\documentclass[12pt]{article}
\usepackage[pdftex]{graphicx}
\usepackage[export]{adjustbox}
\usepackage{amssymb}
\usepackage{amsmath}
\usepackage{xcolor}

\newcommand{\rf}[1]{(\ref{#1})}
\newcommand{\sub}[1]{{(#1)}}
\newcommand{\beq}{\begin{equation}}
\newcommand{\eeq}{\end{equation}}
\newcommand{\bea}{\begin{eqnarray}}
\newcommand{\eea}{\end{eqnarray}}

\newcommand{\e}{\mbox{e}}
\renewcommand{\d}{\mbox{d}}
\newcommand{\g}{\gamma}

\newcommand{\lam}{\lambda}
\newcommand{\Lam}{\Lambda}
\renewcommand{\L}{\Lam}
\renewcommand{\b}{\beta}
\renewcommand{\a}{\alpha}
\newcommand{\al}{\alpha}

\newcommand{\m}{\mu}

\renewcommand{\th}{\theta}
\newcommand{\ep}{\varepsilon}
\newcommand{\eps}{\epsilon}
\newcommand{\om}{\omega}
\newcommand{\del}{\delta}
\newcommand{\Del}{\Delta}
\newcommand{\sg}{\sigma}
\newcommand{\kp}{\kappa}
\newcommand{\vph}{\varphi}

\newcommand{\ga}{{\gamma}}

\newcommand{\oh}{\frac{1}{2}}
\newcommand{\oq}{\frac{1}{4}}

\newcommand{\tr}{\mathrm{tr}\,}
\newcommand{\ra}{\rangle}
\newcommand{\la}{\langle}
\newcommand{\prt}{\partial}
\newcommand{\mi}{\!-\!}
\newcommand{\equ}{\!=\!}
\newcommand{\pl}{\!+\!}

\newcommand{\vg}{\vec{g}}

\newcommand{\cD}{{\cal D}}

\newcommand{\cM}{{\cal M}}

\newcommand{\cT}{{\cal T}}
\newcommand{\cH}{{\cal H}}
\newcommand{\cN}{{\cal N}}

\newcommand{\cO}{{\cal O}}

\newcommand{\cG}{{\cal G}}
\newcommand{\cP}{{\cal P}}
\newcommand{\cB}{{\cal B}}
\newcommand{\cTz}{{\cal T}^{(0)}}
\newcommand{\cTtwo}{{\cal T}^{(2)}}
\newcommand{\cTthree}{{\cal T}^{(3)}}

\newcommand{\tR}{{\tilde{R}}}
\newcommand{\tM}{{\tilde{M}}}

\newcommand{\tW}{{\tilde{W}}}

\renewcommand{\tt}{{\tilde{t}}}
\newcommand{\ttau}{{\tilde{\tau}}}

\newcommand{\hH}{{\hat{H}}}
\newcommand{\hO}{{\hat{O}}}

\newcommand{\hG}{{\hat{G}}}

\newcommand{\hW}{{\hat{W}}}
\newcommand{\hx}{{\hat{x}}}
\newcommand{\hp}{{\hat{p}}}
\newcommand{\hcP}{\hat{\cP}}

\newcommand{\bG}{{\bar{G}}}
\newcommand{\bT}{{\bar{T}}}

\newcommand{\bZ}{{\bar{Z}}}
\newcommand{\bX}{{\bar{X}}}
\newcommand{\bx}{{\bar{x}}}
\newcommand{\bmu}{{\bar{\mu}}}
\newcommand{\bchi}{{\bar{\chi}}}
\newcommand{\bsg}{{\bar{\sg}}}

\newcommand{\no}{\nonumber}

\newcommand{\non}{\nonumber \\}

\newcommand{\SL}{\sqrt{\Lam}}

\newcommand{\dom}{\frac{d \om}{2\pi i}}

\newcommand{\omitt}[1]{ () }
\newcommand{\half}{\frac{1}{2}}

\newcommand{\mus}{{\rm - }}
\newcommand{\plu}{\! +\!}

\newcommand{\ointz}{\oint \frac{dz}{2\pi i \, z}\;}

\newcommand{\SLT}{\sqrt{\L}T}

\newcommand{\Sg}{\Sigma}

\newcommand{\bt}{\beta}
\newcommand{\gm}{\gamma}

\newcommand{\LA}{\left\langle}
\newcommand{\RA}{\right\rangle}
\newcommand{\mf}{\textrm{mf}}
\newcommand{\brak}[1]{\left\langle #1 \right\rangle}
\newcommand{\bra}[1]{\langle #1 |}
\newcommand{\ket}[1]{| #1 \rangle}
\newcommand{\brsh}{\brak{S(h)}}
\newcommand{\dsi}{\delta S_i}
\newcommand{\dsj}{\delta S_j}
\newcommand{\knu}{\kappa_0}
\newcommand{\lnu}{\lambda_0}
\newcommand{\bp}{\textrm{BP}}
\newcommand{\hf}{\hat{F}}
\newcommand{\hd}{\textrm{HD}}
\newcommand{\zm}{Z_{\textrm{matter}}}
\newcommand{\mat}{\textrm{matter}}
\newcommand{\tz}{\tilde{z}}
\newcommand{\tw}{\tilde{w}}

\DeclareMathOperator{\Ei}{Ei}
\DeclareMathOperator{\Li}{Li}

\begin{document}

\begin{titlepage}

\begin{center}
\huge{\bf Elementary Quantum Geometry}\\[2cm]
\Large\bf {\sl Lecture notes by Jan Ambj\o rn}  \\[2cm]
{\large
$$
\hspace{-0.8cm}\int\!\! \cD [g] \, \e^{ -\Lam \int d^2 \xi \,\sqrt{g}} \!\!\int\!\!\! \!\!\int \!\!d^2\xi_1d^2\xi_2 \, \sqrt{\!g(\xi_1)} \sqrt{\!g(\xi_2)}
\; \del\big(D_g(\xi_1,\xi_2) \mi R\big) 
$$}
\end{center}
\begin{figure}[h]
\vspace{-0.5cm}
\centerline{\scalebox{0.15}{\includegraphics[angle = 0]{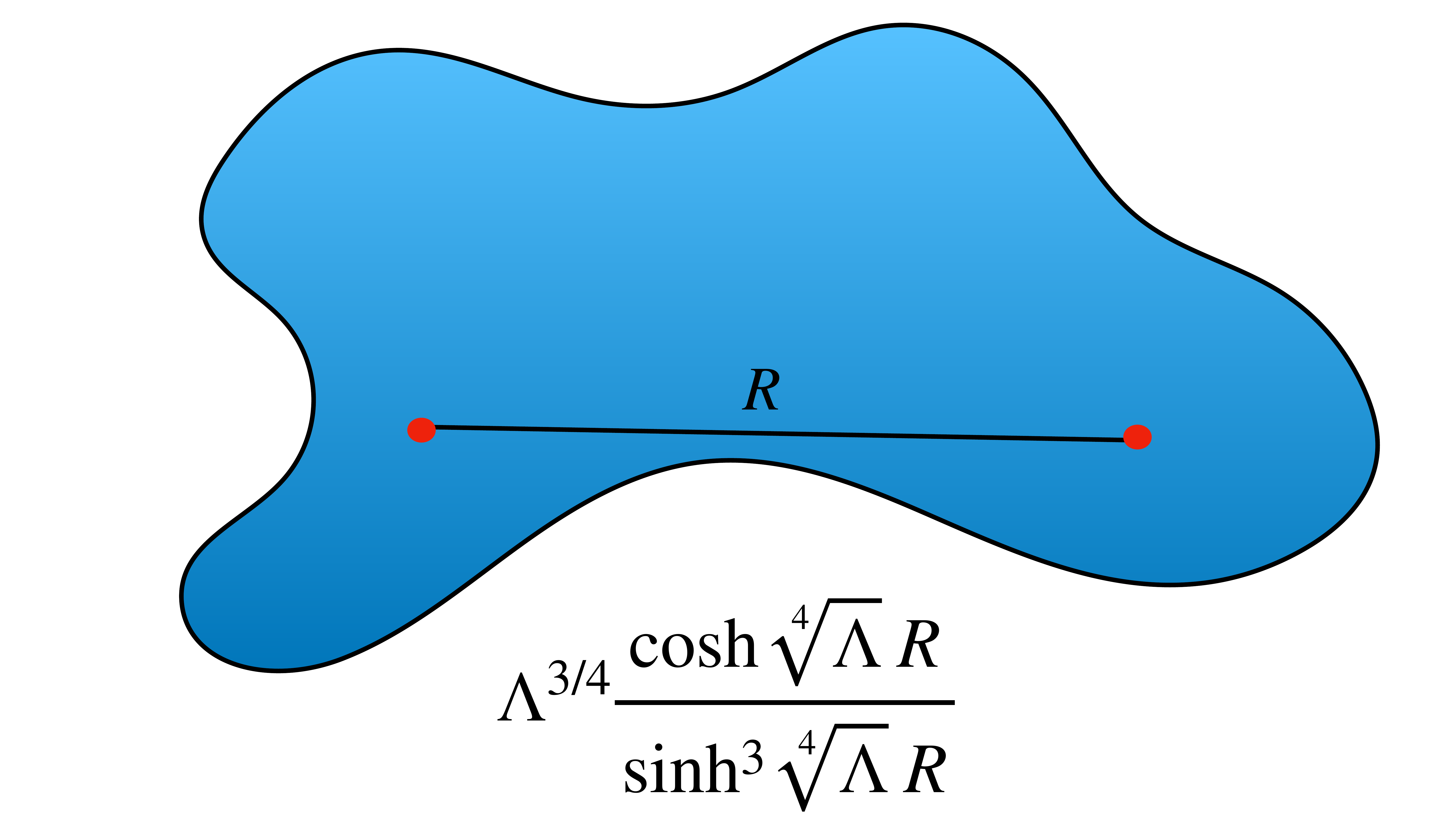}}}
\end{figure}

\vspace{4cm}
\hfill\begin{minipage}[t]{10cm}
\begin{center}
{\it ~~~~The Niels Bohr Institute, University of Copenhagen,} \\
{and} \\
{\it ~~~~Radboud University, Nijmegen, the Netherlands }\\
\end{center}
\hfill April 2022~
\end{minipage}

\end{titlepage}

\section*{Preface}

These Lecture Notes and the related Problem Sets have been used for some years in a course in theoretical physics  
given to Master and PhD students at the Niels Bohr Institute, Copenhagen and at Radboud University, Nijmegen, the Netherlands. 
The idea has been to provide a non-technical introduction to what can be called {\it the statistical theory of geometries}. The theory of General 
Relativity is at present our best attempt to formulate a classical theory of geometry but it has been difficult to quantize the theory, 
and even now it is not entirely clear how to proceed when spacetime is four-dimensional. However, two-dimensional spacetime 
provides an interesting playground for trying to understand what a quantum theory of geometry might entail. If one rotates 
from Lorentzian signature to Euclidean signature of spacetime and use the path integral formalism one arrives at a statistical 
theory of two-dimensional geometries. This theory can be solved by quite elementary methods, and we will do that. 
The solution will also provide us with a beautiful illustration of the Wilsonian view on quantum field theories as associated 
with  universality classes of statistical theories at their critical points. 
If matter fields live on these two-dimensional geometries one easily ends up with string theory, so also the ``surrounding'' 
of two-dimensional quantum geometry is quite rich and important. We will also discuss some elementary aspects of matter
fields coupled to two-dimensional geometries.

As mentioned the notes will often be descriptive rather than providing proofs of the statements. For those interested 
in more technical details I can refer to the book {\it Quantum Geometry, a statistical field theory approach} \cite{book}
by B. Durhuus, T. Jonsson and myself. However, when first preparing the lectures I realized that the book was not really suited for 
students who have just completed their bachelor degree, so the idea of these notes is that the only requirement is some basic knowledge 
of classical analytic mechanics, quantum mechanics and statistical mechanics at bachelor level, and similarly, only simple 
mathematics, like contour integration in the complex plane, is used. Of course, sometimes more advanced concepts are mentioned,
but hopefully never in a way which is essential for an understanding of the topic in question.
No knowledge of quantum field theory or General Relativity is needed (although it does not harm, of course). 
Admittedly, a few concepts
from Riemannian geometry are used, like the concept of a metric 
which describes the geometry and the concept of curvature. However, since we 
only discuss geometry related to at most two-dimensional surfaces, these concepts, in the context we use them, 
can be introduced in an intuitive 
fashion, and in particular, they can be defined for so-called piecewise linear 
geometries where no coordinates are needed for the description.  

Since not much prior knowledge is assumed, the notes have  two sections of ``preliminary material''. One section,
{\it Preliminary material, part A}
is a reminder of the absolute basics in classical mechanics, statistical theory and quantum mechanics, and (this is the primary 
reason it is written) an introduction to the path integral, since the path integral is not standard material  in elementary  
quantum mechanics courses.  This part of the preliminary material is put in front of the real lectures and I usually start the 
lecture series discussing this part. 
 {\it Preliminary material, part B}  reminds the student of the use of Green functions in (classical) physics.
It is a good starting point to know what a classical Green function is, 
as we will basically be calculating (quantum) Green functions of geometry. It is included as an Appendix at the end of the lectures,
intended for self-study for the students for whom the concept of a Green function has become a little hazy.

The Problem Sets are an important integral part of the course.  
Most of the exercises are not really meant to provide the student with 
specific  technical skills, but rather  to supplement or explain in more detail some of the topics discussed in the notes. 
For the same reason detailed solutions are included. Both the Lecture Notes and the Problem Sets existed for a number of years as 
handwritten notes, and I am thankful to Joren Brunekreef , who was my teaching assistant for 
two  years at Radboud University and who
started to tex the Problem Sets and the solutions, despite my advise not to bother to do so. 
This eventually motivated me to also get the Lecture
Notes themselves into a more readable form and to expand the Problem Sets. 

While writing these notes I have benefitted from the insight and skills of my numerous collaborators over the years, and if 
anything deep or ingenious  is present in the notes, they are the ones to be credited. However, when it comes to 
mistakes or conceptual blunders the blame is entirely on me.  In relation to the topics covered in these notes I am, 
apart of course for   the co-authors of the book 
``Quantum Geometry'', Begfinnur Durhuus and Thordur Jonsson,  particularly  indebted to 
Jerzy Jurkiewicz, Renate Loll, Yuri Makeenko, Charlotte Kristjansen, Yoshiyuki Watabiki, Kostas Anagnostopoulos, Timothy Budd,
Leonid Chekhov, Yuki Sato, Stephan Zohren, Willem Westra,  Andrzej G\"{o}rlich, Lisa Glaser, Asger Ipsen, Gudmar Thorleifsson
and Zdzislaw Burda.

\newpage

\newpage

{\bf \Large Table of content}\\

 {\bf \large Preliminary Material, part A: the path integral} \hfill {\bf 6}

\vspace{-12pt}

\begin{itemize}
\item[~] The classical action \hfill  { 6}

\vspace{-6pt}

\item[~]  Statistical mechanics \hfill  {7}

\vspace{-6pt}

\item[]  Classical to quantum \hfill  {11}

\vspace{-6pt}

\item[] The Feynman path integral in quantum mechanics \hfill  { 13}

\vspace{-6pt}

\item[] The Feynman-Kac path integral and imaginary time   \hfill { 17}

\end{itemize}

\vspace{-6pt}

 {\bf \large 1. The free relativistic particle} \hfill   {\bf 21}
 
 \vspace{-12pt}

\begin{itemize}

\item[] The propagator \hfill { 19}

\vspace{-6pt}

\item[] The path integral \hfill  { 22}

\vspace{-6pt}

\item[] Randow walks and universality \hfill 27

\end{itemize}

\vspace{-6pt}

{\bf \large 2. One-dimensional quantum gravity} \hfill  {\bf 31}

\vspace{-12pt}

\begin{itemize}

\item[] Scalar fields in one dimension \hfill { 31}

\vspace{-6pt}

\item[] Hausdorff dimension and scaling relations \hfill {37}

\vspace{-6pt}

\end{itemize}

{\bf \large 3. Branched polymers} \hfill {\bf 41}

\vspace{-12pt}

\begin{itemize}

\item[] Definitions and generalities \hfill { 41}

\vspace{-6pt}

\item[] Rooted branched polymers and universality \hfill {43}

\vspace{-6pt}

\item[] The two-point function \hfill {\bf 45}

\vspace{-6pt}

\item[] Intrinsic properties of branched polymers \hfill {48}

\vspace{-6pt}

\item[] Multicritical branched polymers \hfill {51}

\vspace{-6pt}

\item[] Global and local Hausdorff dimensions \hfill { 52}

\vspace{-6pt}

\end{itemize}

{\bf \large 4. Random surfaces and bosonic strings} \hfill {\bf 55}

\vspace{-12pt}

\begin{itemize} 

\item[] The action, Green functions and critical exponents \hfill  55

\vspace{-6pt}

\item[] Regularizing the integration over geometries \hfill 62

\vspace{-6pt}

\item[] Digression: summation over topologies. \hfill 71

\vspace{-6pt}

\item[] The mass and the string tension \hfill  76

\vspace{-8pt}
 
\begin{itemize}
 \item[]  {\small \sl Scaling of the mass} \hfill 76
 \item[] {\small  \sl Scaling of the string tension} \hfill 84
 \end{itemize}
 
 \vspace{-12pt}
 
 \end{itemize}
 
 \newpage
 
 {\bf \large 5. Two-dimensional quantum gravity}  \hfill 89
 
 \vspace{-12pt}
 
 \begin{itemize}
 
 \item[] {Solving 2d quantum gravity by counting geometries} \hfill 89
 
 \vspace{-6pt}
 
 \item[] Counting triangulations of the disk. \hfill  91

\vspace{-8pt}

\begin{itemize}
 
 \item[] {\small \sl  Branched polymers} \hfill 94
 
 \item[] {\small \sl  Beyond branched polymers: the loop equation} \hfill 96
 
 \end{itemize}
 
 \vspace{-12pt}

 \item[] Multiloop and the loop-insertion operator \hfill 100
 
 \vspace{-6pt}
 
 \item[] Explicit solution for bipartite graphs \hfill 102
 
 \vspace{-6pt}
 
 \item[] The number of large triangulations \hfill 105
 
 \vspace{-6pt}
 
 \item[] The continuum limit \hfill 108
 
 \vspace{-6pt}
 
 \item[] Other universality classes \hfill 112
 
 \vspace{-6pt}
 
 \item[] Appendix \hfill 114
 
 \vspace{-6pt}
 
\end{itemize}

{\bf \large 6. The fractal structure of 2d gravity} \hfill {\bf 116}

\vspace{-12pt}

\begin{itemize}

\item[] Wilsonian universality and the missing correlation length \hfill 116

\vspace{-6pt}

\item[] The two-loop propagator \hfill 115

\vspace{-6pt}

\item[] The two-point function \hfill 124 

\vspace{-6pt}

\item[] The local Hausdorff dimension in 2d gravity. \hfill 126

\vspace{-6pt}

\end{itemize}

{\bf \large 7. The Causal Dynamical Triangulation  model} \hfill {\bf 130}

\vspace{-12pt}

\begin{itemize}

\item[]  Lorentzian versus Euclidean set up  \hfill 130

\vspace{-6pt}

\item[] Defining and solving the CDT model \hfill 131

\vspace{-6pt}

\item[] GCDT: showcasing quantum geometry \hfill 141

\vspace{-6pt}

\item[] GCDT defined as a scaling limit of graphs \hfill 147

\vspace{-6pt}

\item[] The classical continuum theory related to 2d CDT \hfill 150

\vspace{-6pt}

\end{itemize}

{\bf \large References} \hfill {\bf 154}

\vspace{6pt}

{\bf \large Appendix: }  {\bf Preliminary material, part B: Green functions} \hfill {\bf 155}

\vspace{-12pt}

\begin{itemize}

\item[] Basics \hfill 155

\vspace{-6pt}

\item[] Sturm-Liouville boundary conditions \hfill 156

\vspace{-6pt}

\item[] Some higher dimensional Green functions  \hfill 162

\vspace{-6pt}

\item[] Solutions to exercises in preliminary material, part B \hfill 166

\end{itemize}

\newpage

{\bf \large Problem Sets 1-13} \hfill {\bf 170-238}

\vspace{-12pt}

\begin{itemize}

\item[] Set 1: Gaussian integrals and the path integral for the free particle \hfill 170

\vspace{-6pt}

\item[] Set 2: The path integral for the harmonic oscillator. \hfill 173

\vspace{-6pt}

\item[] Set 3: The lattice propagator  and random walks on the lattice \hfill 176

\vspace{-6pt}

\item[] Set 4: Mean field critical exponents for spin systems \hfill 180

\vspace{-6pt}

\item[] Set 5: Various rooted planar trees \hfill 185

\vspace{-6pt}

\item[] Set 6: Branched polymers with hard dimers  \hfill 189

\vspace{-6pt}

\item[] Set 7: Branched polymers coupled to Ising spins and other BPs \hfill 197

\vspace{-6pt}

\item[] Set 8: Asymptotic expansions \hfill 204

\vspace{-6pt}

\item[] Set 9: Branched polymers with loops \hfill 208

\vspace{-6pt}

\item[] Set 10: 2d Graphs with a general even potential \hfill 218

\vspace{-6pt}

\item[] Set 11: Multi-Ising spins coupled to 2d gravity \hfill 225

\vspace{-6pt}

\item[] Set 12: Derivation of  the multiloop formulas in 2d gravity \hfill 233

\vspace{-6pt}

\item[] Set 13: The two-point function and the shape of CDT universes \hfill 235

\end{itemize}

{\bf \large Solutions to Problem Sets 1-13} \hfill {\bf 240-287}

\vspace{-12pt}

\begin{itemize}

\item[] Solutions to Problem Set 1 \hfill 240

\vspace{-6pt}

\item[] Solutions to Problem Set 2 \hfill 244

\vspace{-6pt}

\item[] Solutions to Problem Set 3 \hfill 247

\vspace{-6pt}

\item[] Solutions to Problem Set 4 \hfill 250

\vspace{-6pt}

\item[] Solutions to Problem Set 5 \hfill 253

\vspace{-6pt}

\item[] Solutions to Problem Set 6 \hfill 259

\vspace{-6pt}

\item[] Solutions to Problem Set 7 \hfill 263

\vspace{-6pt}

\item[] Solutions to Problem Set 8 \hfill 266

\vspace{-6pt}

\item[] Solutions to Problem Set 9 \hfill 268

\vspace{-6pt}

\item[] Solutions to Problem Set 10 \hfill 273

\vspace{-6pt}

\item[] Solutions to Problem Set 11 \hfill 276

\vspace{-6pt}

\item[] Solutions to Problem Set 12 \hfill 279

\vspace{-6pt}

\item[] Solutions to Problem Set 13 \hfill 283

\end{itemize}

\newpage

 \setcounter{figure}{0}
 \renewcommand{\thefigure}{A.\arabic{figure}}
\section*{Preliminary material, part A: the path integral}\label{introduction}
\renewcommand{\theequation}{A\arabic{equation}}

\subsection*{The classical action}

\begin{figure}[h]
\centerline{\scalebox{0.18}{\includegraphics{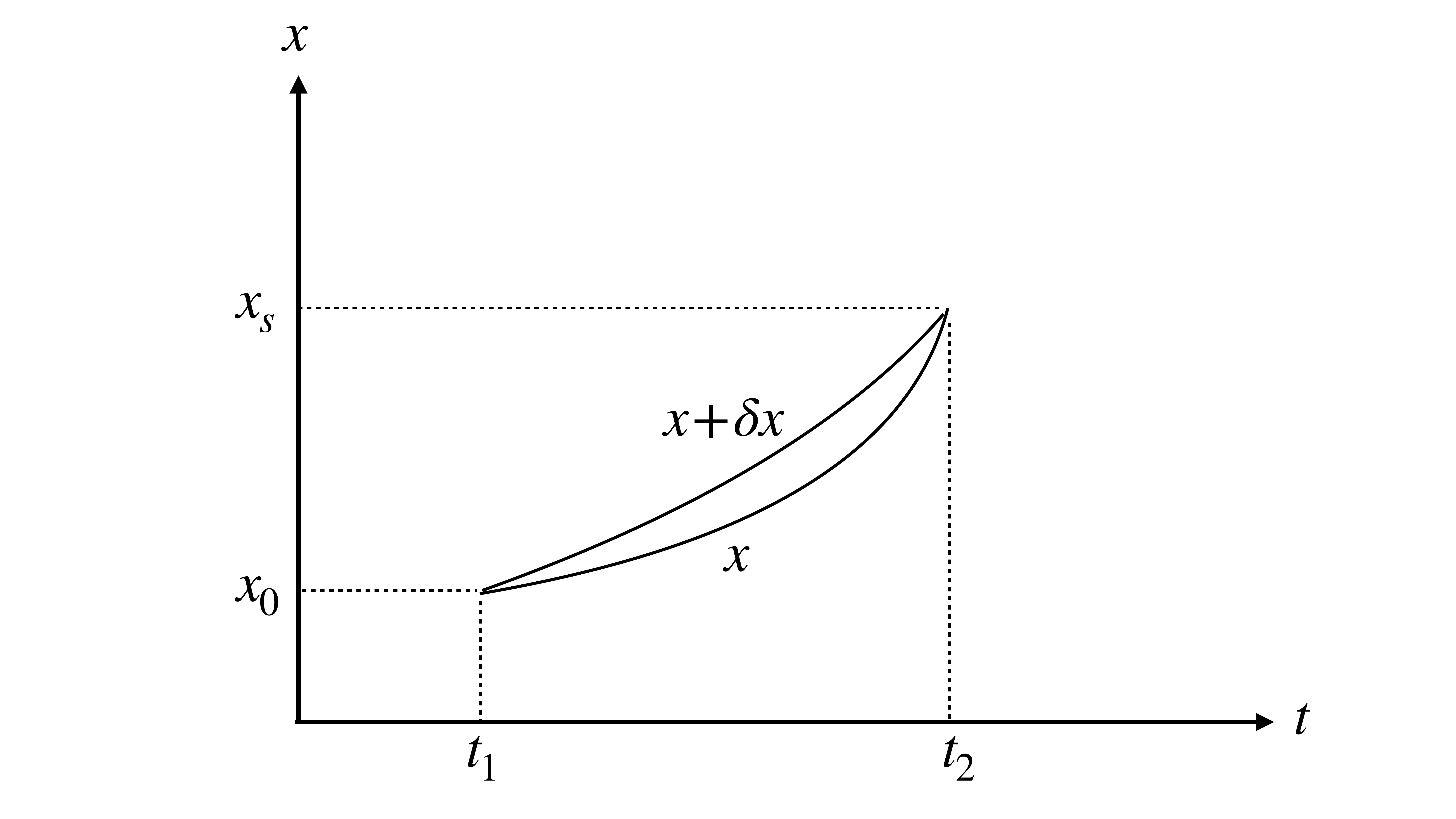}}}
\vspace{-0.3cm}
\caption{{\small  A path $x(t)$ and a path $x(t) + \del x(t)$with a small deformation $\del x(t)$ which vanishes at the endpoints.}}
\label{figpre1}
\end{figure}

Consider a non-relativistic particle with mass $m$ moving in one dimension in a potential $V(x)$. The simplest Hamiltonian and the corresponding equations of motion  (eom) are then,
$p$ denoting the momentum of the particle, 
\beq\label{pre1}
H(x,p) = \frac{ p^2 }{2m}+ V(x)\qquad   \dot{x} = \frac{\prt H}{\prt p}, \quad \dot{p} = -\frac{\prt H}{\prt x}.
\eeq
The Lagrangian $L(x,\dot{x})$ is defined as 
\beq\label{pre2}
\dot{x}\,p -H(x,p) = \oh m \, \dot{x}^2 - V(x) \equiv L(x,\dot{x})
\eeq
and the corresponding eom
\beq\label{pre3}
\frac{d}{dt} \, \frac{\prt L(x,\dot{x})}{\prt \dot{x}} -  \frac{\prt L(x,\dot{x})}{\prt {x}} = 0.
\eeq 
The so-called action $S[x]$ will play a central role in the course. Given a (particle) path $x(t)$ it is defined as 
\beq\label{pre4}
S[x] = \int_{t_1}^{t_2} dt  \; L(x(t),\dot{x}(t)) = \int_{t_1}^{t_2} dt  \; \Big[ \frac{m}{2} \Big(\frac{dx}{dt}\Big)^2 - V(x(t))\Big] .
\eeq
The action should be viewed as a functional on the set of paths $x(t)$, $t \in [t_1,t_2]$. Its relation to the eom is that 
the eom is an extremum of $S[x]$:
\beq\label{pre5}
\frac{\del S[x]}{\del x(t) } = 0 , \quad \Big(\del x(t_1) = \del x( t_2) = 0\Big)  \quad \Rightarrow \quad
\frac{d}{dt} \, \frac{\prt L(x,\dot{x})}{\prt \dot{x}} -  \frac{\prt L(x,\dot{x})}{\prt {x}} = 0.
\eeq
More precisely we consider a path $x(t)$, $t \in [t_1,t_2]$ and an infinitesimal variation $\del x(t)$ away from $x(t)$, with 
the boundary conditions that the variations at the end points of the path are zero as illustrated in Fig. \ref{figpre1}. 
We then define 
\bea\label{pre6}
\del S [x] \!&\equiv& S[x(t) + \del x(t)] -S[ x(t)] \\
\!&=& \!\!\!\int_{t_1}^{t_2} \!\!dt \; \left[
\Big(\frac{\prt L(x,\dot{x})}{\prt {x}} \del x + \frac{\prt L(x,\dot{x})}{\prt \dot{x}} \dot{(\del x)} \Big)  + 
O\Big(\del x^2,\del x \dot{ (\del x)}, (\dot{(\del x)})^2\Big)\right] \nonumber \\
\!\!&=&\!\ \!\! \int_{t_1}^{t_2} \!\!dt \; \left[
\Big(\frac{\prt L(x,\dot{x})}{\prt {x}}  - \frac{d}{d t}\frac{\prt L(x,\dot{x})}{\prt \dot{x}}\Big) \del x(t) + 
O\Big(\del x^2,\del x \dot{(\del x)}, (\dot{(\del x)})^2\Big)\right]. \nonumber
\eea
Thus demanding that $\del S[x] =0$ for all infinitesimal variations (where also the time derivative of $\del x(t)$ can be viewed as 
infinitesimal of the same order) leads to the classical eom for $x(t)$ as indicated in eq.\ \rf{pre5}.

\subsection*{Statistical mechanics}

\begin{figure}[t]
\vspace{-1cm}
\centerline{\scalebox{0.18}{\includegraphics{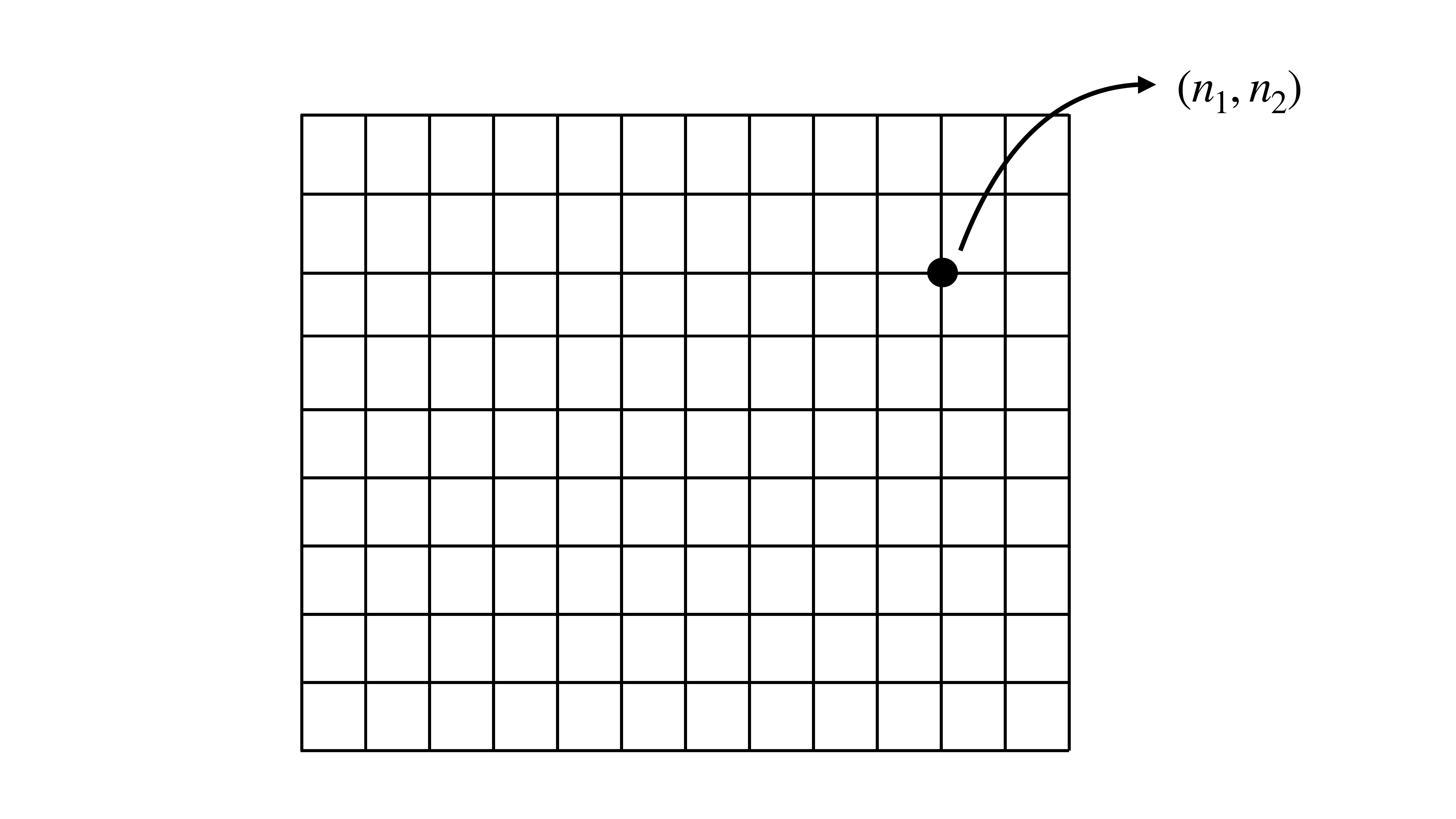}}}
\vspace{-.5cm}
\caption{{\small  A two-dimensional square lattice and a vertex with coordinate $(n_1,n_2)$.}}
\label{figpre2}
\end{figure}

Given a statistical system where the possible energy states $s$ have energies $E_s$, we define the partition function as a
function of the temperature $T$ by
\beq\label{pre8}
Z(T) = \sum_s \e^{-\b E_s},\qquad \beta = \frac{1}{k \, T},
\eeq
where $k$ denotes the Boltzmann constant. The summation is over all states $s$, counting also degeneracies.
An important example is a classical ferromagnetic spin system. We are in $d$ dimensions and consider a hyper-cubic lattice 
(Fig.\ \ref{figpre2} shows a two-dimensional such lattice), where the spins are located at the vertices which we denote with 
integer coordinates $n = (n_1,\ldots,n_d)$. 
The classical spin at site $n$ is then 
represented as a $k$-dimensional vector $\vec{S}(n)$ and a  spin state is then the set $\{\vec{S}(n)\}$ of spins assigned 
to all sites $n$. A model for spin-spin interactions in a crystal  assigns the following classical energy to  a spin-state:
\beq\label{pre9}
E( \{\vec{S}(n)\}) = -J \!\!\!\!\!\!\!\!\!  \sum_{{\rm neighboring}~n,n'} \vec{S}(n) \cdot \vec{S}(n')+ 
\sum_n  a \, \vec{S}(n)^2 + b \, \Big( \vec{S}(n)^2\Big)^2,
\eeq
where $J,a,b$ are coupling constants and $J > 0$ for ferromagnetic system. The partition function is then: 
\beq\label{pre10}
Z(T,J,a,b) = \sum_{\{\vec{S}(n)\}} \e^{- \b E( \{\vec{S}(n)\}) } = \e^{ -\b F(T,J,a,)},
\eeq
where $F(T)$ denotes the free energy of the system.
For the classical spin system the formal summation over the spin states is actually an integration. If the lattice has an extention
$N_i$ in direction $i$ we have 
\beq\label{pre11}
\sum_{\{\vec{S}(n)\}} = \int \prod_{i=1}^d\prod_{n_i =1}^{N_i} \prod_{a=1}^k  d S_a (n_1,\ldots,n_d)
\eeq
 When we have an infinte lattice, i.e.\ $N_i \to \infty$, such a statistical system can have a phase transition as a function 
 of the temperature. In this case, if the phase transition is of order $n$, the $n^{th}$ derivative of $Z(T)$ will be discontinuous at
 the critical temperature $T_c$ where the phase transition takes place. The phase transitions are characterized by certain critical 
 exponents, which we will now define and discuss since they and the associated critical behavior will be  important for our understanding of quantum geometry.
 
 Now we impose an external magnetic field $H$. In the simplest such ferromagnetic system the spins $\vec{S}$ have $d$ components $S_i$, $i=1,\ldots, d$, i.e.\ the same number of components as $\vec{H}$, 
 and we have a partition function which now depends on $H$: 
 \beq\label{1.45}
 Z(T,H) = \e^{-\b F(T,H)} = \sum_{\{\vec{S}(n)\}} \e^{- \b \big(E( \{\vec{S}(n)\}) - \hH \cdot \sum_n  \vec{S}(n)\big)},
 \eeq
 We can now define a number of thermodynamical observables associated with the spin system. Let us write
 $F(T,H) = V f(T,H)$, where $V = \prod_{i=1}^d N_i$ is the volume of the lattice, and $f(T,H)$ thus the free energy density.
 As stated above we have to take $V$ to infinity to have a genuine phase transition in the system. Denote the assumed 
 critical temperature of the system by $T_c$ and the corresponding $\beta$ by $\beta_c$. The specific heat $c_v$ per volume,
 the magnetization $m_i$ per volume and the susceptibility $\chi_{ij}$ are then defined by (we put $k\equ 1$ in the relation
 $\beta = 1/kT$)
 \bea\label{1.70}
 c_v &=&-\b^2 \frac{\prt f}{\prt \b} \quad\propto\quad  | \b \mi \b_c|^{-\alpha} \quad {\rm for} \quad \b \to \b_c\\
 m_i &=&- \frac{\prt f}{\prt H_i}\quad~ \propto \quad |\b \mi \b_c|^\b \quad ~~\,{\rm for} \quad \b > \b_c~~ \b \to \b_c \label{1.71} \\
 \chi_{ij} &=& \frac{\prt m_i}{\prt H_j}\qquad \propto\quad   |\b\mi \b_c|^{-\gamma} \quad \;{\rm for} \quad \b \to \b_c\label{1.72}
 \eea
 where $\alpha$, $\beta$ and $\gamma$ are denoted critical exponents (and the critical exponent $\b$ should not 
 be confused with the inverse temperature $\b$). 

The average value of a quantity $O(n)$ is defined as 
\beq\label{1.48}
\la O(n) \ra = \frac{1}{Z} \sum_{\{\vec{S}(n')\}} O(n) \;\e^{-\b \big(E( \{\vec{S}(n')\})- H_i \sum_n S_i(n')\big) }, 
\eeq
and we can write
\beq\label{1.49}
m_i = \frac{1}{V} \sum_n \la S_i (n)\ra,\qquad \chi_{ij} = \frac{1}{V} 
\sum_{n.n'} \Big\langle  \big(S_i(n) \mi \la S_i(n) \ra\big)\big(S_j(n') \mi \la S_i(n')\ra\big)\Big\rangle.
\eeq
We define the spin-spin correlator as 
\beq\label{1.50}
G_{ij} (n,n') = \Big\langle  \big(S_i(n) \mi \la S_i(n) \ra\big)\big(S_j(n') \mi \la S_i(n')\ra\big)\Big\rangle,
\eeq
and thus 
\beq\label{1.50b} 
\chi_{ij} = \frac{1}{V} \sum_{n,n'} G_{ij}(n,n') = \sum_{n'} G_{ij}(n-n'),
\eeq
where we in the last equation have assumed translational invariance of the system 
(i.e. for a finite lattice periodic boundary conditions). Thus we have 
\beq\label{1.50a}
\boxed{\chi_{ij}(\b) =  \sum_{n} G_{ij}(n\mi n') }
\eeq

 Away from $\b_c$ the 
spin-spin correlation function \rf{1.50} will be short ranged and it falls off exponentially over a few lattice spacings and thus 
$\chi_{ij}(\b)$ will be finite. If the phase transition at $\b_c$ is a second order transition (i.e. the first derivative of $F(\b,H\equ 0)$
wrt $\b$ is finite at $\b_c$, but the second derivative diverges when $\b \to \b_c$) the correlation length of $G_{ij}(n,n')$ 
will diverge when $\b\to \b_c$ and $\chi_{ij}(\b)$ will also diverge as indicated in \rf{1.72} (assuming the critical exponent 
$\gamma > 0$). Denote by $\xi(\b)$ the correlation length of $G_{ij}(n,n')$, defined by the asymptotic exponential fall of of $G$:
\beq\label{1.51}
\frac{-\ln G_{ij} (n,n')}{ |n\mi n'|} \to \xi(\b) \quad \mbox{for} \quad |n\mi n'| \to \infty.
\eeq
Generically the long and short distance behavior of $G(n,n')$  close to the critical (inverse) temperature $\b_c$ is characterized by:
\bea 
G(n,n') &\propto& |n\mi n'|^{2-d - \eta},  ~\qquad \qquad {\rm for}\quad 1 \ll |n\mi n'| \ll \xi(\b), \label{1.81}\\
G(n,n') &\propto& \e^{ - |n-n'|/\xi(\b) + \cO(\ln |n-n'|)} \quad {\rm for}\quad \xi(\b) \ll |n\mi n'| \label{1.82}
\eea
where
\beq\label{1.83}
 \xi(\b) \propto  |\b \mi \b_c|^{-\nu} \quad {\rm for} \quad \b \to \b_c.
 \eeq
In \rf{1.81} and \rf{1.83} we have introduced two  critical exponents, $\eta$ and $\nu$, in addition to 
$\alpha$, $\b$ and $\gamma$. For many systems they are not independent since there exist so-called hyperscaling relations:
\beq\label{1.84}
\alpha = 2\mi  \nu\, d,\qquad \alpha \plu  2\b \plu  \gamma = 2,\qquad \gamma = \nu (2 \mi \eta).
\eeq
Thus there are only two independent exponents. 
The last relation is called {\it Fisher's scaling relation}. Since we will meet it again in our quantum geometry theories, let us 
just show how it can be derived heuristically from the definitions already given. We suppress the indices $i,j$ and assume 
translational invariance of $G$. Approximate $G$ by the asymptotic form \rf{1.81} for $|n| < \xi(\b)$ and put it to zero for 
$|n| > \xi(\b)$, since it according to \rf{1.82} is more or less exponentially suppressed in that region. 
We thus have, replacing summation by integration:
\beq\label{1.85}
\chi(\b) = \int d^d n  \; G(n) \propto \int\limits_{|n| < \xi(\b)}  \frac{d^d n}{n^{d-2+\eta} } \propto \big( \xi(\b)\Big)^{2-\eta} \propto 
\frac{1}{| \b - \b_c|^{\nu (2-\eta)}}, 
\eeq
which implies that  $\gamma = \nu (2\mi \eta)$ according to the definition \rf{1.72} of $\gamma$.

The importance of these exponents is that they are {\it universal}.
Different spin systems can have the same exponents even if their local spin-spin interactions and their critical temperatures 
can be quite different. While the local detail of the spin-spin interactions might be unimportant for the system's critical exponents,
the symmetries are important. Thus spin systems with different symmetries can have different exponents. When the dimension 
of space $d \geq 4$ all spin systems will have the same exponents, the so-called {\it mean-field exponents}. They are:
\beq\label{1.86}
\nu = \oh,\quad \eta =0, \quad \mbox{and from hyperscaling} \quad \gamma =1, \quad {\rm etc.}
\eeq
In Problem Set 4 we will calculate the critical exponents of the above spin system using a so-called mean-field
approximation and we will (not surprisingly....) find the mean field values \rf{1.86}.

 \subsection*{Classical to quantum}
 
 In the transition from classical physics to quantum physics for the simple one-dimensional classical system we considered above
 we first introduce the Hilbert space $\cH$ of square integrable functions $\cH = L^2 (\mathbb{R})$ on the real axis. 
 Next we promote the 
 classical variable $x,p$ to operators  $\hx,\hp$ in the following way:
 \beq\label{pre12}
 \hx: ~\psi(x) \to (\hx \, \psi)(x) = x \psi(x), \qquad \hp:~\psi(x) \to (\hp \, \psi)(x) = \frac{\hbar}{i} \, \frac{d \psi}{dx} 
 \eeq
 where $\psi \in L^2(\mathbb{R})$. Both $\hx$ and $\hp$ are unbounded, but Hermitian  operators. The same is true for the quantum 
 Hamiltonian which is obtained by replacing $x,p$ in the classical Hamiltonian with the  operators $\hx,\hp$:
 \beq\label{pre13}
 \hH = \frac{1}{2m} \, \hp^2 + V(\hx) = - \frac{\hbar^2}{2m} \, \frac{d^2}{dx^2} + V(\hx).
 \eeq
 Any vector $| \psi\ra$ in $\cH$ can be expanded in any orthonormal basis $|e_n\ra$ in $\cH$:
 \beq\label{pre13a}
 | \psi \ra = \sum_n | e_n \ra \la e_n | \psi\ra, \qquad \sum_n | e_n \ra \la e_n | = \hat{{\rm I}}.
 \eeq 
 The eigenvectors $| x \ra$ and $| p\ra$  of the operators $\hx$ and $\hp$, corresponding to the eigenvalues $x$ and $p$, respectively, 
 are defined by 
 \beq\label{pre14}
 \hx \, |x \ra = x \, |x\ra, \qquad \qquad \hp \, | p\ra = p \, |p\ra.
 \eeq 
 These eigenvectors do not belong to $L^2(\mathbb{R})$. Nevertheless, we can still expand
 the vectors in $\cH$ on these vectors, as in eq. \rf{pre13a}:
 \bea\label{pre15}
 |\psi \ra &=& \int dx \, |x\ra \la x | \psi \ra, \qquad \int dx \; |x \ra \la x | = \hat{ {\rm I}}.    \\
 &=&  \int dp \, |p\ra \la p | \psi \ra, \qquad \int dp \; |p \ra \la p | = \hat{ {\rm I}}. \label{pre16}
 \eea
 $\la x | \psi\ra\equiv \psi (x)$ is denoted the wave function of the state $| \psi\ra$ and it is when expanding the 
 states $| \psi\ra$ on the vectors $|x\ra$ that we in \rf{pre12} defined the operators $\hx$ and $\hp$.  
 In particular we have for the state $|p\ra$:
 \beq\label{pre17}
\la x | p\ra = \frac{\e^{i p \,x /\hbar}}{\sqrt{2 \pi \hbar}}, \quad {\rm i.e.} \quad 
| p \ra = \int dx\; |x\ra \la x | p\ra = \int {dx} \; | x\ra \, \frac{\e^{i p \,x /\hbar}}{\sqrt{2 \pi \hbar}} 
\eeq 
The eigenstates and eigenvalues of $\hH$ are of course of particular interest. Denote an eigenstate $|E\ra$ where $E$ is the 
corresponding eigenvalue of $\hH$:
\beq\label{pre17a}
\hH | E\ra = E | E\ra, \quad \psi_E(x) = \la x | E\ra, \quad  \left(- \frac{\hbar^2}{2m} \, \frac{d^2}{dx^2} + V(x)\right) \, \psi_E(x) = 
E \, \psi_E(x).
\eeq
The spectrum (the eigenvalues of the Hamiltonian) 
can be discrete, as when the potential $V(x)= \om^2 x^2/2$ is that of the harmonic oscillator and 
 where the eigenvalues are $E_n = \hbar \om (n+\oh)$. It can also be continuous  as when $V(x)=0$, i.e. the free
particle case,  where the eigenstates of the Hamiltonian are just the states $|p\ra$ and the corresponding eigenvalues of 
the Hamiltonian are $p^2/2m$.

We can now define the {\it quantum partition function} as in \rf{pre8}, just by replacing the classical energies by the quantum energy 
calculated from \rf{pre17a}:
\beq\label{pre18}
Z(T) = \sum_E \e^{-\b \,E} = \sum_E \la E | \, \e^{-\b \, \hH} | E\ra \\
= \tr  \e^{-\b \, \hH} = \int dx \, \la x | \e^{-\b \, \hH} | x\ra,
\eeq
where $\sum_E$ is a summation if the eigenvalues are discrete and a suitable integration if they are continuous.

The {\it  time evolution in quantum mechanics} is simplest described by the Schr\"{o}dinger equation:
\bea\label{pre19}
i \hbar \frac{\prt}{\prt t} \,| \psi(t) \ra &=& \hH \, | \psi(t) \ra, \\\
 i \hbar \frac{\prt}{\prt t}\, \psi(x, t) &=&\! \Big(- \frac{\hbar^2}{2m} \, \frac{d^2}{dx^2} + V(x)\Big) \, \psi(x,t), 
 \quad \psi(x,t) \equiv \la x| \psi(t)\ra.~~~
 \label{pre20}
 \eea 
The formal solution to \rf{pre19} is 
\beq\label{pre21}
| \psi(t) \ra = \e^{-i \hH t /\hbar} | \psi(0)\ra.
\eeq
{\it The basic question asked in quantum mechanics} is the following: given a state $|\psi_0\ra$ at time $t \equ 0$, what is the probability 
amplitude for finding the system in the state $| \psi_s\ra$ at time $t$? The answer is:
\beq\label{pre22}
\la \psi_s | \e^{-i \hH t/\hbar} | \psi(0)\ra.
\eeq
Using the expansion \rf{pre15} in terms of wave functions we can write 
\beq\label{pre23}
\la \psi_s | \e^{-i \hH t/\hbar} | \psi(0)\ra = \int dx_s \int dx_0 \; \psi^*_s(x_s)  \psi_0(x_0)\; \la x_s | \e^{-i \hH t/\hbar} | x_0\ra.
\eeq
In principle we can then answer all questions if we can only calculate 
\beq\label{pre24}
\la x_s | \e^{-i \hH t/\hbar} | x_0\ra,
\eeq
i.e.\ the probability amplitude for a particle which  at time 0 is located  at $x_0$ to be found at $x_s$ at time $t$. This probability 
amplitude can be represented as a {\it Feynman path integral}. The Feynman path integral formalism of quantum mechanics will be 
central in following.

\begin{figure}[h]
\vspace{-0.2cm}
\centerline{\scalebox{0.22}{\includegraphics{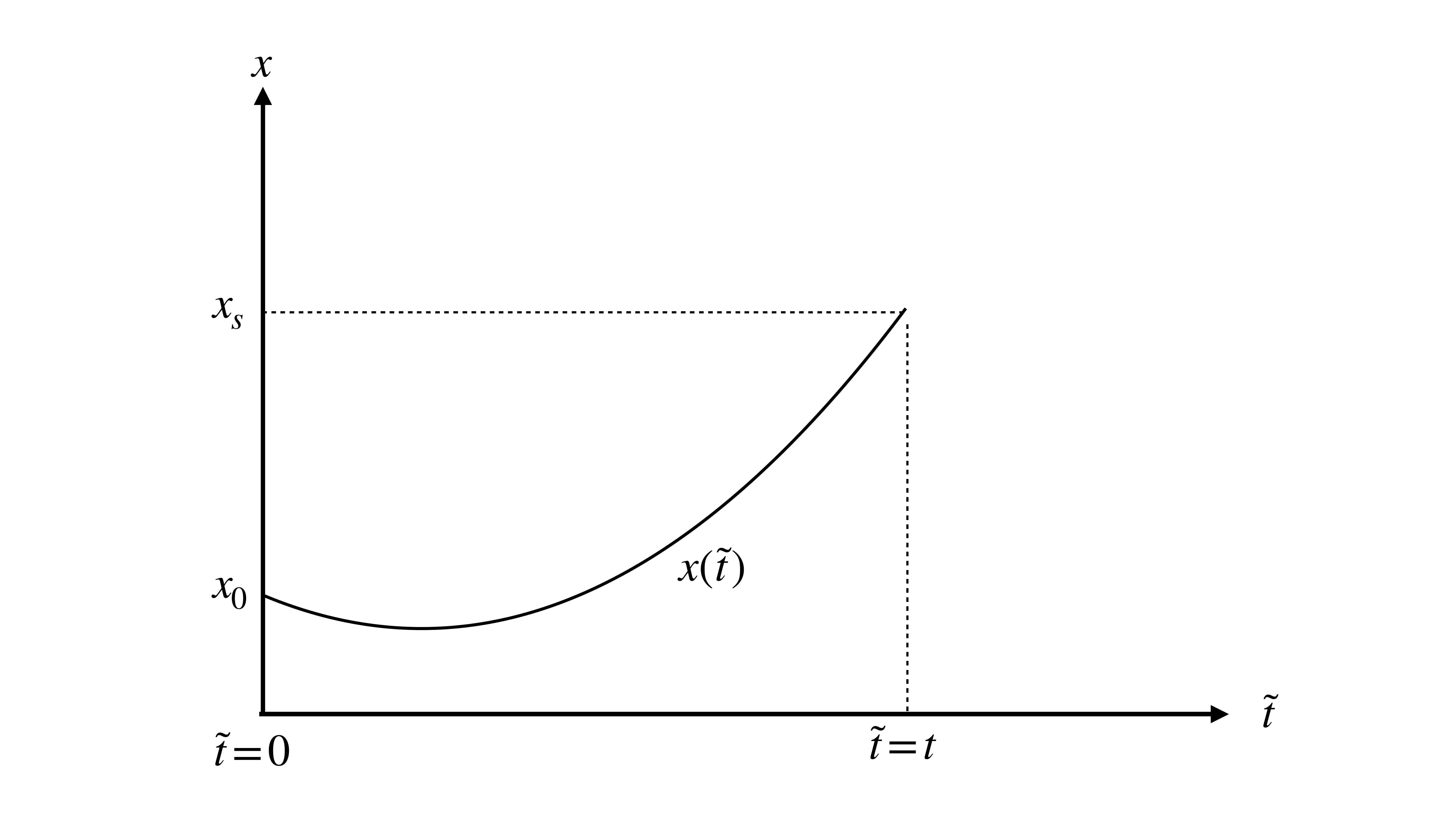}}}
\vspace{-0.2cm}
\caption{{\small  A path $x(\tt)$, $\tt \in [0,t]$.}}
\label{figpre3}
\end{figure}

\subsection*{ The Feynman path integral in quantum mechanics}

\beq\label{pre25}
\boxed{\la x_s | \e^{-i \hH t/\hbar} | x_0\ra = \int\limits_{\substack{x(0) = x_0 \\ x(t) = x_s}} \cD x(\tt) \: \e^{\frac{i}{\hbar} S[x(\tt)]}}
\eeq
The ``integration'' on the rhs of this formula is over the set of continuous path $x(\tt),~ \tt \in [0,t]$ where $x(0) = x_0$ and 
$x(t) = x_s$. For a given path $x(\tt)$, the weight of the integrand is $\e^{i S[x]/\hbar}$ where $S[x]$ is the classical action 
of the path $x(\tt)$. Fig.\ \ref{figpre3} shows a ``typical'' such path (well, as we will see it is not so typical. A ``typical'' continuous 
path is much more ``wild'') . 
Before discussing how the path integral 
 can be derived in a simple way from the standard time evolution of quantum mechanics, let us discuss how the path integral 
provides an intuitive link between quantum and classical physics. Let us consider the limit $\hbar \to 0$ and ask which paths 
contribute to \rf{pre25} in that limit. For small $\hbar$ a little change of a given path $x(\tt) \to x(\tt) + \del x (\tt)$ will 
introduce a large phase shift in $e^{i S[x+\del x]/\hbar}$ compared to  $e^{i S[x]/\hbar}$ and the contributions from 
neighboring paths will cancel unless $S[x+\del x]\approx S[x]$. We have 
\beq\label{pre25a}
S[x+\del x] = S[x] + \int_{0}^t d\tt \, \frac{\del S}{\del x(\tt)} \, \del x(\tt) + \cO\big( (\del x)^2\big).
\eeq
Thus we can obtain  $S[x+\del x]\approx S[x]$ provided that 
\beq\label{pre26a}
\frac{\del S}{\del x(\tt)} = 0 \quad \mbox{i.e. according to \rf{pre5} } \quad x(\tt) = x_{\rm cl} (\tt),
\eeq
and thus the paths which contribute the most in the $\hbar \to 0$ limit to the path integral will be the solutions to the 
classical eom \rf{pre5}. The larger $\hbar$ is compared to the constants which appear in the Lagrangian (the mass of the 
particle, the cyclic frequency $\om$ in the case of a harmonic oscillator, etc), the more ``non-classical'' paths, 
i.e.\  paths which do not satisfy the  classical eom, will be important when calculating the path integral.

Let us now give a precise meaning to the path integral appearing in \rf{pre25}. First we note the following mathematical facts:
\begin{itemize}
\item[(1)] Let $a,b$ be complex numbers. Then 
\beq\label{pre26}
\e^{a+b} = \e^{a} \e^{b}  = \lim_{n\to \infty} \Big( \e^{a/n} \e^{b/n} \Big)^n.
\eeq
\item[(2)] Let $\hat{A}$ and $\hat{B}$ be $N\times N$ matrices. Then in general $\e^{\hat{A} + \hat{B}} \neq  \e^{\hat{A}}\e^{\hat{B}}$
when $[\hat{A},\hat{B}] \neq 0$. However, it is still true that 
\beq\label{pre27}
\e^{\hat{A} + \hat{B}} = \lim_{n\to \infty} \Big( \e^{\hat{A}/n} \e^{\hat{B}/n} \Big)^n, \qquad \mbox{(the Lie-Trotter product formula )}
\eeq

\item[(3)] Let $\hat{A}$ and $\hat{B}$ be Hermitian  operators on a Hilbert
space $\cH$ with eigenvalues which are bounded from below. They can be unbounded operator, but have to be such that $D_A \cap D_B$ is dense in $\cH$, where $D_A$ and 
$D_B$ denote the domains of definition of $\hat{A}$ and $\hat{B}$. Then $\e^{-\hat{A}}$, $\e^{-\hat{B}}$ and 
$\e^{-\hat{A} -\hat{B}}$ are bounded operators with a well-defined norm,
and one has the equivalent of \rf{pre27}:
\beq\label{pre28}
\e^{-\hat{A} - \hat{B}} = \lim_{n\to \infty} \Big( \e^{-\hat{A}/n} \e^{-\hat{B}/n} \Big)^n, \qquad \mbox{(the~Kato-Trotter~theorem)},
\eeq
where the convergence  is in operator norm, i.e.\ a quite strong convergence. The same relation is true if we 
replace $\hat{A}$ by $i \hat{A}$, and similarly for $\hat{B}$ and $\hat{A} + \hat{B}$. In that case $\e^{i\hat{A}}$ etc become 
unitary operators with norm 1. 
\end{itemize}  
With these remarks in mind we now we now apply \rf{pre28} to 
\beq\label{pre28a}
i\frac{t}{\hbar} \hH= i\hat{A} + i\hat{B},\qquad \hat{A} = \frac{t}{2m \hbar} \; \hp^2, \quad \hat{B} =  \frac{t}{\hbar} \; V(\hx).
\eeq
\beq\label{pre29}
\e^{-i t \hH/\hbar} = \lim_{n \to \infty} \Big( \e^{-i\hat{A}/(n+1)} \e^{-i\hat{B}/(n+1)} \Big)^{n+1}  =  
\lim_{n \to \infty}\Big(\hat{O}_\ep\Big)^{n+1}
\eeq 
where 
\beq\label{pre30}
\hat{O}_\ep =  \e^{-i\ep \hp^2/(2m \hbar)} \e^{-i \ep \hat{V}(\hx)/\hbar},\qquad \ep = \frac{t}{n+1}.
\eeq
We can calculate the matrix element $\la x | \hO_\ep | y\ra$:
\beq\label{pre31}
\boxed{\la x | \hO_\ep | y\ra =  \Big(\frac{m}{2\pi i \ep \hbar}\Big)^{\oh}  \; \e^{i (x-y)^2 m/(2 \ep \hbar)}\e^{-i \ep V(y)/\hbar}}
\eeq
Here is the calculation: we use \rf{pre16} and \rf{pre17} to write 
\bea
\la x | \hO_\ep | y\ra &=&\la x| \e^{-i\ep \hp^2/(2m \hbar)} \Big( \int dp \, |p\ra\la p| \Big)  \e^{-i \ep \hat{V}(\hx)/\hbar}|y\ra \nonumber\\
&=& \e^{-i \ep V(y)/\hbar} \int dp \;\la x| p\ra \la p| y\ra \; \e^{-i\ep p^2/(2m \hbar)}  \nonumber\\
&=& \e^{-i \ep V(y)/\hbar}  \int \frac{dp}{2\pi \hbar}  \; \e^{i p(x-y)/\hbar} \; \e^{-i\ep p^2/(2m \hbar)} = {\rm rhs~of~\rf{pre31}},
\nonumber
\eea
where the last equality follows from a Gaussian integral, which will be discussed in  Problem Set 1.   

\begin{figure}[t]
\vspace{-.5cm}
\centerline{\scalebox{0.2}{\includegraphics{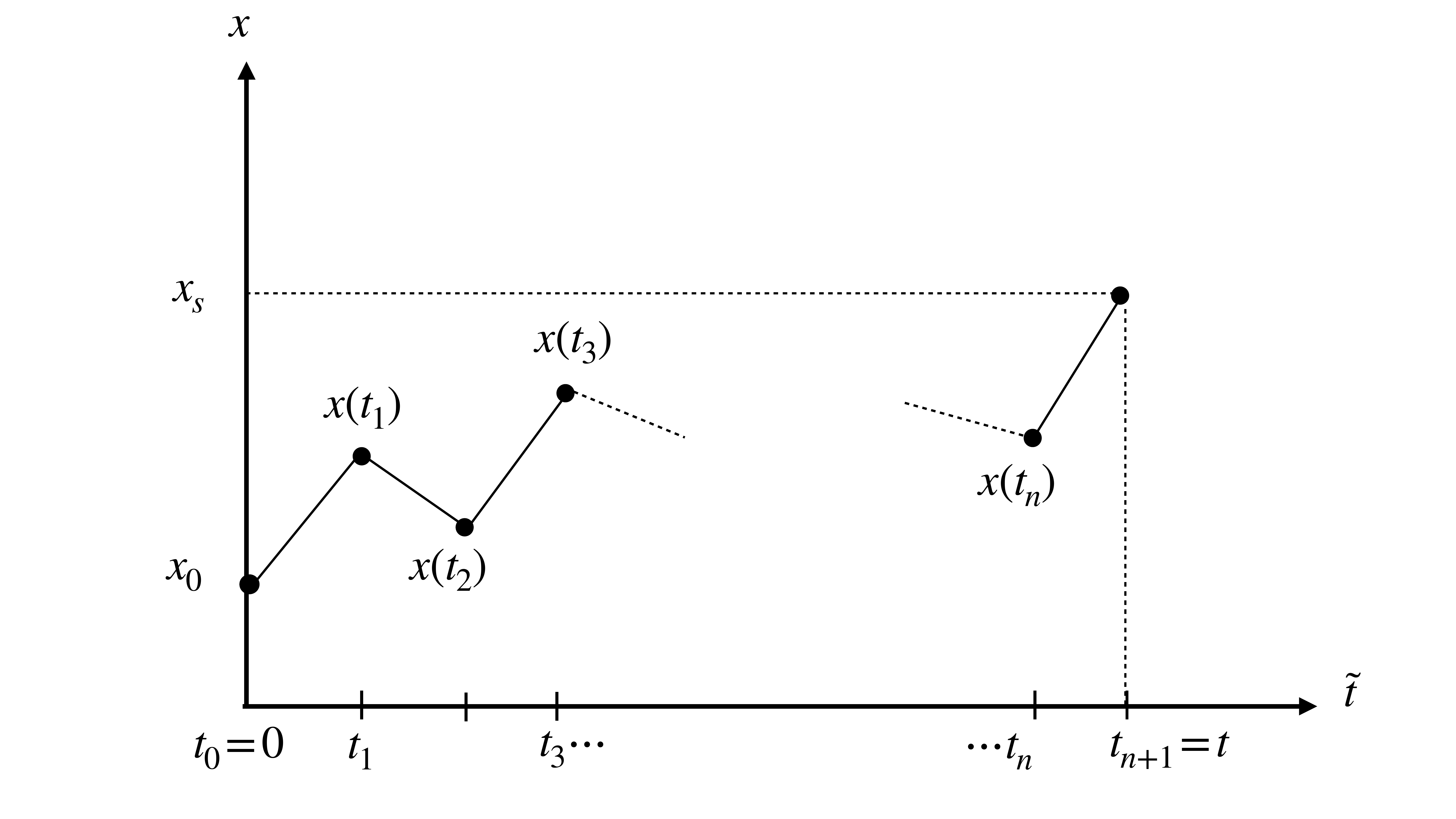}}}
\vspace{-0.5cm}
\caption{{\small  The set of points $\{x(t_0),x(t_1),\ldots,x(t_{n+1})\}$ connected into a piecewise linear path $x(\tt)$, $\tt \in [0,t]$.}}
\label{figpre4}
\end{figure}

We can now write
\beq\label{pre32}
\hat{O}_\ep^{n+1} = \hat{O}_\ep \hat{I} \hat{O}_\ep \hat{I} \cdots  \hat{I} \hat{O}_\ep,\qquad \hat{I} = \int dx \; |x\ra \la x|
\eeq
Thus we can write the matrix elements of $\hat{O}_\ep^{n+1}$ as standard ``matrix" multiplication, where we define
$x_{n+1} = x_s$:
\bea\label{pre33}
\la x_{n+1} | \hat{O}_\ep^{n+1}| x_0\ra &=& \int \prod_{i=1}^n dx_i \; \la x_{n+1} | \hat{O}_\ep |x_n\ra\la x_n\ra 
\la  x_{n} | \hat{O}_\ep |x_{n-1}\ra\cdots \la  x_1 | \hat{O}_\ep |x_0\ra \\
&=& \!\!\Big( \frac{m}{2\pi i \ep \hbar} \Big)^{\frac{n+1}{2}}\!\!\!\int \prod_{i=1}^n dx_i \; 
\exp \Big( \frac{i}{\hbar} \sum_{i=0}^{n} \ep \Big[ \frac{m}{2} \Big( \frac{x_{i+1} \mi x_i}{\ep}\Big)^2 \mi V(x_i) \Big] \Big) \nonumber
\eea
We thus finally have, with $\ep = t/(n\plu 1)$, 
\beq\label{pre34}
\boxed{\la x_s | \e^{-i \hH t/\hbar} | x_0\ra \!=\! \!\lim_{n\to \infty} \Big( \frac{m}{2\pi i \ep \hbar} \Big)^{\frac{n+1}{2}}
\!\!\!\!\int \!\prod_{i=1}^n dx_i 
\exp \Big( \frac{i}{\hbar} \sum_{i=0}^{n} \ep \Big[ \frac{m}{2} \Big( \frac{x_{i+1} \mi x_i}{\ep}\Big)^2\! \mi V(x_i) \Big] \Big)}
\eeq
This gives a precise meaning to the rhs of eq.\ \rf{pre25} via the Kato-Trotter theorem. However, the $x_i$'s which enter in the 
formula are at this point merely integration variables. The BCH  (Baker, Campbell, Hausdorff) formula tells us that\footnote{
One version of the BCH formula useful in this context is 
$\e^{t(\hat{A} + \hat{B})} = \e^{t \hat{A}}\e^{t \hat{B}} \e^{-t^2[\hat{A},\hat{B}] + O(t^3)} $ where the term $O(t^3)$ can 
be expressed in terms of two or more commutators of $\hat{A}$ and $\hat{B}$. Note that the Lie-Trotter (or as it is also called, the 
Suzuki-Trotter) product formula \rf{pre27} follows from this BCH formula.}
\beq\label{pre35}
\hat{O}_\ep   = \e^{-i \eps \hH/\hbar} \;\e^{ \cO(\ep^2)}.
\eeq
It is clear that we can trivially write 
\beq\label{pre36}
\la x_s | \e^{-i \hH t/\hbar} | x_0\ra = \! \int \!\prod_{i=1}^n \!dx_i \, \la x_{n+1} | \e^{-i\frac{ \eps \hH}{\hbar}}  |x_n\ra\la x_n\ra 
\la  x_{n} | \e^{-i  \frac{ \eps \hH}{\hbar}}  |x_{n-1}\ra\cdots \la  x_1 | \e^{-i  \frac{ \eps \hH}{\hbar}} |x_0\ra .
\eeq
In \rf{pre36} the $x_i$'s, $i\equ 1,\ldots,n$ are also 
independent integration variables, but we can now associate the time $t_i \equ i \ep$ to the index $i$ since 
$\e^{-i \eps \hH/\hbar} $ precisely is the evolution operator during a time-interval $\ep$, i.e.\ it 
makes sense to write $x (t_i) \equiv x_i$ in \rf{pre36}. In view of \rf{pre35} we will do the same  \rf{pre34}, although 
strictly speaking there is no strict link of the $i$ in \rf{pre34} to the time $t_i \equ i \ep$ for $1 \leq i \leq n$.
With this assignment $x_i \to x(t_i)$, it is tempting to assign a ``path'' to the sequence of points $x(t_i)$ by joining $x(t_i)$ and 
$x(t_{i+1})$ by a straight line such we can write $x(\tt)$, $\tt \in [0,t]$. 
This is illustrated in Fig.\ \ref{figpre4}. With this assignment we can view \rf{pre34} as a certain 
limit of the integration over the class of ``piecewise linear'' paths shown in Fig.\ \ref{figpre4}.

With the above ``path''-interpretation it is now tempting to perform the following identifications:
\beq\label{pre37}
\frac{x(t_{i+1}) \mi x(t_i)}{\ep} \approx \frac{d x(\tt)}{d \tt},\quad \tt \in ]t_i, t_{i+1}[, \qquad {\rm (wrong!)},
\eeq
\beq\label{pre38}
\sum_{i=0}^{n} \ep \Big[ \frac{m}{2} \Big( \frac{(x_{i+1} \mi x_i)}{\ep}\Big)^2 \mi  V(x_i) \Big]  \underset{n\to \infty}{\to} 
\int_0^t d \tt \;\Big[  \frac{m}{2} \Big( \frac{d x}{d \tt}\Big)^2 \mi V(x(\tt)) \Big] = S[x(\tt)]
\eeq
\beq\label{pre39}
\cD x(\tau) = \lim_{n \to \infty} \cN(n) \; \prod_{i=1}^n dx(t_i), \qquad \cN(n) = \Big(\frac{m}{2\pi i \ep \hbar}\Big)^{\frac{n+1}{2}}.
\eeq
With these identifications the rhs of \rf{pre25} will be precisely the rhs of \rf{pre34} which has a well defined limit. However, 
a number of points should be made clear. First, as explicit written in eq.\ \rf{pre37}, it is somewhat misleading to 
make the identification of the lhs and the rhs of eq.\ \rf{pre37}, even if it is of course  correct for a piecewise linear path
like the one shown in Fig.\ \ref{figpre4}. The reason is that when $\ep \to 0$ the numerical value  of the derivative goes to infinity 
as we will discuss shortly (see eq.\ \rf{pre47}).
Thus \rf{pre38} does not really make much sense for a typical path of the kind shown in Fig.\ \ref{figpre4}.
Rather it should be used moving from the rhs to the lhs when one wants a assign a meaning to the rhs of \rf{pre25}. Nevertheless
the formal use of the classical action in \rf{pre25} is extremely useful in a  number of formal manipulations one can make 
with the path integral as we will also see later. Finally, one could get the idea from the notation \rf{pre39} that $\cD x(\tau)$ 
is (apart from the factor $i$ in $\cN(n)$) a kind of generalization of the finite dimensional Lebesgue measure. However, this 
is not the case. There exists no Lebesgue measure on $\mathbb{R}^\infty$. Below we will discuss what kind of measure one can associate 
with the path integral.

\subsection*{The Feynman-Kac path integral and imaginary time}

Let us make a rotation to imaginary (so-called Euclidean) time $\tau$:
\beq\label{pre40}
t \to -i \tau,\qquad \e^{-i \hH t/\hbar} \to \e^{- \hH \tau/\hbar}
\eeq
As before one proves that 
\beq\label{pre41}
\boxed{\la x_s | \e^{- \hH \tau/\hbar} | x_0\ra = \int\limits_{\substack{x(0) = x_0 \\ x(\tau) = x_s}} 
\cD x(\ttau) \;\e^{-\frac{1}{\hbar} S_E[x(\ttau)]}}
\eeq
\beq\label{pre42}
S_E [x(\ttau)]=\int_0^\tau d \ttau \Big[  \frac{m}{2} \Big( \frac{d x}{d \ttau}\Big)^2 + V(x(\ttau)) \Big] 
 \eeq
 which is called the Feynman-Kac formula and where the rhs of \rf{pre41} should be understood as 
 \beq\label{pre41a}
  \lim_{n\to \infty} \Big( \frac{m}{2\pi  \ep \hbar} \Big)^{\frac{n+1}{2}} \!\!\! \int \prod_{i=1}^n dx_i \; 
\exp \Big( \mi \frac{1}{\hbar} \sum_{i=0}^{n} \ep \Big[ \frac{m}{2} \Big( \frac{x_{i+1} \mi x_i}{\ep}\Big)^2 \plu V(x_i) \Big] \Big)
\eeq

  If we choose $\tau= \b \hbar$ in \rf{pre41} and $x_s =x_0 =x$ and integrate 
 wrt $x$ we obtain a path integral representation of the quantum partition function \rf{pre18}:
 \beq\label{pre42a}
\boxed{ Z(\beta) = \int dx \la x | \,\e^{-\b \hH} | x \ra =  \int dx \!\!\!\int\limits_{\substack{x(0) = x \\ x(\b) = x}} 
\cD x(\ttau) \; \e^{-\frac{1}{\hbar} S_E[x(\ttau)]}.}
\eeq
This formula is also denoted the Feynman-Kac formula and we see that one, loosely speaking, obtains the partition 
function by integrating over all paths which are periodic with period $\b$.

 Note the difference of sign for the potential in \rf{pre42} compared to \rf{pre4}. This can formally be understood 
 by making the replacement $t \to -i \tau$ in the classical action \rf{pre4}, whereby
 \beq\label{pre43}
 iS = i\int d \tt \Big[  \frac{m}{2} \Big( \frac{d x}{d \tt}\Big)^2 \mi V(x(\tt)) \Big]  \to
 -\! \!\int d \ttau \Big[  \frac{m}{2} \Big( \frac{d x}{d \ttau}\Big)^2 \plu  V(x(\ttau)) \Big] = -S_E.
 \eeq
  However, this is the 
 wrong way to think about it. In general a curve $x(t)$ will have no analytic continuation when $t \to -i \tau$. As mentioned 
 one arrives at $S_E[x]$ by following the same steps as before, but since there is no ``i'' in front of the Hamiltonian one 
 obtains instead of \rf{pre31}
 \beq\label{pre44}
 \la x| \hat{O}_\ep^{(E)}| y\ra =  \la x|\; \e^{-\frac{\ep \hp^2}{2m \hbar}} \e^{- \frac{\ep \hat{V}(\hx)}{\hbar}} |y\ra = 
  \Big(\frac{m}{2\pi  \ep \hbar}\Big)^{\oh}  \; \e^{- \frac{(x-y)^2 m}{2 \ep \hbar}}\e^{- \frac{\ep V(y)}{\hbar}}.
  \eeq
  which explains the change of sign. $S_E[x]$ {\it is called the  Euclidean action} because the replacement 
  $t \to -i \tau$, when viewed in a Minkowskian spacetime formally corresponds to changing the signature 
  of the metric to that of Euclidean space:
  \beq\label{pre45}
  ds_M^2 =dx^2 - c^2 dt^2  \to ds_E^2 =dx^2 +   c^2 d\tau^2.
  \eeq
  
Let us finally return to \rf{pre37} and explain why it is wrong and what kind of ``paths'' we are ``integrating'' over in the path integral. 
In this discussion it is more convenient to use the Euclidean version \rf{pre41}-\rf{pre41a} since one can then actually talk about a 
measure on a suitable set of path. When we look at the integrand in \rf{pre41a} it is clear that the dominant terms in the limit 
$\ep \to 0$ will be the kinetic terms proportional $(x_{i+1} -x_i)^2/\ep$ and they will kill any contribution to the integrals unless 
these terms are $\cO(1)$. On the other hand they are not really suppressed any further. Thus we typically expect for $\ep \to 0$:
\beq\label{pre46}
\frac{| x_{i+1}- x_i |}{\ep} \approx \frac{1}{\sqrt{\ep}} \, \sqrt{\frac{2\hbar}{m}}, \quad {\rm i.e.} \quad \frac{| x_{i+1}- x_i |}{\ep} \to \infty.
\eeq
The picture of the piecewise linear path shown in Fig.\ \ref{figpre4} is therefore somewhat misleading. 
In the limit $\ep \to 0$ the derivative of  the curve diverges everywhere and it should be viewed as a continuous 
curve which is nowhere differentiable. We can also estimate the length of such a curve when $\ep \to 0$ (recall 
$\ep = \tau/(n+1)$):
\beq\label{pre47}
\ell (x(\ttau)) = \sum_{i=0}^n |x_{i+1} - x_i | \approx  \sqrt{\frac{2\hbar}{m}} \; n \,\sqrt{\ep} \approx  \sqrt{\frac{2\hbar \tau}{m}} \; \sqrt{n}.
\eeq
 The fact that the length $\ell(x(\ttau)) \propto \sqrt{n}$ for large $n$, rather than going to
 a constant (the length $\ell$ of the given nice continuous curve $x(\ttau)$) signifies that the curve $x(\ttau)$ is {\it not} ``nice'', 
 but actually {\it fractal} with a so-called {\it Hausdorff dimension} equal to 2 (we will discuss this in detail later).
 
 Finally, does it make mathematical sense to view \rf{pre41}-\rf{pre41a} as an integration over continuous path from
 $x_0$ to $x_s$? The answer is yes. We will not go into any detail but just mention a few things.We assume the parameter
 range $[0,\tau]$ is the same for all curves $x(\ttau)$. 
 One can now define the distance between to curves $x_1(\ttau)$ and $x_2(\ttau)$ as
 \beq\label{pre48}
 d(x_1,x_2) = \sup_{\ttau \in [0,\tau]} | x_1(\ttau)-x_2(\ttau)|.
 \eeq
 The existence of this distance turns the space of continuos curves from $x_0$ to $x_s$ into a metric space on which one can 
 define a measure, which again allows us to define integration of functions of curves. It turns out that, loosely speaking, 
 this integration measure is  just our $\cD x(\ttau)$ multiplied by the action of a free particle. This integration measure 
 is denoted the {\it Wiener measure on the set of continuous paths}.   Putting $\hbar =1$ and $m=1$
 we can write
 \beq\label{pre49}
 \cD \mu (x(\ttau)) = \cD x(\ttau) \, \e^{-\oh \int d\ttau  \big(\frac{dx}{d\ttau}\big)^2} = 
 \lim_{n\to \infty} \frac{1}{\sqrt{2\pi  \ep }} \prod_{i=1}^n \frac{dx_i}{\sqrt{2\pi \ep}} \; 
\e^{ -\frac{1}{2} \sum_{i=0}^{n} \ep \big( \frac{x_{i+1} -x_i}{\ep}\big)^2}
\eeq
 Contrary to $\cD x(\ttau)$, $\cD\mu (x(\ttau))$ can be shown to be well defined, and one can now integrate functions
 defined on the set of continuous curves (what we usually call functionals, since the arguments of such a function is itself 
 a function (the curve)). Let $F[x(\ttau)]$ be such a function (functional). The integral of $F[x]$ can now 
 formally  be written as
 \beq\label{pre50}
 \int \cD\mu(x(\ttau)) \; F[x(\ttau)],
 \eeq
 and the precise meaning is obtained by sub-dividing the $\ttau$ 
 parameter range $[0,\tau]$ by $n$ points $\tau_i = i \ep$, as in \rf{pre49}, also
 for the functional $F[x(\ttau)]$. In particular the Feynman-Kac path integral \rf{pre41} is now the integral \rf{pre50} with
 the functional
 \beq\label{pre51}
 F[x(\ttau)] = \e^{-\int d\ttau V(x(\ttau))} = \e^{- \ep \sum_{i=0}^n V(x_i)}.
 \eeq
 The Wiener measure acts as a probability measure on the set of continuous functions and one can then ask interesting 
 questions like: what is the probability that a continuous function is differentiable in a single point? Maybe not surprising 
 from our discussion above the probability is zero! Nevertheless the set of $C^\infty$ functions is dense in the set of 
 continuous functions (much in the same way as set the of rational numbers have measure zero but still is dense in 
 the set of real numbers). 
 
 There are not many potentials $V(x)$  where one can calculate the path integral \rf{pre34} for finite $n$ and then take the 
 limit $n \to \infty$ and in this way obtain $\la x_s | \e^{-i \hH t/\hbar} | x_0\ra$. One is  $V(x) \equ 0$, i.e. the 
 potential for the free particle, where the corresponding path integral 
is discussed in Problem Set 1. Another one is $V(x) = m \om x^2/2$, i.e.\ the harmonic oscillator potential, where 
the path integral is calculated in Problem
Set 2. These path integrals can be performed because they only involve Gaussian integrations. 
From a calculational point of view it is in general easier to solve the Schr\"{o}dinger equation directly. However, there will be 
another class of path integrals which involve geometries, and where the action $S[x]$ is ``geometric'', and where
we will be able to perform the path integral simply by counting geometries.  These are the path integrals 
we will discuss in the  lecture notes.

 \newpage
 
   \setcounter{figure}{0}
 \renewcommand{\thefigure}{1.\arabic{figure}}
 \setcounter{equation}{0}
 \renewcommand{\theequation}{1.\arabic{equation}}
 \section*{1. The free relativistic particle}
 
 \subsection*{The propagator} 

We will now discuss how the Green function for the free {\it relativistic} particle, via the path integral, can be 
described as a scaling limit of a statistical  ensemble of paths and we will encounter the first example of universality
of the scaling limit of geometries. In the following we will use units where $\hbar = c =1$. These constants will then be 
left out of equations, which will simplify the notation. They can of course be reinserted at any point if needed.

In the preliminary notes we discussed the Schr\"{o}dinger equation for a non-relativistic particle, its rotation to ``Euclidean'' time
(the heat- or diffusion-equation), as well as the corresponding Green functions, represented via path integrals. Let us 
just recapitulate, now writing the formulas in $d$ space dimensions. The Schr\"{o}dinger equation reads:
\beq\label{1.1}
\Big( i \frac{\prt}{\prt t} + \frac{1}{2m} \frac{\prt^2}{\prt x_i^2}\Big) \psi(x,t) = 0, \quad i=1,\ldots,d
\eeq
and rotating to Euclidean time $t \to -i \tau$ leads to the diffusion equation:
\beq\label{1.2} 
\Big( - \frac{\prt}{\prt \tau} + \frac{1}{2m} \frac{\prt^2}{\prt x_i^2}\Big) \psi(x,\tau) = 0, \quad i=1,\ldots,d.
 \eeq
 The solution to \rf{1.2} which is zero for $\tau < \tau_0$ and starts out as 
 \beq\label{1.3}
 \psi(x,\tau_0) = \del^{d}(x \mus x_0) \qquad ({\rm i.e.} \quad |\psi (\tau_0) \ra = | x_0\ra)
 \eeq
can be written as  
\beq\label{1.4}
G( x\mus x_0,\tau \mus \tau_0) = \la x |\, \e^{-(\tau-\tau_0) \, \hH} \,| x_0\ra, \qquad \hH = - \frac{1}{2m} \frac{\prt^2}{\prt x_i^2}.
\eeq
$G( x\mus x_0,\tau\mus \tau_1)$ is the Green function of the differential operator 
$\frac{\prt}{\prt \tau} \mi \frac{1}{2m} \frac{\prt^2}{\prt x_i^2} $:
\beq\label{1.5}
\Big( \frac{\prt}{\prt \tau} -\frac{1}{2m} \frac{\prt^2}{\prt x_i^2} \Big) \; G( x\mus x_0,\tau \mus \tau_1) = 
\del(\tau \mus  \tau_0) \del^{d}(x \mus x_0)
\eeq

We want to generalize from a non-relativistic particle to a relativistic particle:
\beq\label{1.6}
\Big(\mi  i \frac{\prt}{\prt t} - \frac{1}{2m} \frac{\prt^2}{\prt x_i^2}\Big) \psi(x,t) = 0 \quad\to\quad  
\Big( \frac{\prt^2}{\prt t^2}  -\frac{\prt^2}{\prt x_i^2}+m^2 \Big) \phi(x,t) =0.
\eeq
We perform again the analytic rotation to Euclidean ``time'' $\tau$:
\beq\label{1.7}
t = -i\tau \equiv -i x_D, \quad D=d+1,\qquad ds^2 = dx_i^2 -dt^2 \to dx_i^2 +dx_D^2 \equiv dx_i^2.
\eeq
where we, with an abuse of notation, have  included the last index $D$ in the sum $dx_i^2$. Thus
\beq\label{1.8}
\Big(  -\frac{\prt^2}{\prt x_i^2} + m^2\Big) \phi(x) =0, \qquad D=d+1.
\eeq
Again the ``propagation'' of the ``Euclidean particle'' is described by a Green function 
\beq\label{1.9}
\Big(  -\frac{\prt^2}{\prt x_i^2} + m^2\Big) \, G(x \mus y) = \del^D(x \mus y).
\eeq
 Introducing the Fourier transformed $\hG$ by
 \beq\label{1.10}
 G(x \mus y) = \int \frac{d^D k}{(2\pi)^D} \; \e^{i k_ix_i} \, \hG(k_i),
 \eeq
 eq. \rf{1.9} can be written as 
 \beq\label{1.11}
 \big( k_i^2 + m^2\big) \hG(k_i)= 1 \quad {\rm i.e.} \quad   \boxed{ \hG(k) = \frac{1}{k_i^2 + m^2}}.
 \eeq
 From \rf{1.10} one now obtains
 \beq\label{1.12}
 \boxed{G(x \mus y) = \frac{1}{(2\pi)^{\frac{D}{2}}} \Big( \frac{m}{|x \mus y|}\Big)^{\frac{D}{2}-1} \; K_{\frac{D}{2}-1} (m | x \mus y|)}
 \eeq
 where $K_\nu(x)$ denotes the second modified Bessel function with index $\nu$. The asymptotic behaviors of $G(x\mi y)$ are:
 \bea \label{1.13}
 G(x \mus y) &\approx&   \frac{\Gamma\big( \frac{D}{2} -1\big)}{4 \pi^{\frac{D}{2}}}\; \frac{1}{|x \mus y|^{D-2}},\hspace{2.1cm}
 {\rm for}\quad m |x-y| \ll 1 \\
 G(x \mus y) &\approx& \frac{1}{(2\pi)^{\frac{D}{2}}}\; \sqrt{\frac{\pi}{2}} \, 
 \frac{m^{\frac{D}{2} - \frac{3}{2}}}{|x \mus y|^{\frac{D}{2} - \frac{1}{2}}}
 \; \e^{-m |x \mus y|} \quad {\rm for}\quad m |x \mus y| \gg 1 \label{1.14}.    
 \eea
 As already mentioned when we discussed the spin-spin correlator on a lattice, this is  
 a generic behavior for our correlators or  Green functions: {\it a power like behavior for small distances 
 and an exponential fall off (with power corrections) for large distances}, measured relative to a parameter, here the mass, which 
 defines  an ``intrinsic'' scale of the physical ``system''.  Here we are discussing a free particle, but as we will see the 
 correlator or Green function or propagator (many names for the same object!) will be described by a 
 statistical ensemble of path, and it will be in this ``system'' that we will define the scale.
 
 There are infinitely many Green functions for a given differential equation and we fix this ambiguity by imposing appropriate 
 boundary conditions. Here the (Euclidean) boundary condition is:
 \beq\label{1.15}
 G(x) \to 0 \quad {\rm for} \quad |x| \to \infty, \quad (D > 2)
 \eeq
 As discussed in the {\it Preliminary Material, part B}, 
 one obtains by analytic continuation $x_D \to i t$ the so-called Feynman Green function
 in Minkowski spacetime. From now on we will stay in Euclidean spacetimes. As long as we deal with quantum field theories 
 in flat spacetimes this procedure is well understood and well defined. However, its status is less clear if we consider quantum 
 field theories in curved spacetime, partly because the concept of an analytic continuation between geometries with 
 Euclidean signatures and geometries with Lorentzian signatures is not well understood or even always well defined. 
 And the status of such a rotation becomes even less clear  when we start discussing systems where the geometry 
 itself is the object of quantization: has a quantum theory of geometries with Euclidean signatures 
 any relation to a quantum theory of geometries with Lorentzian signature? It is a very interesting question and the full
 answer to this question is presently unknown. Here we will perform all calculations using geometries with Euclidean signature
 and we will not discuss the connection to a similar theory of geometries with Lorentzian signatures.
 
 \subsection*{The path integral}
 
 We now want to reproduce the Euclidean Green function \rf{1.11} or \rf{1.12} from a path integral, using a beautiful {\it geometric}
 action for the classical free particle:
 \beq\label{1.16}
 S[P(x,y)] = m_0 \;\ell [P(x,y)].
 \eeq
 In this formula $x$ and $y$ denote {\it spacetime} points in $\mathbb{R}^D$. After rotation to Euclidean spacetime we do 
 no longer work with a separate time coordinate. $P(x,y)$ denotes a {\it geometric} path from $y$ to $x$
 (see Fig.\ \ref{fig1.1}). Let us choose a parametrization of the path:
 \begin{figure}[ht]
\vspace{-1cm}
\centerline{\scalebox{0.2}{\includegraphics{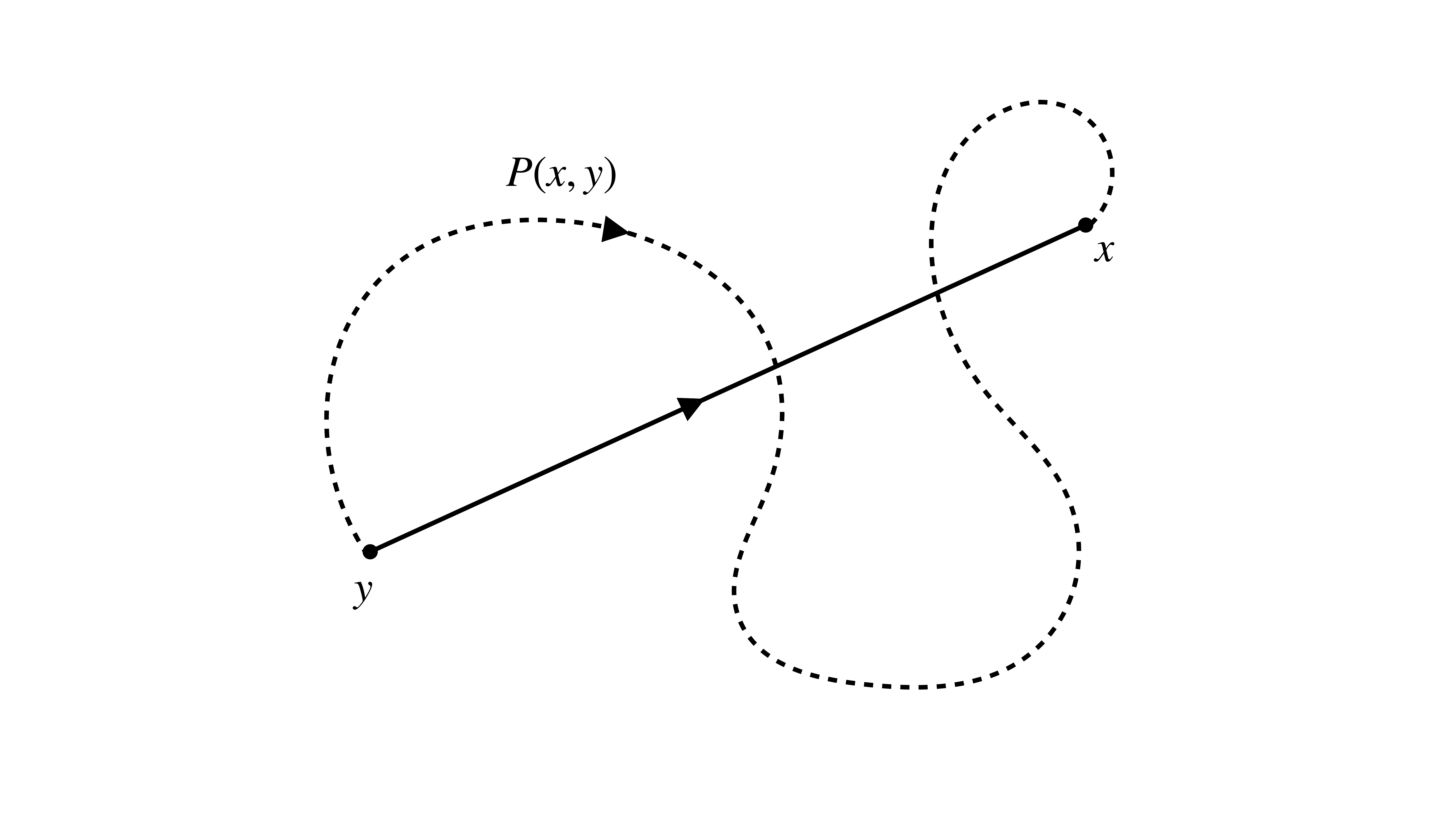}}}
\vspace{-1cm}
\caption{{\small  A path $P(x,y)$ (dashed curve) and the shortest path between $x$ and $y$.}}
\label{fig1.1}
\end{figure}
 \beq\label{1.17}
 P(x,y): \xi \to x(\xi), \quad \xi\in [0,1],\quad x(0) =y,\quad x(1) =x.
 \eeq
 Having chosen such a parametrization we can calculate the length of the path:
 \beq\label{1.18}
\boxed{S_E[P(x,y)] = m_0 \ell[P(x,y)]= m_0 \int_0^1 d\xi \,  \sqrt{ \Big(\frac{d x_i}{d \xi}\Big)^2}},
\eeq
and we can find the eom:
\beq\label{1.19}
\frac{\del S}{\del x_i(\xi)} = - \frac{d}{d \xi} \Big[\frac{d x_i}{d \xi}\Big/\sqrt{ \Big(\frac{d x_i}{d \xi}\Big)^2}\Big] =0
\eeq
Clearly we can find  the minimum of $S[P(x,y)]$ without solving \rf{1.19}, since the minimal length of a path from $y$ to $x$ is 
just $|x\mi y|$. This result is clearly independent of the chosen parametrization of the paths, in agreement with the fact that 
the action \rf{1.18} is {\it reparametrization invariant}: for any twice differentiable function $f: \xi \to f(\xi)$ where $f(0) \equ 0$, 
$f(1) \equ 1$ and $f'(\xi) > 0$ ,the action will be unchanged if we replace $\xi$ by $f(\xi)$. In particular, if $x_{cl} (\xi)$ is a solution 
to \rf{1.19}, then $x_{cl}(f(\xi))$ will also be a solution to \rf{1.19}, and they will both represent geometrically 
the straight line from $y$ to $x$, just with different parametrizations. If we use the action \rf{1.18} in the 
the path integral, it is then natural that we sum only over geometric paths. We thus write

{\underline{Free non-relativistic particle}:}
\beq\label{1.20}
G(x\mus y,\tau_1 \mus \tau_0) = \int\limits_{\substack{x(\tau_0) = y \\ x(\tau_1) = x}} 
\cD x(\ttau) \;\e^{-S_E[x(\ttau)]}=\int\limits_{\substack{x(\tau_0) = y \\ x(\tau_1) = x}} 
\cD x(\ttau) \;\e^{-\int_{\tau_0}^{\tau_1} d \ttau \,\frac{m}{2} \,( \frac{d x}{d \ttau})^2 }
\eeq
{\underline{Free relativistic particle}:}
\beq\label{1.21}
\boxed{G(x\mus y)= \int \cD P(x,y) \; e^{-S_E[P(x,y)]}=\int \cD P(x,y) \; e^{ -m_0 \ell[P(x,y)]}.}
\eeq
Note the difference between the two expressions. In \rf{1.20} $x$ refers to a spatial point, i.e. it has 
coordinates $x_i$, $i=1,\ldots,d$ and the (Euclidean) time $\tau$ appears as the parameter in 
the curve $x(\tau)$ from $x(\tau_0)$ to $x(\tau_1)$. In \rf{1.21} the time-coordinate $x_D$, $D=d+1$ is treated 
on equal footing with the spatial coordinates. To specific a curve from $y$ to $x$ we might have to introduce an 
artificial parameter $\xi$ and write $x(\xi)$. This can clearly be done in many ways, but our summation 
is independent of how we choose such a parametrization since we sum only over geometric paths. However, there
are many more geometric paths joining the two spacetime points in \rf{1.21} than in \rf{1.20}. 

We can label the paths $P(x,y)$ from $y$ to $x$ according to their lengths $\ell[P(x,y)] $. Since all paths of 
the same length have the same action we can formally split the path integral in an integration over paths of a given length $\ell$
followed by an integration over $\ell$:
\beq\label{1.22}
G(x\mus y) = \int_0^\infty d\ell \; \e^{-m_0 \ell} \int\limits_{\substack{\ell[P(x,y)] = \ell}}  \cD P(x,y) \cdot 1 = 
\int_0^\infty d\ell \; \e^{-m_0 \ell}\; \cN_{x,y}(\ell),
\eeq
where $ \cN_{x,y}(\ell)$ denotes the number of paths of length $\ell$ between $x$ and $y$. Thus
{\it the propagator of the free particle is entirely determined by the entropy of paths.}
 We will encounter the same when we study the path integral of higher dimensional geometries than the paths. 
 The ``propagators'' we can define for such ensembles will be entirely determined by the entropy, i.e.\ the number
 of such geometries, and the amazing conclusion
 is that we can quantize geometries, i.e.\ gravity,  and calculate propagators, if we can only count geometries. 

Of course  $ \cN_{x,y}(\ell)= \infty$.

\begin{figure}[t]
\vspace{-1cm}
\centerline{\scalebox{0.20}{\includegraphics{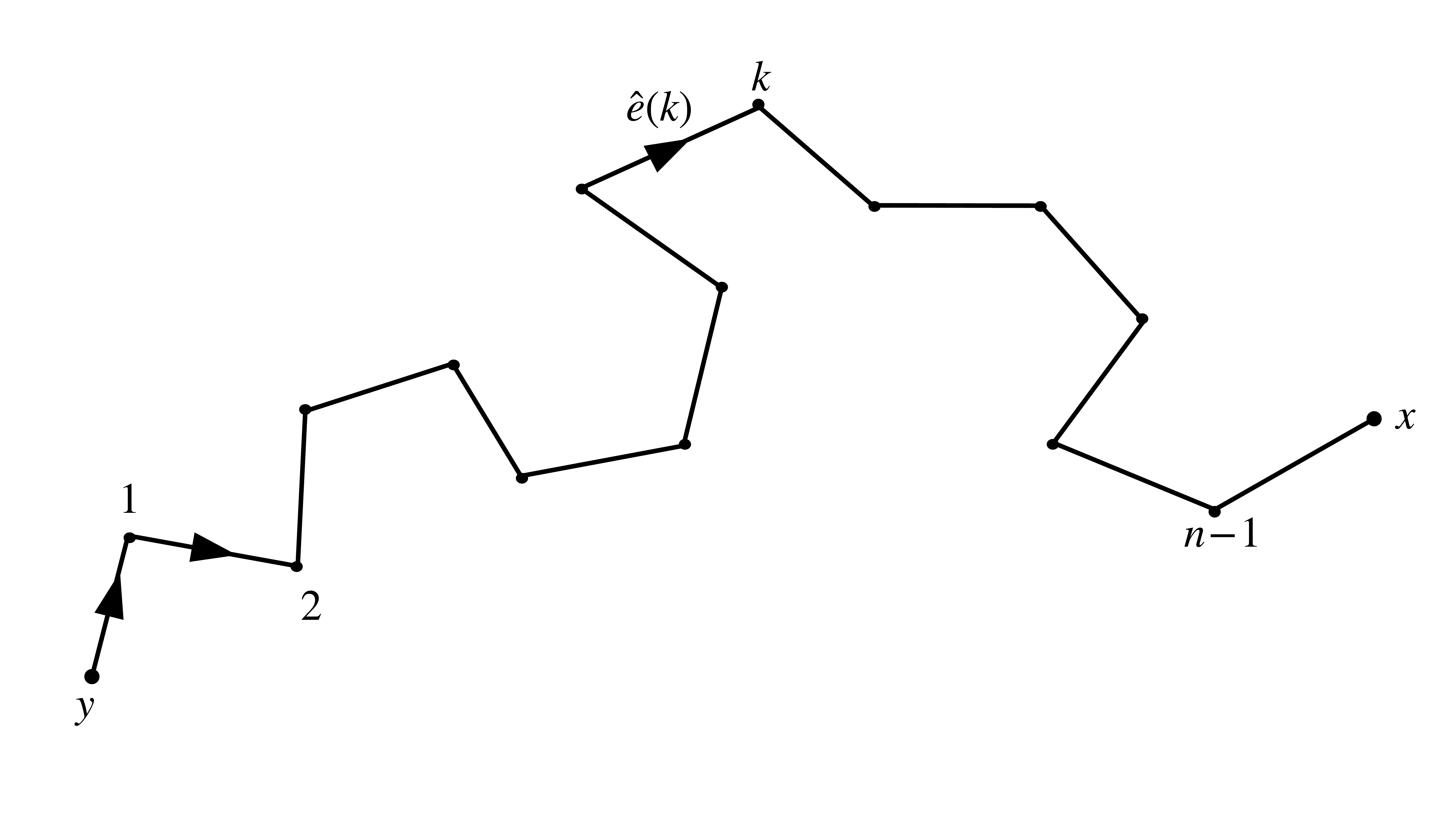}}}
\vspace{-1cm}
\caption{{\small  A piecewise linear curve from $y$ to $x$, 
consisting of $n$ pieces of length $a$ and direction $\hat{e}(k)$, $k=1,\ldots n$.}}
\label{fig1.2}
\end{figure}

In order to make a meaningful counting we have to introduce a cut-off in the same way as we introduced a discretization
of the time interval $[\tau_0,\tau_1]$ in pieces of length $\ep$ for the non-relativistic particle, and considered 
piecewise linear paths in these time intervals. In the case of our (Euclidean) 
relativistic particle we have a formulation where time has no special role. It is thus natural instead  to consider paths in 
$\mathbb{R}^D$ which are piecewise linear and where the {\it length}  of the individual linear pieces  in $\mathbb{R}^D$ 
is $a$ (we use $a$ rather than $\ep$ when refering to distances in $\mathbb{R}^D$). In this way our ``cut-off'' $a$ 
will be independent of a possible chosen parametrization of the path. Let us now 
consider a path  $P_n$ from $y$ to $x$ which consists of $n$ pieces. We then have 
\beq\label{1.23}
\ell[P_n] = n \cdot a,\qquad S[P_n] = m_0 \ell[P_n] = m_0\, a \, n,
\eeq
and the propagator \rf{1.22}, calculated by summing over all such piecewise linear paths, is:
\bea\label{1.24}
G_{a} (x\mus y) &=& \sum_{n=1}^\infty \e^{-m_0\,a \, n} \int_{\{P_n\}} \cD P_n\cdot 1 \\
&=& \sum_{n=1}^\infty \e^{-m_0 a \, n} \int \prod_{j=1}^n d \hat{e}(j) \; \del \Big( a \sum_{j=1}^n \hat{e}(j) \mus (x\mus y)\Big),
\label{1.25}
\eea
where $\hat{e}(j)$ denotes a unit vector along the $j^{th}$ linear segment of a path $P_n$ consisting of $n$ such segments
(see Fig. \ref{fig1.2}).

Note that we have to allow for all $n$ in order not to limit the length of the paths. This is in contrast to our regularization 
used for the non-relativistic particle, where $n$ was linked to $\ep$ and we for a given $n$ could have paths of arbitrary length. 
The propagator now depends on the cut-off $a$ and we want to show how one can obtain the continuum propagator 
in the limit $a \to 0$.
\beq\label{1.26}
\hG_{a} (p)  = \int d^Dx \; \e^{-i p_i (x_i -y_i)} \, G_a(x\mus y) 
= \sum_{n=1}^\infty \e^{-m_0 a \, n} \int \prod_{j=1}^n d \hat{e}(j) \; \e^{ -i a \, p_i \hat{e}_i(j)},
\eeq
and the integration over the unit vectors $\hat{e}$ can be performed\footnote{The integral is equal 
$2 \pi^{\frac{D}{2}} \frac{J_{\frac{D-1}{2}} \big( |p| a\big)}{(|p| a/2)^{\frac{D-1}{2}}} $, where $J_\nu(x)$ is the 
Bessel function with index $\nu$. }:
\beq\label{1.27}
\int d\hat{e} \; \e^{ -i a\, p_i \hat{e}_i} 
= f(|p| a) =f(0) \Big( 1 -\oh \sg^2 a^2 p^2+\cO(a^3)\Big).
\eeq
The only important property of $f$  is that there is no linear term in $a$ and that the $a^2$-coefficient is negative, statements
which follow trivially by expanding the exponential in the integrand in powers of $a$. Thus we obtain from \rf{1.26}
\beq\label{1.28}
\hG_{a} (p)= \sum_{n=1}^\infty \Big(\e^{-m_0 a } f(|p| a)\Big)^n = \frac{\e^{-m_0 a} f(|p| a)}{1-\e^{-m_0 a } f(|p| a)}.
\eeq
We are interested in the limit $a \to 0$. If we can arrange 
\beq\label{1.29}
\e^{-m_0 a } f(|p| a) \to 1-  \oh a^2 \sg^2(p^2 +m_{\rm ph}^2) + \cO (a^3)
\eeq 
then we obtain 
\beq\label{1.30}
\boxed{\hG_{a} (p) \quad \to \quad \frac{2}{a^2\sg^2} \, \frac{1}{p^2 +   m_{\rm ph}^2}=  \frac{2}{a^2\sg^2} G_{\rm cont} (p)},
\eeq
which shows that except for a divergent factor in front (which has a clear interpretation, as we will discuss later), we obtain
precisely the desired continuum result. Is it possible to arrange \rf{1.29}? Yes, by treating $m_0$ as an adjustable parameter
not directly related to the {\it physical} mass of our particle, which we now denote $m_{ ph}$:
\beq\label{1.31}
m_0= \frac{\ln (f(0)}{a} + \oh a \sg^2 \, m_{ ph}^2.
\eeq
Note that $m_0$ actually never appeared in the classical eom, so it seems not disastrous to change it, and \rf{1.31} is 
the simplest example of {\it mass renormalization} in quantum field theory. 

Is this result accidental? At first, a procedure like the one outlined above might seem rather arbitrary. Note however that 
the result did not depend on the detailed form of the function $f$, and we will now show that the {\it scaling limit} 
\rf{1.28}-\rf{1.30} can  indeed be viewed as natural, and it will be our first example of {\it universality}.

\subsection*{ Random walks and universality}

Let us consider a simple model of so-called random walks (RW) in $\mathbb{R}^D$. The $n^{th}$ step in the walk is 
characterized by the initial position $x(n\mus 1)$ and a probability distribution $\cP(v)=\cP(|v|)$ such that $x(n) = x(n\mus 1) \plu v$,
where $v$ is selected to be between $v$ and $v\plu dv$ with probability $\cP(v) d^Dv$. Such a stochastic process is called
Markovian, meaning that the step is independent of any  $x(k)$, $k < n\mus 1$. Further, let $1\mus e^{-\mu}$ be the probability that 
the process stops at $x(n)$ and $e^{-\mu}$ the probability that it continues. The probability that the process will bring us from 
$y$ to $x$ and then stop is 
\beq\label{1.32}
\cG(x\mus y) = \big(1\mus \e^{-\mu}\big) \sum_{n=0}^\infty \e^{-\mu\,n} \int \prod_{i=1}^{n} d^Dx(i) 
\prod_{k=0}^{n} \cP\big(x(k\plu 1)) \mus x(k)\big)
\eeq
where $ x(n\plu 1) \equ x$ and  $x(0) \equ y$. In discussions related to the path integral it is often convenient to work with ``unnormalized'' 
probabilities $P(v)$, i.e.\ we have
\beq\label{1.33}
P(v) = \e^{\mu_c} \cP(v) , \qquad \int d^D v \; P(v) = \e^{\mu_c}.
\eeq
and we then write
\beq\label{1.34}
\boxed{G(x\mus y) =  \sum_{n=0}^\infty \e^{-\mu\, n} \int \prod_{i=1}^{n} d^Dx(i) 
\prod_{k=0}^{n} P\big(x(k\plu 1)) \mus x(k)\big)}
\eeq
In the same way as $\mu \geq 0$ in order for \rf{1.32} to make sense, $\mu \geq \mu_c$ in order for \rf{1.34} to make sense.
Now $G(x\mus y)$ is no longer a normalized probability because $P$ is no longer normalized and because we have 
chosen to drop the factor  $1\mus \e^{-\mu}$ in \rf{1.32}, but
we trivially get back to $\cG(x\mus y)$ with $\mu$ replaced by $\mu \mus \mu_c$ by dividing $G(x\mus y)$ by 
\beq\label{1.35}
\chi(\mu) = \int d^Dx \; G(x\mus y) = \e^{\mu_c} \sum_{n=0}^\infty \big( \e^{\mu_c}\e^{-\mu}\big)^n = 
\frac{e^{\mu_c}}{1-\e^{-(\mu-\mu_c)}}.
\eeq
We denote $\chi(\mu)$ {\it the susceptibility} because of the obvious analogy with \rf{1.50a}. Note that 
\beq\label{1.36}
\chi(\mu) \propto \frac{1}{\mu \mi  \mu_c} \quad {\rm for} \quad \mu \to \mu_c.
\eeq
This singular behavior of the susceptibility will play an important role later. 

Let us analyze the Fourier transform of $G(x-y)$. We denote the Fourier transform of $\cP(v)$, the so-called characteristic
function of the probability distribution, by $\hcP(k)$, since the Fourier transforms of convolutions of functions are the product of 
 the Fourier transforms of the functions (see Problem Set 1 for discussions of this)  we have 
 \beq\label{1.37}
 \hcP(k) = \int d^Dx \; \e^{-i k_j x_j} \, \cP(x),\qquad \hG(k) = \int d^D x \;  e^{-i k_j x_j} \, G(x),
 \eeq
 and
 \beq\label{1.38}
 \hG(k) = \e^{\mu_c} \hcP(k) \; \sum_{n=0}^{\infty} \Big( \e^{-(\mu-\mu_c)} \hcP(k)\Big)^n = 
 \frac{\e^{\mu_c} \hcP(k)}{1- \e^{-(\mu-\mu_c)} \hcP(k)}.
 \eeq
 From the assumption $\cP(x) = \cP(|x|)$ and the assumption that the second and fourth moment 
 of the probability distribution exists, the  characteristic function has the expansion
 \beq\label{1.39}
 \hcP(k) = 1 - \oh \sg^2 \,k^2  + \cO(|k|^4),
 \eeq
 a result which is trivial if we are allowed to  expand $ \e^{-i k_j x_j} $ in powers of $k$ in \rf{1.37}.
 
In our RW model we have viewed all variables as dimensionless. Let us 
now introduce a scaling parameter $a$ with the dimension of length.  
It is now seen that we obtain precisely the results \rf{1.28}-\rf{1.31}  when making the following identifications
\beq\label{1.40}
\mu = a m_0, \quad  \quad k = a \, p, \quad \e^{\mu_c} \hcP(k) = f(k).
\eeq 
In fact our earlier results corresponded to the choice where the general $P(v)$ used here was chosen to be 
proportional to $\delta(|v| \mi 1)$.

Can we give a simple interpretation of the specific scaling we are using to obtain the propagator of a free particle:
\beq\label{1.41}
k = a \,p ,\qquad \mu-\mu_c = \oh \sg^2\, a^2 m_{ph}^2.
\eeq
Yes, it corresponds precisely to the scaling dictated by the central limit theorem for probability distributions. For our purpose 
we can state a simple version of the central limit theorem as follows: under the assumption \rf{1.39} we have
\beq\label{1.42}
\hcP\big({k}/\sqrt{n}\big)^n \underset{n\to \infty}{\to} e^{-\oh \sg^2 k^2},
\eeq
which follows from \rf{1.39} by using the formula $e^z \equ \underset{n\to \infty}{\lim} \big(1\plu z/n\big)^n$. The rhs of 
\rf{1.42} is just the characteristic function of a Gaussian distribution with variance $\sg$ (see again Problem Set 1). Now apply this to 
the first equality in \rf{1.38} and assume that we can substitute \rf{1.42} for all terms\footnote{It is clear that one can start the
summation in \rf{1.38} at any finite $n_0$ rather than at 0, and obtain the same result in the scaling limit, the reason 
being that the propagator diverges in that limit. All terms up to $n_0$ will be finite, and thus not contribute to the 
scaling limit result. Of course this  does  not prove that one can simply replace $\hcP(k)$ by a pure Gaussian distribution,
but it is a strong hint.}, not only for large $n$:
\beq\label{1.43}
\hG(k) = \e^{\mu_c} \hcP(k) \; \sum_{n=0}^{\infty}  \e^{- (\mu-\mu_c)n +\oh \sg^2 k^2} =
\frac{e^{\mu_c}(1\plu \cO(a^2))}{\sg^2 a^2} \sum_{n=0}^\infty \sg^2a^2\; \e^{-\oh \sg^2 a^2 \,n\, (m_{ph}^2+p^2)}
\eeq
where we have made the substitution \rf{1.41}. In the limit $a \to 0$ the sum is converted to an integral
\beq\label{1.44}
\oh\sum_{n=0}^\infty \sg^2a^2 \, \e^{-\oh \sg^2 a^2 n\, (m_{ph}^2+p^2)} \approx 
\oh  \int_0^\infty\!\! ds \; \e^{-\oh s (m_{ph}^2+p^2)} = 
\frac{1}{m_{ph}^2+p^2}~,
\quad s =\sg^2 a^2 \, n.
\eeq
Thus, since $s$ is a finite continuum variable we can view  $a^2 \sim 1/n$ and  the scaling \rf{1.41} is 
basically the same as the scaling in the central limit theorem \rf{1.42}, and the reason that we obtain a universal result
\rf{1.43} can be traced back to the central limit theorem. Finally we remark that the reason we have a  divergent factor
$1/a^2$ in front of the rhs of eq.\ \ref{1.43} can be traced to the use of an unrenormalized $G(x \mus y)$ in \rf{1.34} where 
we have dropped a factor $(1 \mi \e^{-(\mu -\mu_c)})$ compared to $\cG(x \mus y)$ in \rf{1.32}. This normalization factor is 
precisely the susceptibility $\chi(\mu)$, as mentioned above, and we see from \rf{1.36} that we indeed have 
$\chi(\mu) \propto 1/a^2$ when $a \to 0$. 

\begin{figure}[t]
\vspace{-1cm}
\centerline{\scalebox{0.2}{\includegraphics{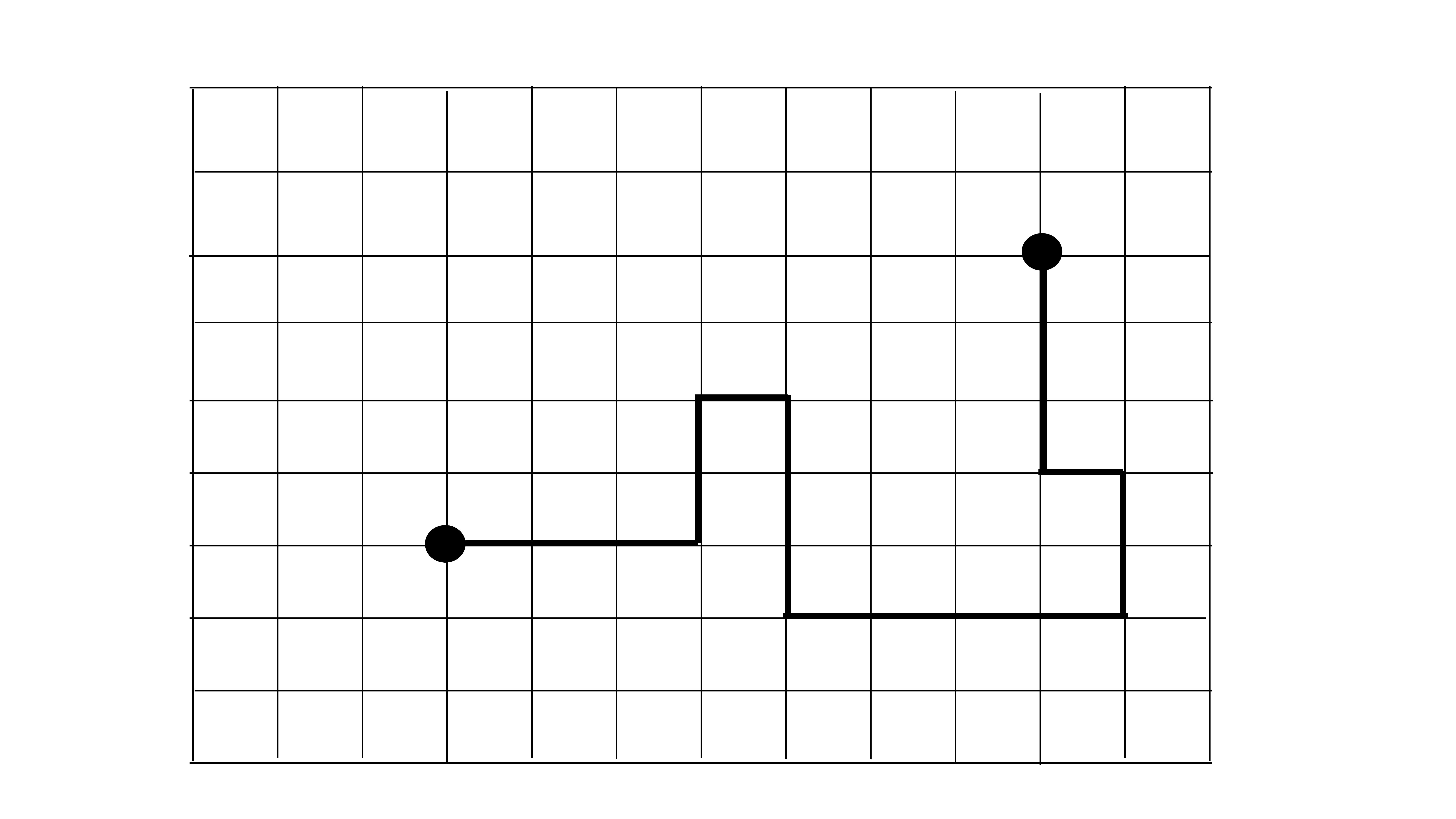}}}
\caption{{\small  A random walk  on a hypercubic lattice}}
\label{fig1.3}
\end{figure}

Since the central limit theorem is valid  under much more general conditions than used here, one can also obtain
the free particle propagator in more general settings than the one discussed above. As an example we mention here (and discuss it
in detail in Problem Set 3) that one can obtain the continuum propagator from RWs on a hypercubic lattice, when the cut-off $a$,
in this case the length of the lattice links, is taken to zero. We have illustrated such a RW on a lattice in Fig.\ \ref{fig1.3}.

 \newpage
 
  \setcounter{figure}{0}
 \renewcommand{\thefigure}{2.\arabic{figure}}
 \setcounter{equation}{0}
 \renewcommand{\theequation}{2.\arabic{equation}}
 \section*{2. One-dimensional quantum gravity}
 
 \subsection*{Scalar fields in one dimension}
 
  If  the RW probability distribution used in Sec. 1 is Gaussian the convolution is ``exact'': the convolution of two Gaussian
  distributions is again Gaussian (for the obvious reason that the characteristic functions are Gaussian and the 
  product of two Gaussians again is a Gaussian, see detailed discussion in Problem Set 1) and we obtain:
  \beq\label{2.1}
\int \prod_{k=1}^{n-1} dx(k)\, \cP\big(x(k) \mus x(k\mus 1)\big) = \frac{\e^{- \frac{|x(n) - x(0)|^2}{2 \sg^2 n}}}{(2\pi  \sg^2 n)^{\oh}} , 
\qquad  \cP(x) =  \frac{\e^{- \frac{|x |^2}{2 \sg^2 }}}{(2\pi  \sg^2)^{\oh} } 
\eeq
Somewhat surprisingly there exists an action which is Gaussian, but still {\it geometric} and still has the 
same classical eom as the geometric action  $S[P(x,y)] = m_0 \ell [P(x,y)]$, where $P(x,y)$ denotes  a path from $y$ to $x$.

In order to describe this action we first make a little digression and describe a few aspects of {\it Riemannian geometry}. Consider
a curved spacetime $\cM$ of dimension $M$. 
Let $\xi =(\xi^1,\ldots,\xi^M)$ be a coordinate system on (a part of) $\cM$. Distances on $\cM$ are 
independent of the chosen coordinate system and described by a symmetric tensor $g_{ab}(\xi)$,
$a,b = 1,\ldots,M$, the so-called metric tensor. $g_{ab}(\xi) $ 
depends on the coordinate system. Let $\xi'$ be another coordinate system and $g'_{ab}(\xi')$ the corrsponding 
metric tensor. The invariant distance $ds$ between between two points with coordinates $\xi$ and $\xi\plu d\xi$ is then 
\beq\label{2.2}
ds^2 = g_{ab}(\xi) d\xi^ad \xi^b = g'_{a'b'}(\xi') d{\xi'}^{a'}  d{\xi'}^{b'},
\eeq
where the relation between $g_{ab}(\xi)$ and  and $g'_{ab}(\xi')$ is 
\beq\label{2.3}
g'_{a'b'}(\xi') = g_{cd}(\xi) \frac{\prt \xi^c}{\prt {\xi'}^{a'}} \frac{\prt \xi^d}{\prt {\xi'}^{b'}} 
\eeq
The transformation property \rf{2.3} defines $g_{ab}(\xi)$ as a tensor and ensures that the distance $ds^2$ 
is coordinate independent. The inverse to $g_{ab}$ is also a tensor and we have $g^{ab} g_{bc} = \del^a_c$. Further, 
the determinant  $g_{ab}$ is denoted $g$: $\det g_{ab} = g$. 
For further reference we note that the so-called Einstein-Hilbert action for the intrinsic geometry of the spacetime $\cM$ is 
\beq\label{2.3a}
S_{\rm{E - H}}[g_{ab}] = K \int d^M \xi \, \sqrt{g(\xi)} \Big( -R(\xi) + 2\Lambda \Big),
\eeq 
where $R$ denotes the intrinsic scalar curvature of $\cM$ (we will discuss the definition and meaning of $R$ later), 
while $1/K$ is proportional to the gravitational coupling constant $G$ and $\Lambda$ is the cosmological coupling constant.

 Let us now consider a scalar fields $X(\xi)$  defined on $\cM$.
That $X(\xi)$ is a scalar field means that  under a change of coordinates $\xi \to \xi'$ it transforms as 
\beq\label{2.4}
X'(\xi') = X(\xi),
\eeq 
which just expresses that the {\it value} of a scalar field in a point $P \in \cM$  is independent of the coordinate system.
Everything we have said above is of course true also if $\cM$ is just flat 
$M$-dimensional space, where we as a natural globally defined coordinate system can use Cartesian coordinates $x^a$ and
$g_{ab}(x)= \del_{ab}$. The change $x =\xi \to \xi'$ will then be a change from Cartesian coordinates to some curvilinear coordinates.
The Gaussian action  for a massless scalar field defined in $\mathbb{R}^M$ is 
\beq\label{2.5}
S[X] =\frac{\kp}{2} \int d^M x \; \del^{ab}\frac{\prt X(x)}{\prt x^a}\frac{\prt X(x)}{\prt x^b} = \frac{\kp}{2}   
\int d^M \xi \,\sqrt{g(\xi)} \, g^{ab} (\xi)
 \frac{\prt X'(\xi)}{\prt \xi^a} \frac{\prt X'(\xi)}{\prt \xi^b},
 \eeq
 where $\kp$ is a coupling constant inserted for dimensional reasons if we want to assign the dimension of length to $X$, 
 and where the rhs is just the action expressed in some curvilinear  coordinates $x \to \xi$ and where $ X'(\xi) = X(x)  $.
One can check that the rhs of \rf{2.5} indeed is invariant under a coordinate change $\xi \to \xi'$ and in particular 
then a change $\xi \to x$ where it reduces to the lhs of \rf{2.5}. In the case where $\cM$ with the metric $g_{ab}(\xi)$ represents
a curved spacetime there exists no coordinate transformation $\xi \to x$, where $x$ is the Cartesian coordinates in $\mathbb{R}^M$,
but the rhs will then be the action for a massless free scalar field defined on the curved spacetime, where by 
``free'' we mean that there is no self-interaction in the action, like a term $X^4$. If we have not only one
scalar field $X(\xi)$, but $D$ scalar fields $X_i(\xi)$, $i=1,\ldots,D$ we can thus write down the corresponding action
for these free massless fields:
\beq\label{2.7}
\boxed{S[X,g_{ab}] =   \frac{\kp}{2} \int d^M \xi \,\sqrt{g(\xi)} \, \Big[ g^{ab} (\xi)
 \frac{\prt X_i(\xi)}{\prt \xi^a} \frac{\prt X_i(\xi)}{\prt \xi^b} + \lam \Big]}
 \eeq
 Several remarks are in order here: as long as the metric $g_{ab}(\xi)$ is fixed, \rf{2.7} just represent $D$ independent
  fields $X_i$.  The last term on the rhs of \rf{2.7} is then irrelevant for the eom since it has no $X$ dependence. Note that 
  \beq\label{2.8}
   \int d^M \xi \,\sqrt{g(\xi)} = {\rm volume}(\cM),
   \eeq
   so the term is  the cosmological term in the Einstein-Hilbert action \rf{2.3a} related to the manifold $\cM$. In the following we are
   going to change the perspective on \rf{2.7} when the dimension $M$ of $\cM$ is 1 or 2, by allowing $g_{ab}$ to 
   be a dynamical variable in addition to the scalar fields $X_i$. This will change everything associated with the interpretation 
   of \rf{2.7}. The scalar fields $X_i$ will no longer be independent since they will interact via the metric field $g_{ab}$ and 
   we will see that when the dimension of $\cM$ is one or two we can view \rf{2.7} as the complete coupled action of 
   $D$ scalar fields and ``gravity''. The reason that \rf{2.7} can be viewed as containing also the action 
   of gravity in these dimensions is, as we will discuss below, that in one dimension there is no intrinsic curvature 
   while in two dimensions the part of the gravitational action \rf{2.3a} which involves the curvature term is a topological 
   term which does not contribute to the eom, and it can consequently be left out as long as we do not consider 
   spacetimes with changing topology.  Thus, for $M=1$ \rf{2.7} will be the classical Lagrangian for 
   one-dimensional gravity coupled to $D$ scalar fields and we will show that this Lagrangian leads to 
   precisely the same eom as the geometric action  \rf{1.18}, i.e.\ it makes sense to view the $D$ scalar fields $X_i(\xi)$ as 
   the coordinates of a particle path $P: \xi \to X_i(\xi)$ in $\mathbb{R}^D$. Also, we will see that $\lam$, the ``cosmological constant''
   in our one-dimensional universe, will be crucial if we want to assign a mass to the free quantum particle.
   When the dimension of $\cM$ is two, 
   we will see that the action \rf{2.7}, which now describes two-dimensional gravity coupled to $D$ scalar fields,
   becomes equivalent to that of  a one-dimensional {\it string} propagating in $\mathbb{R}^D$, and where the {\it geometric} action
   of the string is given by the {\it area} of the worldsheet spanned by the string, in the same way as the geometric action
   of the particle was the length of its worldline. In the rest of 
   this Section we consider the case where the dimension of $\cM$ is $M=1$. 
   
   If $M =1$ we have $\xi = \xi^1$, $g_{ab} = g_{11}$ and we will suppress the index ``1'' and introduce the notation 
   $g_{11}(\xi) = e^2(\xi)$ and \rf{2.2} and \rf{2.3}  read:
   \beq\label{2.9}
   ds = e(\xi) \, d \xi = e'(\xi')\, d \xi', \qquad e'(\xi') = e(\xi) \frac{d \xi}{d \xi'}.
   \eeq 
   Let us assume that $\xi$ (and any other $\xi'$) is normalized to be in the range $[0,1]$. The action \rf{2.7} is then:
   \beq\label{2.10}
    S[X,e] =  \frac{\kp}{2}\int_0^1 d \xi \,e(\xi)\,  \Big[ \frac{1}{e^2(\xi)}  \,
 \Big(\frac{dX_i(\xi)}{d \xi}\Big)^2 + \lam \Big].
 \eeq
  The eom can readily be derived:
  \beq\label{2.11}
  \frac{\del S}{\del e(\xi)} =0 \quad \Rightarrow \quad  -\frac{1}{e^2(\xi)} \Big(\frac{dX_i(\xi)}{d \xi}\Big)^2 + \lam =0,
  \eeq
  \beq\label{2.12}
  \frac{\del S}{\del X_i(\xi)} = 0 \Rightarrow  -\frac{d}{d \xi}\Big( \frac{1}{e} \, \frac{dX_i(\xi)}{d \xi}\Big) =0 
  \Rightarrow   - \frac{d}{d \xi} \Big[ \frac{dX_i(\xi)}{d \xi}\Big/ \Big|\frac{dX_i(\xi)}{d \xi} \Big| \Big] =0.
  \eeq
  The rhs of \rf{2.12} is precisely the eom \rf{1.19}, and  classically \rf{2.10} is thus equivalent to \rf{1.18}
  provided that $\lam > 0$. Note also that if we insert the value of $e(\xi)$ from \rf{2.11} in \rf{2.10} we obtain 
  \beq\label{2.10a}
  S[X] = \kp \sqrt{\lam} \int_0^1 d \xi \,  \left | \frac{dX_i(\xi)}{d \xi}\right | = m_0\, \ell [X], \quad m_0 = \kp \sqrt{\lam},
  \eeq 
i.e. the action \rf{1.18} with the identification $m_0 = \kp \sqrt{\lam}$.

  Of course it is not entirely clear that \rf{1.18} and \rf{2.10} will lead to the same quantum theory, since we in the latter case have 
  two variables, $X_i(\xi)$ and $e(\xi)$. However, as we will now show, even quantum mechanically the two theories are
  identical. 
  
  The action \rf{2.10} is invariant under diffeomorphisms $\xi \to \xi'(\xi)$, $\xi'(0) =0,~\xi'(1) =1,~ d\xi'/d\xi > 0$. We 
  denote the formal number of such diffeomorphisms by ${\rm Vol(diff)}$ (the number is of course infinite). Thus we define 
  \beq\label{2.13}
  \boxed{G(x\mus y) = \int \cD [g_{ab} (\xi)]\!\!\!\! \int\limits_{\substack{X_i(0) = y_i \\ X_i(1) = x_i}}  \!\!\!\!
  \cD X_i (\xi) \;  \e^{-S[X_i(\xi),g_{ab}(\xi)]},\quad  
  \cD [g_{ab}] = \frac{\cD g_{ab}}{{\rm Vol(diff)}}}
  \eeq
  For a given metric $g_{ab}$ the path integral over the fields $X_i$ is essential a straight forward generalization of 
  the non-relativistic path integral for a free particle ($V(x) \equ 0$) to $D$ dimensions, as we will discuss below, the 
  parameter $\xi$ playing the role of the non-relativistic time $t$.
  Many metrics $g_{ab}$ represent the same {\it geometry}, which we denote $[g_{ab}]$. We should only integrate over
  geometries. We have formally represented this on the rhs equation in \rf{2.13}  by dividing the integration over all $g_{ab}$
  by ${\rm Vol(diff)}$. In our one-dimensional case $g_{ab}(\xi)$ has only one component which we have denoted $e^2(\xi)$
  and it is not difficult to find the possible intrinsic geometries of $\cM$. First note that the volume \rf{2.8} of $\cM$, which in the 
  one-dimensional case will be called the length $\ell (\cM)$:
  \beq\label{2.14}
  \ell(\cM) = \int_0^1 d\xi \, \sqrt{g(\xi)} = \int_0^1 d\xi \, e(\xi),
  \eeq
  is of course an invariant under diffeomorphisms. But there are no other invariants, since one can always transform $e(\xi)$ 
  to the constant metric $e'(\xi') = \ell(\cM)$ by a suitable coordinate transformation $\xi \to \xi'(\xi)$:
  \beq\label{2.15}
  \xi' = \frac{1}{\ell} \int_0^\xi d\tilde{\xi} \, e(\tilde{\xi}) \Rightarrow e'(\xi')= \frac{d\xi}{d\xi'} \;e(\xi) = \frac{1}{d\xi'/d\xi}\; e(\xi) = \ell.
  \eeq
  In particular the existence of a constant $e(\xi)$ implies that there is no intrinsic curvature $R(\xi)$ (the expression for 
  $R$ involves the second derivative of $g_{ab}(\xi)$ as we will discuss later), as mentioned above: a curve has no 
  {\it intrinsic} curvature. 
  The integration over intrinsic geometries thus becomes a simple integration over $\ell$, the volume of the geometry:
  \beq\label{2.15a}
  \cD[g_{ab}] = \a \,dl, \qquad 
  \eeq
  where $\a$ is some constant which is not uniquely determined by our formal continuum arguments. In
  principle we have then already calculated the path integral  in \rf{2.13}, 
  since with the choice of $e(\xi) = \ell$ the $X_i$-part of the path integral  
  just becomes the non-relativistic path integral \rf{pre41} of a free particle (potential $V(x) =0$)  
  generalized from 1 to $D$ dimensions. The result of this path integral is just (with suitable normalization) \rf{heat2} 
  with $t=1$ and $b^2 =\ell/2\kp$, and \rf{2.13} becomes:
  \beq\label{2.16}
  G(x \mus y) = \a \int_0^\infty d \ell \; 
  \Big(\frac{\kp}{2\pi \ell}\Big)^{D/2} \; \exp \Big( \mi \frac{   \kp |x \mus y|^2}{2\ell}   \mi \oh \kp\,\lam \, \ell \Big).
  \eeq
  This is the so-called Schwinger proper-time representation of the propagator. By a Fourier transformation we obtain
  \beq\label{2.17}
  \hG(p) = \int d^Dx \, \e^{-i p_j x_j} \, G(x\mus y) = \a \int_0^\infty d\ell \, \e^{- \frac{\ell}{2\kp}  (p^2 +\lam\, \kp^2)} = 
  \frac{1}{p^2+ m_{ph}^2},
  \eeq
  provided we  choose\footnote{Note that in \rf{2.10a} we had $\lam \kp^2 =m_0^2$. The change from $m_0$ to $m_{ph}$ comes
  when we perform the path integral in \rf{2.13}, which we did not actually do here. We only referred to already established results.
  Below we will actually perform the integral and we will see the shift to a renormalized mass.} $\lam \kp^2 =m_{ph}^2$ and $\a =1/2\kp$. It is thus seen that $\lam > 0$ formally seems needed
  in order to obtain the propagator with a mass $m_{ph} > 0$, as remarked earlier. The integral 
  representation \rf{2.17} is the same as we already encountered in formula  \rf{1.44}. 
  
  Rather than appealing, as we did,  to already derived results, when going from eq. \rf{2.13} to \rf{2.16}, 
  it is instructive to derive \rf{2.16} or \rf{2.17} 
  (again) by introducing a cut-off $\ep$, discretizing  and taking the limit $\ep \to 0$. This will bring up the question 
  of how to discretize the space $\cM$ (something which will play a major role later when the dimension of $\cM$
  will be larger than 1). In the case of the non-relativistic path integral we divided the (Euclidean)
  time interval $[0,\tau]$ into sub-intervals of length $\ep$ and at $\tau_k = k\,\ep$ we assigned the variable $x(k)=x(\tau_k)$. 
  We want to do the same thing here, but where $\tau$ had an interpretation as a physical time, the division of the coordinate 
  $\xi$ on $\cM$ is not related to any physical length, and as a minimum we have to require that the cut-off introduced in 
  $\cM$ is invariant under reparametrization (in the limit where $\Del \xi_k \to d\xi_k$):
  \beq\label{2.18}
  \ep = ds = e(\xi_k) \Del \xi_k, \qquad \Del \xi_k = \xi_{k+1} \mus \xi_k.
  \eeq
 The discretized version of \rf{2.10} is now:
 \bea\label{2.19}
 S_\ep [X,e] &=&\frac{\kp}{2} \sum_{k=0}^{n} \Del \xi_k e(\xi_k) 
 \left[ \frac{ \Big(X_i(\xi_{k+1}) \mus  X_i(\xi_k)\Big)^2}{ e^2(\xi_k) \Del \xi_k^2} + \lam\right]  \\
 &=&
  \frac{\kp}{2} \sum_{k=0}^{n} \ep\left[ \frac{ \Big(X_i(\xi_{k+1})\mus  X_i(\xi_k)\Big)^2}{ \ep^2} + \lam\right] .\nonumber
  \eea
  The integration over {\it geometries} $[g_{ab}]$ was reduced to the integration over the volume $\ell$ of these 
  geometries and we have now discretized a geometry of volume $\ell$ into $n\plu 1=\ell/\eps$ pieces. In this way the integration
  over $\ell$ becomes a summation over $n$ and we can finally write for the regularized propagator:
  \beq\label{2.20}
  G_\eps(x\mus y) = \sum_{n=0}^\infty \e^{-\oh \, \lam \, \kp\,\eps\, n}\! \!\!\!\int\limits_{\substack{X(0) = y \\ X(1) = x}} 
  \prod_{k=1}^n \frac{d^DX_i(k)}{(2\pi \ep/\kp)^{D/2}} \; 
  \exp \Big(-\frac{\kp}{2} \sum_{k=0}^n  \frac{(X_i(k\plu 1) \mus X_i (k))^2}{\eps}\Big).
  \eeq
  We have here chosen the normalization factor for path integral wrt the $X_i(\xi)$ variables such that we have 
  a probability distribution $\cP(X)$ as in \rf{2.1}. We are not forced to do that, e.g. one could have omitted the 
  factor $\kp^{D/2}$, in which case one would have worked with a distribution $P(X) = \e^{\mu_c} \cP(X)$, 
  $ \e^{\mu_c} \equ \kp^{-D/2}$.
  As a consequence of \rf{2.1} we obtain by Fourier transformation:
  \beq\label{2.21}
  \hG_\eps(p) = \sum_{n=0}^\infty  \e^{-\frac{\ep n}{2\kp} ( \lam \kp^2 + p^2)}.
 \eeq
 It is thus essentially the same formula as \rf{1.43}, {\it provided} we make the identification $\lam \,\kp^2 = m^2_{ph}$ 
 (or more generally, for unnormalized $P(X)$, $\lam \kp^2 - \frac{\mu_c \kp}{\ep} = m_{ph}^2$), and 
 \beq\label{2.22}
 \boxed{a^2\sg^2 = \frac{\ep}{2\kp}}.
 \eeq
 The parameter $a$ was a cut-off introduced in the space $\mathbb{R}^D$ where  $x$ and $p$  live. We saw 
 explicitly how it was related to distances in this space in the way it was introduced in \rf{1.23} by dividing 
 a path $P$ there of length $\ell[P]$ in $n$ pieces. On the other hand $\ep$ was introduced by dividing the manifold $\cM$ 
 of length $\ell[\cM]$ in $n$ pieces. However, the pieces of size $\ep$ are infinitesimal when measured in units of $a$, 
 or stated differently: if $\ell[P] = n_P a$ is equal to $\ell[\cM] = n_\cM \ep$ then $n_\cM = n_P / (2\kp \sg^2 a)$. This is 
 precisely what we discussed in \rf{pre46}-\rf{pre47}, and is the topic which we will now study in more detail.
 
 \subsection*{Hausdorff dimension and scaling relations}
 
 Let us return to dimensionless units.  Thus in the context of \rf{2.20} we write $x= X \sqrt{\kp/2\ep}$, $\mu = \lam \kp \ep/2$
 and in this way \rf{2.20} becomes a particular simple realization of the general expression \rf{1.34}, where the probability
 distribution $\cP(x)$ is Gaussian. We will now use the general expression \rf{1.34}, which we replicate here for convenience:
 \beq\label{2.23}
G(x\mus y, \mu) =  \sum_{n=0}^\infty \e^{-\mu\, n} \int \prod_{j=1}^{n} d^Dx(j) 
\prod_{k=0}^{n} P\big(x(k\plu 1)) \mus x(k)\big), ~~ \begin{array}{l} x(0) =y \\ x(n\plu 1) =x \end{array}
\eeq
and by Fourier transformation from eq.\ \rf{1.38}
\beq\label{2.24}
 \hG(k,\mu ) = \e^{\mu_c} \hcP(k) \; \sum_{n=0}^{\infty}  \e^{-(\mu-\mu_c)n} \hcP^n(k) = 
 \frac{\e^{\mu_c} \hcP(k)}{1- \e^{-(\mu-\mu_c)} \hcP(k)}.
 \eeq
 Using the central limit theorem we know that $\hcP(k) = 1\mi \oh \sg^2 k^2 +\cdots$ and thus 
 \beq\label{2.25}
 \hG(k, \mu)  \propto \frac{1 + \cdots}{ m^2(\mu) + k^2 + \cdots}, \qquad m^2(\mu) = \frac{2}{\sg^2} \; (\mu -\mu_c).
 \eeq
 where $+ \cdots$ means higher order corrections in $|k|$ and $\mu\mi \mu_c$. By inverse Fourier transformation we obtain
 \beq\label{2.26}
 G(x,\mu) = e^{- m(\mu) |x| + \cdots} 
 \eeq
 where $+\cdots$ indicates logarithmic correction in $|x|$ for large $|x|$. Let us recall the general behavior of the 
 spin-spin correlation function of a statistical spin system near  phase transition point, which $\b_c \sim \mu_c$ and 
 $1/\xi(\b) \sim m(\mu) \sim |\mu \mi \mu_c|^\nu$
 \bea\label{2.27}
 G(x,\mu) &\sim& \frac{c}{|x|^{D-2 +\eta}}, \qquad\qquad \quad |x| \ll 1/m(\mu).\\
 \label{2.28}
 G(x,\mu) &=& \e^{-m(\mu) |x| + \cO (\ln |x|)}, \qquad |x| \gg 1/m(\mu). 
 \eea
For our free particle the central limit theorem ensures that $\nu =1/2$, $\eta =0$  (the mean field 
values). However in the following let us assume that we have an arbitrary $\nu$, since we will later meet such cases. 
We now want to  introduce a scaling parameter with dimension of length in $\mathbb{R}^D$,  $a(\mu) $, a physical length $x_{ph}$ and a 
physical mass $m_{ph}$:
\beq\label{2.30}
m_{ph} a(\mu) = m(\mu)= c (\mu -\mu_c)^\nu,\quad x_{ph} = x \, a(\mu), \quad {\rm i.e.}\quad m_{ph} x_{ph} = m(\mu) x.
\eeq
This ensures that the exponential fall off of the propagator survives in the limit $\mu \to \mu_c$ when expressed in 
terms of ``physical'' distances $x_{ph}$ and a ``physical'' mass $m_{ph}$. A good way to think about this is to consider
that propagator defined  on an infinite  dimensionless lattice. We now introduce the length of the 
lattice links as $a(\mu)$. If $x$ is measured in number of lattice link ``units'', $x_{ph}$ becomes the real physical length. 
When $\mu \to \mu_c$ the correlation length $\xi(\mu) = 1/m(\mu)$, measured in number of lattice links goes to infinity. 
However, we compensate for this by rescaling the physical length of the lattice links $a(\mu)$ such that correlation length 
measured in physical length $x_{ph}$ stays fixed, namely equal to $1/m_{ph}$. This implies, from the assumed behavior 
of $m(\mu)$ that the length $a(\mu)$ of the lattice links scales to zero as $a(\mu) \propto (\mu \mi \mu_c)^\nu$. We
are ``scaling'' the discretize lattice away and recover the continuum.

Let us now consider our ensemble of RWs  (piecewise linear paths in $\mathbb{R}^D$) from $0$ to $x$ defined
by \rf{2.23}. We can view $G(x,\mu)$ as the partition function for this ensemble. The expectation value of an
``observable'' $O$ which takes values on the paths are then defined by
\beq\label{2.31}
\la O\ra_\mu = \frac{1}{G(x,\mu)} \; \sum_{n=0}^\infty \e^{-\mu\, n} \int \prod_{j=1}^{n} d^Dx(j) 
\prod_{k=0}^{n} P\big(x(k\plu 1)) \mus x(k)\big) \; O(\{x_j\}),
\eeq
We now use as an observable $O$ the length $\ell[C(x)]$ of a curve $C(x)$ from $0$ to $x$ and 
we define the Hausdorff dimension $d_H$ 
of the ensemble of paths by:

\beq\label{2.32}
\boxed{\la \ell[C(x)]\ra  \propto | x|^{d_H}} \quad {\rm for } \quad \mu \to \mu_c,  \quad m(\mu)\, |x| = {\rm const}.
\eeq

\begin{figure}[t]
\vspace{-1cm}
\centerline{\scalebox{0.18}{\includegraphics{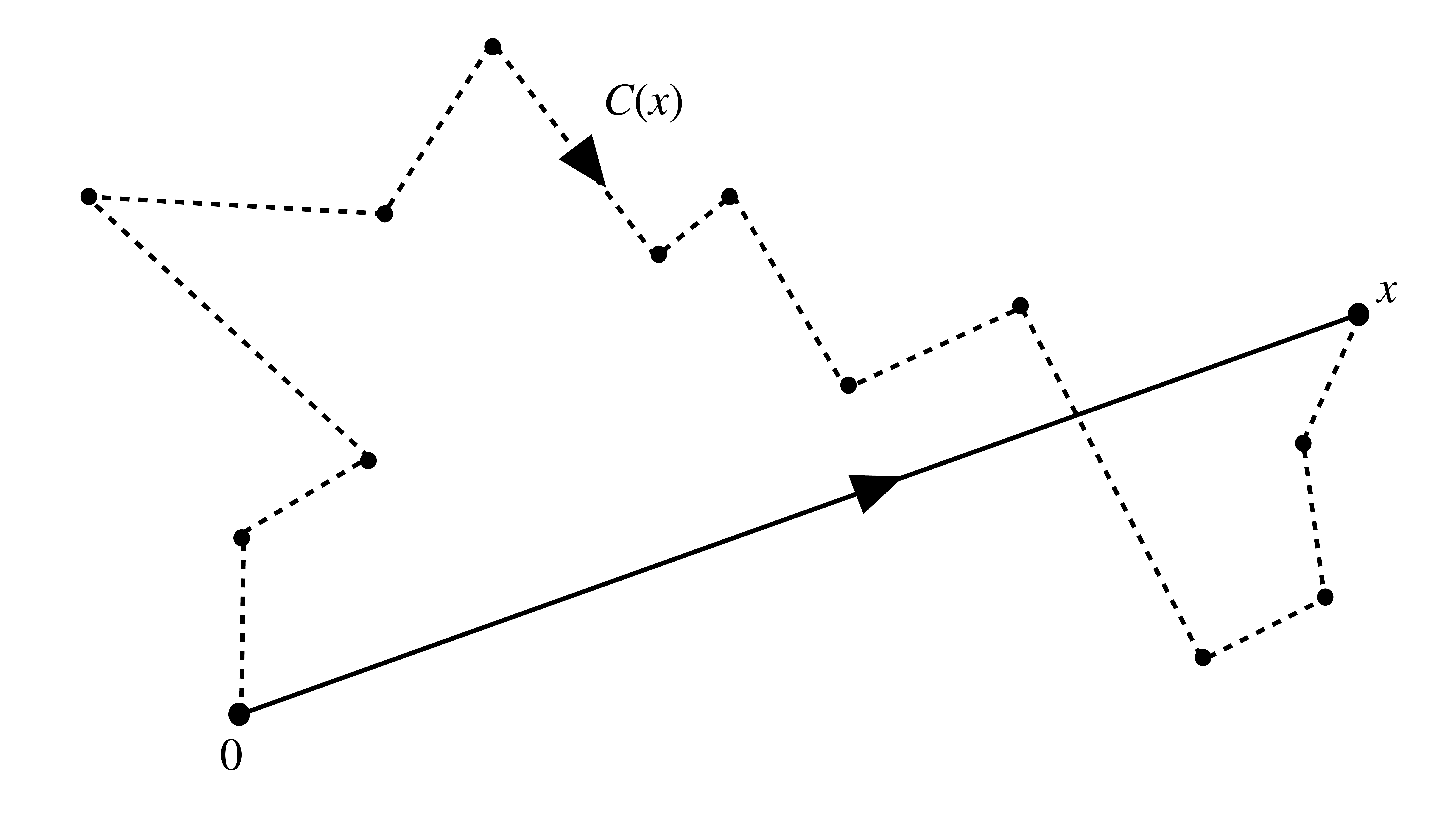}}}
\vspace{-0.5cm}
\caption{{\small  A typical piecewise linear curve $C(x)$ from 0 to $x$.}}
\label{fig2.1}
\end{figure}

The value of $\ell[C(x)]$ on a curve consisting of $n$ pieces is 
\beq\label{2.33}
\ell[C_n(x)]= \sum_{ k=0}^n |x(k\plu 1) \mi x(k)|, \quad x(0) =0~~x(n\plu 1) = x.
\eeq
A typical curve is shown in Fig.\ \ref{fig2.1}. 
When $\m \to \mu_c$ the number $n$ of pieces in a typical 
path will go to  infinity and for such large $n$ it will be approximately true that
\beq\label{2.34}
\la \ell[C(x)]\ra \approx  \sg \la n \ra, \qquad \sg = \int d^Dx\, |x|\, \cP(x),
\eeq
$\sg$ being the average length of a step taken by the random walk, and $\la n \ra$ being the average number of steps.
If we use the discretized action \rf{1.23}-\rf{1.25} this is of course exactly true, since each step in that case has a fixed length.
We can easily calculate $ \la n \ra$ since it follows directly from \rf{2.31} that
\beq\label{2.35}
\la n \ra_\mu = -\frac{1}{G(x,\mu)} \; \frac{\prt G(x,\mu)}{\prt \mu}= - \frac{\prt \ln G(x,\mu)}{\prt \mu} \approx  m'(\mu) |x|,
\eeq
where $m'(\mu)$ denotes the derivative of $m(\mu)$ wrt $\mu$ and where the rhs follows from \rf{2.28}.
From \rf{2.30} we obtain:
\beq\label{2.36}
 m'(\mu) |x| = \frac{\nu}{\mu \mi \mu_c}\; m(\mu) |x| = \frac{ \nu m_{ph} x_{ph}}{\mu \mi \mu_c} \
 \propto m_{ph}x_{ph} \; \frac{|x|^{\frac{1}{\nu}}}{(m_{ph} x_{ph})^{\frac{1}{\nu}}}.
 \eeq
 We thus conclude
 \beq\label{2.37}
 \boxed{ d_H = \frac{1}{\nu}}
 \eeq
 In the case of our RWs we have $\nu=1/2$ and thus $d_H=2$.
 
 In the case of RWs the proof that $d_H=2$ is usually done using instead the ensemble with a fixed number of steps, 
 and then forcing this number of steps to infinity. In this case we have from \rf{2.23}
 \beq\label{2.38} 
 G_n(x) = \int \prod_{j=1}^{n} d^Dx(j) 
\prod_{k=0}^{n} P\big(x(k\plu 1)) \mus x(k)\big), \quad x(0) =0,~ x(n\plu 1) =x.
\eeq
 Before we asked about the average length of a path from 0 to $x$. Now we will instead ask for the distance $|x| $
 travelled by a random walk of $n$ steps (and corresponding average length $\la \ell_n \ra \propto n$), and we define 
 the Hausdorff dimension by 
 \beq\label{2.39}
 \la |x|\ra_n  \propto \la \ell_n\ra^{\frac{1}{d_H}},
 \eeq
 where the averages are calculated wrt $G_n(x)$. 
 If $\cP(x)$ in \rf{2.38} has variance $\sg^2$, then by the central limit theorem $G_n(x)$ will for large $n$ be 
 proportional to  a Gaussian 
 distribution with variance $\sg^2 n$ and for such a distribution one readily shows that
 \beq\label{2.40}
 \la |x|\ra_n  \propto \sqrt{n} \propto \la \ell_n \ra^{\frac{1}{2}},
 \eeq
 Thus $d_H =2$, as expected.
 
 Let us end the Section with the following remark about the significance of $d_H=2$. As we have mentioned it is expected 
 for spin systems that one  has mean-field exponents for $D >4$. As we have studied in Problem Set 4, mean field 
 theory is basically the theory of Gaussian fluctuations, i.e.\ translated to a field theory: free fields. It is believed that one can 
 derive the continuum scalar quantum field theories from lattice spin systems by taking the scaling limit approaching a critical 
 point. If the corresponding critical exponents are mean field exponents it implies that the derived continuum field theory 
 is just a free field theory. Is there a simple explanation why we cannot have interacting  scalar quantum field theories in 
 dimensions $D >4$? Yes: $d_H =2$. One can show that the quantum field theory of a scalar field can be formulated as a 
 theory of particles which interact when their wold lines meet. It is not a very elegant formulation, but it shows that we cannot 
 have interactions for $D >4$. The particles are quantum particles, so their (quantum) paths are two-dimensional since $d_H =2$.
 When $D > 4$ the probability that such two-dimensional objects meet is zero. In dimension $D$ the intersection between 
 a $D_1$- and a $D_2$-dimensional plane is a $(D_1 \plu D_2 \mi D \geq 0)$-dimensional plane, if they meet. Thus the paths will
 not meet for $D>4$ and they will not interact. $D=4$ is marginal (two planes will meet in a point), but it is believed that 
 also here mean field prevails.
  
 \newpage

 \setcounter{figure}{0}
 \renewcommand{\thefigure}{3.\arabic{figure}}
 \setcounter{equation}{0}
 \renewcommand{\theequation}{3.\arabic{equation}}
 \section*{3. Branched polymers} 
  
  \subsection*{Definitions and generalities}
  
\begin{figure}[t]
\vspace{-0.2cm}
\centerline{\scalebox{0.2}{\includegraphics{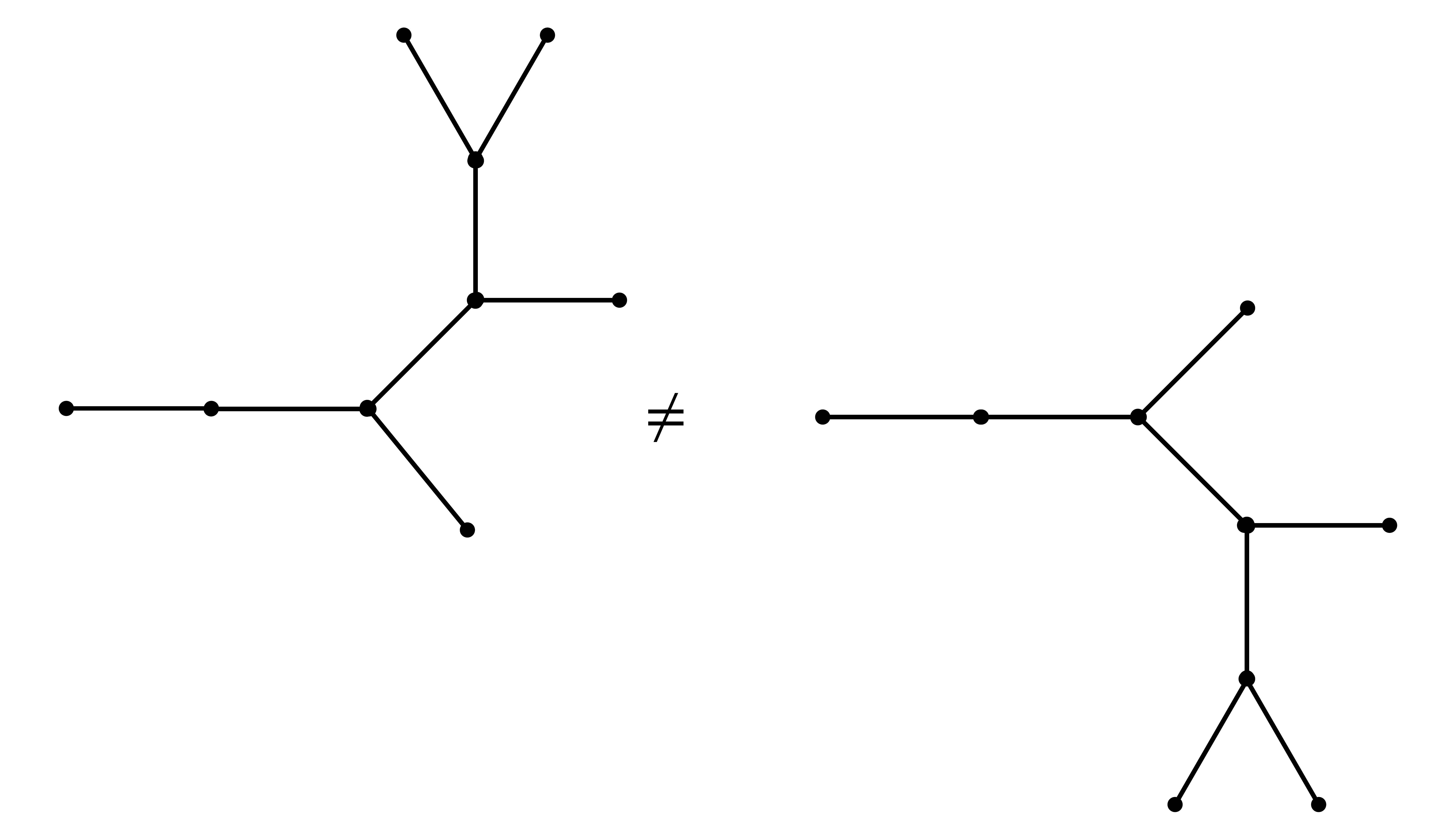}}}
\vspace{-0.2cm}
\caption{{\small  Two BPs which should be seen as inequivalent.}}
\label{fig3.1}
\end{figure}

  We can generalize the random process leading to the RW by enlarging the choices we have when we reach a given vertex:
  before we could stop or continue. Now this last choice is enlarged to branching: the RW is allowed to branch into a number 
  of independent RWs, a process which can be repeated. The process will in this way generate a {\it tree-graph}, i.e. a graph 
  which contains no loops. One can draw the abstract graph in $\mathbb{R}^2$ and we will distinguish graphs which differ by orientation as shown in Fig.\ \ref{fig3.1} (in the actual physical systems to which such trees are approximations, this turns out to
  the natural thing to do). We call these graphs {\it planar branched polymers} or {\it planar trees} (but we will drop the ``planar''
  from now). We will mainly use the notation ``branched polymers'' (BP), since this was the notation used when physicists meet these
  objects in the study of string theory, but in general, and in particular in mathematics, the ``tree'' notation is used.
  Let $v$ denote a vertex on the abstract BP  graph $G$ and $V(G)$ the set of vertices on $G$. We will always assume
  that $G$ is a connected graph. We can assign points $x(v) \in \mathbb{R}^D$ to the vertices, and if $v$ and $v'$ are 
  connected  by a link in $G$ we will also connect $x(v)$ and $x(v')$ in $\mathbb{R}^D$ by a straight line. In this way 
  $G$ is mapped to a graph $G(x)$ in $\mathbb{R}^D$ which we also denote a BP. 
  We now associate an action with this $\mathbb{R}^D$
  graph:
  \beq\label{3.1}
  S[G(x)] = \sum_{\la vv'\ra} \vph ( | x(v) \mi x(v')|) ,
  \eeq
  where $\la vv'\ra$ denotes the link between $v$ and $v'$ if there is any. We are thus summing over all links in \rf{3.1}. $\vph$ is 
  a positive function such that 
  \beq\label{3.2}
  \int d^Dx \; \e^{-\vph(|x|)}  = c_{\vph} < \infty.
  \eeq
  The basic new aspect compared to the RW is that we assign a weight $w_v$ to each vertex, associated with the possibility
  of branching. $w(v)$ will depend only on the {\it order} of the vertex, i.e.\ the number of links to which the vertex belong. Let
  $\sg_v$ denote the order. We will then write $w(\sg_v)$ instead of  $w (v)$. For a RW graph (which is a special BP), 
  one can view the factor the factor $\e^{-\mu}$ as associated with vertices of order 2, 
  while we in the case of unnormalized probabilities
  associated the weight 1 to (the two) vertices of order 1. For the general BP we find it more convenient to associate the 
  weight factor $\e^{-\mu}$ with the links. The final new aspect of BPs compared to RWs is that it is natural to define 
  not only one- and two-point functions but also $n$-point functions, where  $n$ coordinates $\{ x(i)\}$ corresponding to  a
  set of $n$ vertices $\{v(i)\}$, $i=1,\ldots,n$,  are kept fixed while we integrate over the rest:
  \beq\label{3.3}
 \boxed{ G_\mu^{(n)}\big(x(1),\ldots,x(n)\big) =  \sum_{B\in \cB_n} \;\prod_{v \in V(B)}\!\!\! w(\sg_v) 
  \hspace{-3mm}\int\limits_{\{x(v(i)\}} \!\!
  \prod_{v \not\in \{v(i)\} } \hspace{-2mm}   d^D x(v) \; \e^{-S[B(x)]-\mu |L (B)|}}
  \eeq
  In \rf{3.3} $B$ denotes a BP in the set $\cB_n$ of BPs with $n$ marked vertices $\{v(i)\}$, 
  $V(B)$ denotes the set of 
  vertices in $B$, $L(B)$ the set of links in $B$ and $|V(B)|$  and $|L(B)|$ the number of vertices and number of 
  links in $B$, respectively. For a (connected) tree-graph $|V(B)| = |L(B)|+1$. 
  
  If we assume that vertex weights $w_m$ are exponentially bounded, i.e.\ that there exist a constant $c$ such that 
  $w_m \leq c_w^m$, then it is relatively easy to show that $G_\mu^{(n)}\big(x(1),\ldots,x(n)\big)$ exists (i.e.\ the sum in \rf{3.3} 
  is convergent for sufficiently large $\mu$), and in addition that there exists a {\it critical} $\mu_c$, such that the 
  sum is convergent for $\mu > \mu_c$ and divergent for $\mu < \mu_c$, independent of $\{x(v(i))\}$ and $n$. We will 
  not prove this here, but only outline the arguments for $n=1$, i.e.\ the one-point function. The basic observation is 
  that the number of BPs with a given number of links $L$ is exponentially bounded. We will prove this later.
  Let us denote the set of BP with $L$ links $\cB(L)$ and the number of BP graphs with $L$ links $\cN(\cB(L))$ and let us write
  \beq\label{3.4}
  \cN (\cB(L)) \leq c_{bp}^{|L|}.
  \eeq
  Similarly, since each link has two vertices, $\sum_{v \in V(B)} \sg_v = 2 |L(B)|$, and we have 
  \beq\label{3.5}
  \prod_{v\in V(B)} w(\sg_v) \leq   \prod_{v\in V(B)} c_w^{\sg_v} = c_w^{2 |L(B)|}.
  \eeq
  Finally, the one-point function $G_\mu^{(1)}(x(v(1)))$ is by translational invariance of the action  \rf{3.1} independent of $x(v(1))$,
  and we can actually in this case perform the integrals in \rf{3.3} by successive integration, using the tree-nature of 
  the graph $B$, see Fig.\ \ref{fig3.2}:
    \begin{figure}[t]
\vspace{-3cm}
\centerline{\scalebox{0.2}{\includegraphics{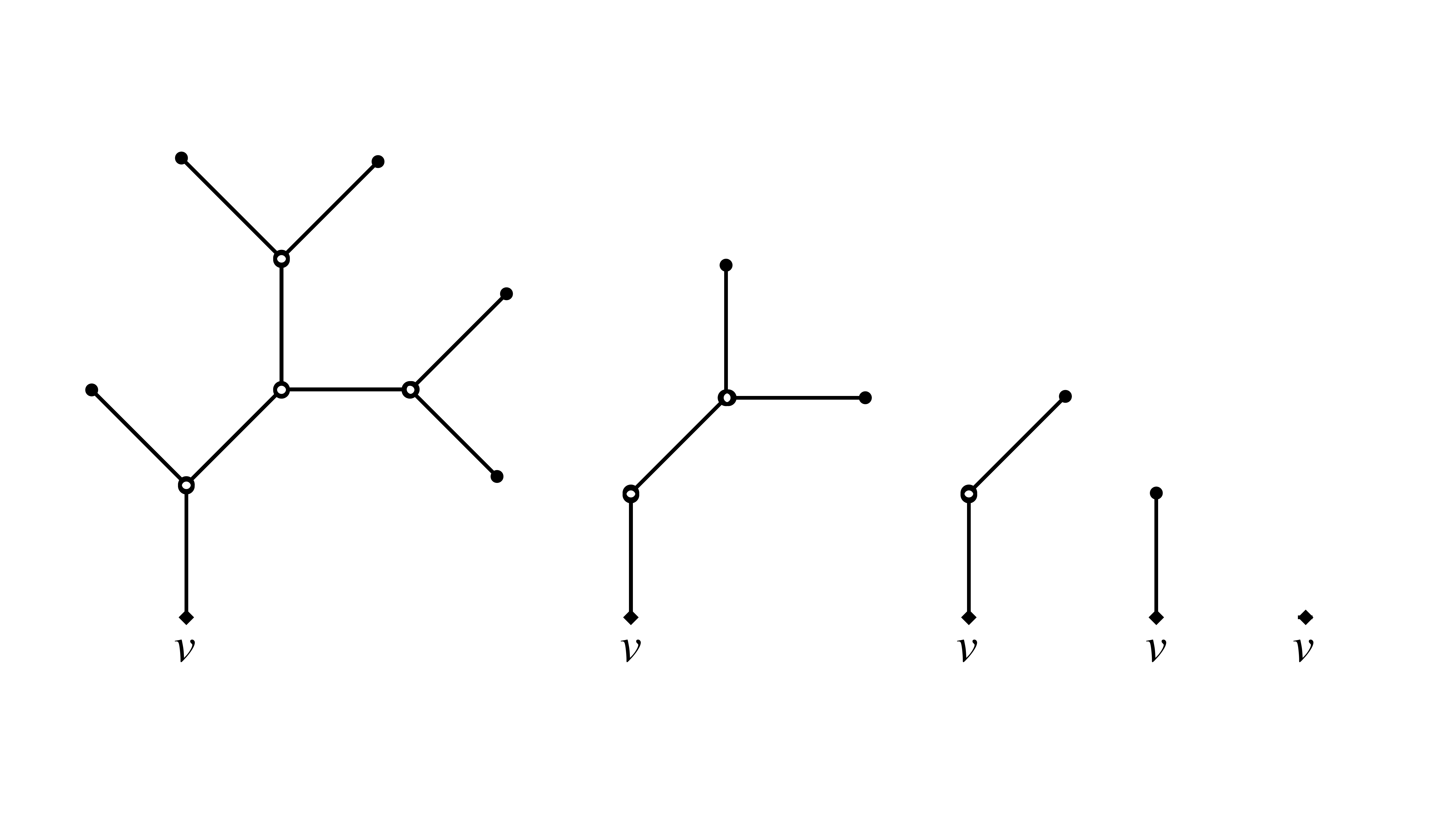}}}
\vspace{-1.2cm}
\caption{{\small  Successive integrations of variables associated with vertices of order 1, except 
the variable associated to the marked vertex $v$ }}
\label{fig3.2}
\end{figure}
  \beq\label{3.6}
  \int_{x(v(1))} \prod_{v \neq v(1)} \d^D x(v)\; \e^{-S[B(x)] }= \left( \int d^D x \; \e^{-\vph(|x|)} \right)^{|V(B)|-1} = c_\vph^{|L(B)|}.
  \eeq
 Thus we can write
 \beq\label{3.7}
 G_\mu^{(1)}(x) \leq  \sum_L \left( c_{bp} c_w^2 c_\vph\right)^L \; \e^{-\mu \, L}.
 \eeq
 We conclude that $G_\mu^{(1)}(x)$ exists for $\mu > \ln ( c_{bp} c_w^2 c_\vph)$ and that the above mentioned 
 $\mu_c \leq \ln ( c_{bp} c_w^2 c_\vph)$. We will later calculate $\mu_c$ more precisely.
 
 \subsection*{Rooted branched polymers and universality}
  
  We now want to show that the one-point function of BPs has 
   a universal critical behavior for $\mu \to \mu_c$, (almost) independent of $\vph(x)$ and 
   the weights $w_m$.  The new aspect compared to the RW is the {\it universality wrt branching }. To simplify the arguments 
   we assume that $c_\vph =1$ and $w_1 =1$ (trivial assumptions), {\it and} we assume in addition that the marked point $v(1)=v_1$
   of the one-point function has $\sg_{v_1} =1$. We call such  one-point function the {\it reduced one-point function $Z(\mu)$} and the 
   corresponding abstract graphs {\it rooted BPs}.  As already remarked $Z(\mu)$ is independent of $x(v_1)$. We write
   \bea\label{3.8}
   Z(\mu) &=& \sum_{B\in \cB_1'} \;\prod_{v \in V(B)}\!\!\! w(\sg_v) 
  \hspace{-1mm}\int\limits_{x(v_1)} \!\!
  \prod_{v \neq v_1 } \hspace{-0.5mm}   d^D x(v) \; \e^{-S[B(x)]-\mu |L (B)|}\\
  &=& \sum_{B\in \cB_1'} \;\prod_{v \in V(B)}\!\!\! w(\sg_v) \;\e^{-\mu \, |L(B)|}.
  \eea
  where $\cB_1'$ denotes the set of the rooted BPs. Each link is assigned a weight $e^{-\mu}$ and we see that $Z(\mu)$ 
  satisfies the graphic equation shown in Fig.\ \ref{fig3.3} which as an algebraic equation reads:
    \begin{figure}[t]
\vspace{-3cm}
\centerline{\scalebox{0.2}{\includegraphics{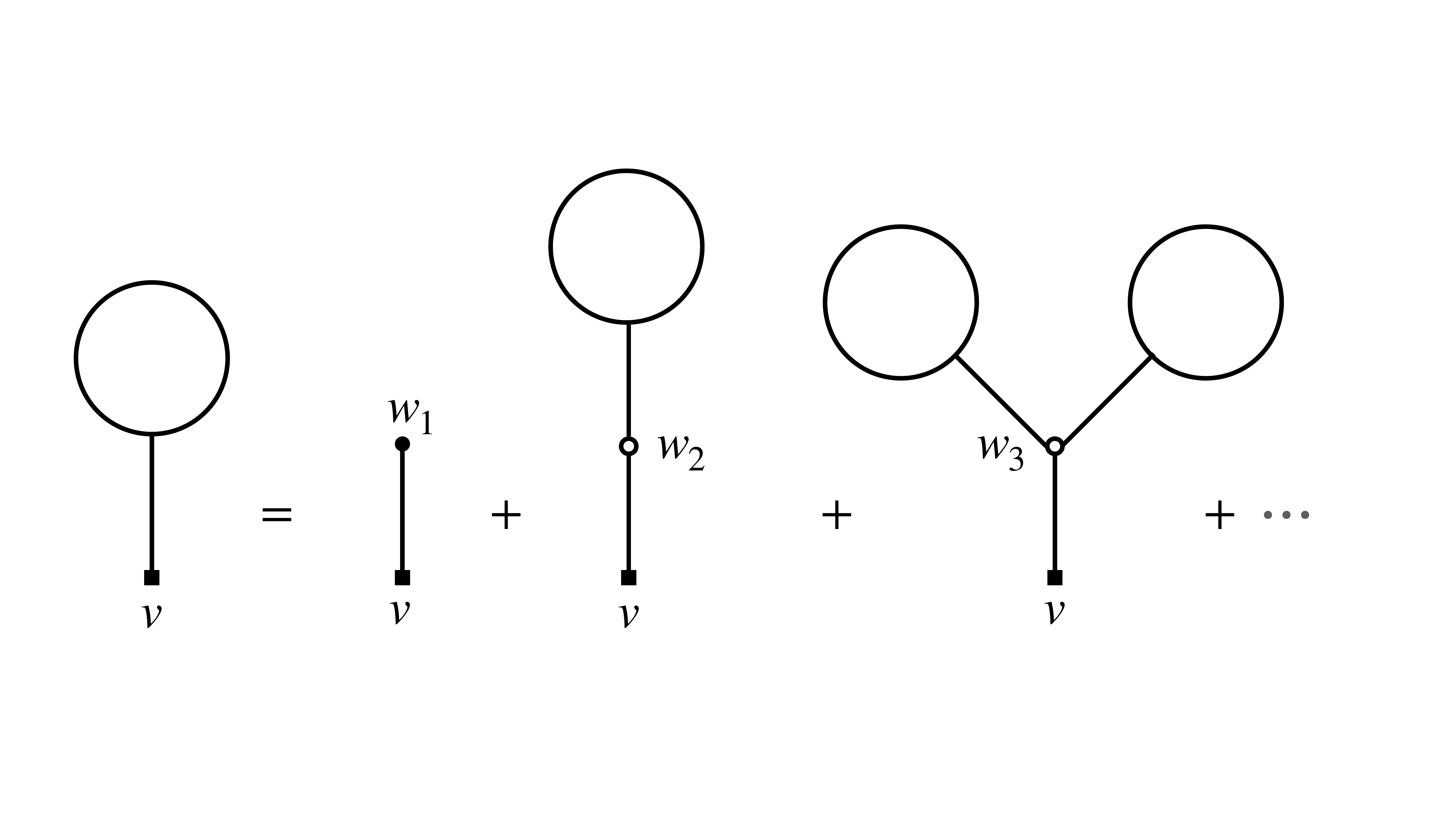}}}
\vspace{-1.7cm}
\caption{{\small  Eq. \rf{3.9} in graphic form, with $w_1=1$. }}
\label{fig3.3}
\end{figure}
  \beq\label{3.9}
  \boxed{Z(\mu) = \e^{-\mu} +   \e^{-\mu} f(Z(\mu)),\qquad f(z) = \sum_{m=2}^\infty w_m z^{m-1}}
  \eeq
  From this we can find $\mu$  as a function of $Z$, shown graphically in Fig.\ \ref{fig3.4}:
  \beq\label{3.10}
  \e^\mu = F(Z), \qquad F(Z) =\frac{1+ f(Z)}{Z},  
  \eeq
  and we can identify the critical point $\mu_c$ as the minimum of the function $F(z)$. To simplify the discussion 
  let us assume that $w_m \geq 0$, $w_m =0$ for $m> m_0$ and that at least one $w_m >0$ for some 
  $m > 2$. Also, we can assume $w_2 =0$ since a $w_2 > 0$ simply adds $w_2$ to $F(Z)$, a constant which will 
  play no role in the arguments to follow.  Thus  $F(z)$ has the shape shown in Fig. \ref{fig3.4} and we obtain:
    \begin{figure}[h]
\centerline{\scalebox{0.18}{\includegraphics{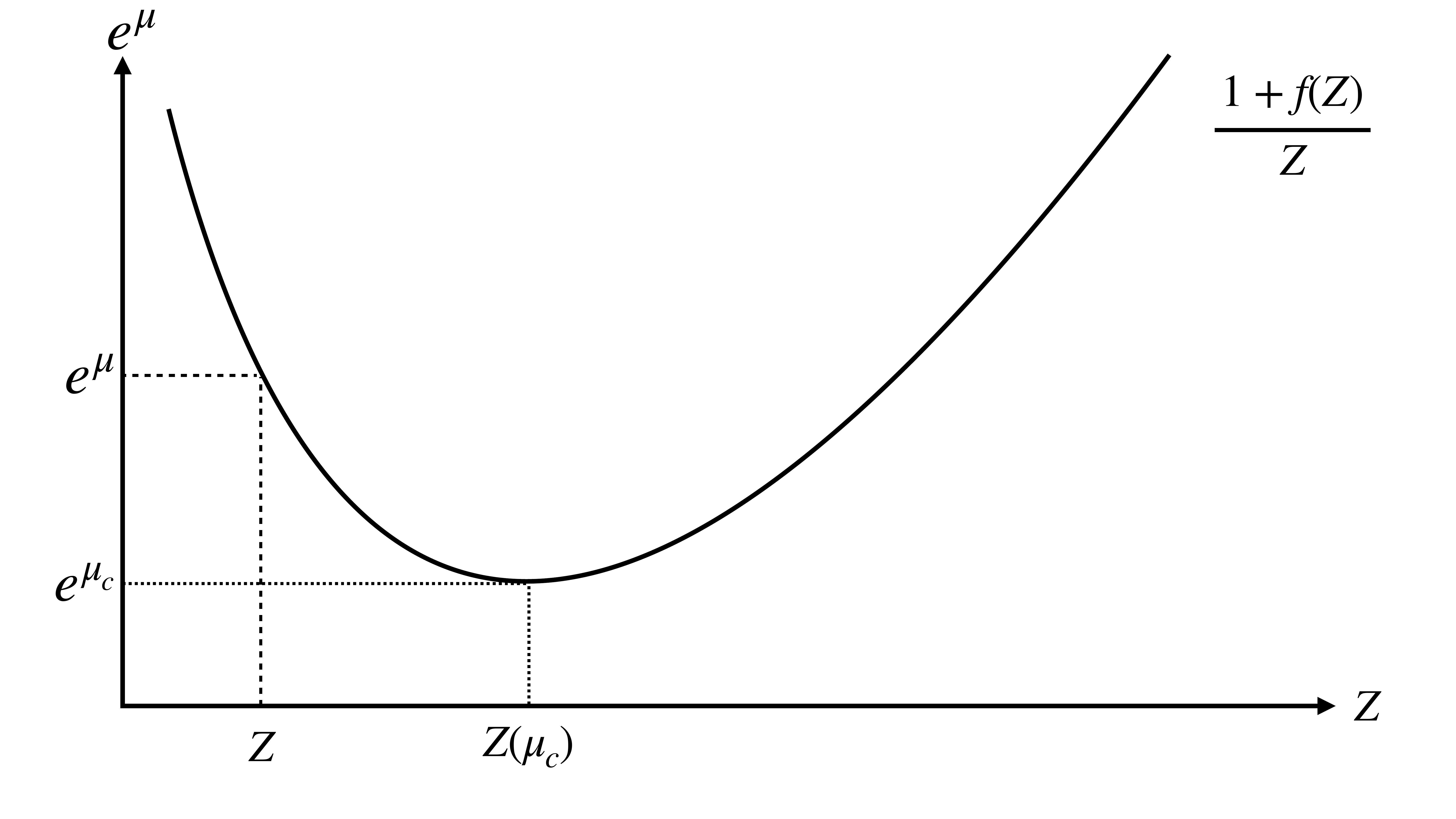}}}
\vspace{-0.5cm}
\caption{{\small  The function $e^\mu = F(Z)$ from eq.\ \rf{3.10}.}}
\label{fig3.4}
\end{figure}
\vspace{-0.4cm}
\beq\label{3.11} 
  \mu \mi \mu_c = c\,(Z(\mu_0) \mi Z(\mu))^2 + \cO\Big( (Z(\mu_0) \mi Z(\mu))^3\Big), 
  \eeq
  or 
  \beq
  \boxed{Z(\mu) \approx  Z(\mu_c) - \tilde{c} \,\sqrt{ \mu \mi \mu_c} = Z(\mu_c) - \tilde{c} \,(\mu\mi \mu_c)^{1 -\gamma}}
  \label{3.12}
  \eeq
  
  We will show below that $\gamma$ can be viewed as the susceptibility exponents for BPs, and we have derived 
  that under the given assumptions about the branching weights $w_m$, 
  {\it the susceptibility  exponent is universal and equal 1/2}. This result is also true if we allow $w_m >0$ for 
  arbitrary large $m$, except 
  in some special situations which we will discuss in  Problem Sets 5 and 7.
  
  \subsection*{The two-point function}
  
  Let us now consider the two-point function, as defined by eq.\ \rf{3.3}. We denote the two marked vertices $v_1$ and $v_2$,
  and the corresponding coordinates $y$ and $x$. For any BP $B \in \cB_2$ there is a unique shortest link-path between 
  $v_1$ and $v_2$. We have indicated that in Fig.\ \ref{3.5}. 
 \begin{figure}[t]
\centerline{\scalebox{0.21}{\includegraphics{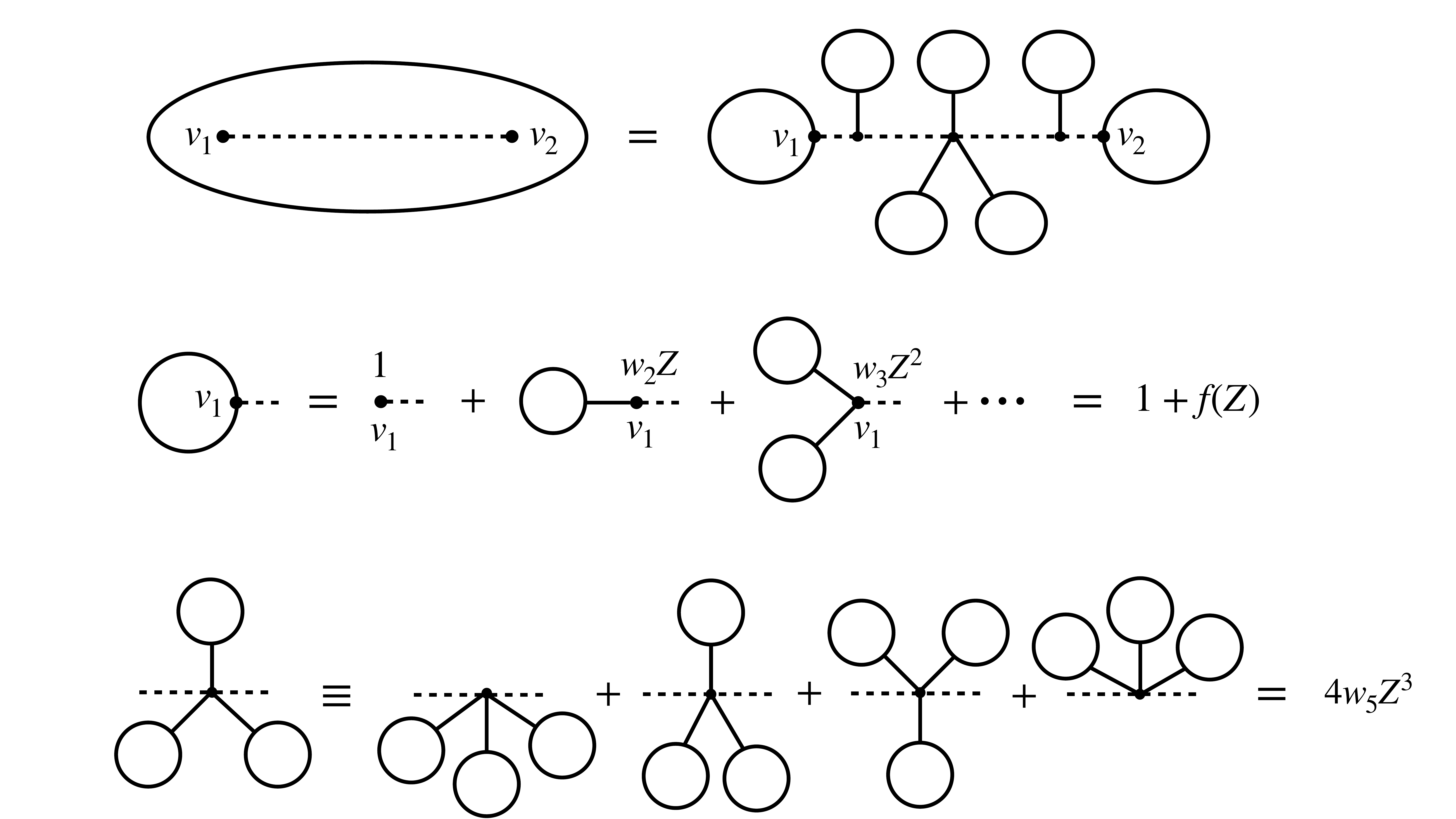}}}
\caption{{\small  The top part shows the decomposition of a graph with two marked vertices $v_1$ and $v_2$
into a shortest link-path (the dashed line) with associated rooted branched polymers, plus the two end graphs, which 
again can be decomposed into rooted branched polymers as shown in the middle graph. Finally the 
bottom graph shows how a contribution from a vertex of order $m$ on the shortest path, really should 
be understood as coming from $m\mi 1$ different terms.  }}
\label{fig3.5}
\end{figure}
 At each vertex along the path, but different from $v_1$ and $v_2$,
  we  can meet a vertex of any order $m \geq 2$, provided the weight $w_m \neq 0$. $m \mi 2$ rooted BPs are then attached 
  to the vertex and they can be arranged in $m-1$ ways as shown in Fig.\ \ref{3.5}. The total contribution from such a vertex will
  thus be:
  \beq\label{3.13}
  \sum_{m=2}^\infty (m \mi 1) \, w_m Z^{m-2}   = f'(Z),
  \eeq
  where $f'$ denotes the derivative of the function $f$.
  If the length of the shortest path is $n$, there will be $n\mi 1$ such contributions. From the $n$ links there will be a contribution
  $e^{-\mu \, n}$. Finally there will be a contribution $(1\plu f(Z))$ from each of the marked vertices $v_1$ and $v_2$ from the
  part of the graph connected to these vertices, but not being part of the shortest path between them, as also 
  illustrated on Fig.\ \ref{3.5}.  Collecting this the two-point function \rf{3.3} can be written as:
  \beq\label{3.14}
  G_\mu^{(2)} (x\mi y) = \big(1\plu f(Z)\big)^2 \sum_{n=1}^\infty \e^{-\mu \, n} \big( f'(Z)\big)^{n-1} \int \prod_{i=1}^{n-1} d^D x_i 
  \; \e^{- \sum_{i=1}^n \vph(|x_i - x_{i-1}|)}
  \eeq
  where $x_0 =y$, $x_n=x$ and $x_i$, $i=1,\ldots,n\mi 1$, denote the coordinates of the $n\mi 1$ vertices of the
  shortest path of length $n$ between the marked vertices $v_1$ and $v_2$. These are the only vertices which 
  we cannot successively integrate over in the way indicated in Fig.\ \ref{fig3.2}. It is seen that the sum in \rf{3.14} is 
  precisely like the sum we encounter in the RW analysis, namely eq.\ \rf{1.34}. Consequently we can write:
  \beq\label{3.15}
 G_\mu^{(2)} (x \mi y) = \frac{\big[1\plu f(Z(\mu))\big]^2}{f'(Z(\mu))} \; G^{(rw)}_{\bmu} (x \mi y),
 \eeq
 where the superscript ``rw'' means the RW propagator and where $\bmu$ is a {\it renormalized} coupling constant:
 \beq\label{3.16}
 \e^{-\bmu} = \e^{-\mu} \, f'(Z(\mu)) = 1\plu \frac{Z(\mu)}{Z'(\mu)}, \qquad\boxed{ \bmu = - \ln \Big( 1\plu \frac{Z(\mu)}{Z'(\mu)}\Big)}
 \eeq
where the expression in terms of $Z'(\mu)$ follows from \rf{3.9} by differentiation wrt $\mu$. From \rf{3.12} it follows that 
when $\mu$ is close to $\mu_c$ we have 
\beq\label{3.17}
\boxed{\bmu  \propto \sqrt{\mu \mi  \mu_c}}
\eeq
 which tells us that the relation between $\mu$ and $\bmu$ is {\it non-analytically} at $\mu_c$.
 
 We know from our RW analysis that $G_\bmu^{(rw)} (x\mi y)$ falls off exponentially with a mass $m(\bmu)$.
 We know that close to the RW critical point $\bmu_c$ we have $m(\bmu) \propto \sqrt{\bmu \mi \bmu_c}$, and we 
 know that $\bmu_c =0$ since we have normalized $\int d^D x e^{-\vph(|x|)} \equ 1$. From 
 \rf{3.15} it is clear that $G_\mu^{(2)} (x\mi y)$ falls off in the same way. Let us call this mass, expressed as a function of 
 $\mu$, for $m_{bp} (\mu)$. We can now write:
 \beq\label{3.18}
 m_{bp}(\mu) = m(\bmu)\;\propto \;\sqrt{\bmu} \;\propto \; (\mu \mi \mu_c)^{\oq} =  (\mu \mi \mu_c)^{\nu_{bp}}.
 \eeq
 
 Similarly , we define the susceptibility of our BPs as 
 \beq\label{3.19}
 \chi_{bp}(\mu) = \int d^Dx \; G_\mu^{(2)} (x \mi y) \;\; \to \;\; \frac{c}{ (\mu\mi \mu_c)^{\gamma}} \quad {\rm for} \quad \mu \to \mu_c.
 \eeq
 Again, from \rf{3.15} we know it will be the same as for the random walk, but the critical behavior will be different 
 because of the non-analytical relation \rf{3.17} between $\bmu$ and $\mu$:
 \beq\label{3.20}
 \chi_{bp}(\mu) = \frac{\big[1\plu f(Z(\mu))\big]^2}{f'(Z(\mu))} \; \chi^{(rw)}(\bmu).
 \eeq
 \beq\label{3.21}
  \chi_{bp}(\mu) \propto \chi^{(rw)}(\bmu) \propto \frac{1}{\bmu}  \propto \frac{1}{\sqrt{\mu -\mu_c}}, \quad {\rm i.e.} \quad 
  \gamma_{bp} = \oh.
  \eeq
  Note that from \rf{3.16} we have for $\mu$ close to $\mu_c$:
  \beq\label{3.22}
  \chi_{bp}(\mu) \;\;\propto \;\;\frac{1}{\bmu} \;\; \propto\;\; -Z'(\mu).
  \eeq
  Thus we have shown that it was justified to use $\gamma$ in formula \rf{3.12}, as promised, and it should be 
  mentioned that this a special case of a more general relation, which we will also use when we discuss string theories.
  We can define a susceptibility function $\chi^{(k)}_{bp}(\mu)$ for the  $k$-point function \rf{3.3} by integrating over 
  all points $x(1),\ldots,x(k)$, except one point. In this way integrals which appear for different $k$ are actually precisely
  the same since we are integrating over all $x$s associated with vertices, except one vertices. The only difference is 
  the way we count the graphs. Let us  consider a graph with $n$ vertices, $n$ very large, where $k$ of them are marked.
  If we want to introduce an additional marked vertex this can essentially be done in $n$ ways (assuming $n \gg k$). Thus 
  there will be $n$ more graphs, but all with the same integral. We can obtain this factor $n$ for each graph by differentiation 
  the $n$-point function from \rf{3.3} wrt $\mi \,\mu$ since the number of links only differs from the number of 
  vertices by 1 for connected tree graphs, and we can thus write 
  \beq\label{3.23}
  \chi^{(k+1)}_{bp} (\mu) \propto - \frac{d}{d\mu} \chi^{(k)}_{bp} (\mu),\quad {\rm i.e.} \quad \gamma_{bp}^{(k+1)} = 
  \gamma_{bp}^{(k)} +1,\quad k \geq 2,
  \eeq
  where we have defined the generalized susceptibility exponent for $G_\mu^{(k)}(x_1,\ldots,x_k)$ in an obvious way.
  The first formula in \rf{3.23}  is {\it almost} relation \rf{3.22} for $k$ =1, but not quite. $Z(\mu)$ is slightly different from 
  $G^{(1)}_\mu (x)$ because in $Z(\mu)$ the marked vertex is of order 1. However, this different does  not really change 
  any critical behavior\footnote{If we assume there are no vertices of order 2, then it is easy to show that 
  $V_1=2 \plu V_3 \plu 2 V_4 \plu 3 V_5 \plu \cdots$, where $V_n$ denotes the number of vertices of order $n$. Thus more
  than half of the vertices are of order one, and when it comes to critical behavior depending on the number of vertices, there will
  be no difference if we consider vertices of order 1 or all vertices.}. 
  
  Finally, the short distance behavior of $G_\mu^{(2)} (x \mi y)$ is of course the same as that of a free particle 
  because of \rf{3.14}:
  \beq\label{3.24}
  G_\mu^{(2)} (x\mi y) \propto \frac{1}{| x \mi y|^{D - 2}}, \qquad |x \mi y| \, m_{bp} (\mu) \ll 1, 
  \eeq
  and the exponent $\eta_{bp} =0$. Summarizing,  the BP critical exponents are 
  \beq\label{3.25}
   \boxed{\nu_{bp} = \oq,\quad \gamma_{bp} = \oh, \quad \eta_{bp} =0, \quad d_H^{(bp)} = 4}
  \eeq
  {\it The Hausdorff dimension of BPs is $d_H^{(bp)} =1/\nu_{bp} = 4$}. A look at the top part of 
  Fig.\ \ref{3.5} makes this result quite natural. In the scaling limit the average number of vertices 
  in each of the rooted branched polymers $Z(\mu)$ diverges as does the number of vertices in the 
  shortest path between the two marked vertices and the divergence of total number of vertices in the BP 
  will be determined by the product of these two numbers. More precisely, when we use the two-point 
  function to derive the Hausdorff dimension we have  
   $|x|^{d_H} \propto 1/(\mu  \mi \mu_c)$ and also that 
  the average number of vertices $\la n \ra_{{\rm bp}} \propto 1/(\mu  \mi \mu_c)$. At the same 
  time the average number of vertices in a rooted BP (derived from $Z(\mu)$) is 
  $\la n \ra_{{\rm rbp}} \propto 1/\sqrt{\mu  \mi \mu_c}$ and 
  the average number of vertices in the shortest path between the two marked points $\la n \ra_{{\rm sp}} \propto 
  1/(\bmu \mi \bmu_c) \propto   1/\sqrt{\mu  \mi \mu_c}$. We can thus write
  \beq\label{3.25a}
  \la n \ra_{{\rm bp}} ~~\propto~~  \frac{1}{\mu  \mi \mu_c} ~~\propto~~ \la n \ra_{{\rm rbp}} \la n \ra_{{\rm sp}}.
  \eeq

  \subsection*{Intrinsic properties of branched polymers}
  
\begin{figure}[t]
\centerline{\scalebox{0.2}{\includegraphics{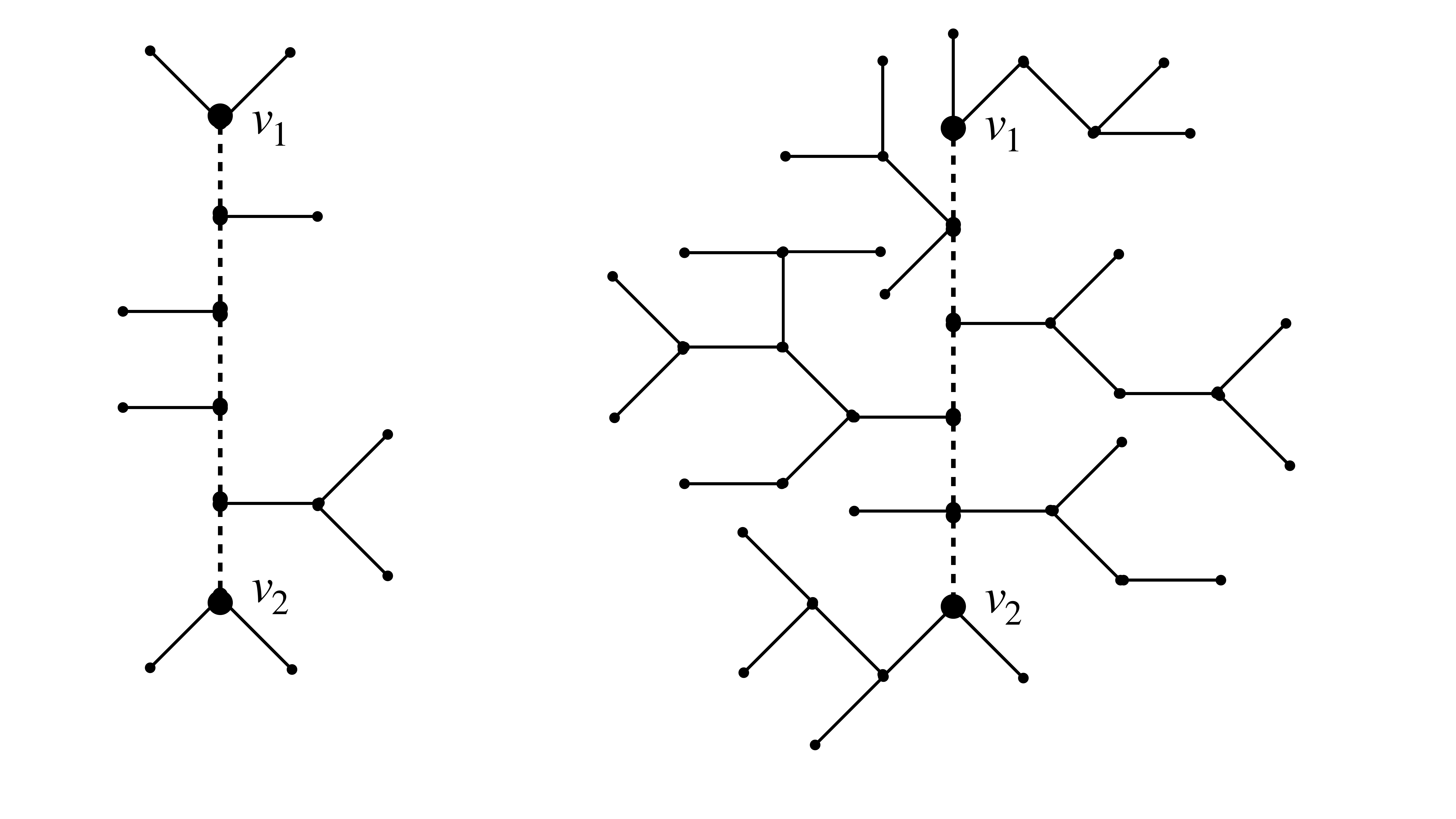}}}
\vspace{-0.5cm}
\caption{{\small  Left:   typical graph where $\mu > \mu_c$ and $r$, the link distance between 
vertices, goes to infinity: we get a linear chain with small outgrowths. Right:  typical graph
when $m(\mu) r$ is constant when $r \to \infty$. In this case $d_H =2 $. Dashed lines show shortest 
paths between vertices $v_1$ and $v_2$.}}
\label{fig3.6}
\end{figure}

  Contrary to RWs, BPs have a non-trivial ``internal life'', independent of the embedding in $\mathbb{R}^D$. 
  We defined the susceptibility $\chi_{bp}(\mu)$ by integration over $x \in \mathbb{R}^D$ as in \rf{3.19}.
  After this integration we obtained
  \beq\label{3.26}
  \chi_{bp}(\mu;v_1,v_2) = \frac{\big[1\plu f(Z(\mu))\big]^2}{f'(Z(\mu))}  \sum_{r=1}^\infty \e^{-\mu r}  f'(Z(\mu))^r,
  \eeq
  where we have explicitly kept the reference to the two marked vertices. 
  We can view this as coming from a partition function for ``abstract'' BPs, where there is no reference to the 
  so-called target space where the $x(v)$ live. In fact, if we define 
  \beq\label{3.27}
 \chi^{(I)}(\mu;v_1,v_2) =   \sum_{B\in \cB_2} \;\prod_{v \in V(B)}\!\!\! w(\sg_v) \; \e^{-\mu \, |L(B)|},
  \eeq
  where $\cB_2$ is the set of BPs with two marked points $v_1,v_2$,
  one obtains precisely \rf{3.26}. Here we have left an explicit reference to the points $v_1$ and $v_2$
  which was left out in \rf{3.19}.  Similarly we would obtain our previously defined $\chi^{(n)}(\mu)$ by defining 
  the equivalent of \rf{3.27} with $\cB_2$ replaced by $\cB_n$ and keeping reference to the marked vertices 
  $v_1,\ldots.v_n$. We now want to introduce the intrinsic {\it link distance} between $v_1$ and $v_2$. Denote
  this distance $r$. Then the decomposition is already given in \rf{3.26}, and \rf{3.26} can be obtained from \rf{3.27}
  by decomposing $\cB_2$ in $\cup_{r=1}^\infty \cB_2 (r)$ where $\cB_2(r)$ denotes the BPs with two marked points 
  separated a link distance $r$. Thus we can write
  \beq\label{3.28}
  G_\mu^{(I)} (r;v_1,v_2) = \sum_{B\in \cB_2(r)} \;\prod_{v \in V(B)}\!\!\! w(\sg_v) \; \e^{-\mu \, |L(B)|}=
  \frac{\big[1\plu f(Z(\mu))\big]^2}{f'(Z(\mu))}  \; \e^{-r \bmu(\mu) },
  \eeq
  \beq\label{3.29}
  \chi^{(I)} (\mu;v_1,v_2) = \sum_{r=1}^\infty  G_\mu^{(I)} (r;v_1,v_2) 
  \eeq
  Since we  know from \rf{3.17} that $\bmu \propto \sqrt{\mu \mi \mu_c}$ it follows that for $\mu \to \mu_c$ 
  we have (suppressing the arguments $v_1,v_2$ in $G$)
  \beq\label{3.30}
  G_\mu^{(I)} (r) = {\rm c} \, \e^{-m_I( \mu) \, r},\quad m_I(\mu) \propto \sqrt{\mu-\mu_c}, 
  \eeq
  Thus 
  \beq\label{3.31}
  \boxed{\nu_I = \oh \quad {\rm i.e.} \quad  d_H^{(I)} = 2}
  \eeq
 From \rf{3.29} we find
  \beq\label{3.31a}
   \chi^{(I)} (\mu) \propto \frac{1}{\sqrt{\mu \mi \mu_c}}, \quad {\rm i.e.} \quad\boxed{ \gamma_I = \oh}
  \eeq
It is instructive to repeat the argument which led to $d_H =1/\nu$ in this new setting. We have our ensemble
of BPs, $\cB_2(r)$ , where two marked vertices are separated a distance $r$, and we ask what is the average volume
(i.e.\ the average number of links) of a graph $B \in \cB_2(r)$. The partition function for these graphs is 
$G_\mu^{(I)} (r) $ and as is clear from \rf{3.28} we obtain the average  value as follows
\beq\label{3.32}
 \la |L(B)|\ra_r = - \frac{1}{ G_\mu^{(I)} (r) } \frac{ \prt G_\mu^{(I)} (r) }{\prt \mu} = m_I'(\mu) \, r  \propto  \frac{r}{\sqrt{\mu\mi \mu_c}}
 \eeq
 Because of \rf{3.30} the formula $m_I'(\mu)\, r$ is actually exact for all $r$, not only valid for large $r$, as \rf{2.35}.
 The formula shows that if $\mu$ is fixed and different from $\mu_c$ the typical BP for large $r$ will just 
 be a linear chain with small outgrowths (see Fig.\ \ref{fig3.6}, left). However we are interested in a limit where 
 $e^{-m_I(\mu) r}$ survives in the limit $r \to \infty$ and $\mu \to \mu_c$, i.e.
 $\sqrt{\mu\mi \mu_c} \; r = {\rm const.}$. In this limit we obtain (see Fig.\ \ref{fig3.6}, right)

 \beq\label{3.33}
 \la |L(B)|\ra_r  \sim r^2\quad {\rm i.e.} \quad d_H^{(I)} = 2.
 \eeq
 Finally we have 
 \beq\label{3.34}
  G_\mu^{(I)} (r) \propto \e^{-m_I(\mu) \,r} \quad \Rightarrow \quad \boxed{ \eta_I =1}
  \eeq
 Recall the notation $G_\mu(x) \sim 1/|x|^{D-2+\eta}$ for $m(\mu) |x| \ll 1$. However, our $r$ in \rf{3.34} should be 
 viewed as the the radial distance from 0 to $x$, i.e. it involves an integration over all point $x$ with $|x| = r$:
 \beq\label{3.35}
 G_\mu(r) \equiv \int d^D x \;\del( |x| -r) \, G_\mu(x) \sim \, \frac{ r^{D-1}}{r^{D-2+\eta}} = r^{1-\eta}, \quad r\, m(\mu) \ll 1.
 \eeq
 Thus \rf{3.34} shows that $\eta_I =1$ and we observe that Fisher's scaling relation is satisfied:
 \beq\label{3.36}
 \boxed{\gamma_I = \nu_I (2 \mi \eta_I)}
 \eeq
 
 \subsection*{Multicritical branched polymers}
 
\begin{figure}[t]
\centerline{\scalebox{0.18}{\includegraphics{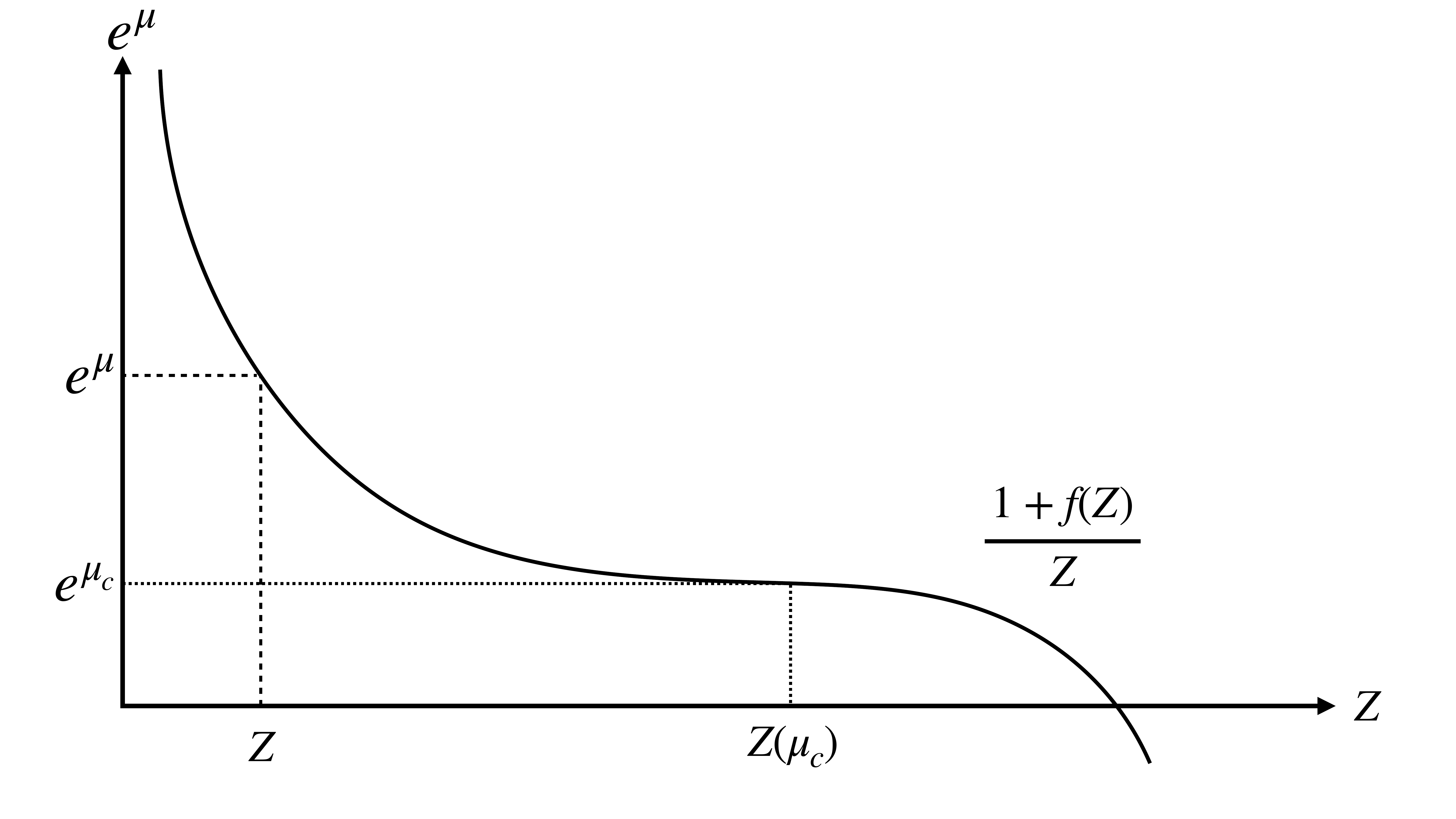}}}
\vspace{-0.5cm}
\caption{{\small  An example of $F(Z)=(1+f(Z))/Z$ where $F'(Z_c) = F''(Z_c) = 0$.}}
\label{fig3.7}
\end{figure}

 We have seen that there is a large universality for BPs: a finite number of positive weights $w_m$, with at least 
 one $w_m$ different from zero for $m >2$, lead to the scaling limit described above. However, by relaxing the requirement 
 that $w_m \geq 0$ we can obtain  different scaling limits. We are thus loosing a strict probabilistic interpretation, but 
 a number of statistical matter systems coupled to BPs will induce such behavior (as we will study in detail in Problem Set 6
 for a specific matter model coupled to BPs) and 
 one will encounter similar situations in two-dimensional gravity systems as we will discuss later. Recall that from a technical 
 point of view the universality came because the function $F(Z)$, defined in \rf{3.9} and \rf{3.10} has a simple minimum, $Z_c$,
 where  $F'(Z_c)  \equ 0$, but $F''(Z_c) > 0$. We can clearly obtain that also $F''(Z_c) \equ 0$ by choosing $w_m$ in a suitable way 
 (see Fig.\ \ref{fig3.7}). As a simple example choose $w_1 \equ 1$, $w_3\equ 1$ and $w_4\equ  -1/12$:
 \beq\label{3.37}
 F(Z) = \frac{ 1\plu  Z^2\mi  \frac{1}{12} Z^3}{Z}, \quad Z_c = \sqrt{2},  \quad F'(Z_c) = F''(Z_c) =0.
 \eeq
This can be generalized to any order $m >2$ by appropriate choices of weights $w_m$:
\bea\label{3.38}
&& F'(Z_c) =\cdots = F^{(m-1)} (Z_c) =0, \quad F^{(m)}(Z_c) \neq 0,  \\
&& \mu - \mu_c \approx c \, (Z_c -Z)^m, \quad \boxed{Z(\mu) \approx Z(\mu_c) -\tilde{c} \, (\mu- \mu_c)^{1/m}, \quad \mu \to \mu_c}
~~~~~~~\nonumber
\eea
We denote a model where this situation is realized an {\it  $m^{th}$-multicritical model} (in Problem Set 5
we will study a number of aspects of multicritical models more closely). 

The graphic equation shown in \rf{fig3.5} is still valid and we obtain:
\beq\label{3.39}
\e^{-\mu} f'(Z(\mu)) = 1 + \frac{Z(\mu)}{Z'(\mu)} \approx 1 - c' \, (\mu-\mu_c)^{\frac{m \mi 1}{m}}\quad {\rm for} \quad \mu \to \mu_c.
\eeq
 As for ordinary BPs we have \rf{3.15}:
 \beq\label{3.40}
 G_\mu^{(2)} (x\mi y) = \frac{\big[1\plu f(Z(\mu))\big]^2}{f'(Z(\mu))} \; G^{(rw)}_{\bmu} (x\mi y),
 \eeq
 where 
 \beq\label{3.41}
 \bmu = - \ln \Big( 1\plu \frac{Z(\mu)}{Z'(\mu)}\Big) \approx c' \, \big(\mu\mi \mu_c\big)^{\frac{m \mi 1}{m}}\quad {\rm for} \quad \mu \to \mu_c.
 \eeq
Thus we obtain 
\beq\label{3.42}
\boxed{m_{bp}(\mu) \propto \sqrt{\bmu} \propto  \big(\mu\mi \mu_c\big)^{\frac{m \mi 1}{2m}}, \quad \nu_{bp} = \frac{m\mi 1}{2m}}
\eeq
and for the intrinsic mass
\beq\label{3.43}
\boxed{ m_I (\mu) \propto \big(\mu\mi \mu_c\big)^{\frac{m \mi 1}{m}},\quad \nu_I = \frac{m\mi 1}{m} = 2\nu_{bp}}
\eeq
 Finally,
 \beq\label{3.44}
 \chi_{bp}(\mu) = \chi_I (\mu)   \propto  \frac{c}{(\mu\mi \mu_c)^{(m\mi 1)/m} }\quad {\rm i.e}\quad 
 \boxed{\gamma_{bp} = \gamma_I = \frac{m \mi 1}{m}}
 \eeq
 Again both set of critical exponents satisfy Fisher's scaling relation since, as for the ordinary BPs, 
 $\boxed{\eta_{bp} =0~{\rm and}~\eta_I =1}$.

\subsection*{ Global and local Hausdorff dimensions}

Until now we have used  \rf{3.32}-\rf{3.33} to define the intrinsic Hausdorff dimension by 
\beq\label{3.45}
\la |L(B)| \ra_r \sim r^{d_H}. 
\eeq
We call this the {\it global Hausdorff dimension} since we can view $r$ as a diameter in the ``B'' universe,
and \rf{3.45} then tell us the  volume of a typical such universe. However, a more geometric definition 
of a Hausdorff dimension is the following: assume our graphs $B\in \cB_1$ have a volume $ |L(B)| \equ L\gg 1$. 
We denote this set of BPs as $\cB_1(L)$. Let $v$ be the marked vertex. 
Let $L_r(B)$ denote the volume of the part of $B$ where the vertices have a link distance less than or equal $r$ to $v$, 
i.e. the volume of a ``ball'' of radius $r$, centered at $v$. We then define the {\it  local Hausdorff dimension $d_h$} by
\beq\label{3.46}
 \la L_r (B)\ra_L \sim r^{d_h} \qquad  1\ll r \ll L^{1/d_h}.
 \eeq 
The average is performed in the ensemble $\cB_1(L)$. 
The idea is that $L^{1/d_h}$ is a typical length scale of a graph of volume $L$ and as long as $r$ is much less than this 
length scale we will have no ``finite size'' effects. We will now show how to use the two-point function to extract $d_h$.
Let is write 
\beq\label{3.47}
G_\mu^{(I)}(r) = \sum_L \e^{-\mu L} G_L^{(I)}(r),
\eeq
 Here $G_L^{(I)}(r)$ denotes the sum over BPs with two marked vertices and volume $L$,  and 
 the marked points separate a link distance $r$. The geometric interpretation of $G^{(I)}_L(r)$ is as follows: we perform 
 the weighted sum over BPs $B$ of volume $L$ and weight $\prod_{v \in V(B)} w_{\sg_v}$.  For each $B$ we mark a vertex 
 $v_1$, then count the number of vertices $v_2$, located a distance $r$ from $v_1$, and finally we sum\footnote{In the 
 definition of the two-point function we are fixing marked vertices $v_1$ and $v_2$, but effectively, for a graph $B$ without 
 the marked vertices we create different triangulations with marked vertices by moving around and marking the 
 vertices as described. There are a few subtleties related to symmetry factors of the graphs, which we will ignore since 
 they are not important for a generic large graph.} over all $v_1$. In this way $G^{(I)}_L(r)$ estimates (up to normalization)
 the average ``area''  $\la S(r)\ra_L$ of a ``spherical'' shell $S(r)$  of radius $r$. 
 For $1 \ll r \ll L^{1/d_h}$ we expect such shells to behave like $r^{d_h-1}$, i.e.
 \beq\label{3.48}
 \boxed{\frac{G^{(I)}_L (r)}{G^{(I)}_L (1)} ~\propto~  \la S(r)\ra_L ~\propto~  r^{d_h-1}, \qquad 1 \ll r \ll L^{1/d_h}}
 \eeq
 Let us use this formula to calculate $d_h$ for (multicritical) BPs. First note that we expect $G^{(I)}_L (r)$ to behave as 
 \beq\label{3.49}
 G^{(I)}_L (r) = \e^{\mu_c L} f(r,L),
 \eeq
 where $f(r,L)$ grows slower than exponential for large $L$. This follows from eq.\ \rf{3.47} since we know that $G_\mu^{(I)}(r)$
 diverges for $\mu < \mu_c$. Close to $\mu_c$ we can write
 \beq\label{3.50}
 G_\mu^{(I)}(r) = \sum_L \e^{- (\mu- \mu_c) L} \, f(r,L) \approx \int dL \;\e^{- (\mu-\mu_c) L} \, f(r,L).
 \eeq
 Using $G_\mu^{(I)}(r) = c \, e^{-m_I(\mu)\, r}$, where $m_I(\mu)$ is given by \rf{3.43}, we find by inverse Laplace transformation:
 \beq\label{3.51}
 f(r,L) \propto \int_{-i\infty+c}^{i\infty +c} d \mu \; \e^{(\mu-\mu_c)L} \e^{-m_I(\mu) r} .
 \eeq
 Expanding $e^{-m_I(\mu)r}$ as $1 \plu (\mu-\mu_c)^{\frac{m\mi 1}{m}} r + \cdots$ we find for small $r$:
 \beq\label{3.52}
 f(r,L) \propto \frac{1}{L^{2-1/m}} \; r
 \eeq
 and thus from \rf{3.48} and \rf{3.49} that 
 \beq\label{3.53}
 \la S(r)\ra_L \propto  \frac{f(r,L)}{f(1,L)} =r \quad \Rightarrow  \quad \boxed{d_h =2 \quad \mbox{for all multicritical models}}
 \eeq 
 We conclude that 
 $$
 \boxed{ \mbox{for ordinary BPs}~ d_H^{(I)} = d_h =2,~~\mbox{but for multicritical BPs}~ 
 d_H^{(I)}  = \frac{m}{m\mi 1} < d_h =2.}
 $$
 
 \newpage
 
 \setcounter{figure}{0}
 \renewcommand{\thefigure}{4.\arabic{figure}}
 \setcounter{equation}{0}
 \renewcommand{\theequation}{4.\arabic{equation}}
 \section*{4. Random surfaces and  bosonic strings}
 
 \subsection*{The action, Green functions and critical exponents}
 
 For the relativistic particle we encountered two actions which were geometric and which were classically (and quantum 
 mechanically) equivalent
 \bea
 S[P(x,y)] &=& m_0\, \ell[P(x,y)]= m_0 \int_0^1 d\xi \,  \sqrt{ \Big(\frac{d X_i}{d \xi}\Big)^2},\quad P: \; \xi \to X_i(\xi),  \label{4.1}\\
S[X,g_{ab}] &=&   \frac{\kp}{2} \int d \xi \,\sqrt{g(\xi)} \, \Big[ g^{ab} (\xi)
 \frac{\prt X_i}{\prt \xi^a} \frac{\prt X_i}{\prt \xi^b} + \lam \Big],  \quad a,b=1 \label{4.2}
 \eea
where $P$ denotes  a path from $y$ to $x$ in $\mathbb{R}^D$. 

\begin{figure}[t]
\centerline{\scalebox{0.15}{\includegraphics{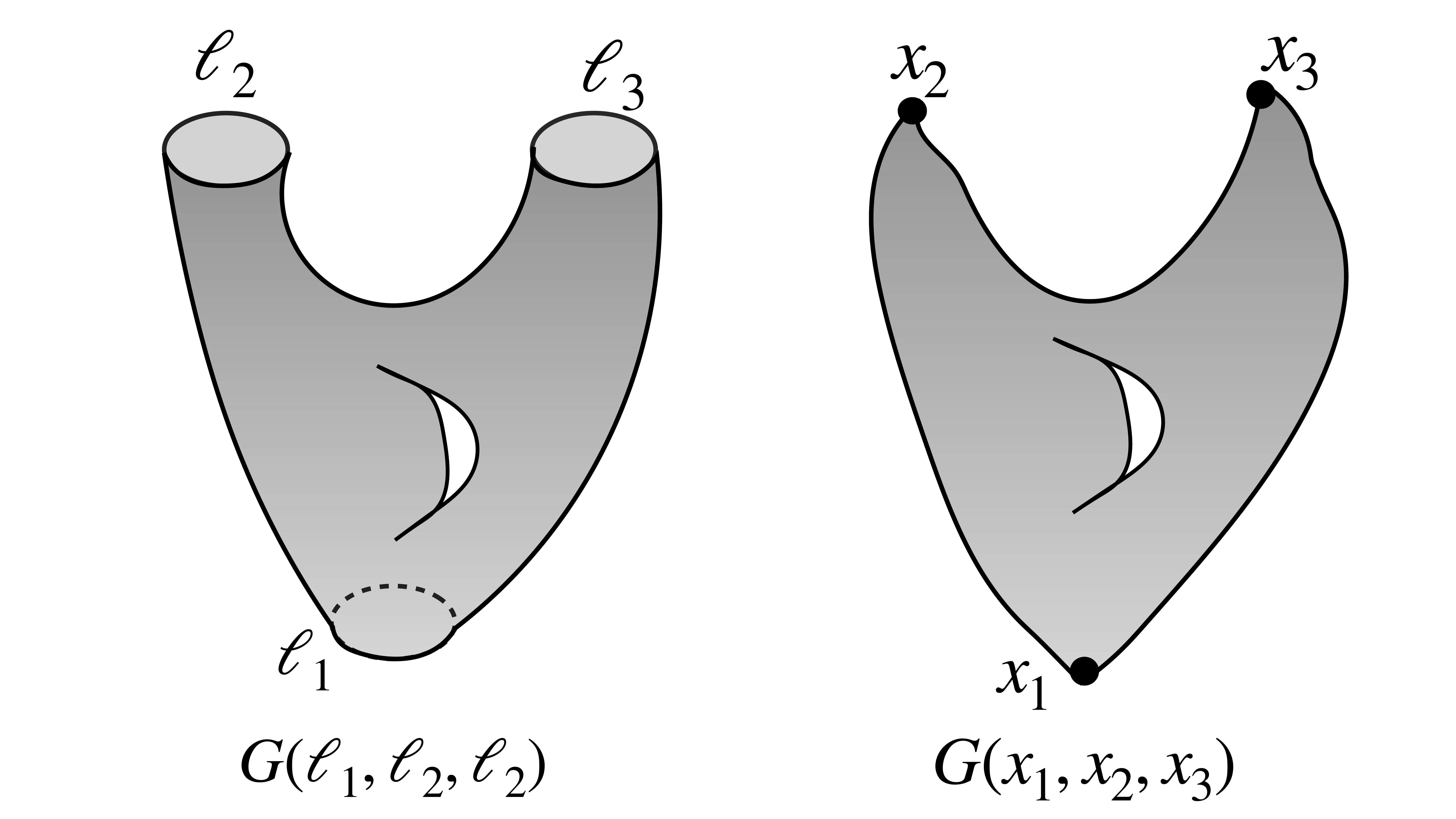}}}
\vspace{-0.2cm}
\caption{{\small  Surfaces with three boundary loops and one handle, or three punctures and one handle, contributing
to $G(\ell_1,\ell_2,\ell_3)$ or $G(x_1,x_2,x_3)$, respectively.}}
\label{fig4.0}
\end{figure}

We now move from one-dimensional geometric objects (paths) to two-dimensional geometric objects (surfaces). We denote
these by  $F$. The two ``boundaries'' of our paths ($y$ to $x$)  were zero-dimensional (points), 
and they are naturally replaced by $n$ one-dimensional boundaries of lengths $\ell_1,\ldots,\ell_n$, and we will 
talk about $n$-loop functions or $n$-loop propagators $G(\ell_1,\ldots,\ell_n)$, 
in the same way as we talked about $n$-point functions 
for BPs. This is illustrated in Fig.\ \ref{fig4.0}. We can contract the loops to points and then we talk about 
$n$-point functions $G(x(1),\ldots,x(n))$ and we say that the surface has $n$ punctures. 
Apart from the boundaries, surfaces also differ from the path by having a non-trivial intrinsic geometry, as we will
discuss. In particular they can be topological distinct and differ (apart from the number of boundaries) by the number
of {\it handles}. In Fig.\ \ref{fig4.0} we have shown a surface with one handle. For reasons which will be clear 
later, we will here mainly consider  surfaces with no handles, i.e.\ surfaces which have the topology of the sphere $S^2$ 
with a number of boundaries.

 Two immediate generalizations of the geometric actions \rf{4.1} and \rf{4.2} for paths to surfaces suggest themselves:
 \bea
 S[F] &=& \kp \, A[F] = \kp\! \int d^2 \xi \; \sqrt{\det h_{ab}}\,, \quad h_{ab}(\xi) = \frac{\prt X_i}{\prt \xi^a}\frac{\prt X_i}{\prt \xi^b},
 \label{4.3}\\
S[X,g_{ab}] &=&   \frac{\kp}{2} \int d^2 \xi \,\sqrt{g(\xi)} \, \Big[ g^{ab} (\xi)
 \frac{\prt X_i}{\prt \xi^a} \frac{\prt X_i}{\prt \xi^b} + \lam \Big],  \quad a,b=1,2 \label{4.4}
 \eea
 where $A[F]$ denotes the area of the surface $F: \xi \to X_i(\xi)$ in $\mathbb{R}^D$. One can write down the
 classical eoms for these actions (treating $X_i$ and $g_{ab}$ as independent variables for $S[X,g_{ab}] $), 
 and they agree for the $X_i(\xi)$ parts, precisely as was the case for the path-actions. From \rf{4.3} it is clear
 that given some boundaries, the minimum of $S[F]$ will be the surface of minimal area connecting these 
 boundaries. Such a situation can be quite singular, as is known from the classical variational theory of minimal 
 area surfaces between boundaries, the standard example being the minimal area surface connecting two circles of radius 
 $r$, the centers separate a distance $R$. When $R$ is sufficiently large compared to $r$ it is clear that the surface with 
 minimal area will consist of the two disks associated with the circles and an infinitely thin tube connecting these disks.
 If we for a moment rotate back to spacetimes with Lorentzian signature we can use the actions
 \rf{4.3} or \rf{4.4} for  {\it relativistic strings}, namely  to describe the time-evolution 
 of a spatial boundary, i.e of a  {\it closed} string. This time-evolution will not necessarily 
 lead to the singular configurations mentioned. However, quantizing the theory we are instructed in the path integral 
 to integrate over all possible surfaces, and as we will see, these singular surfaces will come back and haunt us and
 connect the regularized string theory we consider to BPs.
 
 Thus the actions are the actions for classical relativistic strings and using the path integral we can now define 
 the (Euclidean) $n$-loop (quantum) functions for such strings (and we call it the {\it quantum theory of bosonic strings} 
 or {\it the theory of random surfaces}, this latter notation emphasizing that it is the generalization of RWs to surfaces):
 \bea
 G(\ell_1,\ldots,\ell_n) &=& \int_{\prt F} \cD F \; \e^{-\kp \, A[F]}, \quad \prt F = \{ \ell_1,\ldots,\ell_n\}, \label{4.5}\\
  G(\ell_1,\ldots,\ell_n) &=&  \int \cD [g_{ab}] \int_{\prt F} \cD X \; \e^{-S[X_i,g_{ab}]}
 \label{4.6}
 \eea
 As for the particle the notation $\cD[g_{ab}]$ means that we should only integrate over intrinsic two-dimensional 
 {\it geometries}. Different $g_{ab}$ which just correspond to using different coordinate systems should not be 
 counted as independent. In the following we will use  the one of the two versions \rf{4.5} and  \rf{4.6} that is most 
 convenient for our discussion.

 We will now present some general, formal  arguments, related just to the fact that we have a
 path integral over surfaces. For this purpose it is most convenient to use \rf{4.5}. 
 
 The first thing to note is that interactions between strings seem to be present in the theory without 
 introducing any coupling constants. We can talk about the propagator $G(\ell_1,\ell_2)$ of a string $\ell_1$ to a 
 string $\ell_2$ by summing over all surface in the path integral \rf{4.5} with boundaries $\ell_1$ and $\ell_2$. 
 But without introducing any new coupling constant, $G(\ell_1,\ell_2,\ell_3)$  seems to contain the information
 about a string $\ell_1$ propagating and splitting in two strings $\ell_2$ and $\ell_3$, and $\ell_1$ and $\ell_2$
 joining to $\ell_3$. This is a beautiful aspect of string theory and a feature alien to particle physics.  The situation
 is shown in Fig.\ \ref{fig4.0}
 
 Next, let us consider the two-point function, i.e.\ the two loops $\ell_1$ and $\ell_2$ are contracted to points $x$ and $y$:
 \beq\label{4.7}
 G(x,y) = \int\limits_{\prt F=\{x,y\}}\cD F \; \e^{-\kp \, A[F]},
 \eeq
 We can write
 \beq\label{4.8}
 G(x,y) =\! \int_0^\infty \!\! dA \; \e^{-\kp\, A} \hspace{-5mm}\int\limits_{\prt F=\{x,y\}}\!\!\!\!\cD F \!\; \del(A[F] \mi A) =
  \!  \int_0^\infty \!\!dA \; \e^{-\kp\, A} \; \cN_2(A(x,y))
 \eeq
 where $\cN_2(A(x,y))$ denotes the number of surfaces with area $A$ and two marked points fixed at $x$ and $y$. We thus have 
 the same situation as for the relativistic particle: {\it the propagator is completely determined if we know  the number of surfaces in 
 $\mathbb{R}^D$ with two marked points at $x$ and $y$ and area $A$}. Of course this number is infinite, and we need (as for 
 the particle) to introduce a regularization in order to perform the counting. 
 At the moment we will just assume we have such a regularization.
 The same statement is obviously true if we consider the $n$-point function $G(x(1),\ldots,x(n))$, just with 
 $\cN_n(A(x(1),\ldots,x(n)))$, the number of surfaces with $n$ marked points located at $x(1),\ldots,x(n)$ and area $A$. Note
 also that these surfaces can self-intersect in  $\mathbb{R}^D$. There is nothing in the action which prevents such self-intersection.
 
 We obtain  the {\it susceptibilities} as for RWs and BPs by integrateting the $n$-point functions over $n\mi 1$ of the points:
 \beq\label{4.9}
 \chi^{(n)} (\kp) = \int \prod_{k=1}^{n-1} d x(k) \; G(x(1),\ldots,x(n)) = \int_0^\infty dA \;\e^{-\kp A} \cN_n(A).
 \eeq
 where $\cN_n(A)$ denotes the number of surfaces in $\mathbb{R}^D$ with area $A$ and  $n$ marked points 
 (and one of them kept fixed in order to eliminate translational invariance of $G$). Heuristically we have 
 \beq\label{4.10}
 \cN_n (A) \approx A \, \cN_{n\mi 1} (A),
 \eeq
 for the same reason as discussed for BPs: it is the same integrals over surfaces, the only difference is that 
 one class of surfaces has one more mark than the other, and this mark can put anywhere on the surface, i.e.
 the number of ways this can be done is proportional to $A$. Clearly one needs some kind of regularization to 
 make this into a precise statement, but it should be true for all reasonable regularizations.
 
 The relation between the numbers $\cN_n(A)$ and the so-called susceptibility exponents $\gamma_n$ 
 is the same as we have already encountered for the RWs and BPs. Let us assume we have some regularization
 of our string theory\footnote{\label{lattice} A very simple regularization, like the one mentioned for the particle, is 
 to use a hypercubic lattice. For the particle the paths on the hypercubic lattice would follow the links and the geometric
 action would just be proportional to the number of links. For the string, the surfaces would be made from plaquettes (the sides
 of a minimal lattice hypercube), and again the action would be proportional to the number of the plaquettes constituting 
 the surface. Everything said about counting can be made precise in this setting.} and that the number of surfaces 
 grows exponentially with $A$, up to power like subleading corrections:
 \beq\label{4.11}
 \cN_n (A) \;\propto \;\e^{\kp_c A} A^{\ga_n -1} \Big( 1 + \cO \big(A^{-1}\big)\Big)
 \eeq
 Then
 \beq\label{4.12}
 \chi^{(n)}(\kp) \propto  \int_0^\infty dA \;\e^{-(\kp- \kp_c )A} A^{\ga_n -1} \Big( 1 + \cO \big(A^{-1}\big)\Big) 
\;\;\underset{\kp \to \kp_c}{\to}\;\;
 \frac{c_n}{(\kp \mi \kp_c)^{\ga_n}}\;.
 \eeq
 $\kp_c$ will in general depend on the explicit regularization and in the continuum limit only $\kp -\kp_c$ will 
 survive and correspond to a renormalized $\kp_r= \kp \mi \kp_c$, as we have seen for the particle (and for BPs). Thus the 
 precise exponential growth of the number of surfaces will depend on the regularization, but the subleading 
 power-term relates directly to the continuum limit since this is what determine the divergent power 
 of $\chi^{(n)}(\kp)$ expressed in terms of the renormalized $\kp_r$.  Therefore the subleading power of the number of 
 surfaces should be universal, independent of any (reasonable) regularization, and this turns out to be true.
 
 A trivial consequence of \rf{4.10} is that 
 \beq\label{4.13}
 \ga_n = \ga_{n-1} +1 \qquad   \chi^\sub{n} (\kp) \propto -  \frac{d \chi^\sub{n \mi  1}(\kp)}{d \kp} .
 \eeq
 and if we call $\chi^\sub{2}(\kp)= \chi(\kp)$ (the susceptibility for the two-point function) we can write
 \beq\label{4.14}
\boxed{ \ga \equiv \ga_2, \quad {\rm and} \quad  \chi^\sub{n}(\kp) \;\;\underset{\kp \to \kp_c}{\to}\;\;
  \frac{c_n}{(\kp \mi \kp_c)^{\ga \plu n -2}}}
  \eeq
  
It now follows from simple geometry that  $\boxed{ \ga > 0 \Rightarrow \ga \leq \frac{1}{2}} $ as we will now argue. 
\beq\label{4.15}
 \chi^\sub{n} (\kp)=\int \cD F \; \e^{-\kp A[F]}  \geq \int\limits_{F_1 \cup \cdots \cup F_n} \int \prod_{k=1}^n \cD F_k \;\;\e^{-(A[F_1] +\cdots + A[F_n])} =\chi(\kp)^n
 \eeq
 \beq\label{4.16}
\{F\} \equiv \begin{gathered}\includegraphics[height=2.0cm,valign=c]{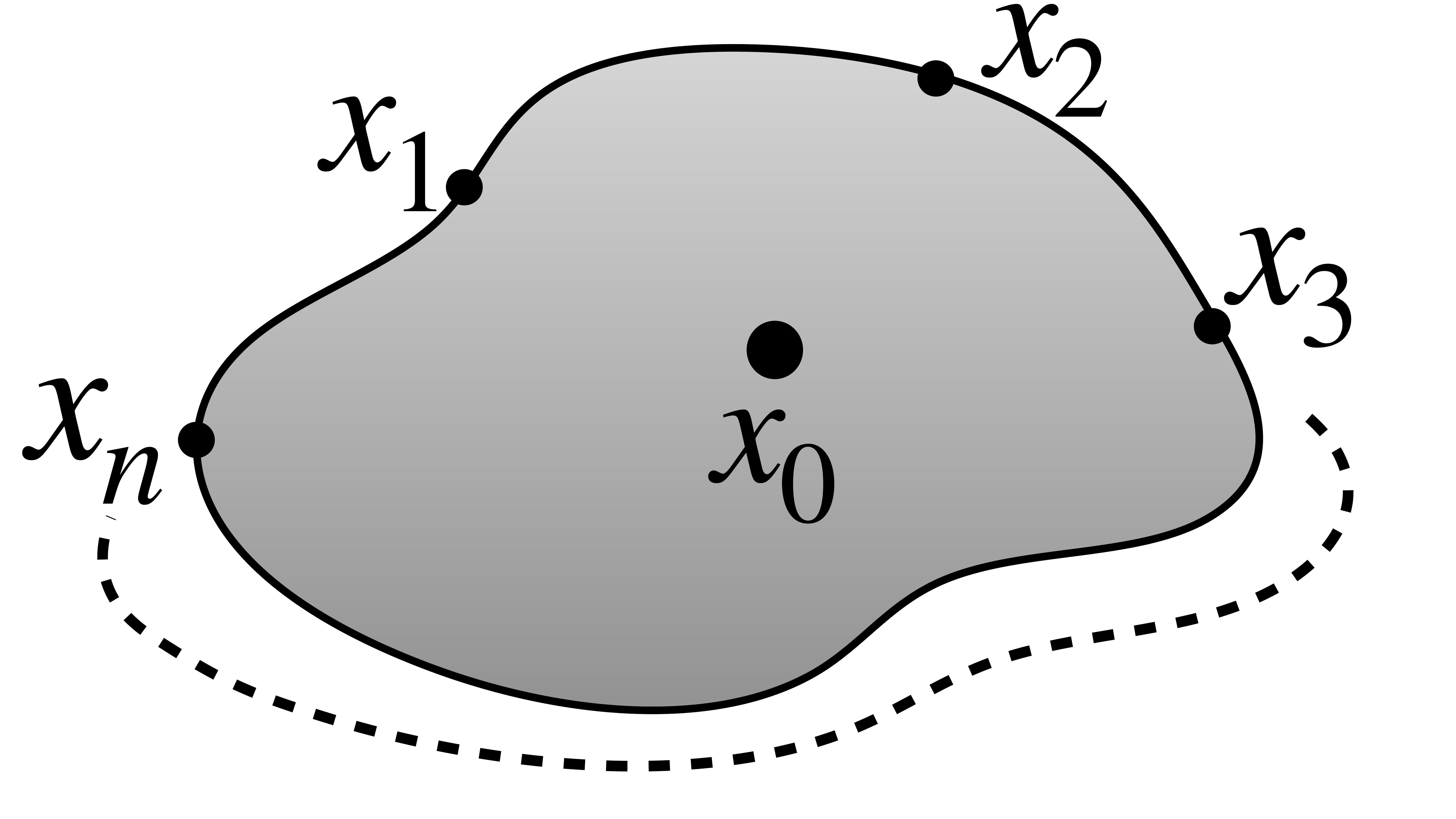}\end{gathered} 
\supseteq
\begin{gathered}\includegraphics[height=2.cm,valign=c]{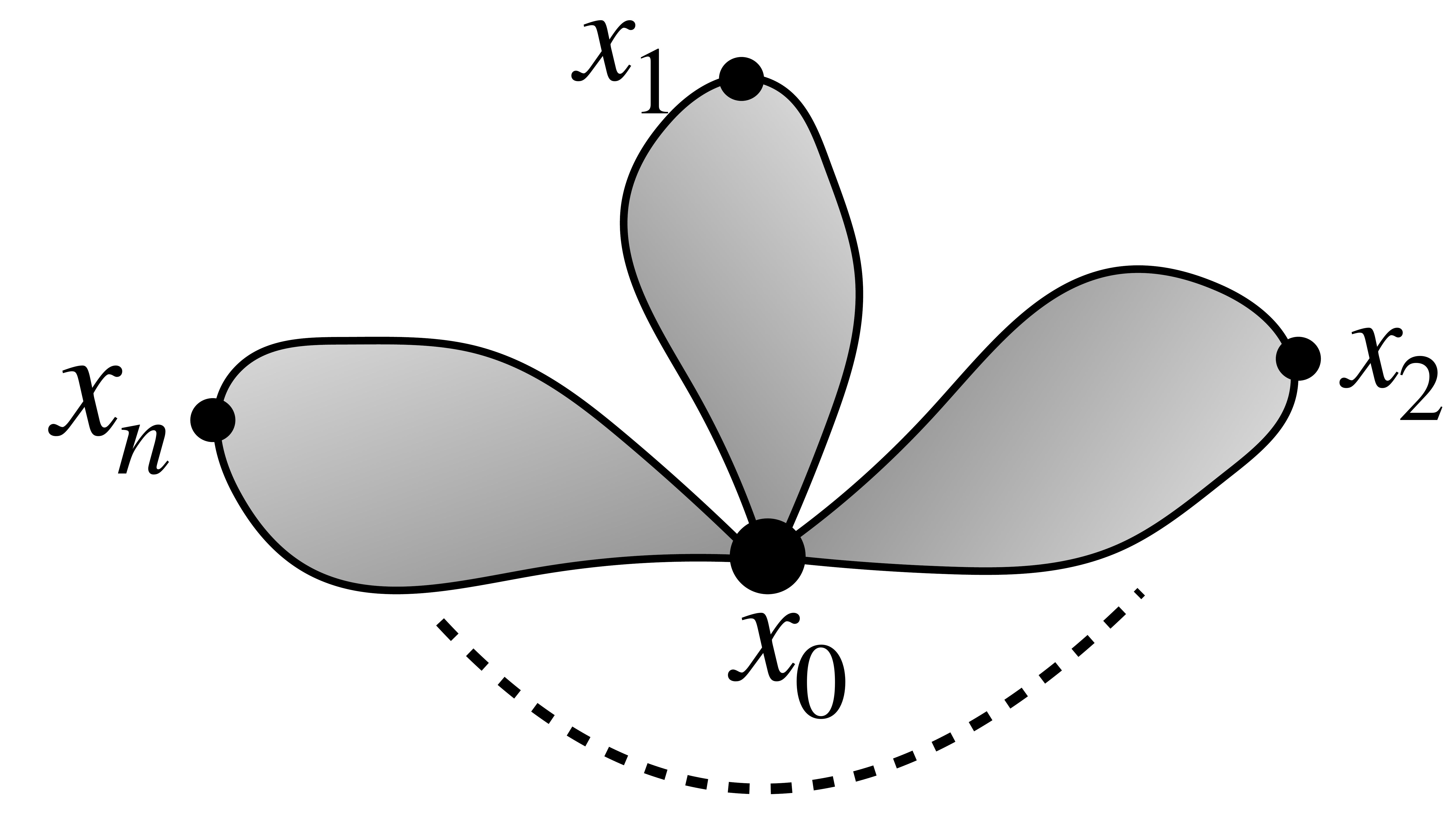}\end{gathered} \equiv  \{F_1\} \cup \cdots \cup \{F_n\} 
 \eeq
The inequality is satisfied simply because there are more surfaces with $n$ marked points $x_i$ than surfaces with 
$n$ marked points of the kind shown on the right hand figure. The $n$ separate surfaces are assumed to join in a 
common ``point'' or little neighborhood around a point $x_0$ which is kept fixed while we integrate over $x_1,\ldots,x_n$,
in this way producing $\chi(\kp)^n $. We have used here the property, special for the geometric action, that 
$A[F_1 \cup \cdots \cup F_n] = A[F_1] +\cdots +A[F_n]$, and also that the decomposition shown to the right in \rf{4.16}
essentially is unique. This is the case of $n>2$, but not for $n=2$, where one cannot define a unique $x_0$ as illustrated here:\\ 
\vspace{-0.7cm}
\beq\label{4.17}
 \begin{gathered}\includegraphics[height=2.5cm,valign=c,rotate = 0]{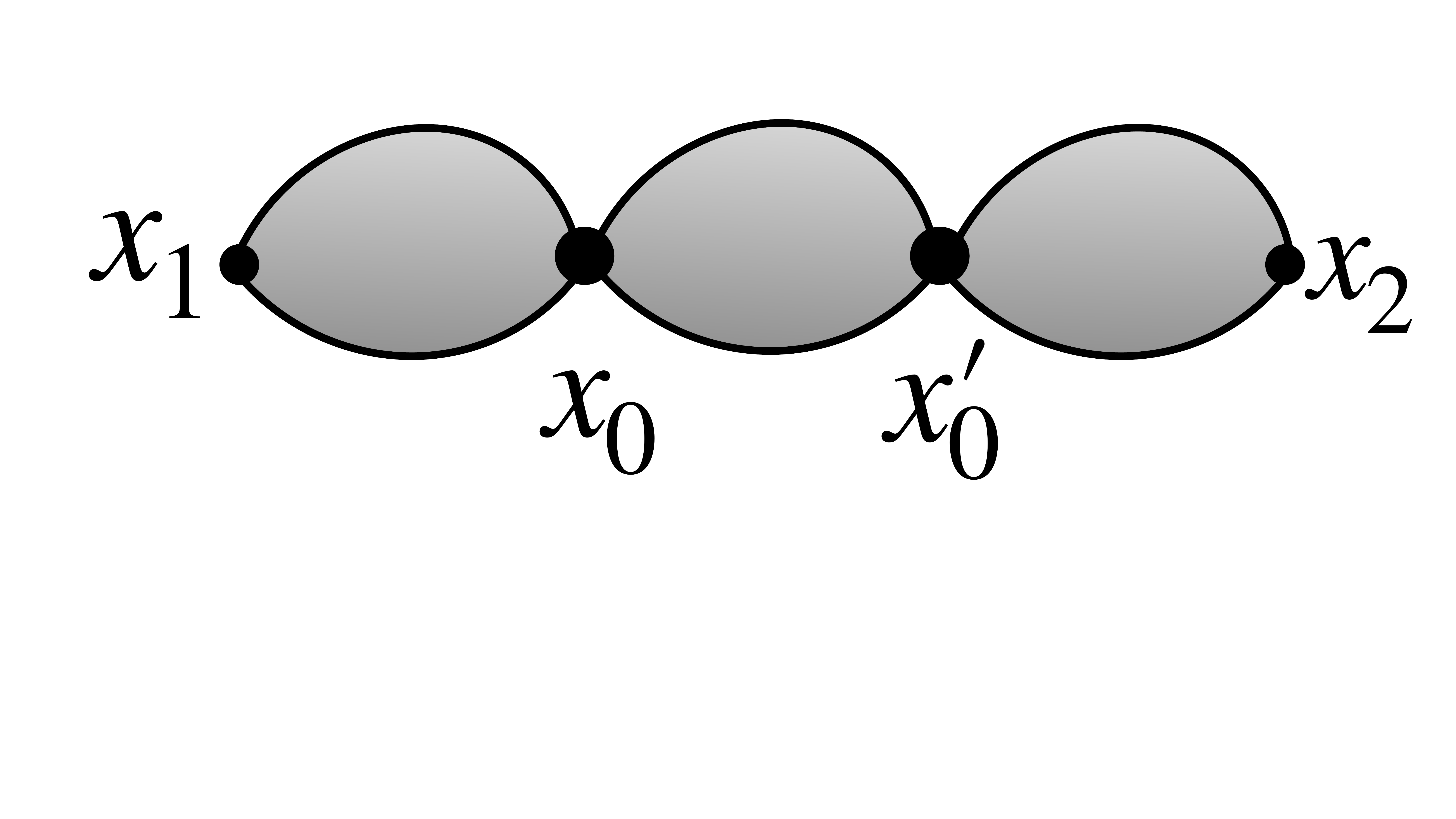}\end{gathered} 
\eeq
\vspace{-1cm}

From \rf{4.15} we conclude 
\beq\label{4.18}
\frac{c_n}{(\kp-\kp_c)^{n-2 +\ga}} \geq \frac{c^n}{(\kp-\kp_c)^{n\ga}}\quad  \Rightarrow\quad  \ga \leq \frac{n\mi 2}{n\mi 1},
\quad n\geq 3,
\eeq
which is the desired result $\ga \leq \oh$. Below we will study the case $n=2$ closer and show that under some universality
assumptions we obtain $\ga = \oh$.

We will now apply the same kind of estimate to the two-point function $G(x\mi y)$ and show that it falls of exponentially.
\beq\label{4.19}
G(x\mi y) = \hspace{-.4cm} \int\limits_{\prt F = \{x,y\}}  \hspace{-0.4cm} \cD F \; \e^{-\kp A[F]}  
\geq   \hspace{-0.4cm}    \int\limits_{\prt F_1 =  \{y,z\}} \hspace{-0.6cm}
\cD F_1 \int\limits_{\prt F_2 = \{z,x\}} \hspace{-0.4cm}  \cD F_2  \;\e^{-(A[F_1]  + A[F_2])}= G(z\mi y) G(x\mi z)
\eeq
\vspace{-0.8cm}
\beq\label{4.20}
\{F\} \equiv \!\!\!\!\!\ \begin{gathered}\includegraphics[height=2.5cm,valign=c]{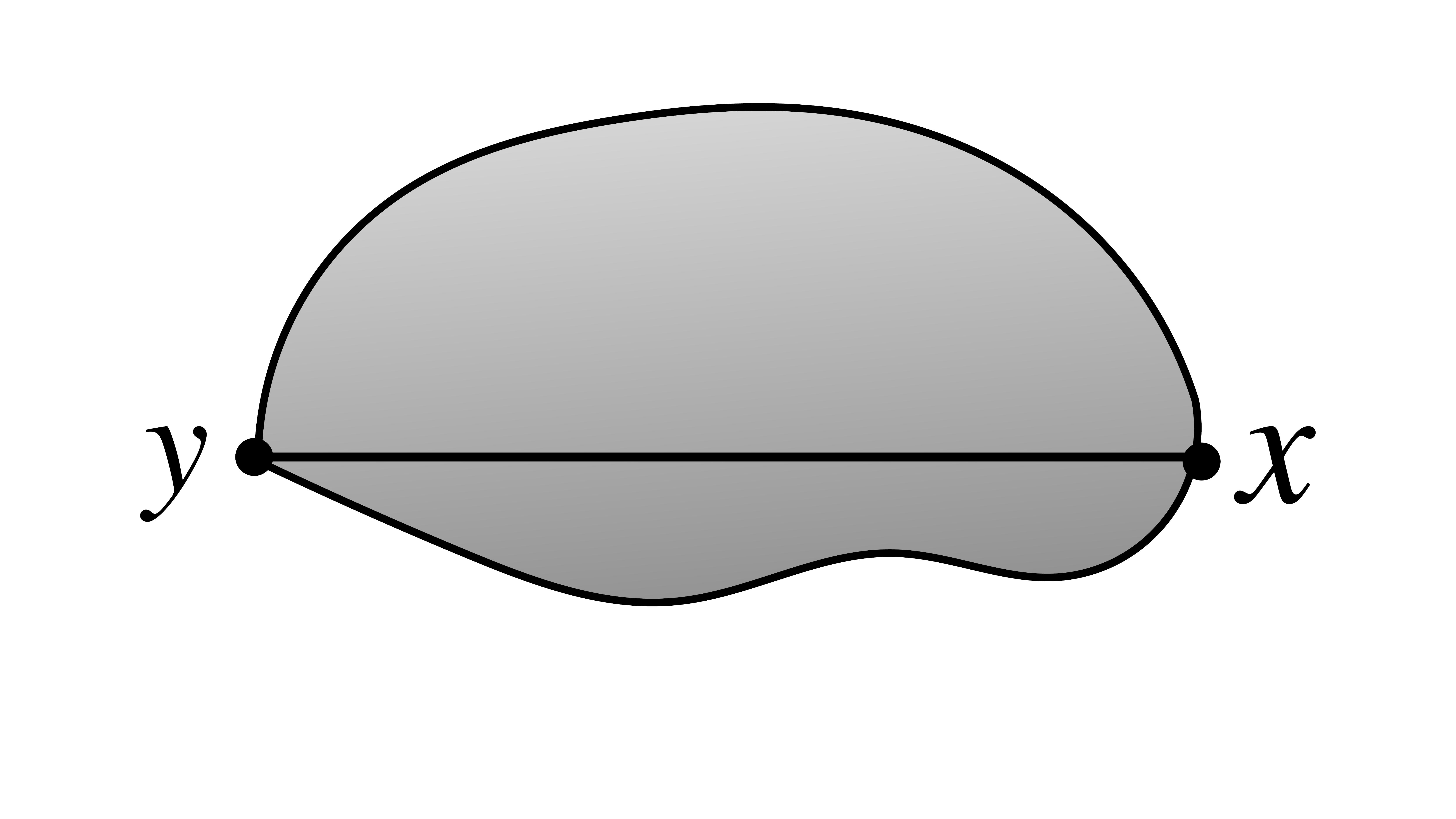}\end{gathered} 
\!\!\!\supseteq
\begin{gathered}\includegraphics[height=2.3cm,valign=c]{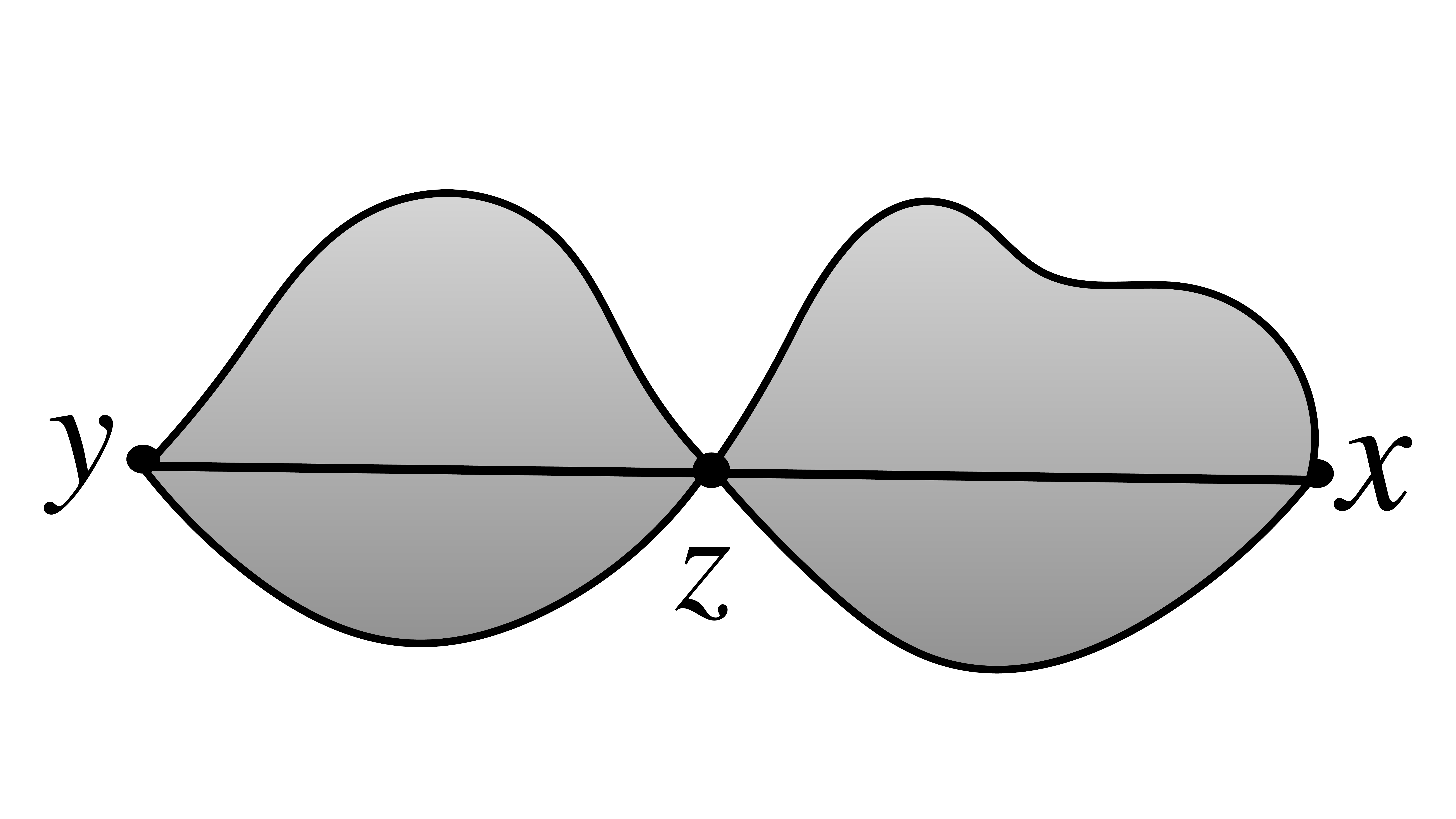}\end{gathered} \equiv \{F_1\} \cup \{F_2\}
\eeq
Again we have the inequality simple because the number of surfaces pinched at $z$ is fewer than the surfaces not pinched.
Since $G(x\mi y) = G(|x\mi y|)$ we can write
\beq\label{4.21}
-\ln G(|x\mi y|) \leq -\ln G(|x-z|) - \ln G(|z\mi y|),\qquad |x\mi y| = |x-z|+|z\mi y|.
\eeq
Eq.\ \rf{4.21} states that $-\ln G(|x\mi y|) $ is a {\it subadditive} function and from general theory (Fekete's lemma) we  
know that for such functions 
\beq\label{4.22}
\boxed{\underset{|x \mi y| \to \infty}{\lim} \frac{\mi \ln G(|x\mi y|)}{ |x\mi y|)} = m(\kp) }
\eeq
If we assume that  $G(x\mi y)  \to 0$ for $|x\mi y| \to \infty$ the mass $m$ has to be non-negative. In addition
$G(x\mi y)$ will be a decreasing function of $\kp$, i.e. $m(\kp_1) \leq m(\kp_1)$ for $\kp_1 < \kp_2$. 
We call $m(\kp)$ the (lowest) mass of the string, and we can write
\beq\label{4.23}
G(x-y) \approx c\;|x\mi y|^\a \, \e^{- m(\kp) |x\mi y|} \quad {\rm for} \quad |x\mi y| \,m(\kp) \gg 1,
\eeq
where the subleading exponent $\a$ in not determined by these general arguments.
It should be noted that we could have applied precisely the same argument in the case of the free particle
to show that the particle propagator, defined by the path integral with the action $S[P] = m_0 \ell[P]$, falls 
off exponentially.

Arguments similar to the ones leading to the existence of the mass $m(\kp)$ also lead to the 
existence of a {\it string tension}. Let us first define the string tension. Consider the one-loop function 
$G(\ell_A)$. Thus the surfaces in the path integral have the boundary $l_A$. In addition  {\it we  assume the surfaces have
no handles, i.e. that all surfaces have the topology of a disk}. We assume the curve defining $\ell_A$ 
is a planar loop in $\mathbb{R}^D$  with area $A$. We now define the string tension $\sg(\kp)$ similarly to the way 
\rf{4.22} defines the mass $m(\kp)$:
\beq\label{4.24}
\boxed{\underset{ A \to \infty}{\lim} \frac{\mi \ln G(\ell_A)}{ A} = \sg(\kp)}
\eeq 
Again we expect from the very nature of the action to have $\sg(\kp_1) \leq \sg(\kp_2)$ for $\kp_1 < \kp_2$.
Why do we call $\sg(\kp)$ the string tension? We can view $G(\ell_A)$ as the partition function for an 
ensemble of fluctuating surfaces (``membranes'', but with very weird properties since they can self-intersect) where the 
boundary is kept fixed. The  Gibb's free energy of these membranes will be $F(A)= - \ln G(\ell_A)$ and the tension of 
the membrane is defined as the change of free energy per unit area when we change the area from $A$ to $A+\Del A$
by changing the boundary:
\beq\label{4.25}
\Del F(A)  = \sg(\kp) \Del A, \quad \mbox{ i.e.\ for large $A$} \quad F(A) \approx \sg(\kp) A ,
\eeq
where we have assumed that the free energy is approximately extensive in the variable $A$ for large $A$.

From the definition of the one-loop function $G(\ell_A)$ with a planar boundary $\ell_A$ enclosing a two-dimensional 
domain of area $A$ in $\mathbb{R}^D$:
\beq\label{4.26}
G(\ell_A) = \int_{\prt F = \ell_A} \cD F \; \e^{-\kp A[F]},
\eeq
we see that there are more surfaces in the set $\ F(\ell_A) \}$ of surfaces with boundary $\ell_A$ than in the two set
of surfaces where we have divided $A$ in subset areas $A_1$ and $A_2$ along some additional boundary in the interior 
of the domain defining $A$. This is illustrate in the case of a rectangle of area $A$ divided into two sub-rectangles of 
area $A_1$ and $A_2$ below:  
\beq\label{4.27}
\{F (\ell_A)\}\equiv \begin{gathered}\includegraphics[height=3.5cm,valign=c]{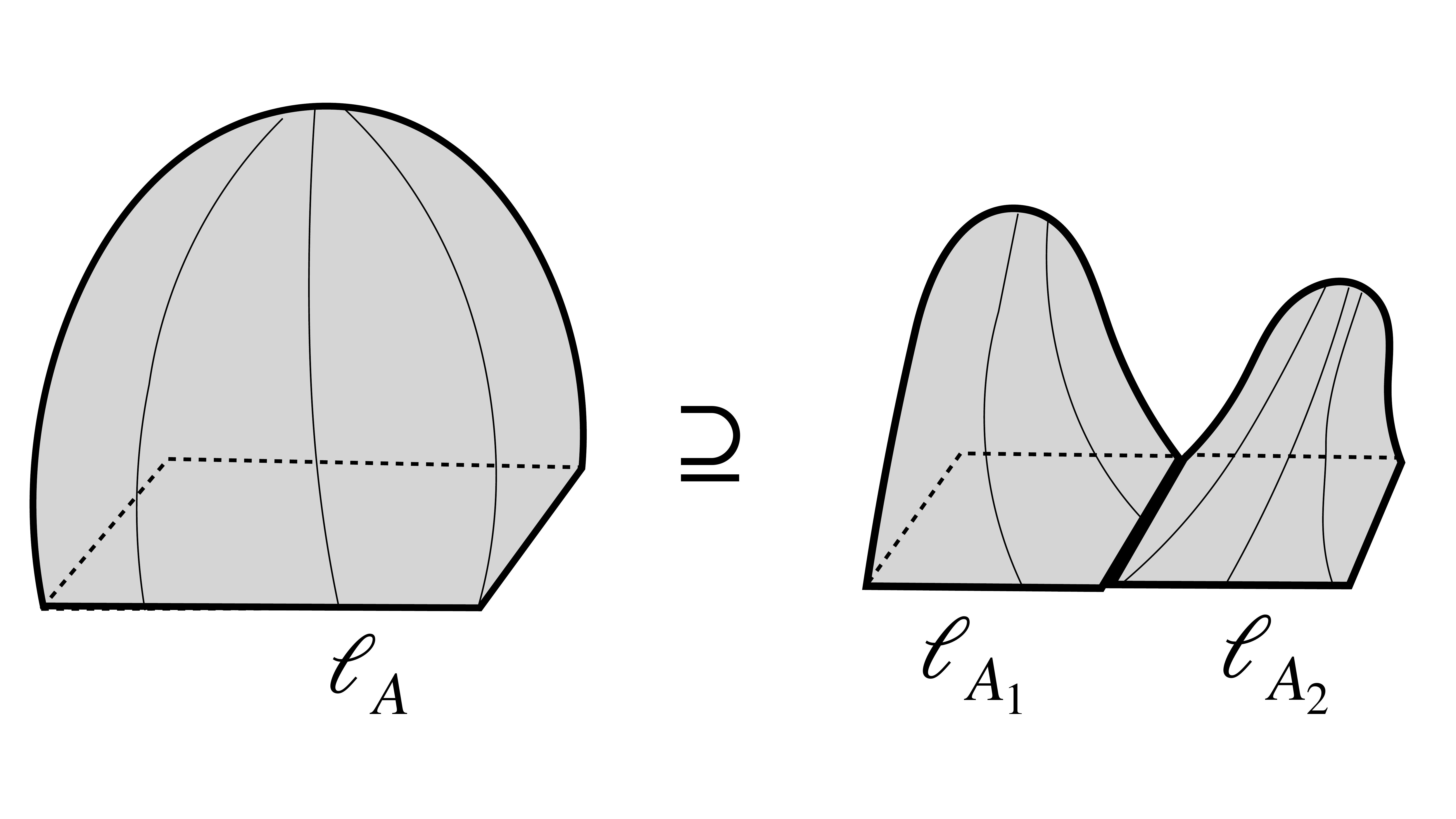}\end{gathered} 
\equiv \{F_1 (\ell_{A_1})\}\cup \{F_2(\ell_{A_2})\}
\eeq
We thus conclude from \rf{4.26}  and \rf{4.27}, in the same way as for the two-point function in \rf{4.19}, that 
\beq\label{4.28}
G(\ell_A) \geq G(\ell_{A_1}) \; G(\ell_{A_2}), \quad A = A_1 +A_2.
\eeq
This leads (again as for the two-point function) to \rf{4.24} and thus 
\beq\label{4.29}
G(\ell_A) \approx A^\a \; e^{-\sg (\kp)\, A} \quad {\rm for } \quad A \, \sg(\kp) \gg 1.
\eeq

Clearly these arguments are very simple and formal, based on counting of surfaces (the number of which is infinite) 
and the simple geometric form of the action. In order to prove them we  need as a starting 
point to {\it define} the path integral over surfaces. As for the particle, a regularization is needed in order 
that we can count. Again there are many ways to introduce such a regularization. If we choose as the action 
$S[F] = \kp A[F]$, \rf{4.3}, a very simple regularization is to use a hypercubic lattice, as already mentioned in footnote \ref{lattice}.
Most of the arguments given above can then be made mathematical rigorous. However, we will here 
use the other geometric action \rf{4.4}, and provide a regularization the path integral using that action.
One reason for this choice is that we can use part of the regularization when we turn to the study of two-dimensional
quantum gravity and so-called ``non-critical strings''. The first thing we have to deal with in that setting is 
how to count two-dimensional geometries $[g_{ab}]$.

\subsection*{\hspace{-1mm} Regularizing the integration over geometries} 

\begin{figure}[t]
\centerline{\scalebox{0.22}{\includegraphics{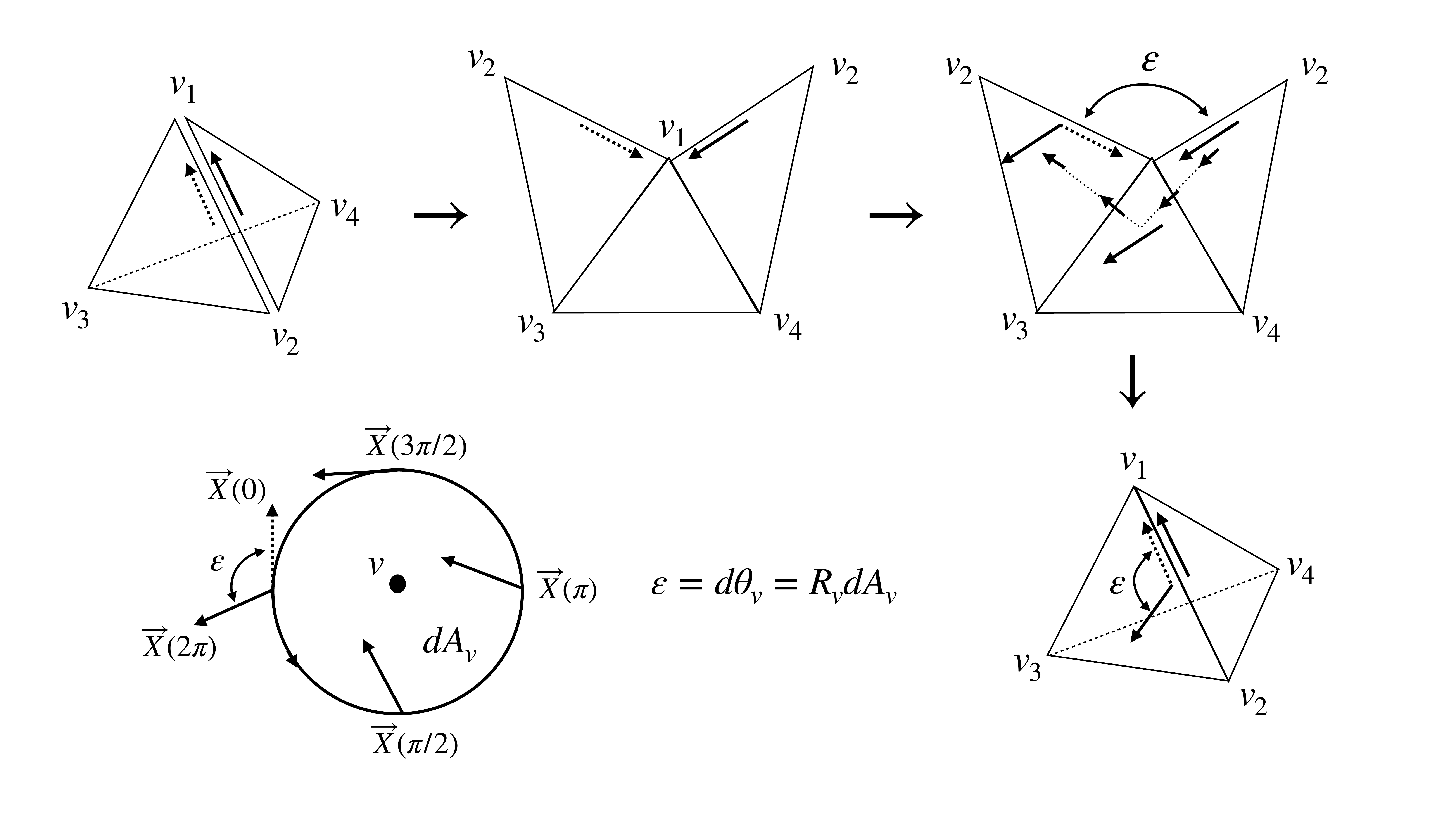}}}
\vspace{-0.5cm}
\caption{{\small  The lower left figure illustrates eq.\ \rf{4.30}. The sequence of four figures illustrates parallel transportation around vertex 
$v_1$ in a triangulation: We cut open the triangulation along link $\la v_1 v_2\ra$ and unfold the triangles 
 neighboring $v_1$ in a plane. The two 
vectors (black and dotted) are parallel since there is no curvature associated with the link   $\la v_1 v_2\ra$. In the plane it is trivial to 
parallel transport the black vector in triangle $\la v_1 v_2 v_4\ra$ to triangle  $\la v_1 v_2 v_3\ra$ 
and compare it to the dotted vector. The 
angle between them is $\ep$, the deficit angle shown on the figure and defined in \rf{4.31}. We can now close the link  $\la v_1 v_2\ra$
which was cut open, and we end with the lower right figure, with the black vector parallel transported around $v_1$.}}
\label{fig4.7}
\end{figure}

\begin{figure}[t]
\vspace{-2cm}
\centerline{\scalebox{0.22}{\includegraphics{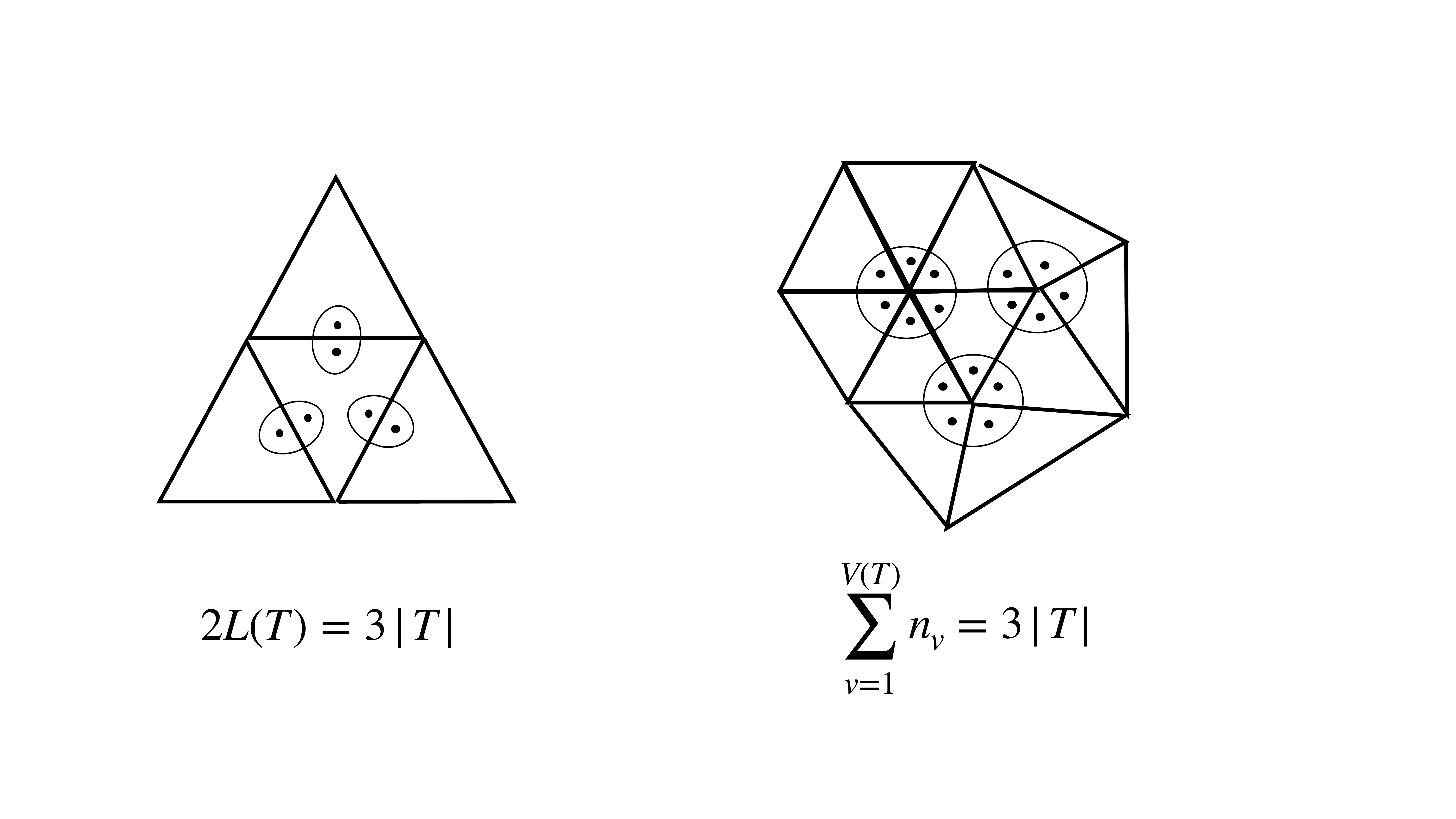}}}
\vspace{-1.2cm}
\caption{{\small  Illustration of eqs.\ \rf{4.33}}}
\label{fig4.8}
\end{figure}

In the case of RWs, piecewise linear paths played an important role. In the case of surfaces it will play an equally important 
role. It will allow us to introduce geometry without having to introduce coordinate systems (and then afterwards have to get rid of 
this freedom by dividing by {\it Vol(diff)}). Stepping one dimension up, the natural replacement of 
a piecewise linear path is a piecewise linear surface, obtained by gluing together triangles.  The lengths of links
are given and each triangle is considered flat in the interior.  In principle we can now calculate the shortest path between
two points on the surface (it will be a certain piecewise linear path on the surface) and thus the intrinsic geometry of the surface
is given. Note that this can be done without ``really'' introducing a coordinate system\footnote{Of course we have to label
the points in the interior of the triangles in some way, but the geometry in the interior of a triangle is defined by the 
length of the links and the statment that the interior is ``flat''.}. Consider the sphere $S^2$ of radius 1 in $\mathbb{R}^3$. 
We know that this sphere has an intrinsic scalar curvature (Gaussian curvature) 1. Now consider a triangulation of the 
kind described above, which approximate the sphere well. One would expect that it is also possible to assign a kind 
of intrinsic curvature to such a triangulation. But where should it be assigned? The interior of the triangles is
declared flat, so one cannot in an intuitive way  assign curvature to an interior point. One property of the intrinsic curvature 
is that it is ``bending invariant'' ( Gauss'  {\it Theorema Egregium}). This makes it unnatural to locate the intrinsic 
curvature on the links, since we can (to some extend) bend the triangulation along the links. We are then left with
the vertices of the triangulation as the place to locate the intrinsic curvature, and this can indeed be done in a 
``natural'' way, as will now be described. Geometrically one can ``detect'' intrinsic curvature by performing a 
parallel transportation along an infinitesimal curve surrounding  a point $v$ on the surface. If the area enclosed by the 
curve is $dA$ and the so-called deficit angle, the angle between the vector before and after being transported around the curve, 
is denoted $d\th$, one has 
\beq\label{4.30}
d\th_v = R_v dA_v + \cO(r^3)
\eeq
where $r$ is a ``typical'' diameter in the domain enclosed by the curve. If $R_v =0$ the surface is locally flat at the point $v$.
This can be understood in a simple way on our piecewise linear surfaces when performing a parallel transportation
around a vertex as illustrated in Fig. \ref{fig4.7}. The deficit angle associated with the parallel transportation around
a vertex $v$ in a triangulation is 
\beq\label{4.31}
\ep_v = 2\pi - \sum_{t  \ni v} \a_t(v), 
\eeq
where the summation is over triangles $t$ which have $v$ as a vertex and $\a_t(v)$ denote the corresponding 
angles in the triangles. Of course this relation is not infinitesimal, like the relation \rf{4.30} for a smooth surface and 
since we have defined the triangulated surface as flat in all other points than the vertices, it is more like 
assigning a $\del$-function-like curvature to the vertices. Writing $dA_v$ in \rf{4.30} as $\sqrt{g(\xi)}d^2\xi$ we can 
integrate the expression \rf{4.30} over the whole surface. Correspondingly we can sum \rf{4.31} over all vertices.
Let us for a moment consider closed surfaces. We would then write 
\beq\label{4.32} 
\mbox{ smooth surfaces:}~~\int d^2 \xi \sqrt{g(\xi)} \; R(\xi)  \sim  \sum_v \ep_v ~~:\mbox{triangulated surfaces}
\eeq 
It is now possible to show that $\sum_v \ep_v$  {\it only depends on the topology of the triangulation}. We will first do
that for a particular class of triangulations which will be of special interest for us,
namely the class of {\it equilateral triangulations}. For such a triangulation $T$ with no boundaries we have 
that $\a_t(v) = \pi/3$ for all triangles and thus 
\beq\label{4.33a}
\sum_v \ep_v = 2\pi \, |V(T)| - \frac{\pi}{3} \sum_v n_v
\eeq
where we have introduced the notation: $T$ denotes a triangulation and at the same  time the set of triangles in the 
triangulation. The number of triangles is denoted $|T|$. $V(T)$ denotes the set of vertices in $T$ and $|V(T)|$ the 
number of vertices. $L(T)$ denotes the set of links in the triangulation and $|L(T)|$ the number of links in the 
triangulation. Finally $n_v$ denotes the order of the vertex in the triangulation, which we here define as the 
number of triangles to which the vertex belongs. From  Fig.\ \ref{fig4.8} we see that 
\beq\label{4.33}
2 |L(T)| = 3 |T| ,\qquad \sum_{v \in V(T)} n_v = 3 |T|
\eeq
and therefore
\beq\label{4.34}
 |V(T)| \mi \frac{1}{6} \sum_v n_v=  |V(T)| \mi \oh |T| =  |V(T)| \mi  |L(T)| \plu |T| \equiv \chi(T)
 \eeq
where $\chi(T)$ is the so-called {\it Euler characteristic} of the triangulation. Also, dropping the assumption that the 
triangles are equilateral, using eq.\ \rf{4.33} it follows immediately that $\sum_v \ep_v$ is still the same simply by using 
that the sum of angles in a (flat) triangle is $\pi$. Thus  
\beq\label{4,34a}
 \sum_v \ep_v = 2\pi |V(T)| \mi  \pi |T| = 2\pi (|V(T)\mi |L(T)| \plu  |T|)= 2\pi \chi(T).
 \eeq
 In general, if a surface $S$ is covered by a set
of polygons then one has 
\beq\label{4.35}
P -L + V \equiv \chi(S) = 2 - 2h -n,
\eeq
where $P$ is the number of polygons, $L$ the number of links and $V$ the number of vertices, $h$ the number of handles
of the surface and $n$ the number of boundaries. Thus the Euler characteristic depends only on the topology and since the 
topology of a surface is characterized completely by the number of handles and number of boundaries, the Euler characteristics
of a surface determines , for a fixed number of boundaries,  the topology of the surface. From the discussion above it is not surprising that we  have
equality in \rf{4.32}, i.e. 
\beq\label{4.36}
\int_\cM d^2 \xi \sqrt{g(\xi)} \; R(\xi)  = 2 \pi \chi(\cM)=  \sum_{v \in V(T)}\ep_v,\quad {\rm (Gauss-Bonnet~theorem)}
\eeq
where $\cM$ is a manifold and $T$ a triangulation with the same topology. The theorem can be extended to surfaces
with boundaries:
\beq\label{4.37}
\int_\cM d^2 \xi \sqrt{g(\xi)} \; R(\xi) + \int_{\prt \cM} ds\; k_g  = \chi(\cM) =  \sum_{v \in  V_I(T)} \ep_v + \sum_{v\in  V_B(T)} \ep_v,
\eeq
where $k_g$ denotes the geodesic curvature of the boundary curve and and $ds$ the line element along the curve. 
$V_I(T)$ denotes the interior vertices while
$V_B(T)$ denotes the vertices at the boundary of the triangulation $T$. 
Finally $\ep_v$ for a boundary vertex $v$ is changed from \rf{4.31} to 
\beq\label{4.38}
\ep_v = \pi - \sum_{t \ni v} \a_t, \quad v \in V_B(T).
\eeq
It is seen from a figure similar to Fig.\ \ref{fig4.7}  that for each boundary vertex $v$, $\ep_v$ in \rf{4.38} is 
just the rotation of the tangent vector
moving around the vertex, when the boundary triangles are put down in a plane. This is of course also the interpretation of the 
infinitesimal  term $k_g\,ds $ on $\cM$.

According to \rf{4.32} the term in the Einstein-Hilbert action for gravity with contains $R$ does not change in 
two dimensions as long as we do not change the topology. This is the reason we could leave it out of a 
two-dimensional theory of gravity as long as we do not consider changing topologies, as already mentioned.

Let us  now consider  manifolds with $h$ handles and $n$ boundaries.  Formally we can write
\beq\label{4.39}
\int_{\cG} \cD [g_{ab}] \;\big( \,\cdot \,\big) = \int_0^\infty dA_{int}  
\int  \cD [g_{ab}] \;\del\Big( \int d^2\xi \, \sqrt{g}     \mi  A_{int} \Big) \; \;\big( \,\cdot \,\big)
\eeq
We denote the space of geometries with the given topology $\cG$ and the subspace with a fixed area $A_{int}$ by 
$\cG_{A_{int}}$, and
the delta-function in the last integral ensures that this integration is over $\cG_{A_{int}}$. We use here the notation $A_{int}$, where 
``int'' is an abbreviation of ``internal'', to signify that the area refers to metric $g_{ab}$, and not to the 
area of the surface measured by the metric induced from the  embedding  in $\mathbb{R}^D$.     
Consider now the space of equilateral triangulations which have the topology of $S^2$ with $n$ boundaries and
where the length of the links is $\ep$. The area of such a triangle 
is $\frac{\sqrt{3}}{4} \;\ep^2$. A triangulation of this kind will belong to $\cG_{A_{int}}$ 
if the number of triangles in such a triangulation satisfies
\beq\label{4.40}
|T| \Big( \frac{\sqrt{3}}{4} \cdot \ep^2\Big) = A_{int}.
\eeq
Denote this set of equilateral triangulations $\cT(A_{int},\ep)$. Clearly the number of triangles for a triangulation in this set 
will go to infinity when $\ep \to 0$.
The main conjecture (which can be proven, but we will not do that here) is that this set of triangulation is sufficiently dense
in the set $\cG_{A_{int}}$ that we can write
\beq\label{4.41}
\sum_{\cT(A_{int},\ep)}  \big(\,\cdot \,\big)   \to \int_{\cG_{A_{int}}} \cD [g_{ab}]  \;\big(\,\cdot \,\big)  \quad {\rm for} \quad \ep \to 0,
\eeq
and integrating  in addition over the area $A_{int}$, and denoting the corresponding set of equilateral triangulations
$\cT(\ep)$, we then formally write
\beq\label{4.42}
\sum_{\cT(\ep)} \big(\,\cdot \,\big)   \to \int_{\cG} \cD [g_{ab}]  \;\big(\,\cdot \,\big)  \quad {\rm for} \quad \ep \to 0.
\eeq
Note that the set $\cT(\ep)$ is {\it independent} of $\ep$, viewed as an abstract set of triangulations. $\ep$ will 
only enter in the implementation of $(\cdot )$ which is some function which is defined on a triangulation and 
may refer explicitly to the length of the links.

One function that has to be included in $(\cdot)$ in \rf{4.42} is the exponential of the action itself. Let us now give a natural definition
of this action on a triangulation $T$. Let us first map this triangulation to a triangulated, piecewise linear surface 
in $\mathbb{R}^D$ by mapping the vertices $v \in V(T)$ to points $X_i(v) \in \mathbb{R}^D$, and defined the corresponding 
surface by declaring the straight line in $ \mathbb{R}^D$ from $X_i(v_1) $ to $X_i(v_2)$ a link if $\la v_1 v_2\ra \in L(T)$,
 and similarly   
$X_i(v_1),X_i(v_2),X_i(v_3)$  for a triangle  in $ \mathbb{R}^D$ if $v_1,v_2,v_2$ defines a triangle in $T$. 
On this piecewise linear surface in $ \mathbb{R}^D$ 
we can now define a coordinate system $\xi$ such that $X_i(\xi)$ are the coordinates on the surface, and we then find 
for a closed surface
\beq\label{4.43}
 \int d^2 \xi \,\sqrt{g(\xi)} \, \Big[ g^{ab} (\xi)
 \frac{\prt X_i}{\prt \xi^a} \frac{\prt X_i}{\prt \xi^b}  \Big]  =
 C \!\!\!\!\!\sum_{\la v v'\ra \in L(T)}  \big(X_i(v)-X_i(v')\big)^2,~~C=\frac{1}{\sqrt{3}}.
 \eeq
 Note that if there are no boundaries this equation can also be written 
 \beq\label{4.43a}
 \int d^2 \xi \,\sqrt{g(\xi)} \, X_i(\xi) \big( \mi \Del_g \big)  X_i(\xi) \Big]  =
   C\!\!\!\!\sum_{ v ,v' \in V(T)}X_i(v)\big(\mi \Del_{vv'}\big) X_i(v'),
 \eeq
 where $\Del_g$ denotes the {\it Laplace-Beltrami operator} on $S^2$ with metric $g_{ab}(\xi)$ and 
 $\Del_{vv'}$ denotes the  {\it combinatorial Laplacian} on triangulation $T$. The combinatorial 
 Laplacian is defined as a $|V(T)|\times |V(T)|$ matrix  where the entries in the diagonal are $-n_v$, 
 the order of the vertex $v$, and the $vv'$ entry is 1 if $v$ and $v'$ are neighbors (i.e.\ belong to the same link),
 and zero otherwise. The Laplace-Betrami operator is defined as
 \beq\label{4.43b}
 \Del_g = \frac{1}{\sqrt{g(\xi)}} \, \frac{\prt}{\prt \xi^a}\, g^{ab}(\xi)\,\frac{\prt}{\prt \xi^b}. 
 \eeq
 
  While the rhs of eqs. \rf{4.43} and \rf{4.43a} indeed look like  reasonable discretizations of the lhs of these equations, 
 we can actually derive  the discretized expressions 
 from our piecewise linear surface picture, which is not really a discretization, but rather 
 a special choice of surface. First consider the given abstract triangulation $T$ where each link has length $\ep$ 
 as embedded in some higher dimensional flat space $\mathbb{R}^k$ such that distances are preserved. Thus we have a
 mapping $v \in V(T) \to y(v) \in \mathbb{R}^k$ such that for all links $\la v v'\ra \in L(T)$ we have $|y(v) \mi y(v')| \equ \ep$.
 There are theorems which ensure that there exists a sufficient large $k$ such that all $T$s can be mapped isometrically
 to $\mathbb{R}^k$ ($k=7$, Nash's theorem). Let us introduce a 
 coordinate system for each triangle (the total coordinate system is then the union of these, including 
 transition functions telling us how to go from one to the other coordinate system in regions of overlap (which will be the links)).
 It is convenient to introduce barycentric coordinates for the triangles. 
 Consider the triangles $t \in T$ defined by the vertices $v_1$, $v_2$ and $v_3$. The
 coordinates of a point in the triangle will then be
 \beq\label{4.44}
 y (\xi) = \xi^1 y (v_1) \plu \xi^2 y(v_2) \plu (1\mi \xi^1 \mi  \xi^2) y(v_3),~~0 \leq \xi^1\plu\xi^2 \leq 1,~~\xi^a \in [0,1]. 
 \eeq
This assigns coordinate $\xi$ to a point in the triangle defined by the three vertices $v_i$ and the corresponding 
values of $X_i(\xi)$ are 
\beq\label{4.45}
X_i(\xi) = \xi^1 X_i(v_1)+ \xi^2 X_i(v_2) + (1\mi \xi^1\mi \xi^2) X_i(v_3).
\eeq
Since the metric is flat and trivially $\del_{\a\b}$ in  $\mathbb{R}^k$ where $y$ lives we have 
\beq\label{4.46a}
g_{ab} (\xi) = \del_{\a\b} \frac{\prt y^\a}{\prt \xi^a}  \frac{\prt y^\b}{\prt \xi^b} = \ep^2 \!
\begin{pmatrix} 1 & \oh \\ \oh &1 \end{pmatrix} , \quad  g^{ab} = \frac{4}{3\ep^2} \!
 \begin{pmatrix} 1 & -\oh \\ -\oh &1 \end{pmatrix} ,\quad \sqrt{g} = \frac{\sqrt{3}\ep^2}{2}.
 \eeq
 Integration over one triangle thus produces (after a little calculation)
 \beq\label{4.46}
 \int_t d^2\xi \;\sqrt{g} \; g^{ab} \,\frac{\prt X_i}{\prt \xi^a} \frac{\prt X_i}{\prt \xi^b}= 
 \frac{1}{\sqrt{3}} \sum_{\la v v'\ra \in L(t)} \big( X_i(v) \mi X_i(v')\big)^2.
 \eeq
 Summing over all triangles then leads to \rf{4.43} (links should only be counted once in neighboring triangles, since
 they will represent the overlap of the two coordinate systems in the triangles).
 For a given triangulation, i.e. a given intrinsic geometry $g_{ab}$ of the corresponding piecewise linear surface,
 the action \rf{4.2} is then
 \beq\label{4.47}
 S[T, X]= \frac{\kp}{2\sqrt{3}} \sum_{\la v v'\ra \in L(T)} \big(X_i(v) \mi  X_i(v')\big)^2 + \frac{\kp \lam \sqrt{3}}{8} \ep^2  |T|.
 \eeq
 As for the particle it is convenience in the following to consider the path integral in terms of dimensionless variables,
 and we thus redefine $ \sqrt{ \kp /\sqrt{3}}\; X \to X$ and $ \frac{\kp \lam \sqrt{3}}{8} \ep^2 \to \mu$ and our 
 dimensionless action is finally
 \beq\label{4.48}
 \boxed{S[X,T] = \oh \sum_{\la v v'\ra \in L(T)} \big(X_i(v) \mi X_i(v')\big)^2 +\mu\, |T|}
 \eeq
 
 We can now define the regularized version of \rf{4.6}
 \beq\label{4.49}
\boxed{ G_\mu(\ell_1,\ldots,\ell_n) = \sum_{T \in \cT(\ell_1,\ldots,\ell_n)} \!\!\!\!\e^{-\mu \, |T|} 
 \int\!\!\!\! \prod_{v \in V(T)/\{ \ell_1,\ldots,\ell_n\}} \!\!\!\!dX(v) \; \e^{-S[X,T]}
 }
 \eeq
 Here $T$ denotes an abstract triangulation with $n$ boundaries and $\cT(\ell_1,\ldots,\ell_n)$ the set of 
 such triangulations. Each boundary consists of a number of vertices and the associated links, connecting the 
 vertices to a loop. These loops have a double meaning in the notation above. They denote at the same time 
 the boundary-loop in the abstract triangulation {\it and} its image in $\mathbb{R}^D$ by the map $v \to X(v)$.
 In \rf{4.49} one does not integrate over the $X(v)$ where $v$ is a boundary vertex in the triangulation: the boundaries 
 are kept fixed in $\mathbb{R}^D$.
 
 Finally, the regularized $n$-point function for surfaces with $h$ handles and $n$ punctures is defined by
\beq\label{4.50}
\boxed{ G_\mu(x(v_1),\ldots,x((v_n)) = \sum_{T \in \cT(v_1,\ldots,v_n)} \!\!\!\!\e^{-\mu \, |T|} 
 \int\!\!\!\! \prod_{v \in V(T)/\{ v_1,\ldots,v_n\}} \!\!\!\!dX(v) \; \e^{-S[X,T]}
 }
 \eeq
 where $\cT(v_1,\ldots,v_)$ denotes the triangulations with $h$ handles and marked vertices $v_1,\ldots,v_n$,  and where the 
 coordinates  $X_i(v_k)$, $k=1,\ldots,n$ on the surface are kept fixed, while the rest are integrated over.
 The regulated susceptibilities are now defined as in \rf{4.9}, except that they now, with the use of the action 
 \rf{4.2} instead of \rf{4.1} and the rescaling of $X$, will be a function of $\mu$:
 \beq\label{4.51}
 \chi^{(n)}(\mu) = \int \prod_{k=1}^{n-1} d x(v_i) \; G_\mu(x(v_1),\ldots, x(v_n)).
 \eeq
 In particular, let us mention that for the one-point function $G_\mu(x(v_1))$  (which is also equal $\chi^{(1)}(\mu)$ and 
 independent of $x(v_1)$ by translational invariance), one can explicitly perform the Gaussian integrals in \rf{4.50}
 by using  \rf{4.43} and \rf{4.43a}. Introducing $Y(v) = X(v) \mi x(v_1)$ we find 
 \beq\label{4.51a}
  \int\!\!\!\! \prod_{v \in V(T)/\{ v_1\}} \!\!\!\!dY(v) \; \e^{-S_g[Y,T]} = \Big(\frac{(2\pi)^{|V(T)|-1}}{\det (-\Del'_{vv'}(T))}\Big)^{\frac{D}{2}},
\eeq 
where $\Del'_{vv'}(T)$ denotes the $(|V(T)\mi 1) \times (|V(T)|-1) $ matrix constructed from the combinatoral Laplacian 
defined for $T$ by deleting the $v_1$th row and column. Thus we have (using $|V(T)| = |T|/2\plu 2\mi 2h$)
\beq\label{4.51b}
\boxed{  G^{(1)}_\mu(x) \equiv \chi^{(1)}(\mu) = \sum_{T \in \cT(v_1)} \!\!\!\!\e^{-(\mu-\mu_0) \, |T|} 
\Big(\frac{(2\pi)^{1-2h}}{\det (-\Del'_{vv'}(T))}\Big)^{\frac{D}{2}}} \quad \e^{\mu_0} = (2\pi)^{D/4},
\eeq
which is a remarkable explicit formula, valid for all $D$ by analytic continuation in $D$.

 Using these regularized functions it is now possible to prove the statements made above for the Green functions.
 We will not given the proofs here, but let us summarized the statements which can be made.

 {\bf Theorem 1:} There exists a critical value $\mu_c$ such that $G_\mu(\ell_1,\ldots,\ell_n)$ is defined by \rf{4.49}
 is convergent for $\mu > \mu_c$ and divergent for $\mu < \mu_c$. This critical value is independent of the number, positions,
 and lengths for the boundary loops as well as the number of handles of the surface. 
 
 {\bf Theorem 2:}  The two-point function $G_\m(x,y)$ falls of exponentially with the distance $|x-y|$ between the two points
 for $\mu > \mu_c$ and the mass 
 $$
 m(\mu) = - \underset{|x-y|\to \infty}{\lim}  \frac{\ln G_\mu(x\mi y)}{|x\mi y|} > 0.$$  
 The mass is independent of the number of handles of the surface. The two-loop
 function $G_\mu(\ell_1,\ell_2)$ has the same exponential fall off  when the distance between the two loops goes to to infinity.
 
 {\bf Theorem 3:}  Consider the ensemble of surfaces with no handles ($h\equ 0$).
 The string tension $\sg(\mu)$, defined as the exponential fall off of the one-loop function $G_\mu(\ell_A)$ 
 for  planar loops $\ell_A$ enclosing a domain of  area $A$ in $\mathbb{R}^D$, exists for any $\mu > \mu_c$ and
 $$
\sg(\mu) =- \underset{A\to \infty}{\lim}  \frac{\ln G_\mu(\ell_A)}{A} > 0.
$$
 
\vspace{6pt}

Theorem 1 implies that if we decompose $G_\mu(\ell_1,\ldots,\ell_n)$ in \rf{4.49} in a sum over Green functions 
constructed from $|T| = N$ triangles
\beq\label{4.52}
G_\mu(\ell_1,\ldots,\ell_n) = \sum_N \e^{-\mu N} G_N(\ell_1,\ldots, \ell_l),
\eeq
then 
\beq\label{4.53}
G_N(\ell_1,\ldots,\ell_n) = \e^{\mu_c N} F(\ell_1,\ldots,\ell_m; N),
\eeq
where $F(\ell_,\ldots,\ell_n;N)$ is subleading in $N$  and
\beq\label{4.54}
G_\mu (\ell_1,\ldots,\ell_n) \sim \sum_N \e^{-(\mu - \mu_c) N} F(\ell_1, \ldots,\ell_n;N).
\eeq
It is now seen that the only way large $N$ can dominate the sum is when $\mu \to \mu_c$. Recall
from \rf{4.40} that $N \ep^2 \propto A_{int}$, the  intrinsic area of a surface. As an order of magnitude 
estimate, \rf{4.54} suggests that $\la N \ra \sim \frac{1}{\mu-\mu_c}$ (and we will later 
prove that this is true for $n \geq 2$, while one has $\la N \ra \sim \frac{1}{\sqrt{\mu-\mu_c}}$ for $n=1$).
Thus it natural to take a limit
\beq\label{4.55}
\mu\mi \mu_c = \Lam \ep^2 \quad \Rightarrow \quad \la A_{int} \ra \propto \la N \ep^2 \ra \sim \frac{1}{\Lam},
\eeq
which is the limit we were aiming for in \rf{4.40}, and a limit which is natural from the way we introduced 
the dimensionless parameter $\mu$ in the first place, namely as $\mu \propto \kp \,\lam \, \ep^2$. The
only new thing in \rf{4.55} is that $\mu$ undergoes an additive renormalization, but that should 
not be a surprise since we have already seen this in the case of RWs and BPs, where the constant $\mu_c$
was related to the exponential growth of the number of RWs or BPs with length or size, respectively. The origin of $\mu_c$
in \rf{4.53} is exactly the same.
Note however that we have not yet made any contact with the actual size of the surfaces $X_i(v)$
embedded in $\mathbb{R}^D$, as we did in the case of RWs or BPs.    
Clearly the behavior of $m(\mu)$ and (and as something new: $\sg(\mu)$) for $\mu \to \mu_c$ 
will of utmost importance when trying to do that,  precisely as it was the case for RWs and BPs. 

Before we study the behavior of $m(\mu)$ and $\sg(\mu)$ in the limit $\mu \to \mu_c$,
we will make a digression and discuss the summation over the number of handles of the surfaces. 
 
 \subsection*{Digression: summation over topologies}
 
    \begin{figure}[t]
\vspace{-1.5cm}
\centerline{\scalebox{0.20}{\includegraphics{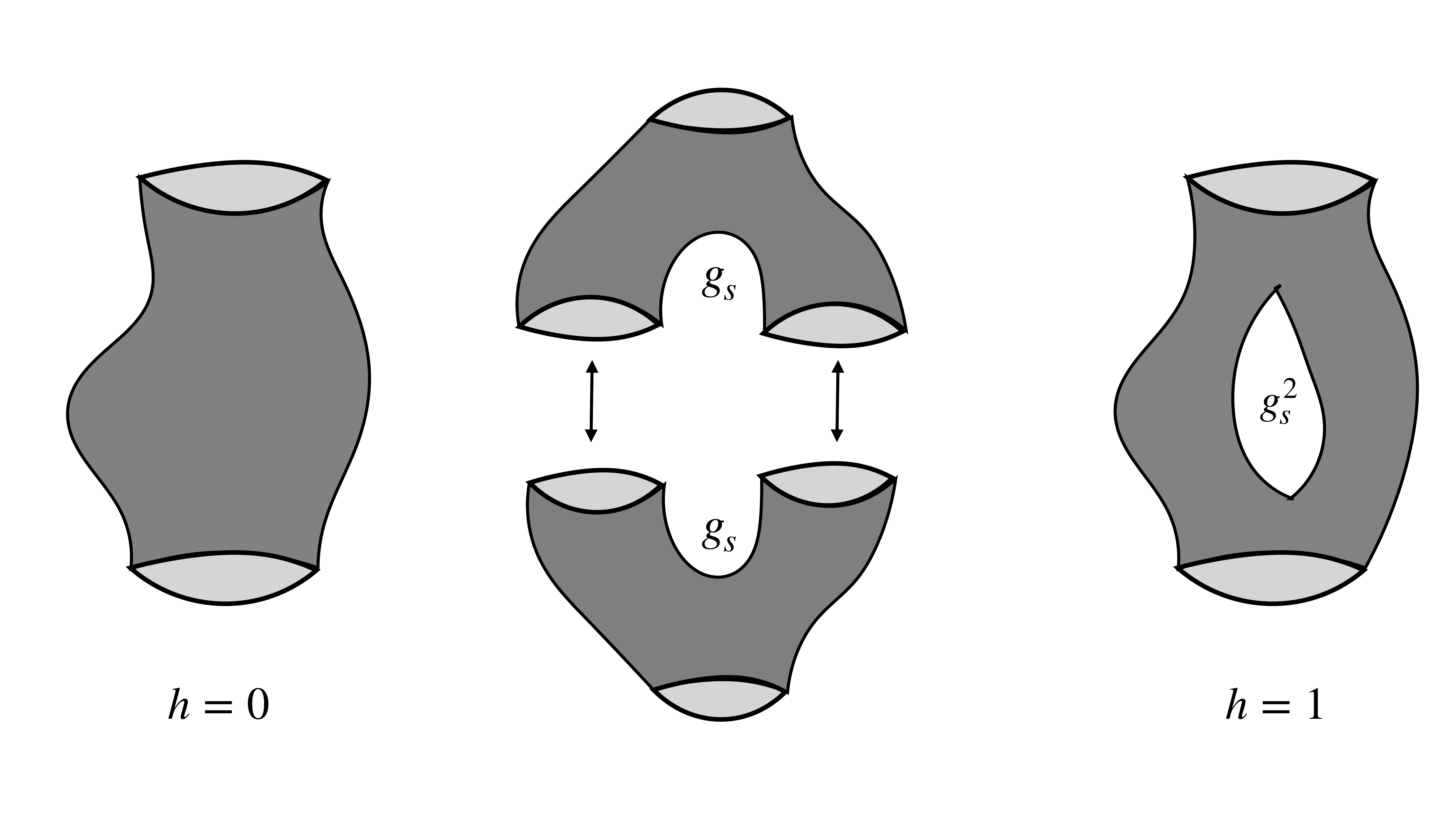}}}
\vspace{-1.0cm}
\caption{{\small  Propagation of a string, $h=0$. A string can split in two during propagation, with associated coupling constant
$g_s$, or two strings can join, again with associated $g_s$. Combined, this lead to propagation of a string, but now spanning a surface with $h=1$.}}
\label{fig4.9}
\end{figure}
 As mentioned above a beautiful aspect of string theory is that viewed from the path integration perspective, as a theory
 of random surfaces in $\mathbb{R}^D$, a $n$-loop function is as natural as a two-loop function. This leads to the 
  inclusion of surfaces with handles, since we can now view a two-loop function with one handles as a ``time''-evolution 
 of the two-loop function, where one loop splits in two loops (made possible because we have three-loop functions), which then 
 ``later'' join again to one loop (again made possible because we have three-loop functions). This is illustrated in Fig.\ \ref{fig4.9}.
 Thus from the figure it is clear that if one includes in the two-loop function the propagation of surfaces with 
 one handle, by iteration one has to include surfaces with an arbitrary high number of handles, i..e.\ we have to 
 sum over all surfaces with two boundaries and all handles:
 \beq\label{4.56}
G_\mu(\ell_1,\ell_2) = \sum_{h=0}^{\infty} G_\mu^{(h)}(\ell_1,\ell_2).
 \eeq
Since we are now discussing the change of topology we have to  step back to the Einstein-Hilbert action \rf{2.3a}.
 As discussed below eq.\ \rf{2.7} the reason we dropped the curvature term in the action in \rf{4.2} is that with 
 no topology change of the surface, it would only contribute with a constant, which we calculated in \rf{4.36} and 
 \rf{4.37}. However, it is natural to include it when we consider the sum in \rf{4.56}:
 \beq\label{4.57}
 \exp\Big\{\frac{1}{2\pi G} \Big(\int_\cM d^2 \xi \sqrt{g} \, R \plu \int_{\prt \cM}  ds \, k_g\Big)\Big\} = 
 \exp \Big\{\frac{2\mi n}{G}\Big\} \; \exp\Big\{-\frac{2h}{G}\Big\}.
 \eeq
 and we then replace \rf{4.56} by
 \beq\label{4.58} 
 G_\mu(\ell_1,\ell_2) = \sum_{h=0}^{\infty}  g_s^{2h}\;G_\mu^{(h)}(\ell_1,\ell_2)\qquad \boxed{g_s = \e^{-\frac{1}{G}}}
 \eeq
 \noindent
 where we  have defined a new coupling constant, the so-called {\it string coupling constant} in terms of 
$G$,  the gravitational coupling constant of two-dimensional gravity. Looking at Fig.\ \ref{fig4.9} there is a 
 factor $g_s$ associated with a splitting of a string in two (the number of boundaries changes from 1 to 2), and again 
 a factor of $g_s$ associated with the joining of two strings to one, and thus a total factor of $g_s^2$ associated with 
 string propagation via a surface with one handle compared to string propagation via a surface with no handles.
 
 Using our regularization we have managed to define $G_\mu^{(h)}(\ell_1,\ldots,\ell_n)$ for any $\mu > \mu_c$. Does 
 the regularization also provide us with a definition of $G_\mu(\ell_1,\ldots, \ell_2)$ by a formula like \rf{4.58} where 
 we sum over all topologies? Clearly \rf{4.58} provides us with a perturbation theory, {\it a string perturbation theory}. 
 We ``just'' have to calculate the contributions for each $h$, and then perform the sum. And for any value of $h$ it is 
 clear that the contribution from order $h\plu 1$ will be small if we choose $g_s$ sufficiently small. First a few general remarks.
 It should be emphasized that the reason it makes sense to talk about an interesting perturbation theory is 
 Theorem 1, which states that $\mu_c$ is the same for all $h$. This is a remarkable result, in particular because $\mu_c$
 is not a universal constant.  It will depend on the way we have chosen to regularize our theory. However, for any reasonable
 regularization, the statement in Theorem 1 is true. We need a common $\mu_c$ for all $h$, since our real interest is 
 in the ``continuum'' limit where the cut-off $\ep$ in \rf{4.55} is taken to zero. Next, viewing \rf{4.58} as a perturbation
 series, can we in principle perform the sum? If the series is convergent, no problem. Below we will show that 
 the series is {\it not} convergent. However, that should not be so surprising, since most perturbation series that 
 one encounters are only asymptotic series. This implies in the wording used above that although it is 
 true that we for a given order $h$ can choose $g_s$ such that the contribution to order $h$ is small, 
 this choice of $g_s$ cannot be made independently of $h$. Eventually, for any fixed $g_s$, the large-$h$ contribution 
 from $G_\mu^{(h)}$ will 
 always be large even if multiplied by $g_s^{2h}$. This does not necessarily mean that the sum cannot 
 be defined and it does not necessarily mean that there is not a well defined answer that one can agree upon. A trivial 
 example of this situation is the perturbative series of the anharmonic oscillator in quantum mechanics. The perturbation 
 theory is in this case only an asymptotic series. However, the summation can be performed by several of the standard 
 methods for summing divergent series, e.g.\  the so-called {\it Borel summation} (which we define and discuss in Problem Set 8). 
 It provides us with an answer. Is this answer
 the correct one (clearly one can get any number by stupid summation of a divergent series)? Yes, we know this because we can 
 define the quantum theory of the anharmonic oscillator in a way which is independent of its perturbation expansion, and we can
 then prove, using this definition, that {\it if} one chooses to perform an perturbative expansion, the Borel sum of the perturbation
 series will give the correct result.   In the case of string theory, we need something similar: we need {\it as a minimum} a 
 regularization which for  a non-zero cut-off provides us with well defined expressions for the $n$-loop functions summed
 over all handles $h$. Since we have finite well defined expressions for each  $h$ and even a perturbation expansion, 
 it is tempting to try to {\it define} our theory including all $h$, by simply using \rf{4.49} and declaring that 
 $\cT(\ell_1,\ldots,\ell_n)$ is the class of all triangulations with $n$ boundaries, independent of $h$. However, as we will 
 show below, it does not work. The expression is simply not well defined except as a formal perturbation series in $h$. 
 
 Nevertheless we might still be able to sum the divergent perturbation series and obtain results which might be interesting
 if they point towards new physics, even if the result in this way is not completely well defined. Let us illustrate this, and the
 nature of the divergent perturbation series by analysing $\chi^{(1)}(\mu)$ given by \rf{4.51b}. Let us start with the simplest
 situation, namely choosing $D=0$. This is then pure two-dimensional quantum gravity (which we will study in some detail
 in the next Section). Since we now have a changing topology we incorporate \rf{4.57} and \rf{4.58} and write:
 \beq\label{4.59a}
 \chi^{(1)}(\mu) = \sum_h g_s^{2h-1} \sum_N \e^{-\mu \, N}\!\!\!\!\sum_{T \in \cT^{(h)}_N(v_1)} \!\!1 =
  \sum_h g_s^{2h-1} \sum_N \e^{-\mu \, N} \cN_1^{(h)}(N)
 \eeq
 where the summation over  triangulations is a  sum over handles $h$ and for given $h$ a sum over the number 
 of triangles, and for given $h,N$ a sum over all such triangulations with one marked vertex $v_1$. Finally,
 $\cN_1^{(h)}(N)$ denotes the number of triangulations with $N$ triangles, $h$ handles and one marked vertex.
 One can calculate the asymptotic form of $\cN_1^{(h)}(N)$. We will do that in the next Section in the simplest case of $h=0$.
 The result is 
 \beq\label{4.59b}
 \cN_1^{(h)}(N) = c_h N^{5(h-1)/2} \, \e^{\mu_c N} \big(1 + \cO(1/N) \big)
 \eeq
 where $c_h$  is  bounded as function of $h$. For any fixed $h$ the number of triangulations grows exponentally
 and for a fixed $h$ the critical $\mu$ in \rf{4.59a} will be the $\mu_c$ which appears in this exponential grows.
 Again, for fixed $h$, large $N$ will dominate in \rf{4.59a} and it makes some sense to use the 
 asymptotic form \rf{4.59b} in \rf{4.59a} if we are interested in the limit $\mu \to \mu_c$. Doing that we obtain
 \bea
 \chi^{(1)}(\mu) &\approx& \sum_h c_h\, g_s^{2h-1} \sum_N N^{5(h-1)/2} \e^{-(\mu \mi \mu_c )\, N} \nonumber \\
 &\approx&
 \frac{g_s}{\mu \mi \mu_c} \sum_h c_h \Gamma\Big(\frac{5h}{2}  \mi \frac{3}{2}\Big)\; 
 \Big( \frac{ g_s}{(\mu \mi \mu_c)^{5/4}}\Big)^{2h-2},
\label{4.59c}
 \eea
where we have replaced the summation with an integration, which is allowed for $h>0$, but not really
for $h=0$. Two aspects are clear: first of all the behavior for $\mu \to \mu_c$ becomes more and more
singular for $\mu \to \mu_c$ with increasing $h$. Thus the sum makes no sense in that limit (which is the 
one we are interested in!) unless we scale the string coupling constant to zero together with taking 
$\mu \to \mu_c$. Such a limit is called {\it the double scaling limit}. We then demand (recalling 
that $g_s = e^{-1/G}$, $G$ the gravitational coupling constant) 
\beq\label{4.59d}
\frac{ \e^{-1/G}}{(\mu \mi \mu_c)^{5/4}} = \e^{-1/G_R}, \quad {\rm or} \quad \frac{1}{G} = \frac{5}{4} \ln \frac{1}{\mu\mi \mu_c} + 
\frac{1}{G_R}.
\eeq
This double scaling limit thus has the intriguing interpretation as a {\it renormalization} of the gravitational coupling constant
$G$: for $\mu \to \mu_c$ the ``bare'' $1/G$ goes to infinity, but leaves behind a renormalized $G_R$. We can now write,
ignoring the term in front of the sum in \rf{4.59c}: 
\beq\label{4.59e}
\chi^{(1)}(G_R)  \propto \sum_h c_h \Gamma\Big(\frac{5h}{2}  \mi \frac{3}{2}\Big)\; \e^{ \frac{\chi(h)}{G_R}}.
\eeq
 So our ``renormalized'' $ \chi^{(1)}(G_R) $ is a sum of contributions for each $h$-sector, given by the 
 Einstein action term for that sector, but with a renormalized gravitational coupling constant, and the 
 ``action contribution'' multiplied by the ``number of geometries'' with handle $h$. This is an amazing
 formula, but unfortunately the series is divergent for any fixed value of $G_R$, since the factorial factor 
 grows too rapidly. It is not even Borel summable. Nevertheless it can be summed! We will not 
 discuss here the methods one can use, but they are discussed in  Problem Set 8\footnote{ Also, in 
  Problem Set 9 it is shown have to carry out the calculations hinted above in detail, in the case
 of BPs, where we enlarge the set of BPs from trees to trees with loops. The number of such polymers 
 will then grow factorially, not exponentially, with the number of links, and we have precisely the 
 problem above. Nevertheless it is possible to perform the summation over such BPs with loops explicitly.}. As already mentioned, unfortunately the result is not unique and  the various results 
 have some  troublesome features, but it is not ruled out that one might find the correct, physical argument 
 to select the ``correct'' sum.
  
 If we return to the surfaces embedded in $D$ dimensions one could hope that the integration over 
 $X$-coordinates could help to ``tame'' the sum over topologies in \rf{4.51b} when $\cT(v_1)$ means
 all triangulations, irrespective of the number of handles. Unfortunately it is not the case as we
 will now argue. The determinant in \rf{4.51b} is the result of the Gaussian integration over $X(v)$s.
 There is a theorem called {\it Kirchoff’s matrix-tree theorem} which states that the determinant 
 is equal to the number of spanning trees in the triangulation, where a spanning tree is 
 a connected tree-subgraph reaching all vertices. Thus we have the following estimate:
 \beq\label{4.59f}
 1 \leq \det (-\Del_{vv'} (T))   \leq \prod_{v \in V(T)} n_v,
 \eeq
 since the product of vertex orders is clearly larger than or equal to the number of spanning trees.
 Let us now be more specific with the class of triangulations we consider (we will also need this in the next 
 subsection). We denote by  $\cT^{(3)}$ the class which satisfies three 
 conditions. (1): the boundary of the triangles sharing a vertex is a circle (i.e.\ locally, around the vertex, 
 we have $\mathbb{R}^2$ topology). (2): a link is uniquely defined by its vertices (i.e.\ we cannot have two links 
 connected to the same two vertices). (3): a triangle is uniquely defined by its three vertices. 
 This implies that $|T| \geq 4$. If a triangulation has $h$ handles it will have $h$ independent 
 non-contractable loops, and for triangulations in class $\cT^{(3)}$ this implies that it contains at least $h$ vertices.
 Recall \rf{4.34}:  $|V(T)| \mi |T|/2 = 2\mi 2h$. Thus
 \beq\label{4.61} 
  |V(T)|  \leq |T| \quad {\rm and} \quad \frac{3 |T|}{|V(T)|} = 6\;\frac{|V(T)| \plu 2h\mi 2}{|V(T)|} \leq 18.
 \eeq
Now we can  estimate that 
 \beq\label{4.63}
 \sum_v \ln n_v \leq |V(T)|\, \ln  \Big(\frac{\sum_v n_v}{ |V(T)| }\Big) =  |V(T)| \,  \ln \Big(\frac{ 3|T|}{ |V(T)| }\Big)  \leq |T|\, \ln 18.
 \eeq
Finally , for $D$ positive 
\beq\label{4.64}
\big(\det (-\Del'_{vv'}(T))\big)^{-D/2}\geq \e^{- \frac{D \ln 18}{2} \; |T|} ,
\eeq
and using the rhs in \rf{4.51b} we are basically getting back to the $D=0$ situation, just with a shifted $\mu$.
For negative $D$ we can use the lower estimate in\rf{4.59f} and replace the determinant by 1 and reach the same 
conclusion. The expression \rf{4.51b} is thus infinite unless we invent some fancy summation procedure, as discussed.

\vspace{6pt}

\subsection*{The mass and the string tension} 

\subsubsection*{Scaling of the mass}

 \begin{figure}[t]
\centerline{\scalebox{0.20}{\includegraphics{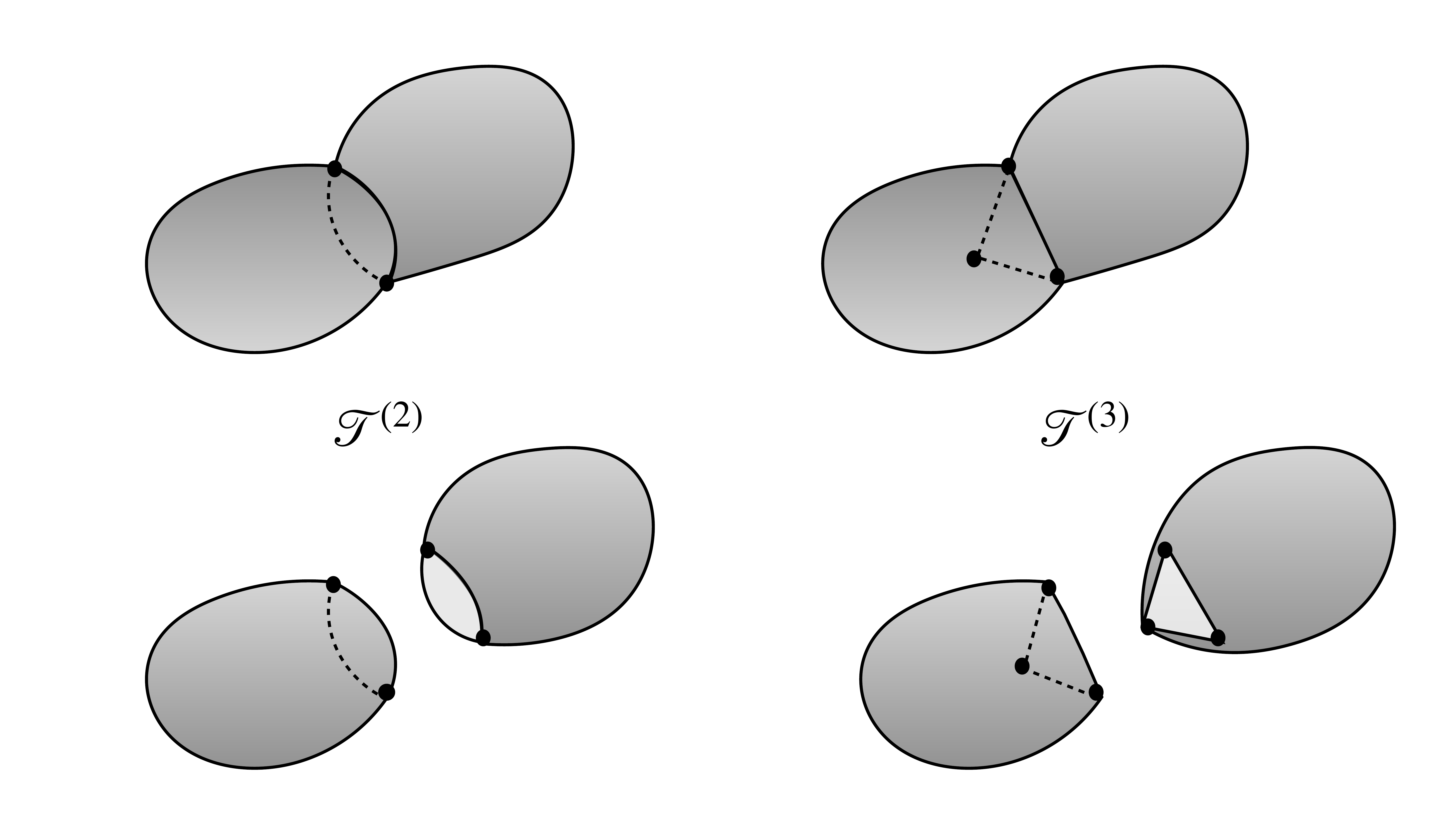}}}
\caption{{\small  A triangulation belonging to $\cT^{(2)}$ can be cut in two disconnected parts along any two-loop which is not a boundary. However,
``ordinary'' triangulations in  $\cT^{(3)}$ may also be cut in two along a three-loop. Such a three-loop is not necessarily a triangle belonging to the 
triangulation, as shown on the right part of the figure.}}
\label{fig4.10}
\end{figure}

\begin{figure}[t]
\vspace{-.5cm}
\centerline{\scalebox{0.20}{\includegraphics{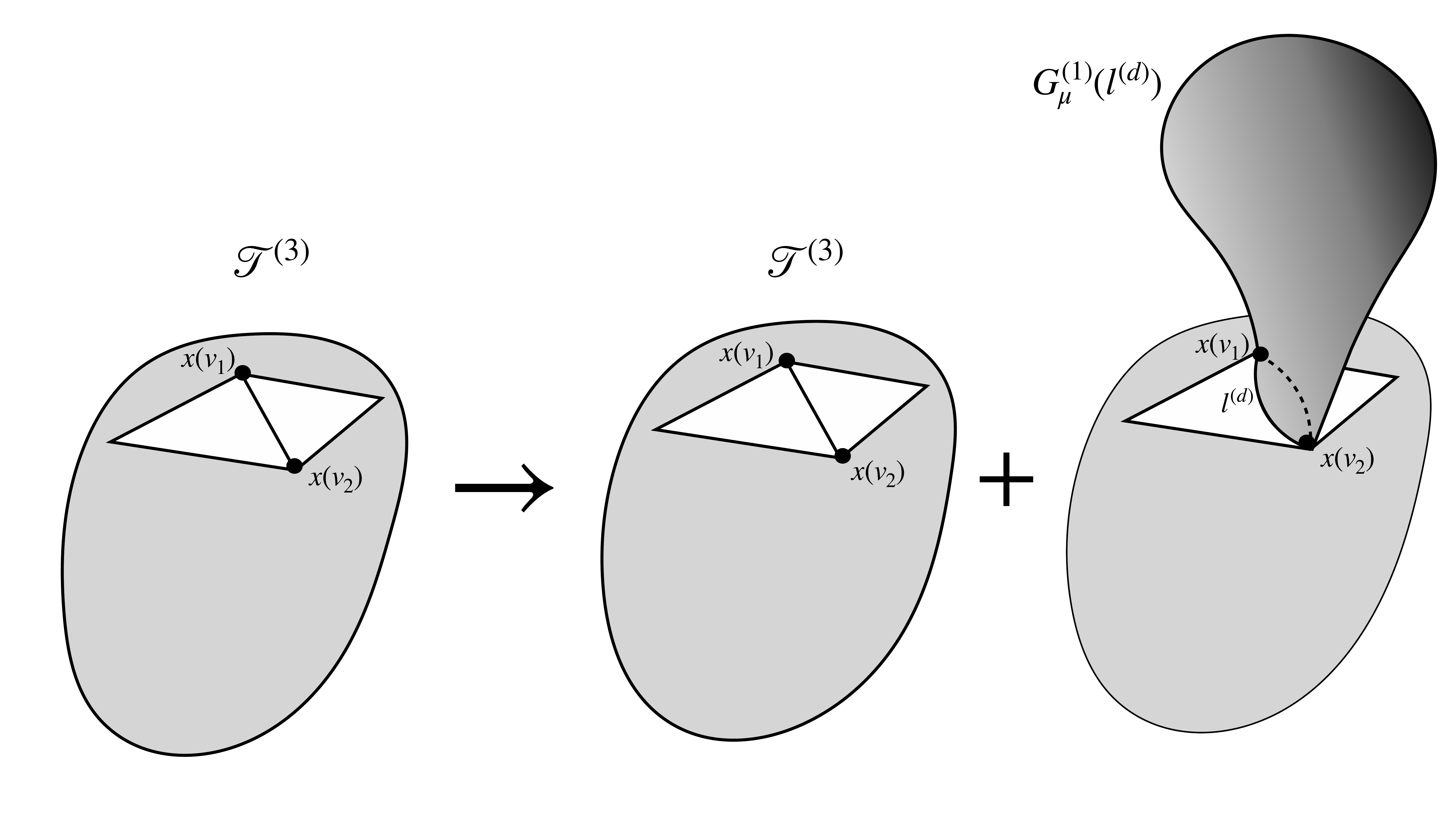}}}
\vspace{-.3cm}
\caption{{\small  Any intrinsic link in a $\cT^{(3)}$ triangulation can be split open to a double link $l^{(d)}$, to which one can attach an 
outgrowth $G^{(1)}_\mu(l^{(d)})$. }}
\label{fig4.11}
\end{figure}

 We have already argued that  the string susceptibility $\g_s  \leq \oh$. We will now show that  
 $\g_s > 0 \Rightarrow \g_s= \oh$. The arguments given here will use triangulations and the action \rf{4.48}
 and will be heuristic in nature, and the basic assumption is {\it universality}: the critical behavior should 
 be independent of the detailed class of triangulations used. A rigorous proof can be given using the hypercubic
 regularization described earlier. We have defined the set of triangulations $\cT^{(3)}$. Let us now specific it 
 further, and define  $\cT^{(3)}_1$ as the set of  $\cT^{(3)}$-triangulations with one boundary which consists of a double-link,
 i.e.\ two links connecting the same two boundary vertices.
 Similarly, $\cT^{(3)}_2$ is defined as the set of  $\cT^{(3)}$-triangulations with two boundaries, where also the other boundary consists of a double-link. By definition there are no interior double-links. 
 Let us   define a  larger class of triangulations denoted by  $\cT^{(2)}$, where we 
  allow double-links (but not triple links etc), but only if cutting the triangulation along  the double-link will separate the 
  triangulation in two disconnected parts. One would not expect our strings defined on 
  this class to exhibit a critical behavior different from the strings defined on  $\cT^{(3)}$ since 
  also in  $\cT^{(3)}$ it might be possible to cut a triangulation in two disconnect parts, not a along
  a ``two-loop'' but along a ``three-loop'', as shown in Fig.\ \ref{fig4.10}. Clearly this difference should not matter when the 
  triangulations are very large, as is the case for those which determine the critical behavior.   
  In class  $\cT^{(2)}$ one now defines  $\cT^{(2)}_1$ and  $\cT^{(2)}_2$ in the same way as for $\cT^{(3)}$. 
  Let now $\ell^{(d)}$ denote such a boundary double link. 
  It will depend on coordinates $x_1$ and $x_2$ which we do usually not  include in the integration. However,
 if we decided to integrate over $x_2$, say, in average it will be at a distance of order 1 from $x_1$, since $x_1$ and $x_2$ 
 interact via a Gaussian term. This distance is very small compared to average distances to most vertices in 
 the triangulation if $|T| \gg1$. We will thus simply approximate the two boundary points by one point  $x$ and 
 in this approximation the one-loop function corresponding the $G_\mu (\ell^{(d)})$ simply becomes the 
 one-point function $G_\mu(x)$, {\it which is independent of $x$}.  In this approximation we see that
 the difference between  between $\cT^{(3)}$ and $\cT^{(2)}$  becomes local as shown 
 Fig.\ \ref{fig4.11}, and summarized by the following change of assignment to each internal link:
 \beq\label{4.66}
 \e^{-\oh (X(v_1) \mi X(v_2))^2} \to \big(1+G^{(1)}_\mu\big)\; \e^{-\oh (X(v_1) \mi X(v_2))^2} ,
 \eeq
since every  link in a   $\cT^{(3)}$ triangulation (which is not a boundary link)  can also be a double-link which serves as the boundary 
for an arbitrary ``outgrowth'' belonging to  $\cT^{(2)}_1$ . 

\begin{figure}[t]
\vspace{-.5cm}
\centerline{\scalebox{0.20}{\includegraphics{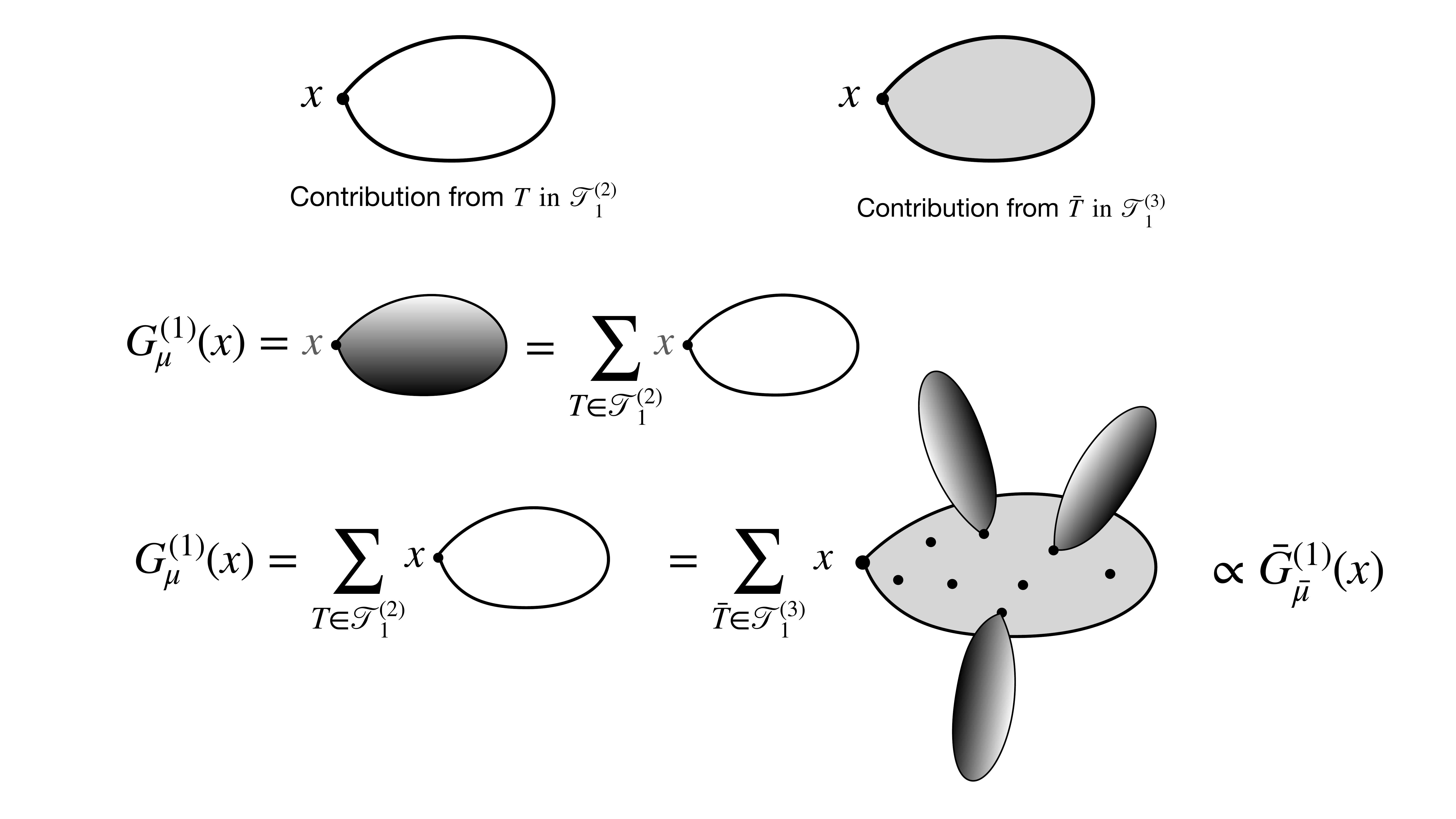}}}
\vspace{-.3cm}
\caption{{\small  The use of the decompostion shown in Fig. \ref{fig4.11} to write the triangulations in $\cT^{(2)}_1$ as triangulations in $\cT^{(3)}_1$
with outgrowths. The dots with an outgrowth attached is really a double link, while the dots without outgrowths symbolise ordinary links in 
$\cT^{(3)}_1$ which have not been cut open.}}
\label{fig4.12}
\end{figure}

In \rf{4.66} $G^{(1)}_\mu$ denotes this 
one-loop or one-point function and we can now make a decomposition:
\bea
G^{(1)}_\mu &=& \sum_{T \in \cT^{(2)}_1} \e^{-\mu |T|} \int \!\!\!\!\prod\limits_{v\in T/\{v_1\}} dX(v) \, \e^{-S_g[X,T]} \nonumber \\
&=& \sum_{\bT \in \cT^{(3)}_1} \e^{-\mu |\bT|}\big( 1+ G^{(1)}_\mu\big)^{|L(\bT)|-2}
\int  \!\!\!\!\prod\limits_{v\in \bT/\{v_1\}} dX(v) \, \e^{-S_g[X,\bT]} \nonumber \\
&=&\!\! \big(1\plu G^{(1)}_\mu\big)^{-1}\!\!\!\sum_{\bT \in \cT^{(3)}_1} \e^{-\bmu |\bT|}\!\!
\int \!\!\!\! \prod\limits_{v\in \bT/\{v_1\}} \!\!\!dX(v) \, \e^{-S_g[X,\bT]}, 
\quad \bmu = \mu \mi \frac{3}{2} \ln \big( 1\plu G^{(1)}_\mu\big), \nonumber  
\eea
 where $\bar{(\cdot)}$ refers to the ensemble $\cT^{(3)}$ and non-bar quantities to the ensemble  $\cT^{(2)}$,
 and where we have used  $ |L(\bT)| -2= \frac{3}{2}|\bT|-1$ when we have two boundary links. Finally $v_1$ is one of
 the vertices in the boundary loop (we integrate over the other vertex, but it is not important in the scaling limit).
 The rhs is just $\bG^{(1)}_\bmu$ except for factor $ (1+ G^{(1)}_\mu)^{-1}$ coming from the 1 in $  \frac{3}{2}|\bT|\mi 1$,
  and the decomposition used is shown in Fig.\ \ref{fig4.12}. Summarizing
 \beq\label{4.67}
 \boxed{ G^{(1)}_\mu =  \big(1\plu G_\mu^{(1)}\big)^{-1} \;\bG^{(1)}_\bmu,\quad 
 \bmu = \mu \mi \frac{3}{2} \ln \big( 1\plu G^{(1)}_\mu\big)}.
 \eeq
 Let us now define the two-loop or two-point function (we will not distinguish when the boundaries 
 are just double-links)  in the ensemble $\cTtwo_2$ as 
 \beq\label{4.67a}
 G^{(2)}_\mu(x\mi y) =   \sum_{T \in \cT^{(2)}_2} \e^{-\mu |T|} \int \!\!\!\!\prod\limits_{v\in T/\{v_1,v_2\}} dX(v) \, \e^{-S_g[X,T]} ,
 \quad y\!=\!x(v_1),~x\!=\!x(v_2),
 \eeq

 where we have two double-link boundary loops, one of which contains vertex $v_1$ with coordinate $y= x(v_1)$ and 
 the other $v_2$ with coordinate $ x = x(v_2)$. 
 We define the susceptibility as 
 \beq\label{4.67b}
 \chi(\mu)  = \int d^Dy \; G^{(2)}_\mu(x\mi y) ,\qquad  \chi(\mu) 
 \propto \frac{1}{(\mu\mi \mu_c)^{\g_s} }\quad {\rm for} \quad \mu \to \mu_c.
 \eeq
 Similarly we define $\bG^{(2)}_\bmu(x \mi y)$ and $\bchi(\bmu)$, by replacing $\cTtwo$ by $\cTthree$, in particular we have
 \beq\label{4.67c}
 \bchi(\bmu)  = \int d^Dy \; \bG^{(2)}_\bmu(x\mi y) ,\qquad  \bchi(\mu) \propto \frac{1}{(\bmu\mi \bmu_c)^{\g_s} }\quad {\rm for} 
 \quad \bmu \to \bmu_c.
 \eeq
 By the assumption of universality we have the same critical exponent $\g_s$ in \rf{4.67b} and \rf{4.67c},
 but $\mu_c$ and $\bmu_c$ will in general be different. The relation between $\bmu$ and $\mu$ is 
 given by \rf{4.67} and in particular we can find the relation for $\mu \to \mu_c$ since we from our general discussion
 around \rf{4.13} expect
 \bea\label{4.67d}
 &&G^{(1)}_{\mu} \to  G^{(1)}_{\mu_c} \plu c \, (\mu \mi \mu_c)^{1-\g_s} \\
 && \bmu(\mu)\mi \bmu(\mu_c) = \tilde{c} \, (\mu\mi \mu_c)^{1-\g_s} \plu (\mu\mi \mu_c) \plu \cdots,\label{4.67e}
\eea
 where $ G^{(1)}_{\mu_c} > 0$ and {\it finite}, since $\g_s \leq \oh$ and thus the critical part goes to zero for $\mu \to \mu_c$.

 \begin{figure}[t]
\vspace{-.5cm}
\centerline{\scalebox{0.25}{\includegraphics{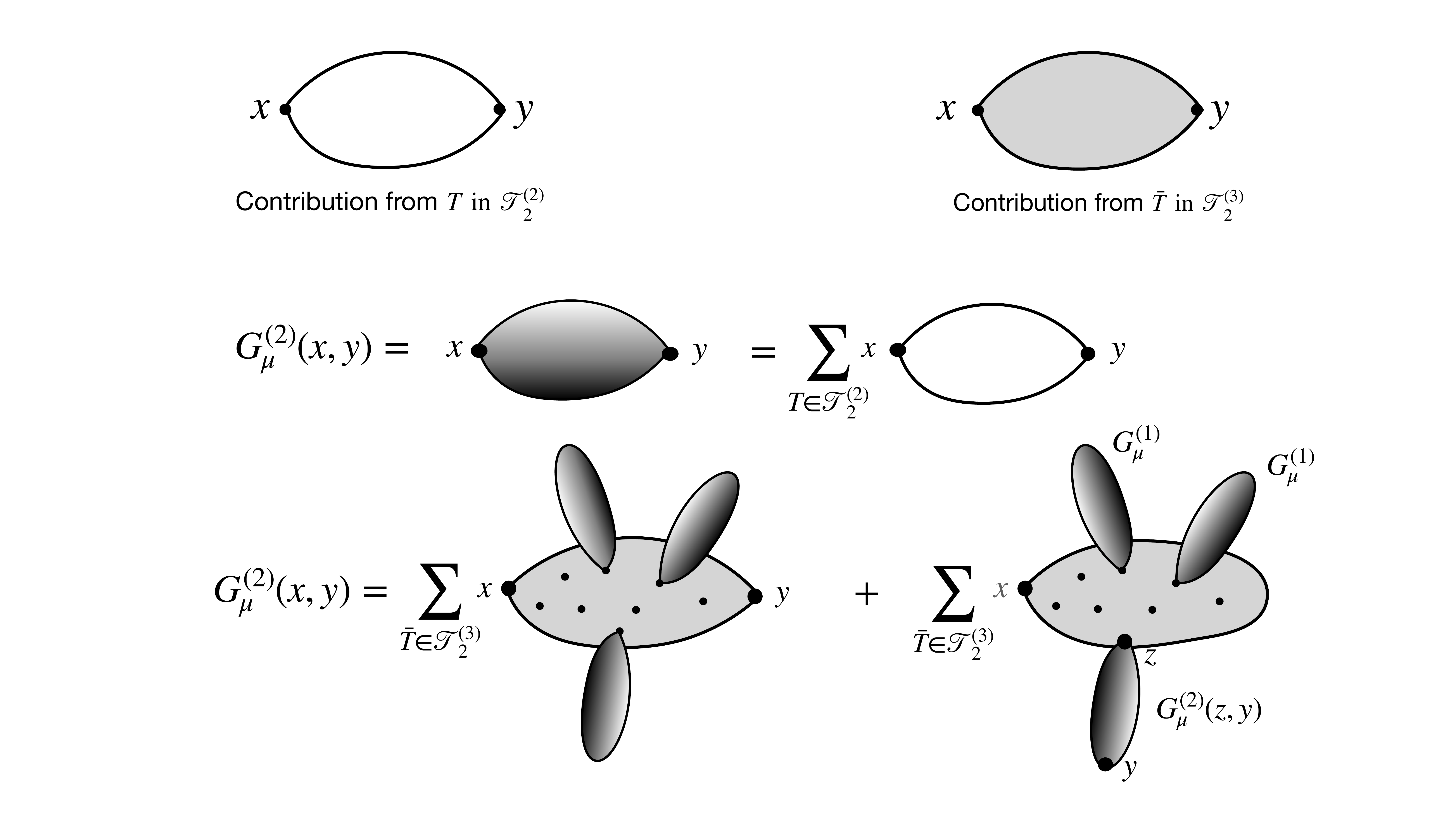}}}
\vspace{-.3cm}
\caption{{\small  The generalization of the one-loop (or one-point) function from Fig.\ \ref{fig4.12} to the two-loop (or two-point) function.}}
\label{fig4.14}
\end{figure}
 We can now  make the same graphical decomposition as we did for $G^{(1)}_{\mu}$ and it is shown in Fig.\ \ref{fig4.14}.
 Transferring it to an equation we obtain the analogue of \rf{4.67}, only for the two-point function
 \beq\label{4.68}
 G_\mu^{(2)}(x\mi y) = \big(1\plu G_\mu^{(1)}\big)^{-2} \Big( \bG^{(2)}_\bmu(x\mi y) \plu 
 \int d^D z\;\bG^{(2)}_\bmu(x\mi z) \, G^{(2)}_\mu(z\mi y)\Big),
  \eeq
 and integrating over $y$ we obtain
 \beq\label{4.69}
 \chi(\mu) =  \big(1\plu G_\mu^{(1)}\big)^{-2} \Big[ \bchi(\bmu) +  \bchi(\bmu) \, \chi(\mu)\Big] ~\Rightarrow ~
 \boxed{\chi(\mu) = \frac{ \big(1\plu G_\mu^{(1)}\big)^{-2} \;\bchi(\bmu)}{ 1 \mi \big(1\plu G_\mu^{(1)}\big)^{-2} \;\bchi(\bmu)}}
 \eeq
 First, the factor $(1\plu G_\mu^{(1)})^{-2}$ in \rf{4.68} comes for the same reason as the factor $(1\plu G_\mu^{(1)})^{-1}$ was 
 present in \rf{4.67}: A $\cTthree$ graph between two boundary loops has $L(\bT)-4 = \frac{3}{2} |\bT| -2$ internal links, 
 the -2 leading to the mentioned factor. As mentioned, \rf{4.67} determines $\bmu$ as a function $\bmu(\mu)$, and 
 for $\mu \to \mu_c$ we have \rf{4.67e}. Now, \rf{4.69} shows that 
 \beq\label{4.69a}
 \boxed{\bmu(\mu_c) > \bmu_c}
 \eeq
 simply because the 
 divergence of $\chi(\mu)$ at $\mu_c$ occurs for a finite value of $\bchi(\bmu)$, namely when the 
 denominator on the rhs of the second equation in \rf{4.69} vanishes. Thus $\bchi(\bmu)$ is a perfectly analytic function around
 the value $\bmu(\mu_c)$ and we can Taylor expand it around this value: 
 \beq\label{4.70}
 \bchi(\bmu(\mu)) = \big(1\plu G_{\mu_c}^{(1)}\big)^{2} + c\, (\bmu(\mu)- \bmu(\mu_c) = 
 \big(1\plu G_{\mu_c}^{(1)}\big)^{2} + c'\, (\mu- \mu_c)^{1-\g_s} 
 \eeq
 where we have used \rf{4.67e}. Inserting this result in \rf{4.69} we obtain a new value $1\mi \g_s$
 for the scaling exponent, which should be compared to the assumed value $\g_s$:
 \beq\label{4.72}
\chi(\mu)  \propto  \frac{1}{(\mu \mi \mu_c)^{1-\g_s}} \propto  \frac{1}{(\mu \mi \mu_c)^{\g_s}} 
\quad   \Rightarrow \quad \boxed{\g_s = \oh}
\eeq
 With this new knowledge let us return to the two-point relation \rf{4.68}. A Fourier transformation leads to 
 \bea\label{4.73}
 &&G^{(2)}_\mu(p) = \big(1\plu G_{\mu_c}^{(1)}\big)^{-2} \Big[ \bG^{(2)}_\bmu(p) + \bG^{(2)}_\bmu(p) \, G^{(2)}_\mu(p) \Big]  
  \\
&&\boxed{ G^{(2)}_\mu(p) = 
\frac{ \big(1\plu G_{\mu_c}^{(1)}\big)^{-2} \,\bG^{(2)}_\bmu(p)}{ 1 - \big(1\plu G_{\mu_c}^{(1)}\big)^{-2}  \bG^{(2)}_\bmu(p)} }
 \label{4.74} 
 \eea
 and we now expand $\bG^{(2)}_\mu(p)$ around $p \equ 0$, remembering that 
 $\bG^{(2)}_\mu(p \equ 0) = \int d^D x \bG^{(2)}_\mu(x) = \bchi(\bmu)$:
 \beq\label{4.75}
 \bG^{(2)}_\bmu(p) = \bG^{(2)}_\bmu(0)\mi \bar{c}(\bmu) \, p^2 + \cdots = \bchi(\bmu) \mi \bar{c}(\bmu) \, p^2 + \cdots ,
 \eeq
 Using this and \rf{4.70} we then obtain
 \beq\label{4.76}
 G^{(2)}_\mu(p) = \
 \frac{\big(1\plu G_{\mu_c}^{(1)}\big)^{-2}(\bchi(\bmu) \mi \bar{c} \, p^2 +\cdots)}{1 \mi  \big(1\plu 
 G_{\mu_c}^{(1)}\big)^{-2}(\bchi(\bmu) \mi \bar{c} \, p^2 +\cdots`)}~~ \underset{\mu \to \mu_c}{\to} ~~
  \frac{1 +\cdots}{c' \sqrt{\mu \mi \mu_c} + \bar{c} \,p^2 + \cdots}
 \eeq
 where the $\cdots$signifies terms of order $p^2$ or $\sqrt{\mu \mi \mu_c}$  in the numerator  and terms of order
 $p^4$, $(\mu \mi \mu_c)$ and  $p^2 \sqrt{\mu \mi \mu_c}$ in the denominator. Thus we can finally write
 \beq\label{4.77} 
\boxed{ G^{(2)}_\mu (p) \approx \frac{1}{\tilde{c} }\; \frac{1}{m^2(\mu) + p^2}, \quad m(\mu) = \check{c} \sqrt[4]{\mu \mi \mu_c}
\quad \nu= \oq, \quad d_H = 4}
 \eeq
 We thus have exponents identical to the ones encountered for BPs. 
 
 We have been working with dimensionless variables since eq.\ \rf{4.48}. In order to understand the relation of the 
 above results to  BPs it is convenient to reintroduce dimensions. We thus write 
\beq\label{4.81}
x \, a(\mu) = x_{ph} , \quad p = p_{ph} \, a(\mu),
\eeq
where $a(\mu)$ is a length-unit in $\mathbb{R}^D$. We can think of $a(\mu)$ as an average length of a 
link of the triangles in the triangulations, when they are mapped into $\mathbb{R}^D$. Thus $a(\mu)$ is not necessarily 
the same as the intrinsic link length $\ep(\mu)$ which appears in \rf{4.55} and which ensures that the 
intrinsic area of a typical triangulation is finite.  The  dependence of $a(\mu)$ on $\mu$ 
will be determined by the requirement that the functions $G_\mu^{(2)}(x,y)$ and
$G^{2)}_\mu(p)$ have a non-trivial behavior for $\mu \to \mu_c$. From \rf{4.77} and \rf{4.81} we see
that the natural way to obtain a non-trivial limit is to define a ``renormalized'' physical mass $m_{ph}$ by 
\beq\label{4.82}
m(\mu) = m_{ph} a(\mu),\qquad a(\mu) = \check{c} \sqrt[4]{\mu\mi \mu_c},
\eeq
which implies that 
\beq\label{4.83}
G^{(2)}_\mu(p)  ~ \underset{\mu \to \mu_c}{\to}~ \frac{1}{a^2(\mu)} \;\frac{1}{ m^2_{ph} + p^2_{ph}},   
\quad {\rm and}  \quad  m(\mu) |x| = m_{ph} |x_{ph}| 
\eeq
This way of taking the scaling limit is similar to the way we did it both for the free particle and for BPs, and we obtain the 
same result! Comparing \rf{4.82} to \rf{4.55} we see that
\beq\label{4.83a}
\boxed{\ep(\mu)  \propto \frac{m_{ph}^2}{\SL} \;a^2(\mu).}
\eeq
This is a relation similar to \rf{2.22} for the RW, and it reflects the same: since $d_H \equ 4$, the average area of
a surface from the path integral embbeded  and measured in $\mathbb{R}^D$ has an area $\la A_{ext}\ra \propto 1/(m_{ph}^4 a^2(\mu))$
if the individual triangles in $\mathbb{R}^D$ have an average area proportional to $a^2(\mu)$. If we insist that the 
average {\it intrinsic} area of a surface,  $\la A_{int}\ra$ , is finite, like in \rf{4.55}, the intrinsic length $\ep(\mu)$ assigned to a link 
in the triangulation has to be much smaller than $a(\mu)$, as is indeed expressed in relation \rf{4.83a}. Let us now
discuss  the BP-picture in more detail.

\begin{figure}[t]
\vspace{-.5cm}
\centerline{\scalebox{0.21}{\includegraphics{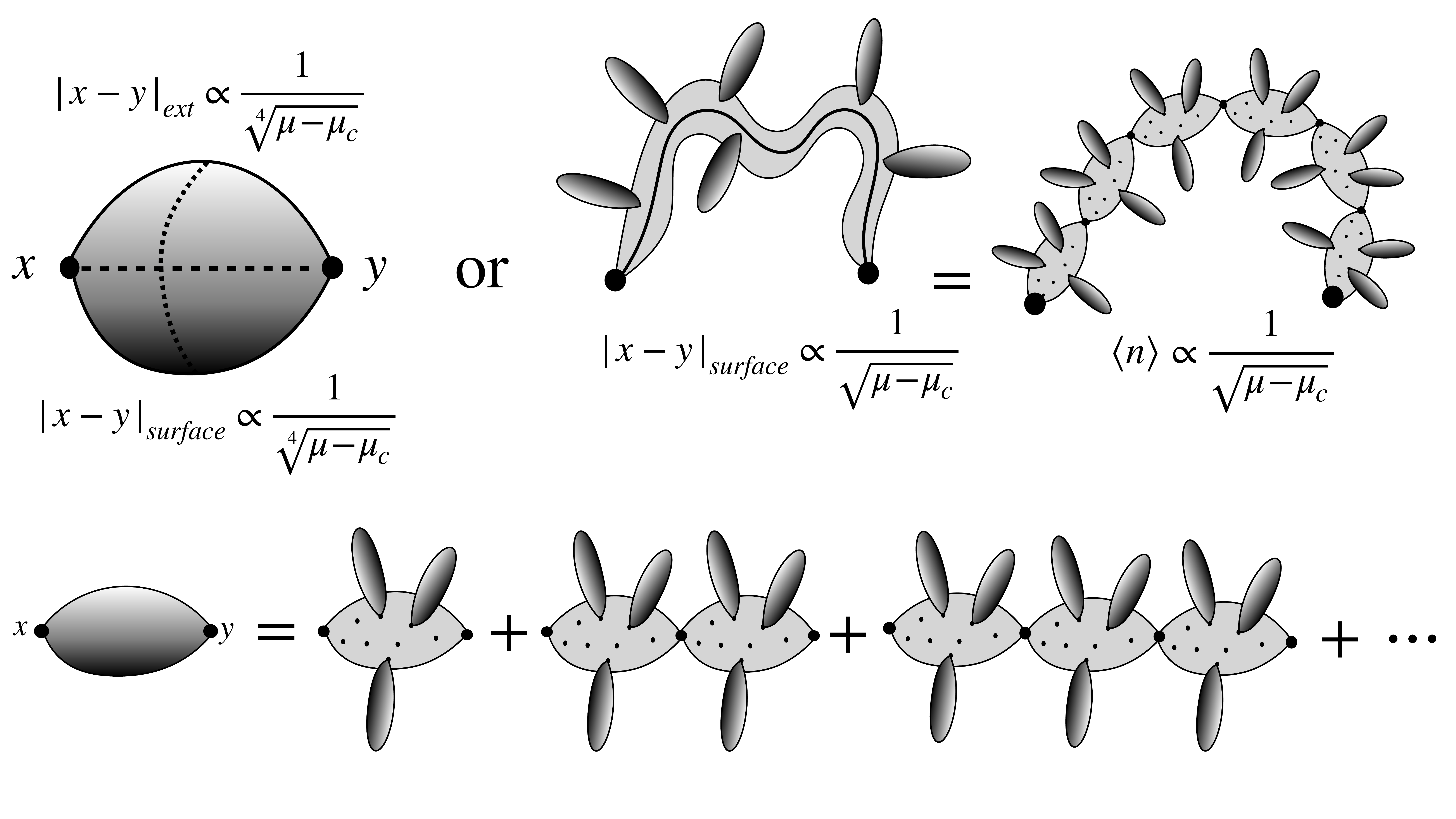}}}
\vspace{-.3cm}
\caption{{\small  $|x-y|_{ext} \propto (\mu \mi \mu_c)^{-1/4}$ in $\mathbb{R}^D$. How does a typical surface in the path integral 
connecting $x$ and $y$ look? The upper left figure shows a surface where the distance between $x$ and $y$ 
within the surface is of the same 
order as  $|x-y|_{ext}$. The middle upper figure shows a surface where the distance between $x$ and $y$ 
is much larger if one 
has to stay within the surface. The upper right figure shows how the middle figure is realized via the $\cTthree$ blobs which 
appear in the decomposition of $G^{(2)}_\mu(x \mi y)$ shown in Fig.\ \ref{fig4.14}, and spelt out in detail in the lower figure 
here. The average number of blobs is $\la n \ra \propto (\mu \mi \mu_c)^{-1/2}$.}}
\label{fig4.15}
\end{figure}

Consider $G_\mu^{(2)}(x\mi y)$ given by \rf{4.67a} as the partition function for surfaces with two marked vertices separated a
distance $|x\mi y|$ in $\mathbb{R}^D$. From the scaling of the Fourier transformed $G_\mu(p)$ given by \rf{4.83} it is 
clear that we have 
\beq\label{4.83b}
G_\mu^{(2)} (x \mi y) \underset{\mu \to \mu_c}{\to} a^{D-2}(\mu) \; G(x_{ph} \mi y_{ph}; m_{ph}),
\eeq
where   $G(x_{ph} \mi y_{ph}; m_{ph})$ is the continuum propagator \rf{1.12} of the free particle. 
From \rf{4.67a} and \rf{4.83b} we have for $\mu \to \mu_c$
\beq\label{4.83c}
\la |T| \ra_{G_\mu^{(2)}(x)} = -  \frac{1}{G_\mu^{(2)}(x\mi y)}\; \frac{\d}{\d \mu}\; G_\mu^{(2)}(x\mi y) \propto
 \frac{1}{\mu \mi \mu_c} \propto \frac{|x|^4}{m_{ph}^4 x^4_{ph}},
\eeq
explicitly showing that the Hausdorff dimension $d_H =4$.
 Similarly, using \rf{4.67d} we have 
 \beq\label{4.83d}
 \la |T| \ra_{G^{(1)}_\mu(x)} = -  \frac{1}{G^{(1)}_\mu(x)} \; \frac{\d}{\d \mu}\; G^{(1)}_\mu (x) \underset{\mu \to \mu_c}{\propto}
 \frac{1}{\sqrt{\mu \mi \mu_c}}
 \eeq
$\la |T| \ra$ is the average number of triangles and we 
have a picture of scaling {\it consistent} with a BP picture where triangles play the role of the 
links in the BPs and $G^{(1)}_\mu(x)$ plays the role 
of the rooted BP partition function $Z(\mu)$. For the ensemble of surfaces defined defined by $G^{(2)}_\mu(x\mi y)$, the smallest
number of triangles needed to connect $x$ and $y$ is of order $|x_{ph} \mi y_{ph}|/\sqrt[4]{\mu \mi \mu_c}$. But how do
we know that the surfaces really look like BPs? The situation is illustrated in Fig.\ \ref{fig4.15}. 

The answer comes from 
the decomposition of $G_\m^{(2)}(x\mi y)$ shown in Fig.\ \ref{fig4.14}. The iteration of the figure or eq. \rf{4.74} leads to 
\beq\label{4.83e}
G^{(2)}_\mu(p) =  \frac{\bG^{(2)}_\bmu(p)}{\big(1 \plu G_\mu^{(1)}\big)^2}+ 
\left(\frac{\bG^{(2)}_\bmu(p)}{\big(1 \plu G_\mu^{(1)}\big)^2}\right)^2 + 
\left(\frac{\bG^{(2)}_\bmu(p)}{\big(1 \plu G_\mu^{(1)}\big)^2}\right)^3 + \cdots 
\eeq
The important point here is that $\bG^{(2)}_\bmu(p)$ is not critical as $\mu \to \mu_c$ and $G_\mu^{(1)} \to G_{\mu_c}^{(1)}$
for $\mu \to \mu_c$, but $G^{(2)}_\mu(p) $ diverges as shown in \rf{4.83}. Note that 
\beq\label{4.83f}
\sum_{n=1}^\infty x^n = a  \Rightarrow \la n \ra = a+1, \qquad \la n\ra := \frac{\sum_n n \,x^n}{\sum_n x^n}.
\eeq
We thus conclude that the number of ``blobs'' in Fig.\ \ref{fig4.15} is $\la n \ra \propto 1/\sqrt{\mu \mi \mu_c}$. 
{\it For each blob we only have a finite number of triangles associated with 
a triangulation $\bT \in \cTthree$ since $\bmu(\mu_c) > \bmu_c$} (see \rf{4.69a}). Thus a shortest path 
in a typical triangulation connecting $x$ and $y$ will be of length proportional to $1/\sqrt{\mu \mi \mu_c}$ since it has to pass 
through all the blobs. This is much longer than the shortest  pass in $\mathbb{R}^D$ between $x$ and $y$, which, as 
mentioned, is of order $ |x_{ph} \mi y_{ph}|/\sqrt[4]{\mu \mi \mu_c}$. It shows that we can consider the 
$ \cTthree$-part of a blob as an effective ``BP-link'' and and these links then perform a RW from $x$ to $y$, increasing the 
length from being proportional to $1/\sqrt[4]{\mu \mi \mu_c}$ to being proportional to the square of this. 
The image of the $\cTthree$-part of a $\cTtwo$-triangulation in $\mathbb{R}^D$ thus  
effectively determines the  geodesic distance between the marked points $x$ and $y$ 
if we are forced to stay within the surface. The analogue for ``real'' BPs is that $x$ and $y$ is connected by 
a {\it unique} shortest link-path in a given BP and this link-path is then mapped to a RW path between $x$ and $y$ 
in  $\mathbb{R}^D$. The typical number of links in this RW will be proportional to $1/\sqrt{\mu \mi \mu_c}$.
Then outgrowths in the form of rooted BPs are attached to the vertices of the shortest path between $x$ and $y$ and 
the average number of links or vertices in such a roooted BP is also proportional to $1/\sqrt{\mu \mi \mu_c}$.  
We have the same situation here for the surfaces. The finite number of outgrowths attached to the $\cTthree$-part of a blob are $G_\mu^{(1)}$-outgrowths, which each contain a number of triangles proportional to $1/\sqrt{\mu \mi \mu_c}$. 
 All together one thus has $1/(\mu \mi \mu_c)$ triangles in  $G^{(2)}_\mu(x) $ as there should be according to \rf{4.83c}. 
This shows that the BP-picture indeed is the correct one for the surfaces which dominate the path integral defining 
$G_\m^{(2)}(x\mi y)$.

We saw in  Section 3 that while the Hausdorff dimension of BPs embedded in $\mathbb{R}^D$ is 4, the intrinsic 
Hausdorff dimension was 2. The reason was simply that the shortest path between two point, staying in the 
BP is in average much longer than the distance between the two points, measured in $\mathbb{R}^D$. As we have just discussed
we have precisely the same phenomenon for our surfaces. In the next Section we will define and discuss in detail the so-called intrinsic Hausdorff dimension for triangulations, but it is clear from the above discussion that for 
the typical triangulations which we meet in the path integral for the bosonic string, we will  find that the ``intrinsic''
Hausdorff dimension is also two because the ``intrinsic distance'' between the points $x$ and $y$, i.e.\ the number
of links or number of triangles one has to pass trough in order to reach from $x$ to $y$ will be of 
order $1/\sqrt{\mu \mi \mu_c}$, while the total number of triangles in a typical triangulation is of  order $1/{(\mu \mi \mu_c)}$.

 \subsubsection*{Scaling of the string tension}
 
Theorem 3, below eq.\ \rf{4.51b}, tells us that  there exists a string tension for $\mu > \mu_c$.
However, it does not tell us if $\sg(\mu)$ scales to zero for $\mu \to \mu_c$. Also, it does not tell us anything about 
subleading corrections to $G_\mu(\ell_A)$ when $A \to \infty$, where $A$ is the area of the planar loop, 
embedded in $\mathbb{R}^D$. There can be many subleading correction, but a generic 
correction comes from the setup, which is such that the length of the boundary  $\ell_A$ has to go to infinity when $A$ goes 
to infinity. This will create a term in the exponential part of $G_\mu(\ell_A)$, depending on the length of the boundary.
Until now we have used the same notation for the set of boundary links $\ell$, viewed  
as a boundary in the triangulation and this boundary mapped 
into $\mathbb{R}^D$. Let us now denote the number of links in $\ell$ by $|\ell |$ and the length of the boundary, mapped 
to $\mathbb{R}^D$ by $L_A$. In general  we expect a behavior
\beq\label{4.78}
G_\mu (\ell_A) = \e^{-\sg(\mu) A- \lam(\mu) L_A+ \cdots}
\eeq
As long as $L_A/A \to 0$, this second term will play no role and does not appear in theorem 3.

\begin{figure}[t]
\vspace{-.5cm}
\centerline{\scalebox{0.20}{\includegraphics{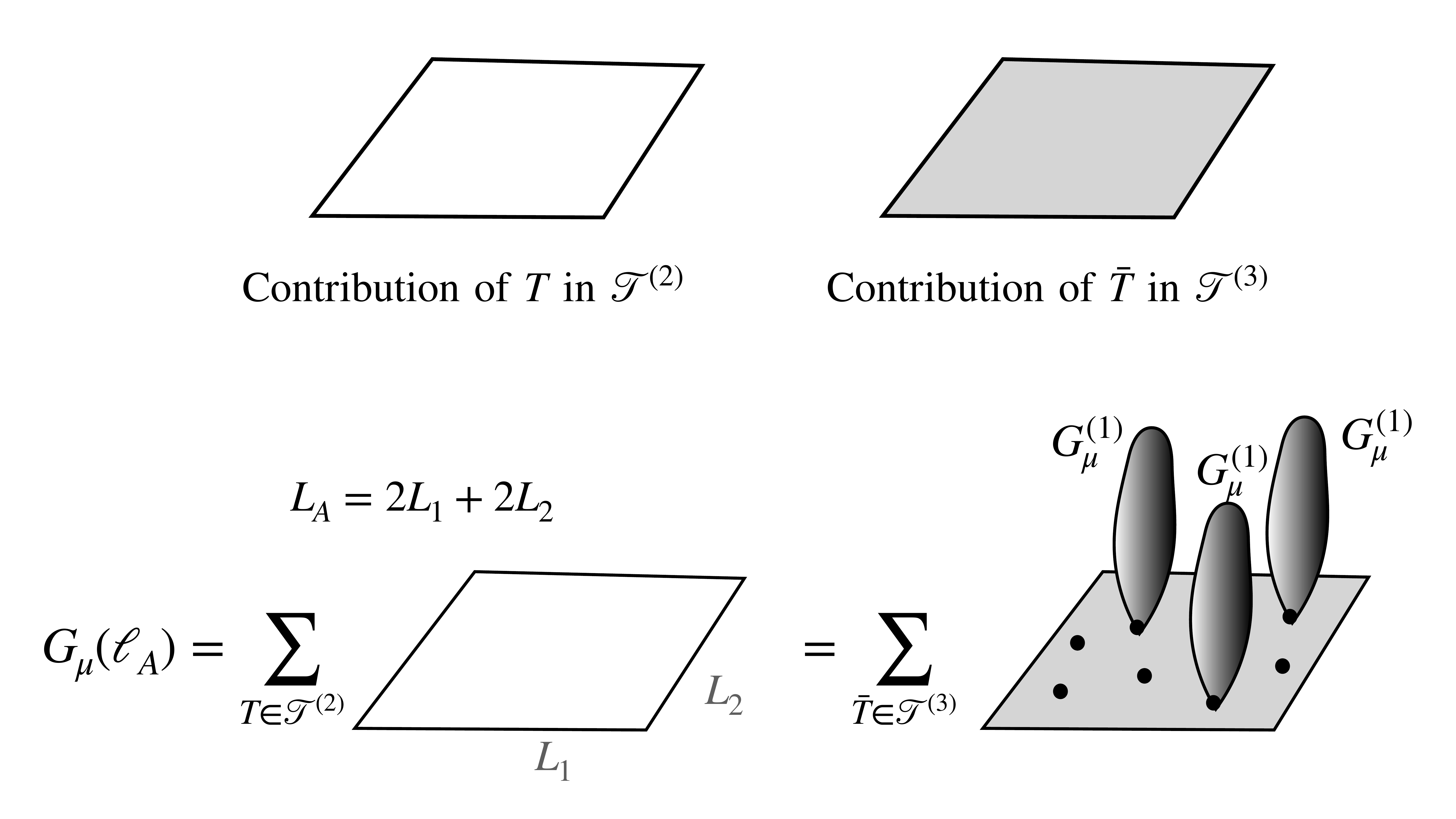}}}
\vspace{-.3cm}
\caption{{\small  The decomposition of triangulations $T \in {\cal T}^{(2)}$ contributing to of $G_\mu(\ell_A)$ into triangulations 
 $\bT \in {\cal T}^{(3)}$ together with outgrowth from some of the links, cut open to double-links. The grey surface 
represents a triangulation $\bT \in {\cal T}^{(3)}$, the dots on $T$ symbolize links, and the dots associated with outgrowths
symbolize links cut open to double-links to which there are $G^{(1)}_\mu$ outgrowths attached.}}
\label{fig4.16}
\end{figure}
Again we can write down the relation between $G_\mu(\ell_A)$ and $\bG_\bmu(\ell_A)$, as illustrated in 
Fig.\ \ref{fig4.16}:
\beq\label{4.79}
G_\mu (\ell_A)  = \bG_\bmu(\ell_A)\; \big(1 \plu G^{(1)}_\mu\big)^{-|\ell_A |} =
 \bG_\bmu(\ell_A)\;\e^{ -\frac{2}{3} (\mu \mi \bmu) |\ell_A |}.
\eeq
As long as $|\ell_A | \leq c \, L_A$ we  conclude from \rf{4.78}  that 
\beq\label{4.79a}
\boxed{\sg(\mu) = \bar{\sg}(\bmu)}
\eeq
but even if that is not satisfied we have $\sg(\mu) \geq \bar{\sg}(\bmu)$.  
Since we have already seen that for $\mu \to \mu_c$ we have $\bmu(\mu_c) > \bmu_c$, 
we know from theorem 3
that $\bar{\sg}(\bmu(\mu_c)) > 0$, and thus that $\sg(\mu_c) > 0$.  Conclusion:  {\it the string tension is not scaling 
to zero for $\mu \to \mu_c$.} 

It is also possible (and relatively easy) to prove directly that the string tension does not scale to zero for $\mu \to \mu_c$. The only assumption 
used is $|\ell_A |  \leq c \, L_A$. The proof is based on the simple estimate that for any triangulation where the $|\ell_A |$ boundary points are 
distributed along the boundary we have for the action \rf{4.48} 
\beq\label{4.80} 
S[X,T]  \geq \sum_{t \in T} A_t \geq A.
\eeq
This follows from the fact that  the area $A_t$ of a triangle spanned by points $X(v_1)$, $X(v_2)$ and $X(v_3)$ is less than or equal to one fourth of the 
squares of the lengths of any two of its sides. Of course the sum of $A_t$'s is larger than or equal to $A$, the minimal area associated with a surface
with planar boundary of length $L_A$. Using \rf{4.80} one can show that $\sg(\mu) \geq 1$ for $\mu > \mu_c$ as long as 
$|\ell_A |  \leq c \, L_A$ (for details consult the book {\it Quantum Geometry, a statisical field theory approach} \cite{book}).

Before discussing the physical consequences of the non-scaling of the string tension, note that the first correction
to this result can easily be calculated from $\sg(\mu) = \bsg(\bmu)$ since we have:
\beq\label{4.80a}
\frac{d \sg(\mu)}{d \mu} =  \frac{d \bmu}{d \mu} \, \frac{d \sg(\mu)}{d \bmu} =
\frac{d \bmu}{d \mu} \, \frac{d \bsg(\bmu)}{d \bmu} \propto \frac{d \bmu}{d \mu} 
\propto \frac{1}{\sqrt{\mu \mi \mu_c}},
\eeq
where we first use that $\sg(\mu) =\bsg(\bmu)$, next that  $\bsg(\bmu)$ is analytical around $\bmu(\mu_c)$, and 
finally  \rf{4.67e} with $\gamma_s = 1/2$. Integrating this relation we obtain
\beq\label{4.80b}
\boxed{\sg(\mu) = \sg(\mu_c) + c \, \sqrt{\mu \mi \mu_c} + \cO(\mu \mi \mu_c), \qquad \sg(\mu_c) > 0}
\eeq  

Let us now discuss the physical consequence of this non-scaling. The basic scaling, already introduced for the 
two-point function in \rf{4.82} and \rf{4.83} ensured that $e^{- m(\mu) |x|}$ survived in the limit $\mu \to \mu_c$ as 
$e^{-m_{ph} |x_{ph}|}$. The natural extension of this is to ensure that $e^{-\sg(\mu) A}$ survives 
in the scaling limit as $e^{- \sg_{ph} A_{ph}}$. Thus we demand
\beq\label{4.84}
\sg(\mu) \, A = \sg_{ph}  A_{ph}, \quad  A_{ph} := A a^2(\mu), \quad {\rm i.e.} \quad \sg_{ph} = \frac{\sg (\mu)}{a^2(\mu)}.
\eeq  
The only way to obtain a finite $\sg_{ph}$ for $\mu \to \mu_c$ is to have a scaling $\sg(\mu) \propto \sqrt{\mu \mi \mu_c}$, but 
from \rf{4.80b} we see it is not the case, although (tantalizing!) the correction to the constant term has the 
right dependence. We conclude from \rf{4.84} and \rf{4.80b} 
that $\boxed{ \sg_{ph} = \infty}$.

\begin{figure}[t]
\vspace{-.5cm}
\centerline{\scalebox{0.18}{\includegraphics{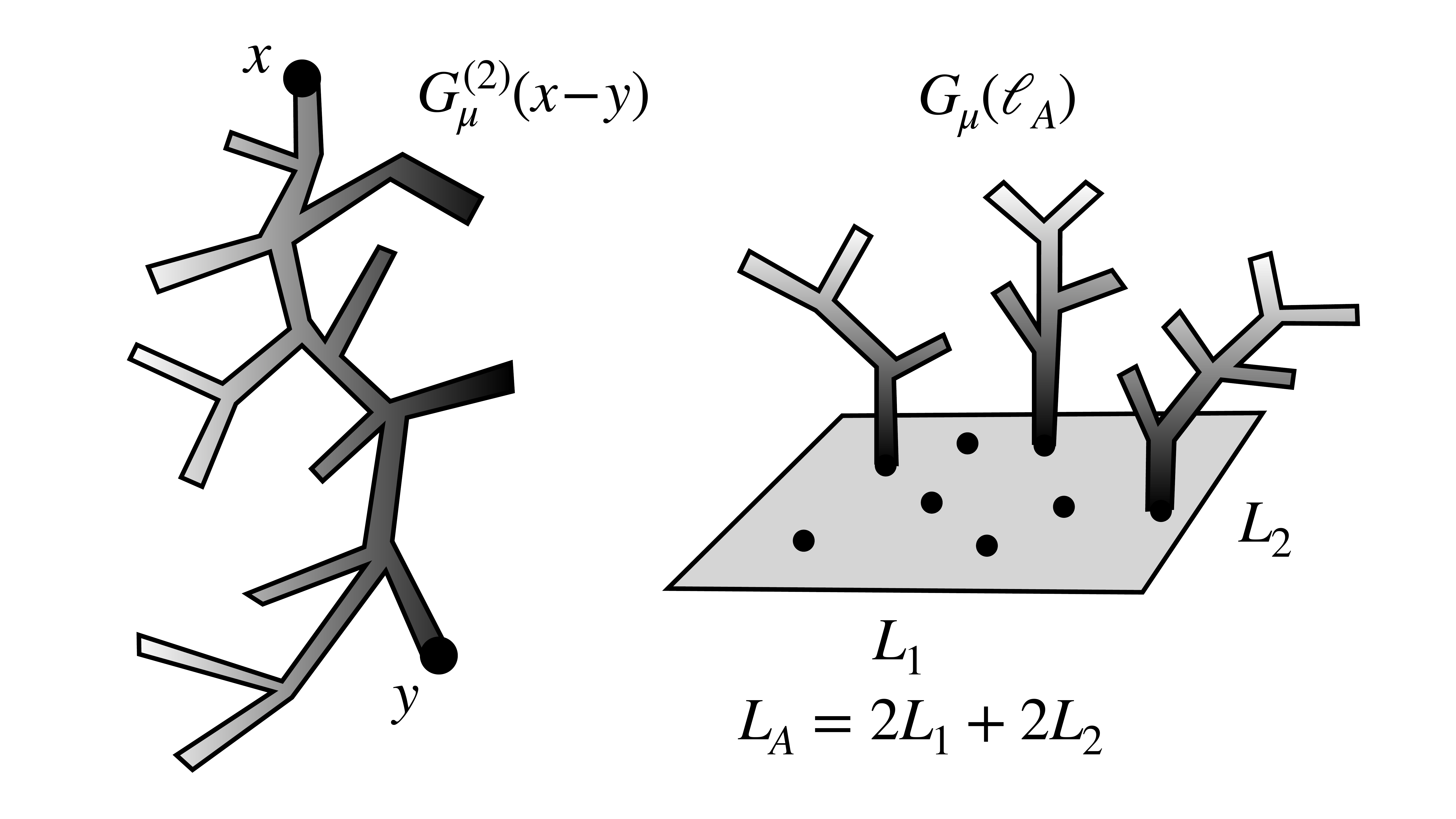}}}
\vspace{-.3cm}
\caption{{\small  The BP-surfaces contributing to the two-point function $G_\mu^{(2)}(x\mi y)$ and the
BP outgrowths from  the minimal surface for $G_\mu(\ell_A)$. Dots symbolize links and dots with outgrowths links
cut open to double links.  }}
\label{fig4.17}
\end{figure}

What does a typical surface in the path integral contributing to $G_\mu(\ell_A)$ look like? The average number of 
triangles in such a surface is 
\beq\label{4.85}
\la |T| \ra_{G_\mu(\ell_A)} = -\frac{1}{G_\mu(\ell_A)} \; \frac{\d}{\d \mu} \, G_\mu(\ell_A) 
\propto \frac{A}{\sqrt{\mu \mi \mu_c}} \propto \frac{A_{ph}}{\mu \mi \mu_c},
\eeq
where we have used \rf{4.78} , \rf{4.80a} and \rf{4.84}. The number of triangles needed to cover a surface with
``dimensionless''  area $A$ is  just $A$ up to a trivial factor, and  $A \equ  A_{ph}/a^2(\mu)$. 
This is basically the number of (blue) $\cTthree$-triangles in a typical 
surface in the decomposition made in Fig.\ \ref{fig4.16}. For each link in that minimal  $\cTthree$-surface we can 
potentially have a $G_\mu^{(1)}$ outgrowth with a number of triangles proportional to $1/\sqrt{\mu \mi \mu_c}$. 
The total number of triangles for such a surface is thus precisely the number calculated in \rf{4.85} and a typical 
surface is therefore well represented by a minimal surface with $G^{(1)}_\mu$-outgrowths everywhere. Since these outgrowths
also are BPs we have a picture like the one presented in Fig.\ \ref{fig4.17}.

This theory of minimal surfaces with BP-like excitations seems to have nothing to do with what is usually known 
as bosonic string theory where one has a finite string tension and where  in addition the lowest mass
excitation is a {\it tachyon}, i.e. a particle with a mass where $m^2 < 0$. The tachyon is a sign of a sick theory 
(all kind of disasters happen when $m^2 <0$, but we will not go into details here). Our statistical theory does 
not have this sickness, since we showed by very general arguments that our propagator has to fall off exponentially,
i.e.\ the lowest mass excitation has $m \geq 0$. Is it possible to make contact to the standard bosonic string theory. The answer
seems to be yes as hinted by the correction term in eq.\ \rf{4.80b} and it seems to be linked to the problem of having 
$| \ell_A |$ fixed boundary points, $| \ell_A | \to \infty$, which at the same time is assigned an intrinsic length 
$\ell_{intrin} =  | \ell_A | \, \ep(\mu)$ and an extrinsic length $\ell_{extrin} =  | \ell_A | \, a(\mu)$, where $\ep$ and 
$a$ are related by \rf{4.83a}. However, since there are still subtleties associated with this resolution of the difference 
between the formally defined continuum bosonic string and our regularized (well defined) bosonic string, we will not 
discuss the topic further.

 \newpage
 
 \setcounter{figure}{0}
\renewcommand{\thefigure}{5.\arabic{figure}}
 \setcounter{equation}{0}
 \renewcommand{\theequation}{5.\arabic{equation}}

 \section*{5. Two-dimensional quantum gravity}
 
 \subsection*{Solving 2d quantum gravity by counting geometries}
 
 We now consider the case where we have no Gaussian matter fields $X_i$ coupled to two-dimensional quantum gravity.
 The Einstein-Hilbert action is given by \rf{2.3a} with $M \!=\! 2$. We have already seen that for two-dimensional
 gravity the curvature term is topological (eqs.\ \rf{4.36} and \rf{4.37}) and thus does not contribute to any dynamics 
 unless we consider processes where the topology changes. On the other hand we have already discussed 
 the problems with two-dimensional quantum gravity and topology changes in Sec.\ 4: ``Digression: summation over
 topologies", so in the following we are going to restrict ourselves to two-dimensional manifolds which have the 
 topology of the sphere ($h \!=\! 0$), but with a number $n$, $n \geq 0$,of boundaries. It is convenient to associate
 independent boundary cosmological constants $Z_i$ to each boundary $i$. In this way our (trivial) action will 
 be (dropping the curvature term in the Einstein-Hilbert action)
 \beq\label{5.1}
 S[g,\Lam] = \Lam \!\int d^2 \xi \, \sqrt{g(\xi)}, \qquad \mbox{no boundary cosmological constants}
 \eeq
 where we denote $\Lam$ the cosmological constant (in our old notation it would be $2\Lam/2\pi G$), and 
 including boundary cosmological constants  
 \beq\label{5.2} 
 S[ g, \Lam, Z_1,\ldots,Z_b] =S[g,\Lam]  + \sum_{i=1}^n Z_i \!\int \!\!d s_i =   
 \Lam\, V_g \plu \sum_{i=1}^n Z_i \,L_{i,g},
 \eeq
 where $V_g$ is the volume of spacetime and $L_{i,g}$ is the length of boundary $i$, calculated using the metric $g$.
 We now define the following partition functions, depending on the boundaries:
 \bea\label{5.3}
 W(\Lam;Z_1,\ldots,Z_n) &=& \int \cD [g] \; \e^{-S[ g, \Lam, Z_1,\ldots,Z_n] },\\
 W(\Lam;L_1,\ldots,L_n) &=& \int \cD [g] \; \e^{-S[ g, \Lam] } \prod_{i=1}^n \del (L_i \mi L_{i,g}), \label{5.4}\\
 W(V;L_1,\ldots,L_n) &=& \int \cD [g] \; \del(V-V_g)  \prod_{i=1}^n \del (L_i \mi L_{i,g}), \label{5.5}
 \eea
 $W(\Lam;Z_1,\ldots,Z_n) $ and $W(\Lam;L_1,\ldots,L_n) $ are related by a Laplace transformation:
 \beq\label{5.6}
 W(\Lam;Z_1,\ldots,Z_n)       =\int_0^\infty \prod_{i=1}^n dL_i\; \e^{-Z_i L_i} \;W(\Lam;L_1,\ldots,L_n) 
 \eeq
 and likewise, $W(\Lam;L_1,\ldots,L_n)$ and $ W(V;L_1,\ldots,L_n)$ are related by a Laplace transformation:
 \beq\label{5.7}
W(\Lam;L_1,\ldots,L_n) = \int_0^\infty dV \; \e^{-\Lam \, V}\; W(V;L_1,\ldots,L_n).
\eeq
 From eq.\ \rf{5.5} it is seen that 
 \beq\label{5.8}
  W(V;L_1,\ldots,L_n) = \mbox{$\#$ of geometries with volume $V$ and boundary-lengths $L_i$}
  \eeq
  It follows that these partition functions of two-dimensional quantum gravity are completely determined if 
  we can {\it count the number of geometries with volume $V$ and boundary lengths $L_i$} and that these partition
  functions in that sense are {\it entirely entropic}. A main result in this Secion will be that we can 
  perform this counting and find
  \beq\label{5.9}
  \boxed{W(V;L_1,\ldots,L_n)  \propto V^{n-7/2}\; \sqrt{L_1\cdots L_n} \; 
  \exp\! \Big(\mi\frac{ (L_1 \plu\cdots \plu  L_n)^2}{4V}\Big) }
  \eeq
  As usual, in order to perform  this counting we first need a regularization of the geometries, and we have it 
  already, namely the one we used when discussing the bosonic string: Dynamical Triangulations (DT), where 
  we consider the subset of geometries defined by equilateral triangles:
  \beq\label{5.10}
  \int \cD [g] \to \sum_{ T \in \cT}
  \eeq
  where $\cT$ denotes a suitable class of equilateral triangulations. As already discussed in the case 
  of the bosonic string, if we have a triangulation $T$ with $|T|$ triangles, and boundaries with $l_i$ links 
  and an assignment of length $\ep$ to the links, we relate  the continuum quantities to the DT quantities 
  by writing
  \beq\label{5.11}
  V \sim |T| \, \ep^2, \qquad  L_i \sim l_i \, \ep
  \eeq
  and we will take a limit where $\ep \to 0$ while $|T|$ and $\ell_i$ go to infinity in such a way that $V$ and $L_i$
  stay fixed. In that limit number of triangulations will be 
  \bea\label{5.12}
  \lefteqn{w(|T|, l_i,\ldots,l_n) =} \\
  && c \, \e^{\mu_c |T|} \; \e^{\lam_c (l_1 + \cdots + l_n)} |T|^{n-7/2} \sqrt{l_1 \cdots l_n}
  \exp\Big(\mi \frac{(l_1 \plu \cdots \plu l_n)^2}{\tilde{c} |T|} \Big) \; \Big[ 1\plu \cdots \Big]\no
  \eea
  where $c$ and $\tilde{c}$, as well as $\mu_c$ and $\lam_c$ depend on the specific set $\cT$ of equilateral
  triangulations we are using, and the $\cdots$ indicate subleading corrections in $|T|$ and $l_i$. 
  The exponential growth, depending on $\mu_c$ and $\lam_c$ will 
  not survive when we convert the counting formula \rf{5.12} to the continuum formula \rf{5.9}, but it is important 
  for being able to make this conversion that the 
  number of triangulations only grows exponentially with $|T|$ and that is only the case if we restrict the 
  topology, i.e. the number of handles $h$ of the two-dimensional manifold. In the following $h \equ 0$.    
  
  \subsection*{Counting triangulations of the disk}
  
\begin{figure}[t]
\centerline{\scalebox{0.2}{\includegraphics{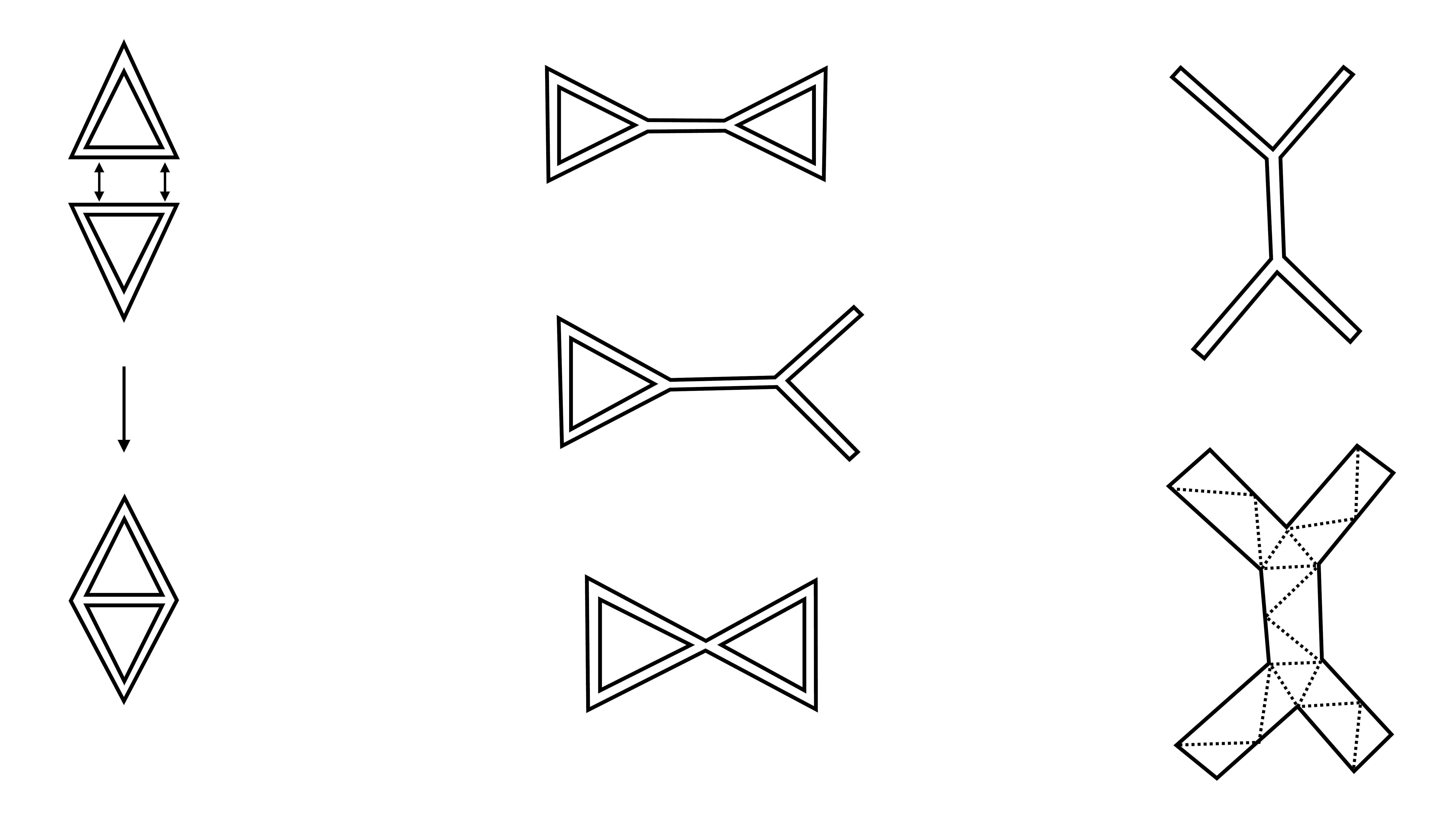}}}
\vspace{-0.5cm}
\caption{{\small  To the left: gluing together two triangles. In the middle: examples of so-called unrestricted triangulations. To the right: double-lines can be 
imitated by regular triangulations.}}
\label{fig5.1}
\end{figure}

  In order to count the triangulations we have to define the class of triangulations we want to count. In Sec.\ 4 we 
  defined two classes of triangulations, $\cTthree$ and $\cTtwo$. In particular $\cTthree$ is a natural class and 
  one can indeed use it (and it has been done). However, we will here choose a somewhat larger class, which at 
  first seems unnatural, but, as we will shortly argue, should be perfectly suitable for extracting a continuum limit 
  when $\ep \to 0$. We will denote the class $\cTz$ and call it {\it unrestricted triangulations}. The main reason 
  for choosing this class is that the counting is easier that for $\cTthree$. Let us consider triangulations with 
  one boundary, i.e.\ triangulations of a disk. Let us use a so-called double-line notation, where we represent
  the triangles as shown in Fig.\ \ref{fig5.1}, and where they are glued together to form a larger triangulation, as
  also shown in the figure. 

  We now allow for more general boundaries as shown, where the boundary can consist of double links (which  
  should  be thought of as having an infinitesimal area between the links if they are not associated with a triangle), and where triangles may only share a single vertex. In all cases the 
  outer boundary lines form a closed curve. Allowing such ``degenerate'' triangulations should make no difference 
  in the $\ep \to 0$ limit, since one can alway imitate such a degenerate triangulation by a  regular one of width 
  $\ep$, again as illustrated in the figure. If it made a difference we should be worried about universality and 
  whether our discretization is a good  one. As mentioned one obtains identical results for $\ep \to 0$ for both regular 
  and unrestricted triangulations, although we are not going to prove this here (for a proof see \cite{book}).
  
\begin{figure}[t]
\vspace{-1cm}
\centerline{\scalebox{0.2}{\includegraphics{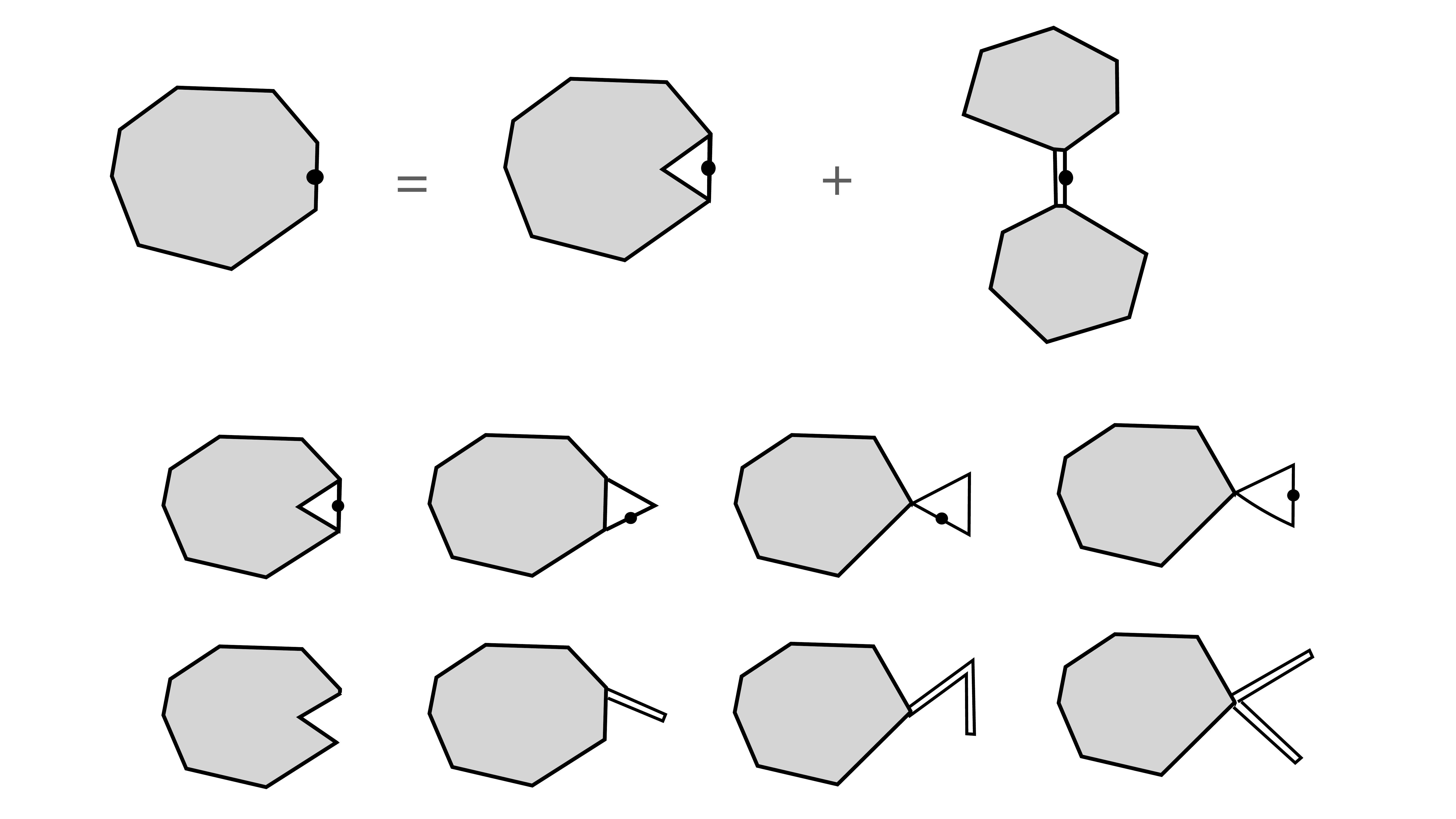}}}
\vspace{-0.2cm}
\caption{{\small  Top: Illustrating that the marked link either belongs to a triangle or a double-link. The two lower rows show  the various ways the triangle can be attached to the rest of the triangulation and some of the ways the mark can 
be placed on the triangle. When the triangle to be removed has more than
one boundary link, removing the triangle results in double-links, such that the new boundary always has one more link than the original boundary. We have 
not specified where the mark should be placed on the new boundary. In principle one is free to invent ones own procedure for that.}}
\label{fig5.2}
\end{figure}

  We now want to count the number of unrestricted triangulations of the disk, where we have marked a link 
  on the boundary. This marking is done in order to avoid some symmetry factor problems in the counting, and it was 
  for the same reason we considered rooted BPs, rather than just BPs, when counting those. We denote 
  the number of triangulations of the disk with $k$ triangles and $l$ boundary links (where one link is marked) by $w_{k,l}$. 
  For convenience we  define $w_{0,0} =1$ (and represents it graphically as a point (a dot)). Further it is natural to define
  $w_{k,0} =0$ for $k>0$ (no triangulations of 
  the disk unless we have a boundary). Finally,
  we denote the generating function for the numbers $w_{k,l}$ by $z\, w(g,z)$:
  \beq\label{5.13}
  z \, w(g,z) = \sum_{k=0}^\infty \sum_{l =0}^\infty g^k \,z^{-l} \, w_{k,l}.
  \eeq  
   A factor $g$ is associated with each triangle and a factor $z^{-1}$ with each boundary link\footnote{In combinatorics 
   one uses the word {\it indeterminate} for the variables like $g$ and $z^{-1}$ in the generating functions. We will not do that here.}. 
   We are using $z^{-1}$ rather than $z$ to enumerate the number of boundary links because the analytic 
   structure of $w(g,z)$ in the complex $z$-plane will be simpler. For the same reason we write 
   $z w(g,z)$ instead of $w(g,z)$ for the generating function to ensure that $w(g,z) \to 0$ for $|z| \to \infty$.
   Thus:
   \beq\label{5.14}
   \boxed{w(g,z) = \sum_{k=0}^\infty \sum_{l =0}^\infty g^k \,z^{-(l+1)} \, w_{k,l} = \sum_{l =0}^\infty 
   \frac{w_l(g)}{ z^{l+1}}, \qquad w_l(g) =  \sum_{k=0}^\infty g^k w_{k,l}.}
   \eeq 
  Here $w_l(g)$ is the generating function for triangulations with $l$ boundary links. In particular we have 
  by definition $w_0(g) =1$ and thus
  \beq\label{5.15}
   w(g,z) \to \frac{1}{z} \quad {\rm for}
   \quad |z| \to \infty   
   \eeq
   Assume now we have a triangulation of the disk with a marked boundary link. The marked link can belong 
   to a triangle or a double link. This is illustrated in Fig.\ \ref{fig5.2}. Removing the triangle and marking a link 
   on the boundary of the remaining triangulation again leads to a marked triangulation of the disk, but with 
   one fewer triangle and a boundary where the boundary length is increased by one. Similarly, removing 
   the double-link the triangulation splits into two disconnected disks where we can also mark the boundaries.
   
   Let the original triangulation $T$ consist of $k$ triangles and $l$ boundary links. In the generating function
   it contributes with a term $g^k/z^{l+1}$. If it is decomposed to a triangle and new triangulation $T'$, we can write 
   this factor as $g \cdot z \cdot ( g^{k-1}/z^{l+2})$ where the factor in the bracket is the weight associated to $T'$
   in the generating function. If $T$ is decomposed into the two disconnected triangulations $T_1$ and $T_2$
   with the number of triangles $k_1$ and $k_2$ and the number of boundary links $l_1$ and $l_2$
   by removing the double-link, we have $k_1 \plu k_2 \equ k$ and $l_1 \plu l_2 \plu 2 \equ  l$. We can now decompose
   the factor $g^k/z^{l+1}$ associated with $T$ into $\frac{1}{z} (g^{k_1}/z^{l_1+1}) (g^{k_2}/z^{l_2+1})$.
   Summing over all triangulations, as done in the generating function, with weights  $g^k/z^{l+1}$, we then arrive 
   at the following equation:
   \beq\label{5.16}
   w(g,z)  ~ \approx ~g \,z \, w(g,z) + \frac{1}{z} \, w^2(g,z).
   \eeq
  This equation can easily be solved.  However, the devil is in the details and this is the reason we do not use 
  ``$=$'' but ``$\approx$'' in \rf{5.16}. The equation is not correct when the triangulations on the lhs have boundary lengths
  $l =0,1$. The $w(g,z)$ on the lhs contains the term $\frac{1}{z}$ coming from $\frac{w_{0,0}}{z}$, but this ``point''
  associated with $l=0$, cannot be found in the decomposition shown in Fig.\ \ref{fig5.2} since 
  the triangulation $T'$ has at least two links, and $T_1$ and $T_2$ are connected by a double-link. We thus have to add
  a term $1/z$ on the rhs of eq.\ \rf{5.16}. Similarly, since $T'$ has at least two links, we should subtract 
  from the $w(g,z)$ on the rhs of  eq.\ \rf{5.16}, associated with summing over $T'$ triangulations, the 
  terms $w_0(g)/z$ and $w_1(g)/z^2$ in notation of eq.\ \rf{5.14}. The corrected eq.\ \rf{5.16} is thus
  \beq\label{5.17}
  w(g,z) = \frac{1}{z} + gz\left[w(g,z) \mi  \frac{1}{z} \mi  \frac{w_1(g)}{z^2} \right] +  \frac{1}{z}\, w^2(g,z)
  \eeq
  or
  \beq\label{5.18}
 \boxed{ w^2(g,z) = \big(z \mi  g\, z^2\big) \, w(g,z) - \big(1\mi g(w_1(g) \plu z)\big)}
  \eeq 
  If we knew $w_1(g)$ this would be a simple quadratic equation for $w(g,z)$. Let us for a moment pretend that 
  we know $w_1(g)$ and define 
  \beq\label{5.19}
  V'(g,z) = z\mi  g z^2,\qquad Q(g,z) = 1\mi g\,\big(w_1(g) \plu z\big)
  \eeq
  Then
  \beq\label{5.20}
 \boxed{ w(g,z) = \oh \left( V'(g,z) - \sqrt{ \big(V'(g,z)\big)^2 \!- 4 Q(g,z)}~ \right)}
 \eeq
 where the square root should be chosen such that it for large $z$ has the expansion
 \beq\label{5.20a}
 V'(g,z) \Big( 1 - 2 \frac{Q(g,z)}{(V'(g,z))^2} - 2  \frac{Q^2(g,z)}{(V'(g,z))^4} - \cdots \Big) 
 \eeq
 ensuring that $w(g,z) \to 1/z$ for $z \to \infty$.
 
 \subsubsection*{Branched polymers}

\begin{figure}[t]
\vspace{-1cm}
\centerline{\scalebox{0.2}{\includegraphics{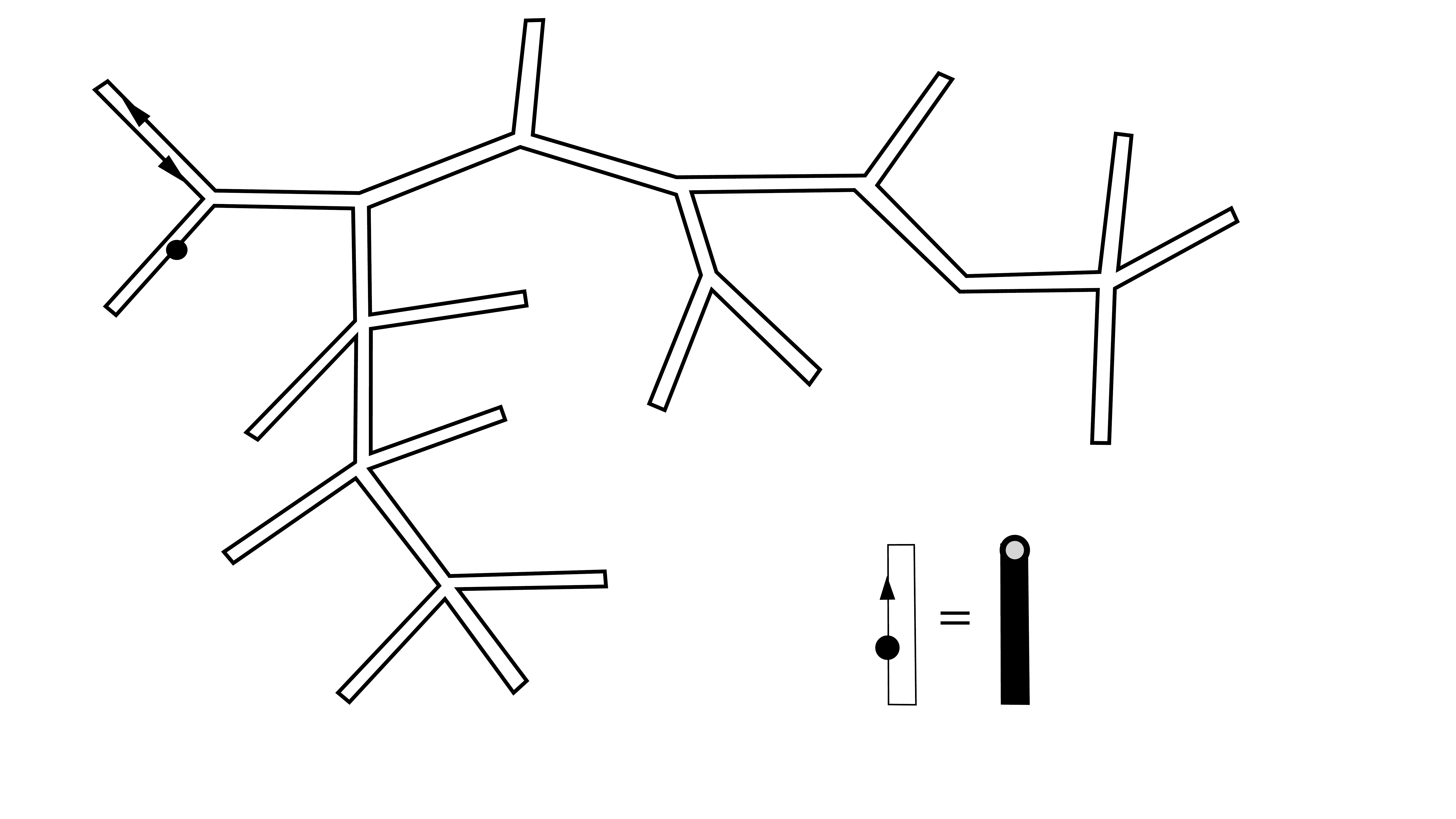}}}
\vspace{-0.9cm}
\caption{{\small Branched polymer of double-links with one link marked.}}
\label{fig5.3}
\end{figure}

\begin{figure}[t]
\centerline{\scalebox{0.2}{\includegraphics{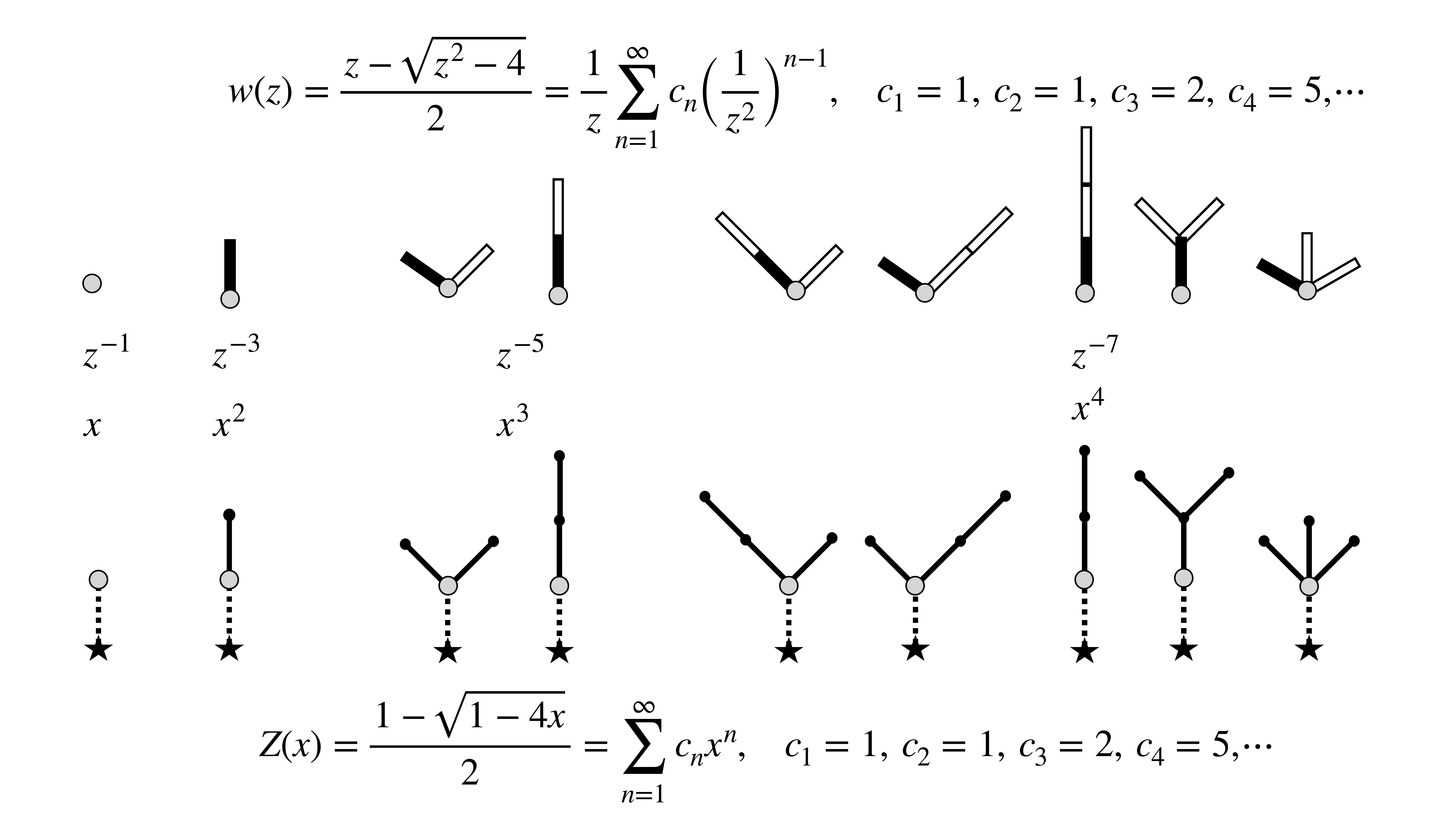}}}
\caption{{\small  Top: BPs (of double-links) with one link marked (black) and corresponding vertex marked (grey).
They are  in bijective correspondance with   rooted BPs with one more link. Bottem: the rooted BPs, where 
the root is shown as a star connected to the marked vertex (grey)  as described in the main text.
}}
\label{fig5.3a}
\end{figure}

 Let us chose $g\equ 0$, i.e.\ we have no triangles. The ``triangulations" are  thus  boundary graphs consisting double links, as 
 illustrated in Fig.\ \ref{fig5.3}. These can clearly be viewed as branched polymers, more or less like the rooted branched polymers, 
 only here the ``root'' is a mark on a double link. In addition, we mark one of the vertices of the double link. In this 
 way we have a vertex relative to which we can define ``height'' (the link distance) of the other vertices, precisely as we can
 define the height of vertices for rooted BPs as the link distance to the root. From \rf{5.20} we obtain the generating function 

 \beq\label{5.21}
 w(z) = \oh \Big(z\mi  \sqrt{z^2 \mi 4}\Big) = \sum_{l=0}^\infty \frac{w_{2l}}{z^{2l+1}} 
 \eeq
\beq\label{5.22}
w_{2l} = \frac{(2l)!}{(l\plu 1)! l!} ~\underset{l\to \infty}{\to}  ~\frac{1}{\pi} \, l^{-3/2} \, 4^l \, \Big(1 \plu \cO\big(\frac{1}{l}\big)\Big)
\eeq
and this is exactly the partition function for BPs where arbitrary branching is allowed with equal weight. Let us spell this out in some 
detail. First the labeling of the BPs. Labeling a link and one of its vertices of BPs is equivalent to introducing a root vertex of 
order one and connecting it to the marked vertex such that the marked link is last link one meets going around the marked vertex 
 counter clockwise, starting with the link connecting the root and the marked vertex. The rooted BP has one more 
link than our BP with a marked link and vertex, and the relation is illustrated in  Fig.\ \ref{fig5.3a} which also shows that 
in the standard rooted BP notation with\footnote{Rather than $x \sim 1/z$ we have $x \sim 1/z^2$ since the links are double
links, each component contributing a factor $1/z$.} $x \sim 1/z^2$ we have instead of $w(z)$ from \rf{5.21} 
the partition function $Z(x)$ determined by 
\beq\label{5.21a}
\frac{1}{x} = \frac{F(Z)}{Z}, \quad F(Z) = \sum_{n=1}^\infty Z^n = \frac{Z}{1\mi Z}, \quad Z(x) = \frac{1\mi \sqrt{1\mi 4x}}{2}.
\eeq
Thus the BPs related to  \rf{5.21} indeed have arbitrary branching with weight  1 and 
we have (of course) precisely the expected behavior for rooted BPs:
\beq\label{5.23} 
w_{2l} \sim l^{\g-2} \, \e^{\lam_c l}\quad {\rm for} \quad l \to \infty, \qquad \g \equ \oh,~~\lam_c \equ \ln 4,
\eeq
i.e.\ the number is growing exponentially. $w(z)$ is an analytic function in the complex $z$-plane 
except for a cut $[\,\mi 2,2]$ on the real 
axis. It has a power expansion in $1/z$, and the radius of convergence is when $z$, coming from infinity, meets the cut, i.e.\ when 
$\frac{1}{z_c} \equ \oh$.  {\it The exponential growth is determined by this point.}

\vspace{6pt}

\subsubsection*{Beyond branched polymers: the loop equation}

The function which appears under the square root in eq.\ \rf{5.20} is a fouth-order polynomial in $z$:
\bea
 f_g(z) &\equ& V'(g,z)\big)^2 \mi 4 Q(g,z) \equ  [z\mi c_1(g)][z\mi c_2(g)] [g z\mi c_3(g)] [g z\mi c_4(g)], \no \\
&& c_1(0) \equ2,~~c_2(0) \equ - 2,~~c_3(0) \equ c_4(0) \equ 1.\label{5.24}
 \eea
 The function $\sqrt{f_g(z)}$ is an analytic function in $1/z$ and the radius of convergence, $1/z_c$,  
 is the largest of the values $|c_1|,~|c_2|,~|c_3|/g,~|c_4|/g$, unless there are special circumstances. There are, as we will now argue. 
 From  \rf{5.21} it is seen that $w_1(0) =0$, and we can to lowest order in $g$ write
 \bea\label{5.25}
 V'(g,z)\big)^2 \mi 4 Q(g,z) &\equ&  z^2 \mi 4 \mi 2g z^3 \plu g^2 z^4 \plu  4g z + \cO(g^2) \\
 &\equ& [z \mi (2\plu 2g) \plu \cO(g^2)] [z \plu (2\mi 2g) \plu \cO(g^2)] [1 \mi g z \plu \cO(g)]^2 \no 
 \eea 
   \begin{figure}[t]
\vspace{-1cm}
\centerline{\scalebox{0.2}{\includegraphics{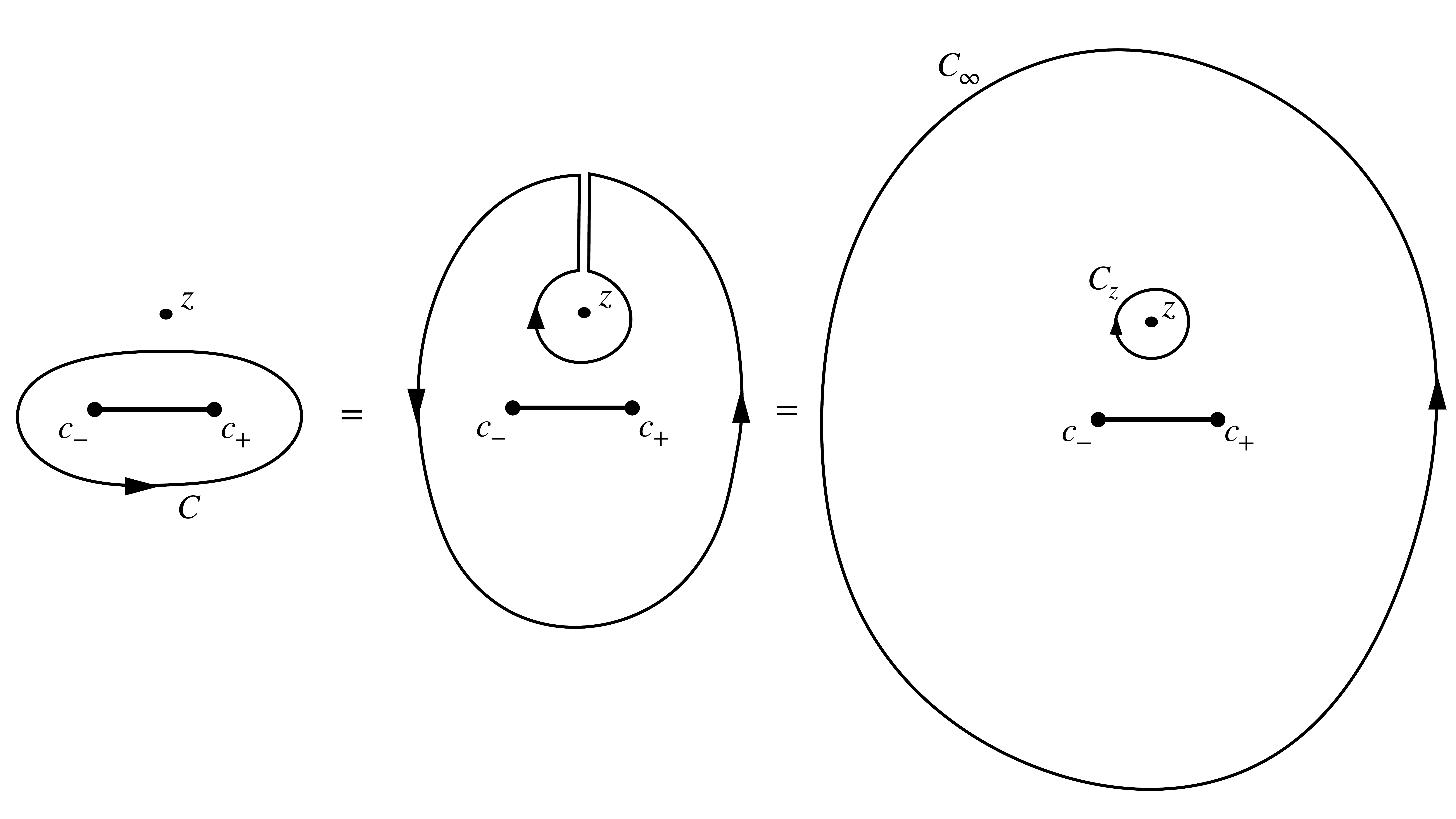}}}
\vspace{-0.2cm}
\caption{{\small  The cut $[c_-.c_+]$ and $z$ and the contours $C$, $C_\infty$ and $C_z$ in the complex plane.}}
\label{fig5.4}
\end{figure}
This implies that $c_i(g)$  change analytically with $g$ from their values \rf{5.24} and the radius of 
convergence of $\sqrt{f_g(z)}$ as a function of $1/z$ will be of order  $g\plu \cO(1)$ 
{\it unless  $c_3(g)$ is  {\it exactly} equal to $c_4(g)$},
in which case  $\sqrt{(g z \mi c_3(g))(g z\mi c_4(g))} = \pm (gz\mi c(g)$, $c(g) \equ c_3(g) \equ c_4(g)$ and 
does not determine the radius of convergence of  $\sqrt{f_g(z)}$. This is needed to avoid that the radius 
of convergence jumps discontinuous from being $1/2$ to $g$ when $g$ changes from being zero to non-zero.
There cannot be such a jump, 
 since the radius of convergence  determines the exponential growth of the 
 number of boundaries for a fixed number of triangles. Having zero triangles or one triangles and then only boundaries, should
 not have a dramatic effect on the number of boundaries. Then\footnote{In \rf{5.26} we have chosen the sign of the square root
 on the rhs to be positive when $z$ is large and positive. This choice is made to ensure  that the term $-g z^2$ in $V'(g,z)$ is cancelled by  the corresponding term coming from the square root, see \rf{5.20a}.}
 \beq\label{5.26}
 -\sqrt{V'(g,z)\big)^2 \mi 4 Q(g,z)} = (gz\mi c(g)) \sqrt{(z\mi c_+(g))(z\mi c_-(g))},
 \eeq
 where we have introduced the notation
 \beq\label{5.27}
 c(g) \equiv  c_3(g) \equ c_4(g),\quad c_1(g) \equiv c_+(g) > c_-(g) \equiv c_2(g),
 \eeq
 where $c(g),~c_\pm(g)$ are analytic around $g\equ 0$. We now conclude that \rf{5.20} can be written as 
 \beq\label{5.28}
 \boxed{w(g,z) = \oh \Big( z-gz^2 \plu (g z \mi c(g) ) \sqrt{(z-c_+(g))(z-c_-(g))} \Big)}
 \eeq 
 The requirement that $w(g,z) \equ  {1}/{z} + \cO(1/z^2) $ leads, by expanding in $1/z$, to three equations. The cancellation 
 of the term $gz^2$  is automatic, the term $z$  has to be cancelled by expanding the square root. 
 Also,  a constant term is not allowed in the expansion, and finally the term which goes like $1/z$ has to have the coefficient 1. 
 These three equations determine $c(g)$ and $c_\pm(g)$ and  lead to a third order equation which can be 
 solved explicitly. However, for our purpose we do not really need the explicit solution. 
 We only need the assumption that $c(g), c_\pm(g)$ are analytic functions in a neighborhood of $g \equ 0$ (which 
 can be checked from the explicit solution).
 
 $w(g,z)$ is  now an analytic function in the complex $z$-plane, except for a cut $[c_-(g),c_+(g)]$ and the radius of convergence 
 of the power series in $1/z$ is detemined by $|c_+(g)|$ (we have $|c_+(g)| \geq |c_-(g)|$) which is an increasing function of $g$, the 
 reason being that with increasing $g$ we have an increased probability of having more triangles and then a larger number of different
 boundaries.  Using eq.\ \rf{5.14} we can now rewrite equation \rf{5.18} in the following way 
 \beq\label{5.29}
 \boxed{\oint_C \dom \; \frac{V'(g,\om)}{z-\om} \; w(g,\om) = w^2(g,z),\quad \mbox{ the loop equation}}
 \eeq
 The contour $C$ encloses the cut $[c_-(g),c_+(g)]$, but not the point $z$, as shown in Fig.\ \ref{fig5.4}. 
 Deforming the contour as 
 also shown in Fig.\ \ref{fig5.4} and using the expansions 
 \beq\label{5.30}
 \frac{1}{z-\om} = - \frac{1}{\om} \sum_{k=0}^\infty \Big( \frac{z}{\om} \Big)^k, \qquad w(g,\om) = \sum_{l=0}^\infty \frac{w_l(g)}{\om^{l+1}},
 \eeq
 to perform the integration along the contour $C_\infty$ at infinity it is seen that the lhs of \rf{5.29} precisely leads to the rhs of 
 eq.\ \rf{5.18}.
 
\begin{figure}[t]
\vspace{-2.7cm}
\centerline{\scalebox{0.23}{\includegraphics[angle = 0]{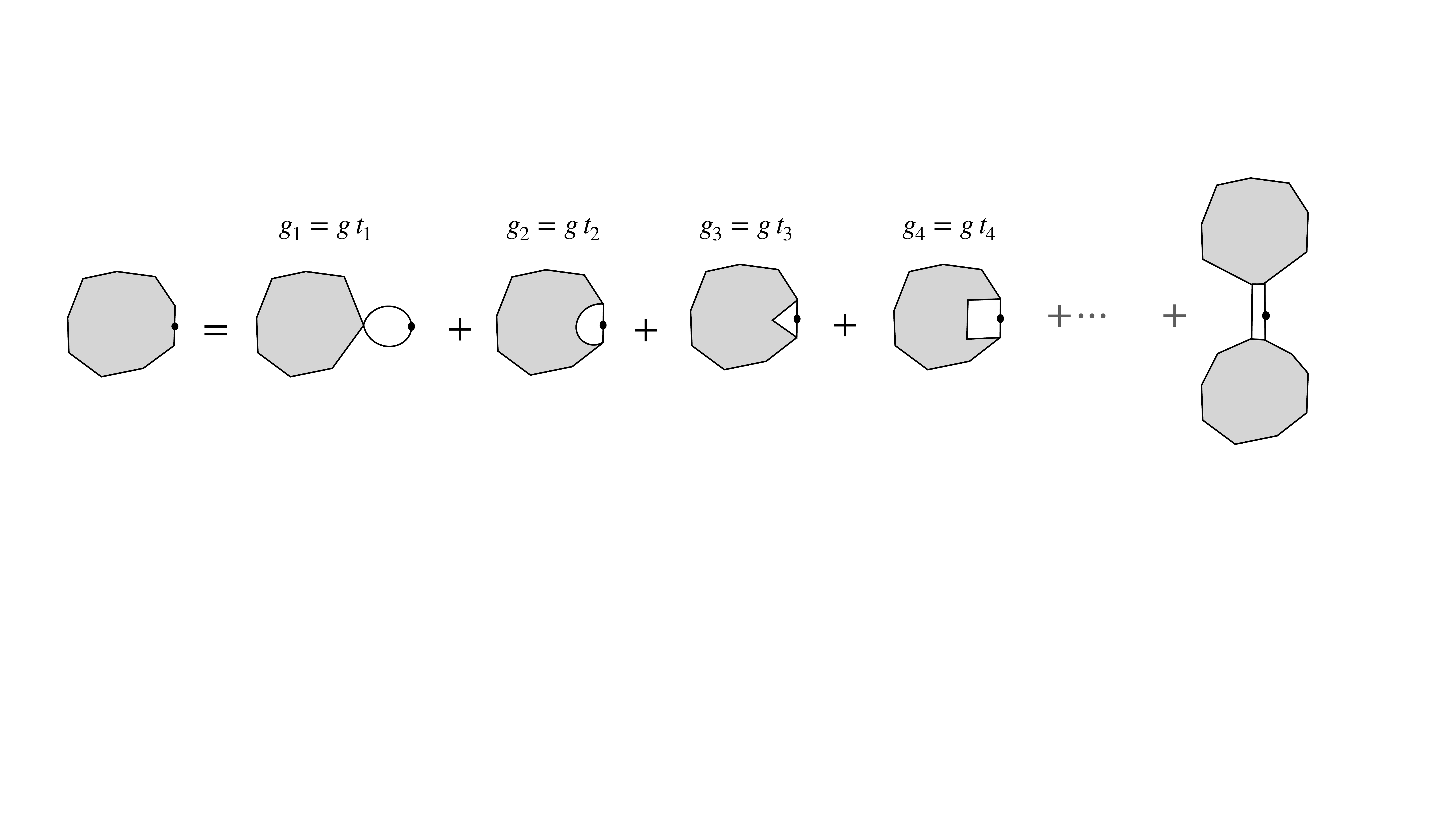}}}
\vspace{-4cm}
\caption{{\small  The general loop equation.}}
\label{fig5.5}
\end{figure}

 The counting can now be generalized, such that we allow not only for triangles and double links, but also for squares, 
 pentagons etc., even ``one-gons'' and ``two-gons'', as illustrated in Fig.\ \ref{fig5.5}.  
 The generalization of \rf{5.18} is then:
 \beq\label{5.31}
 w(g,z) = g \Big( \frac{t_1}{z} \plu \cdots \plu t_n z^{n\mi 2} \Big) \; w(g,z) + \frac{1}{z} \, Q(g,z) + \frac{1}{z} \, w^2(g,z)
 \eeq
 \beq\label{5.32}
 Q(g,z) = 1 -g \sum_{j=1}^n t_j \sum_{l =1}^{j-2} z^l \,w_{j-l-2}(g)
 \eeq
 In these formulas a $j$-polygon is assigned a weight $g_j = g \,t_j$, i.e.\ in the generating function we assign a variable $g$ to each 
 polygon and relative weights $t_j$ to the various kind of polygons. We now use the notation
 \beq\label{5.33}
 V'(g,z) = z - g \big( t_1 \plu t_2z \cdots \plu t_n z^{n-1} \big),
 \eeq
 and we find again \rf{5.20}, but now with $V'(g,z)$ and $Q(g,z)$ generalized to \rf{5.33} and \rf{5.32}: 
 \beq\label{5.34}
 \boxed{w(g,z) = \oh \Big( V'(g,z) - \sqrt{\big( V'(g,z) \big)^2 \mi 4Q(g,z)} \Big)}
 \eeq
 As before we can argue that for fixed $t_j$ and $g$ in a neighborhood of 0, we have 
 \beq\label{5.35}
 \big( V'(g,z) \big)^2 \mi 4Q(g,z)= M^2(g,z) \, \big(z\mi c_+(g) \big) \big(z\mi  c_-(g)\big), 
 \eeq
 \beq\label{5.36}
  M(g,z) = \sum_{k=1}^{n-1} M_k(g) \, (z \mi c_+(g))^{k-1},
 \eeq
 and \rf{5.34} reads, if we introduce the notation $\vg = (g_1,\ldots,g_n)= g(t_1,\ldots,t_n)$ to emphasize the dependence on multiple $g_j$:
 \beq\label{5.37}
 \boxed{w(\vg,z)= \oh \Big( V'(\vg,z) \mi  M(\vg,z) \, \sqrt{(z\mi c_+(\vg) ) (z\mi  c_-(\vg))}}
 \eeq
 We can now solve for $M(\vg,z)$ (where we suppress the $\vg$ dependence)
 \beq\label{5.38}
 M(z)= \frac{ V'(z)}{\sqrt{(z\mi c_+)(z\mi c_-)}} - \frac{ 2 w(z)}{\sqrt{(z\mi c_+)(z-c_-)}}
\eeq
Recall from \rf{5.36} that $M(z)$ is a polynomial of order $n\mi 2$. For any polynomial one can write 
\beq\label{5.39}
M(z) =  \oint_{C_\infty} \dom \; \frac{M(\om)}{\om \mi z},
\eeq
where the contour $C_\infty$ is a contour which can be moved to infinity without crossing $z$. Using the expression
\rf{5.38} for $M(\om)$ in \rf{5.39}, the term with $w(\om)$ will not contribute since  $w(\om) \to 1/\om \pl \cO(1/ \om^2)$ for 
$|\om | \to \infty$ and we have 
\beq\label{5.40}
M(z) =  \oint_{C_\infty} \dom \; \frac{1}{\om \mi z} \; \frac{ V'(\om)}{\sqrt{(\om \mi c_+)(\om\mi c_-)}}.
\eeq
By expanding $\displaystyle{\frac{1}{\om \mi z} = \frac{1}{\om  \mi c_+} \sum_{k=0}^\infty \Big( \frac{z \mi c_+}{\om \mi c_+} \Big)^k}$ we obtain
\beq\label{5.41}
M(z) = \sum_{k=1}^{n-1} M_k (z\mi c_+)^{k-1}, \quad 
M_k (\vg) = \oint_{C_\infty} \dom \;  \frac{ V'(\vg, \om)}{(\om\mi c_+(\vg))^{k+ \oh}(\om\mi c_-(\vg))^{\oh}}
\eeq
The definition of $M_k$ is valid for all integer $k$, also negative $k$, even if only $k > 0$ appears in the sum 
in \rf{5.41}. Note also the contour $C_\infty$ in \rf{5.41} can be deformed to any curve enclosing the cut $[c_-,c_+]$ on the real axis.
For a polynomial \rf{5.33} of order $n \mi 1$, $M_k$ as defined by 
\rf{5.33} will be zero for $k > n\mi 1$. We can now write:
\bea
w(z) &=& \oh V'(z)  - \oh \, M(z) \, \sqrt{(z\mi c_+)(z\mi c_-)}\no \\
&& \oh V'(z) \mi \oh \oint_{C_\infty} \dom \; \frac{V'(\om)}{\om \mi z} \; 
\frac{ \sqrt{(z\mi c_+)(z\mi c_-)}}{\sqrt{(\om\mi c_+)(\om\mi c_-)}}. \no
\eea
Contracting the curve $C_\infty$ back to the curve $C$ shown in Fig.\ \ref{fig5.4} we finally obtain
\beq\label{5.42}
\boxed{w(\vg,z) = \oh 
\oint_C \dom \; \frac{V'(\vg,\om)}{z \mi \om} \; \frac{ \sqrt{(z\mi c_+(\vg))(z\mi c_-(\vg))}}{\sqrt{(\om\mi c_+(\vg))(\om\mi c_-(\vg))}}} 
\eeq
{\it This a solution to the loop equation \rf{5.29}}: We have now a closed expression for $w(\vg,z)$, 
and $c_\pm(\vg)$ are uniquely determined by the condition that $w(z) \to 1/z$ for 
$|z| \to \infty$. Explicitly, expanding the integrand in \rf{5.42} in powers of $1/z$, we obtain:
\beq
w(z) \!=\! \oh \oint\limits_{C_\infty} \dom \;  \frac{ V'( \om)}{\sqrt{(\om \mi c_+)(\om\mi c_-)}} +
 \frac{1}{2z}  \;\oint\limits_{C_\infty} \dom \;  \frac{[\om \mi \oh (c_+ \pl c_-)] V'( \om)}{\sqrt{(\om\mi c_+)(\om\mi c_-)}} +\cO\Big( \frac{1}{z^2} \Big) \no
\eeq
Thus the condition $w(z) \to 1/z$ for $|z| \to \infty$ can be formulated as 
\beq\label{5.43}
\boxed{ M_0(\vg) = 0,\quad  M_{-1} (\vg) = 2}
\eeq
These are two equations which in principle determine $c_\pm(\vg)$.

\subsection*{Multiloops and the loop-insertion operator}

Let us write the generating function for our generalized triangulations (which contain also squares, pentagons etc.) in detail:
\beq\label{5.44}
w(\vg,z) = \sum_{l,k_1, \ldots,k_n}  w_{k_1,\ldots,k_n,l} \, \frac{1}{z^{l+1}} \, \prod_{j=1}^n g_j^{k_j}.
\eeq
{\it In this notation $w_{k_1,\ldots,k_n,l}$ denotes the number of  graphs with the topology of the sphere with one boundary, 
with $k_j$ $j$-sided polygons, 
$j \equ 1,\ldots,n$, and one boundary with $l$ links, where one of the boundary links has a mark.} 

   \begin{figure}[t]
\vspace{-1.3cm}
\centerline{\scalebox{0.18}{\includegraphics[angle = 0]{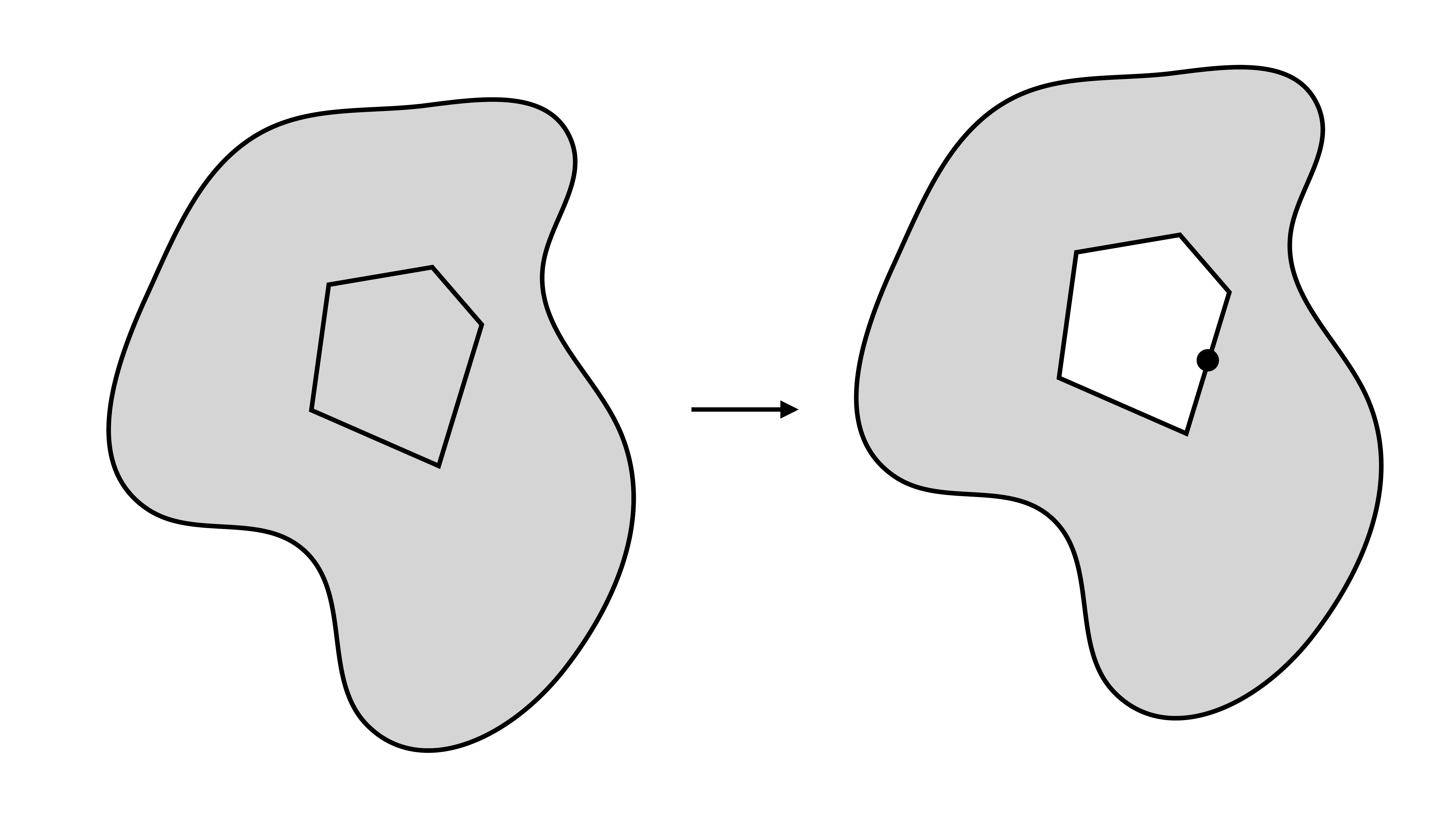}}}
\vspace{-0.7cm}
\caption{{\small  Loop insertion: a polygon is removed, creating a new boundary (a loop) with a marked link.}}
\label{fig5.6}
\end{figure}

Let now $n \to \infty$, i.e.\ we allow polygons with arbitrarily many sides. We introduce the so-called {\it loop insertion operator} as
\beq\label{5.45}
\boxed{\frac{d}{d V(z)} = \sum_{j=1}^\infty \frac{j}{z^{j+1}} \; \frac{d}{dg_j}}.
\eeq 
When $\displaystyle{\frac{d}{d V(z_2)}}$ acts on $w(\vg,z_1)$ it changes
\beq\label{5.46}
g_j^{k_j} \to \frac{j}{z_2^{j+1}}\; k_j\; g_j^{k_j -1},
\eeq
which has the interpretation that it removes a   $j$-polygons, i.e.\ it creates a hole and a corresponding boundary of length
$j$ and it associates a new boundary variable $z_2$ to this boundary. This is the factor $1/z_2^{j+1}$ in \rf{5.46}. The factor $j$ is present 
because we want to mark one of the links on the boundary and it can be done in $j$ ways. The factor $k_j$ is present since there are 
$k_j$ different $j$-polygons and we can choose to remove any one of them. Finally $g_j^{k_j} \to g_j^{k_j-1}$ since we remove 
one $j$-polygon. The process is illustrated in Fig.\ \ref{fig5.6}. Denoting the generating function for graphs with the topologies of the 
sphere with  two marked boundaries by $w(\vg,z_1,z_2)$, where $z_1$ and $z_2$ are used to enumerate the boundaries, we have
\beq\label{5.47}
\frac{d}{d V(z_2)}\; w(\vg,z_1) = w(\vg,z_1,z_2)
\eeq
and by an obvious generalization to $n$ boundaries:
\beq\label{5.48}
\boxed{\frac{d}{d V(z_2)}\cdots \frac{d}{d V(z_n)}\; w(\vg,z_1)= w(\vg,z_1,\ldots,z_n)} 
\eeq
where $w(\vg,z_1,\ldots,z_n)$ is the generating function for the number of spherical graphs with $n$ boundaries, constructed from 
arbitrary $j$-polygons. So if we can calculate the generating function for graphs with one boundary, but constructed with arbitrary 
$j$-polygons,  we in principle have the 
complete information about the graphs with $n$ boundaries.
Note that {\it after} we have constructed $w(\vg,z_1,\ldots,z_n)$ using the loop-insertion operator, we can
put any $g_j \equ 0$. For instance, if we only want to use triangles (as was our starting point) we simply, after having calculated 
the general $w(\vg,z_1,\ldots,z_n)$, choose $g_j \equ 0$ except for $g_3$ (but we need the general expression in order to apply the 
loop-insertion operator, since even if we only use triangles (and double links), a boundary can have any length $l$, and we create 
this boundary by removing a polygon of length $l$. In order to remove it, it has to be present in the first place).


\subsection*{Explicit solution  for bipartite graphs}

Let us consider a special class of ``triangulations'' where there are only loops (and in particular only boundary loops) of even lengths.
Such triangulations are called {\it bipartite} since one can show that the vertices can be divided in two groups, yellow and blue, say,
such that all links only connect different colored vertices. We again assume that the topology of the graphs are spherical with holes,
i.e. boundaries.  The graphs have to be  constructed from $2j$ sided polygons (any odd sided polygon will create a loop
of odd length) and double-links.
The function $V'(z)$ thus has the form 
\beq\label{5.49a}
V'(z) = z \mi g \sum_{j=2}^\infty  t_{2j} z^{2j-1}.
\eeq
For graphs with the topology of the disk and constructed from even sided polygons, loops will automatically be 
of even length. However if the topology of the graph is different from the disk, e.g.\ that of the cylinder,  
one can still have loops with odd length,  
as it is seen by gluing together three squares to form a prism with two boundaries of length 3.
Thus if we restrict the loop inserting operator to act only with $g_{2k}$, it will not create odd loops and thus 
not all graphs one can construct from even sided polygons. It will only construct the subclass of bipartite graphs . 
However, in this subclass  all the operations done on graphs to derive the loop equation  for $w(\vg,z)$ 
and (using the loop insertion 
operator)  the generating functions $w(\vg,z_1,\ldots,z_n)$ for spheres with $n$ boundaries, are still valid. In this way 
we end up with same equations as before, only with \rf{5.49a} instead of the more general potential \rf{5.33}.
We do not expect the continuum limit associated to such a restricted class of triangulations to be different from the 
 continuum limit of our unrestritced triangulations. Since the generating function $z \, w(\vg,z)$ will now be an even function of $z$
we have $c_ - \equ - c_+$ It is convenient to build this into the definitions we have used, and thus we define $c = c_+$ and 
\beq\label{5.49}
w(\vg,z) = \sum_{j=0}^\infty \frac{w_{2j}}{z^{2j+1}} = \oh \Big( V'(\vg,z)  \mi \tM(\vg,z) \, \sqrt{ z^2 \mi c^2(\vg)}\Big),
\eeq
\beq\label{5.50a} 
 \tM(\vg,z) = \oint_{C_\infty}  \dom \; \frac{1}{\om \mi z} \; \frac{ V'(\vg,\om)}{\sqrt{\om^2 \mi c^2(\vg)}} =  
 \oint_{C_\infty} \dom \; \frac{1}{\om^2 \mi z^2} \; \frac{\om \, V'(\vg,\om)}{\sqrt{\om^2 \mi  c^2(\vg)}}, 
 \eeq
 and by expanding 
 $\displaystyle{\frac{1}{\om^2 \mi z^2} = \frac{1}{\om^2 \mi c^2} \sum_{k=0}^\infty \Big( \frac{ z^2\mi c^2}{\om^2\mi c^2} \Big)^k}$ 
 we obtain
\beq\label{5.50b}
\tM(\vg, z) = \sum_{k=1}^\infty \tM_k(\vg) \big(z^2 \mi c^2(\vg)\big)^{k-1} ,\quad  \tM_k(\vg) = \oint_C \dom \; \frac{\om \, 
V'(\vg,\om)}{(\om^2 \mi  c^2(\vg))^{k+\oh}},
\eeq
where the curve $C$ can be any curve enclosing the cut $[-c,c]$ on the real axis.
The requirement $w(z) \to 1/z$ for $|z| \to \infty$ leads to the condition which replaces \rf{5.43}
\beq\label{5.50}
\tM_0 (\vg) = 2.
\eeq
{\it This equation determines $c(\vg)$}: for given $\vg$ one has to perform the integral \rf{5.50b} for $k \equ 0$ and adjust the value 
of $c^2$ such that  the integral is equal 2.  

When applying the loop-insertion operator to functions which 
can be written as  $F(\vg,c^2)$, it is convenient  to write the loop insertion operator as 
\bea\label{5.51}
 \frac{d}{d V(z)}  &=& \sum_{j=1}^\infty \frac{2j}{z^{2j+1}} \; \frac{d}{dg_{2j}} \qquad 
\left( \frac{d}{dg_{2j}} = \frac{\prt}{\prt g_{2j}} + \frac{d c^2}{d g_{2j}} \, \frac{\prt}{\prt c^2}\right) \no \\
 &=&  \frac{\prt}{\prt V(z)} + \frac{d c^2}{d V(z)} \, \frac{\prt}{\prt c^2}
 \eea
Let us now calculate how these operators act on the functions which appear in $w(\vg,z)$:
\begin{itemize}
\item[{1)}] For $|\om | < |z|$ we have 
\beq\label{5.51a} \frac{\prt V'(\vg,\om)}{\prt V(z)} = \sum_{j=1}^\infty \frac{2j}{z^{2j+1}} \frac{\prt}{\prt g_{2j}},
 \Big( \om \mi \sum_{k=1}^\infty g_{2k}\om^{2j-1} \Big) =
\frac{-2\om z}{(\om^2\mi z^2)^2}
\eeq
and by analytic continuation we extend the result to the entire complex $\om$-plane and $z$-plane.
\item[{2)}] We now use this result to write
\beq\label{5.51b}
\frac{\prt \tM_k}{\prt V(z)} = \oint_C \dom \; \frac{ \om }{(\om^2\mi c^2)^{k+\oh}}   \frac{\prt V'(\vg,\om)}{\prt V(z)}= 
\frac{d}{dz} \, \frac{z}{~(z^2\mi c^2)^{k+\oh}}
\eeq
where we first choose $C$ such that it encloses the cut $[- c,c]$, but not $z$. After that, to evaluate the integral, 
we are free to deform  $C \to C_\infty$, picking up the pole term at $z$. The  final contour integral at $C_\infty$ is zero.

\item[{3)}] It follows from the definition \rf{5.50b} of $\tM_k$ that 
\beq\label{5.51c}
\frac{\prt \tM_k}{\prt c^2} = \big( k\plu \oh\big) \, \tM_{k+1}
\eeq

\item[{4)}] Acting with  $d/dV(z)$ on eq.\ \rf{5.50} we obtain:
$$
0 = \frac{ d \tM_0}{dV(z)} = \frac{\prt \tM_0}{\prt V(z)} + \frac{d c^2}{dV(z)} \, \frac{\prt \tM_0}{\prt c^2} = 
\frac{-c^2}{(z^2\mi c^2)^{3/2}} + \oh \tM_1 \, \frac{d c^2}{d V(z)}
$$
This implies that 
\beq\label{5.51d}
 \frac{d c^2}{d V(z)} = \frac{2}{\tM_1(\vg,c^2) }   \; \frac{c^2}{(z^2\mi  c^2(\vg))^{3/2}}
 \eeq

\end{itemize}
From $1) \mi 4)$ it follows that we finally can write
\beq\label{5.52}
\frac{d}{dV(z)} = \frac{\prt}{\prt V(z)} + \frac{2}{\tM_1(\vg,c^2) }   \; \frac{c^2}{(z^2\mi  c^2(\vg))^{3/2}} \; \frac{\prt}{\prt c^2}
\eeq

We can now apply these expression, when calculating 
\beq\label{5.53}
\frac{d}{dV(z_2)} \; w(\vg,z_1) = w(\vg,z_1,z_2)
\eeq
and after an elementary but lengthy calculation (details are discussed in Problem Set 12), one obtains:
\beq\label{5.54}
\boxed{w(\vg,z_1,z_2) =
\frac{1}{2 (z_1^2\mi z_2^2)^2} \left[ z_2^2 \sqrt{\frac{z^2_1 \mi c^2}{z_2^2-c^2}}+ z_1^2 \sqrt{\frac{z^2_2 \mi c^2}{z_1^2-c^2}}- 2z_1z_2 \right]}
\eeq
First of all one can check that the formula is not singular for $z_1 \equ z_2$ by rewriting it as 
\beq\label{5.54a}
w(\vg,z_1,z_2) = \frac{  F^2(z_1,z_2,c^2)}{(\sqrt{z_1^2 \mi c^2} \pl \sqrt{z_2^2 \mi c^2})^2\,\sqrt{z_1^2 \mi c^2} \,\sqrt{z_2^2 \mi c^2}}
\eeq
where 
\beq\label{5.54b}
F(z_1,z_2,c^2) = \frac{ z_2 \sqrt{z_1^2 \mi c^2} - z_1 \sqrt{z_2^2 \mi c^2}}{\sqrt{z_1^2 \mi c^2} -\sqrt{z_2^2 \mi c^2}}
\eeq
Next, it should be emphasized  that \rf{5.54} is a remarkable formula: {\it it does not depend explicitly on $\vg$. The only dependence on 
$\vg$ comes through $c^2$, whose dependence on $\vg$ is obtained by solving \rf{5.50}. $w(\vg,z_1,z_2)$ is said to be universal.}

This simple functional form of the two-loop function makes it easy to obtain the three-loop function by applying $d/dV(z_3)$ in the form given by \rf{5.52} (see Problem Set 12 for details)  and we obtain
\beq\label{5.55}
\boxed{w(\vg,z_1,z_2, z_3) = \frac{c^4}{2 \tM_1}\, \frac{1}{(z_1^2\mi c^2)^{3/2}} \;\frac{1}{(z_2^2\mi c^2)^{3/2}} \; 
\frac{1}{(z_3^2\mi c^2)^{3/2}} }
\eeq
and by induction one can prove (again, details are provided in Problem Set 12)
\beq\label{5.56}
\boxed{w(\vg,z_1,\ldots, z_n) = \left( \frac{2}{\tM_1} \frac{d}{dc^2} \right)^{n-3} \left[ \frac{1}{2 c^2 \tM_1} 
\prod_{k=1}^n \frac{c^2}{(z_k^2\mi c^2)^{3/2}}\right]}
\quad n \geq 3
\eeq
{\it This is the explicit solution to the counting problem for bipartite graphs constructed from even-sided  polygons and double links, 
with the topology of the sphere with $n$ boundaries}. One can find 
the generating function for the number of graphs with a given number of $2j$-polygons and given boundary lengths $l_i$, $i=1,\ldots,n$
(where $l_i$ by construction is even).
Again it is remarkable that the generating function can be completely expressed in a condensed form only depending on factors 
involving $(z_i^2\mi c^2)^{-m-\oh}$ where $1 \leq m \leq n\mi 2$,  and $c^2$ and $\tM_k$, where $1 \leq k \leq n \mi 2$. This follows from rules 
$1) \mi  4)$ above, when using \rf{5.56}. It turns out that in the scaling limit, to be discussed below, the $\tM_k$ 
has a geometric meaning related to so-called
intersection indices on Riemann surfaces. However, we have no time to discuss further this interesting topic 
which bridges between combinatorics and differential  topology (some details can be found in the article \cite{ackm}).

\vspace{6pt}

\subsection*{The number of large triangulations}

Let us now write $g_{2k} \equ g \, t_{2k}$, where we keep the $t_{2k}$ fixed: we are counting $2k$-polygons with relative weight $t_{2k}$ and $g$ enumerates the polygons, no matter how many sides they have, i.e. we write
\beq\label{5.57}
w(g,z_1.\ldots,z_n) = \sum_{k,l_1,\ldots,l_n} w_{k,l_1\ldots,l_n} g^k \prod_{i=1}^n \frac{1}{z_i^{l_i+1}}.
\eeq
In this formula $w_{k,l_1\ldots,l_n} $ is the number of graphs with $k$ polygons, the $2j$-gon counted with relative weight $t_{2j}$, and with 
$n$ boundaries of lengths $l_i$, $i \equ 1,\ldots,n$, where $l_i$ is even for the bipartite graphs we are considering.

Let us first concentrate on the situation where $k \to \infty$ while $l_1, \ldots, l_n$ are kept fixed. We have seen for RWs, BPs, and the 
bosonic string that the large $k$ limit  is associated with non-analytic behavior of the generationg function. We expect the same here.
When we look at the expression for $w(g,z_1,\ldots,z_n)$ there are only two potential sources of such non-analytic behavior: (a) 
$M_1(g,c^2) \to 0$ for $g \to g_c$, or (b) $c^2(g)$ becomes non-analytic for $g \to g_c$ (where $g_c$ is a function $g_c(t_{2j})$ of the 
weights $t_{2j}$. But we keep these fixed, as mentioned above). We will see that (a) and (b) happen at the same point $g_c$.

Assume that $\tM_1(g,c^2(g)) \to 0$ for $g \to g_c$. Introduce the following notation
\beq\label{5.58}
\Del g \equiv g_c \mi g, \qquad \Del (c^2) \equiv c^2(g_c)  \mi c^2(g),\qquad \tM_2^c \equiv \tM_2(g_c,c^2(g_c)).
\eeq 
For a given  $g$, solving \rf{5.50} leads to to a value $c^2(g)$ such that $\tM_0(g,c^2(g)) \equ 2$, i.e.
\bea\label{5.59}
2 = \tM_0(g_c, c^2(g_c)) &=& \tM_0 (g \pl \Del g,c, c^2(g)) + \Del(c^2) \no \\
&=& \tM_0(g,c^2(g)) + \frac{\prt \tM_0}{\prt g} \Del g +\frac{ \prt \tM_0}{\prt c^2} \Del(c^2) + \cdots
\eea
Now use that $ \tM_0(g,c^2(g)) \equ 2$ and $\displaystyle{\frac{ \prt \tM_0}{\prt c^2} = \oh \tM_1(g,c^2(g))}$  (rule  3) above) and expand again
\bea\label{5.60}
0=  \tM_1(g_c,c^2(g_c)) &=&  \tM_1 \big(g \pl \Del g,c, c^2(g) + \Del(c^2)\big) \no \\
&=& \tM_1(g,c^2(g)) + \frac{\prt \tM_1}{\prt g} \Del g +\frac{ \prt \tM_1}{\prt c^2} \Del(c^2) + \cdots
\eea
Using again $\displaystyle{\frac{ \prt \tM_1}{\prt c^2} = \frac{3}{2}\tM_2(g,c^2(g))= \frac{3}{2} \tM_2(g_c,c^2(g_c)) \pl \cdots}$ 
we can finally write
\beq\label{5.61}
0= \frac{\prt \tM_0}{\prt g} \Del g - \frac{3}{4} \tM^c_2 \big( \Del (c^2) \big)^2 + \cO \big((\Del g)^2, \Del(c^2)\Del g\big)
\eeq
In a Appendix to this Section we will show that provided all $t_{2j} \geq 0$ and at least one $t_{2j}$ is not zero, then 
both  ${\prt \tM_0}/{\prt g}$ and $ \tM^c_2 $ are negative. Thus we have reached our conclusion
\beq\label{5.62}
\Del( c^2(g) ) \sim \sqrt{\Del g} \quad {\rm or} \quad c^2(g) = c^2(g_c) -{\rm const.} \, \sqrt{ g_c \mi g} + \cO (g_c \mi g).
\eeq
{\it The point $g_c$ where $\tM_1(g) \equ 0$ is also the point where $c^2(g)$ ceases to be an analytic function of $g$ and the 
singular behavior is a square root singularity.}

From \rf{5.60} and \rf{5.62} we have for $g \to g_c$.
\bea\label{5.63}
\tM_1 (g,c^2(g))  &\equ& -\frac{3}{2}  \tM^c_2\; \Del(c^2) ~~\propto~~ \sqrt{ \Del g}~~ \Rightarrow ~\\
\frac{\prt}{ \prt c^2} \frac{1}{\tM_1(g,c^2)} & \equ&
- \frac{ 3 \tM_2(g,c^2)}{2 \tM_1^2(g,c^2)} ~~\propto~~ \frac{1}{\Del g}\label{5.63a}
\eea
It now follows that the most singular behavior of $w(g,z_1,\ldots,z_n)$ is obtained by differentiating $\tM_1(g,c^2)$ a maximal 
number of times in formula \rf{5.56}, and we obtain a singular behavior
\beq\label{5.64}
\left( \frac{1}{\tM_1} \frac{d}{dc^2}\right)^{n-3} \left[ \frac{1}{\tM_1} \cdots \right] \propto \left[ \frac{1}{\tM_1^{2n-5}} \cdots + lst \right] \propto
\left[ \frac{1}{(\Del g)^{n-5/2}} \cdots +lst \right] ,
\eeq
where {\it lst} means ``less singular terms''.  We obtain now the following behavior for the generating function
\beq\label{5.65}
w(g,z_1,\ldots,z_n) \propto \frac{1}{(\Del g)^{n-5/2}}  \prod_{i=1}^n \frac{1}{(z_i^2 - c^2(g_c))^{3/2}}  +  lst
\eeq
Using the expansion
\beq\label{5.66}
 \frac{1}{(\Del g)^{n-5/2}}  = \frac{1}{g_c^{n-5/2}} \frac{1}{(1- g/g_c)^{n-5/2}} = \frac{1}{g_c^{n-5/2}} \sum_{k=0}^\infty 
 \begin{pmatrix} n\mi 7/2\plu  k\\ k \end{pmatrix}  \, \left(\frac{g}{g_c}\right)^k
 \eeq
 as well as 
 \beq\label{5.67}
 \frac{1}{(z^2-c^2(g_c))^{3/2}} = \frac{1}{z^3} \sum_{j=0}^\infty  \begin{pmatrix} j \plu 1/2\\ j  \end{pmatrix} \, \left(\frac{c^2(g_c)}{z^2}\right)^j
 \eeq
 we obtain from \rf{5.57} and \rf{5.65}, using (as discussed in Problem Set 5) 
 \beq\label{5.67a}
 \begin{pmatrix} m+ k\\ k \end{pmatrix} = \frac{(m \plu k)!}{m!\; k!} ~~\to~~ \frac{k^m}{m!}  
 \Big(1+ \cO\Big(\frac{1}{k}\Big)\Big)
 \quad {\rm for} \quad k \to \infty
 \eeq
 that 
 \beq\label{5.68}
\boxed{  w_{k,l_1,\ldots,l_n} \propto  k^{n-7/2} \sqrt{l_1 l_2 \cdots l_n} \; \left( \frac{1}{g_c}\right)^k \prod_{i=1}^n \big( c(g_c) \big)^{l_i}\; 
 \Big(1+ \cO\Big(\frac{1}{k},\frac{1}{l_i}\Big)\Big)}
 \eeq
 This result is {\it universal}. The only dependence which refers to the choice of $t_{2j}$, the relative weights of the polygons, is $g_c(t_{2k})$ and 
 $c(g_c(t_{2k}))$. These two constants determine the exponential growth of the number of triangulations, both wrt number of polygons
 and wrt lengths of the boundaries. The actual exponential rate of growth is thus not universal, but the fact the growth {\it is} exponential
 is universal, but more importantly:{\it  the leading power corrections to the exponential growth are universal and independent of the
 choice of $t_{2j}$.}
 
 When  deriving  formula \rf{5.12} we have actually assumed that  $l_i \ll \sqrt{k}$. If $l_i \sim  \sqrt{k}$ (which is not 
 unnatural, since it is a graph where the boundary lengths are of the order we would expect for macroscopic boundaries), eq. \rf{5.65} is 
 not quite correct. In \rf{5.65} we should really have used
 \beq\label{5.69}
  \frac{1}{(z^2 \mi  c^2(g))^{3/2}} = \frac{1}{z^3} \sum_{l=0}^\infty  \begin{pmatrix} l \plu 1/2\\ l  \end{pmatrix} \, \left(\frac{c^2(g_c)}{z^2}\right)^l
  \left(\frac{c^2(g)}{c^2(g_c)}\right)^l,
  \eeq
  and the last factor is no longer close to 1 when $l \sim 1/\ln \frac{c^2(g_c)}{c^2(g)} \propto 1/\sqrt{\Del g}$. In the sum \rf{5.66} a typical
  value $\la k \ra$ of $k$ (where the function summed over has a maximum) is likewise of the order $1/\ln (g_c/g)  \propto 1/\Del g$. Therefore,
   precisely  when $l_i$ becomes of the order of $\sqrt{\la k \ra}$ \rf{5.68} needs to be modified and a more careful treatment leads to 
   \rf{5.12}. Instead of doing that we will derived \rf{5.12} in the ``continuum limit'', where the relation between the $l_i$'s and $k$ becomes 
   well defined and given as in \rf{5.11}. We now turns to this.
   
   
   \subsection*{The continuum limit}  
   
   Recall the continuum formulas \rf{5.1} -\rf{5.8} for two-dimensional quantum gravity, which we for convenience repeat here
   \bea\label{5.70}
   S[ g, \Lam, Z_1,\ldots,Z_n] \!\!\!&=&\!\!S[g,\Lam]  + \sum_{i=1}^n Z_i \!\int \!\!d s_i =   
 \Lam\, V_g \plu \sum_{i=1}^n Z_i \,L_{i,g},\\
 \label{5.71}
 W(\Lam;Z_1,\ldots,Z_n) \!\!\!&=& \!\!\!\!\int \cD [g] \; \e^{-S[ g, \Lam, Z_1,\ldots,Z_n] },\\
 \!\!\!&=&\!\! \!\!\!\int_0^\infty \!\! \!\!dV \; \e^{-\Lam \, V} \!\!\int_0^\infty \prod_{i=1}^n dL_i\; \e^{-Z_i L_i} \;W(V;L_1,\ldots,L_n)
 \;\;\; \;\;\;\label{5.71a}
 \eea
The corresponding discretized expressions are 
\bea\label{5.72}
S_T(\mu,\lam_1,\ldots,\lam_n) &=& \mu \, k + \sum_{i=1}^n \lam_i l_i  \\
w(\mu, \lam_1,\ldots,\lam_n) &=& \sum_{T \in \cT(n)} \e^{- S_T(\mu,\lam_1\ldots,\lam_n) }  \label{5.73}\\
&=& \sum_k \e^{- \mu k}\sum_{l_1,\ldots,l_n} \e^{  - \sum_i \lam_i l_i} \; w_{k,l_1,\ldots,l_n}   \label{5.73a}
\eea 
 Here $W(V;L_1,\ldots,L_n)$ denotes the formal number of geometries with volume $V$ and boundary lengths $L_i$ (see \rf{5.8}), 
 and similarly $w_{k,l_1,\ldots,l_n} $ is the number of ``triangulations'' made of $k$ polygons and with boundary lengths $l_i$. 
 Thus we can make the following identification with our generation function:
 \beq\label{5.74}
  z_1 \cdots z_n \, w(g,z_1,\ldots,z_n) \equiv w(\mu, \lam_1,\ldots,\lam_n),\qquad g = \e^{-\mu},\quad \frac{1}{z_i} = \e^{-\lam_i},
  \eeq
  where the factors $z_1\cdots z_n$ just follows the convention \rf{5.13} and where we  identify
  \beq\label{5.75}
  g = \e^{-\mu}, \quad \frac{1}{z_i} = \e^{-\lam_i}, \qquad g_c = \e^{-\mu_c}, \quad \frac{1}{c(g_c)} = \e^{-\lam_c}.
  \eeq
  With these definitions we can write
  \bea\label{5.76}
  \left( \frac{g}{g_c}\right)^k &=& \e^{-(\mu-\mu_c) k} = \e^{-\Lam \, V}  \quad {\rm if} \quad  \boxed{V = k \ep^2, ~~\mu \mi \mu_c = \Lam \ep^2}\\
  \label{5.77}
 \left( \frac{c(g_c)}{z}\right)^l &=& \e^{-(\lam-\lam_c) l} = \e^{-Z \, L }  \quad {\rm if} \quad  \boxed{L = l \ep, ~~\lam \mi \lam_c = Z \ep}
 \eea
 Note that  for fixed $\Lam$ and $\ep \to 0$ we have $\mu \to \mu_c$ and thus $g \to g_c$ and we can write
 \beq\label{5.78}
 \frac{ \Del g}{g_c} = \mu \mi \mu_c = \Lam \ep^2.
 \eeq 
 In agreement with \rf{5.11} $\ep$ can be given the interpretation of the link length in the ``triangulation'' and $V$ and $L$ then represent
 the continuum volume (area) of the triangulation and continuum length of a boundary. Of course the boxed relations 
 in \rf{5.76} and \rf{5.77} only make sense as continuum relations in the limit where $k,l \gg 1$. If we insist on a limit where $V$ and $L$ are
 finite while $\ep \to 0$, we have in this limit $l \sim \sqrt{k}$, exactly the situation discussed above in connection with \rf{5.69}. Recalling 
 the exponential growths of $w_{k,l_1,\ldots,l_n}$ in \rf{5.68}, we see that what appears in the sum over $k$ and $l_i$ in \rf{5.73} 
 is precisely terms $\e^{-(\mu-\mu_c)k}$ and $\e^{-(\lam_i-\lam_c) l_i}$, i.e. with the boxed indentifications $\e^{-\Lam V}$ and $\e^{-Z_i L_i}$, and these 
 allow us to make a very direct translation from \rf{5.73a} to \rf{5.71a} {\it proved we introduce the concept of renormalized cosmological 
 and boundary cosmological constants $\Lam$ and $Z_i$, as done in \rf{5.76} and \rf{5.77}}. We call it a renormalization of the cosmological 
 constant for the following reason: $\mu$ is dimensionless, and the cosmological term in the action \rf{5.72} is $\mu \, k$ which we 
 according the identifications above would write as $\mu k = \Lam_0 V$, where then $\Lam_0 = \mu/\ep^2$. We call $\Lam_0$ the 
 {\it bare} cosmological constant. Now the relation with $\Lam$ in \rf{5.76} can be written as
 \beq\label{5.79}
 \boxed{\Lam_0 = \frac{\mu_c}{\ep^2} +\Lam}, 
 \eeq
 which is a so-called additive renormalization of the bare cosmological constant, 
 needed in order to obtain finite answers from the path integral.
 The constant to be subtracted from $\Lam_0$ can be identified as coming from the entropy of configurations since
 $\mu_c$ determines the exponential growth  of the number of configurations with the same discrete volume 
 (the same number of polygons). The situation is 
 entirely identical to what happened in the case of a free relativistic particle (see \rf{1.31} and \rf{1.41}). 
 A similar interpretation can now be given 
 to the renormalization of the boundary cosmological constant, represented as boxed equations in \rf{5.77}.
 
 With the above indentifications we can now easily take the continuum limit of our multi-loop functions \rf{5.54}-\rf{5.56}. In the limit 
 $\ep \to 0$, i.e.\ $g \to g_c$ we have 
 \beq\label{5.80}
 c^2(g) = c^2(g_c) - \Del(c^2) =  c^2(g_c) - {\rm cnst.} \, \sqrt{\Del g} = c^2(g_c) - {\rm cnst.} \, \sqrt{\Lam} \, \ep.
 \eeq
 \beq\label{5.81} 
 z_i^2 - c^2(g) = {\rm cnst.} \, \big( Z_i + \SL\big) \,\ep 
 \eeq
 after a suitable rescaling of $\Lam$, and from \rf{5.63}
 \beq\label{5.82}
 \tM_1(g,c^2(g)) = {\rm cnst.} \SL \, \ep 
 \eeq
 Thus 
 \bea
 w(\vg,z_1,\ldots, z_n) &=&\left( \frac{2}{\tM_1} \frac{d}{dc^2} \right)^{n-3} \left[ \frac{1}{2 c^2 \tM_1} \prod_{k=1}^n \frac{c^2}{(z_k^2-c^2)^{3/2}}\right]
 \no \\
 \label{5.83} 
 & \to & \frac{{\rm cnst.} }{\ep^{7n/2-5}} 
 \left( - \frac{d}{d\Lam} \right)^{n-3} \left[ \frac{1}{\SL} \prod_{k=1}^n \frac{1}{(Z_k+\SL)^{3/2}}\right]
 \eea
 and we can write for $n\geq 3$ and  in the limit $\ep \to 0$:
 \beq\label{5.84}
 \boxed{w(\vg,z_1,\ldots, z_n) \to   \frac{{\rm cnst.} }{\ep^{7n/2-5}} \;W(\Lam,Z_1,\ldots,L_n)} \hspace{1cm}
 \eeq
 \beq\label{5.85}
 \boxed{ W(\Lam,Z_1,\ldots,Z_n)  = \left( - \frac{d}{d\Lam} \right)^{n-3} \left[ \frac{1}{\SL} \prod_{k=1}^n \frac{1}{(Z_k+\SL)^{3/2}}\right]}
  \eeq
  Finally, we obtain for the two-loop function:
  \bea\label{5.86}
  W(\Lam,Z_1,Z_2) &=& \frac{1}{2 (Z_1-Z_2)^2} \; \Big( \frac{Z_1\pl Z_2 \pl 2\SL}{\sqrt{Z_1 \pl \SL} \, \sqrt{Z_2\pl \SL}} -2\Big)\\
  &=& \frac{1}{2\Big(\sqrt{Z_1 \pl \SL}\pl \sqrt{Z_2 \pl \SL} \Big)^2 \sqrt{Z_1 \pl \SL} \sqrt{Z_2 \pl \SL}}~~~~~~~~\label{5.87}
  \eea
  From \rf{5.6} and \rf{5.7} we can calculate  $ W(\Lam,L_1,\ldots, L_n)$ and $ W(V,L_1,\ldots, L_n)$ by inverse Laplace 
  transformations. The result is for $n \geq 3$
  \bea\label{5.88}
  W(\Lam,L_1,\ldots, L_n)& \propto &  \left( - \frac{d}{d\Lam} \right)^{n-3} \left[ \frac{1}{\SL} \prod_{k=1}^n \sqrt{L_i} \, \e^{-\SL \, L_i}\right]\\
  W(V,L_1,\ldots, L_n) &\propto & V^{n-7/2} \sqrt{ L_1\cdots L_n} \;\; \exp\Big( \mi\frac{(L_1\pl \cdots \pl L_n)^2}{4V}\Big) ~~~~~~~
  \label{5.89}
  \eea
  Please recall that $ W(V,L_1,\ldots, L_n)$ is formally the number of geometries of the sphere with volume $V$ and $n$ boundaries
  of lengths $L_i$. {\it We have managed to ``count'' the number of  these geometries}. 
  
  One can show that \rf{5.89} is actually valid also for $n \equ 0,~ 1$ and $ 2$. 
  For $n \equ 0$ we have, using \rf{5.7}
  \beq\label{5.90b}
  W(V) \propto V^{-7/2}  \quad {\rm i.e.~formally} \quad W(\Lam) \propto \Lam^{5/2},
  \eeq
  where we write ``formally'' since the  Laplace transform \rf{5.7} is singular for $W(V) \propto V^{-7/2}$.  Again, using \rf{5.7}, we find
  \bea\label{5.90}
  W(\Lam,L_1, L_2) &\propto & \frac{\sqrt{L_1L_2}}{L_1+L_2} \; e^{ - \SL (L_1+L_2)} \\
  W(\Lam,L) & \propto & L^{-5/2} \big(1\pl \SL \,L\big) \; \e^{-\SL \, L} \label{5.90a}
  \eea
  We can now use \rf{5.6} to calculate $W(\Lam,Z)$ and obtain
  \beq\label{5.91}
 W(\Lam,Z) = \big(Z\mi \oh \SL\big) \; \sqrt{Z\pl  \SL}  
 \eeq
 Actually, due to the  factor $L^{-5/2}$, the Laplace transform leading to \rf{5.91} is singular for $L \to 0$, so we have thrown away 
 an infinite constant in \rf{5.91}. We would have obtained the same problem had we tried directly to take the scaling limit 
 starting from $w(g,z)$: some constant terms survive (see eq.\ \rf{6.4} below for an explicit formula), which we formally see by looking at the scaling factor $\ep^{5-7n/2}$ in \rf{5.83}. 
 It diverges for $n >1$ and that is why the continuum part dominates any non-universal part. However, for the disk function it
 is opposite. The continuum part is subleading (scales as $\ep^{3/2}$) compared to constant terms. However, these 
 constant terms will go away if we differentiate $w(g,z)$ wrt $g$ or $z$ sufficiently many times and will not play a role for 
 large triangulations. Thus we dismiss the finite part in the limit $\ep \to 0$ as irrelevant for continuum physics. The same statements made for $n\equ 1$
 is even more true in the case of $n\equ 0$, given by \rf{5.90b}, since in this case we do not have the exponential function in \rf{5.89}
 to provide a regularization at $V \equ 0$, and the Laplace transform in $V$ becomes singular at $V \equ 0$.

 The disk function $W(\Lam,L)$ is called the Hartle-Hawking wave function of the (two-dimensional) universe.  An interpretation of 
 this wave function is that it is the amplitude for a universe to evolve from nothing to size $L$. Unfortunate, this evolution
 is in spacetimes with Euclidean signature, and it has never been clear precisely how one should rotate the result back to 
 spacetimes with Lorentzian  signature. 
 
 Finally, considering the limit where $L_i \ll \sqrt{V}$, such that we can ignore the exponential function in \rf{5.89}, we see that 
 $W(V,L_1,\ldots,L_n)$ represents what we in the case of bosonic strings called the susceptibility of a string with $n$ boundaries
 (only are they here entirely intrinsic) and we have (like for bosonic strings)
 \beq\label{5.91a}
 W(V,L_1,\ldots,L_n)  \propto V^{n-2 + (\gamma-1)}, \qquad \gamma = -\oh,
 \eeq  
 valid for $n \geq 3$. 
 Similarly, in the limit where $L_i \ll 1/\sqrt{\Lam}$ we have from \rf{5.88}
 \beq\label{5.91b}
 W(\Lam,L_1,\ldots,L_n) \propto   \frac{1}{\Lam^{n-2 + \gamma}}, \qquad \gamma = -\oh,
 \eeq
We see that a difference between the bosonic strings (which can be viewed as two-dimensional gravity coupled to 
 $D$ scalar fields $X_i$) and pure two-dimensional gravity is that $\gamma$ changes from1/2 to -1/2. This brings up
 the interesting question of how two-dimensional gravity behaves when coupled to other matter fields than scalar fields.
 We have no space to this discussion, except for the few words said in the next subsection.  
 
 \subsection*{Other universality classes}
 
 Let us again emphasize the {\it universality} of the continuum limit. The class of graphs used and the weights associated with the
 different polygons are not important as long as $t_{2k} \geq 0$, $k \geq 2$,  and at least one $t_{2k}$,  is positive. In this sense 
 the situation is quite similar to the one for RWs and BPs. In the case of BPs we saw that allowing some of the weights 
 to become negative, one could reach different universality classes of BPs. The same is true in the case of our two-dimensional
 gravity models. Recall that the critical point $g_c$ was determined by 
 \beq\label{5.84x}
 \tM_1(g_c, c^2(g_c)) = 0, \qquad \tM_2(g_c,c^2(g_c)) \neq 0.
 \eeq
 As shown in the Appendix following this Section,  $\tM_2(g_c,c^2(g_c)) < 0$ follows from $t_{2k} \geq 0$ and one $t_{2k} > 0$, $k \geq 2$.
 If we relax the condition that $t_{2k} \geq 0$ we can 
 obtain a more general scaling at a $g_c$ characterized by 
 \beq\label{5.85x}
 \tM_1(g_c) = \cdots = \tM_{m-1} (g_c) =0, \qquad \tM_m(g_c) \neq 0, \quad m>2.
 \eeq
 Approaching such a point one can show, using rule $1)-4)$ (eqs.\ \rf{5.51a} - \rf{5.51d}), that for $\Del g = g_c-g$ going to zero one has 
 \beq\label{5.86x}
 \Del (c^2) = c^2(g_c) - c^2(g) \propto  (\Del g)^{1/m}.
 \eeq
 The situation is thus very similar to the one encountered for the multicritical BPs and we call the continuum gravity model  obtained
 in this limit the $m^{th}$ {\it multicritical gravity model} (and we will study \rf{5.84x} and \rf{5.85x} in detail in Problem Set 10). 
 It is possible to show that this continuum model corresponds to pure gravity 
 coupled to a so-called $(p,q)$ rational conformal field theory where $(p,q)\equ (2,2m\mi 1)$. One can obtain more general conformal 
 field theories coupled to two-dimensional quantum gravity if we in addition to negative $t_{2k}$ also allow for infinitely many $t_{2k}$
 being different from zero. Again, the situation, from a technical point of view of taking the scaling limit, is very similar to 
 the what happens for BPs (as we will discuss in Problem Set 10).
 
 As hinted in eq.\ \rf{5.86x} we can obtain different critical exponents when matter is coupled to two-dimensional  gravity. 
 The situation is most beautifully
 illustrated in the case of the so-called Ising (spin) model coupled to two-dimensional  gravity.
 The Ising model is the simplest possible spin model,
 where spins are located at the vertices of a lattice, can take values $\pm 1$ and only couple to neighboring spins. If the lattice is 
 a two-dimensional regular lattice the model can be solved analytically (the so-called Onsager solution), and we have the following 
 picture: there exists a critical temperature, $T_c$, where the spin system undergoes a second order phase transition from a high 
 temperature unmagnetized phase to a low-temperature magnetized phase.  At the phase transition the spin-spin correlation 
 length will diverge, and the spin-spin correlation functions can be described by a conformal field theory with so-called central 
 charge $c \equ 1/2$ (in the notation mentioned above it is a $(p,q) \equ (3,4)$ theory. Thus it is not one of the 
 $(p,q) \equ (2,2m \mi 1)$ conformal field theories associated the multicritical models). 
 The Ising model can also be defined on the triangulations used to define two-dimensional gravity, and also 
  the combined model of gravity and Ising spin can be solved analytically.  Again there is critical temperature $T'_c$, separating 
  a magnetized and unmagnetized phase. However, the critical exponents $\alpha$, $\beta$ and $\gamma_m$ for the spin system 
  are different from the Onsager exponents for the Ising model on a regular lattice. The interaction with fluctuating geometries 
  changes the exponents. But even more remarkable, the susceptibility  exponent $\gamma$ of two-dimensional gravity 
  is also changed from $\gamma \equ -1/2$ to $\gamma \equ \mi 1/3$. The change is only at 
  the critical temperature $T'_c$, where  
  the spin-spin correlation length diverges. For $T \neq T_c'$ we have $\gamma \equ -1/2$, as for two-dimensional gravity
  without Ising spins. There is thus an intricate interaction between matter and geometry precisely when the matter interaction 
  becomes long range. Unfortunately there is no space for covering this in these notes, but as a compensation Problem Set 11 
  discusses a mean-field version the Ising model coupled to DT, which captures well this interaction between geometry and matter.

 
 \subsection*{Appendix}
 
 If $t_{2j} \geq 0$ and there exists a $j >1$ such that $t_{2j} > 0$ then $\tM_k (g,c^2) < 0$ for $k > 1$. First we note that 
 \beq\label{5.89x}
 \frac{1}{(\om^2 \mi c^2)^{k+\oh}  } = \frac{1}{\om^{2k+1}} \sum_{i=0}^\infty d_i(k) \frac{c^{2i}}{\om^{2i}}, \quad d_i(k) > 0.
 \eeq 
 Next, we have for $k >1$
 \beq\label{5.90x} 
 \tM_k = \oint_{C_\infty} \dom \; \frac{\om^2 \mi g \sum_{j=2}^\infty t_{2j} \om^{2j}}{(\om^2 \mi c^2)^{k+\oh}} = 
 -g \sum_{j=0}^\infty t_{2(k+j)} d_j (k) c^{2j}< 0.
 \eeq
 For $k =1$ we get
 \beq\label{5.90y}
 \tM_1 = 1 -g \sum_{j=0}^\infty t_{2(1+j)} d_j (1) c^{2j},
 \eeq
 It is thus seen that $\tM_1(0,c^2(0)) =1$ (it is the case of BPs and $c^2(0)=4$).
 The same kind of calculation shows that 
 \beq\label{5.91x}
 \frac{\prt \tM_0}{\prt g} =  \oint_{C_\infty} \dom \; \frac{\frac{\prt}{\prt g} \big(\om^2 \mi g \sum_{j=2}^\infty t_{2j} \om^{2j}\big)}{(\om^2 -c^2)^{\oh}}
 = -g  \sum_{j=2}^\infty t_{2j} d_j (0) c^{2j}< 0
 \eeq
 
 We can use this information to show that $c^2(g)$ is an increasing function of $g \in [0,g_c]$ ($c^2(0) \equ 4$ (BPs)). We have already 
 argued for that intuitively, since increasing $g$ implies an increasing number of polygons in an average ``triangulation'' which consists of 
 double-lines and polygons. Thus one can have more boundaries of different types and the same length than if we have fewer polygons,
 and $c^2(g)$ determines the exponential growth of the number of boundaries as a function of the length of the boundaries.  However
 using \rf{5.91x} we have directly
 \beq\label{5.92x}
 2 = \tM_0(g,c^2(g)) \Rightarrow 0 =  \frac{\prt \tM_0}{\prt g} +  \frac{\prt \tM_0}{\prt c^2 }\frac{d c^2}{dg} = 
  \frac{\prt \tM_0}{\prt g} +  \oh M_1(g,c^2(g)) \, \frac{d c^2}{dg} 
\eeq 
Since $M_1(0,c^2(0))  =1$ 
and the first zero of $M_1(g,c^2(g))$ is at $g=g_c$,  $M_1(g,c^2(g)) > 0$ for $g \in [0,g_c[$ and consequently $d c^2/dg > 0$
in the same interval. 

Finally note that one has explicitly 
\beq\label{5.94x}
\frac{\prt \tM_0}{\prt g} = \frac{1}{g}   \oint_{C_\infty} \dom \; \frac{(\om V'(\om) -\om^2)}{(\om^2 -c^2)^{\oh}} =  \frac{1}{g} \Big( \tM_0 -\oh c^2(g)\Big) 
= - \frac{1}{2g} \big( c^2(g) -4\big)
\eeq
Thus it is possible to write the important relation \rf{5.61} explicitly as 
\beq\label{5.95x}
\boxed{\Del g = \frac{3 g (- \tM^c_2) }{2(c^2(g) - 4)} \; \big( \Del(c^2)\big)^2 }
\eeq

 \newpage
 
 \setcounter{figure}{0}
 \renewcommand{\thefigure}{6.\arabic{figure}}
 \setcounter{equation}{0}
 \renewcommand{\theequation}{6.\arabic{equation}}
 
 \section*{6. The fractal structure of 2d gravity}
 
 \subsection*{Wilsonian universality and the missing correlation length}
 
 In the last Section we saw how universal scaling limits describing aspects of two-dimensional quantum gravity could
 be obtained. While we found critical points, critical surfaces and approached them in various ways, which provided a 
 wonderful realization of the Wilsonian point of view, where the continuum quantum theory is related to the approach 
 to critical surfaces, somehow the most important and intuitive part of this picture was missing. The primary intuitive reason 
 for the Wilsonian universality is the existence of a  correlation length which diverges 
 when we approach the critical surface. It is this 
 divergence of a correlation length which makes the underlying lattice structure irrelevant 
 and allows us to define a continuum theory with no reference to the lattice. But where is this correlation
 length when we consider two-dimensional quantum gravity? A priori it is not so clear how to define a correlation 
 length in a theory of quantum gravity. In the path integral we have to integrate over all geometries, but at the same time 
 a correlation length, being a ``length'', has a refer to a geometry.
 Still, we will show in this section that one can define a two-point 
 function on the triangulations with a correlation length $\xi(\mu)$ which diverges as $|\mu \mi \mu_c|^{-\nu}$ when we approach the critical point $\mu_c$, and where  the scaling exponent $\nu \equ 1/4$ determines the Hausdorff dimension ($d_H \equ 4$)
 and where the susceptibility exponent $\gamma$ calculated from this two-point function precisely is the $\gamma \equ - 1/2$
 already determined in the former Section.

 \subsection*{The two-loop propagator}
 
\begin{figure}[t]
\vspace{-1cm}
\centerline{\scalebox{0.2}{\includegraphics[angle = 0]{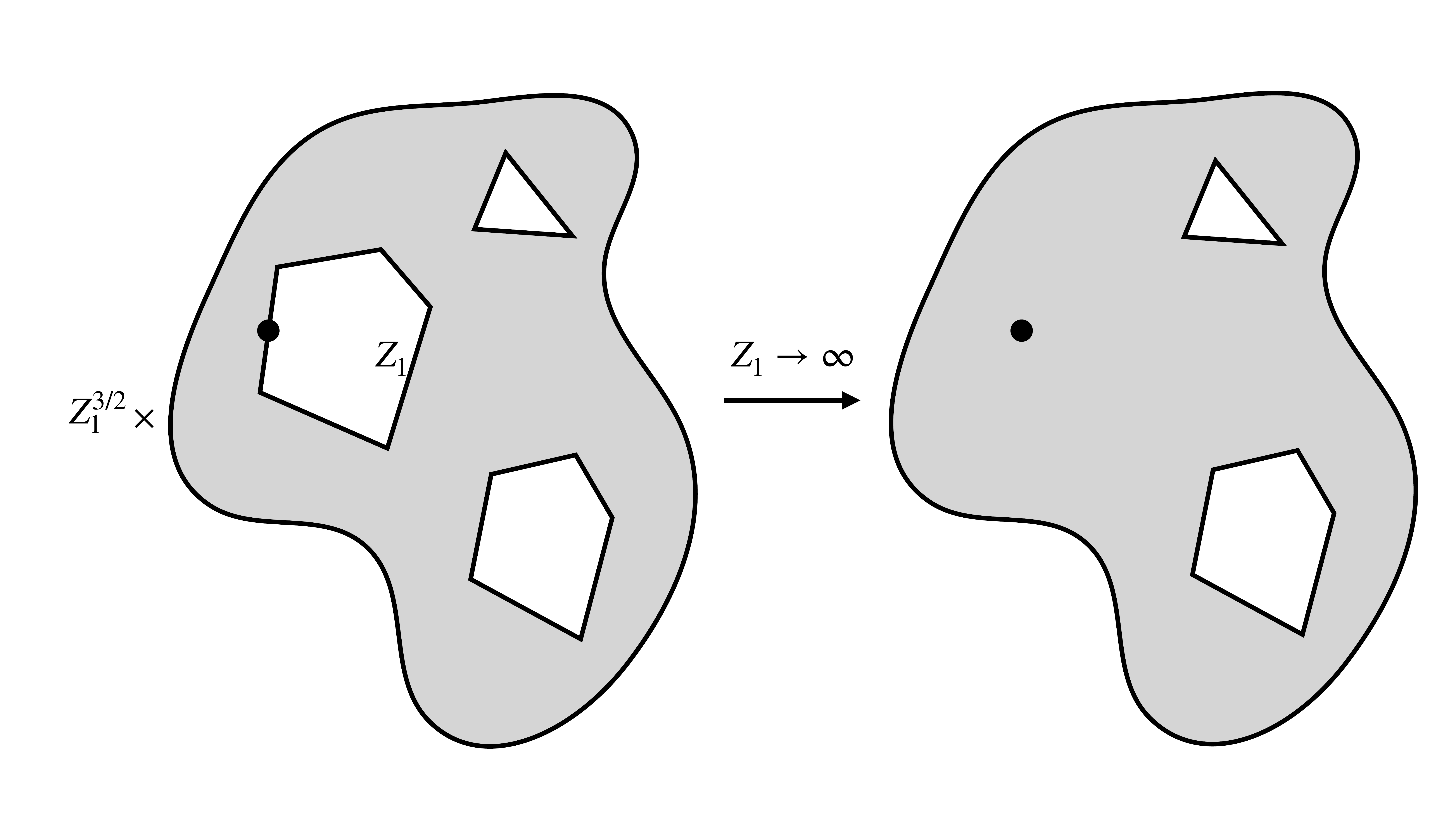}}}
\vspace{-0.4cm}
\caption{{\small  Illustration of eq. \rf{6.8}}}
\label{fig6.1}
\end{figure}

Let us return to the set up where we only allow for triangles and double-links. The disk function was given by \rf{5.28}, which 
we repeat here for convenience:
\beq\label{6.1}
w(g,z) = \oh \Big( z-gz^2 \plu (g z \mi c(g) ) \sqrt{(z-c_+(g))(z-c_-(g))} \Big)
\eeq
Taking the continuum limit for  this model is slightly different from the situation for the bipartite triangulations. since we have 
that $c_+ (g) \neq - c_-(g)$. The critical behavior is still obtained when $M_1(g)  \to 0$ and we have that $c_+(g)$ at this point 
becomes a non-analytic function of $g$. Let us here list the behavior for $\Del g =g_c \mi g \to 0$, defining $z_c = c_+(g_c)$:
\bea
c(g) &=&  g_c z_c \big( 1 + \oh \alpha \sqrt{\Del g} \big) + \cO(\Del g) \label{6.2a}\\
c_+(g) & = & z_c\big( 1 - \alpha \sqrt{\Del g}\big) + \cO(\Del g) \label{7.2b}\label{6.2b}\\
c_-(g) &=& c_-(g_c) + \cO(\Del g) \label{6.2c}
\eea
Here $\alpha$ is  a constant which can be calculated like we did in \rf{5.95x} for the bipartite graphs. We now define the continuum
cosmological constant $\Lam$  and boundary cosmological constant $Z$ as above:
\beq\label{6.3}
z=z_c(1+ \ep Z), \quad \Del g = \alpha^{-2} \ep^2 \Lam,
\eeq
 These relations are the equivalent to \rf{5.80} and \rf{5.81}, and the factor $\alpha^{-2}$ in \rf{6.3} is a rescaling 
 of the cosmological constant, done in order  to ensure that 
 $z-c_+(g) \propto Z+\SL$. A  similar rescaling was performed in \rf{5.81} and it is just to obtain nice-looking formulas.
 Taking the limit $\ep \to 0$ we write $w(g,z)$ as 
 \beq\label{6.4}
 w(g,z) = \oh \Big( z - g z^2 + \ep^{3/2} g_c z_c^{3/2} \; W(\Lam,Z)  + \cO(\ep^2)\Big),
 \eeq
 where
 \beq\label{6.5}
 W(\Lam,Z) = \Big( Z \mi \oh \SL\Big) \sqrt{Z +\SL}
 \eeq
 This is precisely \rf{5.91}, which we derived by a  Laplace transformation of  $W(V,L)$. As discussed below eq.\ \rf{5.91}
 the part $z-gz^2$ has no continuum limit, but will play no role in the following\footnote{In certain situations, not discussed 
 in these notes it {\it can} play a role, see  footnote \ref{footnote19}.}. The  continuum multiloop functions 
 $W(\Lam, Z_1, \ldots,Z_n)$ are given by \rf{5.85} and \rf{5.87}. What one observes is that the following formula is valid
 \beq\label{6.8}
 \lim\limits_{Z_1 \to \infty} Z_1^{3/2} \,W(\Lam,Z_1,\ldots,Z_n) \propto \Big( \!\mi \frac{d}{d \Lam} \Big)\, W(\Lam,Z_2,\ldots, Z_n)
 \eeq
 It is even true for $n \equ 2$ and for $n \equ 1$, although in particular the $n \equ 1$ case requires some additional arguments (which we will
 not present here). The interpretation is as follows: from the relation \rf{5.6} between $W(\Lam,Z_1, \ldots,Z_n)$ and 
 $W(\Lam,L_1,\ldots,L_n)$ it is seen that $Z_1 \to \infty$ corresponds to $L_1 \to 0$. Thus we are contracting the marked loop with 
 boundary cosmological constant $Z_1$ to a ``marked point'', and  we multiply by the factor $Z_1^{3/2}$ to get rid of a remaining 
 allover $Z_1$ factor associated with  the marked point. But this marked point can be anywhere on the surface. Thus the number of surfaces
 with a marked point is related to the number of surfaces without a marked point by multiplying with the volume (area) of the surface. This 
 is precisely implemented by $\big( \mi \frac{d}{d \Lam}\big)$. Recall that $\Lam$ only appears in the continuum action \rf{5.2} as $\Lam V$ 
 and thus differentiating  $W(\Lam,Z_1, \ldots,Z_n)$  defined as the path integral \rf{5.3} brings down a factor $V$. Eq. \rf{6.8} is illustrated 
 in Fig.\ \ref{fig6.1}. We will later use this procedure to contract  marked loops to  marked points.
 
\begin{figure}[t]
\vspace{-1.5cm}
\centerline{\scalebox{0.2}{\includegraphics[angle = 0]{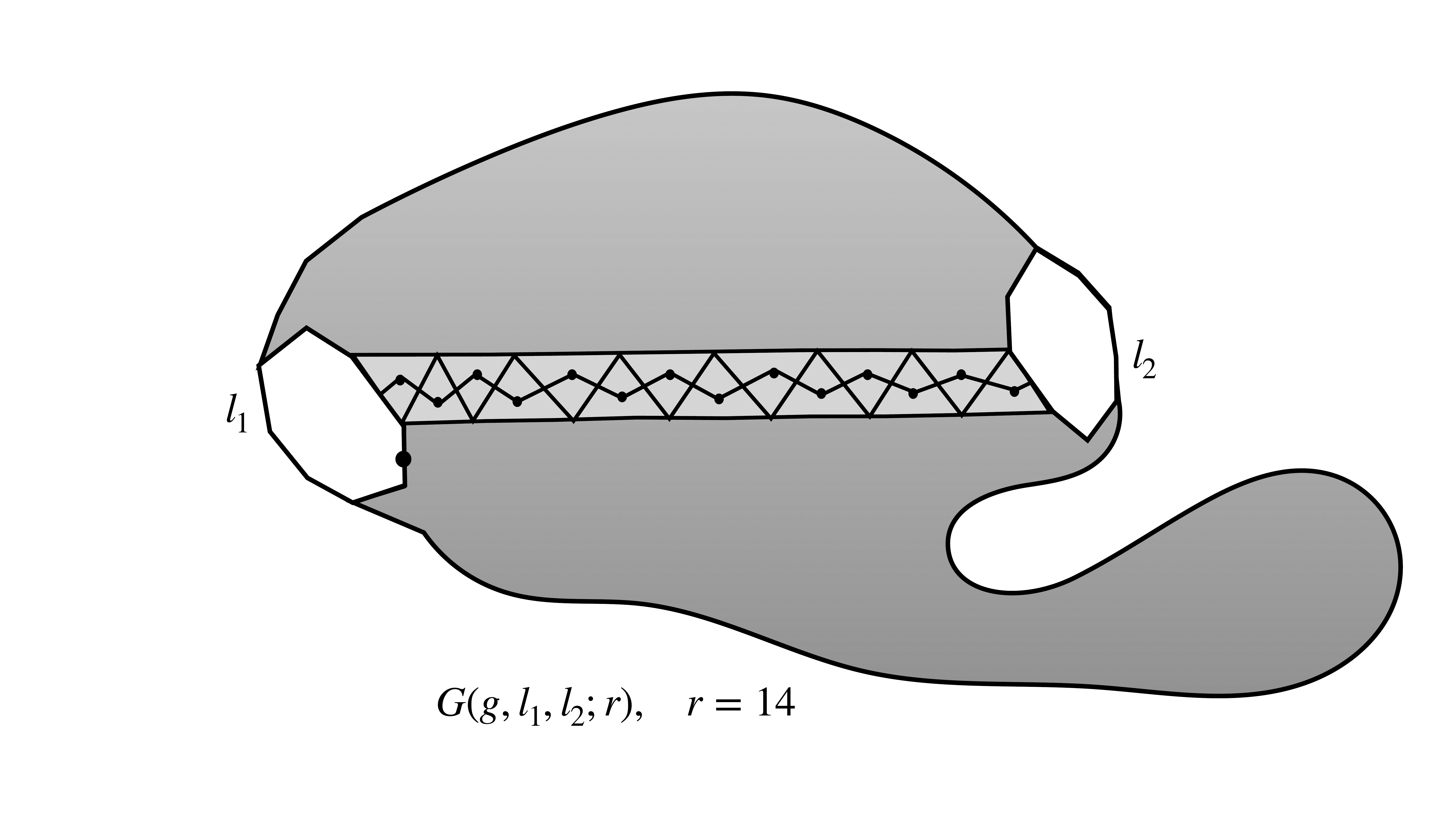}}}
\vspace{-0.8cm}
\caption{{\small  Illustration of a configuration with boundary loops of $l_1$ and $l_2$, separated a graph distance $r= 14$.}}
\label{fig6.2}
\end{figure}

Let us now consider the two-loop function, but with the additional constraint that every point on the ``exit'' loop has a fixed distance to 
the ``entrance'' loop. In order to formulate this in a precise way, we return to the discrete formulation in terms of triangles and double-links.
We mark a link on the entrance loop, but not on the exit loop. These choices are made for convenience, as we will explain later. We denote
the two-loop function by 
\beq\label{6.10}
G (g, l_1,l_2; r), \qquad g = \e^{-\mu}.
\eeq
 The relation between $g$ and $\mu$ is the standard one we have been using, given by \rf{5.75}. $l_1$ denotes at the same 
 time the entrance loop and  the number of links in the 
 entrance loop (one link marked), and similarly $l_2$ denotes the exit loop and the number of links in the exit loop. 
 $r$ denotes the {\it graph distance} between 
 $l_2$ and $l_1$ in a given triangulation with the two boundary loops. Given a link in $l_2$ and a link in $l_1$ we define the graph distance 
 between these links as shortest path through neighboring  triangles which connects the two links, the length of the path counted as 
 as number of triangles present in the path. One can imagine the path as a piecewise linear path passing through the centres of the 
 neighboring triangles. The situation is illustrated in Fig.\ \ref{fig6.2}. 
 The distance between a given link in $l_2$ and the boundary $l_1$ is defined
 as the minimum of the distances between the given link in $l_2$ and the links in $l_1$.  We now require that {\it each link in $l_2$ has 
 the same distance $r$ to $l_1$} (note that this does not ensure that each link in $l_1$ has the same distance to $l_2$, but there is 
 of course at least one link in $l_1$ which has the distance $r$ to $l_2$). This way of defining the distance between the boundaries ensures 
 that we have the composition law:
 \beq\label{6.11}
 G(g, l_1,l_2; r_1 + r_2) = \sum_{l=2}^\infty G (g, l_1,l;r_1) \; G (g, l, l_2; r_2).
 \eeq
 This is where the choice of marking and non-marking of the boundary loops comes into play. 
 With our choice there is no additional $l$ weight factor in the sum, related to the way one can ``turn'' the two cylinders relative to each other, when gluing them together to one cylinder.

 We will think of the graph distance defined this way as the {\it geodesic distance} between the two loops for a given triangulation
 in the piecewise linear geometry defined by the equilateral triangulation. Of course this is not strictly speaking correct, but we expect
 for very large generic triangulations and very large distances that the real geodesic distance will be proportional to the graph distance and we will in the 
 following not distinguish between the two.

\begin{figure}[t]
\vspace{-1.5cm}
\centerline{\scalebox{0.2}{\includegraphics[angle = 0]{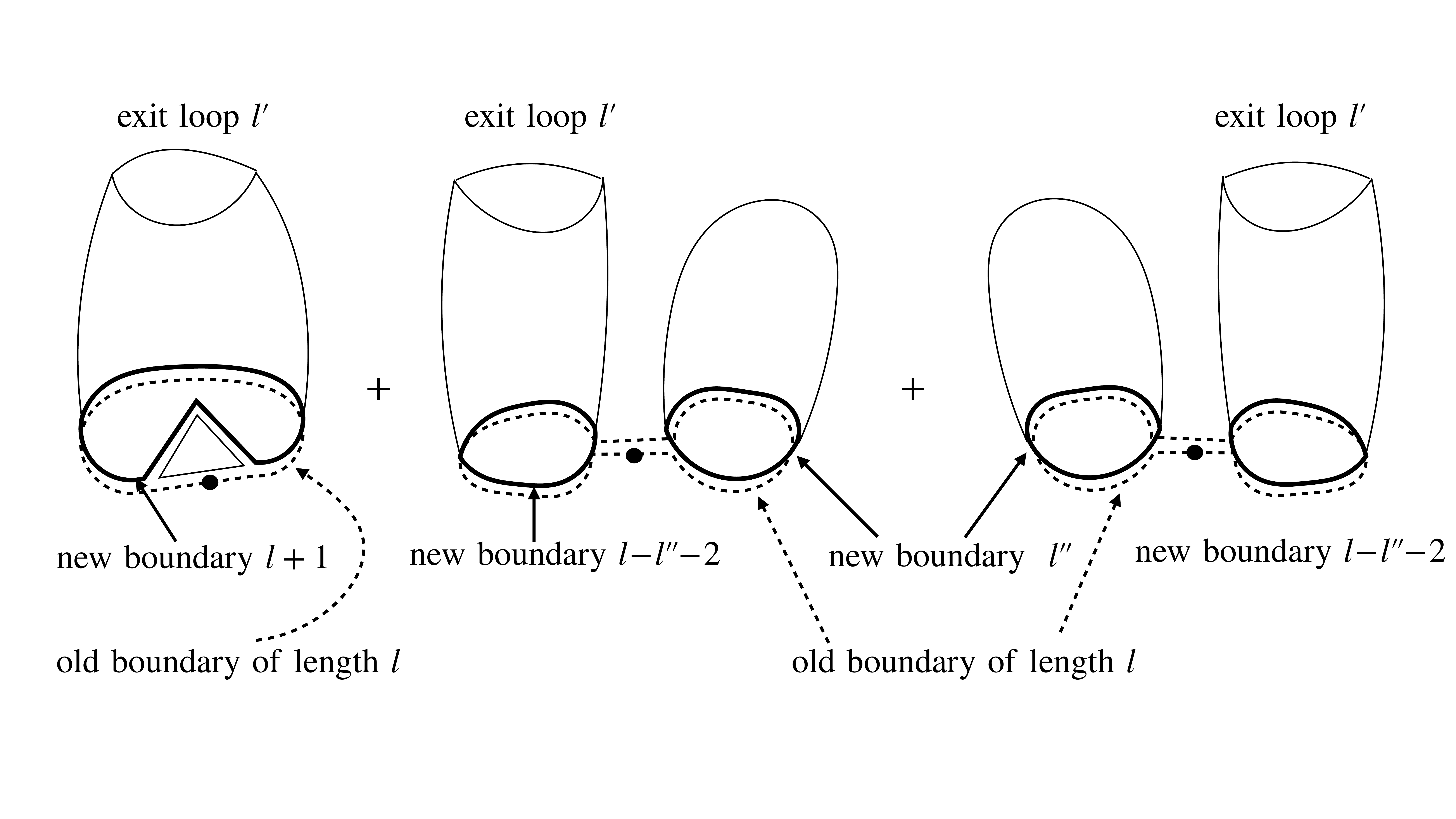}}}
\vspace{-1.0cm}
\caption{{\small  Graphically illustration of equation \rf{6.12}.}}
\label{fig6.3}
\end{figure}

\begin{figure}[t]
\vspace{-1.0cm}
\centerline{\scalebox{0.15}{\includegraphics[angle = 0]{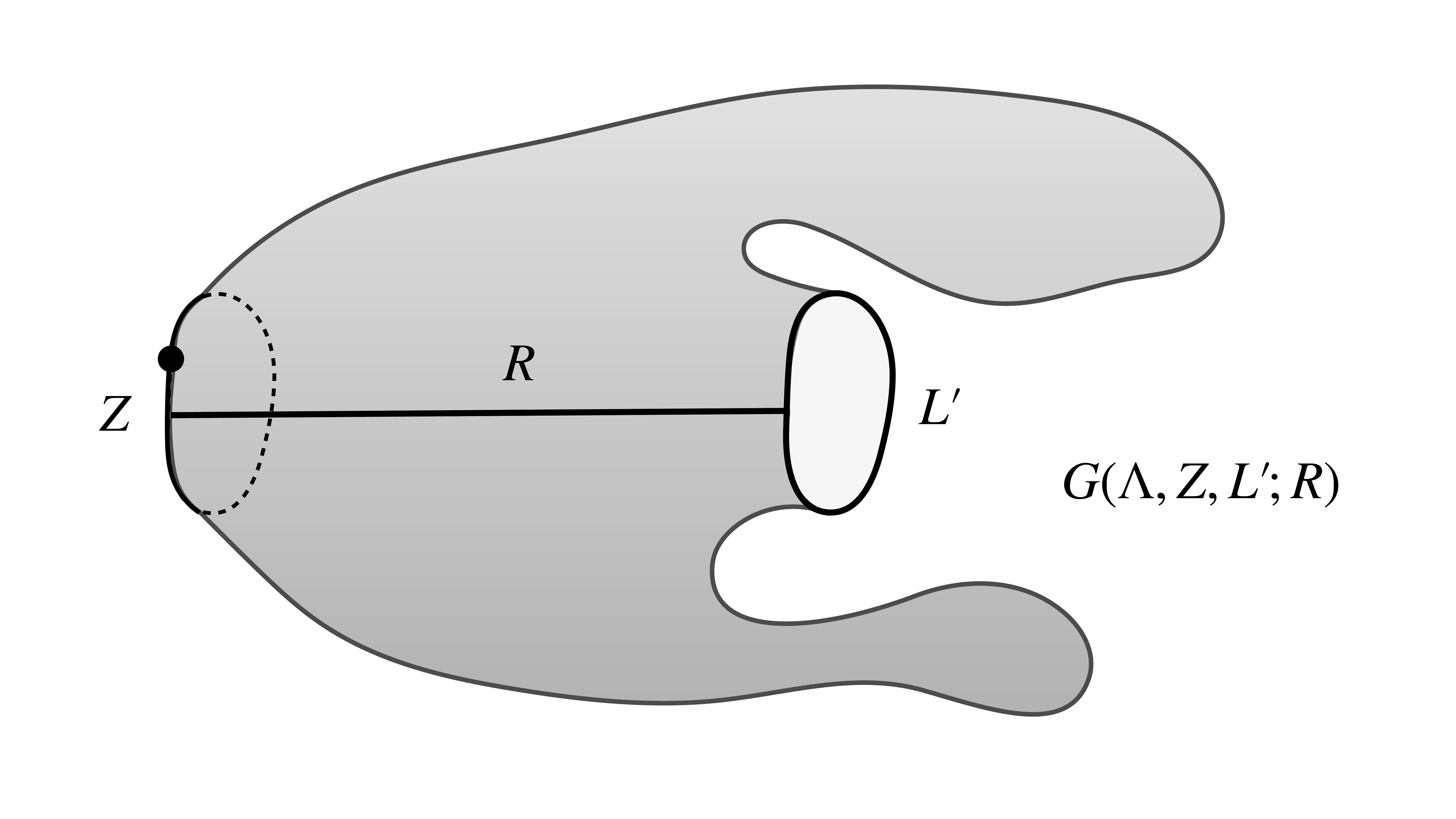}}}
\vspace{-.5cm}
\caption{{\small  Graphically illustration of $G(\Lambda,Z,L';R)$.}}
\label{fig6.4}
\end{figure}

 We now use the same ``moves'' as shown in Fig.\ \ref{fig5.2}, except that we apply them to the cylinder surfaces used in the
 calculation of $G (g, l_1,l_2;r)$. 
 We ``peel'' away a triangle from the entrance loop, moving ``closer'' to the exit loop, or, when we meet 
 a double-link, we chop away a ``baby'' universe, as shown in Fig.\ \ref{fig6.3}. Algebraically, we can write the operation as follows:
 \beq\label{6.12}
 G (g, l,l';r) = g \tilde{G} (g, l\pl 1,l';r) + 2 \sum_{l'' =0}^{l-2} w_{l''}(g) \, G(g, l\mi l'' \mi 2,l';r),
 \eeq
 where $w_l(g)$ as usual denotes the disk function with a boundary consisting of $l$ links. $g \tilde{G} (g, l \pl 1,l';r)$ (the left graph in 
 Fig.\ \ref{fig6.3}) is not really of the form $g G(g,l\pl 1,l';r)$ since removing a triangle in general will spoil the property that exit  links have 
 a distance $r$ to the entrance loop. However, applying the removal of triangles $l$ times will ``in average'' get us one step 
 closer to the exit loop, i.e.\ to $G(g,l, l',r\mi 1)$. Thus we write
 \beq\label{6.13}
 g \tilde{G} (g,l\pl 1, l'; r) = g G (g, l\pl 1,l';r) - \frac{1}{l} \, \frac{\prt G (g, l,l';r)}{\prt r},
 \eeq
 where the factor $1/l$ in front of the derivative term refers to $1/l^{th}$ of the $l$ times we in average have to remove a triangle to get
 from $r$ to $r\mi 1$. Clearly this is not a rigorous result the way it is presented here. The operation $- \frac{1}{l} \, \frac{\prt}{\prt r}$ is 
 rather intuitive, at best.  However, it can be made rigorous, but it is rather tedious\footnote{For the really dedicated 
 readers we can refer to the article \cite{ab1}.}, and we will simply accept \rf{6.13}. Inserted in \rf{6.12}
 we obtain:
 \bea\label{6.14}
 \frac{\prt G (g,l,l';r)}{\prt r} &=& -l \, G(g, l,l';r) + g l \, G (g, l \pl 1,l';r) +\\
 && 2 l \sum_{l'' =0}^{l-2}  G(g, l\mi l'' \mi 2,l';r)\, w_{l''}(g). \no 
 \eea  
  The last term is a kind of convolution. We thus introduce the (discrete) Laplace transformation, which turns convolutions into 
 products:
 \bea\label{6.15}
 \hG(z) := \sum_{l=0}^\infty  \frac{G(l)}{z^{l+1}}, \qquad G(l) = \oint \frac{dz}{2\pi i} \, z^l \, \hG(z).\\
 H(l) = \sum_{l'=0}^l G(l') F(l\mi l' ) ~~ \implies ~~z\hH(z) = z\hG(z)\cdot z\hat{F}(z) \label{6.15a}
 \eea
 The inversion formula in \rf{6.15} assumes the contour is in the region where $\hG(z)$ is analytic, and 
 the convolution formula uses the rearrangement 
 \beq\label{6.15b}
 \sum_{l =0}^\infty\sum_{l'=0}^l \frac{ G(l\mi l')F(l')}{z^l} = \sum_{l'=0}^\infty \sum_{l-l' =0}^\infty \frac{G(l-l')}{z^{l-l'}} \; \frac{F(l')}{z^{l'}}
 \eeq
 We now introduces the discrete Laplace transform for  variable $l$  in  $G(g,l,l';r)$ and by an abuse of notation we still denote 
 it $G(g,z,l';r)$:
 \beq\label{6.15c}
 G(g,z,l';r) = \sum_{l=0}^\infty \frac{G(g,l,l';r)}{z^{l+1}.}
 \eeq 
 From \rf{6.15} and \rf{6.15a}  one obtains after a little algebra
 \beq\label{6.17}
 \frac{\prt G (g,z,l';r)}{\prt r} = \frac{\prt}{\prt z} \Big[ \big( z - g z^2 -2 w(g,z) \big) \; G(g,z,l';r) \Big]
 \eeq
 Recall from  \rf{6.1} and \rf{6.4} that the term in $( \cdot )$ is precisely the part that scales:
 \beq\label{6.17a}
 z -gz^2 -2 w(g,z) \propto - \ep^{3/2} W(\Lam,Z)
 \eeq
 This makes it possible directly to take the continuum limit of \rf{6.17}. Assume the scaling \rf{6.3} and in addition
 \beq\label{6.18}
 \ep \, l = L, \quad \ep\, l' = L', \quad \ep^\del r \propto  R.
 \eeq
 Naively, we would expect that $\del \equ 1$. However, that will not be the case. 
 First note that 
 \beq\label{6.18a}
 \sum_l  = \frac{1}{\ep} \sum_l \ep \to    \frac{1}{\ep}  \int dL
 \eeq
 Next, it then follows from the composition rule \rf{6.11} that the continuum limit of $G(g,l,l';r)$ has to scale as 
 \beq\label{6.19}
 G(g,l,l';r) ~\propto ~\ep\, G(\Lam,L,L';R)
 \eeq
 Then \rf{6.15} leads to, for $\ep \to 0$,
 \beq\label{6.20}
 G(g,z,l';r) ~\propto~  \int_0^\infty dL \; \e^{-L Z} G(\Lam,L,L';R) = G(\Lam,Z,L';R)
 \eeq 
 From \rf{6.17} we then have
 \beq\label{6.21}
 \ep^\del \frac{\prt }{\prt R} \, G(\Lam, Z,L';R) = - \frac{1}{\ep} \, \frac{\prt}{ \prt Z} \Big( \ep^{3/2} W(\Lam,Z) \, G(\Lam,Z,L';R) \Big)
 \eeq
 Thus ${\del =\oh}$: {\it the ``geodesic'' distance $R$ scales anomalously} :
 \beq\label{6.22}
 \boxed{{\rm dim} [R] =  \oh  {\rm dim} [L] = \oq  {\rm dim} [V]}
 \eeq
 and we have the equation 
 \beq\label{6.23}
\boxed{ \frac{\prt }{\prt R} \, G(\Lam, Z,L';R) = -  \frac{\prt}{ \prt Z} \Big(  W(\Lam,Z) \, G(\Lam,Z,L';R) \Big).}
 \eeq
 The function $G(\Lam, Z,L';R) $ describes the ``propagation'' of a spatial universe, where the length distribution 
 is dictated by the boundary cosmological constant $Z$ (and the boundary has a mark),  a distance $R$ to a spatial 
 universe where the boundary has length $L'$. The distance $R$ is an intrinsic geodesic distance for each 
 geometry which contributes to path integral defining $G(\Lam, Z,L';R) $. The situation is illustrated in Fig.\ \ref{fig6.4}.
 
 Eq.\ \rf{6.23} is an ordinary 1st order partial differential equation of the form
 \beq\label{6.24}
 \frac{\prt f(x,y) }{\prt y} = - \frac{ \prt ( w(x) f(x,y))}{\prt x}, \qquad f(x,y_0) = h(x),
 \eeq
 where the last equation serves as a boundary condition.
 The solution to this equation is 
 \beq\label{6.25}
 f(x,y) = h( \bx(y;x)) \; \frac{w(\bx(y;x))}{w(x)},
 \eeq
 where $\bx(y)$ is a solution to the ordinary differential equation (the so-called characteristic equation for the partial differential equation):
 \beq\label{6.26}
 \frac{d \bx}{dy} = - w(\bx), \quad \bx(y_0) = x ,
 \eeq
 In eq.\ \rf{6.25} we have written $\bx(y_0) \equiv \bx(y_0;x)$ to emphasize the dependence on $x$ via the boundary condition. The solution to 
 the characteristic equation is 
 \beq\label{6.27}
 y- y_0 = \int_{\bx(y)}^x \frac{dx'}{w(x')}.
 \eeq
 
 We can directly apply this to \rf{6.23} if  we impose the natural boundary condition
 \beq\label{6.28}
 G(\Lam,L,L';R \equ 0) = \del(L-L').
 \eeq
 This implies that 
 \beq\label{6.29}
 G(\Lam,Z,L'; R \equ 0) = \int_0^\infty dL \; \e^{-ZL} \, G(\Lam,L,L';   R \equ 0) = \e^{-Z L'}
 \eeq
 which will serve as our boundary condition for \rf{6.23}. Corresponding to \rf{6.26} and \rf{6.27} we have
 \beq\label{6.30}
 \frac{d \bZ}{d R} = - W(\Lam,\bZ), \qquad \bZ(0) = Z,
 \eeq
 \beq\label{6.31}
 R =  \int^Z_{\bZ(R;Z)} \frac{dZ'}{W(\Lam,Z')} =  \int^Z_{\bZ(R;Z)} \frac{dZ'}{(Z'-\oh \SL)\sqrt{Z'\pl \SL}},
 \eeq
 which is easily integrated using the substitution $\xi = \sqrt{Z'\pl \SL}$. One obtains:
 \beq\label{6.32}
  R = \frac{1}{\sqrt{\frac{3}{2}\SL}} \ln \frac{H(\bZ)}{H(Z)} ,\qquad H(X) = 
 \frac{\sqrt{X \pl \SL} \pl \sqrt{\frac{3}{2} \SL}}{\sqrt{X \pl \SL} \mi \sqrt{\frac{3}{2} \SL}},
\eeq
From this we can find $\bZ(R;Z)$ and $W(\Lam,\bZ)$
\beq\label{6.33}
\bZ(R;Z) = \oh \SL + \frac{3}{2} \SL 
\left[ \frac{ \Big( H(Z) \,\e^{ \sqrt{\frac{3}{2} \SL} \, R} + 1\Big)^2}{ \Big( H(Z) \,\e^{ \sqrt{\frac{3}{2} \SL} \, R} - 1\Big)^2} - 1\right]
\eeq
\beq\label{6.34}
W(\Lam,\bZ(R;Z)) = \big(\bZ(R;Z) - \oh \SL\big) \sqrt{ \frac{3}{2} \SL} \left[  
 \frac{  H(Z) \,\e^{ \sqrt{\frac{3}{2} \SL} \, R} + 1}{  H(Z) \,\e^{ \sqrt{\frac{3}{2} \SL} \, R} - 1}\right]
 \eeq
 Our final solution is thus 
 \beq\label{6.35}
 \boxed{G(\Lam,Z,L';R) = \frac{W(\Lam, \bZ(R;Z))}{W(\Lam,Z)} \; \e^{-\bZ(R;Z) L'}}
 \eeq
 and by a Laplace transformation in $L'$:
 \beq\label{6.35a}
 G(\Lam,Z,Y;R) = \frac{W(\Lam, \bZ(R;Z))}{W(\Lam,Z)} \; \frac{1}{Y \plu \bZ(R;Z)}
\eeq
 
 \subsection*{The two-point function}
 
\begin{figure}[t]
\vspace{-1.5cm}
\centerline{\scalebox{0.15}{\includegraphics[angle = 0]{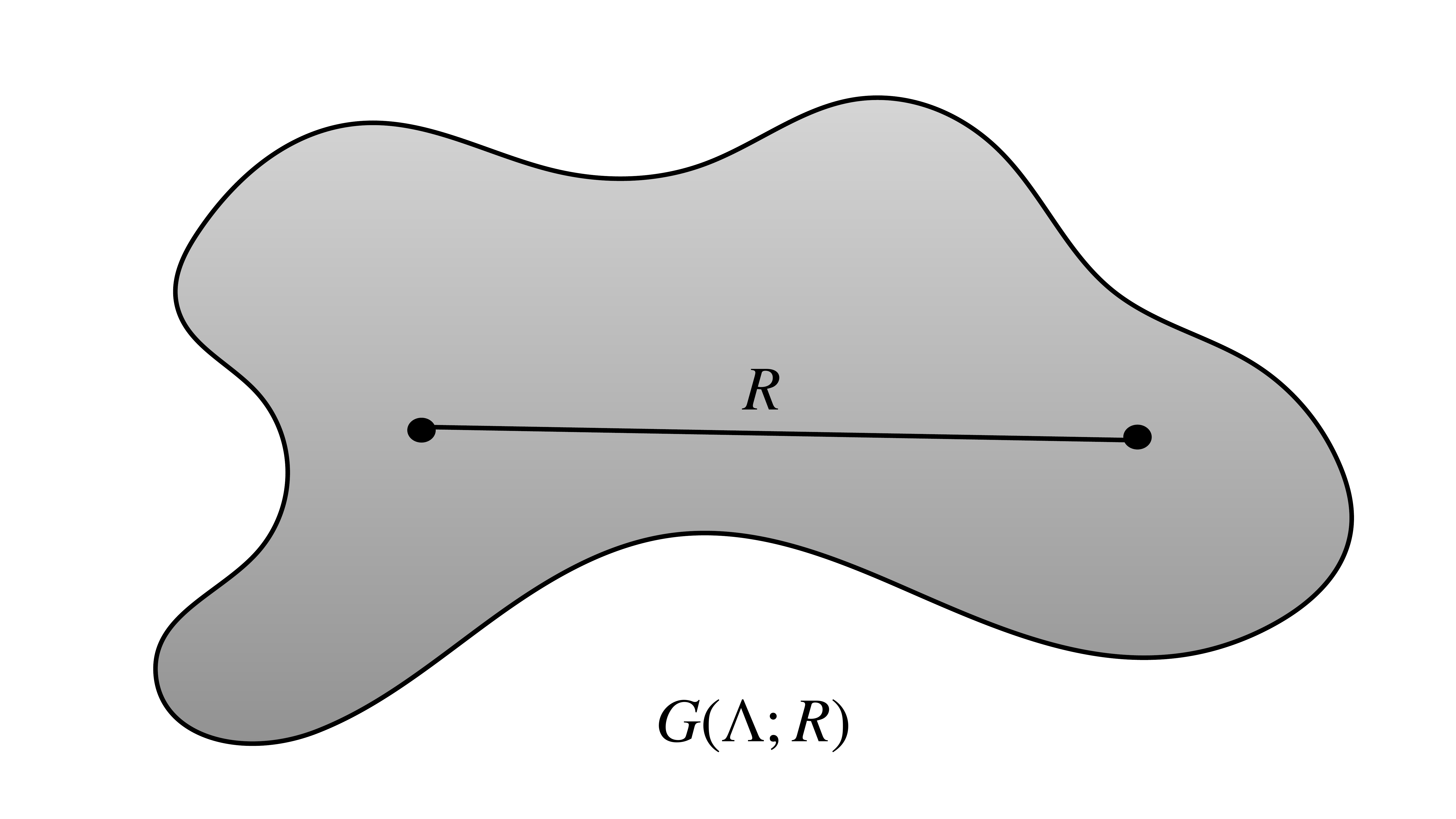}}}
\caption{{\small  Graphically illustration of $G(\Lambda;R)$.}}
\label{fig6.7}
\end{figure}

 However, we are more interested in the limit where the exit and entrance loops are contracted to points. For the exit loop this is easy,
 we just take $L' \to 0$ in \rf{6.35}. From \rf{6.8} we also know how to contract the entrance loop to a marked point, namely 
 by multiplying with $Z^{3/2}$ and taking $Z\to \infty$. Since $Z^{3/2}/W(\Lam,Z) \to 1$ in this limit we obtain
 \beq\label{6.37}
 G(\Lam;R) = W(\Lam;\bZ(R;Z\equ \infty))
 \eeq
 Since $H(Z) \to 1$ for $Z \to \infty$ the expression for  $W(\Lam;\bZ(R;Z\equ \infty))$ becomes quite simple:
 \beq\label{6.39}
 \boxed{G(\Lam;R) = c\, \Lam^{3/4} \; \frac{ \cosh \sqrt[4]{\Lam} \, \tR}{\sinh^3  \sqrt[4]{\Lam} \, \tR}} \qquad \tR = \oh \Big(\frac{3}{2}\Big)^{1/2} R,
 \quad c= \Big(\frac{3}{2}\Big)^{3/4}  
 \eeq
We denote $G(\Lam,R)$ {\it the two-point function and it has the interpretation as the partition function for universes where two marked points
are separated a geodesic distance $R$}, as shown in Fig.\ \ref{fig6.7}. It has the following definition in terms of a path integral over geometries:
\beq\label{6.38}
\boxed{G(\Lam;R) = \int \cD [g] \; \e^{ -\Lam \int d^2 \xi \,\sqrt{g}} \int\!\!\! \int d^2\xi_1d^2\xi_2 \, \sqrt{\!g(\xi_1)} \sqrt{\!g(\xi_2)}
\; \del\big(D_g(\xi_1,\xi_2) \mi R\big)}
\eeq
where $D_g(\xi_1,\xi_2)$ is the geodesic distance between $\xi_1$ and $\xi_2$, measured in a metric $g_{a b}(\xi)$ defining a given 
geometry in the path integral.  

Formula \rf{6.39} is quite amazing. 
It is simple to derive from the dynamical triangulation formalism by counting triangulations, simple to 
define in the scaling limit (the continuum limit), but impossible to calculate directly from the continuum definition \rf{6.38} because
the geodesic distance  $D_g(\xi_1,\xi_2)$ is an immensely complicated function of the metric $g$ for a general 
geometry.  We also see  
that the ``quantum average'' of $D_g(\xi_1,\xi_2)$ indeed is very ``quantum'' since the dimension of $R$ is different from 
the dimension of $D_g(\xi_1,\xi_2)$ one would expect for the geodesic distance of a nice smooth geometry. This is of course
only possible if a typical geometry appearing in the path integral is not at all nice and smooth at the scale set by $R$. But since 
$R$ was arbitrary, this has to be true for geometries at all scales. We will return to discuss this further below.

It is seen that $G(\Lam;R)$ behaves very much like an ordinary 2-point function:
\bea\label{6.40}
G(\Lam;R) &\propto& \frac{1}{\tR^3} \qquad  \hspace{1.7cm} {\rm for} \quad \tR \ll \frac{ 1}{\sqrt[4]{\Lam}} \\
G(\Lam;R) &\propto&  \Lam^{3/4} \; \e^{ -2\sqrt[4]{\Lam} \, \tR}   \qquad {\rm for} \quad \tR \gg \frac{ 1}{\sqrt[4]{\Lam}} \label{6.40a}
\eea
 In addition the 2-point function has a ``stringy'' feature: it has an infinity set of equidistance  mass excitations (for a real string it is actually
 the not the mass excitations $m_n$, but $m_n^2$ which are equidistancely separated). If we use
 \beq\label{6.40b}
 \frac{\cosh x}{\sinh^3 x} = 4 \e^{-2x} \, \frac{1\plu \e^{-2x}}{(1\mi \e^{-2x})^3},\qquad \frac{1\plu z}{(1\mi z)^3} = 
 \frac{d}{dz} z \frac{d}{dz} \frac{1}{1\mi z} =\sum_{n=1}^\infty n^2 \,z^{n-1},
 \eeq
 we can write
 \beq\label{6.41}
 G( \Lam;R) \propto \sum_{n=1}^\infty n^2 \, \e^{-2n \sqrt[4]{\Lam}\, \tR}
 \eeq
 and we have  mass excitations $m_n \propto 2n \sqrt[4]{\Lam}$.
 
 Returning to the dimensionless variables variables $r$ and $\mu$ used before taking the continuum limit 
 ( $ R \propto \ep^{1/2} r$ and $ \Del \mu \equ\m \mi \m_c \propto \ep^2 \Lam$), we can write 
 \rf{6.38}, \rf{6.40} and \rf{6.40a} as 
 \beq\label{6.43} 
 G_\m (r) \propto \Del \mu^{\frac{3}{4}} \;\frac{ \cosh \Del \mu^{\oq} \, r}{\sinh^3 \Del \mu^{\oq} \, r}  
 \quad\approx \quad
 \begin{matrix}
 r^{-3} & {\rm for} &  r \ll  \Del \mu^{-\oq} \\
   \Del \mu^{\frac{3}{4}} \; \e^{-2 \Del \mu^{\oq} r}   & {\rm for} & r \gg \Del \mu^{-\oq}
 \end{matrix}
 \eeq
 Following our discussion of intrinsic critical exponents  for BPs (see \rf{3.35}),  we see from the short distance 
 behavior of the  two-point function, $G_\mu(r) \propto r^{1-\eta}$, that the exponent $\eta = 4$. This is a quite 
 unusual exponent! \footnote{In ordinary quantum field theory in flat spacetime one considers $\eta \equ 2$ 
 to be an upper bound on the 
 anomalus scaling dimension, the reason being that the propagator then behaves like $1/|p|^{2-\eta}$ for  large momentum. 
 If $\eta > 2$ the propagator is growing with large momentum, making any probabilistic interpretation of scattering processes 
 in quantum field theory problematic. This is also the reason we in the discussion of bosonic string theory, keeping an eye 
 on Fisher's scaling relation $\gamma = \nu(2-\eta)$, said that having a $\nu >0$ and a $\gamma >0$ goes
 hand in hand. However, surprisingly, for the two-point function of instrinsic 2d gravity the situation is  different.}. 
 Also we read off from the exponential decay of $G_\mu(r)$ that the exponent $\nu = 1/4$. We also 
 know from \rf{5.91a} that $\gamma =-1/2$ and thus the unusual value of $\eta$ ensures that Fisher's scaling 
 relation is satisfied for $G_\m(r)$:
 \beq\label{6.43a}
 \boxed{\gamma = \nu (2-\eta), \qquad \gamma =-\oh,~~\eta = 4,~~ \nu = \oq}
 \eeq  
 While we appealed to  general considerations when using  eq.\ \rf{5.91a} to argue that $\gamma = -1/2$, 
 it can also be shown directly from
 \rf{6.43} using the elementary definition of susceptibility in terms of  the two-point function:
 \beq\label{6.43b}
 \chi(\Del \mu) = \sum_{r=1}^\infty G_\mu(r) = {\rm cnst.} - \frac{1}{6} \Del \mu^{\oh} + \cdots
 \eeq
 In the discussions related to spin systems, RWs and BPs the critical exponent $\gamma$ of the susceptibility was defined 
 by the divergence of $\chi(\Del \mu)$ for $\Del \mu \to 0$, namely $\chi(\Del \mu) \propto (\Del \mu)^{-\gamma}$. 
 That of course assumes that $\chi(\Del \mu)$ {\it is} divergent for $\Del \mu \to 0$, which was the case. Here we have 
 $\gamma \equ \mi 1/2$ and we will define the susceptibility by the leading non-analytic term, which in this case is $\sqrt{\Del \mu}$.
 Also, one should not be surprised that the coefficient multiplying $\sqrt{\Del \mu}$ is negative (despite $\chi(\Del \mu)$ of course 
 being a positive function). If $\gamma \equ \mi 1/2$, the susceptibility exponent  of the three-point function 
 will be $\gamma \pl 1 > 0$, 
 i.e.\ the three-point function will diverge for $\Del \mu \to 0$, and it is of course positive. But we essentially get the 
 three-point function by $\chi^{(3)}(\Del \mu) \propto - \frac{d}{d\mu}  \chi(\Del \mu)$, as discussed in the case of the bosonic string.
 Thus the coefficient multiplying $\sqrt{\Del \mu}$  in \rf{6.43b} has to be negative\footnote{Note that we still have an 
 equation like \rf{4.69}, if we considered different classes of triangulations $\cT^{(2)}$ and $\cT^{(3)}$ as was the case for 
 the bosonic string. However, if $\gamma < 0$ the susceptibility does not go to infinity when $\mu \to \mu_c$. Thus we cannot 
 conclude that \rf{4.69a} is valid, i.e.\ that $\bmu(\mu_c) > \bmu_c$, which was the main reason we could 
 argue for the BP picture of 
 bosonic strings.}.

 Maybe the most important consequence of the  exponential behavior shown for the two-point 
 function $G_\mu(r)$ is, using the now standard arguments from RWs and BPs,  that 
 {\it the global Hausdorff dimension, $d_H$,  of the set of spherical triangulations is  4}.  Let us now show that 
 also the {\it local} Hausdorff dimension $d_h  \equ 4$, by studying the geometric meaning of $G(\Lam;R)$.

\subsection*{The local  Hausdorff dimension in  2d  gravity} 

The two-point function $G(\Lam;R)$ defined for a given cosmological constant $\Lam$ is related to the two-point function 
$G(V;R)$ defined for a given volume $V$ by a Laplace transformation
\beq\label{6.44}
G(\Lam;R) = \int_0^\infty dV \; \e^{-\Lam \, V} \; G(V;R).
\eeq
The continuum definition of $G(V;R)$ is then (from \rf{6.38} and \rf{6.44})
\beq\label{6.45}
G(V;R) = \!\!\int \!\!\cD [g] \del \Big( \!\int \!\!d^2\xi \sqrt{g} \mi V\Big) \!\!\int\!\!\!\! \!\int \!\!d^2\xi_1d^2\xi_2 \, \sqrt{\!g(\xi_1)} \sqrt{\!g(\xi_2)}
\; \del\big(D_g(\xi_1,\xi_2) \mi R\big)
\eeq
Thus $G(V;R)$ is proportional to the number of geometries with volume $V$ and where in addition 
two marked points are separated a geodesic distance $R$. 
We will now provide a more precise picture of this, which relates $G(V;R)$ to the {\it local Hausdorff dimension}, exactly 
as we did in the case of BPs.

For a given point with coordinates $\xi_1$ we define the ``area'' (in 2d, like here, the length) of a spherical shell located a geodesic 
distance $R$ from $\xi_1$ as
\beq\label{6.46}
S_V(\xi_1,R; g) = \int d^2\xi \sqrt{g(\xi)} \; \del\Big( D_g(\xi,\xi_1) -R\Big)
\eeq
The average of $S_V(\xi_1)$ over the whole manifold is 
\beq\label{6.47}
S_V(R;g) = \frac{1}{V} \int d^2\xi \sqrt{g(\xi)} \; S_V(\xi,R;g)
\eeq
The quantum average of $S_V(R;g)$ over all geometries with volume $V$ is now 
\beq\label{6.48}
\la S_V (R) \ra = \frac{1}{W(V) }
\int \cD [g] \; \del\Big(   \int d^2\xi \sqrt{g} - V\Big) \; S_V(R;g) 
\eeq
where the partition function $W(V)$ according to \rf{5.90b} is given by 
\beq\label{6.49}
W(V) =  \int \cD [g] \; \del\Big(   \int d^2\xi \sqrt{g} - V\Big) ~ \propto~ V^{-7/2}
\eeq
From \rf{6.46} - \rf{6.49} we obtain
\beq\label{6.50}
\boxed{\la S_V (R) \ra = \frac{G(V;R)}{V W(V)} ~\propto~ V^{5/2} G(V;R) }
\eeq
So $G(V;R)$ {\it has a simple geometric interpretation: for a fixed $V$ it is proportional to the quantum average area of a spherical shell 
of radius $R$.}

For a smooth two-dimensional geometry $g$ we have for $R$ sufficiently small
\beq\label{6.51a}
S_V (R;g) \propto R \quad {\rm for} \quad R \ll \frac{1}{\sqrt{V}}
\eeq
For a smooth $d$-dimensional geometry  $g$ we have 
\beq\label{6.51}
S_V (R;g) \propto R^{d-1}  \quad {\rm for} \quad R \ll \frac{1}{V^{1/d}}
\eeq
If the space is fractal with Hausdorff dimension $d_h$ we have (this is the {\it definition} of $d_h$)
\beq\label{6.52}
\la S_V (R) \ra \propto  R^{d_h-1}  \quad {\rm for} \quad R \ll \frac{1}{V^{1/d_h}}
\eeq
Let us now calculate $\la S_V (R) \ra$ using \rf{6.50}. From \rf{6.44} we have by an inverse Laplace transformation:
\beq\label{6.53}
G(V;R) = \int_{-i \infty}^{i \infty} \frac{d \Lam}{2\pi i} \; \e^{V \Lam} \, G(\Lam;R)
\eeq
We expand $G(\Lam;R)$ in powers of $\Lam$:
\beq\label{6.53a}
G(\Lam;R) \propto \frac{1}{\tR^3} - \frac{\Lam \tR}{15} + \frac{4}{189} \Lam^{3/2} \tR^3 +c_5 \Lam^2 \tR^5 + c_7 \Lam^{5/2} \tR^7 + \cdots
\eeq
Now use 
\beq\label{6.53b}
 \int_{-i \infty}^{i \infty} \frac{d \Lam}{2\pi i} \; \e^{V \Lam} \Lam^n = \frac{d^n}{dV^n} \del (V), \quad 
  \int_{-i \infty}^{i \infty} \frac{d \Lam}{2\pi i} \; \e^{V \Lam} \Lam^{n-1/2} = \frac{1} { \Gamma(- n \plu \oh) V^{n+\oh}}.
  \eeq
  We discard the contributions from $\Lam^n$ terms since they corresponds to zero volume $V$ and obtain
  \beq\label{6.54}
  \boxed{\la S_V (R) \ra \propto R^3 \Big(1 + \cO\big( \frac{R^4}{V} \big) \Big)}, 
  \eeq
and comparing with \rf{6.52} we conclude that $\boxed{d_h \equ 4}$.

  \begin{figure}[t]
\vspace{-1.0cm}
\centerline{\scalebox{0.25}{\includegraphics[angle = 0]{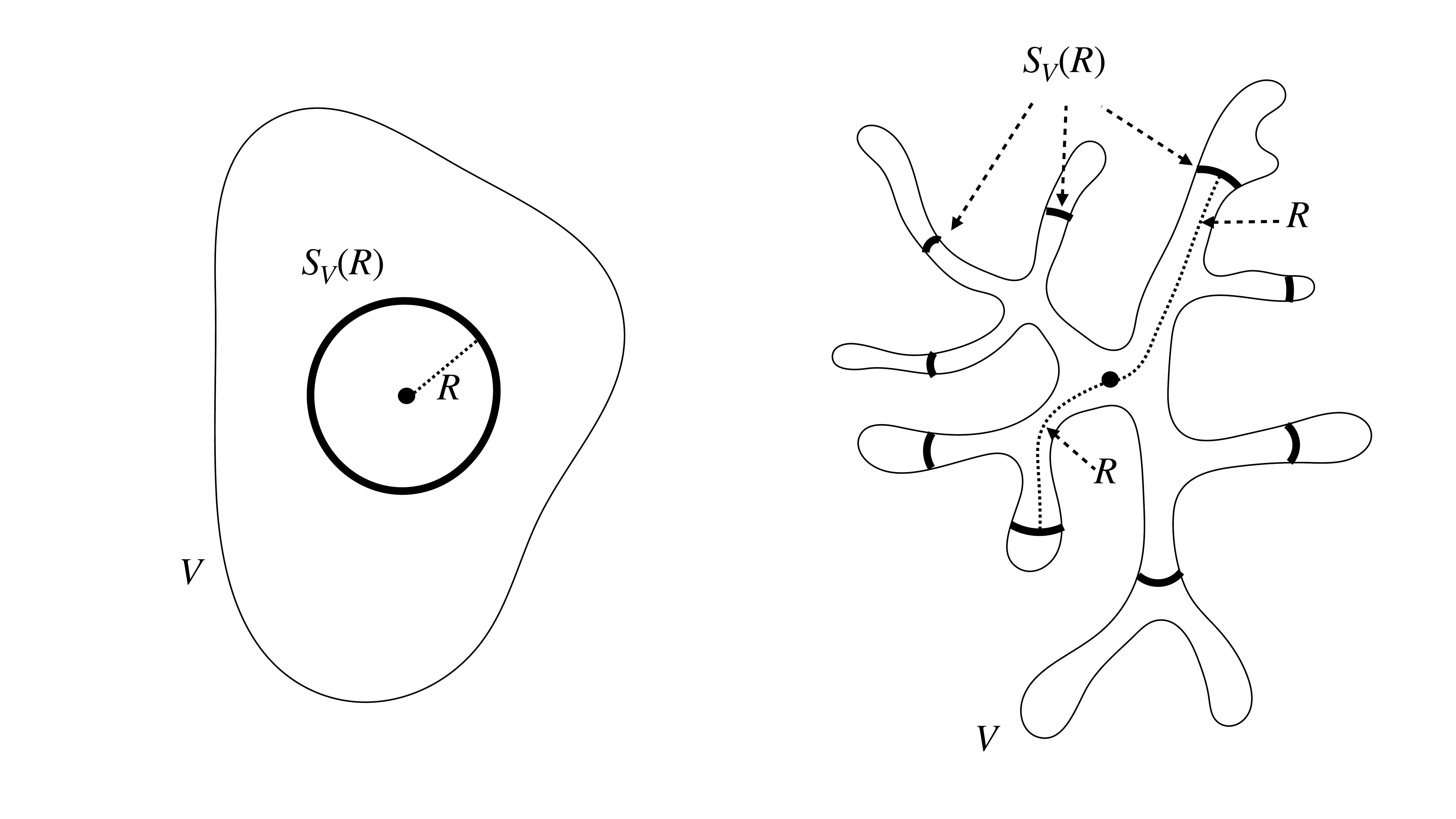}}}
\vspace{-1.0cm}
\caption{{\small  Left: situation when $R \leq R_g$ for a given point. Right: situation when $R \gg R_g$. In the scaling limit, starting out 
with triangulations, there will be infinitely many disconnected parts of $S_V(R)$ with probability one, no matter how small $R$ is. This is 
the reason we can have $S_V(R)\propto R^3$.}}
\label{fig6.8}
\end{figure}

How is it possible that $\la S_V(R)\ra $ is not proportional to $R$ for small $R$? Let us assume that the smooth geometries 
constitutes a dense set in the set of all continuous geometries entering in the path integral. In order to talk about this in 
a meaningful way one has to have a measure defined on the set of continuous geometries, much like we have the Wiener measure in 
the case of RWs. Important progress has been made in this direction in mathematics in the recent years, but we have no 
space to discuss it here \footnote{The interested reader can consult \cite{brownian-map,math1,math2,math2} for a review.}. 
Let us just assume that we have such a measure. Now for each smooth geometry $g$ we use in the path integral 
\rf{6.45} we can find an $R_g$ such that
\beq\label{6.55}
S_V(R;g) \propto R\quad {\rm for} \quad R \leq R_g.
\eeq
This situation is illustrated on the left part of  Fig.\ \ref{fig6.8}. However, in \rf{6.45} the $R$ is chosen independent of $g$ and is a parameter outside 
the integration and the result $d_h \equ 4$ shows that for any $R$, no matter how small, there will be 
many more smooth geometries $g$ for with $R \gg R_g$ than there will be smooth geometries $g$ where $R \leq R_g$.
For the geometries where $R \gg R_g$ the geometry looks more  like the one shown on the right part of Fig.\ \ref{fig6.8}, and 
in such a situation there is no reason why \rf{6.55} should be valid. In fact, with probability 1, if we pick  randomly  a smooth geometry $g$, we 
will obtain $R \gg R_g$. The same statement would be ``even more true'' if applied to the full set of continuous geometries 
which enters in the path integral \rf{6.45}. It is a  beautiful result  that this set of rather unwieldy  geometries has a well defined 
Hausdorff dimension, namely $d_h = d_H =4$, and in a sense it is the generalization of the RW result, where the set of 
random walks has $d_H \equ 2$, the double of the dimension expected for a smooth path, only are we in the case of geometries 
talking about entirely intrinsic properties, while we in the case of RWs talked about properties of the RW embedded in $\mathbb{R}^D$.

\newpage
 
\setcounter{figure}{0}
 \renewcommand{\thefigure}{7.\arabic{figure}}
 \setcounter{equation}{0}
 \renewcommand{\theequation}{7.\arabic{equation}}
 
 \section*{7. The Causal Dynamical Triangulation  model}

\subsection*{ Lorentzian versus Euclidean set up}

The two-dimensional Euclidean gravity model we have studied satisfies the Wilsonian criterium for universality: to a large extent
it is independent of the details of the short distance regularization. We were not restricted to use triangulations as building 
blocks, but could use any (finite) combination of  polygons as building blocks, as long as the weights of polygons were 
all positive, and we would obtain the same continuum multi-loop functions when the dimensionless cosmological 
coupling constant $\mu \to \mu_c$ in such a way that $\mu = \mu_c + \ep^2 \Lam$, where the link length in the graphs went to 
zero while the {\it continuum} cosmological constant $\Lam$ survived. In that limit the average number $N$ of polygons in the 
graphs also diverged for multi-loop functions with three or more loops and we could talk about a finite continuum 
limit of the volume $V \propto N \ep^2$, where $N$ denoted the number of polygons. We had $\la V \ra\propto \Lam^{-1}$ if 
we did not fix the volume of spacetime, but considered the model with a fixed cosmological constant $\Lam$. 
Further, by studying the two-point function as a function of the so-called geodesic distance, 
we identified the  correlation length which diverged when we approached the critical point $\mu_c$. 
In this sense the two-point function acted precisely as the two-point function of a spin system and the universality of the 
results could be understood as a result of the divergent correlation length, in the same way as universality of phase transitions 
of spin systems can be understood as the result of a divergent correlation length between the spins, which makes many details of 
the short distance lattice structure and interactions irrelevant for scaling limit. 

We also stated that when we coupled  matter to 2d gravity we could change the critical behavior of the ensemble of polygons when 
the matter system itself had long range interactions (and in addition the long range correlation in geometry would change the 
critical properties of the matter system). We  mentioned that this change of critical behavior could many times be obtained 
by assigning negative weights to some of the polygon building blocks. This whole complex of systems
provides a  lattice regularization of two-dimensional Euclidean quantum gravity coupled to conformal field theories, and is denoted
 {\it Euclidean Dynamical triangulations} (EDT) or {\it Quantum Liouville Theory}.

 We will now introduce a new, and different universality class of 2d models, 
 denoted {\it Causal Dynamical Triangulations} (CDT).
 Historically, the motivation was that time and space might be more different than it appears in the truly Euclidean approach 
 we have pursued so far. One could emphasize this by insisting that the starting point was to consider  Lorentzian
 geometries with a global proper time (and in particular thus a local causal structure, which gave rise to the name CDT, when 
 one implemented this via triangulations) , and then perform the rotation to Euclidean signature by rotating this global time to 
 imaginary global ``time''. In this way we arrive with a set of Euclidean geometries which are more restricted than the ones 
 we have studied so far. We triangulate these geometries as before, using equilateral triangles, and using the corresponding
 Regge action we arrive at a new statistical system of two-dimensional  geometries which we still denote CDT, despite 
 the rotation to Euclidean signature. By our choice of 
 geometries we have broken the symmetry between space and (Euclidean) time and from a Wilsonian point of view 
 it is then a distinct possibility that our statistical system of geometries (CDT) will be in a different universality class\footnote{
 The way we defined EDT above, it was not really a single universality class, since the universality classes were labeled by a 
 continuum parameter, the central  charge $c$ of the conformal field theory coupled to the two-dimensional geometry. 
 When comparing
 EDT to CDT, we will from now on have in mind the specific model where there is no conformal matter coupled to two-dimensional 
 geometries, i.e.\ in the labeling mentioned, the case $c \equ 0$.} 
  than the statistical systems of geometries denoted EDT where this symmetry is manifest. We will see that it is indeed the case.  

\subsection*{Defining and solving the CDT model}

\begin{figure}[t]
\vspace{-1cm}
\centerline{\scalebox{0.20}{\includegraphics{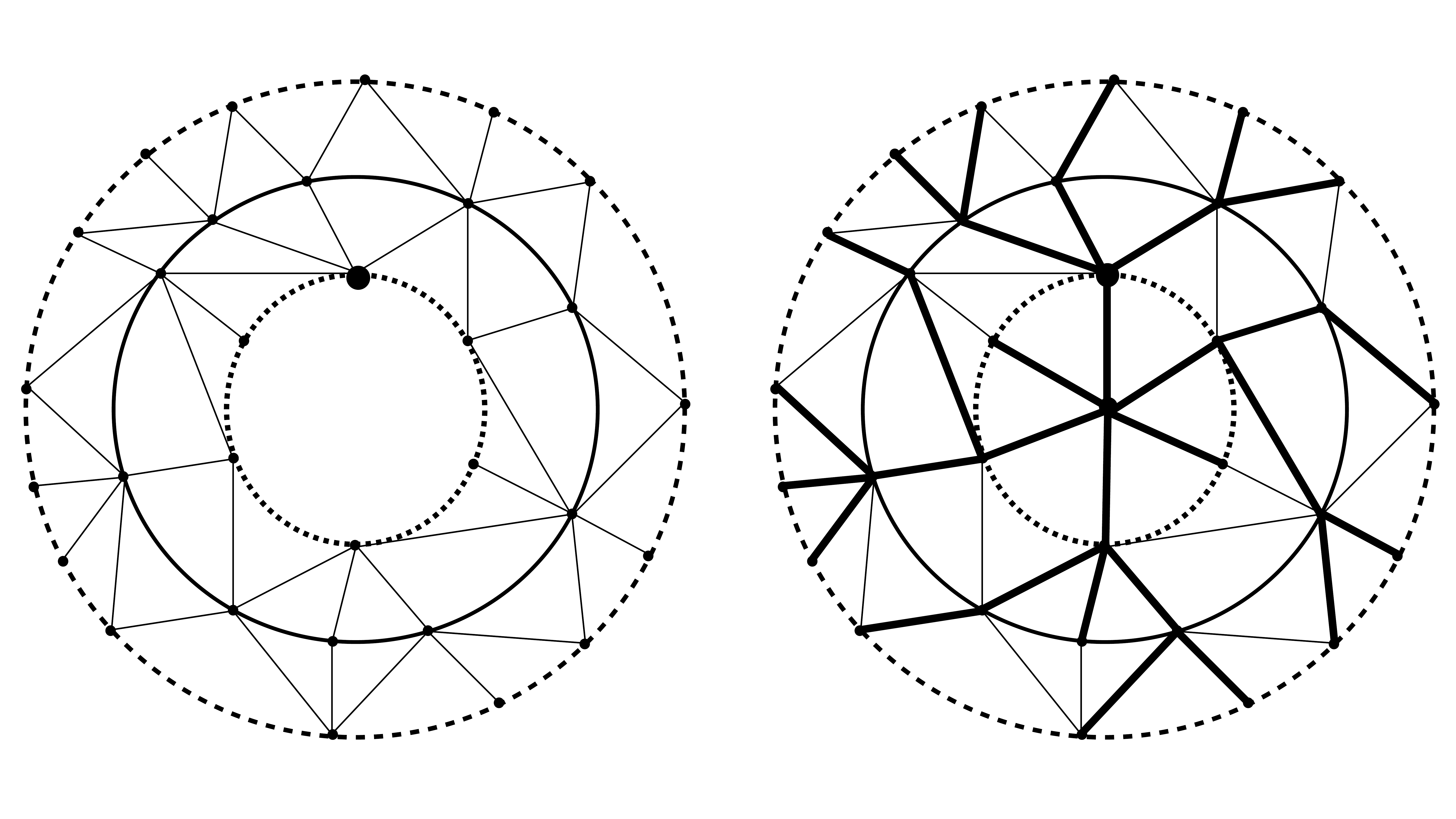}}}
\vspace{-1cm}
\caption{{\small  Left figure: A triangulation of the cylinder (represented as an annulus). Constant time slices corresponding to 
$t=0,1,2$ are circles. The boundary loops have dashed and dotted lines.  
A vertex (or the spatial link to the right of it) on the entrance loop $t\equ 0$ is marked. Right figure: 
the corresponding branched polymer (thick black links). 
An artificial vertex at $t \equ  \mi 1$ connected to each vertex at the $t \equ 0$ loop 
ensures a bijection between the CDT triangulations with boundaries at times 0 and $t$ 
and rooted  branched polymers of {\it height $t$} (the root connects the vertex at $t \equ \mi 1$ to the marked 
vertex at $t \equ 0$).}}
\label{figcdt1}
\end{figure}

We label ``time'' by an integer coordinate $t$. For each time coordinate $t$ ``space'' will be assumed to have 
the topology of $S^1$. Space at time $t$ will consists of $l_t$ links glued together via $l_t$ vertices such that 
the topology of space is $S^1$. Given space at $t$ and space at $t \plu1$ we now fill out  the ``slab'' in between by equilateral 
triangles, such that a triangle has two vertices with time coordinate $t$ and one with time coordinate $t\plu 1$ or oppositely has 
one vertex with time coordinate $t$ and two vertices with time coordinate $t\plu 1$. The triangles are glued together such that 
they form a triangulation with the topology of a cylinder where one boundary consists of $l_t$ links and the other boundary 
consists of $l_{t\plu 1}$ links. The total number of triangles is $l_t+ l_{t\plu 1}$, but for given $l_t$ and $l_{t\plu 1}$ there are of course 
many ways we can glue together the triangles to form a cylinder with the given length of boundaries. 
Continuing this  way we construct triangulations which have slices of constant time  labelled 
by $i \equ 0,1,2, \ldots, t$ and 
link-lengths $l_i > 0$ (we do not allow slices of constant time without at least one link). This triangulation $\cT$ has the 
topology of the cylinder  with  boundaries  of lengths $l_0$ and $l_t$. The left panel in Fig.\ \ref{figcdt1} 
shows such a triangulation with the
cylinder presented as an annulus where the circles represent the spatial slices at times 0,1 and 2. For the 
purpose of combinatorics it is convenient to mark one of the  link (and its ``first'' vertex if the loop is oriented counter clockwise) 
on the spatial boundary loop  
with  $l_0$ links (the ``entrance'' loop), but have no marked link on the boundary loop with $l_t$ links (the ``exit'' loop). 
This choice of labeling is similar to the one we used in the last Section when we considered the two-loop function 
$G(g,l_1,l_2;r)$, and we choose it for the same reason:  from a combinatoral point of view it makes the gluing of 
two cylinders easier and we will have a composition law similar to \rf{6.11}, only with $r$ replaced by $t$.
The total number of triangles and the total number of vertices in $\cT$ will be
\beq\label{cdt2}
N_{\cT} = 2 \sum_{i=1}^{t-1}   l_i + l_0+l_t,\quad V_{\cT} =  \sum_{i=0}^{t}   l_i, \quad {\rm i.e.} \quad 
2V_{\cT} = N_{\cT} + l_0 + l_t
\eeq
where the first equation simply reflects that each internal spatial link is the spatial link of two neighboring triangles, 
while a boundary link is a spatial link for just one triangle. 
The action associated with such a triangulation will as usual just be the cosmological term:
\beq\label{cdt3}
S[\cT] = \mu N_{\cT},
\eeq
where $\mu$ is the dimensionless cosmological constant. 

It is possible to make a bijective map from this class of triangulations to branched polymers as shown  in the right panel of 
 Fig.\ \ref{figcdt1} (see \cite{djw} for details). 
 Let a  vertex at a loop at time $i < t$ be connected  with links to $k$ vertices at the loop at time $i\plu 1$. Moving
counter clockwise around the vertex,  we declare that all links except the last one will belong to the BP. In addition we have 
added a marked vertex which we connect to all vertices at the entrance loop corresponding to $i \equ 0$. The BP defined 
in this way has thus a marked vertex and a corresponding marked link, which is the link connected to the marked vertex on
the entrance loop. If we define the {\it height} of a vertex in the BP as the link distance from the marked vertex, it 
follows by construction that  the vertices at height $i$ are precisely the vertices at the loop at time $i-1$ in the triangulation.
Given a BP with a marked vertex and a corresponding marked link, one can reconstruct the triangulation and in this way
prove the bijection. The number of links in the BP, $L_{BP}$, is equal to the number of vertices 
$V_{\cT}$ in the triangulation, and from \rf{cdt2}
in follows that if we ignore boundary contributions we have 
\beq\label{cdt4}
S[\cT] = \mu N_{\cT} \approx 2 \mu L_{BP}.
\eeq
Thus, if we define the {\it partition function of CDT} by summing over all triangulations constructed as described above, 
using as weights $e^{-\mu N_{\cT}}$, 
we expect to obtain the same result as if we performed the summation of all BPs using weights $e^{-(2\mu) L_{BP}}$ (apart
from the above mentioned boundary terms).   The set of BPs is precisely the set of BPs we encountered in Chapter 5 
(see eq.\ \rf{5.21} and \rf{5.21a}) and we can thus expect a critical $\mu_c$ given by
\beq\label{cdt5}
 \e^{-2 \mu_c} = \oq,\quad{\rm i.e.} \quad \mu_c = \ln 2.
 \eeq
 Below we will verify, using very simple arguments, that this is indeed true. Because of the strong link to BPs we also 
 expect that the CDT theory of geometry belongs to a different universality class than the EDT theory, and as we will see
 this is indeed the case. 

We will be interested in a continuum limit of the above lattice construction, where, somewhat similar to what we did in  EDT, 
we take the number of boundary links, the number of triangles and the time steps to infinity in such a way that 
one one can take the link length $\ep$ to zero while keeping the continuum boundary lengths, the continuum areas and 
the continuum time finite. In order to implement this we start by  keeping $l_0$ and $l_t$ fixed and sum   over all  
``cylindrical'' triangulations $\cT(l_0,l_t;t)$ of the kind described above, with fixed $l_0$, $ l_t$ and $t$, using the action \rf{cdt3}.
 Let us denote this amplitude
\beq\label{cdt4a}
G_\mu(l_0,l_t;t)\equiv G(g,l_0,l_t;t) = \!\!\!\sum_{ \{\cT(l_0,l_t;t)\}} g^{N_{\cT(l_0,l_t;t)}},\quad g = \e^{-\mu}
\eeq 
Having two cylindrical triangulations $\cT(l_0,l_{t_1};t_1)$ and $\cT(l_{t_1},l_{t_2};t_2\mi t_1)$, they can be glued together
to a single cylindrical triangulation $\cT(l_0,l_{t_2};t_2)$  along the boundaries of 
lengths $l_{t_1}$.  No additional symmetry factor is related to this gluing because we have chosen to mark the entrance 
loop and not the exit loop, as already mentioned. Thus we can write
\beq\label{cdt6}
G(g,l,l';t_1\plu t_2) = \sum_{l''} G(g,l,l'';t_1)\; G(g,l'',l';t_2).
\eeq
As a special case of \rf{cdt6} we can write
\beq\label{cdt7}
 G(g,l,l';t\plu 1) = \sum_{l''} G(g,l,l'';1)\; G(g,l'',l';t).  
 \eeq
{\it Thus it is clear that we can find $G(g,l,l')$ by iteration if we only know  $G(g,l,l';1)$}\footnote{
One could have used the same argument in the case of EDT, eq.\ \rf{6.11}, and one can indeed find 
$G(g,l.l';1)$ in the EDT case and in this way find the two-loop EDT function. We refer to \cite{book} for details.}. In order to find 
$G(g,l,l';1)$ we introduce (as usual) the generating function for $G(g,l,l';t)$ and define
\beq\label{cdt8}
G(g,x,y;t) = \sum_{l,l'} x^ly^{l'} G(g,l,l';t), \quad  x = \e^{-\lam_{en}},~y= \e^{-\lam_{ex}}
\eeq
where we have indicated that the indeterminate $x,y$, if positive and real, also can be given an interpretation as 
boundary cosmological constants for the entrance and exit loop (as was also the case in EDT where we just used 
 $z_1 \equ 1/x$ and $z_2 \equ 1/y$ as the indeterminate in the generating function instead of $x$ and $y$).
 In this way we can invert \rf{cdt8} if needed:
\beq\label{cdt8a}
 G(g,l,l';t) = \oint \frac{dx}{2\pi i x^{l+1}}   \oint \frac{dy}{2\pi i y^{l'+1}} \; G(g,x,y;t),
\eeq  
where the integration contours enclose $x\equ y \equ 0$ and lie within the convergence radii of the power series in 
$x$ and $y$. It follows from 
\beq\label{cdt11}
\oint \frac{dz}{ 2 \pi i \;z^{n+1}}= \del_{0,n} \qquad n\in \mathbb{Z}.
\eeq

The relation \rf{cdt6} can now be written as
\beq\label{cdt10}
G(g,x,y;t_1\plu t_2) = \ointz G(g,x,z^{-1};t_1) G(g,z,y;t_2),
\eeq
where the integration contour encloses $z \equ 0$ and for fixed $g,x,y$  lies inside the radius of convergence $r(g,y)$
for $G(g,z,y;t_2)$ and inside  the radius of convergence  $r(g,x)$ for $G(g,x,z^{-1};t_2)$ as a power series in $1/z$, 
i.e. in the region $z \geq 1/r(g,x)$. This is possible when we consider $g,x,y$ less than their critical values, given by 
eq.\ \rf{cdt15} below, since then $r(g,x) > 1$.

It is now easy to find $G(g,x,y;1)$, just looking at Fig.\ \ref{figcdt1}:
\beq\label{cdt12}
G(g,x,y;1) =\sum_{k=0}^\infty \left( gx \sum_{l=0}^\infty
 (gy)^l \right)^k -
\sum_{k=0}^\infty (gx)^k = \frac{g^2 xy}{(1\mi gx)(1\mi gx \mi gy)}.
\eeq
Formula \rf{cdt12} is simply a book-keeping device for all possible
ways of evolving from an entrance loop of any length in one step to
an exit loop of any length. The subtraction of the term $1/(1 \mi gx)$ 
has been performed to 
exclude the degenerate cases where either the entrance or the exit 
loop is of length zero. 

We now use \rf{cdt12} in \rf{cdt10} with $t_1 \equ 1$ and $t_2 \equ t$.
\beq\label{cdt13}
G(g,x,z^{-1};1)= \frac{g^2 x}{(1\mi gx)^2} \;\frac{1}{z \mi  \frac{g}{1\mi gx}},
\eeq
and the integration contour in \rf{cdt10} should include $z=g/(1\mi gx)$ and $z \equ 0$, but $z \equ 0$ does not contribute since
$G(g,z,y;t)/z$  is finite for $z\to 0$ (the entrance loop has length $l \geq 1$). We thus  obtain
\beq\label{cdt14}
\boxed{G(g,x,y;t \plu 1) = \frac{gx}{1 \mi gx}\; G\Big(\frac{g}{1\mi gx},y;g;t\Big).}
\eeq
This equation can be solved by iteration (see \cite{al} for details). However, rather than doing that and then deriving 
the continuum limit of $G(g,x,y;t)$, we will use it to  directly ``guess'' the continuum limit.
Let us assume that there are critical points $g_c$, $x_c$ and $y_c$ like in 2d EDT, such that we can write
\beq\label{cdt15}
g = g_c \e^{-\Lam \ep^2/2},\quad x = x_c \e^{-X \ep},\quad y= y_c \e^{-Y \ep},\quad T = \ep\, t ,
\eeq
where $\ep$ denotes the link length, while $\Lam$, $X$ and $Y$ are the continuum cosmological constant and the 
continuum boundary cosmological constants.
The last relation, $T\equ \ep \,t$,  
was absent for 2d EDT since we had no time slicing like here in CDT. In 2d EDT we could have used the geodesic distance
from a point or an entrance loop as ``time'', but the corresponding links  at a distance $r$  from the entrance loop did not form
a connected loop, but branched out in many loops, reflecting that the Hausdorff dimension of 2d EDT is 4  
and that the geodesic distance has an  anomalous dimension (recall Fig. \ref{fig6.8}). 
Here in CDT the situation is seemingly different and 
successive time slices labeled by the integer $t$ stay connected by construction. Thus the relation $T \propto \ep \,t$
 between the continuum time $T$ and the dimensionless integer lattice time $t$ is reasonable. We will adjust $T$ such that 
 the constant of proportionality is 1. We will now assume that $G(g,x,y;t)$ has a limit $\ep^\eta  G_\Lam(X,Y;T)$
 when $\ep \to 0$. This is similar to the situation we encountered in 2d EDT for $G(g,x,y;r)$ and from 
\rf{cdt10} it folows that $\eta \equ  - 1$, like in the EDT case. An assigment 
\beq\label{cdt16}
G(g,x,y;t) \to  \ep^{-1} G_\Lam(X,Y ;T)\quad {\rm for} \quad \ep \to 0
\eeq
is only meaningful if we  in eq.\ \rf{cdt14}  have 
\beq\label{cdt17}
\frac{g_cx_c}{1 \mi g_cx_c} =1,\quad \frac{g_c}{1 \mi g_cx_c} =1, \quad {\rm i.e.} \quad x_c =1,~g_c = \oh.
\eeq
It is seen that we indeed have confirmed the prediction \rf{cdt5}.
Inserting \rf{cdt17}  back in \rf{cdt14} and using \rf{cdt15}  we obtain to lowest order in $\ep$
\beq\label{cdt18}
G\big(g,1\mi \ep X, y; T \plu \ep\big) = (1\mi 2\ep X)\, G\big (g,1\mi  \ep [X \plu  \ep(\Lam \mi X^2)], y;T\big),
\eeq
i.e.\ (suppressing the arguments $y$ and $g$ in $G(g,x,y;t)$)
\beq\label{cdt19}
\ep \frac{\prt G\big(1\mi \ep X;T)\big)}{\prt T} = \mi 2\ep  X \, G\big(1\mi  \ep  X;T\big) + 
\ep(\Lam \mi X^2) \,\frac{\prt G\big(1\mi \ep X;T\big)}{\prt X}
\eeq
or 
\beq\label{cdt20}
 \boxed{\frac{\prt G_{\Lam}(X,Y;T)}{\prt T} =-\frac{ \prt}{\prt X} \Big(  (X^2 \mi \Lam)  G_\Lam(X,Y;T) \Big)}
 \eeq
This partial differential equation has the same structure as the one we already meet in 2d EDT and we can solve it 
precisely in the same way (see \rf{6.35a}):
\beq\label{cdt21}
G_\Lam(X,Y;T) = \frac{\bar{X}^2(T;X) \mi \L}{X^2 \mi \L}\; \frac{1}{\bar{X}(T;X) \plu Y}, 
\eeq
where $\bar{X}(T;X)$ is the solution to the characteristic equation
\beq\label{cdt22}
\frac{d \bar{X}}{dT} = -(\bar{X}^2 \mi\L),~~~~\bar{X}(T\equ 0;X)=X.
\eeq
i.e.
\beq\label{cdt23}
\bar{X}(T;X) = \SL \; 
\frac{(\SL\plu X)\mi \e^{-2\SLT}(\SL\mi X)}{(\SL+X)\plu \e^{-2\SLT}(\SL\mi X)}.
\eeq

We can now introduce the continuum boundary length $L$ as we did for EDT 
\beq\label{cdt26}
L \equiv \ep \, l,\qquad x^l = x_c^l\;\e^{-\ep l X} = \e^{-L X}, 
\eeq 
and provided we, like in \rf{cdt16},  make the identification
\beq\label{cdt29a}
G(g,l_1,l_2;t) \to \ep \, G_\Lam(L_1,L_2;T) \quad {\rm for} \quad \ep \to 0.
\eeq 
 it is seen that with a change of variables from
$x,y$ to $X,Y$  the integration contours in \rf{cdt8a} change from circles to integration along the imaginary $X,Y$ axes
in the limit when $\ep \to 0$ and the continuum limit of \rf{cdt8a} reads:
\beq\label{cdt27}
G_\Lam( L_1,L_2;T) = 
\int_{-i \infty}^{i \infty} \frac{dX}{2\pi i}  \int_{-i \infty}^{i \infty} \frac{dY}{2\pi i} \; \e^{L_1 X+ L_2 Y} G_\Lam(X,Y;T)
\eeq
This is just the inverse Laplace transformation of the continuum limit of \rf{cdt8}:
\beq\label{cdt28}
G_{\Lam}(X,Y;T) = \int_0^\infty \!\!\!dL_1 \!\! \int_0^\infty \!\!\!dL_2\; \e^{-X L_1-Y L_2} G_{\Lam}( L_1,L_2;T),
\eeq
From the solution \rf{cdt21} we can easily perform the inverse Laplace transformation wrt $Y$ as in \rf{cdt27} and we obtain
(corresponding to \rf{6.35})
\beq\label{cdt32}
G_\Lam( X,L;T) = 
\frac{\bar{X}^2(T;X)\mi \L}{X^2-\L}\; \e^{-\bar{X}(T;X) L}.
\eeq
Using \rf{cdt23} we see that we have the following large $T$ behavior
\bea \label{cdt29}
G_\Lam(X,L;T) &\buildrel{T\rightarrow\infty}\over\longrightarrow &
\frac{4\L \,\e^{-\SL L}}{(X+\SL)^2}\;\e^{-2\SLT},  \\
G_\Lam(L_1,L_2;T) &\buildrel{T\rightarrow\infty}\over\longrightarrow& 
4\L \,L_1\e^{-\SL (L_1\plu L_2)}\,\e^{-2\SLT}\label{cdt29b}
\eea
i.e. the two-loop functions fall off exponentially. The factor $L_1$ in \rf{cdt29b} is present only because 
we have chosen to mark a point on the entrance loop.

The disk amplitude played an important role in our EDT theory. It had the interpretation of the Hartle-Hawking wave function 
of the universe and it was the building block for all higher loop functions. Looking at Fig.\ \ref{figcdt1}, the right panel defines 
a triangulation with the topology of the disk. It is a special configuration in the sense that it has a special point in the center and 
at least we should in addition sum over all times $t$. In the continuum we formally achieve this by starting with 
 $G_\Lam(X,L;T)$ and contracting the loop $L \to 0$ and then integrating wrt $T$. We thus {\it define} the CDT 
 disk function as 
\beq\label{cdt33}
W_\Lam(X) \equiv \int_0^\infty \!\!\! dT \, G_\Lam(X,L\equ 0;T) =
 \int_{\SL}^X \frac{ d \bar{X} }{X^2\mi  \Lam} = \frac{1}{X\plu \SL}.
\eeq 
The integral can be performed by using \rf{cdt22} to change integration variable from $T$ to $\bar{X}(T)$. 
The limit $T \to \infty$ corresponds according to \rf{cdt23} to $\bar{X} \equ \SL$.  
Taking the Laplace transform we obtain
\beq\label{cdt33a}
W_\Lam(L) = \int_{-i \infty}^{+i \infty} \frac{dX}{2\pi i} \; \e^{L X} W(\Lam,X) = \e^{-\SL \,L},
\eeq
i.e.\ the CDT disk amplitude falls off exponentially with $L$, the decay determined by the square root of the 
cosmological constant. 

\begin{figure}[t]
\vspace{-1cm}
\centerline{\scalebox{0.12}{\includegraphics{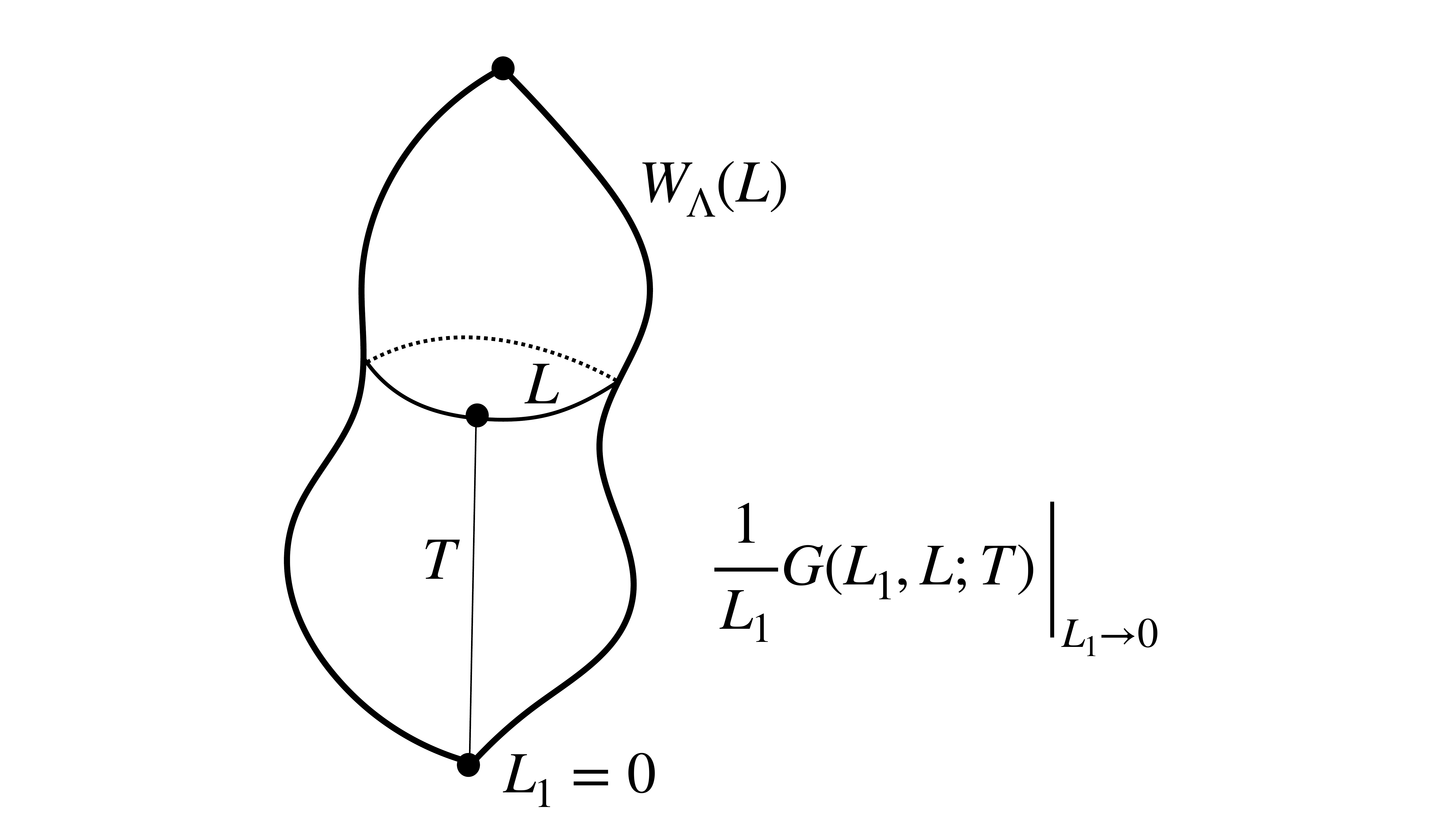}}}
\caption{{\small First the propagation from a contracted entrance loop (length $L_1 \equ 0$) to a point at distance $T$. This point
belongs to a loop of length $L$, where all points have distance $T$ from the entrance loop. After that the universe continues 
to evolve until it eventially disappears (the upper black dot). That continued evolution is described by $W_\Lam(L)$.}}
\label{figcdt10}
\end{figure}

We now {\it define} the following two-point function in CDT: The starting point is the two-loop function
$G_\Lam(L_1,L_2;T)$. We  contract the length $L_2$ of the exit  loop to zero as we did for $W_\Lam(X)$.
We also contract the entrance loop to zero, but to compensate for the factor $L_1$ which we ``artificially'' introduced 
by marking a point on the entrance loop, we divide by $L_1$ before taking the limit $L_1 \to 0$. For functions $F(L)$
where $F'(0)$ exists it can be done by writing:
\beq\label{cdt32a}
G(L) = L F(L), \quad G'(0) = F(0),\quad{\rm i.e.} \quad F(0) = \int_{-i \infty}^{i \infty} \frac{dX}{2\pi i} \; \e^{LX} X G(X)\Big |_{L=0}.
\eeq
We now define the two-point function $G_\Lam(T)$ as the  sum over all CDT cylinder surfaces where the 
entrance and exit loops are contracted to points as described above, and where {\it a marked point has a distance 
$T$ to the entrance loop (point)}. A typical such surface is shown in Fig.\ \ref{figcdt10}. From the figure it follows that
\bea\label{cdt32b}
G_\Lam(T)\!\! &=& \!\! \lim_{L_1 \to 0} \frac{1}{L_1} \int \!dL \;G_\Lam(L_1,L)\, L \,W(L) =  
\mi \int_{-i \infty}^{i \infty} \frac{dX}{2\pi i} \; \frac{X (\bX^2\mi \Lam)}{X^2 \mi \Lam} \frac{d W_\Lam(\bX)}{d \bX} \hspace{4mm}
 \nonumber\\
&=&- (\bX^2 \mi \Lam)\left. \frac{d W_\Lam(\bX)}{d \bX}\right|_{|X| = \infty} \!\!= \frac{d W_\Lam (\bX(T;\infty))}{ dT} 
= \e^{-2 \SL T}.
\eea
The first equality follows from the figure: the loop at distance $T$ has a length $L$ and the marked point can be anywhere. 
The surface can now continue in all possible ways compatible with CDT surfaces until a spatial loop contracts to a point, i.e.\ 
precisely as $W_\Lam(L)$. The next equality uses \rf{cdt32a} and \rf{cdt32}.  
The third equality follows from deforming the integration contour to a circle at infinity, where $\bX(T;X)$ is independent
of $X$, as seen from \rf{cdt23} which also leads to the final result \footnote{\label{footnote19} The result differs from \rf{6.37} where there 
is no differentiation wrt  $T$ (or better $R$ in eq. \rf{6.37}). 
The difference can be traced back the non-scaling part of the disk amplitude $w(g,z)$ given by \rf{6.4}, although it 
seemingly cancels out in the differential equation \rf{6.17} which leads  to \rf{6.23}.  It would nevertheless enter if one 
tried to define the two-point function as in Fig.\ \ref{figcdt10}, starting out at a discretized level. In EDT the 
non-scaling part of $w(g,z)$ cannot be ignored because of the fractal nature of the geometries. The chance
that the loop where the black dot in the figure  is located has a macroscopic length is simply zero and for a microscopic 
loop, the corresponding dominating contribution from $w(g,z)$ will be the non-scaling part of \rf{6.4}. A detailed 
discussion can be found in \cite{al}.}.

If we return to discrete variables we have
\beq\label{cdt31}
G_\mu(t) \propto \e^{-2 \sqrt{\mu - \mu_c} \, t}, \quad \mu_c = -\ln g_c = \ln 2,
\eeq
i.e.\ the two-point function behaves precisely as the two-point function for intrinsic 
BPs, and it has the same critical exponents: 
\beq\label{cdt31a}
\boxed{\nu_{cdt} = \oh \quad ({\rm i.e.} \quad d_H \equ 2), \qquad \gamma_{cdt} = \oh,\qquad \eta_{cdt} = 1.}
\eeq
Of course it is not surprising, given the bijective mapping between BPs and CDT configurations, but we have 
now shown it by explicit calculations.  Also, the result is manifest  different from the 
EDT result where we had a propagator behavior $G_\mu(t) \propto e^{- \sqrt[4]{\mu - \mu_c} \, t}$ for large $t$
and $G_\mu(t) \propto t^{-3}$ for small $t$ (where $t$ denoted the link distance between two marked points), and where 
the corresponding critical exponents were 
\beq\label{cdt31b}
\nu_{edt} = \oq,\qquad \gamma_{edt} = - \oh,\qquad \eta_{edt} =  4.
\eeq
{\it Our conclusion is that the CDT ensemble of 
2d geometries belongs to a different universality class.}

\begin{figure}[t]
\vspace{-1cm}
\centerline{
\scalebox{0.4}{\includegraphics[angle=90]{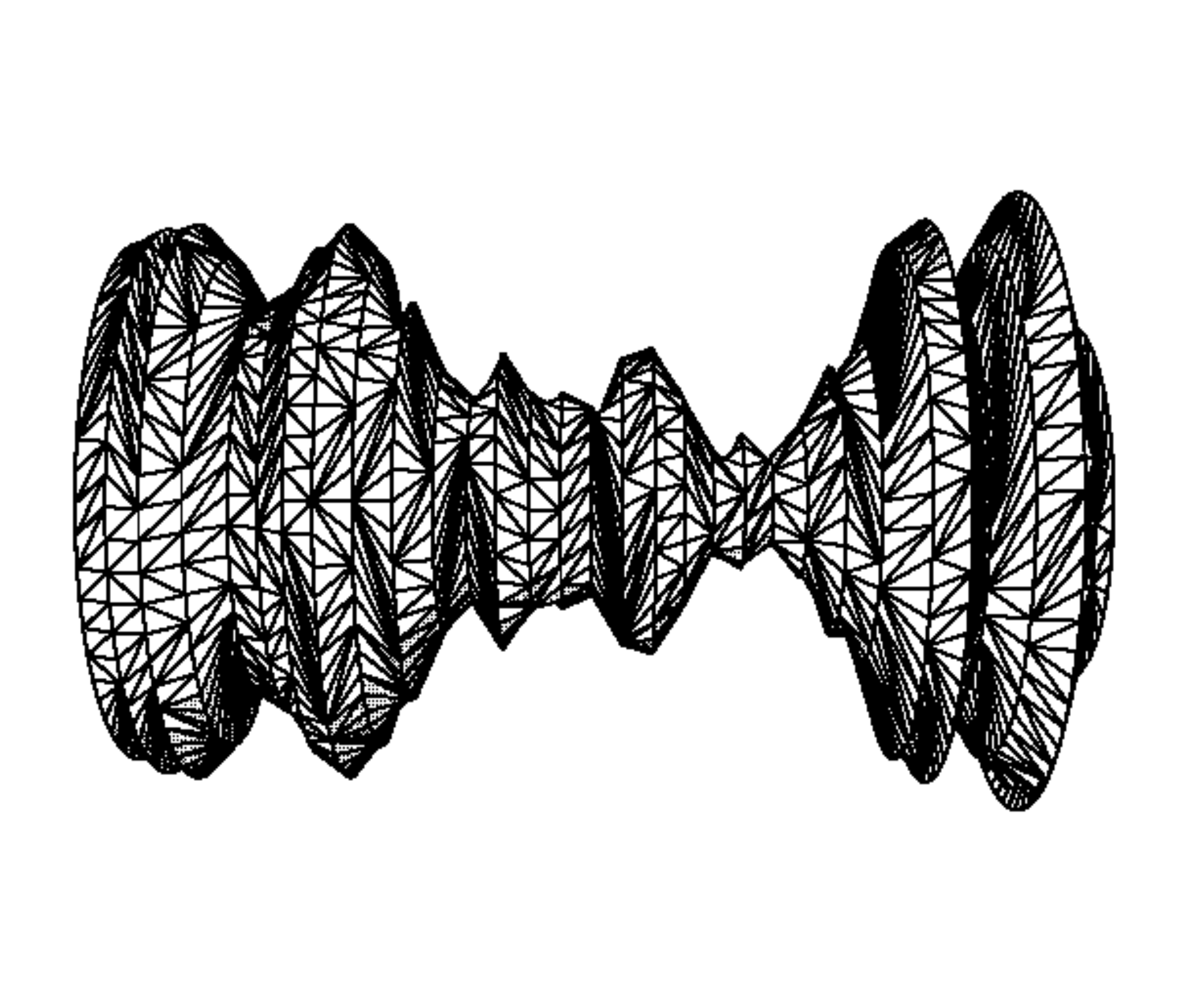}}}
\vspace{-1cm}
\caption{{\small A ``typical'' configuration contributing to the path integral defining the amplitude \rf{cdt35}. Time is 
in the horizontal  direction and the configuration is a triangulation which at times $t_n$  
consists of a number of links $l (t_n)$ which are drawn as a circle of length $l(t_n)$. The circles are then connected by triangles.
The triangulation is generated by a so-called Monte Carlo simulation of the CDT system. In such simulation one
 computer-generate the CDT triangulations with the relative probabilities with which they are represented in the path integral.}}
\label{figcdt11}
\end{figure}

By taking the Laplace transform of  eq.\ \rf{cdt20} we obtain
\beq\label{cdt34}
\frac{\prt}{\prt T} G_\Lam(L_1,L_2;T) = - \hat{H}(L_1)\, G_\Lam(L_1,L_2;T) ,\qquad \hat{H}(L) = - L \frac{ d^2}{d L^2} + \Lam \, L 
\eeq
Thus we can write 
\beq\label{cdt35}
\boxed{G_\Lam(L_1,L_2;T) = \langle L_2 |\, \e^{-\hat{H} \,  T} | L_1 \rangle},
\eeq
{\it where $\hat{H}$ is the Hamiltonian for the evolution of our spatial universe of length $L$}. It is an Hermitian operator 
on the positive real axis ($L$ has to be non-negative) with the scalar product 
\beq\label{cdt36a}
\la \Psi_2 | \Psi_1\ra = \int_0^\infty \frac{d L}{L} \; \Psi_2^*(L) \Psi_1(L).
\eeq
One can find the eigenfunctions and eigenvalues of $\hat{H}$:
\beq\label{cdt37}
\hat{H} \Psi_n = E_n \Psi_n, \quad E_n =  2n \SL, \quad \Psi_n(L) = P_n(L) \e^{-\SL L},\quad n=1,2,\ldots
\eeq
where $P_n(L)$ is a polynomial of order $n$ such that $P_n(0) \equ 0$ for $n \geq 1$ (we leave it as an exercise to show this). 
Formally ``the wave function of the universe'' $W_\Lam(L) \equ \Psi_0(L)$ is also an eigenfunction
of $\hat{H}$, corresponding to $E_0 \equ 0$. However, it is not a normalizable eigenfunction when using the scalar 
product \rf{cdt36a}. In Fig.\ \ref{figcdt11} we have shown a typical configuration contributing to the 
path integral defining the propagator \rf{cdt35}. If $T$ is sufficiently large
the ground state of $\hat{H}$, $\Psi_1$, will dominate the expression \rf{cdt35} and in that approximation we find that
\beq\label{cdt35a}
\la L(t) \ra \propto \frac{1}{\SL},  \qquad P(L(t)) \equ \frac{ \Psi^2_1(L)}{L} \equ  4\Lam L \e^{-2 \SL L}, \qquad 0 \ll t \ll T,
\eeq
where $P(L(t))$ denotes the probability distribution for the length $L(t)$ of the spatial universe at time $t$.
Thus looking at Fig.\ \ref{figcdt11} and making a normalized 
histogram for the lengths of  the spatial circumferences shown for the discrete times $t_n$ 
should reproduce the $P(L)$ in eq.\ \rf{cdt35a},
assuming that the time $T$ is large enough for the ground state of $\hat{H}$ to dominate in the region $0 \ll t \ll T$.

\subsection*{GCDT: showcasing quantum geometry}
\begin{figure}[t]
\vspace{-1cm}
\centerline{\scalebox{0.18}{\includegraphics{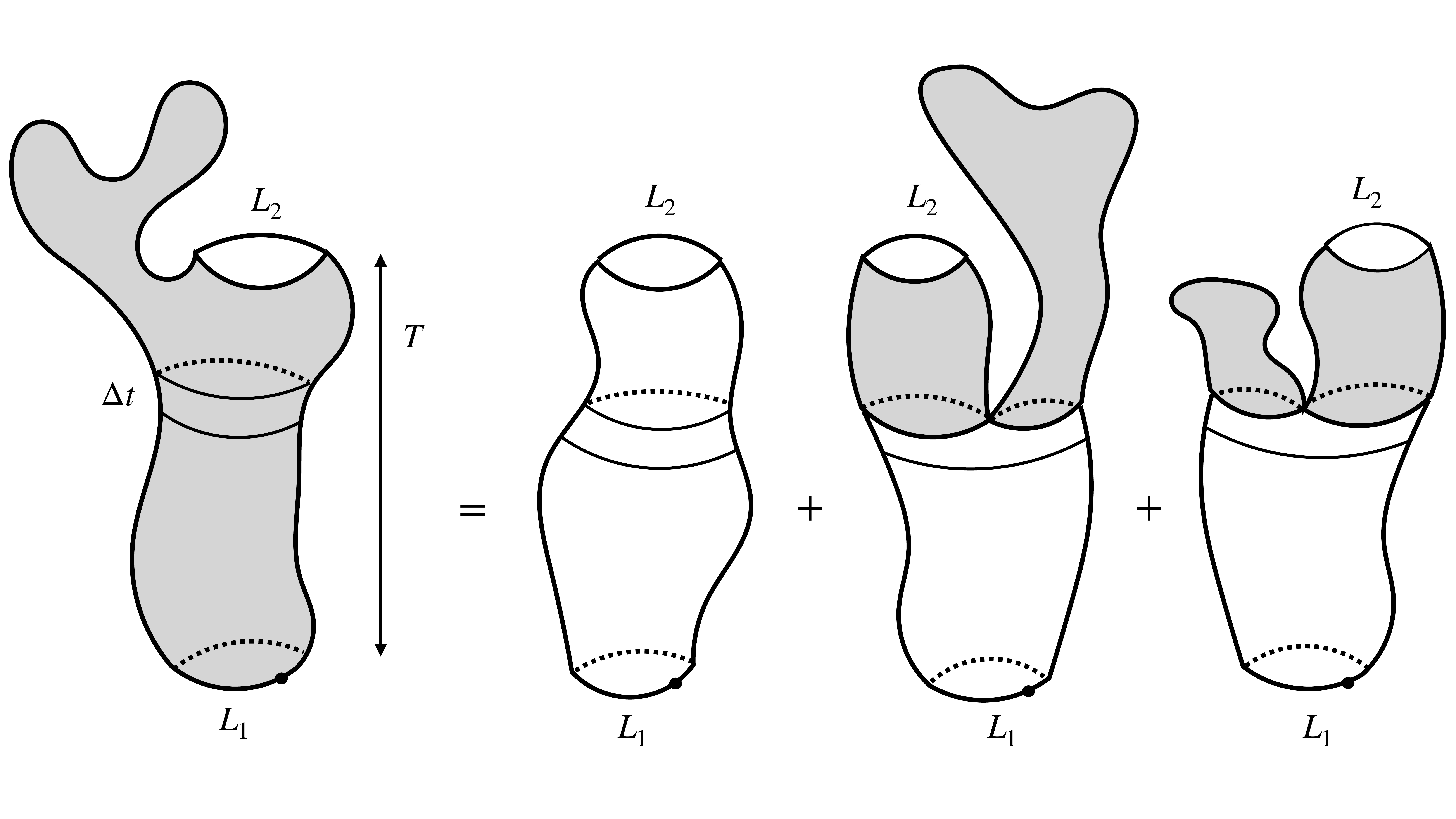}}}
\vspace{-0.5cm}
\caption{{\small  The GCDT  cylinder amplitude $G_\Lam(L_1,L_1;T)$, 
expressed in terms of the CDT cylinder amplitude and the GCDT disk
function. As time progresses either no split occurs (the CDT cylinder amplitude) or no split happens before time $t$ and then
a split happens between $t$ and $t+dt$ with ``probability'' $g_s dt$. After that one part develops as a baby universe, the other
part develops as the GCDT cylinder amplitude. To count all such configurations we have to integrate over $t$. Note that
baby universes can develop in time which extends beyond $T$! }}
\label{figcdt2}
\end{figure}

Above we defined the CDT model. One can ask if it is possible to generalize the model without leaving the universality class.
If we consider the cylinder amplitude, a natural generalization is still to have a time foliation but allow outgrows like 
shown in Fig.\  \ref{figcdt2}. In this way the topology is still that of a cylinder, 
but we allow the creation of baby universes, which have the topology of a disk. 
We denote this theory {\it Generalized} CDT (GCDT). At this point we 
have actually not defined the GCDT disk amplitude starting from any triangulation. Thus the figure involves both an unknown cylinder amplitude and an unknown disk amplitude. However we will show that consistency of 
quantum geometry allows us to determine the amplitudes before actually providing a definition 
via triangulations and taking a scaling limit!

We already allowed the creation of baby universes in  EDT, so are we not just getting back to EDT? The difference is that 
in the case of EDT the creation of baby universes were allowed already at the discretized level without any constraint, 
and when we took the continuum limit the number of such baby universes became quite dominant, and strictly 
infinite in the continuum limit, as illustrated in Fig.\ \ref{fig6.8}. Here we are already in the continuum and for Fig.\ \ref{figcdt2} to make sense in the continuum there should only be a finite number of baby universes for a finite  continuum time $T$.

\begin{figure}[t]
\vspace{-1cm}
\centerline{\scalebox{0.2}{\includegraphics{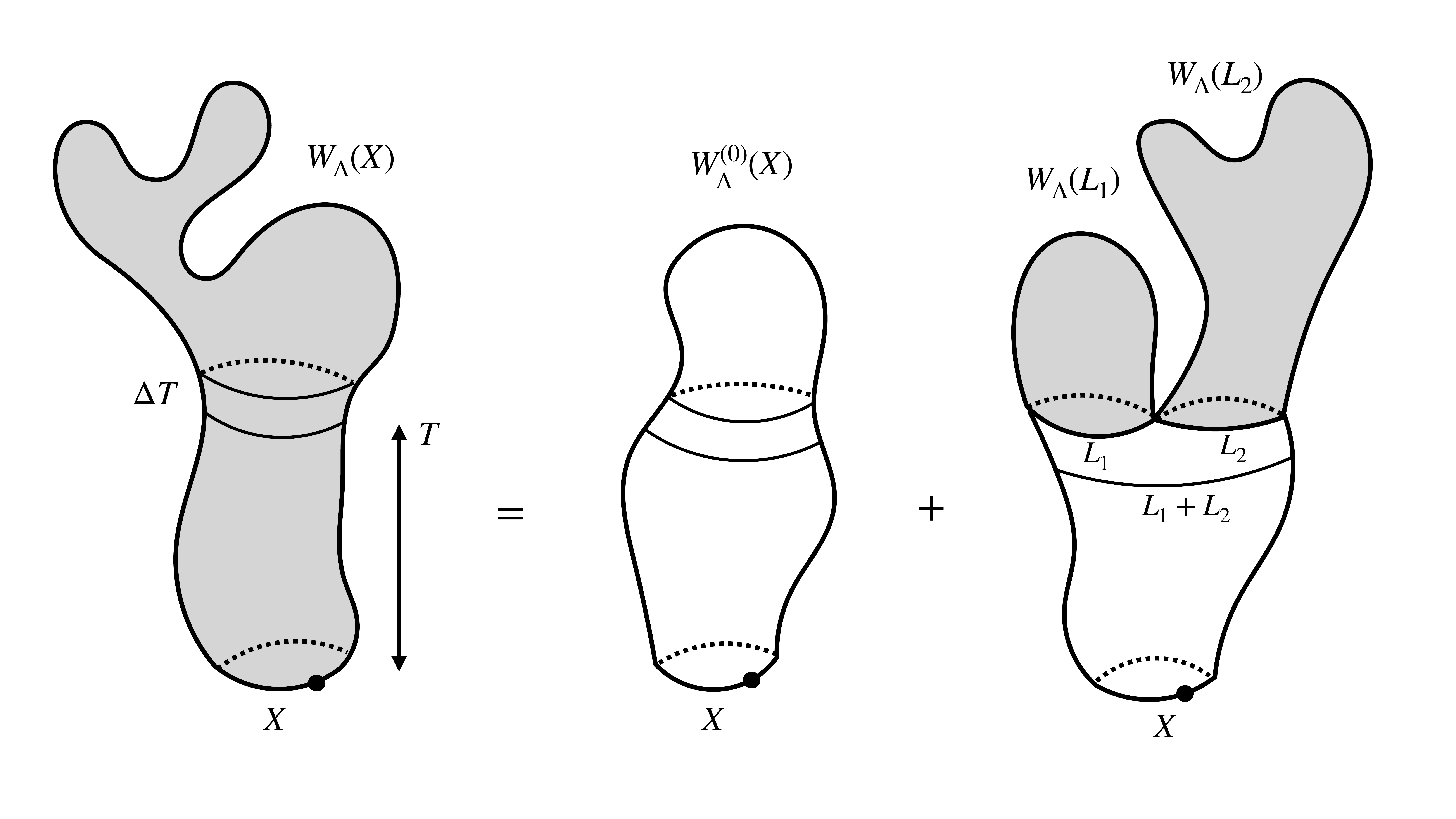}}}
\vspace{-0.8cm}
\caption{{\small  The disk amplitude $W_\Lam(X)$ for GCDT: either no split takes place, or at a time interval between $T$ and 
$T\plu \Del T$ it splits with ``probability'' $g_s \Del T$. 
After that anything can happen at future times, except that the two universes have to have 
the topology of a disk, i.e.\  we will have $W_\Lam(L_1)$ and $W_\Lam(L_2)$ if the loop of length $L_1\plu L_2$ splits in two loops 
of lengths $L_1$ and $L_2$.  We thus integrate wrt $T$, $L_1$ and $L_2$ to count all possible configurations.}}
\label{figcdt3}
\end{figure}
In the same way as Fig. \ref{figcdt2} is a kind of consistence relation  if we allow for the creation of baby universe, we can find 
a consistence relation for the (undefined) disk amplitude itself. It is shown in Fig.\ \ref{figcdt3}. It involves the cylinder 
amplitude and disk amplitude from CDT and  we know these. We can thus  write down an actual equation
corresponding to Fig.\ \ref{figcdt3}
\bea
W_\Lam(X)\!\!\! &=& \!\!\!W^{(0)}_\Lam(X) + g_s
\!\!\int_0^\infty \hspace{-4mm}dT \!\!\int_0^\infty  \hspace{-4mm} dL_1\!\!  \int_0^\infty  \hspace{-4mm} dL_2\;
(L_1\plu L_2) G^{(0)}_\Lam(X,L_1\plu L_2;T) W_\Lam(L_1)W_\Lam(L_2)\nonumber \\
&=& W^{(0)}_\Lam(X) + g_s\int_0^\infty \hspace{-4mm}dT  \; \frac{ \bar{X}^2(T) \mi  \Lam}{X^2 \mi \Lam} \; 
\frac{\d W^2_\Lam(\bar{X})}{\d \bar{X}} \Big|_{\bar{X} = \bar{X}(T)} \nonumber \\
&=& W^{(0)}_\Lam(X) + g_s \frac{W^2_\Lam(\sqrt{\Lam}) \mi W^2_\Lam(X)}{X^2 \mi \Lam}. \label{cdt36}
\eea
The meaning of the coupling constant $g_s$ is explained in the figure captions of Figs.\ \ref{figcdt2} and \ref{figcdt3}.
The superscript $^{(0)}$ refers to the CDT functions, which are explicitly given by \rf{cdt32} and \rf{cdt33}. Further, the factor $L=L_1\plu L_2$ is present because the loop at time $T$ is pinched at  a point and that can be at $L$ different points, 
morally speaking (in a discretized 
version the loop would have  $l$ links  and $l$ vertices and could be pinched in $l$ ways). The second line follows from inserting
\rf{cdt32} and \rf{cdt33} and performing the $L$ integrals, which just lead from $W(L)$ to the Laplace transform $W(\bar{X})$.
Finally the third line follows from \rf{cdt22} which allows us to replace the $T$ integration by an integration over $\bar{X}$,
as was also done in \rf{cdt33}.
We can now solve for $W_\Lam(X)$:
\beq\label{cdt40}
2g_s W_\Lam(X) \equ\Lam \mi X^2 \plu  \hW_\Lam(X),\quad 
\hW_\Lam(X) \equ \sqrt{(X^2\mi \Lam)^2 \plu 4 g_s \big(g_s W^2(\SL) \plu X \mi \SL\big)}.
\eeq
 
 \begin{figure}[t]
\vspace{-2cm}
\centerline{\scalebox{0.22}{\includegraphics{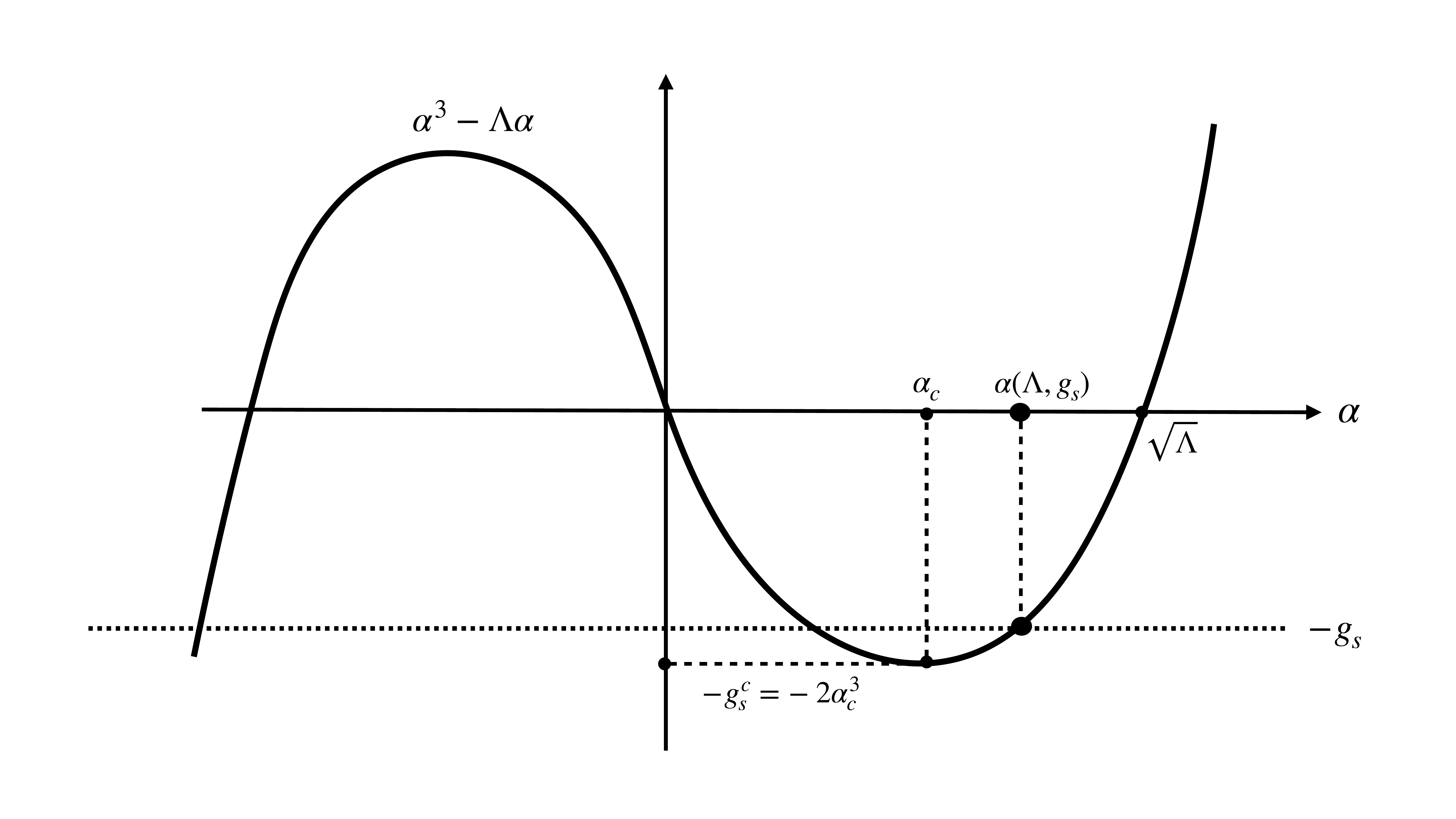}}}
\vspace{-1cm}
\caption{{\small The graphic solution $\alpha(\Lam,g_s)$ to eq. \rf{cdt41}. For given $\Lam$ there 
is no solution when $g_s >  g_s^c = -2 \alpha_c^3 =\frac{ 2}{3 \sqrt{3}} \Lam^{3/2}$. For $g_s = g_s^c$ 
the solution $\alpha(\Lambda,g_s^c) = \alpha_c = \sqrt{\Lam/3}$. This implies that   $\alpha(\Lam,g_s)$ will decrease 
from $\SL$ for $g_s =0$ to its minimum value $\al_c$ when $g_s = g_s^c$.}}
\label{figcdt4}
\end{figure}

We want $W_\Lam(L)$ to be a continuous deformation of $W^{(0)}_\Lam(L)= e^{-\SL L}$ when $g_s$ is close to zero. In particular we want it to fall off exponentially with $L$. For $g_s \equ 0$ the 4th-order polynomial under the square root in \rf{cdt40} has double 
zeros at $X = \pm \SL$. For small $g_s$ these double zeros will split, unless $W_\Lam(\SL)$ is fine tuned, and this will 
result in two cuts in $\hW_\Lam(X)$, one cut close to $-\SL$ and the other cut close to $\SL$. We cannot allow the cut close to 
$\SL$, since by inverse Laplace transformation it will result in an exponential growing $W_\Lam(L)$. Thus we have to insist
that   $W_\Lam(\SL)$ is fine tuned such that the fourth-order polynomial has a double zero on the positive real axis. As simple
calculation then leads to 
\beq\label{cdt41}
\hW_\Lam(X) = (X-\al) \sqrt{(X+\al)^2 -\frac{2g_s}{\al}},  \qquad  \al^3\mi \Lam \al  \plu  g_s \equ 0.
\eeq 
One has to choose the solution $\a(\Lam,g_s)$ to the third order equation which is closest to $\SL$, 
as illustrated in Fig.\ \ref{figcdt4} and we then have a power 
expansion of $\al$  in powers of $g_s/\Lam^{3/2}$:
\beq\label{cdt41a}
\al(\Lam, g_s) = \SL \,\Big(1 -  \frac{g_s}{2\Lam^{3/2}}+ \cdots \Big) = \SL\;F\Big( \frac{g_s}{\Lam^{3/2}}\Big), \quad F(0) \equ 1,
\eeq
and a corresponding expansion of $W_\Lam(X)$. 

\begin{figure}[t]
\vspace{-1cm}
\centerline{\scalebox{0.22}{\includegraphics{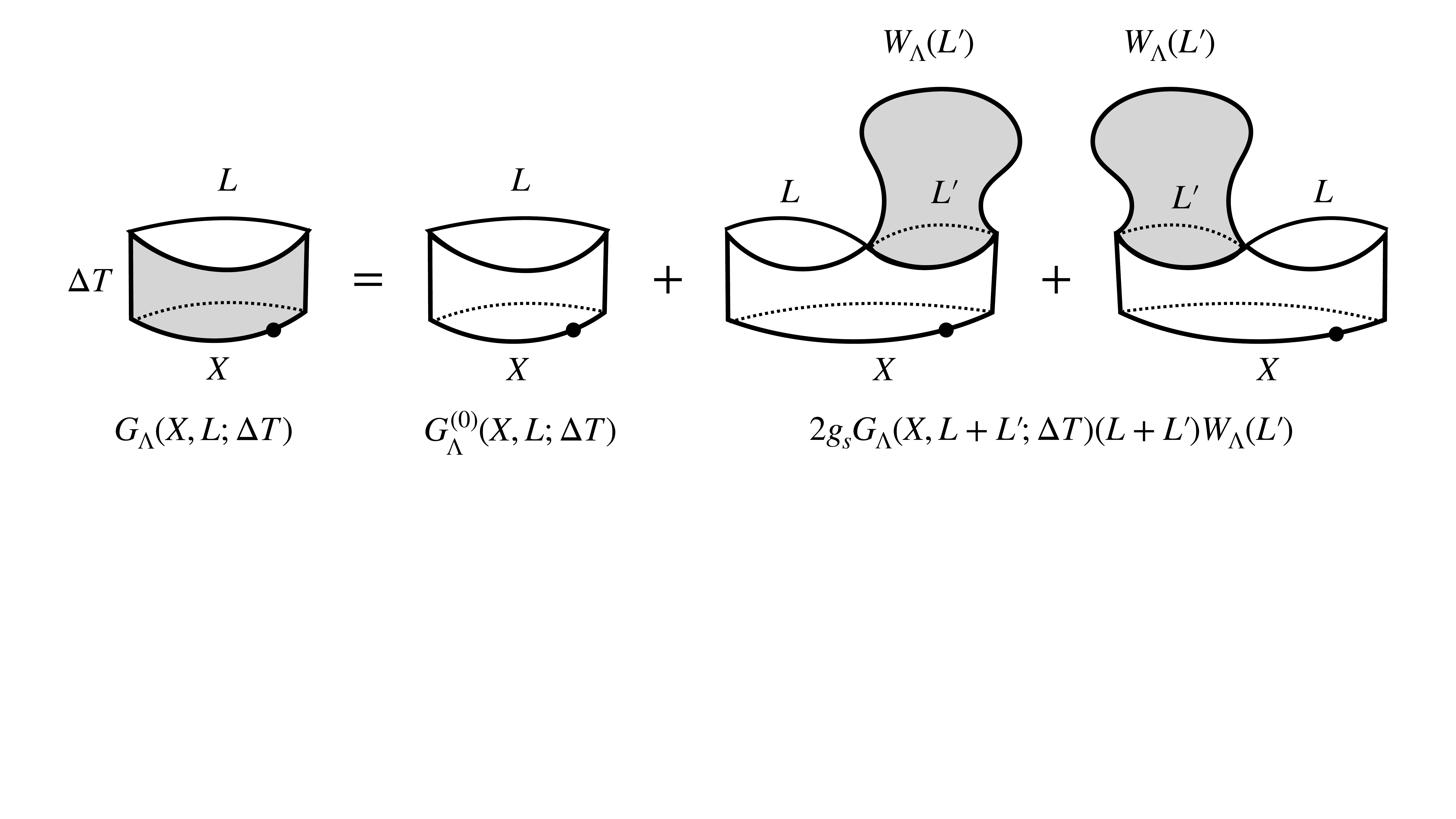}}}
\vspace{-4cm}
\caption{{\small  The infinitesimal propagation $G_\Lam(X,L;\Del T)$ for GCDT: either no split takes place or 
a split will occur  with ``probability'' $g_s \Del T$. After the split one part will be a baby universe with the topology of a disk.
For the split to occur, the exit loop (with length $L\plu L'$)  has to pinch at a point. This can happen in $L\plu L'$ ways. 
We have to integrate wrt $L'$ to count all possible configurations.}}
\label{figcdt5}
\end{figure}

Knowing $W_\Lam(X)$ we can now return to Fig.\ \ref{figcdt2} and find an equation for $G_\Lam(X,L;T)$. Rather than using 
Fig.\ \ref{figcdt2}, which involves a time integration from 0 to $T$,  it is more convenient to use the infinitesimal version of it, 
where time only changes by $\Del T$ as indicated on  the figure and shown in detail  in Fig.\ \ref{figcdt5} . This figure 
leads to the following equation for $G_\Lam(X,L;\Del T)$:
 \bea
G_\Lam(X,L,\Del T)\! \!\!&=&\!\!\! 
G_\Lam^{(0)}(X,L,\Del T) \plu 2g_s \Del T \!\!\int \!dL' (L\plu L') G_\Lam^{(0)}(X,L\plu L',\Del T) W_\Lam(L')\no \\
&=&\! \!\!\e^{-XL} \mi  \Del T \frac{\prt}{\prt X} \Big(\big[(X^2 \mi \Lam)  \plu
 2g_s   W(X) \big] \e^{-XL} \Big)\plu \cO(\Del T^2) \label{cdt42}
 \eea 
Here we have used \rf{cdt32} and \rf{cdt22} to write
\beq\label{cdt43}
G_\Lam^{(0)}(X,L;\Del T) = \e^{-XL} - \Del T  \frac{\prt}{\prt X} \Big((X^2\mi \Lam) \e^{-X L}\Big) + \cO(\Del T^2) ,
\eeq
Thus we see that the only change going from CDT to GCDT is the replacement
\beq\label{cdt44}
X^2 \mi \Lam\;\; \to\;\; X^2 \mi \Lam \plu 2g_s W_\Lam(X) = \hW_\Lam(X).
\eeq
We can finally write 
\bea
G_\Lam(X,Y; T\plu \Del T)\!\!\! &= &\!\!\!\int dL  \; G_\Lam(X,L,\Del T) \;G_\Lam(L,Y; T)\no \\
& =& \!\!\!G_\Lam(X,Y; T) - \Del T  \frac{\prt}{\prt X} \Big( \hW_\Lam(X) G_\Lam(X,Y; T) \Big)
\hspace{1.0cm} \label{cdt45}
\eea
which leads to the generalization of \rf{cdt20}
\beq\label{cdt46}
\boxed{\frac{\prt G_{\Lam}(X,Y;T)}{\prt T} =-\frac{ \prt}{\prt X} \Big(   \hW_\Lam(X)   G_\Lam(X,Y;T) \Big)}
 \eeq
 
 The solution is by now standard and generalizes \rf{cdt21} and \rf{cdt22}:
 \beq\label{cdt47}
 G_\Lam(X,L;T) = \frac{\hat{W} (\bX(T,X))}{\hW(X)} \; \e^{-\bX(T) L}, \qquad \frac{d \bX}{dT} = -\hW(\bX),\quad \bX(0)= X.
 \eeq
 One can find $\bX(T;X)$ (and thus $G_\Lam(X,L;T)$) expressed in terms of elementary functions (see Problem Set 13).  
 Here we will only  provide the expression for $X \to \infty$, i.e.\ when we contract the entrance loop 
 to a point:
 \ \beq\label{cdt48}
 \bX(T;X\equ \infty) \mi \al = \frac{\Sg^2}{ \sinh (\Sg T) \big(\Sg \cosh (\Sg T) + \al \sinh(\Sg T)\big)}, 
 \quad \Sg\equ \sqrt{\al^2 - \frac{g_s}{2\al}},
 \eeq
 where we, using \rf{cdt41a}, can write
 \beq\label{cdt48a}
 \Sigma(\Lam,g_s) = \SL \; H\Big( \frac{g_s}{\Lam^{3/2}}\Big), \quad H(0) \equ1, \quad H\Big(\frac{2}{3 \sqrt{3}}\Big) \equ 0
 \eeq
 It can be shown that $\bX(T,X) \mi \al$ falls off exponentially as $e^{-2 \Sg T}$, 
 not only for $X \equ \infty$ as shown in \rf{cdt48}, but for all 
 $X > \al$ (see Problem Set 13). It follows from \rf{cdt47} that the cylinder amplitude falls off as $e^{-2 \Sg T}$, 
 but the coefficient $\Sigma$ decreases from the CDT value $2 \SL$ towards
 zero when $g_s$ increases to the critical value $g_s^c= 2 \al_c^3 =2\Lam^{3/2}/(3 \sqrt{3})$.
 
 We can define the two-point function for GCDT precisely as we did for CDT, eq.\ \rf{cdt32b}, and we obtain
 in the same way (only replacing $X^2 \mi \Lam$ with $\hW_\Lam(X)$)
 \beq\label{cdt48b}
\boxed{G_\Lam(T) =  \frac{d W_\Lam (\bX(T;\infty))}{ dT} 
 =   \frac{\Sg^3}{\al} \frac{\Sg \sinh (\Sg T) + \al \cosh(\Sg T)}{ \big(\Sg \cosh (\Sg T) + \al \sinh(\Sg T)\big)^3}} 
 \eeq
 It is somewhat tedious, but straight forward, to derive the formula using \rf{cdt48} (see Problem Set 13 for some details). 
 The formula itself is 
 remarkable and it looks like a simple generalization of the formula derived for the two-point function in EDT.  However, the 
 consequences are very different, and lead to the critical exponents of CDT, so {\it  we have achieved our goal: to 
 find a non-trivial generalization of CDT, which still belong to the same universality class}. Let us discuss this in the same 
 way as we did for the two-point function of CDT. Firstly it falls of exponentially as $\e^{-2 \Sg T}$ when $T \to \infty$. 
 This indicates critical exponent $\nu_{gcdt} = 1/2$ provided we have a discretized theory where we can write
 $T = \ep \, t$ and $\sqrt{ \mu \mi \mu_c} \propto  \Sg \, \ep$. We will discuss such theories in the next subsection. Next we have 
 $G_\Lam( T \to 0) = 1$, as for CDT, and this indicates an anomalous dimension $\eta_{gcdt} = 1$. Finally 
 \bea
 \chi (\Lam) \!\!\! &=&\!\! \int_0^\infty dT \; G_\Lam(T) = W_\Lam(\bX(T \equ \infty;\infty)) -  W_\Lam(\bX(T \equ 0;\infty))\no \\
 & =& W_\Lam(\al)  = \frac{\Lam - \al^2}{2g_s} = \frac{1}{2\al}, \label{cdt48c}
 \eea 
and again this indicates a susceptibility exponent $\gamma_{gcdt} \equ  \oh$ provided we can write 
$ \sqrt{\mu \mi \mu_c} \propto \al \, \ep$ in a discretized theory. 
 
 $W_\Lam(X)$ can be expanded in powers of $g_s$
 \beq\label{cdt49}
 g_sW_\Lam(X) = \sum_{n=0}^\infty  W^{(n)}_\Lam(X) g^{n+1}_s
 \eeq
 where $W^{(n)}_\Lam(X)$ is the disk amplitude with $n\plu 1$ ``CDT disk components'': 
 $W^{(0)}_\Lam(X)$ is the CDT amplitude with 
 one component, the CDT universe, $W^{(1)}_\Lam(X)$ is the universe where at some time $T$ it split in two components
 which then continue their propagation in time independently  without splitting any further, i.e. as two CTD universes,
 and higher powers of $g_s$ capture the iteration of this splitting process. Let us (for reasons to be clear later) mark one 
 of the components. One can think of the mark as associated with the end point in time of that particular 
 CDT component (where the length of the boundary loop is contracted to a point). We obtain the corresponding disk amplitude 
 by differentiating $ g_sW_\Lam(X)$ wrt $g_s$ since the marking of a component in $W^{(n)}_\Lam(X)$ can be done in $n\plu 1$
 ways. A short calculation, using \rf{cdt40} and \rf{cdt41} leads to the remarkably simple result\footnote{This result is the equivalent
 to the result in EDT that $d W_\Lam^{\rm edt}(X)/d\Lam \propto 1/\sqrt{X + \SL}$, which one easily proves by differentiating 
 $W_\Lam^{edt}(X)$ wrt $\Lam$ and which we discuss in more detail in Exercise 10. The difference is that we in EDT differentiate 
 wrt $\Lam$, not $g_s$. In the continuum limit of EDT there is no strict equivalent to $g_s$ since an infinite number of baby 
 universes will be created in a continuum time $\Del T$, even if $\Del T$ is small. Alternatively one can say the baby universes 
 are everywhere and thus  marking the ``top'' of a baby universe will be ``proportional'' to just marking a point, which is 
 exactly the counting provided in EDT by differentiating wrt $\Lam$. }
 \beq\label{cdt50}
 \tW_\Lam (X) :=  \frac{d (g_s W_\Lam(X))}{d g_s} 
 = \frac{1}{\sqrt{(X\plu \al)^2 \mi 2g_s/\al)}}
 \eeq 
 In some sense $\tW_\Lam(X)$ {\it is} the natural generalization of the CDT disk amplitude $W^{(0)}(X)$. Recall that 
 the CDT disk amplitued was defined by \rf{cdt33}, and since we now have the cylinder amplitude for GCDT we 
 could use a similar definition of the disk amplitude (which will then differ from $W_\Lam(X)$). A calculation like 
 \rf{cdt33}, just using \rf{cdt47} instead of \rf{cdt21} and \rf{cdt22}, leads to 
 \beq\label{cdt50a}
 \int_0^\infty dT\; G_\Lam(X,L \equ 0;T) = -\int_X^\al \; \frac{d\bX}{\hW_\Lam(X)} = \frac{X\mi \al}{ \hW_\Lam(X)}  = \tW_\Lam(X).
 \eeq
 If we look at Fig.\ \ref{figcdt2}, it is seen that contracting $L$ to a point and integrating wrt $T$ can be viewed as labeling 
 one of the CDT components in Fig.\ \ref{figcdt3} (the component which contains the contracted loop), and we have indeed
 agreement between \rf{cdt50} and \rf{cdt50a}.

 Let us use $\tW_\Lam(X)$ as a partition function and calculate the average number of CDT components (baby universes) 
 in this ensemble. Again we obtain this number by differentiating $g_s\tW_\Lam(X)$ wrt $ g_s$
 \beq\label{cdt51}
 \la n \ra_{g_s} =\frac{1}{ \tW_\Lam(X) } \frac{d(g_s \tW_\Lam(X))}{d g_s}= 
 1 +\frac{g_s}{ 3\al^2-\Lam}\; \Big[\frac{X \plu 2\al}{(X\plu \al)^2 \mi  2g_s/\al}\Big]  
 \eeq
 It is seen that $\la n \ra_{g_s} \to \infty$ for $g_s \to g_s^c$ simply 
 because $ \frac{d \al}{dg_s} = \frac{1}{3\al^2 \mi \Lam}$ diverges at that point.
 So for $g_s > g_s^c$ the GCDT theory breaks down and since the number of baby universes proliferates, it is natural 
 to conjecture that the EDT picture will take over in any sensible extension of GCDT. This is indeed the case as we will
 explain  in the next subsection. 
 
 \subsection*{GCDT defined as a scaling limit of graphs}
 
 Above we defined GCDT via pictures! It is possible to define GCDT in the same way as we have defined EDT and CDT, 
 via triangulations (and also on  much  more general classes of graphs), 
 where the links have length $\ep$, and then take the scaling limit 
 $\ep \to 0$. The {\it continuum} expression for the GCDT disk amplitude was 
 \beq\label{cdt60}
 W_\Lam(X) = \frac{\Lam \mi X^2 \plu (X \mi C) \sqrt{ (X \mi C_+) (X \mi C_-)}}{2g_s},
 \quad
 C = \al,~ C_{\pm} = -\al \pm \sqrt{\frac{2g_s}{\al}}.
 \eeq
 This expression is formally quite similar to the expression for the disk amplitude in EDT {\it before} one takes the scaling limit:
 \beq\label{cdt61}
 w(z) = \frac{ - \kp \plu z \mi \kp z^2 \plu t(z \mi c) \sqrt{(z\mi c_+)(z\mi c_-)}}{2 \th}, \quad V'(z) = \frac{1}{\th}\,(z \mi \kp (1 \plu z^2))
 \eeq
 \begin{figure}[t]
\vspace{-1cm}
\centerline{\scalebox{0.2}{\includegraphics{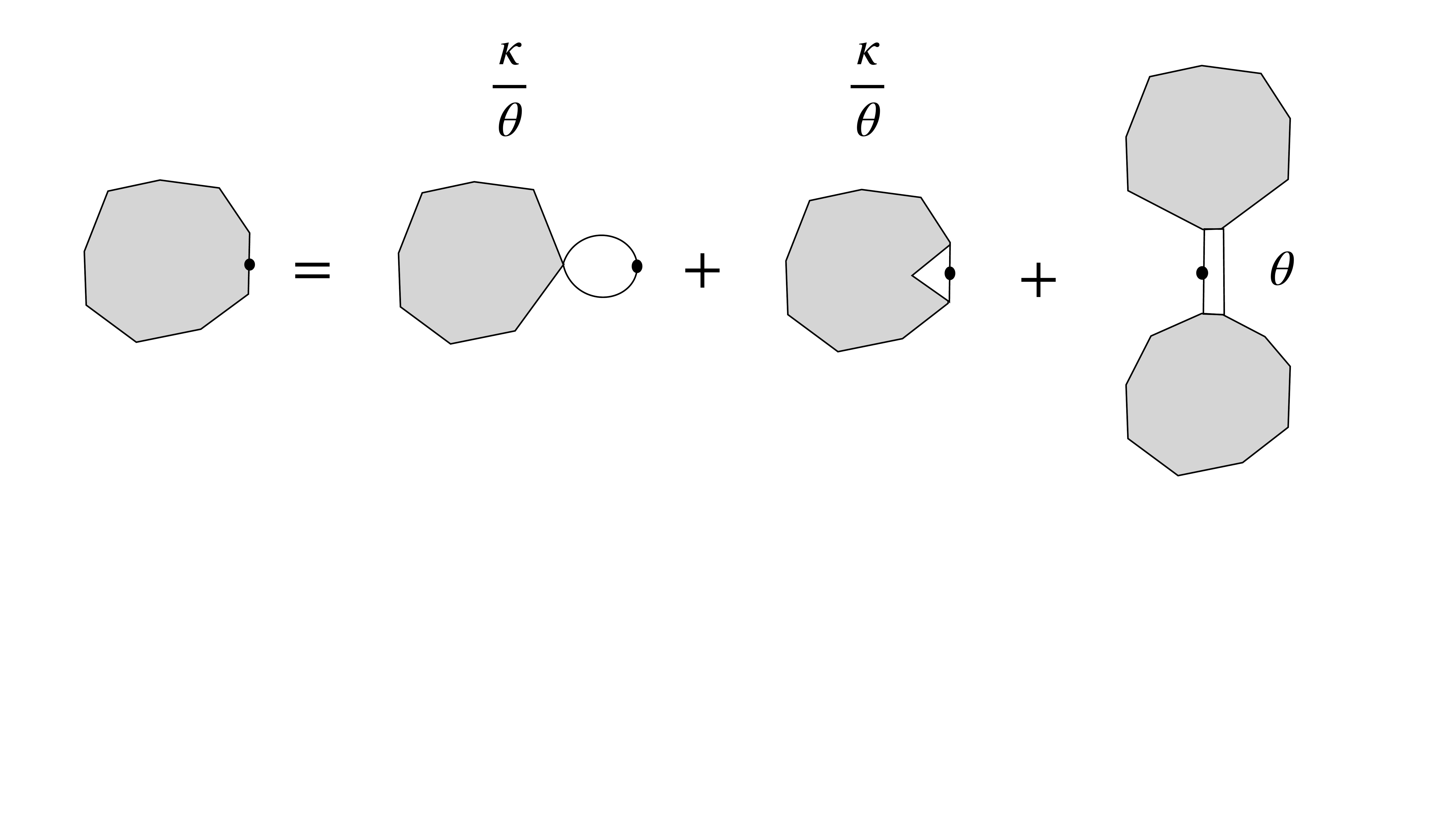}}}
\vspace{-3.5cm}
\caption{{\small  The graphic equation for $w(z)$ in the case there $V'(z)$ is given by \rf{cdt61}.}}
\label{figcdt7}
\end{figure}
A simple shift $z \to z \plu 1/2\kp$ will eliminate the linear term which is present in \rf{cdt61} but not present in \rf{cdt60}.
 However, the standard scaling limit of EDT is such that $c_-$ does not scale and therefore, in the EDT scaling 
 limit the cut from $c_+$ to $c_-$ will develop into a cut from a $C_+$ to $\mi \infty$, a situation distinctly different from 
 what is seen in \rf{cdt60}. For this reason  
 we have introduced here a new (allover) coupling constant $\th$, which was formerly chosen to be 1 in the EDT case.
 In the defining picture for the EDT disk amplitude this new coupling constant will appear as a coupling contant
 for the splitting of $w(z)$ in two $w(z)$ as shown in Fig.\ \ref{figcdt7}, as one can check recalling the arguments
 leading to the picture in the first place. For a fixed $\th$ the critical EDT point $\kp_c$ was the point where 
 $c_+(\kp_c) \equ c(\kp_c)$ and approaching this point, which depends on $\th$, according to 
 \beq\label{cdt62}
  \kp = \kp_c(\th)\big(1 \mi  \oh \ep^2 \, \Lam_{edt}\big), \quad z = z_c(\th) ( 1 \plu \ep X_{edt}) 
  \eeq
  leads to the standard EDT disk amplitude. The only way to obtain something different is to scale $\th \to 0$ at the same time.
  It is intuitive clear that this is what we have to do in order to make contact to GCDT. Recall that Fig.\ \ref{figcdt7} was also
  used to derive the two-loop propagator in EDT, viewing it as the discretized version of Fig.\ \ref{figcdt5}. It is then clear that if 
  we want to prohibit the creating of baby universes, such that only finite many appear in the continuum limit, we have to scale 
  $\th$ to zero. In fact, if we write $\th \equ g_s\, \ep^3$ then a calculation leads to 
  \beq\label{cdt63}
   w(z) ~\buildrel{\ep \rightarrow 0}\over\longrightarrow ~  
   \frac{1}{\ep}\, W_\Lam(X), \quad \kp \equ \kp_c \,\big(1 \mi  \oh \ep^2 \Lam\big), 
   \quad z\equ z_c ( 1 + \ep X), \quad \kp_c \equ \oh, ~z_c \equ 1.
   \eeq
   $\kp_c$ and $z_c$ are just the critical values for CDT. It is now clear from  \rf{cdt41a} and \rf{cdt48a} that 
   if we have a scaling limit $\th \equ g_s \,\ep^3$ and $\kp_c\mi \kp \equ \ep^2 \Lam$ then we indeed can write 
   \beq\label{cdt64}
   \sqrt{\kp_c \mi \kp} ~\propto~ \al \, \ep ~\propto~ \Sigma \, \ep \qquad \mbox{ for $\ep \to 0$ and fixed $\Lam$ and $g_s$.} 
   \eeq
   This shows, as already remarked, that the critical exponents in this scaling limit are the ones of CDT and not 
   the EDT exponents.
  \begin{figure}[t]
\vspace{-1.3cm}
\centerline{\scalebox{0.17}{\includegraphics{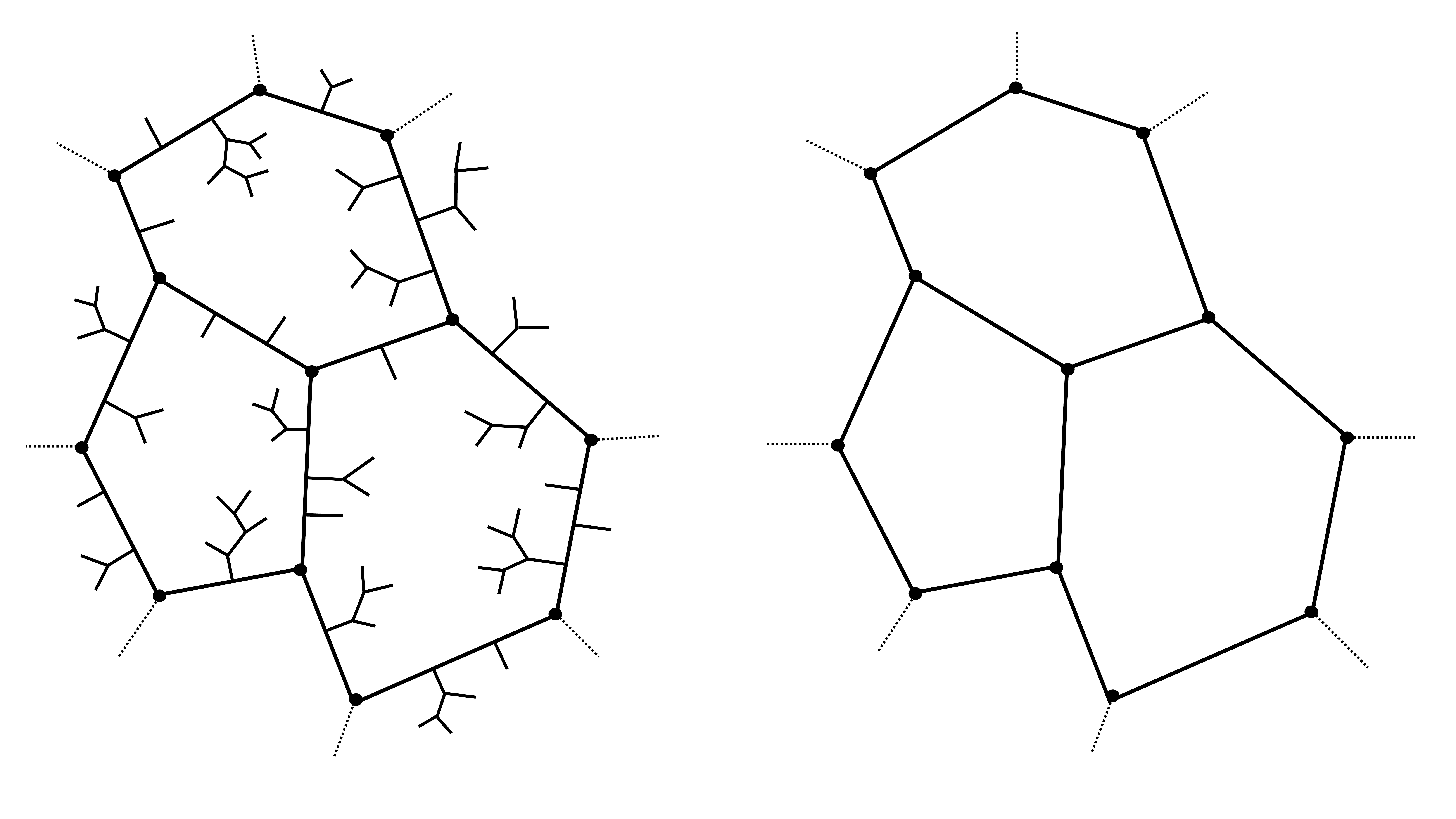}}}
\vspace{-0.7cm}
\caption{{\small  A planar graph constructed from vertices of order 1 and 3 (the left graph in the picture) can be viewed as 
a ``skeleton'' graph (the right graph) consisting only of vertices of order 3,  decorated with tree-graphs (rooted BPs).}}
\label{figcdt8}
\end{figure}

 The graphs generated by $V'(z)$ given in \rf{cdt61} will be made of triangles and ``one-gons''. Let $T_1$ and $T_3$ denote
 the number of one-gons and triangles in a connected planar graph (a ``triangulation'' $T$) constructed from these objects. 
 The factor $\th$ associated with $T$ will be $\th^{-T_1 + T_3}$.
 It follows from  writing $z  \equ \th \tilde{z}$ in \rf{cdt61}, i.e.\ we obtain the standard form 
 $V'(\tilde{z}) \equ \tilde{z} \mi \kp( \th^{-1} \plu \th \tilde{z}^2)$
 and thus a factor $\kp/\th$ and a factor $\kp \th$ associated with one-gons and triangles, respectively.
 Let us consider the graph dual to $T$, i.e.\ the graph $\tilde{T}$
 where a vertex is put in the center of each triangle and each one-gon, and the vertices in  neighboring triangles or one-gons
 are connected by (dual) links. In this way we obtain a planar graph which consists of vertices of order 3 or 1. Such a graph is 
 shown in Fig.\ \ref{figcdt8}. Let $F$ denote the number of faces in $\tilde{T}$, $L$ the number of links and $V \equ T_1\plu T_3$ the 
 number of vertices in $\tilde{T}$. From Euler's relation we can write
 \beq\label{cdt65}
 F\mi L\plu V =1, \quad {\rm where} \quad 2L = T_1\plu 3T_3, \quad {\rm i.e.} \quad 2F \mi 1 = -T_1 \plu T_3.
 \eeq
 The factor $\th$ associated with $T$ can then be written as the factor 
 \beq\label{cdt66}
 \th^{2 F(\tilde{T}) -1}
 \eeq
 associated with  the dual graph $\tilde{T}$. Eq.\ \rf{cdt66} shows that the number of faces in the dual graphs will be 
 suppressed when $\th \to 0$. The following picture then emerges: as long as
 $g_s < g_s^c$ (for fixed $\Lam$) the criticality of the ensemble of $\phi^3$ and $\phi$ graphs, 
 exemplified in Fig.\ \ref{figcdt8}, is determined 
 by the criticality of the BPs dressing the $\phi^3$ skeleton graphs and the average number of faces, links and vertices in the 
 skeleton graphs will be {\it finite} in the scaling limit. This is the GCDT limit. However, for $g_s$ larger than $g_s^c$ a different
 scaling limit will prevail, where the BPs will not be critical (and they are thus not important in the scaling limit), but now
 the skeleton graphs will define the criticality. This is the EDT limit.
 
 Seemingly, insisting on only a finite number of baby universes being present in the scaling limit of the triangulated surfaces 
 showed up has a scaling limit on the set of dual graphs where the number of faces were finite. This is not  a coincidence, 
 as will now be discussed. One can formulate a more detailed relationship between
 such classes of graphs, which also keeps track of graph distances. It is most easily done, not starting with triangulations, but 
 with quadrangulations. We will end this section discussing this, without providing many details, not to mention proofs (they 
 can be found in \cite{ab2}, which also contains a combinatorial definition of GCDT and 
 discusses how to take the scaling limit in detail).
 Let us define {\it a planar map} as a connected graph which can be projected on the sphere without any links crossing. It will 
 consist of a number of faces, links and vertices. On such graphs one can mark a  vertex and  
 then define the graph distance  from the marked vertex to other vertices, i.e. one can define a distance function on the 
 graphs. It can then be shown 
 that there exists a bijection $\Phi$ from the planar quadrangulations with a marked 
 vertex\footnote{Strictly speaking one has to make the marking somewhat more precise but we will not go into the
  technical details of how to do that.} to the planar maps with a marked vertex, 
 such that if $Q$ is a planar quadrangulation with  $N$ faces and $n$ local maxima of the distance function, then $\Phi(Q)$ is 
 a planar map with $N$ links and $n$ faces, and the distance labelling of $Q$ is mapped to the distance labelling of $\Phi(Q)$.
 The scaling limit of GCDT is one where we (loosely speaking) keep the number $n$ of baby universes  
 fixed while taking $N$ to infinity.
 Starting out with the marked vertex, or more generally with a marked entrance loop, the distance function to 
 the marked point or to the marked entrance loop on the  quadrangulations serves  as the common time $T$, and 
 we have a picture where the vertices at a given distance first form a connected loop which develops in time and then 
 can split in two baby universes which again can split  as time progresses. Each baby universe will eventually vanish 
 ``in the vacuum''. The points where the baby universes vanish are the points where the distance function has local maxima.
 The bijection $\Phi$ sends these GCDT quadrangulations into general planar graphs with $N$ links and $n$ faces, and 
 the volume of a baby universe (i.e.\ its number of quadrangles) will be proportional to  the degree of the face (i.e.\ the number 
 of links constituting the boundary of the face). More precisely a baby universe of volume $V$ is mapped to a face of 
 degree $2V$ by $\Phi$. In this way the GCDT scaling limit can be understood also as a scaling limit on the set of planar maps
 where the average number of faces is finite.
 
 \subsection*{The classical continuum theory related to 2d CDT}
 
  We have now  theories, CDT and GCDT, which we have defined as quantum theories. It is natural to ask if there exist
  {\it classical} theories, which  lead to  CDT or GCDT when quantized. For CDT there exists an obvious 
  candidate where the symmetry imposed naively agrees with the symmetry imposed on CDT configurations: 
  {\it Ho\v{r}ava-Lifshitz gravity theory} (HLG). It is a modification of General Relativity where time is given a special role. 
  In so-called {\it projectable} HLG it is assumed that spacetime has a time foliation and that the theory 
  is invariant under spatial diffeomorphisms and time re-definitions, also called foliation preserving diffeomorphisms. This clearly 
  restricts the class of geometries and it agrees with the set of geometries we used to define CDT. In the same way as 
  the Einstein-Hilbert action is  characterized as being the unique action invariant under diffeomorphisms and containing at most
  second derivatives of the metric, one can find the action which contains at most second derivatives and is invariant 
  under foliation preserving diffeomorphisms. In spacetime dimensions larger then 2, what is usually denoted HLG is 
  a theory which actually contains higher spatial derivatives which are added to the theory to make it perturbatively renormalizable.
  However, in two dimensions renormalizability of gravity is not an issue (we have precisely quantized 2d gravity in these 
  lectures!) and we will not add such terms to two-dimensional HLG.  We will not go into any detail, but only mention that the 
  invariance under spatial diffeomorphisms implies that starting out with the metric variables 
  $g_{\mu\nu}(x,t)$, $\nu,\mu \equ 0,1$, where 0,1 signifies time and space, the only remaining  variables will be  
  \beq\label{cdt80} 
  N(t)  \equ \sqrt{|g_{00}(t)|} \quad {\rm and} \quad L(t) = \int dx \, \sqrt{|g_{11}(x,t)|} .
  \eeq 
  $L(t)$ is the length of a spatial universe at time $t$ and it cannot be changed by a spatial diffeomorphism. Further, one 
  one assumes in  projectable HLG that $g_{00}(x,t)$ is  a function only of $t$. In projectable HLG 
  the so-called proper time:
  \beq\label{cdt81}
  t_p(t) = \int_0^t dt' N(t'), 
  \eeq
  is invariant under time redefinitions, and thus a physical observable and corresponds to  $N_p(t_p) \equ 1$.
  The classical HLG action  rotated to Euclidean signature can now be written (choosing proper time)
  \beq\label{cdt82}
   S_E[L] = \int dt_p \left( \frac{\dot{L}^2(t_p)}{4L(t_p)} + \Lam L(t_p)\right)
   \eeq
   and the corresponding quantum amplitude will be
   \beq\label{cdt83}
   G_\Lam(L_1,L_2;T) = \int \cD L(t_p) \; \e^{ - S_E[L]},\quad L(0) \equ L_1,~~L(T) \equ L_2.
   \eeq 
   If we write
   \bea\label{cdt84}
   G_\Lam(L_1,L_2;T) &=& \la L_2 | \e^{-\hH T} |L_1\ra\\
   G_\Lam(L_2,L_1;\ep) &=& \la L_2 | (I - \ep \hH + \cO(\ep^{3/2}) | L_1\ra,\label{cdt85}
   \eea
   we obtain, by discretizing the proper time interval in steps of $\ep$, from \rf{cdt82} and \rf{cdt83}:
   \beq\label{cdt86}
     G_\Lam(L_2,L_1;\ep) = \frac{L_1}{\sqrt{4\pi \ep L_2}} \;\exp \left(- \frac{(L_2\mi L_1)^2}{4 \ep L_2} \plu
     \ep \Lam L_1\right),
     \eeq
     where the origin of the factor $L_1$ comes from the marking of the entrance loop of the cylinder amplitude 
     $G_\Lam(L_1,L_2;T)$. Integrating \rf{cdt85} with a wave function $\Psi(L_1)$ we have
   \beq\label{cdt87}
   \int \frac{dL_1}{L_1}  G(L_2,L_1;\ep) \Psi(L_1) = \Psi(L_2) - \ep \big( \hH \Psi\big) (L_2) + \cO(\ep^{3/2})
   \eeq
   and Taylor expanding $\Psi(L_1)$ appearing in the integral on the lhs of \rf{cdt87} around $L_2$ one finally obtains
   \beq\label{cdt88}
  \big( \hH \Psi\big) (L_2) =\left( -L_2 \frac{d^2}{d L_2^2} + \Lam L_2 \right) \Psi(L_2).
  \eeq
  Thus $\hH$ is precisely the CDT Hamiltonian   \rf{cdt34}. We have thus shown that the classical two-dimensional HLG
  when quantized leads  to CDT.

  \begin{figure}[t]
\vspace{-1cm}
\centerline{\scalebox{0.9}{\includegraphics{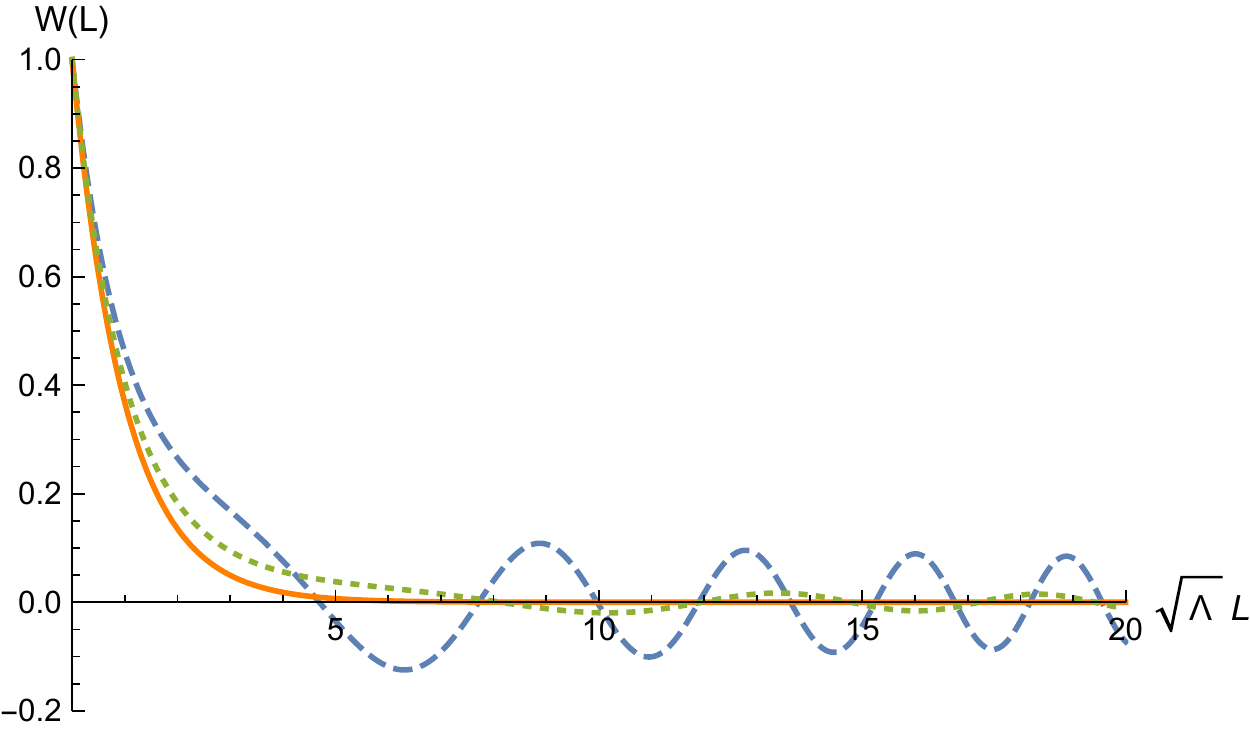}}}
\caption{{\small  The Hartle-Hawking wave function \rf{cdt90} which $c\equ 0$ plotted for $g_s \equ 0$ 
(orange curve and pure CDT, i.e.\ $W(\L) \equ \e^{-\SL L}$),
for $g_s/\Lam^{3/2} \equ 1/6$ (dotted green curve) and for $g_s/\Lam^{3/2} \equ 1/3$ (dashed blue curve). The oscillatory behavior 
starts only for $\SL L > \Lam^{3/2}/g_s$ and then the  fall off the changes from exponential to $1/L^{1/4}$. }}
\label{figcdt20}
\end{figure}

 It is less clear how to associate a classical  continuum theory to GCDT.  Looking at Fig.\ \ref{figcdt2} it seems
 difficult to associate a classical Hamiltonian  to the propagation of 
 space in (proper) time $T$. The creation of baby universes is not a natural part of a classical theory 
 and the attempts to define a classical theory leading to the GCDT when quantized have so far been forced to 
 put in some kind of  ``baby universes'' by hand in the classical theory,
 and are at best ``unusual" classical theories. 
 Rather, it seems more natural to view GCDT as a  quantum generalization of the quantum theory we have denoted 
 CDT, namely a quantum generalization where we allow 
 for the creation of baby universes. In this way one appeals to the idea that in a quantum theory  
 everything which is not protected by some symmetries (and corresponding 
 conserved charges) should be allowed. That would then lead not to CDT but to GCDT. 
 It  is remarkable that we can actually solve this generalized theory explicitly and perform
 the summation over all possible baby universes. Once we have allowed for the creation of baby universes, it is also natural 
 to allow for the creation of {\it wormholes}, i.e.\ a baby universe is created, but rather than vanishing in the vacuum it is 
 allowed to connect back to the ``parent universe'', in this way changing the spacetime topology (for an illustration see Fig.\ \ref{fig4.9}
 in the case of a propagating string). Even more remarkable than being able to sum over all baby universes is the fact that 
 one can perform this added summation which also includes the summation over all wormholes. We will not go into any 
 detail here, just mention two things. Firstly, the underlying technical 
 reason one is able to perform this summation is the bijection between
 BPs and CDT, which then can be generalized to a mapping between the generalized surfaces 
 (with baby universes and wormholes) and 
 BPs with loops. As we saw in Problem Set 9, one can sum these BPs with loops.  Secondly, one can write down an 
  ``effective Hamiltonian'' where the effect of this summation is taken into account and it is a very simple generalization of 
  the CDT Hamiltonian 
 \beq\label{cdt89}
 \hH^{\rm eff}  = -L \frac{d^2}{dL^2} + \Lam L  - g_s L^2.
\eeq
It is seen that the potential is unbounded from below, a reflection of the fact that the pertubation series in $g_s$ is not 
even Borel summable because of the numerous wormhole configurations of higher spacetime genus. In this sense the 
situation is somewhat similar to the situation we described in string theory and also in EDT if one included geometries 
with arbitrary high genus. Nevertheless, one can make $ \hH^{\rm eff}$ a selfadjoint operator with a discrete energy spectrum
where the eigenvalues $E_n \to E_n^{\rm cdt}\equ 2n \SL$ for $g_s \to 0$ and $n \ge 1$. In particular, $E_0 \equ 0$ is still 
an eigenvalue, and the ``disk'' function $W_{\Lam,g_s}(L)$, which now includes all possible wormholes, satisfies 
\beq\label{cdt90}
 \hH^{\rm eff}  W_{\Lam,g_s}(L) = 0, \qquad W_{\Lam,g_s}(L) =
 \frac{ {\rm Bi}\Big(\frac{\Lam - g_s L}{g_s^{2/3}} \Big)}{{\rm Bi}\Big(\frac{\Lam }{g_s^{2/3}}\Big)} + 
 c \cdot {\rm Ai}\Big(\frac{\Lam - g_s L}{g_s^{2/3}} \Big) 
 \eeq
 Bi and Ai are the standard Airy functions (which we also met in Problem Set 9). 
 One can check that for $g_s \to 0$ then $W_{\Lam,g_s}(L) \to \e^{-\SL L}$ for $L \leq \Lam/g_s$. A couple
 of solutions is shown in Fig.\ \ref{figcdt20}. Eq.\ \rf{cdt90}
 contains the term ${\rm Ai}(x/g_s^{2/3})$ which for $x >0$ falls off like $\e^{-2 x^{3/2}/3g_s}$. It is thus not 
 part of a perturbative expansion in $g_s$ and undetermined by the requirement that  $W_{\Lam,g_s}(L) \to \e^{-\SL L}$ for 
 $g_s \to 0$. {\it $W_{\Lam,g_s}(L)$ is the non-perturbative Hartle-Hawking wave function of our quantum GCDT universe. }

\newpage

  \setcounter{figure}{0}
 \renewcommand{\thefigure}{B.\arabic{figure}}
 \setcounter{equation}{0}
 \renewcommand{\theequation}{B\arabic{equation}}

 \section*{Appendix}

\subsection*{Preliminary material, part B:  Green functions}
 
 \subsubsection*{Basics}
 
The purpose here  is to remind the reader about Green functions 
as they are used in classical physics. No proofs will be given,
no mathematical rigor is attempted and the details of calculations 
are left as exercises. Our starting point will be a simple inhomogeneous second order differential
equation:
\beq
\left( - \frac{d^2}{dt^2} + V(t) \right) \psi(t) = J(t)\label{basic1}
\eeq
Assume that we can solve the following equation
\beq
\left( - \frac{d^2}{dt^2} + V(t) \right) G(t,s) = \delta(t-s)\label{basic2}
\eeq
This generates a solution to \rf{basic1}:
\beq
\psi(t) = \int ds \;G(t,s) J(s) \label{basic3}.
\eeq
$G(t,s)$ is a Green function for the differential equation \rf{basic1}.

Viewed as operators we can write (very formally, no discussion of 
domains etc):
$$
\hat{D} : \psi \mapsto \hat{D}\psi, ~~~~~
(\hat{D}\psi)(t) = -\psi^{''}(t) + V(t)\psi(t)
$$
$$
\hat{G}  : \psi \mapsto \hat{G}\psi,~~~~~  (\hat{G}\psi)(t) =
\int ds \,G(t,s) \,\psi(s).
$$
Equation (\ref{basic2}) can now be written as a formal 
operator identity
\beq
\hat{D} \hat{G} = \hat{I} \label{oper1}
\eeq
which suggests that 
\beq
\hat{G} = \hat{D}^{-1}.\label{oper2}
\eeq
In general $\hat{D}^{-1}$ is not well defined unless we restrict the function
space for $\hat{D}$ since $\hat{D}\psi = 0$ might have many non-trivial
solutions, namely the solutions to the homogeneous equation:
\beq
\left(-\frac{d^2}{dt^2} + V(t)\right) \psi(t) = 0\label{homogenous}
\eeq
{\it Sometimes} we can eliminate these solutions by imposing
{\it boundary conditions}. We usually have to do that anyway if we want  
$\hat{D}$ to be  {\it Hermitian}. Let us consider $\hat{D}$ defined in the interval $t \in [t_i,t_f]$.

\subsubsection*{Sturm-Liouville Boundary conditions at $\mathbf{t_i, t_f}$:}
\bea\label{B1}
       \alpha \psi(t_i) + \beta \psi'(t_i) & = & 0\\
       \gamma \psi(t_f) + \delta \psi'(t_f) & = & 0\label{B2}
\eea
These boundary conditions ensure that $\hat{D}$ is Hermitian since:
$$
\int_{t_i}^{t_f} \phi \frac{d^2}{dt^2} \psi = (\phi \psi' - 
\phi' \psi)
\mid_{t_i}^{t_f} + \int_{t_i}^{t_f} \Bigl( \frac{d^2}{dt^2}\phi
\Bigr) \psi
$$
For generic choice of $\alpha, \beta, \gamma, \delta$  (\ref{homogenous}) will 
{\it not} have a solution, i.e. $\hat{D}$ will be invertible.
Let $\phi_1(t)$ be a solution to (\ref{homogenous}) 
which satisfies \rf{B1} (but not \rf{B2} since we assume that no 
such solution exists). 
Let $\phi_2(t)$ be
solution to \rf{homogenous} which satisfies \rf{B2} (but not \rf{B1} ....). Let
$\psi_n(t)$ be an {\it eigenfunction} of the operator $\hat{D}$ (with the
given boundary conditions \rf{B1} and \rf{B2}):
\beq
\left[ -\frac{d^2}{dt^2} + V(t) \right] \psi_n(t) = \lambda_n \psi_n(t)
\label{eigenvalue}
\eeq
The functions $\psi_n(t)$ form a complete set since $\hat{D}$ is Hermitian: 
Any $L^2[t_i,t_f]$ function can be expanded as:
\beq
f(t) = \sum_n c_n \psi_n(t), ~~~c_n = \int_{t_i}^{t_f} \psi_n^*(t) f(t)
\eeq

There are two ways to construct $G(t,s)= \hat{D}^{-1}$:
\bea
 {\rm (I):}~~~ G(t,s) &=& \frac{-1}{w} \left\{ \begin{array}{cl}
                             \phi_1(t) \phi_2(s), & t<s \\
                             \phi_2(t) \phi_1(s), & t>s
                            \end{array} 
                          \right.
\label{GI}\\
{\rm (II):}~~G(t,s)& = &\sum_n \frac{\psi_n^*(s) \psi_n(t)}{\lambda_n}
\label{GII}
\eea
where
\beq\label{wronski}
w(\phi_1,\phi_2) \equiv  \phi_1 \phi'_2 - \phi_2 \phi'_1
\eeq
is called the {\it Wronskian} of $\hat{D}$. $w(\phi_1,\phi_2) =const.$
if $\phi_1,\phi_2$ satisfy (\ref{homogenous}).

{\sl Exercise 1}:~~~Show that (\ref{GI}) and (\ref{GII}) solves 
(\ref{basic2}) and that (\ref{basic3}) will satisfy the correct 
boundary conditions.

Let us now consider the special situations where $V(t)$ is {\it independent}
of time and $t_i \rightarrow -\infty, t_f \rightarrow +\infty$, i.e.\
\bea
{\rm  case~(a):} \quad &&\left[- \frac{d^2}{dt^2} + \tilde{\omega}^2 \right] \psi(t) = J(t)
\label{eqnoa}\\
{\rm  case~(b):} \quad &&  \left[- \frac{d^2}{dt^2} - \omega^2 \right] \psi(t) = J(t)
\label{eqnob}
\eea

{\sl Case (a)}

\begin{figure}[ht]
\centerline{\scalebox{0.20}{\includegraphics{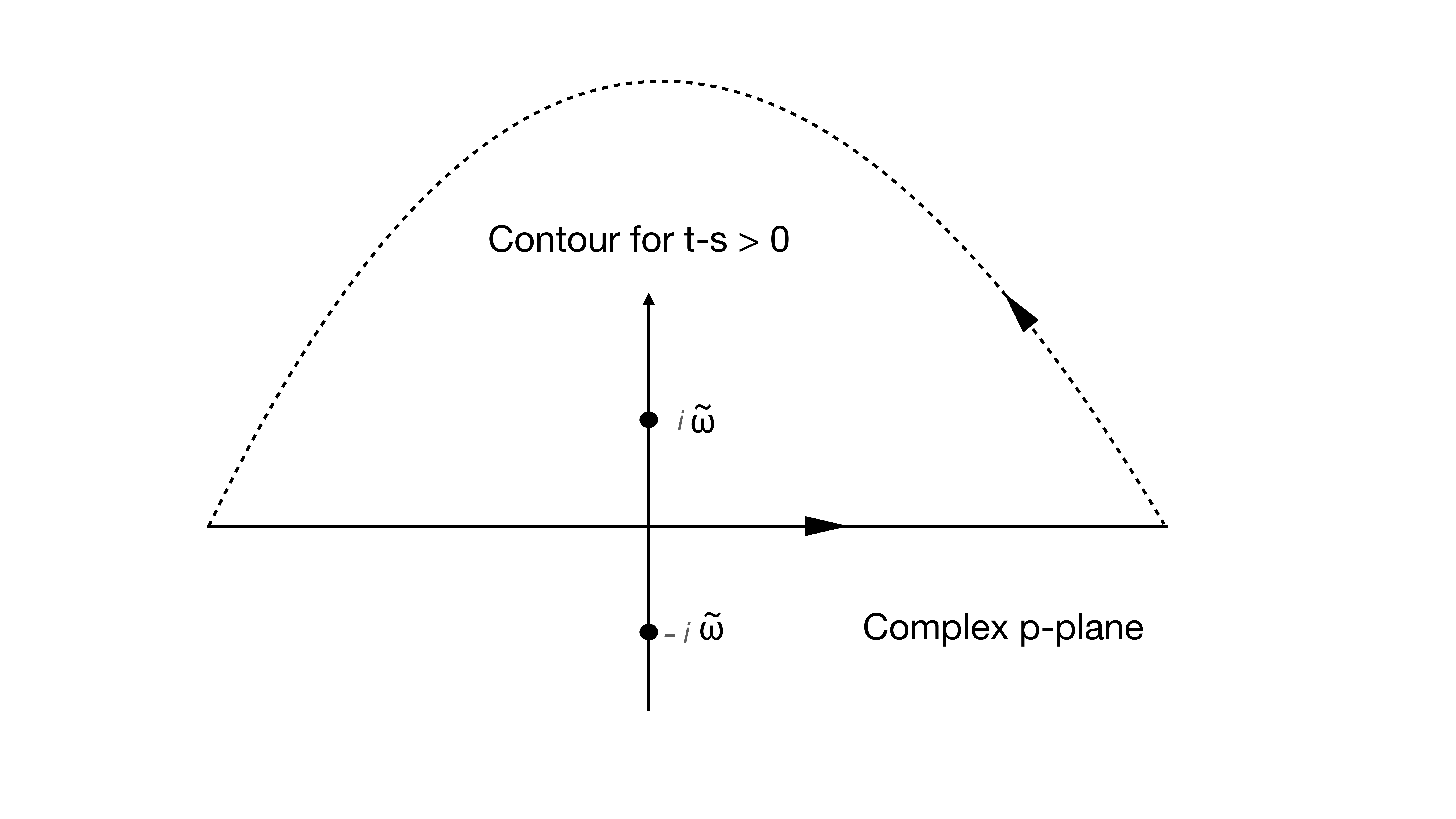}}}
\vspace{-1cm}
\caption{{\small  The integration along the real axis closed by a contour in the upper complex half-plane  for $t-s >0$. 
For $t-s < 0$ the integration contour should lie in the lower complex half-plane}}
\label{figgreen1}
\end{figure}

Impose {\it the boundary conditions:} 
\beq
\psi(t) \rightarrow 0 ~~{\mbox{\rm for}}~~ t \rightarrow \pm \infty
\label{B3}
\eeq
With these boundary condition $\hat{D}$ becomes Hermitean on $L^2(\mathbb{R})$.
The complete solution to the homogenous equation  corresponding to \rf{eqnoa} (i.e.\ $J(t) =0$) is:
\beq
\psi(t) = a\, \e^{-\tilde{\omega}t} + b\, \e^{\tilde{\omega}t} \label{homo1}
\eeq
It follows that $\lambda \equ 0$ is {\it not} an eigenvalue for \rf{B3}.
We see  that $\phi_1(t) \equ e^{\tilde{\omega}t}$ and  $\phi_2(t) \equ e^{-\tilde{\omega}t}$ and this implies by (\ref{GI}) that 
\beq
G(t,s) = \frac{1}{2\tilde{\omega}} \left\{ \begin{array}{cl}
                              \e^{\tilde{\omega}t}\, \e^{-\tilde{\omega}s}, & t<s\\
                              \e^{-\tilde{\omega}t} \,\e^{\tilde{\omega}s}, & t>s
                             \end{array}
                      \right\}    
       = \frac{\e^{-\tilde{\omega}|t-s|}}{2\tilde{\omega}}
\eeq
Let us use construction (\ref{GII}). The solution to the eigenvalue equation
(\ref{eigenvalue}) is:
\beq
\left( - \frac{d^2}{dt^2} \plu  \tilde{\omega}^2 \right) \psi_p(t) =
\lambda_p \psi_p(t) ~~ \Rightarrow ~~\psi_p(t) \equ \e^{ipt}, ~~~
\lambda_p \equ p^2 \plu \tilde{\omega}^2.
\eeq
Strictly speaking these are {\it generalized} eigenfunctions since they
do not belong to $L^2(\mathbb{R})$ and do not satisfy the imposed boundary 
conditions (but they stay bounded at least, contrary to the functions 
in eq.\ (\ref{homo1})). Obviously, they  form a complete set.
\beq
G(t,s) = \sum_p \frac{\psi_p^*(s) \psi_p(t)}{\lambda_p} ~\rightarrow
\int \frac{dp}{2\pi} \;\frac{\e^{ip(t-s)}}{p^2 + \tilde{\omega}^2}
= \frac{\e^{-\tilde{\omega}|t-s|}}{2\tilde{\omega}}
\label{9a}
\eeq
{\sl Exercise 2:} Show this using residue calculus (see Fig.\ \ref{figgreen1})\\

\vspace{12pt}



{\sl Case (b)}

Let us impose the boundary condition
\beq
\psi(t) \sim \e^{i \omega t} ~~{\mbox{\rm for}}~~t \rightarrow -\infty, 
~~~\psi(t) \sim \e^{-i \omega t} ~~{\mbox{\rm for}}~~ t 
\rightarrow +\infty
\label{B4}
\eeq
The functions which enter in these boundary 
conditions can be viewed as the analytic continuation of solutions 
$e^{\tilde{\omega}t}$ and  $e^{-\tilde{\omega}t}$ to the 
homogeneous equation of case (a) which satisfy one of the conditions in 
\rf{B3} for  $t \rightarrow -\infty$ and $t \rightarrow +\infty$.
Rotate the $\tilde{\om}$ from case  (a) as  
$\tilde{\omega}(\th) \equ e^{-i \th} \tilde{\om}$ where $\th \in [0,\pi/2[$. For any such $\tilde{\om}(\th)$ we still have that the 
solutions $\phi_1(t)\equ e^{\tilde{\om} (\th) t}$ and $\phi_2(t)\equ  e^{-\tilde{\om} (\th) t}$ to the  homogeneous equation
of case (a), with $\tilde{\om}$ replaced by  $\tilde{\omega}(\th)$, go to zero when $t \to \mp \infty$, respectively. 
We now view the $\omega$ of case (b) as the limit of $\tilde{\omega}(\th)$ for $\th \to \pi/2$, i.e. we write 
$\omega \to \omega (1\mi i\ep)$ and the solutions to the homogeneous equations in case (b) can then be viewed as 
$$
\phi_1(t) =\e^{i(\omega-i \epsilon)t}, \qquad\phi_2(t) = \e^{-i(\omega-i\epsilon)t}
$$
$$
\phi_1(t) \rightarrow 0 \quad {\rm for}\quad t \rightarrow -\infty, 
\qquad \phi_2(t) \rightarrow 0 \quad {\rm for}\quad t \rightarrow \infty
$$
We will use this interpretation of boundary conditions below, i.e. we will replace $\om$ with $\om \mi i\ep$ if 
ambiguities arise.

The complete solution of the homogeneous 
equation (\ref{homogenous}) is now in  case (b):
\beq\label{homeq}
\psi(t) = a\, \e^{-i \omega t} + b\, \e^{i \omega t},
\eeq
i.e. $\lambda \equ 0$ is {\it not} eigenvector of (\ref{eigenvalue}) with 
boundary condition \rf{B4}.
We obtain $\phi_1(t) \equ e^{i \omega t}$ and $\phi_2(t) \equ e^{-i \omega t}$, 
i.e. the Green function constructed according to (\ref{GI}) is:
\beq
G(t,s) = \frac{1}{2 \omega i} \left\{ \begin{array}{ll}
       \e^{i \omega t}\, \e^{-i \omega s}, & t<s\\
      \e^{-i \omega t}\, \e^{i \omega s}, & t>s
\end{array} \right\}
= \frac{\e^{-i \omega |t-s|}}{2i \omega}.
\eeq
Can we use (\ref{GII}) to construct $G(t,s)$? 
The answer is yes, with some care!
The eigenvalue eq.\ (\ref{eigenvalue}) leads to:
\beq
\left( - \frac{d^2}{dt^2} - w^2 \right) \psi_p(t) \equ \lambda_p(t)\psi_p(t) 
~~ \Rightarrow ~~ \psi_p(t) \equ \e^{ipt},~~~
\lambda_p \equ p^2 \mi  {\omega}^2.
\eeq
Again these are generalised eigenfunctions in the sense that 
they do not belong to $L^2(\mathbb{R})$ and do not satisfy the boundary conditions.
Compared to (\ref{9a}) we have a problem for $p = \pm w$ where
$\lambda_p = 0$,
i.e.\ for the solutions \rf{homeq} to the homogeneous equation.
However, we now resolve this ambiguity by replacing $\om \to \om \mi i\ep$, as mentioned above.
With this prescription we have $\lambda_p \equ p^2 \mi (\om \mi i\ep)^2$ and assuming  $\om >0$ and $\ep$ infinitesimal
we can write $\lambda_p \equ p^2\mi  \om^2 \plu i\ep$ (where we have redefined $2\om \ep \to \ep$). Finally we can then write:
\beq
G_F(t,s) = \sum_p \frac{\psi^*(s) \psi_p(t)}{\lambda_p} ~\rightarrow
\int \frac{dp}{2\pi} \: \frac{\e^{ip(t-s)}}{p^2-\omega^2+i \epsilon} =
\frac{\e^{-i \omega |t-s|}}{2i \omega}.
\label{feynman1}
\eeq
{\sl Exercise 3:}~Perform the p integration using residue calculus.\\[2.0ex]
By Fourier transformation we get (where we with an abuse of notation will use the 
same symbol $G$ also for the Fourier transformed Green function):
\beq
G_F(p) \equiv \int dt \;G_F(t)\; \e^{-ipt} = \frac{1}{p^2-{\omega}^2+i\epsilon}
\label{feynman2}
\eeq
This Green function is called the {\it Feynman Green function} or the Feynman propagator.

\begin{figure}[ht]
\vspace{-1cm}
\centerline{\scalebox{0.2}{\includegraphics{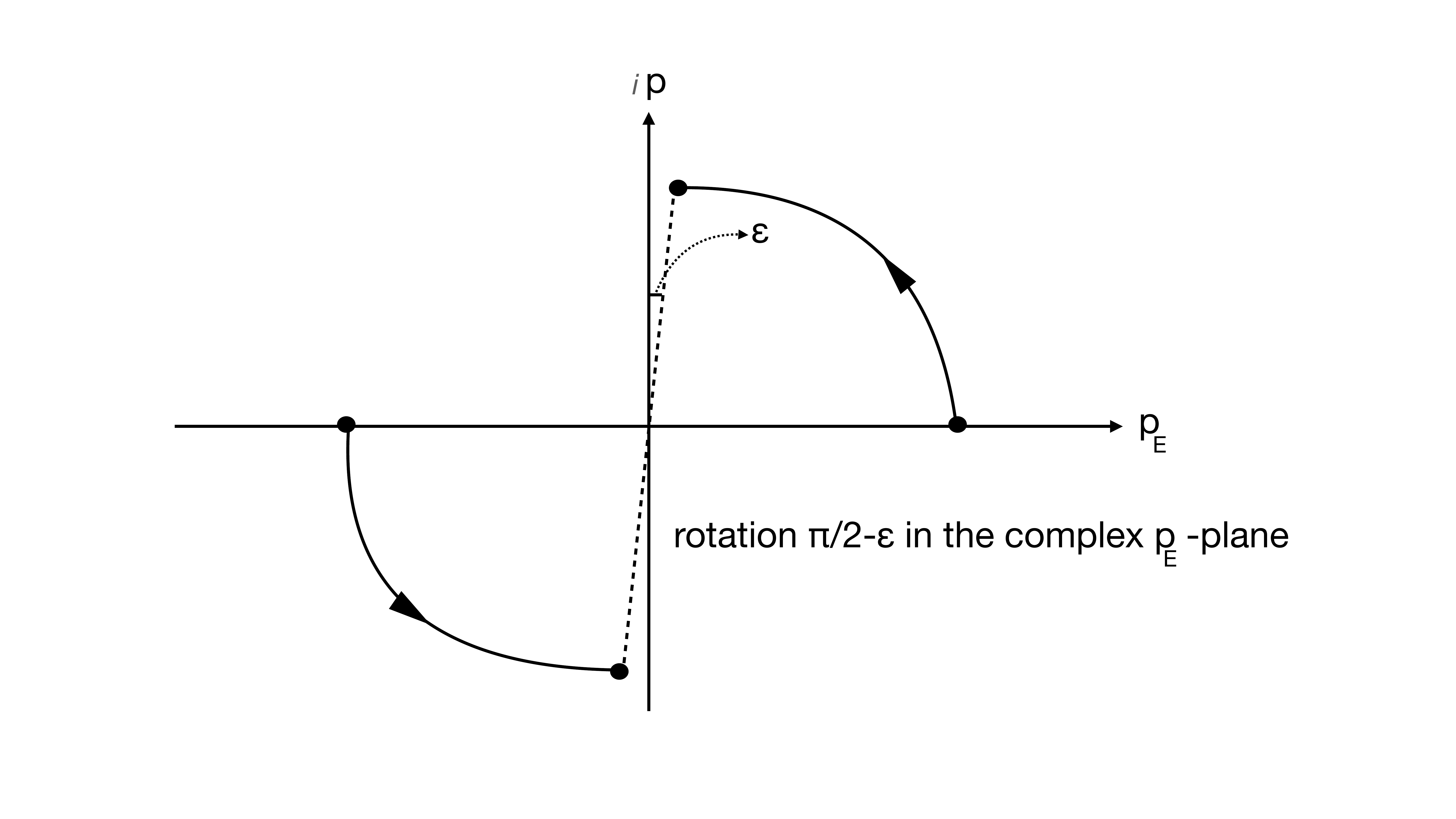}}}
\vspace{-1cm}
\caption{{\small  The rotation $p_E \to e^{i(\pi/2 -\ep)}p_E$ in the complex $p_E$ plane. In this way a real value $p_E$ 
appearing in the propagator $G_E(p_E)$ is transformed into $ip(1-i\ep)$ where $p$ 
is the real value appearing in the propagator $G_F(p)$.}}
\label{figgreen2}
\end{figure}

Rather than viewing the $i\epsilon$ prescription in (\ref{feynman2})
as originating from the analytic continuation in $\tilde{\omega}$ 
we can view it as an analytic continuation in the complex $p$-plane.  In this way the Green function
$G_F(p)$ is seen as (minus\footnote{The minus sign relating the two Fourier transformations is a triviality related to 
the definitions \rf{eqnoa} and \rf{eqnob}. Had we in case (b) used two + signs instead of the two - signs, 
this minus sign would be absent. And it 
is not unnatural to use the opposite sign in case (b). It comes if we think of case (b) as arising from case (a) by an 
analytic continuation $ t \to i t$ rather than $\tilde{\om} \to i \om$.}) the analytic continuation of the Fourier 
transformed $G_E(p)$ of (\ref{9a}), where
\beq 
G_E(p_E) = \frac{1}{p_E^2 + {\omega}^2} \label{euclidian}
\eeq
under a rotation $\pi/2\mi \ep$ of in the complex $p_E$-plane, as show in the Fig.\ \ref{figgreen2}.
Explicitly we have, defining $p_E(\th) = e^{i \th} p_E$, $\th \in [0,\pi/2]$ and noting that $p_E(\th) \to i p (1-i\ep)$ for 
$\th \to \pi/2-\ep$,  the follow behavior under the rotation
\beq\label{e-f}
\frac{1}{p_E^2+ \om^2} \;\;\to \;\;\frac{1}{(p_E(\th))^2+ \om^2} \;\;\; \underset{\th =\pi/2-\ep}{\to}
\;\;\; \frac{1}{-p^2-i\ep + \om^2}.
\eeq
{\it Thus we can obtain the Feynman Green function  (\ref{feynman2}) from (\ref{euclidian}) by a rotation of 
$\pi/2\mi \ep$ in the complex $p_E$-plane without encountering any singularities.}. We are spelling this out in detail
because it turns out to be a general principle in quantum field theory that one can obtain the so-called time-ordered 
Green functions (the Feynman Green functions) from the Euclidean Green functions by such an analytic continuation without 
encountering any singularities. This will be important for us since we will be working in spacetimes with Euclidean signatures
and we will calculate the Green functions there. Thus, should we want to rotate back to spacetimes with Lorentzian signature, 
we expect to obtain the Feynman Green functions, not the standard retarded Green functions we will discuss below.

We can now write down the general solution to the harmonic oscillator 
problem (b) with an {\it external} force $J(t)$:
\beq
\psi(t) = \int_{-\infty}^{\infty} dt' \;G_F(t \mi t') J(t')
\eeq
This is {\it not} the solution we will usually consider in a classical
problem where $\psi(t) \equ x(t)$ (the coordinate of the particle). In this case
we are interested in a {\it causal} Green function, rather than $G_F$. The 
response to $J(t)$ should not influence $x(t)$ at earlier times $t'<t$,
i.e.
\beq
G_R(t,t') = 0 ~~~{\mbox{ for}}~~~ t < t'
\eeq
We denote this Green function {\it the retarded Green function}.
\beq
G_R(p) = \int_{-\infty}^{\infty} dt \;\e^{-ipt}\, G_R(t) = \int_0^{\infty} dt
\;\e^{-ipt}\,G_R(t) 
\eeq
$G_R(p)$ is analytic in the lower complex $p$-plane if $G_R(t) = 0$ for $t<0$.

{\sl Exercise 4:} ~Show this, assuming $G_R(t)$ is bounded for $t \in \mathbb{R}$.

The retarded Green function is given by 
\beq
G_R(t) = -\theta(t)\, \frac{\sin \omega t}{\omega},
 ~~~~~~~\theta(t): \left\{ \begin{array}{cl}
                        0 & {\rm for}~~~ t<0\\
                        1/2 & {\rm for}~~~ t=0\\
                        1 & {\rm for}~~~t>0
                    \end{array}
            \right.      
\label{retardeda}
\eeq

{\sl Exercise 5:}~Show that $G_R(t)$ satisfies (\ref{basic2}) with 
$V(t)\equ -\om^2$.

Since $G_R(t)$ and $G_F(t)$ satisfy the same inhomogenous 
second order differential equation, the
{\it difference} is a solution of the  homogenous equation
(\ref{homogenous}):
\beq
G_F(t) = G_R(t) +  \frac{\e^{i \omega t}}{2i \omega}
\eeq
In the same way we can define {\it the advanced Green function} $G_A(t)$ such
that $G_A(t) = 0$ for $t>0$
\beq
G_A(t) = \theta(-t)\,\frac{\sin \omega t}{\omega} 
\eeq
If we denote $e^{-i \omega t}$ as {\it positive frequency oscillations} and 
$e^{i \omega t}$
as {\it negative frequency oscillation} we can say that $G_F(t)$ propagate
{\it positive frequencies forward in time} and {\it negative frequencies
backwards in time}. $G_R(t)$ and $G_A(t)$ do not allow a split, but denote by
$G_R^{(+)}(t)$ the positive frequency part, $G_A^{(-)}(t)$ the negative
frequency part of these functions. then,
\beq
G_F(t) = G_R^+(t) + G_A^-(t) = \frac{1}{2 i\omega} \big[\theta(t)\, \e^{-i \omega t}
 + \theta(-t)\, \e^{i \omega t}\big].
\eeq

{\it Is this discussion relevant for higher dimensions?  Yes!}

{\bf Example:~ Relativistic free massive scalar particle:}
\beq
\left[ -\frac{\partial^2}{\partial t^2}+\frac{\partial^2}{\partial x_i^2}-m^2 
\right] \phi(x_i,t) = 0 ~~~~~~~~i=1,...,n
\eeq
Translational invariance invites to a Fourier transformation:
\beq
\phi(x_{i},t) = \int \frac{d^3k}{{(2\pi)}^3}\; \e^{ik_ix_i}\, \phi(k_i,t)
\eeq
\beq
\left[ -\frac{d^2}{dt^2}-w_k^2 \right] \phi(k,t) = 0~~~~~~~~~ 
\begin{array}{lcl}
 {\omega}_k^2 & = & k_i^2+m^2\\ 
 {\omega}_k & = & \sqrt{k_i^2+m^2}
\end{array}
\label{massive}
\eeq
Eq.\ (\ref{massive}) is identical to the harmonic oscillator problem 
already considered, i.e.
\beq
G_F(t,k_i) = \frac{1}{2 i \omega_k} \left[ \theta(t)\, \e^{-i {\omega}_kt} + 
\theta(-t)\, \e^{i {\omega}_kt}\label{feynman3}
\right]
\eeq
\bea
G_F(k_0,k_i) &= &\int dt~ \e^{ik_0t} G_F(t,k_i) \nonumber \\
&=& \frac{1}{k_0^2-\omega_k^2-m^2
+i \epsilon} 
= \frac{-1}{k^{\mu}k_{\mu}+m^2-i \epsilon}
\label{feynman4}.
\eea
\beq\label{retarded}
G_R(t,k_i) = -\theta(t)\, \frac{\sin {\omega}_kt}{{\omega}_k},~~~~~
G_A(t,k_i) = \theta(-t) \,\frac{\sin {\omega}_kt}{{\omega}_k}.
\eeq

\subsubsection*{ Some of the higher dimensional Green functions:} 

{\bf 1. Electrostatics:} 
\beq
\nabla \cdot E = \rho, ~~~E = - \nabla \phi~~ \Rightarrow~
- \Delta \phi=\rho  ~~~\mbox{ (Poissons eq.)}
\eeq
The Green function of this problem satisfies:
\beq
\Big(\mi \frac{\partial}{\partial x_i^2}\Big)\; G(x,y) = \delta^{n} (x-y), 
\label{gf}
\eeq
where $n$ is the dimension of space. If $n >2$ we expect $\phi(x)$
to fall of at infinity for $\rho(x)$ having compact support,
i.e. we have the  "Euclidean" boundary condition discussed in case (a).
In particular, $G(x,y) \rightarrow 0$ for $|x-y| \rightarrow \infty$,
since $G(x,y)$ has the interpretation as $\phi(x)$ for a $\delta$--function
source located in $y$. This is 
opposite to the expectations for a  wave equation, 
recall the difference between (B3) and (B4).
We can solve (\ref{gf}) by Fourier transformation:
$$
G(x\mi y) = \int \frac{d^n p}{(2\pi)^n} ~\e^{ip(x-y)} G(p)\qquad
\delta^n(x\mi y) = \int  \frac{d^n p}{(2\pi)^n} ~\e^{ip(x-y)} 
$$
and \rf{gf} can be written as 
$$
p^2 \, G(p) = 1
$$
We conclude that the Green function in $p$ space is 
\beq
G(p) = \frac{1}{p^2},
\eeq
and by Fourier transformation we find for $n >2$
\beq
G(x) = \int \frac{d^n p}{{(2 \pi)}^n}\;  \frac{\e^{ipx}}{p^2} =
\frac{\Gamma(\frac{n}{2} \mi 1)}{4 {\pi}^{\frac{n}{2}}\,|x|^{n-2}}.
\label{electros}
\eeq

{\sl Exercise 6:}~Show (\ref{electros}) 
by using $\displaystyle{\frac{1}{p^2} \equ \int_{0}^{\infty} d\alpha \, \e^{-\alpha p^2}}$
and interchanging the  $\alpha$ and the $p$ integrations (not really allowed, but .....).

From \rf{electros} we see that $G(x) \rightarrow 0$ for $|x| \rightarrow \infty$ as 
desired. For $n = 3$ we get:
\beq
G(x,y) = \frac{1}{4\pi} \frac{1}{|x \mi  y|}
\eeq
\beq\label{coulomb}
\phi(x) = \frac{1}{4\pi} \int d^3y\; \frac{\rho(y)}{|x \mi  y|}~~~ 
\mbox{ (Coulombs Law!)}
\eeq

\vspace{12pt}

{\bf 2. Retarded wave functions:}

$$
\left( - \frac{\partial^2}{\partial t^2} + \frac{\partial}{\partial x_i^2} 
\right) \phi(x_i,t) = - \rho(x_i,t)
$$
From \rf{coulomb} we can guess the solution since we ``know" that disturbances caused by
$\rho$ travel with velocity 1 (= c):
\beq
\phi(x,t) = \frac{1}{4\pi} \int d^3y \;\frac{\rho(y, t \mi |x \mi  y|)}{|x \mi y|}
\eeq
Let us derive this from the retarded Green function \rf{retarded}:
\beq
G_R(k,t) = - \theta(t)\, \frac{\sin |k| t}{|k|} , ~~~\omega_k = |k| 
~~\mbox{for}~~ m = 0
\eeq
\begin{eqnarray*}
G_R(x,t) & = & -\int \frac{d^3k}{(2\pi)^3} \;\theta(t)\, \frac{\sin |k| t}{|k|}
\,\e^{ik_ix_i}\\ 
 &  = & -\theta(t) \frac{1}{4\pi^2} \int d |k| |k|^2 \frac{\sin |k| t}{|k|}
\int_{0}^{\pi} d \theta \, \sin \th \;\e^{i |k| |x| \cos \theta}\\
 & = &- \frac{\theta(t)}{2{\pi}^2} \frac{1}{|x|} \int_{0}^{\infty} d|k|\; 
\sin |k|t\; \sin |k| |x| = -\frac{\theta(t)}{4\pi} \frac{1}{|x|} \delta(t - |x|)
\end{eqnarray*}
 Conclusion: 
$$
G_R(x,t) = -\frac{\theta(t)}{4\pi|x|} \delta(t - |x|) = -\frac{\theta(x^0)}
{2\pi} \delta(x_{\mu}x^{\mu})
$$
and using the retarded Green function we can find any wave 
propagation caused by the source $\rho(x,t)$':
\bea
\phi(x,t) &= &\int dt' d^3x' ~G_R(x\mi x',t\mi t')\; \rho(x',t')\nonumber\\
& = &\frac{1}{4\pi} \int d^3y\; \frac{\rho(y,t \mi  |x\mi y|}{|x\mi y|}
\label{wave}
\eea

\vspace{12pt}
\noindent
{\bf 3. Classical electrodynamics}
\vspace{6pt}

Maxwell's equations can be written as 
\beq
\partial^{\mu}F_{\mu \nu} = - j_{\nu};~~~~~F_{\mu\nu}= \partial_\mu A_\nu
-\partial_\nu A_\mu ,
\eeq
or (imposing the Lorentz gauge condition $\partial^{\nu}A_{\nu}=0$ 
on $A_\nu$):
\beq
\left( -\frac{\partial^2}
{\partial t^2} + \frac{\partial^2}{\partial x_i^2} \right) A_{\nu} = -j_{\nu},
~~~~~~~~~\partial^{\nu}A_{\nu}=0
\eeq
The electromagnetic field triggered by a
current distribution $j_{\nu}$(x,t) is then given by 
\beq
A_{\nu}(x,t) = \int d^4y\; G_R(x^{\mu} \mi  {y}^{\mu}) \,
j_{\nu}({y}^{\mu})
\label{Asol}
\eeq
{\sl Exercise 7:}~Why is $\partial^{\nu}A_{\nu} = 0$ if $A_{\nu}$ 
is given by (\ref{Asol})?
(what is the physical requirement for $j^{\nu}(x,t)$?)

\vspace{12pt}

{\bf 4. Heat conduction or diffusion}
\vspace{6pt}

The equation for heat conduction (or diffusion) in a medium with a source $J(x,t)$ is 
\beq
\frac{\partial\, \varepsilon(x,t)}{\partial t} =b^2\, {\nabla}^2 \varepsilon(x,t) +
J(x,t), ~~~x_i = 1,...,n
\eeq
The Green function satisfies
\beq
\left( \frac{\partial}{\partial t} - b^2{\nabla}^2 \right) G(x,t ; y,t') =
\delta^{n}(x\mi y) \delta(t \mi  t')
\label{heat}
\eeq
and the boundary conditions are: causal  propagation and the 
requirement that $G(x,t) \rightarrow 0$ for $|x| \rightarrow
\infty$.\\[1.0ex]
Again we find the Green function by Fourier transformation:
\beq
G_R(k,t) = \int d^n x ~\e^{-ik_{i}x^{i}} \,G_R(x,t).
\eeq
If we use this in eq.\ (\ref{heat}) we get 
\beq
\left( \frac{\partial}{\partial t} + b^2k_{i}^2 \right) G(t,k) = \delta(t)
\eeq
which has the solution
\beq
G(k,t) = \theta (t)\; \e^{-b^2k_{i}^2t}\label{heat1}
\eeq
{\sl Exercise 8:}~ Show (\ref{heat1}).\\[1.0ex]
We can now transform back to $x$ variables
\beq
G(x,t) = \int \frac{d^n k}{{(2 \pi)}^n} \;\e^{ikx}\, G(k,t) =
\frac{\theta (t)}{(4 \pi b^2  t)^{n/2}} \;\e^{-\frac{|x|^2}{4b^2t}}
\label{heat2}
\eeq
{\sl Exercise 9:}~Show (\ref{heat2}).

Note
\beq
G(x,0^+) = \delta^{n}(x).
\label{47.a}
\eeq
We can now solve for the heat conduction from a
source  $J(x,t)$:
\beq
\varepsilon(x,t) = \int d^n x' dt' \;G_R(x \mi x', t \mi t')\,J(x',t').
\label{heat3}
\eeq

\vspace{12pt}
{\bf 5. The Schr\"{o}dinger equation}
\vspace{6pt}

Note that $t_{heat} \rightarrow it_{schr}$ turns the heat equation
into the Schr\"{o}dinger equation:
\beq
i \frac{\partial}{\partial t} \psi = -\frac{\hbar}{2m} \nabla^2 \psi ,
~~~~~~~b= \sqrt{\frac{\hbar}{2m}} 
\label{sch1}
\eeq
The Green function and the solutions are obtained by 
the same analytic continuation of (\ref{heat1})-(\ref{heat3}).
In particular, we find
\beq
\psi (x,t) =\Big(\frac{m}{2\pi i\, \hbar  \,  t}\Big)^{\frac{n}{2}} \;
\exp{\Bigl(\frac{i |x\mi x_0|^2m}
{2 \hbar t}\Bigr)}.
\label{schr2}
\eeq
$\psi(x,t)$ is the solution for $t>0$ to (\ref{sch1}) 
such that 
$$
\psi (x,0^+) =\delta^n(x\mi x_0)
$$ 
(see (\ref{47.a})), i.e. $\psi (x,0^+) = \la x | x_0\ra$ and has 
the quantum mechanical interpretation as the eigenvector of the 
operator $\hat{x}$ corresponding to the eigenvalue $x_0$
and $\psi(x,t)$, $t \geq 0$, is the wave function of the free particle 
located at $x_0$ at time $t=0$, i.e. 
\beq
\psi(x,t) \equiv \la x|e^{-it \hat{H}/\hbar}|x_0\ra = \Big(\frac{m}{2\pi i\, \hbar  \,  t}\Big)^{\frac{n}{2}} \;
\exp{\Bigl(\frac{i |x\mi x_0|^2m}
{2 \hbar t}\Bigr)}.
\eeq

\newpage

\setcounter{figure}{0}
 \renewcommand{\thefigure}{SB.\arabic{figure}}
 \setcounter{equation}{0}
 \renewcommand{\theequation}{SB\arabic{equation}}

\centerline{{\large \bf{Solutions to Preliminary Material, Part B}}}

\vspace{6pt}

{\bf Exercise 1}

Recall the following:
\beq\label{1}
\frac{\d}{\d x} | x| = 2 \th (x) -1,\qquad \frac{\d}{\d x} \th (x) = \del (x).
\eeq
where $\th(x)$ is the function which is 1 for $x > 0$ and 0 for $x < 0$. Thus, differentiating a  continuous function $f(x)$ where 
$f'(x^-_0) = a$ and $f'(x_0^+) =b$ one obtains 
\beq\label{2}
\frac{\d^2}{\d x^2} f(x)  = (b-a) \delta(x-x_0) + f''(x)
\eeq
where $f'(x)$ and $f''(x)$ denote the first and second derivative of $f$ for $x \neq x_0$. 

Assume now that $\phi_1$ and $\phi_2$ both satisfy (B6). Then we find that the Wronskian $w(\phi_1,\phi_2)$ is constant:
$$
\frac{\d}{\d t} w = \phi_1 \phi_2'' - \phi_1'' \phi_2 = \phi_1 V(t) \phi_2 - V(t) \phi_1 \phi_2 =0
$$
In (B11) we can then consider $w$ as a constant. $G(t,s)$ is a continuous function of $t$, but it is not differentiable in $t \equ s$.
For $t \neq s$ it satisfies (B2) since both $\phi_1(t)$ and $\phi_2(t)$ satisfy (B6). Thanks to \rf{2} the second derivative of 
$G(t,s)$ wrt $t$ for $t=s$ gives a contribution 
$$
\frac{-1}{w} \Big[\phi_2'(s_+)\phi_1(s)- \phi_1'(s_-)\phi_2(s)\Big] \; \delta(t-s) = - \delta(t-s)
$$
Thus $G(t,s)$ satisfies (B2).\\

Let us now turn to (B12). Acting with $D(t)$ on $G(t,s)$ we obtain from (B9), assuming that we can 
commute differentiation and summation (we will not deal with such subtleties....) that 
$$
D(t) G(t,s) = \sum_n \frac{\psi_n^*(s) D(t) \psi_n(t)}{\lam_n} = \sum_n  \psi_n^*(s) \psi_n(t) = \del(t-s),
$$
where we have used that the eigenfunctions $\psi_n(t)$ are assumed to be a complete set of normalized basis vectors in $L^2[t_i,t_f]$.
Such a choice can be made since $\hat{D}$ is an Hermitian operator. \\

Finally, let us look at the solution (B3) and use the representation (B11) for $G(t,s)$. We have 
$$
\psi(t_i) = \phi_1(t_i) \int \d s \, \phi_2(s) J(s), \qquad \psi(t_f) = \phi_2(t_f) \int \d s \, \phi_1(s) J(s),
$$
and it is then clear that $\psi(t)$ satisfies the boundary conditions since $\phi_1(t)$ and $\phi_2(t)$ were chosen to satisfy the boundary conditions at $t_i$ and $t_f$, respectively.   \\

\vspace{6pt}

{\bf Exercise 2} 

In its simplest form the residue theorem states that
\beq\label{3}
\oint_C \frac{\d z}{2\pi i} \; \frac{f(z)}{z-z_0}  = f(z_0)
\eeq
where $C$ is a simple closed curve, oriented anti-clockwise, enclosing $z_0$,  
and $f(z)$ is a holomorphic function in an open simple connected 
region of the complex plane containing $C$. In the case where $t \mi s > 0$ we apply the theorem as shown in Fig.\ B.1, writing 
$$
\frac{\e^{i z (t-s)}}{z^2 + \tilde{\om}^2} = \frac{f(z)}{ z - i \tilde{ \om}},\qquad f(z) = \frac{\e^{i z (t-s)}}{z + i \tilde{\om}},
$$
and using that the part of line-integral in the upper half-plane will vanish when the curve is at infinity and $t \mi s > 0$ since the 
function will then vanish exponentially there because ${\rm im}\, z > 0$.
If $t\mi s < 0$ we use a contour integral as in Fig.\ B.1, only with the arc in the lower complex plane, 
such that ${\rm im} \, z < 0$ on the arc.\\

\vspace{6pt}

{\bf Exercise 3}

We choose the same contours as in Exercise 2, only are the poles now located at $p= \pm (\om  \mi i\ep)$. For $t \mi s > 0$ we
have to choose the upper half-plane contour and the pole enclosed by the contour is located at $z_0 = -\om \plu i \ep$, i.e. we have 
$$
\frac{\e^{i z (t-s)}}{z^2 -(\om-i\ep)^2} = \frac{f(z)}{ z + { \om}\mi i \ep},\qquad f(z) = \frac{\e^{i z (t-s)}}{z - {\om}+i\ep}.
$$
The contour integral leads to 
$$
\oint \frac{\d z}{2\pi i} \; \frac{\e^{i z (t-s)}}{z^2 -(\om-i\ep)^2} = \frac{ \e^{-i \om |t-s| } }{-2 \om} = \frac{1}{i} G(t,s).
$$
The calculation for $t\mi s < 0$ and the choice of contour in the lower half-plane leads to the same result.\\

\vspace{6pt}

{\bf Exercise 4}

Let us assume $G_R(t) \equ 0$ for $t < 0$ and bounded for $t \geq 0$. (B.31) then shows that $G_R(p_R  + i p_I)$ is well defined 
and holomorphic for $p_I < 0$. $G_R(p_R + i p_I)$ is {\it holomorphic} since 
$$
i \frac{\prt G_R(p)}{ \prt p_R} =  \frac{\prt G_R(p)}{ \prt p_I} \quad \left( = \int_0^\infty G_R(t) \; t \,\e^{-i (p_R + i p_I) t} \right)
$$

\vspace{12pt}

{\bf Exercise 5}

Recall \rf{1} which tells us that differentiating $\th(x)$ we obtain $\del(x)$. We need yet another rule:
\beq\label{4}
\del(x) \, f(x) = \del(x) \, f(0), \quad {\rm i.e.} \quad \del(x) f(x) = 0 \quad {\rm if} \quad f(0) \equ 0.
\eeq
We now use this rule differentiating $G_R(t)$:
$$
-\frac{\d G_R(t)}{\d t} = \th(t) \cos \om t + \del(t) \, \frac{\sin \om t}{\om} =  \th(t) \cos \om t .
$$
$$
-\frac{\d^2 G_R(t)}{\d t^2} = -\th(t) \,\om \sin \om t + \del(t) \, \cos \om t = \om^2 G_R(t)  + \del(t)
$$
$$
 \left(-\frac{\d^2}{\d t^2} - \om^2\right) G_R(t)  = \del(t).
 $$
 
 \vspace{12pt}
 
 {\bf Exercise 6}
 
 Let us use that we have the one-dimensional integrals
 \beq\label{5}
 \int \frac{\d p}{2 \pi} \, \e^{-\alpha p^2} = \frac{1}{2 \sqrt{\alpha \pi}},\qquad
 \int \frac{\d p}{2 \pi} \, \e^{-\alpha p^2 - ixp} = \frac{\e^{-x^2/4\alpha}}{2 \sqrt{\alpha \pi}},
 \eeq
 The first integral we will calculate in a separate exercise dealing with Gaussian integrals. The second integral follows 
 from the first by completing the square: $\alpha (p^2+ ipx) = \alpha (p+i x/2\alpha)^2 + x^2/4\alpha$. The $p$-integral can now be performed since the integration along the horizontal line with imaginary coordinate $ix /2 \alpha$ gives the same result as 
 integrating along the real axis with $x \equ 0$, again because of the residue theorem (no poles between the two horizontal lines).
 
 From \rf{5} we obtain, simply by choosing the $p_1$-axis parallel to the vector $x$:
 \beq\label{5a}
 \int \frac{\d^n p}{(2 \pi)^n} \, \e^{-\alpha p^2 - ix\cdot p} = \frac{\e^{-x^2/4\alpha}}{2^n (\alpha \pi)^{n/2}},
 \eeq
 since we have $n \mi 1$ integrals like the left hand integral in \rf{5} and one integral (the $p_1$ integral) like the right hand side 
 of \rf{5}.
 
 We can now write, assuming that we can change the order of integration:
 $$
\int \frac{\d^n p}{(2 \pi)^n} \, \frac{\e^{-ipx}}{p^2} = \int_0^\infty \!\!\!\!\d \alpha \int \frac{\d^n p}{(2 \pi)^n} \, \e^{-\alpha p^2 - ix\cdot p}=
 \int_0^\infty \!\!\!\!\d \alpha \; \frac{e^{- x^2/4\alpha}}{2^n (\pi \alpha)^{n/2}} =  \int_0^\infty\!\!\!\! \d \beta \;
 \frac{\beta^{n/2-2} \; \e^{-\beta}}{4 \pi^{n/2} |x|^{n-2}},
 $$
 where we in the last integral has made the substitution $ |x|^2 /4\alpha = \beta$. The wanted result now follows from 
 the following formula for the $\Gamma$-function:
 \beq\label{6}
 \Gamma(x) = \int^\infty_0 \d \beta \; \beta^{x-1} \e^{-\beta}.
 \eeq

 \vspace{12pt}
 
 {\bf Exercise 7}
 
 A physical current will safisfy $\prt^\mu j_\mu (x) \equ 0$. This is just the continuity equation. Let now $A_\mu(x)$ be given 
 by (B.53). We then have 
 \bea
 \frac{\prt}{\prt x_\mu} A_\mu (x) &=& \int \d^4 x' \, \frac{\prt}{\prt x_\mu} G_R(x\mi x') \, j_\mu(x') = 
 \int \d^4 x' \, \Big(\mi \frac{\prt}{\prt x'_\mu}\Big) G_R(x\mi x') \, j_\mu(x') \nonumber \\
 &=& \int \d^4 x' \,  G_R(x\mi x') \,\frac{\prt}{\prt x'_\mu}  j_\mu(x') = 0,\nonumber
\eea
where we have assumed that boundary terms vanish at spatial or temporal infinity when we perform the partial integration 
to go from line one to line two in the equations.\\

\vspace{6pt}

{\bf Exercise 8}

From the rules \rf{1} and \rf{4} we obtain 
$$
\frac{\prt}{\prt t} G(k,t) = - b^2 k^2 \th(t) \e^{-b^2k^2 t} + \del(t) \, \e^{-b^2k^2 t} =  - b^2 k^2 G(k,t) + \del(t).
$$

\vspace{12pt}

{\bf Exercise 9}

The calculation is identical to the one done in eq.\ \rf{5a}.

\newpage

\renewcommand{\theequation}{\arabic{equation}}

\newpage

\setcounter{equation}{0}
\setcounter{figure}{0}
 \setcounter{figure}{0}
 \renewcommand{\thefigure}{PS1.\arabic{figure}}
\renewcommand{\theequation}{ps-1.\arabic{equation}}

\centerline{{\bf \Large  Problem Sets 1-13}}

\subsection*{Elementary Quantum Geometry Problem Set 1}
In this problem, we treat the free non-relativistic particle in the framework of the path integral.

\subsubsection*{Gaussian integrals}
Prove the following identities (assume the integration range to be $(-\infty, \infty)$ unless specified otherwise):
\begin{enumerate}
    \item \hspace{5cm}{$\displaystyle{\int  \e^{-(x^2+y^2)} \,dxdy = \pi.}$}\\
    \item \hspace{5cm} $\displaystyle{\int \e^{-\frac{1}{2}a x^2}\, dx = \sqrt{\frac{2\pi}{a}}.}$\\
    \item Let $A_{ij}$ be a real symmetric $n\times n$ matrix with positive eigenvalues.
        \[\int d^n x \,\e^{-\frac{1}{2} x_i A_{ij} x_j} = \frac{(2\pi)^{\frac{n}{2}}}{\sqrt{\det{A}}}. \]
        Note that summation over repeated indices is implied all throughout the exercise.
    \item Let $z_i \in \mathbb{C} = x_i+i y_i$ and let $A$ be a Hermitian matrix with positive eigenvalues.
        \[ \int \prod_{i=1}^n dx_i dy_i \, \e^{-\frac{1}{2} z^\dagger A z} = \frac{(2\pi)^n}{\det{A}}. \]
        Here 
    \[ z = \left(\begin{array}{c} z_1 \\ \vdots \\ z_n \end{array}\right), \quad \quad z^\dagger = \left(z_1^*, \cdots, z_n^*\right). \]
    \item Let $x, b$ be real $n$-vectors, and $A$ a real symmetric matrix with positive eigenvalues. Furthermore, let
        \[ S(x) = \frac{1}{2} x^T A x + b^T x. \]
        We now define the so-called ``classical solution'' $x_c$, for which 
        \[ \frac{\partial}{\partial x_i} S(x) = 0, \quad i = 1, \cdots, n. \]
        Show that 
        \[ S(x_c) = -\frac{1}{2} b^T A^{-1} b, \quad x_c = -A^{-1} b. \]
        Then write $x = x_c + \Delta x$, where $\Delta x$ denotes the ``fluctuations'' around the classical solution. Then show that
        \[ S(x) = S(x_c) + S(\Delta x, b=0). \]
        Subsequently, show that
        $$
            \int d^n x\, \e^{-S(x)} = \e^{-S(x_c)} \int d^n (\Delta x) \;\e^{-S(\Delta x, b=0)} 
            =\e^{-S(x_c)} \frac{\left(2\pi\right)^{\frac{n}{2}}}{\sqrt{\det{A}}},
        $$
       and thus  
        $$
            \int d^n x\, \e^{-\left(\frac{1}{2} x^T A x + b^T x \right)} = \frac{\e^{\frac{1}{2}b^t A^{-1} b} (2\pi)^{\frac{n}{2}}}{\sqrt{\det{A}}}.
        $$
        This result is still valid if we make the substitutions $A \to i A$, $b \to i b$. We then encounter the Fresnel integrals
        $$
        \int dx\, \e^{-\frac{i a x^2}{2}} = \sqrt{\frac{2\pi}{i a}}, \qquad 
        \int d^n x\, \e^{-i\left(\frac{1}{2} x^T A x + b^T x \right)} = \frac{\e^{\frac{i}{2}b^t A^{-1} b} (2\pi)^{\frac{n}{2}}}{\sqrt{i^n\det{A}}}.
        $$

    \item  The Fourier transformed of a function $f(x)$ is defined by 
        $$
        {\cal F}(f)(k) := \int d^nx f(x) \; \e^{-i k^T x}\quad {\rm and~then} \quad f(x) = \int \frac{d^n k}{(2\pi)^n} \e^{i x^T k} {\cal F} (f) (k).
        $$
        Show that the Fourier transformation of a Gaussian function is still a Gaussian:
        $$
        {\cal F}( \e^{-\frac{1}{2} x^T A x}) =  \frac{(2\pi)^{\frac{n}{2}}}{\sqrt{\det A}} \; \e^{- \frac{1}{2} k^T A^{-1}k }
        $$
        
        \item
        The convolution $(f\!*\!g)(x)$ of two functions $f(x)$ and $g(x)$ is defined as 
      $$
      (f \!*\! g) (x) = \int dy\, f(x\!-\!y) g(y) = \int dy \, g(x\!-\!y) f(y)
      $$
      It is a property of convolutions that the Fourier transformed of a convolution is the product of the Fourier transformed of the 
      functions:
      $$
      {\cal F}  (f \!*\! g) (k) =   {\cal F}  (f ) (k) \cdot  {\cal F} ( g) (k)
      $$
      Let $f_A(x)$ denote the Gaussian function 
      $$
      f_A (x) = \frac{1}{ \sqrt{\det ( 2\pi A)}} \;  \e^{-\frac{1}{2} x^T A^{-1} x}\quad {\rm i.e.} \quad {\cal F} (f_A)(k) =  
      \e^{- \frac{1}{2} k^T A k }
      $$
       Use this to show that 
        $$
        (f_A * f_B) (x) = f_{A+B} (x), \quad f_A^{*N} (x) = f_{NA}(x), \quad f^{*N} = f*f*f \cdots *f \quad N~{\rm times}.
        $$
 
 \end{enumerate}
         
\subsubsection*{The free non-relativistic particle}

   In the lectures, we saw that
    $$
            \bra{x_{n+1}} \hat{O}_\epsilon^{n+1} \ket{x_0} = \left(\frac{m}{2\pi i \epsilon \hbar}\right)^{\frac{n+1}{2}} \int \prod_{i=1}^n dx_i \; \e^{\frac{i}{\hbar} \sum_{i=0}^n \epsilon \left[ \frac{m}{2} \left(\frac{x_{i+1}-x_i}{\epsilon}\right)^2 - V(x_i)\right]}.
 $$
     For the free particle, we have $V(x_i) = 0$. In that case, we can perform the integrals for all the $x_i$ successively, starting with $x_1$ since they are all Gaussian integrals. In addition they are convolutions. We can thus use what we have just learned about 
  convolutions of Gaussians. 
  
  \begin{itemize}
  
  \item[8.]
  Show, using the convolution of Gaussians mentioned above,  that    
      $$
            \int dx' \frac{1}{\sqrt{  2\pi ia}} \;\e^{\frac{i }{ 2a} \left(x''-x'\right)^2} \frac{1}{\sqrt{2\pi  i b}} \;\e^{\frac{i }{2 b} \left(x'-x\right)^2} = 
            \frac{1}{\sqrt{  2 \pi i(a\!+\! b)}} \;\e^{\frac{i }{2(a+b)} \left(x''-x\right)^2}
        $$
        and use this result to prove, by successive convolutions,  that for $V(x) = 0$ we have
        $$
            \bra{x_{n+1}} \hat{O}_\epsilon^{n+1}\ket{x_0} = \sqrt{\frac{m}{2\pi i \hbar t}}\; \e^{\frac{i}{\hbar} \frac{m}{2} \frac{\left(x_b-x_a\right)^2}{t}} = \Psi (x_b \!- \! x_a,t),
        $$
        where 
        \[ t = (n+1)\epsilon, \quad \quad x_b = x_{n+1}, \quad \quad x_a = x_0. \]
        
         Why is this result independent of $n$?

         {\it $\Psi(x,t) \equiv \langle x| \Psi(t) \rangle$ is the wave function of the free particle}.
      
\item[9.]
Show that the Fourier transformed of $\Psi(x,t)$ wrt $x$ is 
$$
{\cal F}(\Psi) (k) = \e^{-i \frac{k^2 \hbar}{2m} \,t} =  \e^{-\frac{i}{\hbar} \frac{p^2}{2m} \,t} = \tilde{\Psi}(p,t), \quad p = \hbar k.
$$
Here $\tilde{\Psi}(p,t) \propto \langle p | \Psi(t)\rangle$  is of course the solution to the Schr\"{o}dinger equation for the free particle in momentum basis:
$$
i \hbar \frac{\partial}{\partial t} \tilde{\Psi}(p,t) = \frac{p^2}{2m}\;  \tilde{\Psi}(p,t)
$$ 

\end{itemize}

\newpage

\setcounter{equation}{0}
\setcounter{figure}{0}
\renewcommand{\thefigure}{ps-2.\arabic{figure}}
 \renewcommand{\theequation}{ps-2.\arabic{equation}}

\subsection*{Elementary Quantum Geometry Problem Set 2}

In this problem, we treat the next-simplest case of the path integral: the harmonic oscillator. The action is written 
\begin{align*}
    S[x] &= \int_{t_a}^{t_b} dt \;L\left(x(t)\right), \qquad
    L(x(t)) = \frac{m}{2} \left( \dot{x}^2(t)-\omega^2 x^2(t)\right).
\end{align*}

\begin{itemize}
    \item[1.] Show that the classical solution to the eom with boundary conditions 
    $$ 
    x(t_a) = x_a\quad {\rm  and} \quad    x(t_b) = x_b 
    $$
 takes the form
  \bea 
  x_c(t) &=& \frac{x_b \sin{\omega (t\mi t_a)}+x_b \sin{\omega(t_b\mi t)}}{\sin{ \omega(t_b\mi t_a)}} ,\qquad 
  \nonumber \\
  S[x_c(t)] &=& \frac{m\omega}{2\sin{\omega(t_b\mi t_a)}} \big[(x_a^2\plu x_b^2)\cos{\omega(t_b\mi t_a)}-2 x_a x_b\big]
  \label{p2j1}\eea
\
\item[2.] Make the decomposition
 \[ 
 x(t) = x_c(t) + \Delta x(t), \quad \Delta x(t_a) = \Delta x(t_b) = 0.
  \]
  and show that one has 
  \beq\label{p2j2}
S[x(t)] = S[x_c(t)] + S[\Delta x (t)] 
\eeq  
\end{itemize}

Now recall that the propagation amplitude can be expressed in terms of the path integral as
 $$
 \langle{x_b}| \e^{-i(t_b-t_a) \hat{H}}|{x_a}\rangle = 
 \int_{\substack{x(t_a)=x_a \\ x(t_b)=x_b}} \mathcal{D}x(t) e^{\frac{i}{\hbar} S\left[x(t)\right]}. 
 $$
 \begin{itemize}
  
  \item[3.] Show that 
  \[ 
  \int_{\substack{x(t_a)=x_a \\ x(t_b)=x_b}} \mathcal{D}x(t) e^{\frac{i}{\hbar} S\left[x(t)\right]} = e^{\frac{i}{\hbar} S\left[x_c(t)\right]} \int_{\substack{\Delta x(t_a)=0 \\ \Delta x(t_b)=0}} \mathcal{D}\Delta x(t) e^{\frac{i}{\hbar} S\left[\Delta x(t)\right]} 
  \]
\end{itemize}
 Set $y(t) = \Delta x(t)$.  We now have to compute 
  \[
  \int_{\substack{y(t_a)=0 \\ y(t_b)=0}} \mathcal{D}y(t) e^{\frac{i}{\hbar} S\left[y(t)\right]} = 
        \lim_{\epsilon \to 0} \big\langle {y(t_b)}\equ 0\big| \hat{O}^{n+1} \big|{y(t_a)}\equ 0\big\rangle , 
        \quad \quad t_b-t_a = (n+1)\epsilon. 
\]
        As shown in the lectures, this corresponds to the $n \to \infty, \epsilon \to 0$ limit of the integral
  \[ 
        \left(\frac{m}{2\pi i \epsilon \hbar}\right)^{\frac{n+1}{2}} \int \prod_{i=1}^n d y_i \exp{\left[\frac{i}{\hbar} \sum_{i=0}^n \epsilon \left( \left(\frac{y_{i+1}\mi y_i}{\epsilon}\right)^2-\omega^2 y_i^2\right)\right]}, \quad \quad y_{n+1}=y_0 = 0.
 \]
        We can write the exponent in terms of a matrix product:
  \[ 
        \exp{\left[\frac{i}{\hbar} \sum_{i=0}^n \epsilon \left( \left(\frac{y_{i+1}\mi y_i}{\epsilon}\right)^2-\omega^2 y_i^2\right)\right]} =        \exp{\left[\frac{i}{\hbar \epsilon} y^T A_{n\times n} y\right]}.
 \]
 \begin{itemize}
 
 \item[4.]
        Show that the matrix $A_{n\times n}$ can chosen as the following symmetric matrix
        \[ A_{n\times n} = \left\{
            \begin{array}{ccccc}
                2- \omega^2\epsilon^2 & -1 & 0 & \cdots & 0 \\
                -1 & 2-\omega^2\epsilon^2 & -1 & & 0 \\
                0 & -1 &\ddots & \ddots &\vdots \\
                \vdots & & \ddots & & -1 \\
                0 & 0 & \dots &-1 & 2-\omega^2 \epsilon^2
            \end{array}
            \right\}
        \]

    \item[5.] Let $ D_n \equiv  \det A_{n\times n}$. Prove that 
   \beq
            D_n = \left(2-\epsilon^2 \omega^2\right)D_{n-1} - D_{n-2},  \qquad D_0 = 1, ~~ D_{-1} = 0
            \label{eq:recur} \\ 
   \eeq     
        
        \end{itemize}

 \subsubsection*{Generating functions} 
        
        We will use the method of generating functions to solve this recursion relation for the determinant. Let $a_n$ be a sequence of numbers. We call 
        \[ f(x) = \sum_n a_n x^n \]
        the \emph{generating function} for that sequence.

        In combinatorics and probability theory, generating functions are very useful since a given problem is often formulated in a less restrictive way using the generating function, and consequently easier solved. Once $f(x)$ is known, we can recover the coefficients $a_n$ straightforwardly as $a_n = \left.\frac{1}{n!} \left(\frac{d}{dx}\right)^n f(x) \right|_{x=0}$. We will use generating functions all the time! 

 For  $D_n$ we found the recursion relation \eqref{eq:recur}. Define the generating function
 $$
 D(x) = \sum_{n=0}^\infty D_n x^n .
 $$
 \begin{itemize}       
        \item[6.]  Show that it satisfies the equation
 \[ 
 D(x) =1+ \left(2-\omega^2\epsilon^2\right) x D(x)-x^2 D(x), \quad {\rm i.e.} \quad  D(x) = \frac{1}{1-(2-\omega^2\epsilon^2) x + x^2}.
  \]
\end{itemize}
  Now introduce a new variable $\tilde{\om}$ such that $\omega\epsilon/2 = \sin( \tilde{\om}\epsilon/2)$. 
  \begin{itemize}
  \item[7.]      
        Show that, in terms of this new variable, we have
   $$
            D(x) = \frac{1}{1-2 x\cos\left( \tilde{\om}\epsilon\right) +x^2} =
       \frac{1}{\e^{i\tilde{\om} \epsilon} \mi   \e^{-i\tilde{\om} \epsilon}}  \Big(\frac{1}{\left(e^{-i \tilde{\om} \epsilon}\mi x\right)} -
   \frac{1}{\left(e^{i \tilde{\om} \epsilon}\mi x\right)}\Big)
   $$
   \item[8.]   
        Next, show that
        \[D_n = \frac{\sin\left((n\plu 1)\epsilon\tilde{\om}\right)}{\sin \epsilon \tilde{\om}} \]
        \item[9.]
        Finally use this to show that
   $$
            \int_{\substack{y(t_a)=0 \\ y(t_b)=0}} \mathcal{D} y(t) e^{\frac{i}{\hbar} S\left[y(t)\right]} = 
            \lim_{\epsilon\to 0} \sqrt{\frac{m}{2\pi i \epsilon \hbar} 
            \;\frac{\sin \tilde{\om} \epsilon}{\sin \left((n\plu 1)\epsilon\tilde{\om}\right)}} 
            =\sqrt{  \frac{m\omega}{2\pi i \hbar\sin \left(\omega(t_b\mi t_a)\right)}}
    $$
        \end{itemize}

        To summarize:
 $$
        \boxed{    \langle x_b| \;\e^{-\frac{i}{\hbar} (t_b-t_a) \hat{H}} |x_a\rangle  =
         \sqrt{\frac{m\omega }{2\pi i \hbar \sin \omega(t_b\mi t_a)}} \;\e^{\frac{i}{\hbar} S\left[x_c\right]} }
 $$    
 $$       
   \e^{\frac{i}{\hbar} S\left[x_c\right]} = 
   \e^{\frac{i}{\hbar} \frac{m \omega}{\sin\omega(t_b-t_a)} \left(\left(x_b^2+x_a^2\right)\cos\omega(t_b-t_a)-2x_ax_b\right)}.
  $$

    Now recall that
        \[ Z = \tr \;\e^{-\beta \hat{H}} = \int dx\; \langle {x}|\; \e^{-\beta\hat{H}}|{x}\rangle , \quad \quad \beta = \frac{1}{k_B T}.\]
   Define 
   $$
   z(x) :=     \langle {x}|\; \e^{-\beta\hat{H}}|{x}\rangle,\qquad \rho(x) := \frac{z(x)}{Z} 
    $$ 
\begin{itemize}
\item[10.]
        Show from the previously found result that
    $$
    z(x)  = 
        \sqrt{\frac{m \omega}{2 \pi \hbar \sin \hbar \beta \omega}} \;\e^{-\frac{m \omega}{\hbar} \tanh \frac{\beta \hbar \omega}{2} x^2}
 \quad {\rm and} \quad   
      Z = \int dx\; z(x) = \frac{1}{2\sinh \frac{\beta \hbar \omega}{2}}.
$$

\item[11.] Let $\psi_0(x)$ denote  the ground state wave function of the harmonic oscillator.
     Show that 
   $$
   \rho(x) \to \left|\psi_0^2(x)\right|, \quad {\rm for} \quad  T \to 0 ~({\rm i.e.}~\beta \to \infty)
    $$
    and explain why we can obtain this result without any detailed calculation.

\end{itemize}

\newpage

\setcounter{equation}{0}
\setcounter{figure}{0}
\renewcommand{\thefigure}{ps-3.\arabic{figure}}
 \renewcommand{\theequation}{ps-3.\arabic{equation}}

\subsection*{Elementary Quantum Geometry Problem Set 3}

In this exercise, we discuss the lattice propagator represented as a random walk (RW) on the lattice. 
\subsubsection*{The lattice propagator}
The continuum Laplace operator takes the form
\[ \Delta f(x) = \frac{\partial^2}{\partial x^2} f(x). \]
Now we replace the continuum by a $D$-dimensional hypercubic lattice, $(a \mathbb{Z})^D$. The lattice sites are given by
\[ x_i(\vec{n}) \equ a \cdot n_i, \quad \vec{n} = (n_1, \cdots, n_D). \]
Functions on the lattice are functions of the sites, i.e. $f(x_i(n))$. The lattice Laplacian then takes the form
\[ 
\left(\Delta_L f\right)\left(x_i(n)\right) = 
\sum_{j=1}^D \frac{f\left(x_i(n)\plu e_j a \right)\plu f\left(x_i(n)\mi e_j a\right) \mi 2 f\left(x_i(n)\right)}{a^2}.
 \]
Here $e_j$ is a unit vector in the $\hat{j}$ direction.
    We will first derive the lattice propagator in momentum space. The Fourier transform $\hat{F}(p)$ of a function $F(x)$ on the lattice is written
        \[ \hat{F}(p) = \sum_{x_n} a^D \cdot \e^{i p \cdot x_n} F(x_n). \]
        \begin{enumerate}
            \item Prove that
                \[ \hat{F}(p_i) = \hat{F}\left(p_i\plu \frac{2\pi}{a}e_i\right). \]
                Thus, $\hf$ is periodic with period $\frac{2\pi}{a}$. Therefore, we assume from now on that $p_i \in \left[ -\frac{\pi}{a}, \frac{\pi}{a}\right]$. This is called the (first) \emph{Brillouin zone}.

            \item Now ``prove"  the \emph{inversion formula} (use basics from Fourier series):
                \[ F(x_n) = \int_{-\frac{\pi}{a}}^{\frac{\pi}{a}} \frac{d^Dp}{(2\pi)^D}\, \e^{-i p x_n} \hf(p). \]
\end{enumerate}                

\noindent                In order to obtain the lattice propagator, we want to solve
                \begin{align}
                    \left(-\Delta_L(x_n)\plu m^2 \right)G(x_n\mi x_m) = \delta(x_n\mi x_m). 
                    \label{eq:greens}
                \end{align}
                The lattice delta function is defined as
                \[ \delta(x_n\mi x_m) \equiv \frac{1}{a^D} \delta_{nm} \quad {\rm thus:}
                \quad  \sum_{x_n} a^D \delta(x_n\mi x_m) f(x_n) = f(x_m), \]
                where the sum is over all lattice sites. Note that this precisely corresponds to the ``shifting'' property of the continuum Dirac delta function.

\begin{enumerate}
\setcounter{enumi}{2}
            \item Now transform equation \eqref{eq:greens} to momentum space, and show that it results in 
                \[
                \left[\sum_{i=1}^D \frac{2}{a^2} \left(1\mi \cos a p_i\right) \plu  m^2\right] G(p) = 1. \]
            \item Show that this is equivalent to 
                \[ G(p) = \frac{a^2}{4 \sum_{i=1}^D \sin^2 \frac{a p_i}{2} \plu m^2 a^2 }. \]
            \item Subsequently show that in the limit $a \to 0$ this reduces to
                \[ G(p) = \frac{1}{p^2 \plu m^2}, \]
                consistent with the analogous result in continuum field theory.
        \end{enumerate}

        \subsubsection*{Calculation of the lattice propagator in $x_n$-space}
        
We write $\Delta_L$ in matrix form $\left(\Delta_L\right)_{nm}$ as follows:
        \[ \left(\Delta_L f\right)(x_n) = \sum_m \left(\Delta_L\right)_{nm} f(x_m). \]
        \begin{enumerate}
 \setcounter{enumi}{5}      
            \item Show that
                \[ - \left(\Delta_L\right)_{nm} = a^{-2} \left[ 2 D \,\delta_{nm} \mi  Q_{nm} \right], \]
                where $Q_{nm} \equ 1$ if $n,m$ are neighbouring lattice sites, and $Q_{nm} \equ 0$ otherwise.

            \item Now show that $\left( \left(-\Delta_L\right) \plu m^2  \right)$
                is invertible for $m^2 > 0$ (e.g.\ by using the Fourier transformed operator), 
                and that  $\left(-\Delta_L\plu m^2\right)^{-1}$ allows an expansion in   the Neumann series
                \[ \frac{a^{2-D}}{2D\plu m^2 a^2} \sum_{k=0}^\infty \left(\frac{Q}{2D\plu m^2 a^2} \right)^k. \]
                Next, show that
                \[ \left(-\Delta_L\plu m^2\right)_{nm}^{-1} = 
                \frac{a^{2-D}}{2D\plu m^2a^2} \sum_{P(x_n,x_m)} \left(2D\plu m^2 a^2\right)^{-L(P)/a}, \]
                where $P(x_n,x_m)$ is a lattice path from $x_m$ to $x_n$ and $L(P)$ is the length of this path. Such a lattice path follows the links of the lattice, and the path length is equal to its number of links times the link length $a$.

\end{enumerate}

 \noindent               We now consider the path integral for the free relativistic particle and we use the classical action
                \[ S(P) = m_0 L(P). \]
                We want to calculate
                \[ G(x,y) = \int \mathcal{D} P(x,y) e^{-S[P(x,y)]}\]
                and we provide a regularization by restricting the paths to a hypercubic lattice:
                \[ G_a(x_n,x_m) = \sum_{P(x_n, x_m)} e^{-m_0(a) L[P(x_n,x_m)]}. \]
                As a side note: by this definition, $G_a(x_n, x_m)$ is dimensionless, contrary to the ``real'' $G(x, y)$.

\begin{enumerate}
 \setcounter{enumi}{7}   

            \item Show that we can choose $m_0(a)$ as a function of the lattice spacing $a$, such that
                \[ \frac{a^{2-D} }{2D \plu  m^2a^2}\, G_a(x_n, x_m) =  \left(-\Delta_L\plu m^2\right)_{nm}^{-1}. \]
                i.e.\ the (dimensionless) path integral Green function $G_a$ goes to the continuum Green function
                $G_{cont}$ as follows for $a \to 0$
                $$ 
                \frac{a^{2-D} }{2D}\, G_a(x_m,x_n) \to G_{cont} (x_m,x_n)
                $$

            \item Then compare $m_0(a)$ with the form of $m_0(a)$ for the free particle regularized by piecewise linear paths constructed from building blocks of length $a$ as in the notes:
                \[ m_0(a) = \frac{\log f(0)}{a} \plu  c^2 a m^2 \plu O(a^3). \]
                Here on the lattice we find 
                \[ m_0(a) = \frac{\log 2D}{a} \plu  \frac{1}{2D} \,a m^2 \plu  O(a^3)\]
            \item Give a simple interpretation of $\log 2D$ in terms of the number of paths on the lattice.
        \end{enumerate}

\newpage

\setcounter{equation}{0}

\setcounter{equation}{0}
\setcounter{figure}{0}
\renewcommand{\thefigure}{ps-4.\arabic{figure}}
 \renewcommand{\theequation}{ps-4.\arabic{equation}}

\subsection*{ \hspace{-3mm} Elementary Quantum Geometry Problem Set 4}

This exercise centers around mean-field critical exponents for the simplest ferromagnetic model of classical spins. 

Consider a hypercubic lattice in $\mathbb{R}^D$. We set the link length $a=1$ (it can always straightforwardly be re-introduced later on if necessary). The Hamiltonian is written
\begin{align}
    \label{eq:ham}
    H = -J  \sum_{\brak{ij}} S_i S_j - h \sum_i S_i.
\end{align}
The sum over $\brak{ij}$ indicates that we sum over all pairs of lattice sites $i,j$ (all links of the lattice). For a 
hypercubic lattice in $D$ dimensions the number of links is $D$ times the number of vertices. 
The external magnetic field strength is given by $h$ and $S_i \in \mathbb{R}$ is the spin at lattice site $i$.

The partition function is given by 
\begin{align}
    \label{eq:part}
    Z(\beta;h;J) = \int \prod_i \left(d S_i \rho(S_i, \beta)\right) e^{-\beta H(S_i)}.
\end{align}
The function $\rho(S, \beta)$ describes the spin properties of the individual lattice sites (or ``atoms''). We will assume
\begin{align}
    \label{eq:rho}
    \rho(S,\beta) = e^{-\kappa(\beta)S^2 -\lambda(\beta) S^4}, \quad \quad \kappa(\beta) > 0.
\end{align}
Here $\kappa(\beta)$ and $\lambda(\beta)$ are ``material'' constants that have only a weak $\beta$-dependence, i.e. $\kappa(\beta) = \kappa_0,\, \lambda(\beta) = \lambda_0$ as a first approximation (which we will use).

We thus have the ``single atom'' partition function 
\begin{align}
    \label{eq:atompart}
    Z_{s.a.}(\beta, h) = \int dS \;e^{-\kappa_0 S^2 - \lambda_0 S^4+\beta h S}.
\end{align}
For the expectation value of the spin we have for this "single atom" partition function
\begin{equation}
    \label{eq:spin-ex}
    \left\langle S(h)\right\rangle_{s.a.} = \frac{1}{Z_{s.a.}(\beta,h)} \int dS \;S \;e^{-\kappa_0 S^2-\lambda_0 S^4} 
    e^{\beta h S} 
\end{equation}
$$    
    \left\langle S(h \equ 0) \right\rangle_{s.a.} = 0.
$$

We will now use $Z(\beta)$ in an approximation where we write
\begin{align}
    S_i = \left\langle S(h) \right\rangle + \delta S_i, \quad \quad \left\langle S_i \right\rangle = \left\langle S(h)\right\rangle 
\end{align}
and assume that terms of order $(\delta S_i)^3$ can be ignored.

{\it  (1) Show that in this approximation we can write} 
        \begin{align}
            \label{eq:zbh}
            Z(\beta,h) = \left(\prod_{i=1}^V Z_{\textrm{mf}}(\beta,h)\right) \times \int 
            \left(\prod_i d(\delta S_i)\right)\; e^{-\beta H_F(\delta S_i)}. 
        \end{align}

 \noindent
  In (\ref{eq:zbh}) $V$ denotes the ``volume'' of $\mathbb{R}^D$, i.e. the number of lattice sites. Furthermore, $Z_{\textrm{mf}}$ 
  is the  ``mean field'' partition function per ``atom site'' $i$:
        \begin{align}
            Z_{\textrm{mf}}(\beta,h) &= e^{-\beta f_{\textrm{mf}}(\beta,h)} \\
            \beta f_\mf &= (\kappa_0 - D J \beta) \left\langle S(h)\right\rangle^2 + \lambda_0 \left\langle S(h)\right\rangle^4 -\beta h \brak{S(h)}\label{p4jx1} \\
            \beta H_F(\delta S_i) &= \frac{1}{2} \beta J \sum_{\brak{ij}} \left(\delta S_i-\delta S_j\right)^2 + 
            \left(\kappa_0-D J \beta+ 6 \lambda_0 \brak{S(h)}^2\right) \sum_i (\delta S_i)^2  \label{eq:bhf}
        \end{align}
         Finally , $\brak{S(h)}$ is determined by the equation
        \begin{align}
            \label{eq:sh}
            2(\kappa_0 - D J \beta) \brak{S(h)} + 4 \lambda_0 \brak{S(h)}^3 = \beta h. 
        \end{align}
        This is the condition that ensures that $\brak{\delta S_i} = 0$ when we only keep terms up to quadratic order in $(\delta S_i)$.

{\it (2) Prove that }
  $$
            \frac{d f_\mf}{d h} = - \brak{S(h)}
  $$

 This is the standard result for the free energy density. Why is it called the ``mean field approximation''?
To see this, consider the partition function for this system with interactions turned off, i.e. 
        \[ Z_{J=0}(\beta, h) = \prod_{i=1}^V Z_{s.a.}(\beta,h). \]
        where 
        \begin{align}
            Z_{s.a.}(\beta,h) &= e^{-\beta f_{\textrm{free}}(h)} \label{eq:zsa} \\
            \beta f_{\textrm{free}}(h) &= \knu S^2 + \lnu S^4 - \beta h S \label{p4jx2}
        \end{align}
        If we then ignore the fluctuations, i.e.\ put $\delta S =0$, we see that the only difference 
        between $Z(\beta,h,J)$ and $Z_{J=0}(\beta,h)$ is the shift
        \[ \knu \to \knu - D \beta J. \]
   This can be understood in the following way. Let us assume $S_i = \brak{S(h)}$. Then we can write
        \[ H = -J \sum_{\brak{ij}} S_i S_j - h \sum_i S_i =- (h +DJ\brak{S(h)}) \sum_i S_i \]
        Thus each spin feels not only the external field $h$, but also the local field from the neighbours,
        and we can formally write $H$ as sum of single spins interacting with an effective magnetic field $h_{\rm eff}$: 
        \[ h \to h_{\rm eff} = h+   D J \brak{S(h)}. \]

 {\it (3)  Check the consistency of this picture by showing that  $f_{\textrm{free}}(h_{\rm eff},S)$ in eq.\ (\ref{p4jx2}) agrees
 with $ f_\mf(h,\brak{S(h)})$ in eq.\ (\ref{p4jx1}) when we identify $S = \brak{S(h)}$}\\
  
 {\it (4) Assume now that $h=0$ in \eqref{eq:sh}. For given $J$ draw $\brak{S(h\equ 0)}_\beta$ as a function of $\beta$.} \\

Spontaneous magnetization starts at 
  $$
  \beta_c = \frac{\kappa_0}{D J},\quad \quad T_c = \frac{1}{k_b \beta_c}.
  $$
 Define the critical exponent for magnetization by
   $$
   \brak{S(h\equ 0)}_T \sim (T_c-T)^{\beta}, \quad \quad T \approx T_c,\quad T < T_c. 
   $$
        Note that the exponent $\beta$ is \emph{not} the inverse of $T$ here.

 {\it (5) Now convince yourself that $\beta = \frac{1}{2}$.}

 The susceptibility is defined as 
        \[ \chi(T) = \left.\frac{d \brak{S(h)}}{d h}\right|_{h=0}. \]

 {\it (6) Use \eqref{eq:sh} to show that 
        \[ \frac{\partial \brak{S(h)}}{\partial h} = \half \frac{\beta}{(\knu-DJ\beta)+6\lambda \brak{S(h)}^2}, \]
        and thus:
   $$
            \chi(T) = \half \frac{\beta}{\knu-DJ\beta} \quad T > T_c , \qquad\quad
            \chi(T) = \frac{1}{4} \frac{\beta}{DJ\beta - \knu} \quad T < T_c.
   $$
}

 The critical exponent of susceptibility is defined as 
  $$
  \chi(T) \to \frac{c}{\left|T-T_c\right|^\gamma} \quad \textrm{for } T \to T_c. 
  $$

 {\it (7)  Convince yourself that $\gamma = 1$ in the mean-field approximation.}
 
 \vspace{10pt}

 Finally consider the spin-spin correlation function:
        \[ \brak{(S_i-\brak{S_i})(S_j-\brak{S_j})} = \brak{\delta S_i \delta S_j} \]
        for $h=0$.
        The correlation length $\xi(T)$ is defined as the exponential fall-off of $\brak{\delta S_i \delta S_j}$, i.e. by
        \[ \xi (T) =- \frac{\log (\brak{\delta S_i \delta S_j})}{|i-j|}\quad {\rm for}\quad |i-j| \to \infty \]
       From \eqref{eq:bhf} we know:
        \[ \brak{(\delta S)_i (\delta S)_j}  = \frac{\int \prod_k d(\delta S_k) \;\delta S_i \delta S_j \;e^{-\beta H_F(\delta S_k)}}{\int \prod_k d(\delta S_k)\; e^{-\beta H_F(\delta S_k)}}, \]
        where $H_F(\delta S)$ is quadratic in $\delta S$. Recall from Gaussian integration:
        \[ \frac{\int \prod_k d x_k \;e^{-\half x_i A_{ij} x_j + J_i x_i}}{\int \prod_k dx_k \;e^{-\half x_i A_{ij} x_j}} = e^{\half J_i A_{ij}^{-1} J_j}. \]
        Furthermore, by definition
        \[ \brak{x_i x_j} = \frac{\int \prod_k dx_k 
        \; (x_i x_j ) \; e^{-\half x_l A_{lm} x_m }}{\int \prod_k dx_k \;e^{-\half x_l A_{lm} x_m}}. \]
     
 {\it  (8)     Use the last two equations to  prove that }
        \[ \brak{x_i x_j} = \left(A^{-1}\right)_{ij}. \]

 {\it (9) Let  $\Delta_L$ denote the lattice Laplacian. Use \eqref{eq:bhf} to show that}
        \bea
            \brak{\delta S_i \delta S_j} &=& \left(A^{-1}\right)_{ij}, \quad
            A_{ij} = \beta J \left[ \Delta_L + m^2 \right]_{ij}, \nonumber \\
            m^2 &=& \frac{1}{\beta J} \left[ \knu - D J \beta + 6 \lnu \brak{S(0)}^2\right] \label{p4-mass}
        \eea

We know  the long distance behavior of  $[ \Delta_L + m^2 ]_{ij}^{-1}$. It is 
   $$    
    -\frac{\log \left[ \Delta_L + m^2 \right]_{ij}^{-1}}{ | i-j |} = m   \quad {\rm for}\quad |i-j| \to \infty               
$$

{\it (10) Show that} 
        \[ m(T) \propto  \sqrt{|T-T_c|} \quad \textrm{for } T \to T_c. \]

        For a spin system
        \[ \brak{\delta S_i \delta S_j} \sim e^{-\frac{|i-j|}{\xi(T)}} \]
        defines the correlation length $\xi(T)$. Close to the phase transition it might have a non-analytic behavior:
        \[ \xi(T) \sim \frac{1}{|T-T_c|^\nu}, \]
        signifying long-range correlations.

 {\it  (11)  Convince yourself that $\nu = \half$.}

Finally we know 
        \[ \left(\Delta_L + m^2 \right)_{ij}^{-1} \sim \frac{1}{|i-j|^{D-2}}.   \quad {\rm for} \quad
        1 \ll |i-j| \ll \xi(T) .\]
For our spin system one defines the \emph{anomalous scaling exponent} $\eta$:
        \[ \brak{\delta S_i \delta S_j} \sim \frac{1}{|i-j|^{D-2+\eta}}, \quad 1 \ll |i-j| \ll \xi(T). \]
        Thus, $\eta = 0$ in mean-field theory.

\newpage

\setcounter{equation}{0}

\setcounter{equation}{0}
\setcounter{figure}{0}
\renewcommand{\thefigure}{ps-5.\arabic{figure}}
 \renewcommand{\theequation}{ps-5.\arabic{equation}}

\subsection*{Elementary Quantum Geometry Problem Set 5}

\subsubsection*{Rooted planar trees}
Recall first:
\begin{align}
    z(\mu) &= \sum_{\bp} e^{-\mu |\bp|} \prod_{v \in \bp} w_v \\
    &= \sum_L e^{-\mu L} \sum_{\left\{\bp : |\bp | = L\right\}} \prod_{v \in \bp} w_v
\end{align}
Write 
$$
g = e^\mu, \quad f(z) = \sum_{n=2}^\infty w_n z^{n-1}, \quad w_1 = 1
$$
and $w_v \equiv w_{n(v)}$, where $n(v)$ is the order of the vertex $v$.

Furthermore, let us write
$$
{\cal N } (L) :=  \sum_{\left\{ \bp: |\bp| = L\right\}} \prod_{v \in \bp} w_v = g_c^L \cdot h(L), \quad {\rm i.e.} \quad
z(g) = \sum_L  h(L) \; \left(\frac{g_c}{g}\right)^L.
 $$
where for $L$ large we have $\frac{\log h(L)}{L} \to 0$. {\it We call $\mathcal{N}(L) = g_c^L h(L)$ the number of branched polymers with weight $w_{n(v)}$ and length $L$}. We have
\[ g = \frac{1+f(z)}{z} \]
and $g_c \equiv e^{\mu_c}$ is determined by 
$$
\left.\frac{dg}{dz}\right|_{z_c} = 0. 
$$
\begin{enumerate}

\item Show that for $0 < \gamma < 1$ we have 
        $$
       \boxed{ h(L) \propto L^{\gamma -2}\Big(1 + \cdots\Big) ~~  \Longleftrightarrow ~~  z(g)\mi z(g_c)  
       \propto \left(1 \mi  \frac{g_c}{g}\right)^{1-\gamma} + \cdots }
        $$

    \item Assume all $w_n = 1$ for $n = 2,3, \cdots$, i.e. all branchings are allowed and have the same weight. Show that
        $$
        g_c = 4 \quad {\rm and} \quad \mathcal{N}(L) \to  4^L L^{-\frac{3}{2}}\quad {\rm for} \quad L \to \infty 
        $$
     \item

        Show that the explicit expression for $z(g)$ in this case is 
        $$
        z(g) = \frac{1 - \sqrt{ 1- \frac{4}{g}}}{2}
        $$
        This ensemble of trees is called the ensemble of {\it uniform random rooted trees}.
        
    \item Assume $w_3=1$ and all other $w_n = 0$. Show that 
    $$ 
    g_c =2 \quad {\rm and} \quad  \mathcal{N}(L) \to  2^L L^{-\frac{3}{2}} \quad {\rm for} \quad L \to \infty.
    $$
    \item Show that  for $w_n = 1, w_k = 0, k \neq n, n > 2$ we have 
      $$
      g_c(n) = (n-2)^{\frac{1}{n-1}} + (n-2)^{-\frac{n-2}{n-1}}, \qquad \mathcal{N}(L) \to g_c^L(n) L^{-\frac{3}{2}} \quad {\rm for } \quad
      L\to \infty
      $$
        Why does $g_c(n) \to 1$ for $n \to \infty$?
    \item Discuss the case $w_2 = 1, w_n =0, n > 2$. (Solve for $z$)

\end{enumerate}

{\it Note that in all the cases we have discussed so far we have $\gamma = \frac{1}{2}$. } (except for the case discussed 
in question 6, which was not really a BP). We will now discuss when $\gamma \equ 1/2$ and how to obtain 
BPs with $\gamma \neq 1/2$.

\subsubsection*{Criticality of branched polymers}

The basic equation is
\[ e^\mu = \frac{1+f(z)}{z}, \quad f(z) = \sum_{m=2}^\infty w_m z^{m-1} \]
 Let us assume $w_2 = 0$ and denote $e^\mu = g$. Also, let us assume that $f(z)$ is a polynomial of order $n$. 
\begin{enumerate}
\setcounter{enumi}{6}

    \item Show from the very definition of $z$ in terms of $\mu$ that for $g \to \infty$ we have $z \to 0$.
 \end{enumerate}   

        For decreasing $g$, the value of $z$ will increase. Criticality is encountered at the \emph{first} extremum of $g(z)$ for increasing $z$. Let this point be $z_c$. By assumption: $w_{n+1} \neq 0$ and $ w_{n+2}, w_{n+3}, \cdots = 0$. Assume that 
        \[ g'(z_c) = \cdots = g^{(n-1)}(z_c) = 0, \quad g^{(n)}(z_c) \neq 0 \]
        
\begin{enumerate}\setcounter{enumi}{7}
        
    \item Show that 
    \beq
            \label{eq:gz}
        g(z) \mi g(z_c) = \frac{\Big(1\mi \frac{z}{z_c}\Big)^n}{z}, \quad {\rm and~that}~
       ( w_2 \equ 0) \quad g(z_c) = \frac{n}{z_c} 
        \eeq
        Hint: Taylor expand and use that $z g(z)$ is a polynomial of order $n$.

    \item Find the explicit branching weights $w_m$ corresponding to this function. 
    Note that they alternate in sign starting out with $w_3 > 0$.

            \item Define $g_c = g(z_c)$.
Show that eq.\ \rf{eq:gz} leads to 
        $$
            z(g) = z_c-z_c^{1+\frac{1}{n}} \left(g\mi g_c\right)^{\frac{1}{n}} \plu  O\left((g\mi g_c)^{\frac{2}{n}}\right) 
     \quad {\rm  i.e.} \quad  \gamma = 1\mi \frac{1}{n}.
     $$

    \item Show that if all weights $w_3, \cdots, w_{n+1}$ are positive then the \emph{only} critical behavior is 
      $$
      z(g) = z_c - \kappa (g\mi g_c)^{1/2} + O(g\mi g_c) \quad 
       {\rm  i.e.} \quad  \gamma = 1/2
       $$
\end{enumerate}

        Let us now consider the situation where we allow arbitrarily high branching, 
        i.e.\ $w_m$ can be different from zero for arbitrarily high $m$.
        
\begin{enumerate}\setcounter{enumi}{11}
     
    \item Assume that $f(x)$ is at least two times differentiable, that $f(0),f'(0) \geq 0$, that $f''(x) >0$ and that 
    $f''(x) > 1/x^2$ for large $x$.  Show that  $\gamma = 1/2$ for such an $f(x)$.

    \item  Some examples of such functions: $f(x) = e^x-1-x$, $f(x) = x^2 \tanh x$. 
Find the corresponding $w_m$ for these functions. Note that in the second example 
we have an oscillating sign of $w_m$, but nevertheless $\gamma = 1/2$.

 \end{enumerate}

        We want to generalize \eqref{eq:gz} in a non-trivial way:
        \begin{align}
            g(z)-g(z_0) = \frac{\left(1-\frac{z}{z_0}\right)^s}{z}, \quad z < z_0, \quad n-1 < s < n, 
        \end{align}
 where we assume $s> 1$ and the weight $w_2=0$, i.e. $g(z_0) \equ \displaystyle{\frac{s}{z_0}}$ (like in (\ref{eq:gz}))  
        
 \begin{enumerate}\setcounter{enumi}{13}       
    \item Show that  weights for $m> 2$ are:
        \begin{align}
            w_m &=   \frac{1}{z_0^{m-1}} (-1)^m \frac{\Gamma(s+1)}{\Gamma(s-m+2)\Gamma(m)}\\
            &= \frac{1}{z_0^{m-1}} \cdot \frac{\Gamma(m-1-s)}{\Gamma(-s)\Gamma(m)} \underset{m \to \infty}{\propto} \frac{(-1)^n}{m^{s+1}} 
             \end{align}
    \item Show that the sign of $w_m$ is oscillating for $m < s+2$, like in the situation for integer values of $s$, 
    but is constant for $m > s+2$.

    \item Show that for $1 < s < 2$ all weights are positive.
        
    \item Show that for $s > 1$ we have
        \[ z-z_c \underset{g\to g_c}{\sim} \kappa (g-g_c)^{\frac{1}{s}} \left(1+O\left((g-g_c)^{\frac{1}{s}}\right)\right) \]
        Thus, $\gamma_s = 1-\frac{1}{s}$.

\end{enumerate}

        For $1 < s < 2$ we have an example of a situation where all weights are positive but $\gamma \neq 1/2$. It requires infinite branching \emph{and} that $w_n$ should not be suppressed too much. For example, a power law $\frac{1}{n^{1+s}}$ rather than $\frac{1}{n!}$, and only $1<s<2$.

\newpage

\setcounter{equation}{0}
\setcounter{figure}{0}
\renewcommand{\thefigure}{ps-6.\arabic{figure}}
 \renewcommand{\theequation}{ps-6.\arabic{equation}}

\subsection*{Elementary Quantum Geometry Problem Set 6}
\subsubsection*{Branched polymers with ``matter''}
Consider a regular lattice in two dimensions. On such a lattice one can put down ``dimers'' (rods), illustrated by wiggly lines. One can create a statistical model of these dimers on the lattice by associating with each dimer a fugacity $\xi$. We will be interested in so-called \emph{hard dimers}, where the dimers are not allowed to touch each other. The partition function for these hard dimers is then
\[ z(\xi) = \sum_{\{\hd\}} \xi^{\left|\hd\right|} \]
where the summation is over all possible ways one can put down the hard dimers on the lattice, and $\left|\hd\right|$ is the number of dimers in the particular dimer configuration.

\begin{figure}[!ht]
\centerline{   \includegraphics[height=5cm]{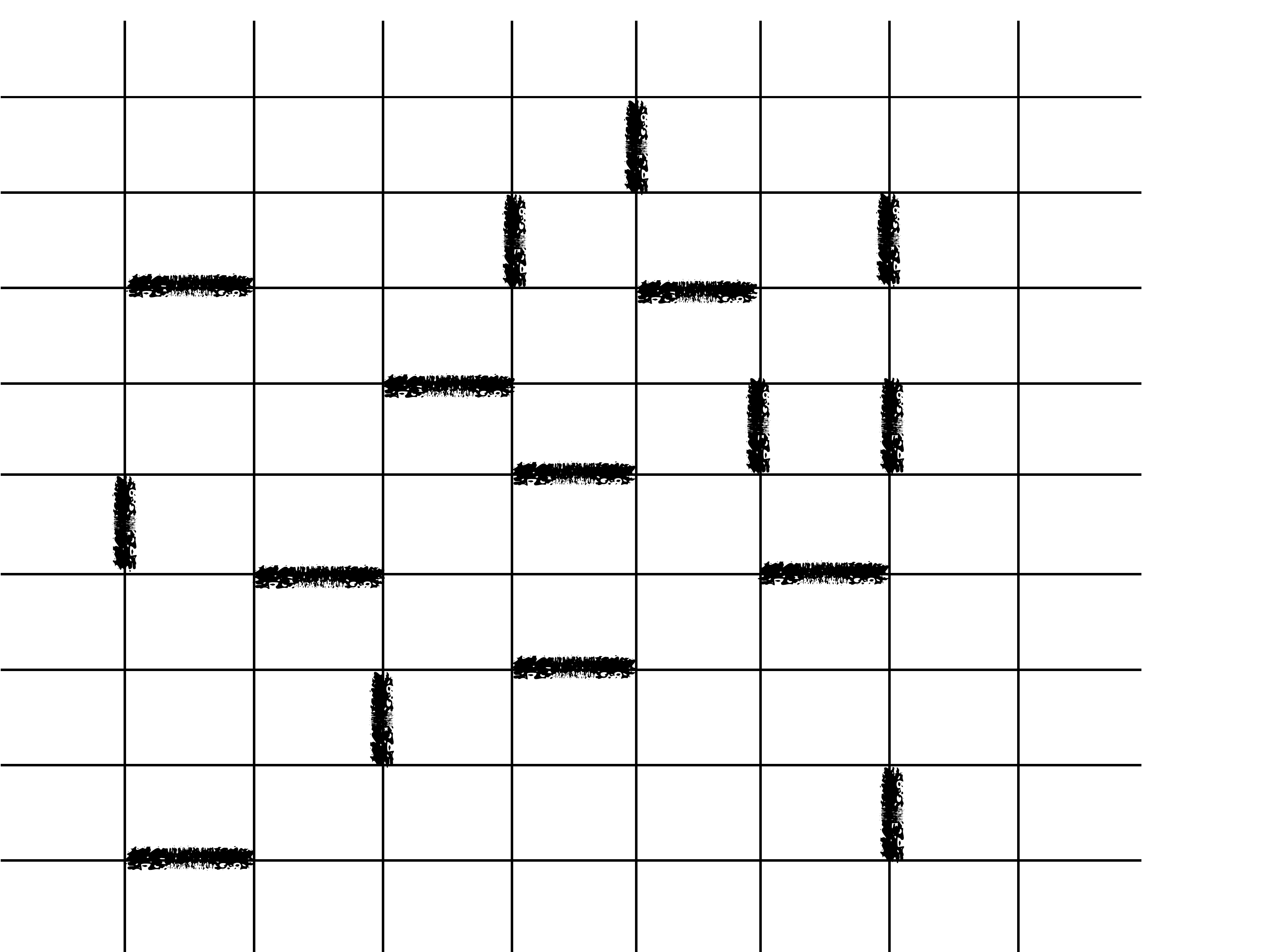}}
 \caption{{\small A regular square lattice with (hard) dimers on some of the links (the wiggly lines).}}
 \label{dimerlattice}
\end{figure}

For two-dimensional lattices these hard dimer models play an important role for exactly solvable lattice spin systems (related to the high-temperature expansion of the spin systems) and the interesting critical behavior of a dimer model is actually obtained for a somewhat ``unphysical'' \emph{negative} value of the fugacity.

Let $G$ be a connected graph. It is now clear how to define a hard dimer model on $G$. Let us now consider the ensemble of 
  planar trees or  BPs. 
On each of these trees we can put down dimers and we can consider the partition function of the combined system:
\[ z(\mu,\xi) = \sum_{\bp} \left(\prod_v w_v\right) \left(\prod_\rho e^{-\mu}\right)\sum_{\hd(\bp)}  \xi^{\left| \hd (\bp) \right|}\]
Here $i$ is a vertex in a BP and $v(i)$ the weight of that vertex. Furthermore, $\rho$ is a link in a BP and $e^{-\mu}$ the usual weight. On each BP we have a statistical system of HDs.

Such an average over both lattices and matter systems on these lattices is called an \emph{annealed average} (contrary to another kind of average: a \emph{quenched average} where we first calculate the free energy of the matter system (i.e. $\log Z$) on a lattice and \emph{then} average over lattices).

Here we consider the simplest BP system:
\begin{align*}
    w_v &= 1 \quad \textrm{if } \mathrm{ord}(v) = 1,3 \\ 
    w_v &= 0 \quad \textrm{if } \mathrm{ord}(v) \neq 1,3 
\end{align*}
where $\mathrm{ord}(i)$ is the order of the vertex $i$.

\begin{figure}[t]
\vspace{-1cm}
\centerline{\includegraphics[height=6cm]{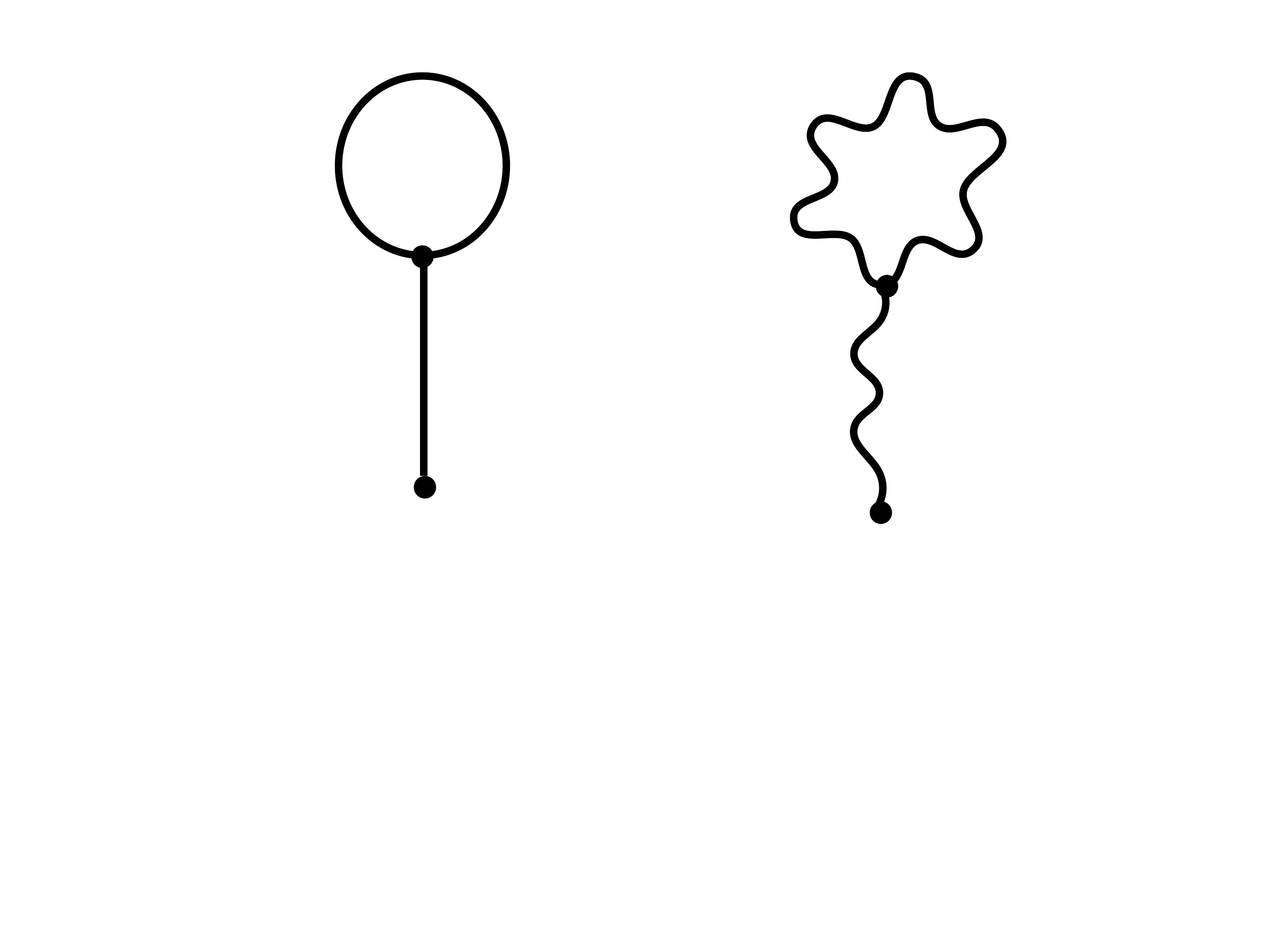} }
\vspace{-3cm}
\caption{{\small Graphical representation of $z(\mu,\xi)$ where no dimer touches the root (left figure) and 
$\tilde{z}(\mu,\xi)$ where a dimer {\it does} touch the root (right figure)}}
\label{figxx2}
\end{figure}

Let us now consider the equations of \emph{rooted} BPs. We have two situations: the link touching the root \emph{does not} have a dimer and the link touching the root \emph{does} have a dimer, indicated graphically in Fig.\ \ref{figxx2}.

\begin{enumerate}
    \item Convince yourself that the graphical representation shown in fig.\ \ref{dimerequations} is correct and leads to the following equations:
        \[ e^{\mu} = \frac{1+z^2+2z \tz}{z}, \quad e^{\mu} = \xi \frac{1+z^2}{\tz} \]
        Then write $g = e^\mu$ and

\begin{figure}[!ht]
 \vspace{-1cm}
 \centerline{ \includegraphics[height=8cm]{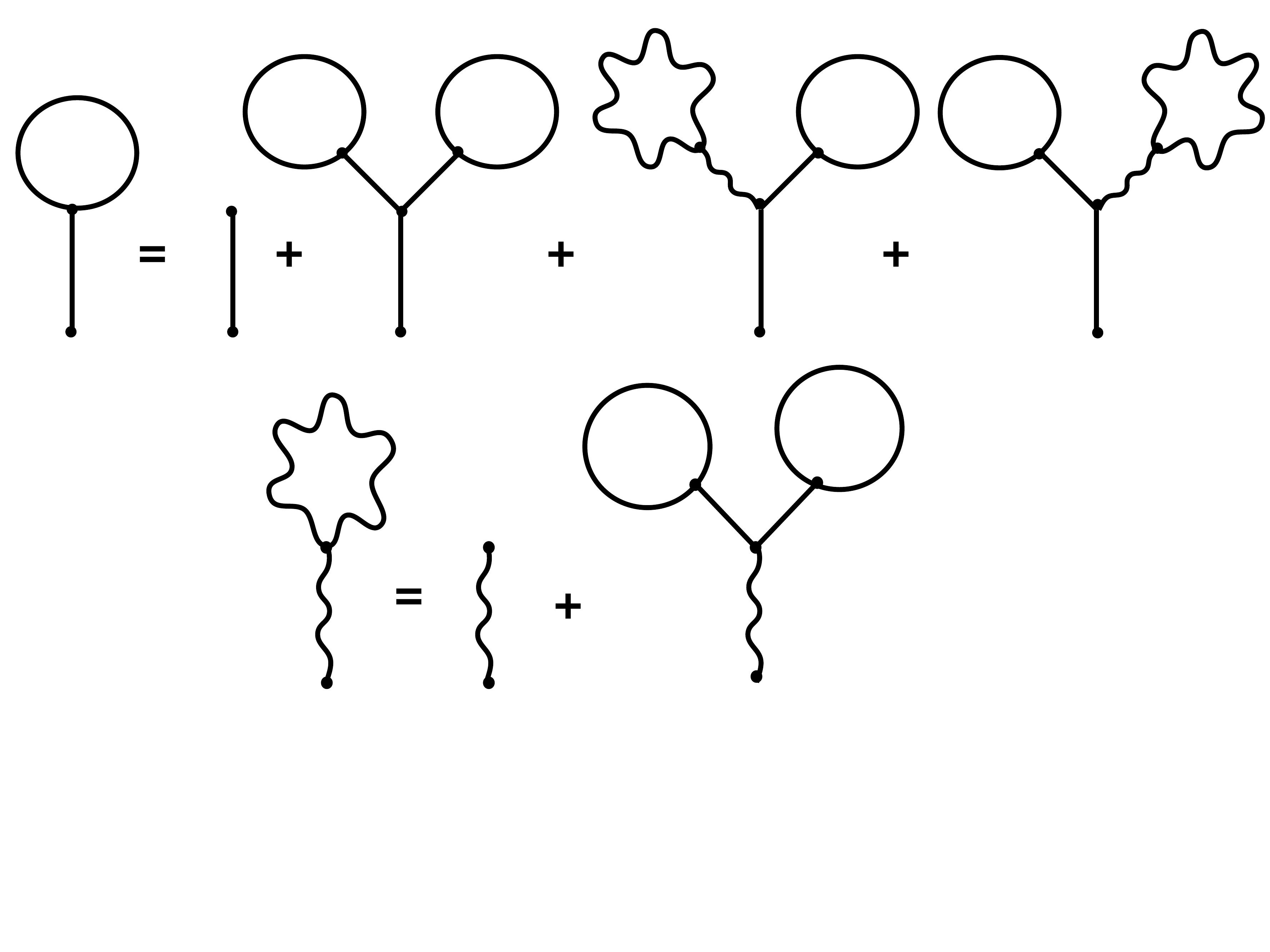}}
 \vspace{-2.3cm}
    \caption{{\small The graphical representation of the equations for rooted hard dimers.}}
    \label{dimerequations}
\end{figure}

    \item Show that
    \beq\label{p6j1}
        g = \frac{1+z^2}{z} + \frac{2\xi}{g}(1+z^2). 
     \eeq
        and that the solution is 
          \beq\label{p6j2}
          g(z,\xi) = \half \left[ \frac{1+z^2}{z} + \sqrt{\frac{(1+z^2)^2}{z^2} + 8 \xi (1+z^2)}\right] 
          \eeq
  The function is plotted in Fig.\ \ref{fig1z} for various $\xi$s, and the minima of the curves
  indicated by the dashed line.         

\begin{figure}[!ht]
\vspace{0.2cm}
\centerline{ \includegraphics[height=7cm]{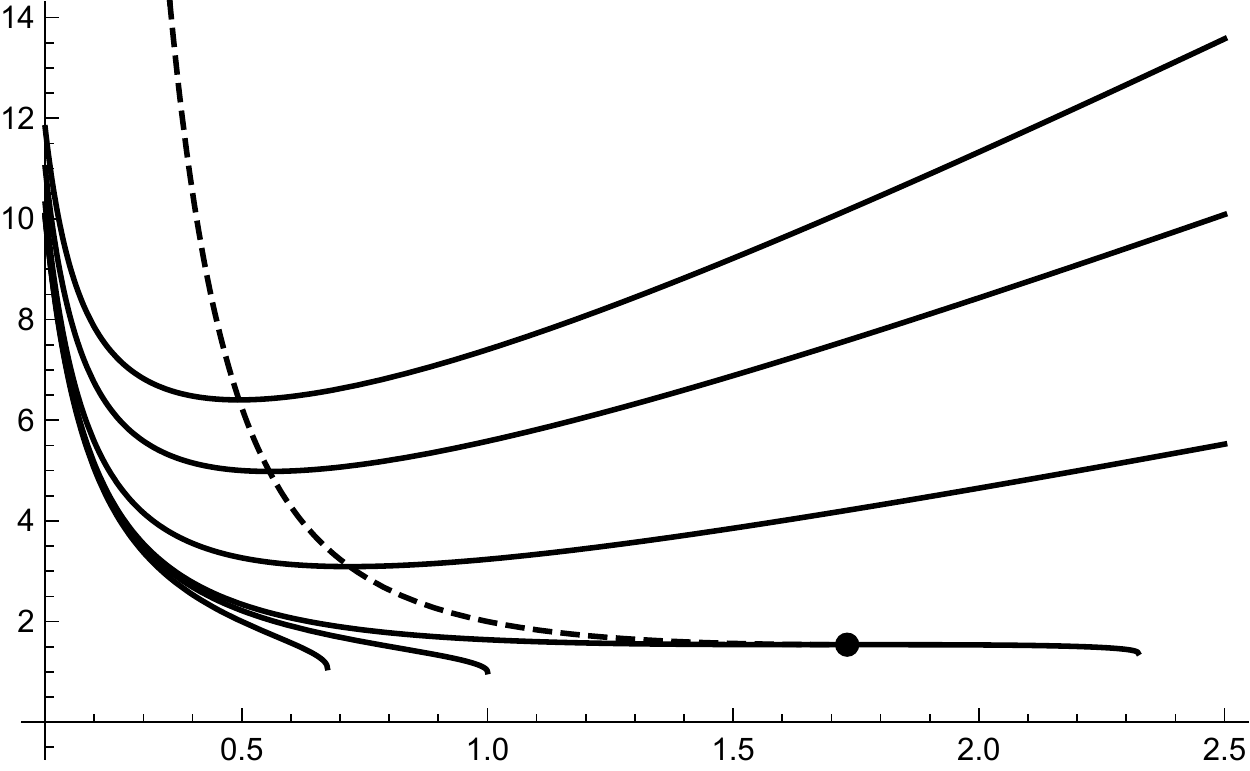}}
\caption{{\small The various curves $z \to g(z,\xi)$ for a number of values of $\xi$. For $\xi > \xi_c = -4/27$
the curves have a minimum which is also shown as the dashed curve. last curve having a minimum is 
the one with $\xi =-4/27$, and the endpoint is indicated by a dot.}}
  \label{fig1z}
  \end{figure}

\end{enumerate}

 \begin{enumerate}\setcounter{enumi}{2}
    \item Differentiate eq.\ \rf{p6j1}  with respect to $z$ (while keeping $\xi$ fixed) to find equations for
        \[ \frac{dg}{dz} = 0 \quad \textrm{and } \quad \frac{d^2g}{dz^2} = 0 \]

        There is only one $\xi$ (which we denote $\xi_c$) where both equations are satisfied.
    \item Find $\xi_c$, $g_c$, and $z_c$, and subsequently argue that $\left.\frac{d^3g}{dz^3}\right|_{z_c} \neq 0$.
\end{enumerate}

\begin{figure}[!ht]

\centerline{ \includegraphics[height=6cm]{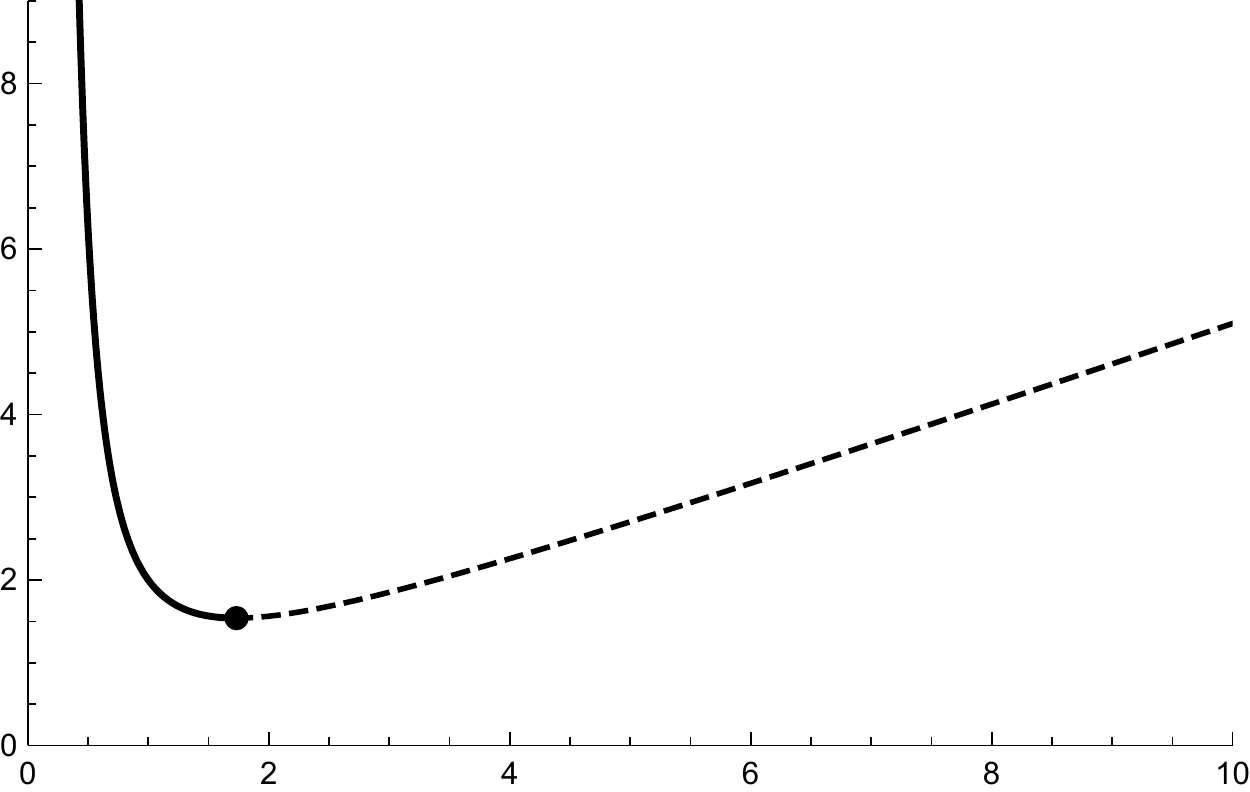}}
\caption{{\small The critical curve is shown in thick black in the figure and the end point with a dot. The endpoint is $(z_c,g_c)$.
The continuation of the curve is the same as in Fig.\ \ref{fig2h}.}}
            \label{fig3s}
        \end{figure}

For a given $\xi$ (larger than $\xi_c=-4/27$) we have a point $(z_k(\xi),g_k(\xi))$ where ${dg}/{dz} = 0$. It is
the minimum of the curve $z \to g(z,\xi)$ given by eq.\ \rf{p6j2}. We denote the curve $\xi \to  (z_k(\xi),g_k(\xi))$
{\it the critical curve}. This is where we can take the continuum limit for a given $\xi$. The critical curve is 
is shown on Fig. \ref{fig3s}. Expanding around the point $z_k(\xi)$ we have  ($\xi > \xi_c$ is kept fixed)
 \beq\label{p6j5}
 g(z,\xi)\mi g(z_k(\xi),\xi) \propto c_2(\xi) (z\mi z_k(\xi))^2 + O((z\mi z_k(\xi))^3) \quad 
{\rm i.e.} \quad  \gamma = \frac{1}{2}.
\eeq
However, for $\xi \to \xi_c$ we have $c_2(\xi) \to 0$ since $g''_{zz}(z_c,\xi_c)\equ 0$ and we obtain for $\xi=\xi_c$
\beq\label{p6j6}
g(z,\xi_c)\mi g(z_c(\xi_c),\xi_c) \propto (z\mi z_c(\xi_c))^3,\quad
       {\rm  i.e.}\quad  \gamma = 1\mi \frac{1}{3} = \frac{2}{3}.
\eeq       
 Thus, the multicritical behavior can actually be reproduced by having a matter system on BPs. The negative weight comes from the matter system.


 {\bf To summarize:} We have a curve of criticality (in this case the curve where $\frac{dg}{dz}=0$) as the 
 matter coupling constant $\xi$ varies. For all points on the curve we have the same 
 critical behavior, $\gamma \equ \frac{1}{2}$,  except at the endpoint of the critical curve, 
 where $\gamma \equ \frac{2}{3}$.
 This is typical in critical phenomena for statistical systems: one has a phase transition line where all 
 points on the line have the same critical behaviour except at  
 the endpoint, where the order of the transition, and therefore the corresponding  critical exponents,  can change.

\begin{enumerate}\setcounter{enumi}{4}
\item 
Show that the critical curve is given by 
\beq\label{p6j3}
g_k(z_k) = \frac{(1+z_k^2)^2}{2z_k^3}.
\eeq

\item 
Understand how the critical curve, shown on Fig.\ \ref{fig2h}, is related to the functions $z \to g(z,\xi)$ 
 given by eq.\ \rf{p6j2} (also shown on the figure), not only for $z < z_c$ (the {\it real} critical curve), but also 
 for $z > z_c$.
\begin{figure}[!ht]
\vspace{0.5cm}
\centerline{ \includegraphics[height=7.5cm]{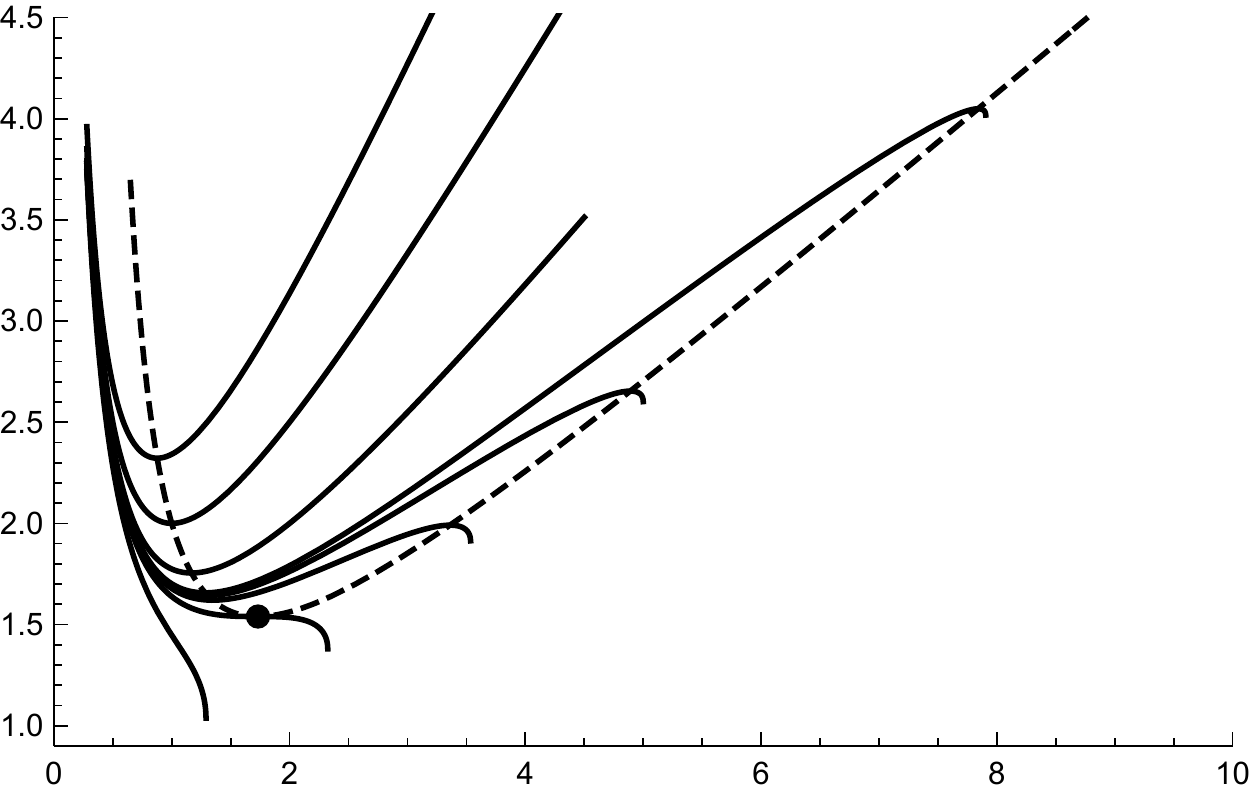}}
\caption{{\small The various curves $z \to g(z,\xi)$ for a number of values of $\xi$. For $\xi \in ]-4/27,-1/8]$
the curves have a local maximum to the right of the local minimum and the curve $g_k(z_k)$ given by eq.\ \rf{p6j3} 
passes through these local maxima for $z_k > \sqrt{3}$, as seen on the figure.}}
            \label{fig2h}
        \end{figure}

\end{enumerate}
 

        Let us couple matter to a BP (the dimer model we have just considered is a particular example). We then have a partion function:
        \[ \zm(BP) \equiv \sum e^{-S(\textrm{matter}, \textrm{BP})} \]
        where the sum is over all possible matter configurations on the given BP. We now consider the situation where we sum over all BPs with a given number of vertices $V$ or links $L$ (note that $V = L\plu 1$). We can view $L$ (or $V$) as the ``volume'' of the BP, and we have
        \[ \zm(L) \equiv \sum_{\bp, \left|\bp\right| = L} \left( \prod_v w_v\right) \zm(\bp). \]
        For large $L$ we expect
        \[ \zm(L) = e^{-L\left(f(\textrm{matter})\right)+o(L)} \]
        where $f(\mat)$ is the free energy per unit volume (or free energy density).
 The total (grand canonical) partition function of matter on the ensemble of BPs is then:
        \begin{align*}
            Z(\mu,\mat) &= \sum_\bp e^{-\mu |\bp|} \zm(\bp) \\
            &= \sum_L e^{-\mu L} \zm(L)
        \end{align*}
        We thus see that the critical point $\mu_c$ is precisely
        \[ \boxed{\mu_c = -f(\mat)} \]
        Consider the model $Z(\mu,\mat)$ as a BP model where the weight for each BP is 
        \[ \prod_v w_v \to \left(\prod_v w_v\right) \cdot \zm(\bp) \]

        {\it Matter changes the weight of each BP, but if we can calculate $\mu_c$ (which will be a function of the matter couplings) we have automatically calculated the free energy density of the matter on BPs.}

\vspace{12pt}

        Let us now apply this to our dimer model. Our matter coupling is the fugacity. We had $e^{\mu} = g$, 
        and thus $e^{\mu_c(\xi)} = g_k(\xi)$ where $g_k(\xi)$ was determined by the condition 
        $g'_z(z,\xi) = 0$.     Recall from a magnetic system that
        \[ \frac{df}{dH} = -m(H) \]
        It turns out that for the Ising model in two dimensions, at high temperature $f(H)$ has a singularity for an imaginary magnetic field $H = i \tilde{H}$ at a certain critical value $\tilde{H}_c(T)$, called the Lee-Yang edge singularity:
 \beq\label{p6jj3}
 \frac{\partial f}{\partial \tilde{H}}\Big|_{\rm singular} \propto \left(\tilde{H}\mi \tilde{H}_c(T)\right)^\sigma, \quad 
 \frac{\partial^2 f(\tilde{H})}{\partial \tilde{H}^2}\Big|_{\rm singular} \propto (\tilde{H}\mi \tilde{H}_c(T))^{\sigma-1},
  \quad \sigma = -\frac{1}{6} .
 \eeq

        Our fugacity is similar: the coupling of the magnetic field to the spin was
        \[ e^{H S} \to e^{i H S}\qquad {\rm when~the~magnetic~field~is~imaginary} \]
        so for aligned spins it is $e^{iH}$ per site. One might even consider diluted spin models where not every lattice site has a spin variable. Similarly, the fugacity is $\xi$ per link where there is a dimer. Let us write $\xi = e^h$. For \emph{negative} fugacity we have 
        \[ h \sim i \pi \plu \log |\xi|. \] 
        Let us now ask if we have a singular (i.e. non-analytic) behavior of $f(\xi)$ for $\xi \to \xi_c$.
 We want to calculate 
  \beq\label{p6jj4} 
  \left.\frac{\partial^2 f(\xi)}{\partial \xi^2}\right|_{\xi\to\xi_c} \sim (\xi\mi \xi_c)^{\sigma-1},\quad
   f(\xi) = -\mu_0(\xi) = -\log g_k(\xi) 
  \eeq
  First show the following 
  \begin{enumerate}\setcounter{enumi}{6}
  \item 
  Show that on the critical curve $\xi \to (z_k(\xi),g_k(\xi))$ we have 
  \beq\label{p6jj5}
   \xi = \frac{(1\mi z^4)(1\plu z^2)}{8 z^6}
   \eeq
   which in principle determines $z(\xi)$.
   
   \item
   Show that 
  \beq\label{p6jj6} \left.\frac{d\xi}{dz}\right|_{z_c} = 0, 
  \eeq
  which implies that $\xi(z)$ has a quadratic minimum:
   \beq\label{p6jj7}
   \xi-\xi_c = \half \left.\frac{d^2 \xi}{dz^2}\right|_{z_c} (z\mi z_c)^2 + \cdots, \quad 
   {\rm provided}\quad \left.\frac{d^2 \xi}{dz^2}\right|_{z_c} \neq 0
   \eeq
\end{enumerate}
 In order to determine the behaviour of $g_k(\xi)$ around $\xi_c$ we expand 
  $g(z,\xi)$ around $z_c, \xi_c$ (the reason we do not directly try to expand $g_k(\xi)$ around $\xi_c$ is that 
  if it has a critical behaviour, we expect it to be singular around this point. However $g(z,\xi)$ is itself 
  perfectly regular around $z_c,g_c$)
        \begin{align}
            g(z,\xi) &= g(z_c, \xi_c)+\left.\frac{\partial g}{\partial z }\right|_{z_c, \xi_c} \hspace{-4mm}(z\mi z_c)+ 
            \half \left.\frac{\partial^2 g}{\partial z^2}\right|_{z_c, \xi_c} \hspace{-4mm} (z\mi z_c)^2 + 
            \left.\frac{\partial g}{\partial \xi }\right|_{z_c, \xi_c}  \hspace{-4mm}(\xi\mi \xi_c)\nonumber\\
            &\quad + 
            \half \left.\frac{\partial^2 g}{\partial \xi^2}\right|_{z_c, \xi_c} \hspace{-4mm} (\xi\mi \xi_c)^2  + \left.\frac{\partial^2 g}{\partial \xi \partial z} \right|_{z_c, \xi_c} \hspace{-4mm} (z\mi z_c)(\xi\mi \xi_c) + 
            \left.\frac{1}{6} \frac{\partial^3 g}{\partial z^3}\right|_{z_c,\xi_c} \hspace{-4mm} (z\mi z_c)^3\nonumber \\
      &=       g(z_c, \xi_c) + \left.\frac{\partial g}{\partial \xi }\right|_{z_c, \xi_c} \!\!\!\!\!\!\!(\xi\mi \xi_c)+ \half \left.\frac{\partial^2 g}{\partial \xi^2}\right|_{z_c, \xi_c} \!\!\!\!\!\!\!(\xi\mi \xi_c)^2  + 
      \left.\frac{\partial^2 g}{\partial \xi \partial z} \right|_{z_c, \xi_c} \!\!\!\!\!\!\!(z\mi z_c)(\xi\mi \xi_c) \nonumber\\
&\quad  +
    \left.\frac{1}{6} \frac{\partial^3 g}{\partial z^3}\right|_{z_c,\xi_c} \!\!\!\!\!\!\!(z\mi z_c)^3 \label{p6jj8}
        \end{align}
when we use that 
        \[ \left.\frac{\partial g}{\partial z}\right|_{z_c,\xi_c} = \left.\frac{\partial^2 g}{\partial z^2}\right|_{z_c,\xi_c}  = 0 \]

\begin{enumerate}\setcounter{enumi}{8}

 \item Use \rf{p6jj7} and \rf{p6jj8} to show that 
 \beq\label{p6jj9}
 g(z_k(\xi), \xi) = g(z_c,\xi_c) + \left.\frac{\partial g}{\partial \xi}\right|_{z_c,\xi_c}  \hspace{-4mm}
 (\xi\mi \xi_c)+\kappa \cdot (\xi\mi \xi_c)^{3/2} + O\left((\xi\mi\xi_c)^2\right) 
 \eeq

  \item 
  show that 
  \beq\label{p6jj10}
  \frac{d^2 g(z_k(\xi),\xi)}{d\xi^2} \propto \frac{1}{(\xi\mi \xi_c)^\half} \implies \sigma = \half
  \eeq
 so we indeed have a critical ``magnetization'' for $\xi \to \xi_c$:
 \end{enumerate}

\newpage

\setcounter{equation}{0}
\setcounter{figure}{0}
\renewcommand{\thefigure}{ps-7.\arabic{figure}}
 \renewcommand{\theequation}{ps-7.\arabic{equation}}

\subsection*{Elementary Quantum Geometry Problem Set 7}

In this Problem Set we consider various aspects of BPs: BPs with infinite Hausdorff dimension, BPs coupled to Ising spins and 
the relation between BPs with dimers and Ising spins (this last relation is valid for any reasonable lattice system on which one can 
put Ising spins and dimers).

\subsubsection*{BPs with infinite Hausdorff dimension}

Let us consider BPs with $w_{m+1} = \frac{1}{m^{s+1}}$. Again taking $g = e^\mu$, we have:
\beq\label{p7j1} g = \frac{1+f(z)}{z}, \quad\quad f(z) = \sum_{m=1}^\infty \frac{z^m}{m^{s+1}} = \Li_{s+1}(z) 
\eeq
Here we encounter the so-called \emph{polylogarithm} function $\Li$. For $s > 0: f(1) = \Li_{s+1}(1) = \zeta(s+1)$, which is the Riemann zeta function.
\begin{enumerate}
    \item Show that $f'(1) = \infty$ for $s \leq 1$, and use this to argue that $z_c < 1$ (the radius of convergence 
    of $f(z)$) and therefore $\gamma = 1/2$ (so we have ordinary BPs).
    Actually, the argument can be extended and $z_c < 1$ for $s < 1.5915....$ and correspondingly $\gamma =1/2$
\end{enumerate}
Let $s \in ] n, n\plu 1 [$, with $n$ an integer $ \geq 2$. One can show that for $z \to 1$:
        \[ \Li_s(z) = \Li_s(1)+\Li_s'(1)(1\mi z)+ \cdots + \Li_s^{(n-1)}(1)(1\mi z)^{n-1} + c \cdot(1\mi z)^{s-1} + \cdots \]
        Thus we have:
        \begin{align}
            \label{eq:ggc}
            g-g_c = c_1(1\mi z) + \cdots + c_n(1\mi z)^n + c_s(1\mi z)^s + \cdots
        \end{align}
 \begin{enumerate}\setcounter{enumi}{1}       
    \item Argue (without giving detailed calculations) that one can add weights
   \beq\label{p7j3}
    \tilde{w}_2, \cdots, \tilde{w}_{n+2} \quad  \textrm{i.e. }\quad w_{k+1} = \frac{1}{k^{s+1}} + \tilde{w}_{k+1}, 
    \quad k = 1 \cdots, n+1
    \eeq
        such that in this model we have:
      \beq\label{p7j2}
      g-\tilde{g}_c = \tilde{c}_s(1\mi z)^s + \cdots
   \eeq
\end{enumerate}  
              This is precisely the same scaling as we encountered before, and an example of universality: in an earlier exercise 
              we found the weights $w_{m+1}$ which produced the  relation $g-g_c \propto  (z_c-z)^s$, without any corrections and we saw 
              that these coefficients asymptotically behaved like $1/m^{s+1}$. Here we have chosen in \rf{p7j1} weights $w_{m+1}$  
              which are exactly $1/m^{s+1}$. When correcting these coefficients in a minimal way (which does not affect the asymptotic 
              behavior), like in \rf{p7j3} we obtain the critical behavior $g-g_c \propto  (z_c-z)^s$, but there are corrections to this 
              expression, as indicated with $+ \cdots$ in \rf{p7j2}, but corrections which do not influence the critical behavior.
 \emph{However:} if we do \emph{not} add $\tw_2, \cdots, \tw_{n+2}$ we have \eqref{eq:ggc} (and note:
 if we do not allow negative weights (which do not have a straight forward probability interpretation), we 
 cannot get rid of the first $n$ terms in \eqref{eq:ggc}). Note also that if we do not add the terms  we have 
 \beq\label{p7j4}
 \frac{d g}{d z} < 0  \quad z \in [0,1],
 \eeq
 while if we add the terms $\tilde{w}_m$ in \rf{p7j3} we obtain $ \frac{d g}{d z}\big|_{z =1}  \equ 0$.
 
\begin{enumerate} \setcounter{enumi}{2} 
  \item Show that by inverting \eqref{eq:ggc} we obtain:
   \begin{align}
            \label{eq:zone}
            z\mi z_c = z\mi 1 = d_1(g\mi g_c)+d_2(g\mi g_c)^2+\cdots + d_n(g\mi g_c)^n + d_s(g\mi g_c)^s + \cdots
        \end{align}
\end{enumerate}   
     The situation is thus very different from
        \[g\mi g_c = (1\mi z)^s \implies z\mi 1= - (g\mi g_c)^{1/s} \]
        How do we define the critical exponent $\gamma$ for the case \eqref{eq:zone}?
 Recall that $\gamma$ was defined by
        \[ \frac{dz}{d\mu} \to \frac{c}{(\mu\mi \mu_c)^\gamma} \quad \quad \textrm{for } \mu \to \mu_c. \]
        But this definition assumed that $(\mu\mi \mu_c)^{-\gamma}$ was the dominating  term, 
        i.e. $\gamma > 0$. We can write:
        \[ \frac{d^{n+1}z}{d \mu^{n+1}} \to \frac{1}{(\mu\mi \mu_c)^{\gamma + n}}. \]
        Let us apply this to \eqref{eq:zone}:
        \[ \frac{d^{n+1}z}{d \mu^{n+1}} \to \frac{1}{(\mu\mi \mu_c)^{n+1-s}}, \quad {\rm and~ thus} \quad 
       \boxed{ \gamma = 1\mi s < 0   ~~ {\rm for}~~ s> 2}.
        \]

\noindent
        These  BPs are very different from the ones where $\gamma > 0$. They are dominated by configurations where a few vertices have very high order and the rest have order 1. We will not prove that here, 
        but there are simple arguments pointing in that direction.
 
 Recall that if 
        $ 0 < \gamma < 1: {dz}/{d\mu} \propto (\mu\mi \mu_c)^{-\gamma} \to \infty$ and thus  $d\mu /dz \to 0$.         One important consequence of \eqref{eq:zone} is that $d\mu/{dz} \neq 0$ for $z \to z_c$ .              
        \begin{enumerate}\setcounter{enumi}{3}        
    \item 
    Use this in the expression for 
        $G^{(I)}(\mu)$, the intrinsic two-point function, to 
show that the mass $m_I(\mu) \to c > 0$ as $\mu \to \mu_c$. 
 \end{enumerate}   
    Thus the mass does not scale to zero. Recall that we have $m(\mu) \equiv \left|\mu\mi \mu_c\right|^\nu$. 
    If $m(\mu)$ does not scale to zero we formally have $\nu = 0.$
        Furthermore, for the Hausdorff dimension $d_H$ we had $d_H = \frac{1}{\nu}$, so formally 
    $\nu =0$ implies that $d_H = \infty$.  Effectively, one can reach all vertices in just a few steps!
Intuitively this is possible if we have vertices of very high order, such that many vertices can be connected via these 
high order vertices.

However, there is more to be said, since differentiating the partition function $z(\mu)$ $n$ times, it  becomes divergent
for $\mu \to \mu_c$. Thus the $m$-point functions, $m \geq n$ {\it are} critical, but we will not discuss the interpretation of this 
any further here.

\vspace{6pt}

\subsubsection*{The Ising model coupled to BPs}

        Let BP be a branched polymer:
        \[ Z_\bp (\beta, h) = \sum_{\{\sigma_i\}} e^{\beta \sum_{\langle i j \rangle} \sigma_i \sigma_j + h \sum_i \sigma_i} \]
        where $\langle i j \rangle$ is the link between neighbouring vertices $i$ and $j$ in the BP. We take $\sigma_i = \pm 1$ and $h$ is an external magnetic field. This spin model is the so-called Ising model, and one can put the model on any graph consisting 
        of vertices and links. In particular one can put the Ising model on regular lattices. More physics related to the Ising model is 
        discussed in Problem Set 11. Here we consider the Ising model on BPs. The total partition function is then
 $$
            Z(\mu,\beta,h) = \sum_\bp e^{-\mu |\bp|} \rho(\bp) Z_\bp(\beta, h), \qquad 
            \rho(\bp) = \prod_i w_i
  $$
        We consider rooted BPs and use the convention that the root vertex has no magnetic field attached. Denote $Z_+$ the partition function where $\sigma_{\textrm{root}} \equ 1$ and $Z_-$ the partition function where $\sigma_{\textrm{root}} \equ -1$. As usual we define $f(Z) \equ \sum_{n=2}^\infty w_n Z^{n-1}$ and assume $w_1 \equ 1$. We assume until stated 
        differently that $f(Z) $ is such that $\gamma \equ 1/2$.

\begin{enumerate}
    \item Show:
        \begin{align}
            Z_+ &= e^{-\mu} \left[ e^{\beta+h}+e^{-\beta - h} + e^{\beta + h}f(Z_+) + e^{-\beta - h} f(Z_-) \right] \label{eq:zp} \\
            Z_- &= e^{- \mu} \left[ e^{\beta-h}+e^{-\left(\beta - h\right)} + e^{-(\beta - h)}f(Z_+) + e^{\beta - h} f(Z_-) \right]  \label{eq:zm}
        \end{align}
    \item Show that when $h = 0: Z_+ = Z_-$ and 
 \begin{equation}\label{p7jj1}
 e^{\mu} = 2\cosh \beta \; \frac{1+ f(Z)}{Z}   
 \end{equation}
 and thus that 
        \begin{equation}
            \label{eq:mucb}
            \mu_c(\beta) = \hat{\mu}_c  + \ln(2 \cosh \beta),
            \quad  {\rm where} \quad 
   e^{\hat{\mu}} = \frac{1+f(Z)}{Z}, \quad (Z=Z_+=Z_-).
 \end{equation}
\end{enumerate}  
So for a fixed $\beta$ we see that the critical $Z_c$ is determined by the same equation as the BPs without 
Ising spins, namely 
\begin{equation}\label{p7jj2}
 \frac{d}{dZ} \left(\frac{1+f(Z)}{Z}\right)\Big|_{Z=Z_c} = 0.
  \end{equation}
 \begin{enumerate}\setcounter{enumi}{2} 
 \item      Show that    
 \begin{equation}\label{p7jj3}
 \frac{1+f(Z_c)}{Z_c} = f'(Z_c) \quad {\rm and~thus}\quad e^{\mu_c(\beta)} = 2\cosh \beta \; f'(Z_c).
 \eeq
      
 \end{enumerate}
For $\beta \to 0$ (or $T \to \infty$): $\mu_c(\beta) = \hat{\mu}_c+\ln 2$.
\begin{enumerate}\setcounter{enumi}{3}
 \item Explain the $\ln 2$ as coming from the entropy of Ising spins.
\end{enumerate}
        Recall from the discussion of the dimer model that the partition function for a fixed ``volume'' $L$ (number of links)
        \[ Z_L(\beta) = e^{-f_L(\beta) \cdot L} = e^{-F_L(\beta)} \]
        and for $ L \to \infty$ we have that $f_L(\beta) \to f(\beta) + O(1/L)$, where the free energy per volume, $f(\beta)$
        is related to the critical point by 
        \[ f(\beta) = - \mu_c(\beta) \]
A critical temperature in the spin model is a $\beta_c$  where $f(\beta_c)$ is  non-analytic. 

\vspace{6pt}

\noindent
{\it            Thus there is {\bf no} critical temperature $\beta_c$ for Ising models on BPs, since $\mu_c(\beta)$ is 
analytic for all $\beta$}

\vspace{6pt}

\noindent
            For the Ising model the situation on a regular lattice is the following:
            
            \vspace{6pt}

            $d=1$: no phase transition.

            $d=2$: the Onsager phase transition, the most famous phase transition in physics!

            $d \geq 2$: a phase transition
            
            \vspace{6pt}
            
\noindent
            For BPs, we have $d_H \equ 2$ for $\gamma \equ \half$, which was what we assumed above. But contrary to the 
            situation for a regular lattice we have no magnetic phase transition. 
            It is possible to check that also in the cases discussed above, 
            where   $\gamma < 0$ and $d_H \equ \infty$ we have no magnetic phase transition.
  \emph{Therefore $d_H$ is not a good indicator of dimension in all situations} (the linear structure of the trees seems
   more important in this case). 
 
 \vspace{12pt}

\noindent 
 Let us finally ask whether we have spontaneous magnetization. Recall for regular lattices:

   $d=1$: no spontaneous magnetization and  $d \geq 2$: spontaneous magnetization.

  We define spontaneous magnetization as
  $$ 
  \langle m \rangle = -\lim_{h \to 0^+} \frac{\partial f(\beta, h)}{\partial h} = 
            \lim_{h \to 0^+} \frac{\partial \mu_c(\beta, h)}{\partial h}
   $$
where $\mu_c(\beta,h)$ is the critical value of $\mu$ obtained by solving eqs.\ (\ref{eq:zp}) and (\ref{eq:zm})
 for $\mu$ and then finding the smallest value of $\mu$ for given $\beta$ and $h$.

            First we consider BPs with $\gamma = \half$. After that we will analyze  $\gamma < 0$ separately.

{\it Case 1: $\gamma = \half$}
$$
Z_\pm(\beta, h) = Z_c(\beta)+\Delta Z_\pm \quad \quad \textrm{for } h \to 0: \Delta Z_\pm = c_\pm h + O(h^2) 
$$
\begin{enumerate}\setcounter{enumi}{4}
        \item Show from \eqref{eq:zp} and \eqref{eq:zm} by expanding to linear order in $h$ that:
 \begin{eqnarray}
                \label{eq:ppm}
        &&   (\Delta Z_+ \plu \Delta Z_-)\left(e^{\mu_c(\beta)}\mi 2 \cosh \beta f'(Z_c)\right) + 2Z_c e^{\mu_c(\beta)} \Delta \mu = 0
  \hspace{3cm}         \\
            &&\nonumber\\
                \label{eq:pmm}
  &&     (\Delta Z_+ \mi \Delta Z_-)\left(e^{\mu_c(\beta)} \mi 2 \sinh \beta f'(Z_c) \right) = 4h(1\plu f(Z_c)) \sinh \beta
\end{eqnarray}
\item Show, using (\ref{p7jj3}) that $\Delta \mu \equ O(h^2)$ and thus that $\langle m \rangle \equ 0$.
\end{enumerate}

 \emph{For $\gamma \equ \frac{1}{2}$ we have no spontaneous magnetization and the Hausdorff dimension $d_H  \equ 2$
 is not a good guidance.}

{\it Case 2: $\gamma < 0$}

To be specific, let us consider the case where $w_{n+1} \equ {1}/{n^{s+1}}$, $s >2$ which we analyzed above and which 
has $\gamma \equ  1 \mi s$. 
 Here we have $Z_c(\beta) \equ 1$ (the radius of convergence of $\sum_{n=1}^\infty w_{n+1}Z^n \equ f(Z)$).
 If $h > 0$  one expects  that $\Delta Z_+ \geq 0$ since in average each vertex will have more $\plu$ spins than $-$
 spins. Since the spin at the root is fixed to be +, the spin interaction between the root and its neighbor vertex will thus in average contribute positively to $Z_+$ (and similarly negatively to $Z_-$) compared to the situation where $h\equ 0$. 
 However, since $Z_+(\beta,h \equ 0))$  already assumes the  maximum
 value 1, $Z_+(\beta,h)$ cannot increase further and thus  $\Delta Z_+ \equ 0$. Also, note that we no longer 
 have $e^{\mu_c(\beta)} = 2 \cosh \beta f'(Z_c)$, since  $\frac{d}{dZ} \frac{1+f(Z)}{Z} \neq 0$ in \eqref{eq:zone} for $s>2$.
\begin{enumerate}\setcounter{enumi}{6}

        \item Use this to show that based on \eqref{eq:ppm} we get $\Delta \mu \propto \Delta Z_-$ and from \eqref{eq:pmm} we get $\Delta \mu \propto h$.

        \item Show that we have spontaneous magnetization and find   $\langle m \rangle $ as a function of $\beta$. 
\end{enumerate}        

 \noindent          {\it  Now for $\gamma < 0$ we have $d_H = \infty$, leading to spontaneous magnetization. Therefore in this case the Hausdorff dimension is a good guidance.}

\subsubsection*{The relation between the Ising model and hard dimers}

            Let $G$ be a connected graph. It can be a regular lattice, a BP or another kind of random
            graph (we are later going to consider so-called two-dimensional random graphs). 
            We can place an Ising spin model on this graph by assigning the spins to the vertices, and the interaction
            between spins will be between neighboring vertices connected by a link in the graph. We have as before:
            \[ Z_G(\beta, h) = \sum_{\{\sigma_i\}} e^{\beta \sum_{\langle i j \rangle} \sigma_i \sigma_j + h \sum_i \sigma_i} \]
\begin{enumerate}

 \item Let  $V$ be the number of vertices in $G$ and $L$ the number of links. Use the identity
            \[ e^{\sigma X} = \cosh X + \sigma \sinh X, \quad \quad \sigma = \pm 1 \]
            to show that
            \[Z_G(\beta,h) = \cosh^V h \cosh^L \beta \cdot \sum_{\{\sigma_i\}} \prod_j \big(1\plu \sigma_j \tanh h\big) 
            \prod_{\langle k l  \rangle} \big(1\plu \sigma_k \sigma_l \tanh \beta\big) \]
            
\end{enumerate}

\noindent            Expanding the products and summing over $\sigma_i$, it is clear that terms with an odd number of $\sigma_{i_1} \sigma_{i_2} \cdots \sigma_{i_{2n+1}}$ will average to zero. We want to use this and let $\beta \to 0$ (the high temperature
expansion) 
            
 \begin{enumerate}  \setcounter{enumi}{1}         
        \item  Let  $\theta(n)$ denote the number of ways one can put down $n$ hard dimers on $G$.   Show that
            \begin{eqnarray*}
                Z(\beta, h) &= &(2 \cosh h)^V \cosh^L \beta \times\\
                && 
                \Big(1+\tanh^2 h\big[\theta(1) \beta \plu O(\beta^2)\big] + \tanh^4 h \big[\theta(2) \beta^2 \plu O(\beta^4)\big] + \cdots \Big)
            \end{eqnarray*}

\end{enumerate}

\noindent            Define $\xi \equ \beta \tanh^2 h$. We now take the limit $\beta \to 0,~h \to i \pi/2$ while $\xi$ is fixed.
\begin{enumerate}\setcounter{enumi}{2}
        \item Show that the partition function for the hard dimer on $G$ with \emph{negative} $\xi$ is 
            \[ \tilde{Z}(\xi ) = \lim_{h \to i \frac{\pi}{2}, ~\beta \to 0,~ \xi~{\rm fixed}} \;\; 
            \frac{1}{(2 \cosh h)^V} Z(\beta, h) \]

\end{enumerate}

\noindent  {\it This  relates the dimer model and the Ising model with an imaginary magnetic field.}

\newpage

\setcounter{equation}{0}
\setcounter{figure}{0}
\renewcommand{\thefigure}{ps-2.\arabic{figure}}
 \renewcommand{\theequation}{ps-8.\arabic{equation}}

\subsection*{Elementary Quantum Geometry Problem Set 8}

\subsubsection*{Asymptotic expansions}

Most perturbation expansions are only so-called asymptotic expansions. This is true even for the perturbative 
expansion of the  ground state energy $E_0$ of the quantum mechanical anharmonic oscillator:
$$
    \hat{H} = \frac{1}{2m} \hat{p}^2 + \half m \omega^2 \hat{x}^2 + g \hat{x}^4 ,\quad \qquad 
    E_0 = \sum_{n=0}^\infty c_n g^n.
$$
The coefficients $c_n$ in the expansion can be calculated to any order  using textbook perturbation 
theory. However, the coefficients $c_n$ grow so fast that the radius of convergence in the power series is zero. 
Note that this is not surprising: if there was a radius of convergence, 
the theory for $g$ and $-g$ (for small $g$) would essentially be the same since 
everything would be analytic in $g$ for small $g$, but that is clearly not the case. The dynamical system 
above is well defined for positive small $g$, and is just a small deformation of the harmonic oscillator. However, for small
negative $g$ it is a very unhealthy system. For sufficient large energy we have classical run-away solutions accelerating 
to infinity and quantum mechanically there will for every energy always be a finite probability for tunnelling to 
such a situation. This implies that it is even non-trivial to define $\hat{H}$ as an Hermitian operator 
for negative $g$ (and the possible definitions are non-unique).

The  non-convergence of the perturbative series of $E_0$ for any $g$ leads to the question: assume that we have calculated all the $c_n$. Do we have a way to calculate $E_0$?
One method is {\it Borel summation}. Let $f(x)$ be ``defined'' by its formal power series. The Borel transform 
of $f$, $B(f)$, is then also defined as a formal power series
\begin{equation}\label{p8j1}
f(x) =  \sum_{n=0}^\infty a_n x^n, \qquad B(f) (x) :=  \sum_{n=0}^\infty \frac{a_n}{n!} x^n
\end{equation}
Assume now that the power series for $B(f)$ has  radius of convergence $r >0$ and that the corresponding function
$B(f) ( x)$  can be analytically continued into a wedge region $|\arg z| < \epsilon$  of the complex plane, and 
that it grows slower than exponential in this region. Then one can write formally write, interchanging 
summation and integration, which might or might not be allowed from a mathematical point of view,
\begin{align*}
    f(x) &= \sum_{n=0}^\infty a_n\, x^n = \sum_{n=0}^\infty n!  \; \frac{a_n}{n!}\, x^n, \qquad
    n! = \int_0^\infty dt \,t^n e^{-t} \\
    f(x) &= \int_0^\infty e^{-t} \left( \sum_{n=0}^\infty \frac{a_n}{n!} (xt)^n\right) = \int_0^\infty e^{-t} B(f) (xt).
\end{align*}
This integral now exists and it is called {\it the  Borel sum of the formal power series $f(x)$}.
\begin{enumerate}
    \item Assume $x> 0$ and apply this procedure to 
        \begin{align}
            f(x) = \sum_{n=0}^\infty n! (-1)^n x^{n+1} = x \sum_{n=0}^\infty n! (-1)^n x^{n} ,
            \label{eq:borelfx}
        \end{align}
     to obtain
 \begin{equation}\label{p8j2}
  f(x) = x \int_0^\infty dt \, \frac{\e^{-t}}{1\plu xt} 
  \end{equation}
\end{enumerate}

 The perturbation series of the anharmonic oscillator is divergent like \eqref{eq:borelfx}, so it has zero radius of convergence. Two questions arise. (a) Assume the perturbation series can be Borel summed, like the series  
\eqref{eq:borelfx}. Of course we know that $E_0(g)$ exists in quantum mechanics. How can we 
be sure that the Borel sum actually gives the correct value of $E_0(g)$.  To be sure of that 
one has to appeal to properties of $E_0(g)$, which have to be proven outside perturbation theory, 
i.e.\ using general theorems for unbounded Hermitian operators like $\hat{H}$, known from
functional analysis. and combine these with other general conditions which a function  $f(x)$ has to satisfy 
in order that the Borel sum of its asymptotic series actually is equal to $f(x)$. We will not discuss these 
mathematical issues. (b) At a much more mundane level one can ask the following: even if we know 
that one, by some fancy method, is able to sum a series like  \eqref{eq:borelfx} to the correct answer, to what 
extent does a perturbation expansion which is only an asymptotic expansion  help us at all? Clearly, given a value 
of $x$, e.g.\ the coupling constant $g$ of the anharmonic oscillator, it makes no sense to continue calculating 
to very high order since $n! x^n \to \infty$ for $n \to \infty$. In fact, in general this observation also 
contains the practical answer we  know from perturbation theory, here formulated for the 
asymptotic series: for a given $x$, higher order terms will only improve the approximation to 
the function $f(x)$ we are looking for if $n < |x|$ (such that $|a_n x^n| <1,~|a_n| \sim n!$). Of course there exist
many methods by which one can do much better than just naively summing the series up to a given $n$, 
but as with the Borel summation, to be sure that they work, one has to know something more about the function $f(x)$.

        There are other (less general) methods to sum divergent series:
\begin{enumerate}\setcounter{enumi}{1}        
    \item Show that the formal power series $f(x)$  defined in \eqref{eq:borelfx} satisfies the following differential equation:
        \[ \frac{df}{dx} + \frac{1}{x^2} f(x) = \frac{1}{x} \]
    \item Solve this to find $f(x)$ explicitly.   (Hint: change variables to $u = \frac{1}{x}$ if you do not remember 
    the general solution to a linear differential equation.) 
\end{enumerate}        
\noindent        The exponential-integral function $\Ei(u)$ is defined by
\begin{equation} 
\Ei(u) = - \int_{-u}^\infty dt \, \frac{\e^{-t}}{t} \qquad {\rm for}\quad u< 0 \label{p8j3}
\end{equation}
It is convenient to define the following function
\begin{equation}
\Ei_c(u) = -\int_{-u}^{-c} dt \, \frac{\e^{-t}}{t}  
 \qquad {\rm for}\quad u,c > 0 \label{p8j4}
\end{equation}
  \begin{enumerate}\setcounter{enumi}{3}
    \item Show that
        \[ f(x) = -\e^{\frac{1}{x}} \Ei\Big(\mi \frac{1}{x} \Big) , \qquad x >0\]
        and that $f(x)$ can be written in the form \eqref{p8j2}.
  \end{enumerate}
  Thus we have seen that  starting from the formal power series \eqref{eq:borelfx} we obtain the 
  Borel sum \eqref{p8j2} and solving the differential equation which the formal power series obeys, we obtain the 
  same function. (and it is easy to show that the function $f(x)$ we have found precisely has the asymptotic
  expansion \eqref{p8j2}, e.g.\ by partial integration in representation \eqref{p8j3} of $\Ei (-1/x)$). 
   \begin{enumerate}\setcounter{enumi}{4}
   
   \item
   Perform the partial integrations and convince yourself that it is correct....
    \end{enumerate}
  \begin{enumerate}    \setcounter{enumi}{5} 
    \item Now repeat the same  steps for the function 
  \begin{equation}\label{p8j5}
  g(x) = \sum_{n=0}^\infty n!\, x^{n+1} 
\end{equation}
and Borel sum to obtain (formally)
\begin{equation}\label{p8j6}
 g(x) = x \int_0^\infty dt \, \frac{e^{-t}}{1 \mi xt}
 \end{equation}
 and show that $g(x)$ satisfies the differential equation:
  \begin{equation}\label{p8j7}
  \frac{dg}{dx} - \frac{1}{x^2} g(x) = -\frac{1}{x}, \
 \end{equation}  
 and finally that for all positive $c$ we have  solutions 
 \begin{equation}
    g_c(x) = -e^{-\frac{1}{x}} \Ei_c\Big(\frac{1}{x} \Big) , 
 \end{equation}
 to eq.\ \eqref{p8j7},  which {\it all} have the asymptotic expansion \eqref{p8j5}.
 (Hint for the asymptotic expansion: partial integrate $\Ei_c (1/x)$ and use that 
 the asymptotic expansion of $e^{-1/x}$ is zero!) 
  
\end{enumerate}

\noindent
 The series \eqref{p8j5}  is \emph{not} Borel summable ( $B(g/x)(x) = 1/(1\mi x)$ has a singularity on the positive real axis
 and the integral \eqref{p8j6} does not exist).  However there is no problem  solving the differential equation for 
 positive $x$  and the corresponding solutions $g_c(x)$ all have  the correct  asymptotic expansion.
        Notice also that $e^{-{1}/x}$ is a solution to the homogeneous differential equation. This is why
        we found a whole family of solutions:  
 $$ 
 g_{c_1}(x) -g_{c_2} (x)= {\rm const.} \, \e^{-\frac{1}{x}},
 $$
        and since the Taylor expansion around  $0^+$ of $e^{-\frac{1}{x}}$ is zero, they all have the same asymptotic
        expansion. {\it So we here see a simple example where the asymptotic series does not fix the 
        function uniquely. For that we need more information.} Of course we also have homogeneous solutions 
        we could add to our function $f(x)$ which was Borel summable. However, in this case the solution 
        to the homogeneous differential equation would be $e^{1/x}$, which blows up at $x\to 0^+$, and 
        there might be good physical arguments to discard this contribution.

\noindent
        We say that the $e^{-\frac{1}{x}}$ is a ``non-perturbative'' contribution: if $x\equ g$, the coupling constant, 
        the contribution $e^{-\frac{1}{g}}$ will never be seen in a simple perturbation expansion. 
        Nevertheless, there are many examples in physics where such contributions are important. Maybe the simplest one is the energy shift between the two lowest energy levels for the anharmonic oscillator with double well:
        \[ V(x) = -\frac{\omega^2 x^2}{4} + \frac{g}{4} x^4 \]
        where the energy shift is 
        \[ \Delta E \propto \hbar \omega \, \e^{-\left(\frac{\omega^4}{3\hbar \omega} \frac{1}{g}\right)} \]

\noindent
In the next exercise we will count BPs with loops and see that the partition function is only defined as an asymptotic series which is \emph{not} Borel summable. Nevertheless, we can find a differential equation for it, which we can subsequently solve.

\newpage

\setcounter{equation}{0}
\setcounter{figure}{0}
\renewcommand{\thefigure}{ps-9.\arabic{figure}}
 \renewcommand{\theequation}{ps-9.\arabic{equation}}

\subsection*{Elementary Quantum Geometry Problem Set 9}
\subsubsection*{Branched polymers with loops}

\noindent
We start with the simplest branching, where
\[ w_3 = 1, \quad w_1 = 1+j, \quad e^\mu = g. \]
Usually we have always chosen $w_1 = 1$, but for later use it is convenient to choose $w_1 = 1+j$ where we will set $j=0$ in the end.

Now the rooted BP partition function satisfies the following graphical equation,\\
\begin{figure}[!ht]
    \centering
    \vspace{-0.4cm}
   \centerline{ \includegraphics[width =0.6\linewidth]{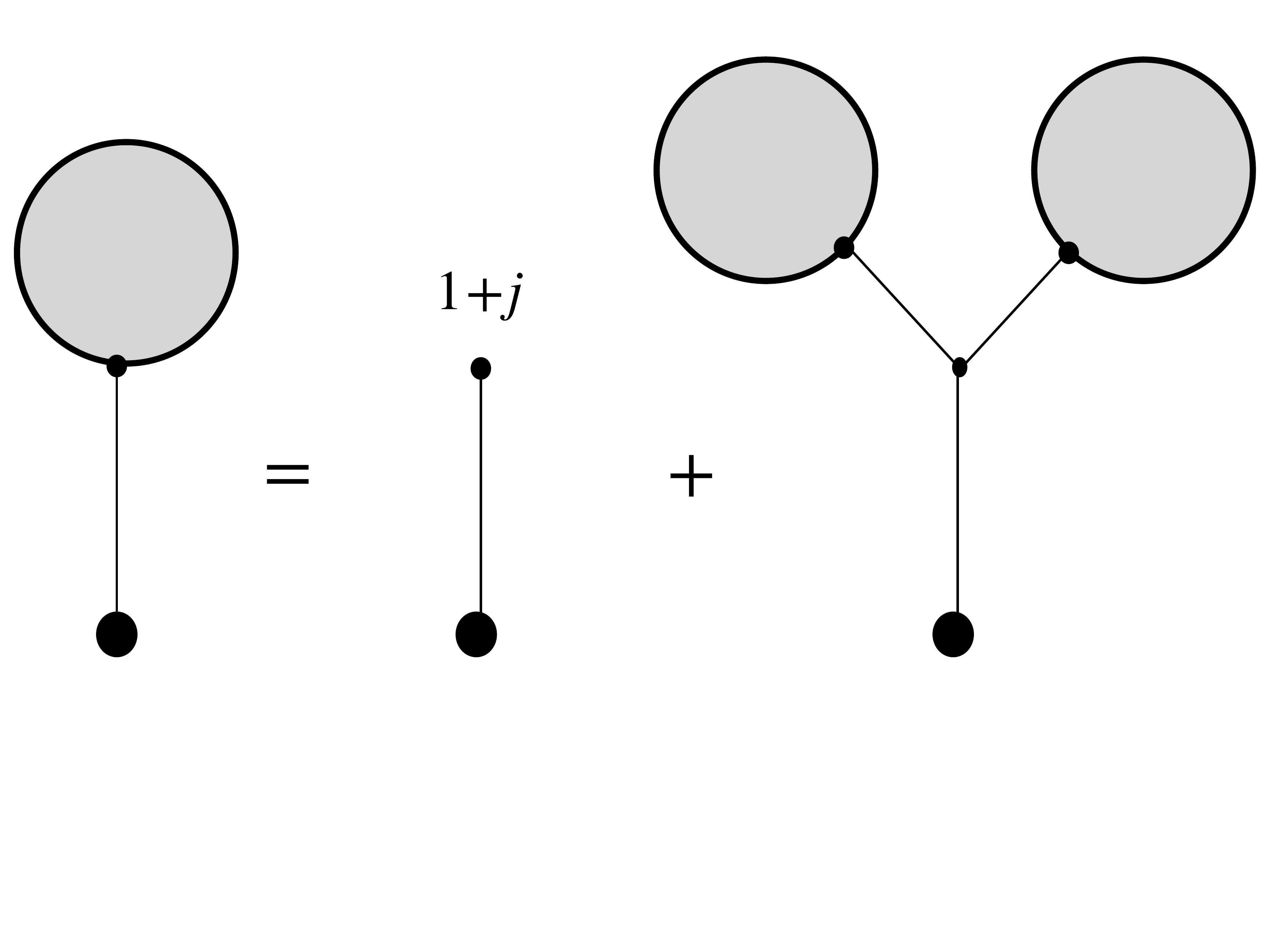}}
   \vspace{-2cm}
    \caption{{\footnotesize Equation for the rooted branched polymers. The bottom big dot is the root and the free little
    dot of order 1 has weight $w_1 = 1+j$.} }
    \label{figx1}
\end{figure}

 which when formulated in terms of $g$ and $z$ reads:
\begin{align}
    z&= \frac{1}{g} (w_1+w_3 z^2),\quad {\rm i.e.} \quad 
    z= \frac{1}{g} (1+j+z^2) \label{p9jy}
\end{align}
\begin{enumerate}
    \item Solve this for $z(g,j)$ and determine whether a plus or minus sign should be used in the process.
    \item Show that $\gamma \equ \half$, $z_c \equ \sqrt{1\plu j}$, and $g_c \equ 2 \sqrt{1\plu j}$.
\end{enumerate}
 Define the susceptibility (almost) like in the notes:
 \beq\label{p9j1}
 \chi(g,j) = \frac{dz(g,j)}{dj}.
 \eeq
 The difference between this susceptibility and the one we used in the notes is that there we differentiated
 wrt $g$ rather than $j$. The power with which $j$ appear in a graph is $V_1 \mi 1$, were $V_1$ is the 
 number of vertices of order 1 (the -1 is the root). The power with which $g$ appears in a graph 
 is $L = V\mi 1$ in a BP. However, for the graphs we consider we have $V_1\equ L/2\mi 3/2$. Thus there will 
 not be any difference in the critical behavior of this $\chi$ and the one in the notes. Graphically 
 the difference is that the two marked vertices in the graph for the present $\chi$ will be vertices
 of order 1, while for the $\chi$ in the notes it can be any vertex in the graph. The graphic definition 
 for the present $\chi$ is shown below. Note the following: in order to obtain the correct number of graphs 
 it is important to be aware that {\it $\chi$ has two marked vertices, and these marks can be distinguished: }
 one mark is on the root, while the other marked vertex came from removing a 
 $j$ from one of the vertices of order 1 when differentiating wrt $j$. 
 The present definition will be more convenient when we consider graphs which are not tree graphs. 
 The graphical representation of $\chi(g,j)$ is shown in fig.\ \ref{figx2}.
             \begin{figure}[!ht]
            \vspace{-0.4cm}
             \centerline{\includegraphics[height=8cm]{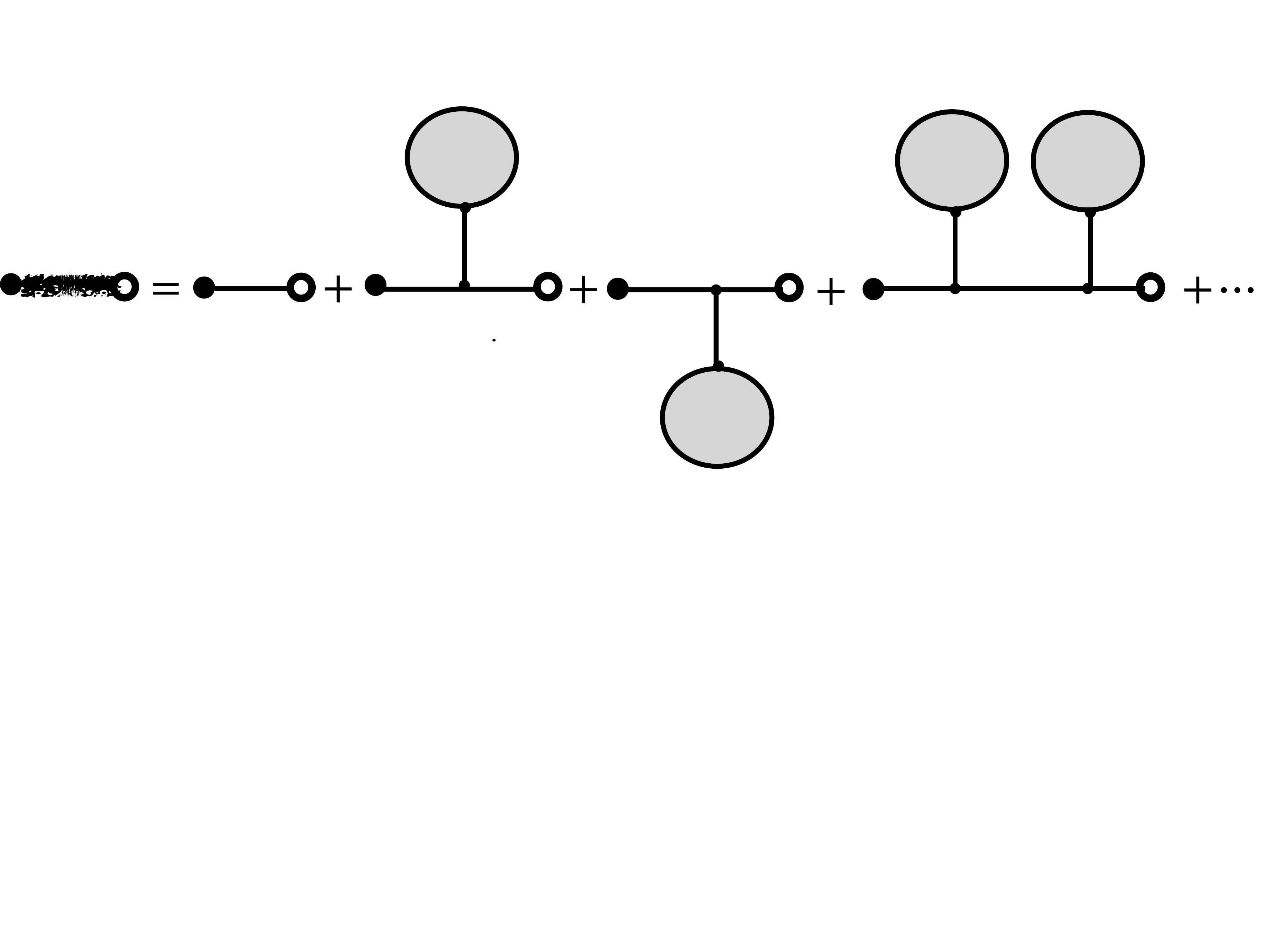}}
             \vspace{-4cm}
            \caption{{\footnotesize The susceptibility $\chi(g,j)$ defined by eq.\ \rf{p9j1}. The graph on the lhs of the equality sign, 
            the susceptibility, is defined as the sum of the graphs on the rhs of the equality sign. The large black
            dot denote the root (coming from $z(g,j)$ in eq.\ \rf{p9j1}) and the large dot to the right is the vertex on which 
            $d/dj$ has acted. The weight $1\plu j$ of such a vertex is then replaced by weight 1.} }
            \label{figx2}
        \end{figure}

        Now let us define 
  \beq\label{yy}\Delta \equiv \frac{g^2}{4} - (1+j), \quad \quad z = \frac{g}{2} - \sqrt{\Delta}
   \eeq
          At the critical point we have $\Delta_c = \Delta(g_c, j) = 0$.        
\begin{enumerate}\setcounter{enumi}{2}
   \item Show from the graphical representation of $\chi$ in fig.\ \ref{figx2} that 
       \begin{align} \label{p9jh1}
           \chi(g, j) = \frac{1}{2\sqrt{\Delta}}  = \frac{1}{g-2z},
       \end{align}
       which is of course what one obtains just by differentiating the $z(g,j)$ which we have already found explicitly.
       Expand $\chi(g, j)$ in inverse powers of $g$. The coefficient of $(1\plu j)^{n-1}/g^{2n-1}$ gives the number 
       of BPs with $2n\mi 1$ links and a root and another marked vertex and $n\mi 1$ vertices of order 
       one (each associated with a factor $(1\plu  j)$). The first coefficients are 1,2 and 6. Draw the 
       corresponding BPs with 1, 3 and 5 lines.
\end{enumerate}
     
\begin{figure}[t]
 \vspace{-1cm}
\centerline{\hspace{2.5cm}\includegraphics[width=1.0\linewidth, angle = 0]{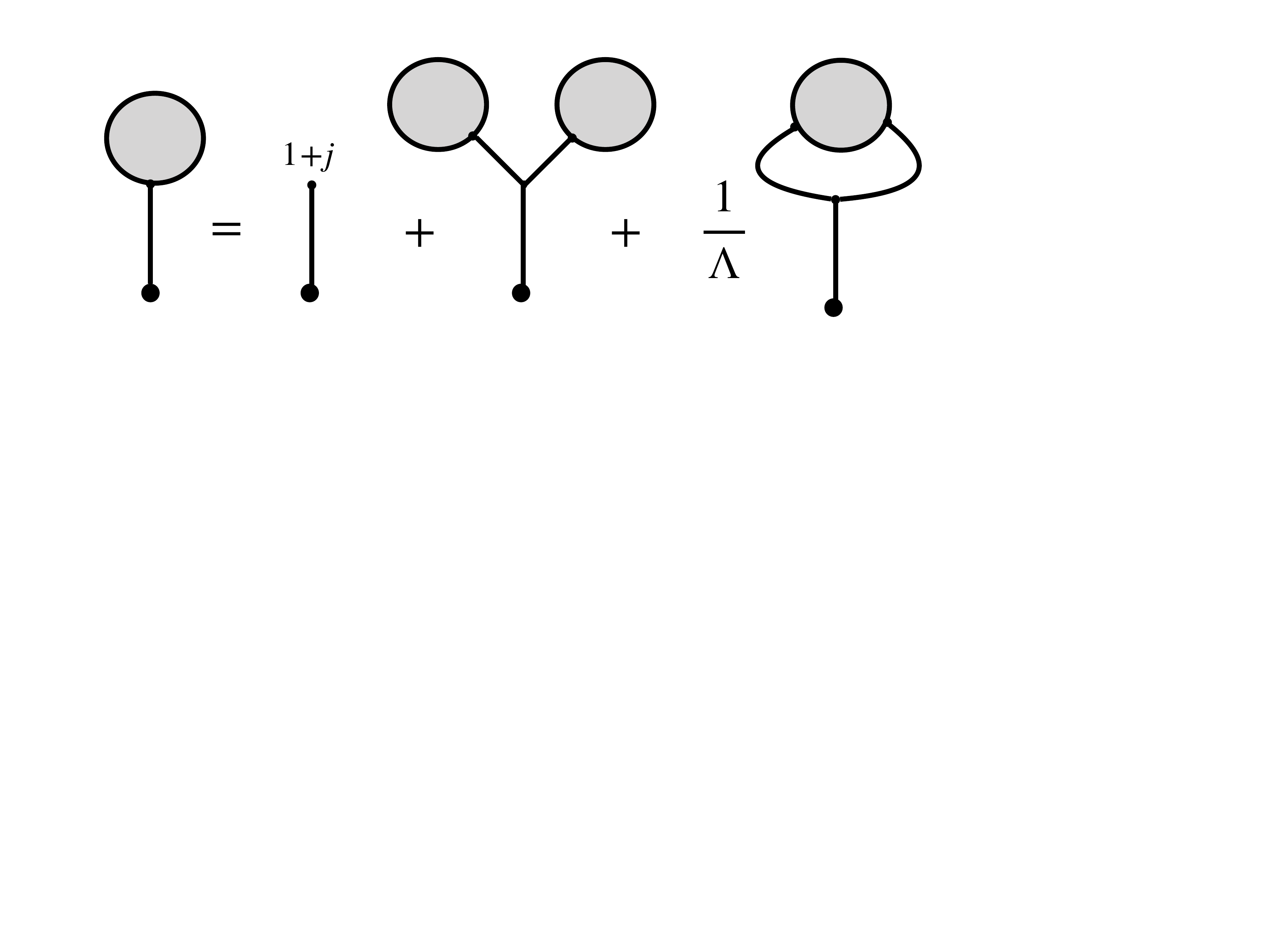}}
\vspace{-7.5cm}
\caption{{\footnotesize The graphical equation  is not an algebraically closed equation since it involves the 
 two-point function appearing in the last graph in the figure. A new coupling constant $\Lam$ is introduced, such that 
 $\Lam^{-n}$ will be muiltiplying graphs with $n$ loops when we solve the graphical equation iteratively in the number of loop.}  }
\label{figy1}
\end{figure}

      
       Consider now rooted BPs with loops. We have to define these. We choose to use a graphic definition which is 
       a generalization of Fig.\ \ref{figx1}, shown in Fig.\ \ref{figy1}. 
In Fig.\ \ref{figy1} we have generalized the definition of the two-point function (or susceptibility) from $z$ in eq.\ \rf{p9j1} which 
had no loop, to the $Z$ which has loops, i.e. the equation corresponding to Fig.\ \ref{figy1} reads:
 \beq\label{p9jh3}
Z = \frac{1}{g} \Big( (1\plu j)  + Z^2 + \frac{1}{\Lambda}  \chi^{(2)} \Big), \qquad \chi^{(2)}(g,j,\Lambda) = \frac{d Z(g,j,\Lambda)}{d j}.
\eeq
Thus we now have a differential equation in $Z$, rather than an algebraic equation as for the BPs without loops. However, we can 
turn eq.\ \rf{p9jh3} into an infinite set of algebraic equations by first introducing the $k$-point functions $\chi^{(k)}$:
\beq\label{p9jhx1}
\chi^{(k+1)}(g,j,\Lambda) =
   \frac{d \chi^{(k)}(g,j,\Lambda)}{d j} = \frac{d^k Z(g,j,\Lambda)}{d j^k},
 \eeq  
and then differentiation eq.\ \rf{p9jh3} wrt $j$:
\bea
\chi^{(2)} &=& \frac{1}{g} \Big( 1 + 2 Z \, \chi^{(2)}  + \frac{1}{\Lambda}  \chi^{(3)} \Big), \label{ds1}\\
\chi^{(3)} &=& \frac{1}{g} \Big( 2 Z \, \chi^{(3)} + 2 \chi^{(2)}\chi^{(2)}+ \frac{1}{\Lambda}  \chi^{(4)} \Big), \label{ds2}\\
\chi^{(4)} &=& \cdots\cdots \label{ds3}
\eea
These equations allow us to make a systematic double expansion in powers of 
$1/g$ and $1/\Lambda$, such that for $j\equ 0$ the number
 of graphs  of the $k$-point function with $n$ lines and $\ell$ loops is the coefficient to the power $g^{-n} \Lambda^{-\ell}$
 when we make the expansion   
 \bea\label{p9jx1}
 Z(g,j,\Lambda)\!\! &=& \!\! \sum_{\ell,n} \frac{Z_{n,\ell}(j)}{\Lambda^\ell \; g^n} = 
 \sum_{\ell=0}^\infty   \frac{Z_\ell(g,j)}{\Lambda^\ell}, \quad 
 Z_\ell (g,j)  = \!\!\! \sum_{n=|3\ell -1| }^{\infty}\!\!  \frac{Z_{n,\ell}(j)}{g^n}, \hspace{1.8cm}\\
 \chi^{(k)}(g,j,\Lambda)\! \!&=&\!\!  \sum_{\ell,n} \frac{\chi^{(k)}_{n,\ell}(j)}{\Lambda^\ell \; g^n} = 
 \sum_{\ell=0}^\infty   \frac{\chi^{(k)}_\ell(g,j)}{\Lambda^\ell}, \quad \chi^{(k)}_\ell (g,j) =  \!\!\!\!\!
 \sum_{n=|3\ell -1| }^{\infty} \!\! \frac{\chi^{(k)}_{n,\ell}(j)}{g^n}
 \eea
 Thus  $Z_{n,\ell}(j\equ 0)$ is the number of graphs with $\ell$ loops and $n$ lines. A natural starting point of the iteration of 
 these equations is the BPs, i.e. $Z_0$ and $\chi_0^{(k)}$ since we know these functions explicitly:
 \bea\label{p9jh5}
 Z_0 &=&\frac{g}{2} -\sqrt{\Delta} = \frac{1}{2} \Big( g \mi \sqrt{g^2 \mi 4(1\plu j)}\Big)\\ 
 \chi^{(2)}_0  &=&\frac{1}{g\mi 2Z_0} = \frac{1}{2\sqrt{\Delta}}  \label{p9jh5a}\\ 
 \chi^{(k+2)}_0 &=&     {2^k \,(2k-1)!!} \;\Big(\chi^{(2)}_0\Big)^{2k+1}.\label{p9jh5b}
 \eea
       \begin{figure}[t]
            \vspace{-0.5cm}
       \centerline{\includegraphics[width=1.0\linewidth]{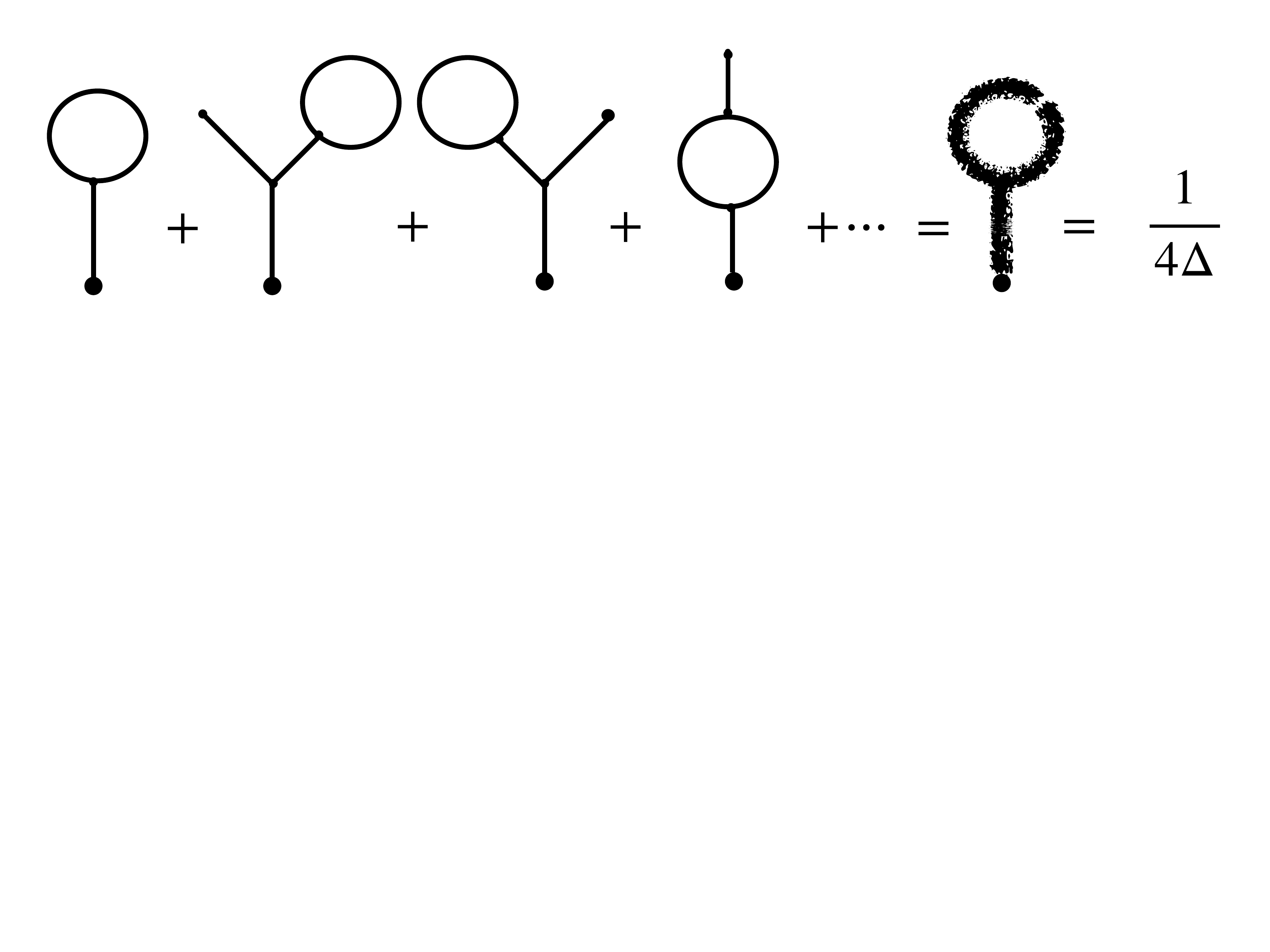}}
       \vspace{-7cm}
       \caption{{\footnotesize The summation of one-loop  BPs}}
       \label{figx3}
        \end{figure}
\begin{enumerate}\setcounter{enumi}{3}

\item Show that the first iterations are 
\bea\label{p9jh6}
Z_1 &=& \frac{1}{g} \Big( 2Z_0 Z_1 + \chi_0^{(2)}\Big),  \\
 \label{p9jh6a}\\
Z_2 &=&  \frac{1}{g} \Big( 2Z_0 Z_2 + Z_1^2  + \chi^{(2)}_1\Big), \label{p9jh6b}
\eea
and show that it can be written
\beq\label{p9jh7}
 Z_1 =  \Big(\chi_0^{(2)}\Big)^2 ,\quad  \chi_1^{(2)} = 4 \Big(\chi_0^{(2)}\Big)^3, \quad Z_2 = 5  \Big(\chi_0^{(2)}\Big)^5.
 \eeq
 \item Argue that the one-loop diagrams corresponding to $Z_1$  can be presented as in  Fig.\ \ref{figx3}. 
 Find the coefficients to $1/g^2$, $1/g^4$ (and, if you are energetic, $1/g^6$
   in the expansion of $1/(4 \Delta)$. These are the number of one-loop diagrams with  2, 4 and 6 lines.
   Draw them.

   \item Argue (no proof...) that the graphs representing $Z_2$ are of the form shown in Fig.\ \ref{figx4},
       \begin{figure}[!ht]
            \centering
            \includegraphics[width=0.9\linewidth]{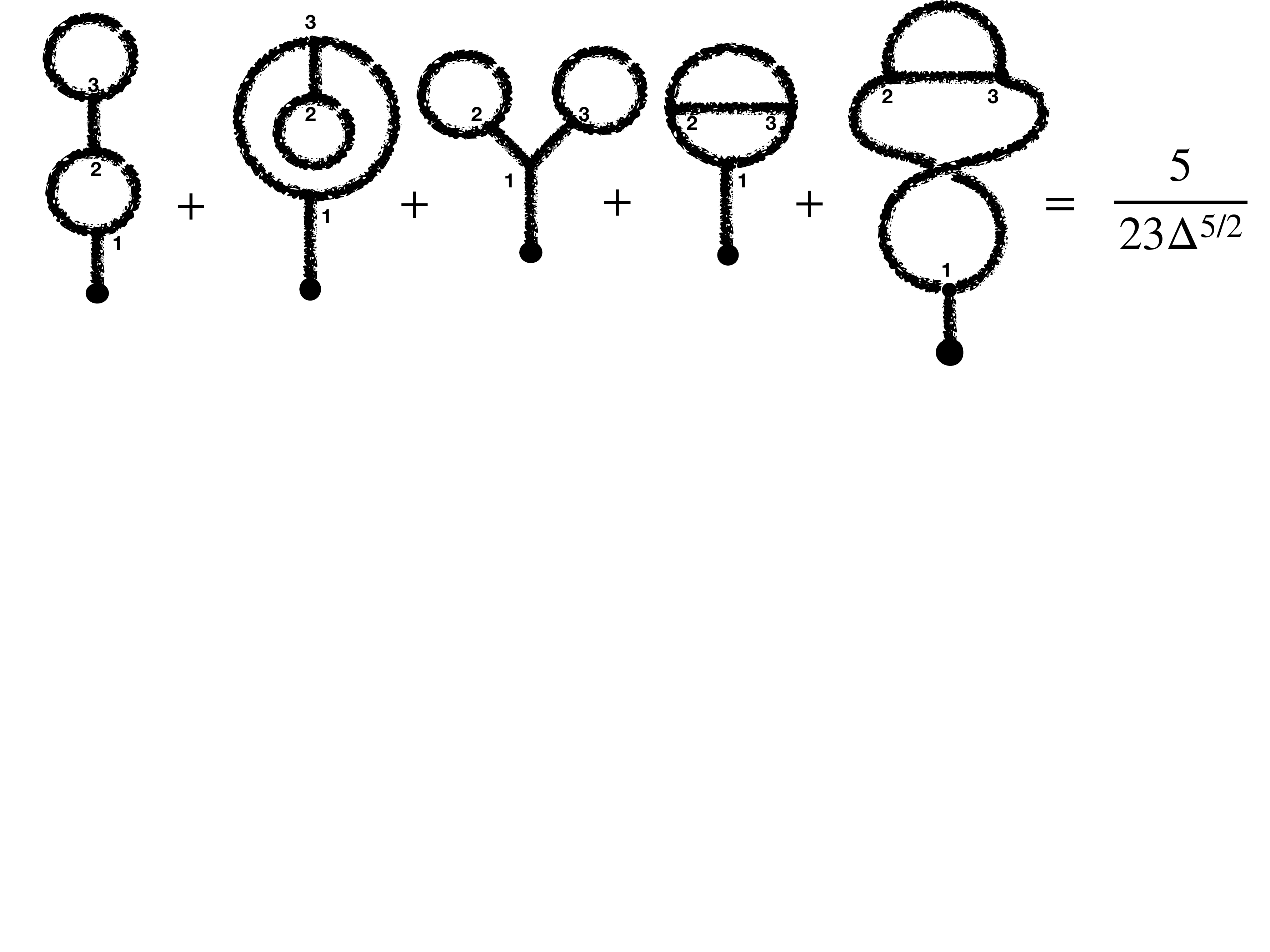}
            \vspace{-6.5cm}
            \caption{{\footnotesize The two-loop  diagrams. Note that the coefficient here, 5/32, and the coefficient 1/4
            in the former figure, are the coefficients reproduced in the asymptotic expansion given by eq.\ \rf{p9janx1}.}}
            \label{figx4}
        \end{figure}
i.e. all two-loop  $\phi^3$ graphs with one ``external'' line and one ``external''  vertex. 
Extend the arguments to  $Z_\ell$ with $\ell$  loops, which  can then be represented as 
dressed $\ell$-loop $\phi^3$  ``tadpole'' graphs, and argue  that generation function of these behave like 
       \begin{align}\label{p9jk1}
         Z_\ell(g,j)= \frac{C_\ell}{\Delta^{{3}\ell/2-1/2}} 
       \end{align}
       \end{enumerate}
       
 \noindent      Returning to the full generating function for rooted BPs, also including loops, we can now write
       \begin{align}
           Z(\Lambda, g, j) &= \sum_{\ell=0}^\infty \frac{Z_\ell(g,j)}{\Lambda^\ell} = 
            \frac{g}{2} +\sqrt{\Delta} \sum_{\ell = 0}^\infty \frac{C_\ell}{\left(\Lambda \Delta^{\frac{3}{2}} \right)^\ell}, \quad C_0 = -1.
       \end{align} 
       Introduce the notation
       \begin{align}
           \Lambda \Delta^{\frac{3}{2}} &= \frac{3}{2} t \\
           Z(g,j,\Lambda) &= \frac{g}{2} - \Delta^\half F(t) \label{eq:bigf}
       \end{align}
We then insert this $Z$ in the defining equation \rf{p9jh3} for $Z$. Keeping $g$ fixed we have 
$$
\Delta = \frac{g^2}{4}-(1\plu j) \implies \frac{d}{dj} = -\frac{d}{d\Delta}
$$ 
and thus, using   $F$ from \eqref{eq:bigf} instead of $Z$ 
       \begin{align}
           \label{eq:diff-f}
           F^2+\frac{F}{3t}+\frac{dF}{dt} = 1. 
       \end{align}
 In this equation we finally put $j=0$.  
 
 \begin{enumerate}\setcounter{enumi}{6}
 \item Show that eq.\ \rf{p9jh3} implies eq.\ \rf{eq:diff-f}.
 \end{enumerate}

\noindent        Eq.\ \rf{eq:diff-f} a so-called \emph{Riccati equation} and it can be solved. The solution can be expressed in terms of Airy functions Ai, Bi and their derivatives. The Airy functions $w(z)$ are solutions to following differential equation
\beq\label{p9jg1}
 \frac{d^2 w}{dz^2} = z\, w(z),\qquad w(z) = a \, {\rm Ai}(z) + b\, {\rm Bi}(z),
 \eeq
 and the solution to eq.\ \rf{eq:diff-f} in terms of Airy functions is then 
\beq\label{p9jg2}
F(t) = \frac{1}{\sqrt{z}} \; \frac{w'(z)}{w(z)}, \quad t = \frac{2}{3} \, z^{3/2}.
\eeq
 where Ai, Ai', Bi and Bi' have the following asymptotic expansions for large $|z| \leq \pi/3$, expressed in terms of $t = 2 z^{3/2} /3$: 
 \beq\label{ai}
  {\rm Ai}(z) = \frac{ e^{-t}}{2 \sqrt{\pi}\; z^{-1/4}} \sum_{k=0}^\infty \frac{(-1)^k a_k}{t^k} ,\quad   
  {\rm Bi} (z)= \frac{ e^{t}}{\sqrt{\pi}\; z^{-1/4}} \sum_{k=0}^\infty \frac{a_k}{t^k} .
  \eeq
 \beq\label{bi}
  {\rm Ai'}(z) = -\frac{ e^{-t}}{2 \sqrt{\pi}\; z^{1/4}} \sum_{k=0}^\infty \frac{(-1)^k d_k}{t^k} ,\quad   
  {\rm Bi}' (z)= \frac{ e^{t}}{\sqrt{\pi}\; z^{1/4}} \sum_{k=0}^\infty \frac{d_k}{t^k} .
  \eeq
  One has 
  \beq\label{ci}
  a_k = \frac{\Gamma(3k\plu \half)}{54^k k! \, \Gamma(k\plu \half)},\qquad d_k = - \frac{6k\plu 1}{6k\mi 1} a_k.
  \eeq
These expansions are only asymptotic since the coefficients $|a_k|$, $d_k$ grow like $k!$. 

From the asymptotic expansions \rf{ai}  we see that the requirement that the asymptotic expansion of $F(t)$ starts
out as $1 + O(1/t)$ only fixes $F(t)$ up to exponential corrections of order $e^{-2t}$. We have 
\bea\label{di}
\lefteqn{F(t) = \frac{{\rm Bi}'(z)}{{\rm Bi}(z)} \; \left( \frac{1 + c \, {{\rm Ai}'(z)}/{{\rm Bi'}(z)}}{1+  c \,{{\rm Ai}(z)}/{{\rm Bi}(z)}} \right)}\\
&=&\!\!\!1 1 \mi \frac{1}{6 \,t}  \mi\frac{5}{72 t^2} \plu\cdots \plu e^{-2t} \big(k_0 \plu \frac{k_1}{t} \plu \cdots\big) 
\plu  e^{-4t}\big(h_0 \plu \frac{h_1}{t} \plu  \cdots\big) \plu  \cdots\hspace{18mm}
\eea 
So there is a one-parameter class of solutions, depending on the constant $c$, which have the same leading asymptotic expansion 
        \begin{align}\label{p9janx1}
            F(t) = 1-\frac{1}{6t}-\frac{5}{72t^2} - \cdots 
        \end{align}

\begin{enumerate}\setcounter{enumi}{7}
    \item Write
        \[ F(t) = \sum_{n=0}^\infty \frac{c_n}{t^n} \]
        and use \eqref{eq:diff-f} to find a recursive relation for the $c_n$ and find the first two coefficients shown in \rf{p9janx1}. 
    \item Show that 
        \[ c_n \propto - \frac{\Gamma(n)}{2^n} \quad \textrm{for } n\to\infty, \]
        up to factors $n^{-\alpha}(1+ O(1/n))$ where $\alpha$ is not determined.
\end{enumerate}


\noindent        Finally we have achieved our goal:
 \beq\label{final} 
 Z(g,\Lambda) = \frac{g}{2} - \sqrt{\Delta} + \sqrt{\Delta} \sum_{\ell \geq 1}^\infty \left(\frac{3}{2}\right)^\ell \frac{c_\ell}{\left(\Delta^{\frac{3}{2}} \Lambda\right)^\ell}
 \eeq
 where 
 \beq 
 \sqrt{\Delta} = \sqrt{\frac{g^2}{4}-1} = \half \sqrt{g-g_c}, \qquad g_c = 2 
\eeq

\vspace{6pt}  

Let us now discuss if we can associate any critical behavior to this partition function when $g \to g_c \equ 2$. Our starting point 
was that $Z_0(g) \equ g/2 \mi  \sqrt{\Delta}$ was the partition function for BPs and that it has $\gamma \equ  1/2$,  i.e.  
$ Z_0(g) = c_0 \plu  c_1 (g\mi g_c)^{1-\gamma}  + \cdots$, where $\gamma =1/2$.  Similarly we have for $Z_\ell(g)$, the partition 
function for PBs with $\ell$ loops, that $Z_\ell(g) \propto \Delta^{(1 -3\ell )/2}$.

\noindent      
\begin{enumerate} \setcounter{enumi}{9}       
    \item Show that $ Z_\ell(g) \sim (g \mi g_c)^{1-\gamma_\ell }$ where $\gamma_\ell =  \frac{3}{2} \ell\plu \half$.
\end{enumerate}

It is now clear that we cannot associate an ordinary critical behavior to the function $Z(g,\Lambda)$ given by eq.\ \rf{final}
since for fixed $\Lambda$ it becomes more and more singular for increasing number of loops when $g \to g_c$. However,
one can try to take a so-called {\it double scaling limit}, where we together with a scaling $g\to g_c$ also scale the 
``coupling constant for loops", $1/\Lambda$ to zero such that 
\beq\label{double}
\Lambda \Delta^{3/2} = \frac{3}{2} t\quad {\rm is~fixed}.
\eeq 
The ``physics'' of this double scaling limit is the following: all partition functions $Z_\ell(g)$ have the same critical point 
$g_c=2$. This is actually quite remarkable. The number $Z_\ell(k)$ of BPs with $k$ links and $\ell$ loops 
has a leading asymptotic behavior
\beq\label{number}
Z_\ell (g) = \sum_k \frac{Z_\ell (k)}{g^k}, \quad Z_\ell (k) \sim k^{\gamma_\ell-2} \; 2^k\big( 1\plu  O(1/k)\big)  
\quad {\rm for} \quad k\gg 1.
\eeq
Thus there is an exponential growth $2^k$ of the $\ell$-loop BPs with the number of links. However, the sub-leading,
but universal, factor $ k^{\gamma_\ell-2}$
grows with $\ell$, and since $\ell$ in principle can be of an order proportional to $k$ this factor can 
actually end up being more important than the exponential growth. This has two consequences: (1) eventually, for large $k$ 
the number of BPs with a very large number of loops (proportional to $k$) will completely dominate in numbers those of 
small $\ell$ and (2) this rapid growth (factorial, not exponential) is the reason that the partition function $Z(g,\Lambda)$ 
given by eq.\ \rf{final} is only given by an asymptotic expansion which is not convergent, and (as we argued)  thus does not 
uniquely define $Z(g,\Lambda)$. The double scaling limit is an attempt to take a limit which tries to make a compromise
between allowing the number of links to go to infinity (which is needed if we want to associate any continuum physics
to BPs), i.e. to let $g \to g_c$ and thus $\Delta \to 0$, and at the same time allow BPs of arbitrary high loop 
number $\ell$ to play a role. Clearly taking $\Lambda \to \infty$ suppress graphs with a large number of loops, but 
the double scaling limit is the only one where we in principle can have graphs with an infinite number of links co-existing 
together with graphs having an infinite number of loops.
\begin{enumerate}\setcounter{enumi}{10}   
    \item Finally: show that starting out with any BP (weights $w_3, w_4, \ldots$) where $\gamma = \half$, the leading higher loop diagrams reduce precisely to the $\phi^3$ diagrams we have already considered (just with changed $w_3 \neq 1$).
\end{enumerate}

\vspace{12pt}

\noindent        {\bf Lesson:}
 We have given a perturbative definition of $Z(g,\Lambda)$ and we have found the expansion, and even explicit functions (the Airy functions) which reproduce this expansion. However, the $Z(g,\Lambda)$ is not uniquely fixed by its asymptotic expansion. In order to completely fix it we need a {\it non-perturbative} definition (which we do not have).

This example illustrates in a quite precise way the problem encountered in string theory, 
where one has a well-defined expansion in genus of the worldsheet (the equivalent to our expansion in loops), 
but is lacking a non-perturbative definition of string theory itself. We can even go one step further 
and study the BP equivalence to the attempts in string theory to find a  
non-perturbative definition of the theory.

The graphs we have studied from a combinatorial point of view are basically $\phi^3$-graphs. It should thus not come as 
a surprise that the  defining equation can be derived from a  $\phi^3$ ``field theory''. We write ``field theory'' because we will 
only keep the zero-dimensional real number $\phi$ in the path integral, not the real field $\phi(x)$. In this way the 
path integral will just generate the graphs, but propagators will be trivial equal to a number (which we choose to match 
the $1/g$ we assigned to each link in the combinatorial approach.

We define the partition function of the $\phi^3$-graphs to be 
\beq\label{p9js1}
\Omega(g,j,\Lambda) = \int d\phi  \; e^{-S(\phi)},\qquad  
S(\phi) = \Lambda\Big( \frac{1}{2}\, g\, \phi^2 \mi \frac{1}{3}\,\phi^3 \mi(1+j)\phi \Big)  
\eeq 
We are already here facing the problem that the integral is ill-defined if we simply integrate $\phi$ along the real axis. 
However, for the moment we will ignore this.
We obtain the Feynman graphs corresponding to $\Omega$ by expanding 
$e^{\Lambda( \phi^3/3+(1+j)\phi)}$ in powers of $\phi$ and performing
 the remaining Gaussian integral using Wick's theorem. This procedure {\it is} well defined to any finite order.
 If we consider a connected {\it tadpole} Feynman graph  (i.e. a graph coming from the interaction term  
 $\Lambda( \phi^3/3+(1+j)\phi)$ with one ``external'' vertex of order 1) 
 where the total number of vertices is $V$  and the number of links is $L$, it will have $\ell$ loops, where 
 \beq\label{p9js2}
 \ell = L -V+1
 \eeq
 Since a factor $\Lambda$ is associated to each vertex, except the external vertex, and a factor $1/\Lambda$ to each link, 
 we see that the total $\Lambda$-factor
 associated to a connected tadpole Feynman  graph with $\ell$ loops will be $1/\Lambda^\ell$.
The other coupling constants are chosen such that the connected {\it inequivalent} tree-graphs are assigned a weight 1 for each vertex
 of order 3 and weight (1+j) for each vertex of order 1, while each link is assigned a weight $1/g$, 
 such that we reproduce the standard BPs.  
 
 \vspace{6pt}
 
 In this field theoretical language we have:
 \beq\label{p9js3}
 \langle \phi \rangle := \frac{1}{\Omega}  \;  \int d\phi \; \phi \; e^{-S(\phi)} = 
 \frac{1}{\Lambda} \, \frac{1}{\Omega}  \frac{d\Omega}{dj} :=Z\qquad 
 \Big({\rm i.e.} \quad \frac{1}{\Lambda}\,\frac{d \Omega}{dj}=  Z\, \Omega\Big)
 \eeq 
Our partition function $Z$ for BPs with one external  vertex is precisely $\langle \phi \rangle$. Had we been in higher 
dimensions the vertex would have a coordinate $x$ and we would have $\langle \phi(x) \rangle$. Similarly we can write
\beq\label{p9js4}
 \langle \phi^2 \rangle := \frac{1}{\Omega}  \;  \int d\phi \; \phi^2 \; e^{-S(\phi)} =  
 \frac{1}{\Lambda^2} \frac{1}{\Omega}  \frac{d^2\Omega}{dj^2} =Z^2 + \frac{1}{\Lambda} \, \frac{d Z}{d j}.
 \eeq 
The so-called Dyson-Schwinger equation states that the expectation value of the classical eom is zero. The classical eom is 
\beq\label{p9js5}
0= \frac{ d \, S}{d \phi} = \Lambda \, \Big( g \phi \mi \phi^2 \mi (1\plu j) \Big) .
 \eeq
 Using eqs.\ \rf{p9js3} and \rf{p9js4} we obtain
 \beq\label{p9js6}
 g Z -\Big(Z^2 \plu \frac{1}{\Lambda} \, \frac{d Z}{d j}\Big) - (1\plu j) = 0,
 \eeq 
 which is precisely our fundamental graphical equation \rf{p9jh3} for BPs. Let us for completeness derive the DS equation:
 \beq\label{p9js7}
 0 = \int d\phi \; \frac{d}{d\phi}  e^{-S(\phi)} = -  \int d\phi \; \Big(\frac{d S}{d\phi}\Big) \; e^{-S(\phi)} ,
 \quad {\rm thus} \quad \left \langle \frac{d S}{d\phi} \right \rangle =0.
 \eeq
 
 We now have the following situation: we have a perturbative expansion of graphs defined by  the integral \rf{p9js1}. The integral 
 itself is ill defined when the integration contour is along the real axis, but the perturbative expansion makes sense to 
 any (finite) order, by expanding the interaction  $e^{\Lambda( \phi^3/3+(1+j)\phi)}$ in powers of $\phi$ and performing the 
 remaining Gaussian integration (as already remarked above). Is it possible to make the complete integral well defined 
 and in this way arrive at a non-perturbative definition of the theory? Yes, in fact it is easy: rotate the integration contour 
 by $\pi/6$ in the complex $\phi$ plane. We thus make the substitution $\phi \to e^{i \pi/6} \phi$. In this way we still integrate 
 over real $\phi$, but the action is changed to 
 \beq\label{p9js8}
 S_{\rm mod}(\phi) = \Lambda \Big( \frac{g}{4}(1+i \sqrt{3}) \phi^2 - \frac{1+j}{2} (  \sqrt{3} + i) \phi -\frac{i}{3} \phi^3\Big)
 \eeq
 and 
 \beq\label{p9js9}
 \Omega_{\rm mod} = e^{i \pi/6}\int_{-\infty}^{\infty} d\phi \; \e^{-S_{\rm  mod}(\phi)}
 \eeq
 is well defined. Further it is easy to show that the  expectation value of $\phi^n$ calculated {\it perturbatively} to a finite 
 order is unchanged, since in such a calculation we have just performed a well defined rotation of the contour of integration.
 One can also directly check that the factors of $e^{i \pi/6}$ cancel between vertices and links. 

\vspace{6pt}
 
 Thus it seems as if we have managed to define the summation over BPs non-perturbatively. However, it turns out that 
 $ \Omega_{\rm mod}$ defined in this way is complex, and the non-perturbative contributions to $\langle \phi^n \rangle$ will 
 typically be complex. Clearly we do not really want complex contributions and it reflects that the non-perturbative definition mentioned
 is not really based on any physical principle. Such a principle is presently missing, both for our BPs and for string theory.

\newpage

\setcounter{equation}{0}
\setcounter{figure}{0}
\renewcommand{\thefigure}{ps-10.\arabic{figure}}
 \renewcommand{\theequation}{ps-10.\arabic{equation}}

\subsection*{Elementary Quantum Geometry Problem Set 10}

The purpose of this exercise is to show that the characterization of criticality for ensembles of polygon graphs 
can be done in a way very similar to what we did for BPs and also to generalize the generic 
behavior in the notes  to so-called multicritical behavior, again as for BPs. 

\subsubsection*{A general even potential $\mathbf{V(x)}$}

We use the notation
\begin{align}
    \label{eq:vx}
    V(x) = \frac{1}{g} \sum_n t_n x^{2n} = \frac{1}{g} \tilde{V}(x)  
\end{align}
and we will, for a start, consider the $t_n$ as fixed such that we can only vary $g$. {\it 
Note that the notation of this $g$ is somewhat  different from the $g$ in the notes.}
Recall that the generic behavior was obtained if $t_1 >0$ and $t_n \leq 0$ for $n>1$ (and at least one of 
these $t_n < 0$).

Consider 
\begin{align}
    \oint_C \frac{d\omega}{2\pi i} \frac{f(\omega)}{\sqrt{\omega^2 \mi a^2}} 
\end{align}
where $f(\omega)$ is analytic in a region $\Omega$ including the cut $[-a,a]$ and the contour $C$ encircles the 
cut and is located in $\Omega$.
\begin{enumerate}
    \item Show that
        \begin{align}
            \oint_C \frac{d \omega}{2\pi i} \frac{f(\omega)}{\sqrt{\omega^2\mi a^2}}=\int_{-a}^a \frac{dx}{\pi} \frac{f(x)}{\sqrt{a^2\mi x^2}} = \int_{-1}^1 \frac{dy}{\pi} \frac{f(ay)}{\sqrt{1\mi y^2}} 
        \end{align}
        Hint: contract $C$ to be just above and below the cut and use that
        \[ \omega = x \pm i \epsilon \implies \frac{1}{\sqrt{\omega \mi a}} = \mp i \frac{1}{\sqrt{a\mi x}}\quad {\rm for} \quad x \in ]-a,a[ \]
 \end{enumerate}

 \noindent       For integer $k > 0$  in 
        \[ \oint_C d\omega \frac{f(\omega)}{\left(\omega^2 \mi a^2\right)^{k+\half}} \]
        one cannot simply contract the contour to the cut because the integral 
        \[ \int_{-a}^a \frac{dx}{\pi} \frac{f(x)}{(a^2\mi x^2)^{k+\half}} \]
        is singular. However, we have
        \begin{align*}
            \oint_C d\omega \frac{f(\omega)}{\left(\omega^2\mi a^2\right)^{k+\half}} &= \frac{1}{k\mi \half} \cdots \frac{1}{\half} \left(\frac{d}{da^2}\right)^k \oint_C d\omega \frac{f(\omega)}{\left(\omega^2\mi a^2\right)^{\half}}  \\
          &  = \frac{1}{k\mi \half} \cdots \frac{1}{\half} \int_{-1}^1 dy \frac{\left(\frac{d}{da^2}\right)^k f(ay)}{\sqrt{1\mi y^2}}
        \end{align*}

\begin{enumerate}\setcounter{enumi}{1}
    \item Show from the formulas in the notes for the disk amplitude $W(z)$,  
    that for an even potential we have
        \begin{align}
            W(z) = \left(\int_0^a \frac{dx}{\pi} \frac{x V'(x)}{(z^2\mi x^2)\sqrt{a^2\mi x^2}}\right) \sqrt{z^2\mi a^2} 
        \end{align}
        
        \item 
        Show that for $V(x)$ given by \eqref{eq:vx} the condition $W(z) \to \frac{1}{z}$ for $|z| \to \infty$ leads to 
        \begin{align}
            g(a^2) = \int_0^a \frac{dx}{\pi} \frac{x \tilde V'(x)}{\sqrt{a^2\mi x^2}} 
        \end{align}
 \end{enumerate}       
 
 \noindent
        {\it This determines $g$ as a function of the position of the cut $a$.}

\begin{enumerate}\setcounter{enumi}{3}
    \item Show that
        \begin{align}\label{p10jg1}
      g(a^2)=      \int_0^a \frac{dx}{\pi} \frac{x \tilde{V}'(x)}{\sqrt{a^2\mi x^2}} = \sum_n \frac{t_n a^{2n}}{B(n, \half)}
        \end{align}
        Hint: use the integral below and the definition of the beta-function:
        \[ \int_0^{\frac{\pi}{2}} d\theta \sin^{2n}\theta = \frac{\pi}{2} \cdot \frac{(2n\mi 1)!!}{(2n)!!} ,
     \quad   \qquad B(x,y) := \frac{\Gamma(x)\Gamma(y)}{\Gamma(x\plu y)}\]
  \end{enumerate}
   
 \noindent      {\it This fixes $g(a^2)$ as a polynomial, knowing the poynomial $\tilde V(x)$.}
 
 \vspace{6pt}

\noindent        Consider the simplest situation where
        \[ t_1 \equ \half, \quad t_2 \equ - \frac{1}{4} \quad \quad \implies \tilde V'(x) = x \mi x^3  \]

\begin{enumerate}\setcounter{enumi}{4}
    \item Show that
        \[ g(a^2) = \frac{1}{4} a^2 \mi  \frac{3}{16} a^4 \]
        \[ g(a^2) = g(a_c^2) \mi  \frac{3}{16}(a_c^2\mi a^2)^2 =\frac{1}{12}\mi \frac{3}{16}\Big(\frac{2}{3}\mi a^2\Big)^2\]
 \begin{figure}[!ht]
\centerline{\includegraphics[width=0.6\linewidth, angle = 0]{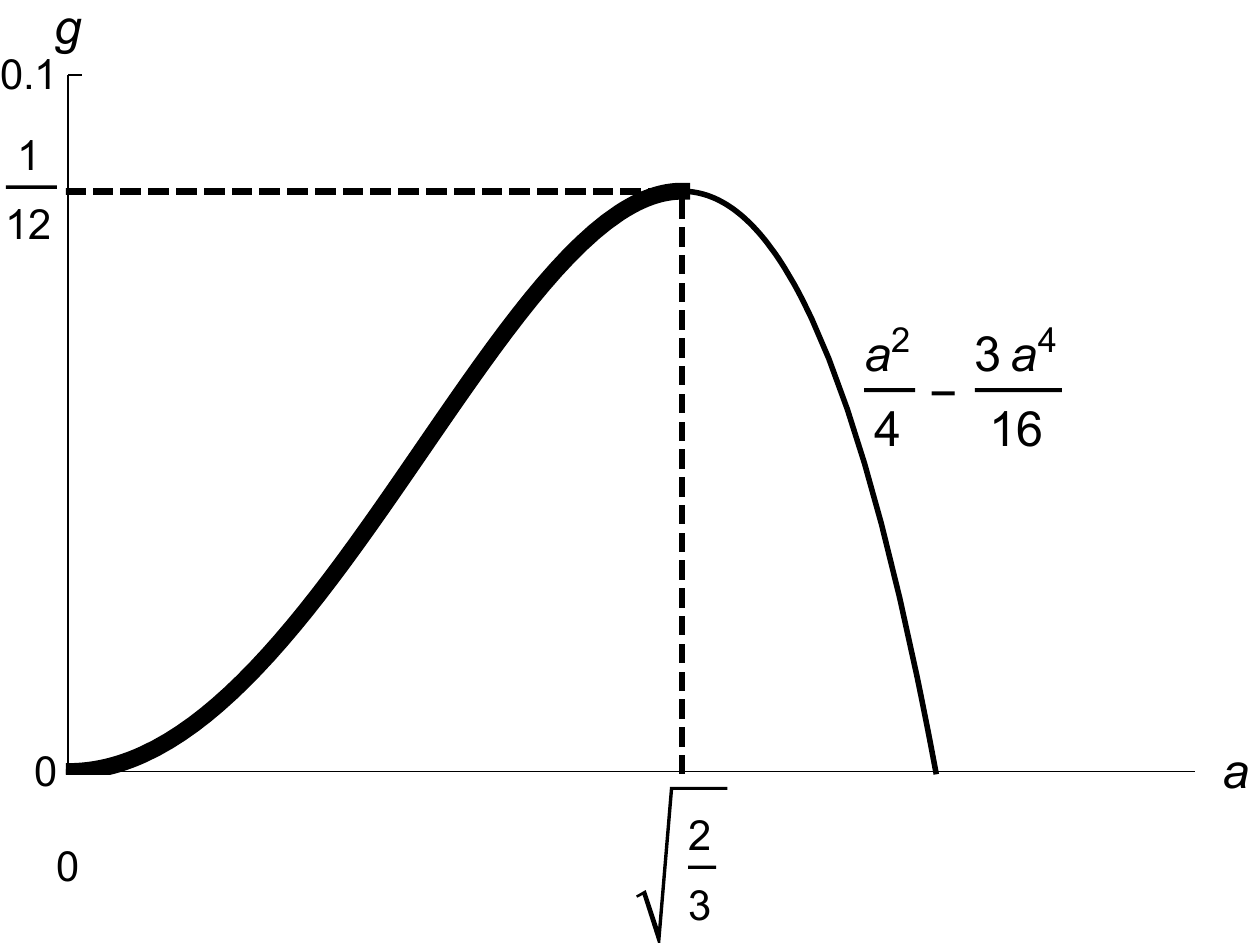}}
\caption{\footnotesize{The Physical region of the curve is from $(a,g) = (0,0)$ to $(\sqrt{2/3},1/12)$, shown in thick black.}}
\end{figure}
\end{enumerate}

\noindent
Thus we   start at $g \equ0$ for $a^2\equ 0$ and reach $g_c$ where $\frac{dg}{da^2} \equ  0$ for $a_c^2 \equ \frac{2}{3}$.
For larger $g$ we have no solution $g(a^2)$. The situation is thus somewhat similar to the BP case (except 
that the $g$ used here is more like the $e^{-\mu} \equ 1/g_{BP}$: 
$g \to 0$ corresponds to the partition function
going to zero, as does $\mu \to \infty$ for BPs. This is why the curve turns downwards on the figure rather 
than upwards as in the BP case): the critical behavior is obtained when $g$ is approaching it maximal 
value $g_c$ and the continuum limit is obtained by expanding around that maximum.

\begin{enumerate}\setcounter{enumi}{5}
    \item Show that we have the same qualitative behavior for $t_1 > 0, t_n \leq 0$ for $n> 1$ and at least one $t_n < 0$.
\end{enumerate}

 \noindent  {\bf Thus: universality!} Close to $g_c$ we 
  have $g_c\mi g \approx (a_c^2\mi  a^2)^2$, or equivalently $a^2 \approx a_c^2 \mi  c \cdot \sqrt{g_c\mi g}$. 

\vspace{6pt}

\noindent        We see that the situation is very similar to the BP case, and inspired by BPs we can now define multicriticality by dropping the requirement that $t_n \leq 0$ for $n > 1$. We thus lose a strict probabilistic interpretation of the random triangulations. However, like for BPs it is often possible to view the negative weights as coming from some matter interacting with the random geometry.

\vspace{6pt}

\noindent
        Let us in the same way as for BPs define a multicritical point by
\beq\label{p10j1}       
            \frac{dg}{da^2}\Big|_{a^2 = a_c^2} = 0, \cdots, \left(\frac{d}{da^2}\right)^{m-1} g \Big|_{a^2=a_c^2} = 0, 
            \qquad 
            \left(\frac{d}{da^2}\right)^m g \Big|_{a^2=a_c^2} \neq 0
\eeq
        To satisfy this we require a polynomial of order at least $2m$ (we consider only even potentials $V(x)$).
        
 \begin{enumerate}\setcounter{enumi}{6}       
    \item Show that {\it if} the polynomial is of order $2m$ and we assume $t_1 \equ 1/2$, then we have:
        \beq\label{p10j2}
            g(a^2) = g(a_c^2) \mi c\,(a_c^2\mi a^2)^m, \qquad 
            g(a_c^2)=c\cdot a_c^{2m}, \quad c = \frac{1}{4m\, a_c^{2m-2}}
        \eeq
\end{enumerate}

\noindent        
The value $a_c^2 > 0$ can be chosen arbitrarily, but after that the coefficients $t_n, n > 1$ are completely fixed
(we already assumed $t_1 \equ 1/2$). 
{\it We choose $a_c^2 \equ 1$ from now on.} 
        
\begin{enumerate}\setcounter{enumi}{7}
    \item Show that with the choice $a_c\equ 1$ we have (in the case of the $2m$'th order polynomial)
        \beq
            \label{eq:tn}
            t_n = \frac{(-1)^{n-1}}{4m} \binom{m}{n} B\big(n,\frac{1}{2}\big) \quad \textrm{for } n \leq m \qquad
            t_n = 0 \quad \textrm{for } n > m
        \eeq
\end{enumerate}

\noindent {\it We thus have the multicritical behavior ($a_c^2 =1$)}
        \begin{align}\label{p10jh1}
            a^2 = 1\mi \Big(1\mi \frac{g}{g_c}\Big)^{\frac{1}{m}}, \quad g_c \equ \frac{1}{4m}
        \end{align}

\vspace{12pt}

\noindent        Define 
        \begin{align}
            \tilde M_k(a^2) = \oint_C d \omega \frac{\omega \tilde V'(\omega)}{(\omega^2\mi a^2)^{k+\half}} 
        \end{align}
        where $M_k \equ \frac{1}{g} \tilde M_k$, with $M_k$ as in the notes.
        
 \begin{enumerate}\setcounter{enumi}{8}
       
    \item Show that
        \begin{align}
            \tilde M_0(a^2) = 2 g(a^2)
        \end{align}
\end{enumerate}        
        
\noindent         Define $\displaystyle{\epsilon = a_c^2\mi a^2 \sim \left(1\mi \frac{g}{g_c}\right)^{\frac{1}{m}}}$.
 
 \begin{enumerate}\setcounter{enumi}{9}

    \item Show that $ \epsilon \to 0$
        \beq
            \label{eq:mk}
            M_k(a^2) = \mu_k \epsilon^{m-k}+O\left(\epsilon^{m-k+1}\right), 0 < k < m ,\qquad
            M_m(a^2)  \neq 0, \qquad 
        \eeq
\end{enumerate}

\noindent        The potential corresponding to the choice \eqref{eq:tn} of $t_n$ (where only $g$ is allowed to vary) is called the {\it Kazakov potential}, and varying $g$ we have a behavior like \eqref{eq:mk}. Taking $\epsilon \to 0$ (or $g \to g_c$) we approach the {\it $m$th-multicritical point} in a specific way.

\vspace{6pt}

 \noindent       A more general approach to the $m$th-multicritical point is obtained by also allowing $t_n \to t_n \plu \delta t_n$ 
 but in such a way that \eqref{eq:mk} is satisfied (with $\mu_k$ depending on $\delta t_n$). 
 We say that the choice $\mu_1, \cdots, \mu_{m-1}$ defines the approach to the $m$th-multicritical point. 
 One can show that the $\mu_k$'s are related to so-called intersection indices on Riemann surfaces.
 
 \vspace{6pt}
 
 We have now defined the so-called multicritical behavior, if we have a situation like (\ref{p10jh1}), with $m=2,3,\ldots$:
\begin{equation}\label{p10jh2}
a^2 = 1 \mi  \Big(1\mi \frac{g}{g_c}\Big)^{1/m},\quad g_c = \frac{1}{4m}, \qquad {\rm or} \quad  
g(a^2) = \frac{1}{4m} \mi  \frac{1}{4m}\;(1\mi a^2)^m.
\end{equation}
Let us now generalize the critical behavior, for $m < s < m\plu 1$ to 
\begin{equation}\label{p10jh3}
a^2 = 1 \mi \Big(1\mi \frac{g}{g_c}\Big)^{1/s},\quad g_c = \frac{1}{4s}, \qquad {\rm or} \quad  
g(a^2)=  \frac{1}{4s}\mi   \frac{1}{4s}\;(1\mi a^2)^s.
\end{equation} 
From eq.\ (\ref{p10jg1}) we can now find potential $\tilde{V}(x)$  by an expansion of $(1\mi a^2)^s$ in powers of $a^2$.
We have already made this expansion in problem 5, dealing with multicritical BPs.

\begin{enumerate}\setcounter{enumi}{10}

\item
Show that we have 
\begin{equation} 
t_n \sim \frac{ 1}{n^{s + 3/2}} \quad {\rm for} \quad n \to \infty.
\end{equation}
\end{enumerate}

Thus the potential $\tilde{V}(x)$ is given by an infinite power series in $x^2$. The corresponding function 
is a hypergeometric function. So whenever $s$ is non-integer we need triangulations which have vertices 
of arbitrarily high order, if the model shall reproduce a critical behavior like (\ref{p10jh3})

 \vspace{6pt}

\noindent        We end this exercise by proving an amazing universal result:
        \begin{align}
            \label{eq:uni}
            \frac{d gW(z)}{dg} = \frac{1}{\sqrt{z^2-a^2}} 
        \end{align}

 \noindent       This result is true for {\it any} even potential $V(x)$ of the form \eqref{eq:vx}. We call it an amazing 
        universal result, but it should maybe not come as big surprise, considering that we have already in the notes 
        proven that the two-loop function is universal in the sense that it only depends on $a$, and eq.\ \rf{eq:uni} 
        is essential the disk amplitude, differentiated after $g$. This differentiation corresponds to 
        putting a mark everywhere on the disk, and this is combinatorially the same as contracting one of the loops
        to a point. Thus one should be able to obtain eq.\ \rf{eq:uni}  from the expression for the two-loop 
        function and one can indeed do that (after some work...). In the notes we only dealt with positive 
        probabilities, i.e.\  $t_n \leq 0$ for $n \geq 2$, but the combinatorial argument would also be valid 
        if we dropped that restriction on the $t_n$'s.  Eq.\ \rf{eq:uni} has a direct translation to conformal field theories coupled to 2d quantum gravity in the scaling limit $g \to g_c$, $a \to a_c$ as we will discuss later.

\begin{enumerate}\setcounter{enumi}{11}

    \item Show that
        \begin{align}
            \label{eq:m1}
            \tilde M_1(a^2) = 4 \frac{dg}{da^2}, \quad \tilde M_{k+1}(a^2) = \frac{1}{k+\half} \;\frac{d}{da^2}\tilde M_k(a^2)
        \end{align}
    \item Show that we can write
        \beq
            g W(z) = \half \left[ \tilde V(z) - \tilde M(z) \sqrt{z^2\mi a^2} \right],\qquad
            \tilde M(z) = \sum_{k=1}^m \tilde M_k (z^2\mi a^2)^{k-1}.
        \eeq
        
    \item Use \eqref{eq:m1} to show \eqref{eq:uni}. Hint: 
        \bea
            \frac{dgW}{dg} &=& \frac{da^2}{dg} \frac{dgW}{da^2} 
            = - \half \frac{da^2}{dg} \left(\frac{d\tilde M(z)}{da^2} \sqrt{z^2-a^2} -  \frac{\tilde M (z)}{2\sqrt{z^2-a^2}}\right)
          \nonumber \\ 
            &=& \frac{1}{\tilde M_1} \frac{1}{ \sqrt{z^2-a^2}}   \left( \tilde M(z) - 2(z^2-a^2) \frac{d \tilde M(z)}{da^2} \right)
        \eea
\end{enumerate}

 \noindent       The continuum limit of \eqref{eq:uni} reads
        \begin{align}
            \frac{dW(Z)}{d\Lambda} = \frac{1}{\sqrt{Z\plu \sqrt{\Lambda}}} 
        \end{align}
        where 
        \begin{align*}
            z^2 &= a_c^2 + \epsilon Z, \quad a^2 = a_c^2 - \epsilon \sqrt{\Lambda} \\
            g-g_c  &\sim (a_c^2-a^2)^m \sim \epsilon^m \Lambda^{m/2}
        \end{align*}
        $W(z)$ is the disk amplitude with one marked point at the boundary and $Z$ the boundary cosmological constant. 
        Differentiating with respect to the cosmological constant $\Lambda$ corresponds to an insertion anywhere.
       It turns out that one obtains the same result in so-called quantum Liouville theory coupled to a 
       $(p,q)$ conformal field theory where the insertion is a specific so-called  conformal operator, 
       namely the so-called primary operator with largest negative dimension in the $(p,q)$ conformal theory.

\newpage

\setcounter{equation}{0}
\setcounter{figure}{0}
\renewcommand{\thefigure}{ps-11.\arabic{figure}}
 \renewcommand{\theequation}{ps-11.\arabic{equation}}

\subsection*{\hspace{-2mm} Elementary Quantum Geometry Problem Set 11}

\subsubsection*{Multi-Ising spins coupled to 2d gravity}

\subsubsection*{Physics of the Ising model on a regular lattice}
The partition function of the Ising model is  
\beq\label{p11-1}
Z(\bt) = \sum_{\{\sg_i\}} \e^{\frac{\bt}{2} \sum_{\LA ij  \RA} (\sg_i\sg_j -1)}, \quad \sg = \pm 1.
\eeq
where the Ising spins are placed at the vertices $i$  of the lattice, and neighboring spins interact. $\LA ij \RA$
denotes the link between site $i$ and $j$ if they are neighbors.   $\{\sg_i\}$ denotes the set of all spin configurations
and the summation in the action is over all links.

For an infinite lattice there exists a so-called critical $\bt_0$, such that for $\bt > \bt_0$ we have magnetization while 
for $\bt < \bt_0$ we have no magnetization. The phase transition at $\bt_0$ is a second order phase transition.

For very large $\bt$ (small temperatures, $T \equ 1/\bt$) almost all $\sg_i \equ1$ (or almost all $\sg_i \equ -1$). The excitations around the configuration where all $\sg_i \equ1$ are {\it small} spin clusters with $\sg_i \equ -1$. The reason that the spin clusters are small is that the energy $E=\bt/2 \sum_{\LA ij  \RA} (\sg_i\sg_j -1)$ is only different from zero when 
two neighboring spins are different, i.e.\ along the boundaries between regions of $\sg_i \equ1$ and $\sg_i \equ -1$.
For a regular lattice, starting with all spins $\sg_i \equ1$, a large region with $\sg_i \equ -1$ will also have a long boundary
(at least like $\sqrt{A}$, $A$ being the area of a region where $\sg_i =-1$). Thus large regions of $\sg_i \equ -1$ will be
suppressed for large $\bt$ if we start out in a state (the ground state) where all $\sg_i \equ 1$.  

As $\bt $ decreases towards $\bt_0$ the size of $\sg_i = -1$ spin clusters as well as the number of them will grow and 
at $\bt_0$ there is an equal number of $\sg_i =1$ and $\sg_i =-1$ spins, and the distributions of $\pm$ spin clusters 
will be the same and the cluster sizes can be large.

\subsubsection*{The Ising model on dynamical triangulations}  

\begin{figure}[!ht]
\begin{center}
\vspace{-1cm}
\includegraphics[width=14cm]{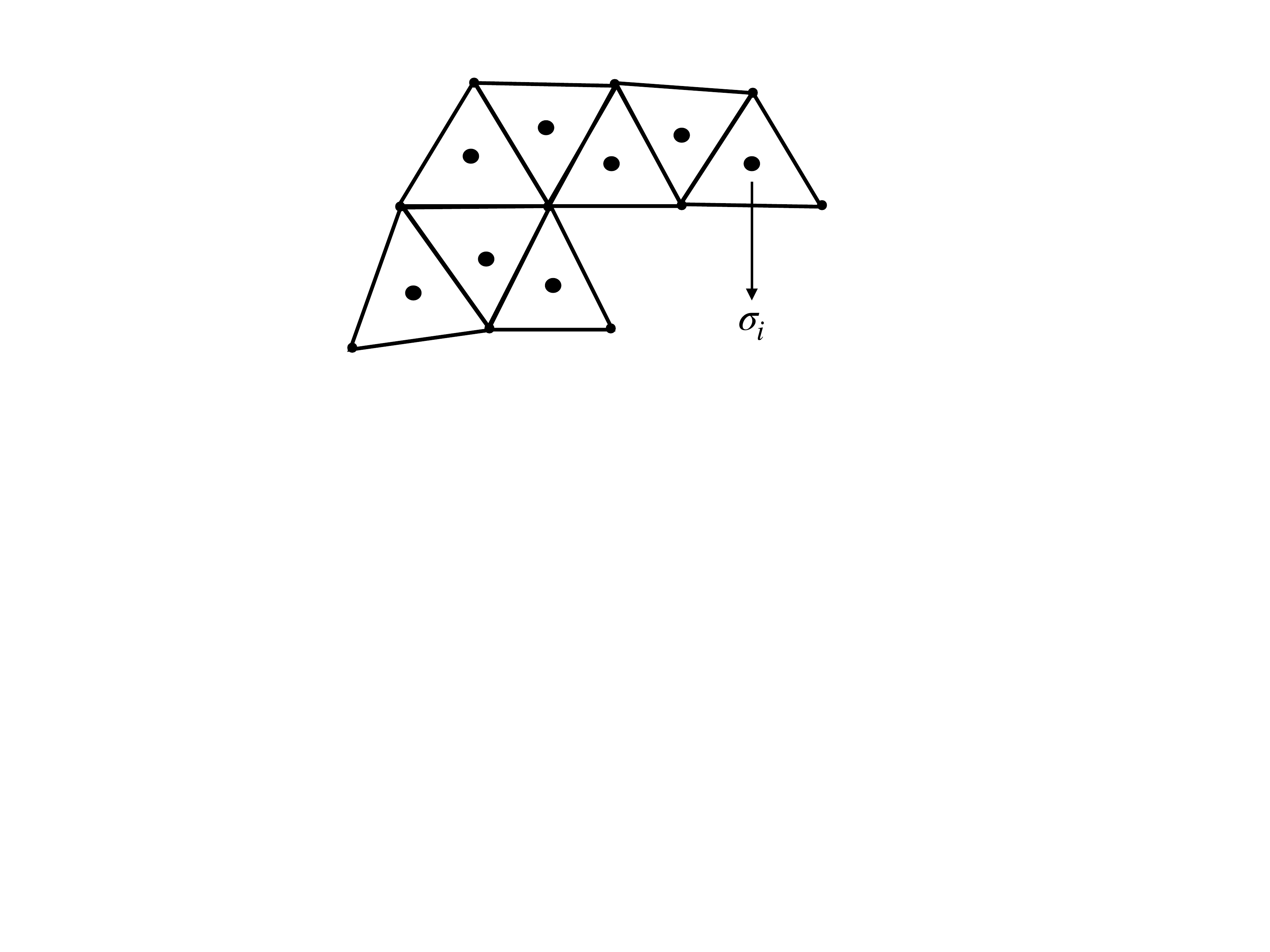}
\end{center}
\vspace{-7cm}
\caption{{\footnotesize Now the index $i$ refers to a triangle and the interaction is between neighboring triangles.}}
\label{fig1}
\end{figure}

Let us now consider Ising spins coupled to dynamical triangulations (DT). We put the spins at the center of the 
triangles (this is not essential, but convenient here) as illustrated in fig.\ \ref{fig1}.

The partition function is defined as 
\beq\label{p11-2}
Z(\mu,\bt) = \sum_T \e^{-\mu N_T} Z_T (\bt).
\eeq
where the summation is over a suitable class of triangulations. $Z_T(\bt)$ refers to the partition 
function \rf{p11-1}, defined as mentioned on the graph corresponding to the triangulation $T$.

A few facts about this model (which we are not going to prove). For each $\bt$ there is a critical $\mu_0(\bt)$.
In addition there is a $\bt_0$ such that for $\bt > \bt_0$ there is magnetiztion and for $\bt < \bt_0$ there is 
no magnetiztion. The phase transition at $\bt_0$ is {\it third order} on the ensemble of DT. Also the critical 
exponents $\alpha,\bt_m,\gm_m$ for the magnetic system at $\bt_0$ are different from the famous Onsager
exponents on a regular lattice.  {\it So the DT ensemble of geometries influences the critical properties of the 
spin system.} One can express the size of the spin clusters at the critical $\bt_0$ as a function of the critical 
exponents, i.e.\ the fractal properties of spin clusters of the Ising model change on dynamical triangulations.

{\it In addition the spin system influences the critical properties of geometry, but only for $\bt= \bt_0$, i.e.\ when the 
spin system itself is critical.} We know that for pure gravity (DT without matter), the susceptibility exponent is 
$\gm = -1/2$. For $\bt \neq \bt_0$ this is still true for the combined system. However, for $\bt= \bt_0$ one finds 
$\gm (\bt_0) = -1/3$. Thus the long range interactions of the large spin clusters also change the fractal structure 
of the dynamical triangulations. 

\subsubsection*{The mean field model}

\begin{figure}[!ht]
\vspace{-1.0cm}
\begin{center}
\includegraphics[width=14cm]{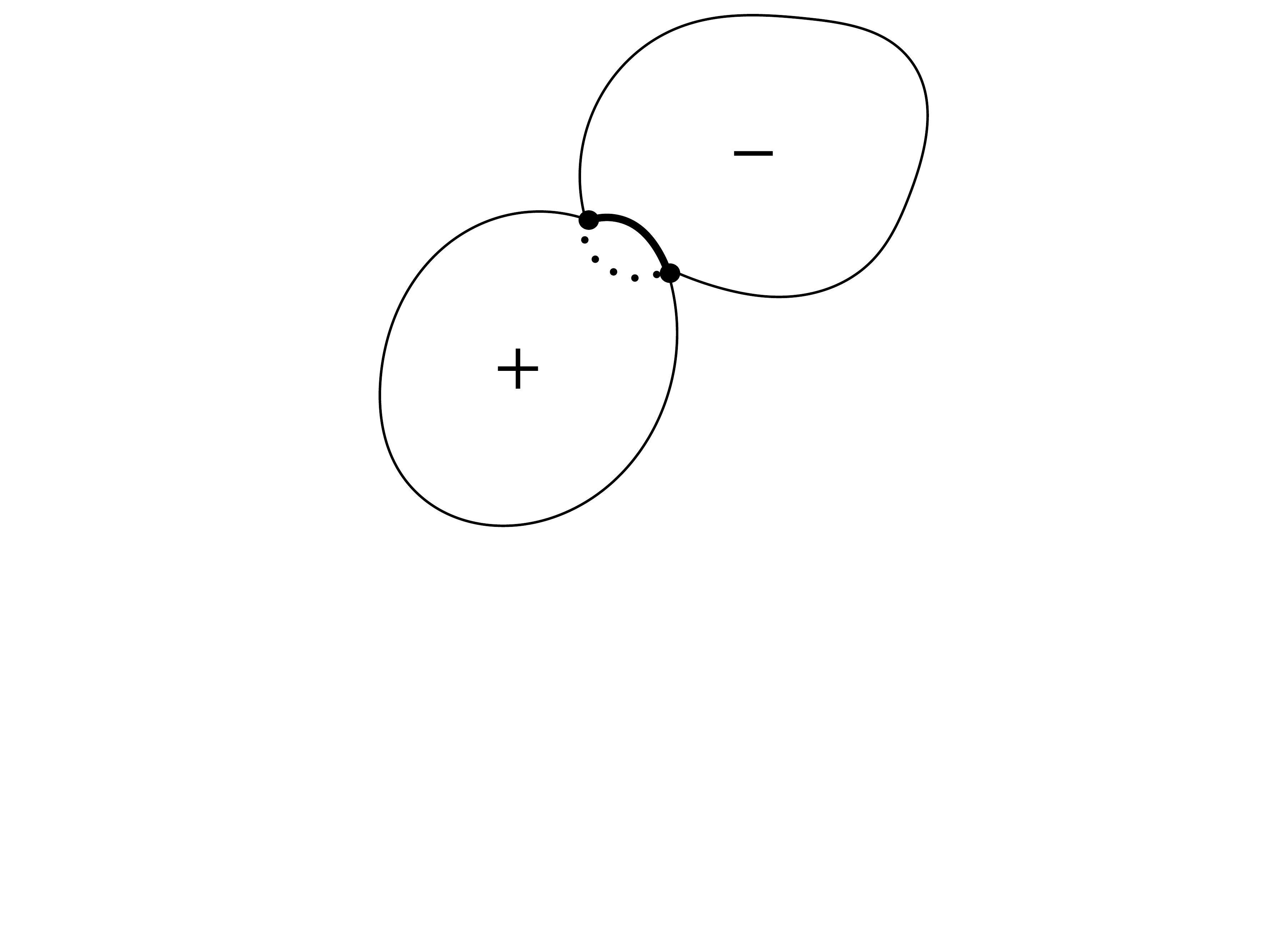}
\end{center}
\vspace{-5cm}
\caption{{\small Two macrospic spin clusters of $\pm$ spin, separated by a minimal boundary.}}
\label{fig2g}
\end{figure}

The purpose of this exercise is to understand this interplay between geometry and matter in a simple ``mean-field'' 
model. The starting point is that the minimal boundaries separating $\pm$ spin clusters can be very different 
from those on a regular lattice as illustrated in fig.\ \ref{fig2g}. So for {\it some} geometries it is possible to have
huge spin clusters separated by small boundaries, i.e.\ small energy. Conversely, the relative weight of these 
geometries will be enhanced in the combined matter-geometry ensemble relative to 
more regular triangulations which do not have such a ``pinching'',  simply because the 
energy of the matter part will be small. 

\begin{figure}[!h]
\vspace{-1cm}
\begin{center}
\centerline{\includegraphics[width=14cm, angle=0]{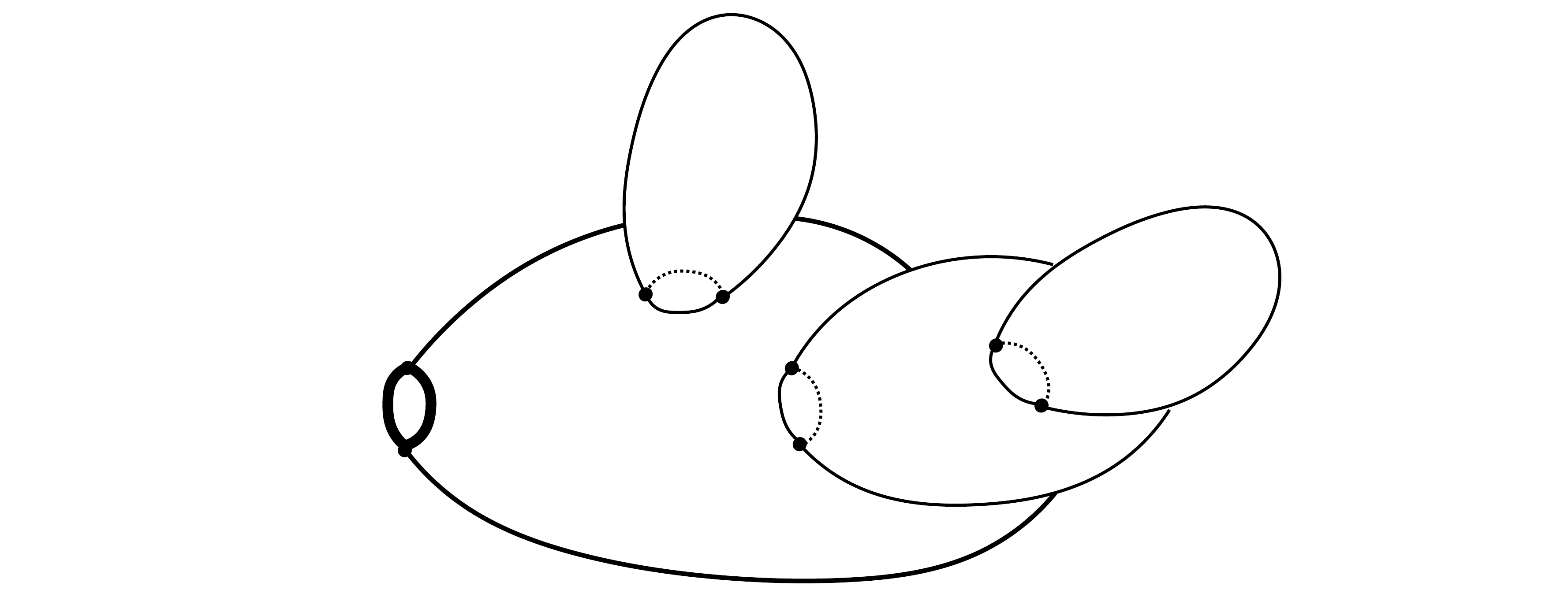}}
\end{center}
\vspace{-1cm}
\caption{{\footnotesize The hierarchical structure  of baby universes separated from their parent universe by only two links,
relative to the boundary.}}
\label{fig3}
\end{figure}

Let us now consider the following toy model designed to capture this: we allow two links to be connected to 
the same two vertices, but only if cutting the triangulation  along the two links separates the triangulation in 
two disconnected parts (as illustrated in fig.\ \ref{fig2g}). We now consider triangulations which have one boundary,
and this boundary consists only of two links. We can now make a decomposition of the triangulation by peeling 
away ``baby universes'' connected to the rest of the surface by only two links and then closing the links. The fact 
that we have a boundary makes this a systematic procedure, the ultimate ``parent'' universe being connected
to this boundary (see fig.\ \ref{fig3}). This class of triangulations is denoted ${\cal T}^{(2)}$ in the notes. 
We now only sum over spin configurations which are such that the spin of a baby universe component is either 
$+$ or $-$. This approximation is inspired by fig.\ \ref{fig2g} and is expected to be a good approximation for 
large $\bt$. It is also expected to be good for somewhat smaller $\bt$ if we have many ``independent'' Ising models
coupled to the ensemble ${\cal T}^{(2)}$. We write ``independent'' because different copies of the Ising spins 
do not interact directly, but they interact indirectly via the common geometry which they influence. The model is 
expected to be correct all the way down to a critical $\bt_0$ if $n$, the number of independent Ising spins, 
is sufficiently large. For $n\equ 1$ it is only an approximation, which does not give the correct critical exponents
(like ordinary mean-field theory in 2 and 3 dimensions for spin systems). 

\begin{figure}[!h]
\vspace{-1cm}
\begin{center}
\centerline{\includegraphics[width=12cm, angle=0]{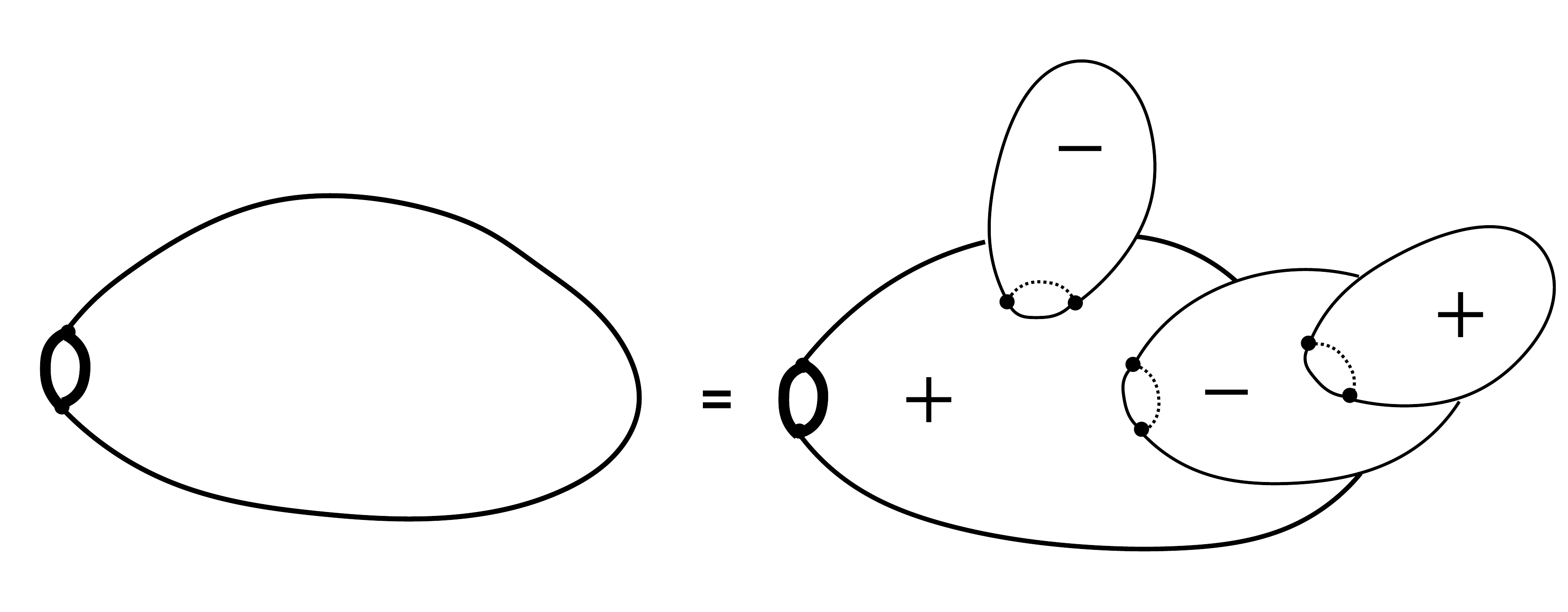}}
\end{center}
\vspace{-1.5cm}
\caption{{\footnotesize Illustration of the spin configurations included in the summation in eq.\ \rf{p11-1a}.}}
\label{fig4}
\end{figure}

We can now write down the one-loop function 
\beq\label{p11-1a}
G(\mu,\bt) = \sum_{ T \in {\cal T}^{(2)}(2)} \e^{-\mu N_T} {\sum_{\{\sg_i\}}}' \e^{\frac{\bt}{2} \sum_{\LA ij  \RA} (\sg_i\sg_j -1)}
\eeq
where ${\cal T}^{(2)}$ refers to the class of triangulations discussed and ${\cal T}^{(2)}(2)$ refers to this class with 
a boundary consisting of 2 links. The $\sum'$ refers to the summation over the restricted class of spin configurations 
we mentioned above. Eq.\ \rf{p11-1a} is illustrated in Fig.\ \ref{fig4}.

For a given $\bt$ the model \rf{p11-1a} has a critical $\mu_0(\bt)$, such that the sum is convergent for $\mu > \mu_0(\bt)$
and divergent for  $\mu < \mu_0(\bt)$.

We now define:
\beq\label{p11-2a}
\chi(\mu,\bt) = - \frac{\prt G(\mu,\bt)}{\prt \mu}
\eeq

If there is no Ising spin, i.e.\ we have our original pure gravity model from the notes, we define
\beq\label{3a}
G_0(\mu) = \sum_{ T \in {\cal T}^{(2)}(2)} \e^{-\mu N_T},\qquad
\chi_0 (\mu) =  - \frac{\prt G_0(\mu)}{\prt \mu}
\eeq
and we denote the corresponding critical $\mu$ by $\mu_0$. We know that since $\gm_0= -1/2$ 
for the model without Ising spin that 
\bea 
\chi_0(\mu) &=& a_1 + a_2 (\mu\mi \mu_0)^{1/2} + \cdots \label{p11-4a}\\
G_0(\mu) &=& b_1 + b_2 (\mu\mi \mu_0) + b_3 (\mu\mi \mu_0)^{3/2} + \cdots \label{p11-5a}
\eea
and we write 
\beq\label{6a}
\chi(\mu,\bt) = c_1 + c_2 (\mu\mi \mu_0(\bt)) +\cdots + c (\mu \mi \mu_0(\bt))^{-\gm(\bt)} + \cdots
\eeq

\vspace{6pt}
{\bf We want to determine {\boldmath${\gm(\bt)}$}}
\vspace{12pt}

\begin{itemize}
\item[1)] Show, by first summing over the spin of baby universes, that 
\bea 
 G(\mu,\bt)\! \!&=& \!\!\sum_{ T \in {\cal T}^{(2)}(2)} \e^{-\mu N_T}\Big( 1 \plu  \e^{-2\bt} G(\mu,\bt ) \Big)^{N_L(T)-2} \nonumber\\
&&\!\! \sum_{ T \in {\cal T}^{(2)}(2)} \e^{-\mu N_T}\Big( 1 \plu \e^{-2\bt} G(\mu,\bt ) \Big)^{3N_T/2-1}\label{p11-7a}
\eea
We will ignore the $-2$ and $-1$ in the powers, since we will solve the model close to criticality where $N_T$ 
and $N_L$ are large.
\item[2)] show using \rf{p11-7a} that we have 
\beq\label{p11-8a}
G(\mu,\bt) = \sum_{ T \in {\cal T}^{(2)}(2)}  \e^{-\bmu N_T} = G_0(\bmu),
\eeq
\beq\label{p11-9a}
\bmu= \mu - \frac{3}{2} \,\log \Big ( 1 \plu  \e^{-2\bt} G(\mu,\bt ) \Big),
\eeq

\item[3)] Show that 
\beq\label{p11-10a}
\mu= \bmu + \frac{3}{2} \,\log \Big ( 1 \plu \e^{-2\bt} G_0(\bmu ) \Big).
\eeq

\item[4)] Show that 
\beq\label{p11-11a}
\frac{\prt \mu}{\prt \bmu} = \frac{1-\frac{3}{2} \e^{-2\bt} \chi_0(\bmu) + \e^{-2\bt} G_0(\bmu)}{1+\e^{-2\bt} G_0(\bmu)}
=  \frac{ \e^{2\bt}-\big(\frac{3}{2} \chi_0(\bmu) - G_0(\bmu)\big)}{\e^{2\bt}+ G_0(\bmu)}
\eeq
and thus that 
\beq\label{p11-12a}
\frac{\prt \bmu}{\prt \mu} = \frac{\e^{2\bt} +G_0(\bmu)}{\e^{2\bt} -\big(\frac{3}{2}  \chi_0(\bmu) \mi  G_0(\bmu)\big)}
\eeq

\item[5)] Show, using $G(\mu,\bt) \equ G_0(\bmu)$ (see \rf{p11-8a}), that 
\beq\label{p11-13a}
\chi(\mu,\bt) = \chi_0 (\bmu) \; \frac{\prt \bmu}{\prt \mu}
\eeq

\item[6)] Show from the definition of $G_0(\mu)$ and $\chi_0(\mu)$ that $\frac{3}{2} \chi_0(\bmu) \mi G_0(\bmu)$ 
is a decreasing function of $\bmu$, with its maximum for the smallest possible value one can use in 
$G_0$ and $\chi_0$, namely the critical value $\mu_0$, i.e. $\bmu \equ \mu_0$. Note that both $\chi_0(\mu_0)$ and 
$G_0(\mu_0)$ are finite at $\mu_0$, according to \rf{p11-4a} and \rf{p11-5a}.

\item[7)] Show that this implies that there exists a $\bt_0$ such that the denominator in \rf{p11-12a} is $>0$ for all 
$\mu > \mu_0(\bt)$ provided $\bt > \bt_0$.

\item[8)] Let now $\bt > \bt_0$ be fixed. We have a critical point $\mu_0(\bt)$. Let us now decrease $\mu$ towards this 
critical point. The susceptibility satisfies \rf{p11-13a}, which we write in detail as 
\beq\label{p11-13b}
\chi(\mu,\bt) = \chi_0 (\bmu(\mu,\bt)) \; \frac{\prt \bmu(\mu,\bt)}{\prt \mu}.
\eeq
Use this formula and the arguments above to argue that $\chi(\mu,\bt)$ can only be critical if 
\beq\label{p11-14a}
\bmu(\mu_0(\bt),\bt) = \mu_0,
\eeq
and that this implies that 

\beq\label{p11-15a}
\framebox{ $\gm(\bt) = \gm_0 = -1/2~~ {\rm for} ~~\bt > \bt_0$.}
\eeq

\end{itemize} 

$\bt_0$ is the largest $\beta$ for which 
\beq\label{p11-16a}
\e^{2\bt} = \frac{3}{2} \chi_0(\bmu) -G_0(\bmu)
\eeq
has a solution. For $\bt \leq \bt_0$ we now define $\bmu_0(\bt)$ by
\beq\label{p11-17a}
\frac{\prt \mu}{\prt \bmu}\Big|_{\bmu_0(\bt)} = 0.
\eeq
Thus, according to \rf{p11-11a}, $\bmu_0(\bt)$ satisfies eq.\ \rf{p11-16a}.

\begin{itemize}

\item[9)] Show that  we have 
\beq\label{p11-18a}
\bmu_0(\bt_0 )= \mu_0,\qquad \quad \bmu_0(\bt) > \mu_0\quad {\rm for } \quad \bt < \bt_0.
\eeq

\item[10)] Show that \rf{p11-17a} implies that 
\beq\label{p11-19a}
\mu \mi  \mu_0(\bt) = c\, \big(\bmu \mi  \bmu_0(\bt)\big)^2 + O\big((\bmu \mi \bmu_0(\bt))^3\big).
\eeq

\item[11)] {\it Assume now that $\bt < \bt_0$.} Use \rf{p11-13a}, \rf{p11-18a} and \rf{p11-19a} to show, by Taylor expanding 
around $\bmu_0(\bt)$ ( $> \mu_0$, so one {\it can} Taylor expand), that for $\mu \to \mu_0(\bt)$
 we have 
 \beq\label{p11-20a}
 \chi(\mu,\bt) \sim \frac{1}{ \bmu\mi \bmu_0(\bt)} \sim \frac{1}{ \sqrt{\mu\mi \mu_0(\bt)}} \qquad \bt < \bt_0.
 \eeq

\end{itemize}
 Thus 
  \beq\label{p11-20b}
\framebox{ $\gm(\bt) = 1/2~~ {\rm for} ~~\bt < \bt_0$}, \quad {\rm like~BPs!}
\eeq

We have now seen that there is a phase transition at $\bt_0$, where the critical exponent $\gm(\bt)$ jumps 
from $-1/2$, the value for pure gravity without Ising spins, for $\bt > \bt_0$, to $\gm(\bt) \equ 1/2$, the value 
for BPs, for $\bt < \bt_0$.

We will finally determine $\gm(\bt_0)$. We can no longer Taylor expand for $\mu \to \mu_0(\bt)$ because 
$\bmu_0(\bt_0) \equ \mu_0$ (see \rf{p11-18a}), and $\chi_0(\mu_0)$ and $G_0(\mu_0)$ are not analytic in that point. 
However, we know their behavior there, see \rf{p11-4a} and \rf{p11-5a}.

\begin{itemize}

\item[12)] Show that 
\beq\label{p11-21a}
\frac{\prt \mu}{\prt \bmu} \sim (\bmu \mi  \bmu_0(\bt_0))^{-\gm_0} + O(\bmu-\mu_0)
\eeq

\item[13)] Show 
\beq\label{p11-22a}
\mu \mi  \mu_0(\bt_0) = c \,(\bmu \mi  \mu_0)^{1-\gm_0}+ O\big((\bmu\mi \mu_0)^2\big)
\eeq

\item[14)] Show, using \rf{p11-13a}, that this implies 

\beq\label{p11-23a}
\chi(\mu,\bt_0) \sim \frac{1}{ (\bmu\mi \mu_0(\bt_0))^{-\gm_0}}\sim \frac{1}{ (\mu\mi \mu_0(\bt_0))^{-\gm_0/(1-\gm_0)}}
\eeq
Thus
 \beq\label{p11-24a}
\framebox{ $\displaystyle{\gm(\bt_0) = \frac{\gm_0}{\gm_0\mi 1} = \frac{1}{3}}$}
\eeq

\end{itemize}
{\bf Summary:} For large $\bt$ ($\bt > \bt_0$), i.e.\ for low temperature, we have $\gm(\bt) \equ \gm_0$ (=  - 1/2). This 
is the magnetized phase where spin fluctuations are small, and the geometry is not affected by the spin. 
At $\bt \equ \bt_0$ there is a phase transition. At the transition $\gm(\bt)$ jumps to 1/3. For $\bt < \bt_0$ (high temperature)
there are many baby universes and $\gm(\bt) \equ 1/2$, like for BPs. In this phase there is no spontaneous magnetization
in accordance with the fact that BPs have no spontaneous magnetization. 

\vspace{6pt}

{\bf General remarks:} the high temperature phase of our model, where $\bt < \bt_0$, 
does not represent well a single Ising spin coupled to DT (as already mentioned). In the real, full model 
one has $\gm(\bt) = -1/2$ for $\bt < \bt_0$. Also, in the full model $\gm(\bt_0) = -1/3$ (and not $1/3$). However, again
as already mentioned, the model represents very well many Ising spins coupled to DT. 
The models with many Ising spins coupled to DT cannot be solved analytically, 
but have been studied by computer simulations. 

\vspace{6pt}

{\bf Finally:} Note that the physics of the magnetized baby universes seems amazingly similar to the physics of 
real magnets, the baby universes playing the role of magnetized domains.

\newpage

\setcounter{equation}{0}
\setcounter{figure}{0}
\renewcommand{\thefigure}{ps-12.\arabic{figure}}
 \renewcommand{\theequation}{ps-12.\arabic{equation}}

\subsection*{\hspace{-2mm} Elementary Quantum Geometry Problem Set 12}

\subsubsection*{Deriving the multiloop formulas}

The purpose of this problem set is to derive the the multiloop formulas \rf{5.62}, \rf{5.65} and \rf{5.66} using \rf{5.61}. We will simply 
use the representation \rf{5.52} for the loop insertion operator and act  on the disk function $w(\vg,z)$ written in the form 
\rf{5.49}, using the results \rf{5.51a}-\rf{5.51d}. Let us write \rf{5.62} in the following way
\beq\label{p12-1a}
w(\vg,\om,z) = \frac{1}{(z^2\mi \om^2)^2} \left( -2 z\om + 
\frac{2z^2\om^2 \mi c^2(z^2\plu \om^2)}{(z^2\mi c^2)^{1/2} (\om^2\mi c^2)^{1/2}}\right)
\eeq

\begin{itemize}

\item[(1)] Show that 
\beq\label{p12-1}
\frac{2}{\tM_1} \frac{d}{dc^2} \sum_{k=1}^\infty \tM_k (\om^2\mi c^2)^{k-1/2} = - \frac{1}{(\om^2\mi c^2)^{1/2}}.
\eeq

\item[(2)] Show that 
\beq\label{p12-2}
\frac{\prt}{\prt V(z) } \sum_{k=1}^\infty \tM_k (\om^2\mi c^2)^{k-1/2} = \frac{d}{dz} \Big[ \Big( \frac{\om^2\mi c^2}{z^2\mi c^2} \Big)^{1/2} 
\frac{z}{z^2-\om^2}\Big].
\eeq

\item[(3)] Use now \rf{5.51a} to write 
\beq\label{p12-3}
\frac{d w(\vg,\om) }{d V(z)} = \frac{-2\om z}{(z^2\mi \om^2)^2}  + \frac{c^2}{(z^2\mi c^2)^{\frac{3}{2}} (\om^2\mi c^2)^{\oh}} -   
\frac{d}{dz} \Big[ \Big( \frac{\om^2\mi c^2}{z^2\mi c^2} \Big)^{\oh} 
\frac{z}{z^2\mi \om^2}\Big]
\eeq
and show that the last two terms, after differentiation, can be reorganized in the following way:
\beq\label{p12-4}
\frac{1}{(z^2\mi \om^2)^2 }\left(  \frac{(z^2\mi \om^2) (z^2\mi c^2) \om^2}{(z^2\mi c^2)^{3/2} (\om^2\mi c^2)^{1/2}} + 
\frac{(\om^2\mi c^2) (z^2\plu \om^2)}{(z^2\mi c^2)^{1/2} (\om^2\mi c^2)^{1/2}}   \right)
\eeq

\item[(4)]
Use the above to prove formula \rf{p12-1a}.

\end{itemize}

We now turn to the proof of the three-loop formula \rf{5.65}. Since the two-loop function only depends on the 
coupling constants $\vg$ via the position of the cut, $c(\vg)$, the loop insertion operator becomes very simple 
in the form \rf{5.49}
when acting on the two-loop function
\begin{itemize}
\item[(5)] Prove that 
\beq\label{p12-5}
\frac{d}{dc^2} \left( \frac{2 z^2 \om^2 -c^2(z^2\plu \om^2)}{ (z^2\mi c^2)^{1/2} (\om^2\mi c^2)^{1/2}}\right)
= \oh \frac{c^2 (z^2\mi \om^2)^2}{(z^2\mi c^2)^{3/2} (\om^2\mi c^2)^{3/2}}
\eeq

\item[(6)] Use this to prove formula \rf{5.65} for the three-loop function

\end{itemize}

Let us next prove the 4-loop formula. What we have to show is that
\beq\label{p12-11}
\frac{d}{d V(z)} \frac{f(c)}{\tM_1} = \frac{2}{\tM_1} \frac{d}{dc^2} \frac{ f(c)}{\tM_1 (z^2 \mi c^2)^{3/2}}.
\eeq

\begin{itemize}
\item[(7)] Show that
\beq\label{p12-11a} 
\frac{\prt \tM_1}{\prt V(z)}  = -\frac{d}{d c^2} \frac{2c^2}{(z^2 \mi c^2)^{3/2}}
\eeq
and use this to show \rf{p12-11}.
\end{itemize}

Finally, let us turn to the $n$-loop formula, which we have just proven for $n=3,4$. Assume it is correct up to $n\mi 1 \geq 4$.

\begin{itemize}

\item[(8)] Use \rf{p12-11a} to prove the following
\beq\label{p12-6}
\Big[ \frac{d}{d V(z)} , \frac{2}{\tM_1} \frac{d}{dc^2}\Big] =0
\eeq

\item[(9)] Use this and \rf{p12-11} to prove the multiloop formula \rf{5.66}

\end{itemize}

\newpage

\setcounter{equation}{0}
\setcounter{figure}{0}
\renewcommand{\thefigure}{ps-13.\arabic{figure}}
 \renewcommand{\theequation}{ps-13.\arabic{equation}}

\subsection*{Elementary Quantum Geometry Problem Set 13}

In this Problem Set we will solve the characteristic equation \rf{cdt47}, use the solution to find the two-point function as well as 
to calculate the ``average shape'' of the quantum universe, also in a situation where the universe is ``expanding'' to infinity for 
$T \to \infty$. 

\subsubsection*{The characteristic function and the two point function}

\begin{itemize}

\item[(1)] Show that the solution to 
\beq\label{p13-1}
\frac{d \bX(T)}{d T} = - \hW(\bX), \qquad \bX(T \equ 0;X) \equ X,
\eeq
can be written as 
\beq\label{p13-2}
T = \int_{X(T;X)}^X \frac{dx}{\hW(x)} = \int_{\bX(T;X)}^X \frac{dx}{(x-\al) \sqrt{ (x\plu \al)^2 \mi  \frac{2g_s}{\al}}}
\eeq

\end{itemize}
Here $\bX(T)$ and $X$ are both larger than $\al$ and it is seen that $T \to \infty$ implies that $\bX(T) \to \al$.
Let us now introduce the notation 
\beq\label{p13-3}
\Sigma = \sqrt{\al^2 \mi \frac{g_s}{2\al}},  \quad \cosh \beta = \frac{\al}{\sqrt{g_s/2\al}},\quad \sinh \beta = \frac{\Sigma}{\sqrt{g_2/2\al}},
\eeq
The integral in \rf{p13-2} can be written as 
\beq\label{p13-4}
T = - \frac{1}{\Sigma} \sinh^{-1} \sqrt{F(x)}\Big |^X_{\bX(T)}, \qquad
 F(x) =  \sinh^2(\beta/2) + \sinh^2\beta \frac{ \sqrt{g_s/2\al}}{ x-\al} 
 \eeq
\begin{itemize}
\item[(2)] Shown that \rf{p13-4} is the integral appearing in \rf{p13-2} by differentiation.

\item[(3)] Show that \rf{p13-4} leads to 
\beq\label{p13-5}
\bX(T) - \al = \frac{\Sigma^2}{\sqrt{\displaystyle{\frac{g_s}{2\al}}}} \;\, 
\frac{1}{ \sinh^2\Big( \Sigma T \plu \sinh^{-1} \sqrt{F(X)}\Big) \mi \sinh^2( \beta/2)}
\eeq

\item[(4)] Show that the large $T$ behavior of $\bX(T)$ is 
\beq\label{p13-6} 
\bX(T) \mi  \al  \to \frac{4\Sigma^2}{\sqrt{\displaystyle{\frac{g_s}{2\al}}}}  \;\; 
\Big( \sqrt{ 1\plu F(X)} \mi  \sqrt{F(X)} \Big)^2 \; \e^{ -2 \Sigma T}
~~ {\rm for } ~~ T \to \infty
\eeq
\item[(5)] Show that for $X \to \infty$ we have 
\bea\label{p13-7}
\bX(T)\mi \al &=& \frac{\Sigma^2}{\sqrt{\displaystyle{\frac{g_s}{2\al}}}} \;\; \frac{1}{\sinh (\Sigma T) \; \sinh ( \Sigma T + \beta)}\nonumber \\
&=&  \frac{ \Sigma^2}{ \sinh(\Sigma T) \Big( \Sigma \cosh ( \Sigma T) + \al \sinh (\Sigma T) \Big)}.
\eea

\item[(6)] Show that for $g_s \to 0$ \rf{p13-7} agrees with the corresponding CDT expression \rf{cdt23} in the Lecture 
Notes (for $X \to \infty$)

\end{itemize}

We now turn to the two-point function $G_\Lam(T)$. We have seen that it can be expressed as 
\beq\label{p13-8}
G_\Lam(T) = \frac{d\, W_\Lam( \bX(T, X\equ \infty))}{d T} , \qquad 2g_s W_\Lam(x) = \Lam -x^2 + \hW_\Lam(x).
\eeq
\begin{itemize}

\item[(7)] Show that 
\bea\label{p13-9}
2g_sG_\Lam(T) \!\!\!&=& \!\!\! -2 \bX(T) \frac{d \bX(T)}{dT} \mi \frac{d^2 \bX(T)}{dT^2}, 
\quad \bX(T) \equiv \bX(T, X \equ \infty) \hspace{2cm} \\
&= & \!\!\! 
\frac{4 \Sigma^3 ( \al^2 \mi \Sigma^2) \;\Big( \al \cosh (\Sigma T) + \Sigma \sinh(\Sigma T)\Big) }{\Big( \Sigma \cosh (\Sigma T) 
+ \al \sinh (\Sigma T)\Big)^3}\label{p13-10a}
\eea
You should not try to perform the (trivial) differentiations leading from \rf{p13-9} to \rf{p13-10a} by hand unless you really love 
calculations. Rather, convince yourself that the end result is precisely the formula for $G_\Lam(T)$ given in the Lecture Notes.
 
\end{itemize}

\subsubsection*{The average shape of CDT and GCDT universes}

Until now we have mainly considered the two-loop function in the form $G_\Lam(X,L;T)$, given by \rf{cdt47}. However, here 
it will convenient to consider the situation where we have a boundary cosmological constant $Y$ associated with the exit loop 
at $T$. The corresponding two-loop function $G_\Lam(X,Y;T)$ is obtained by a Laplace transformation:
\beq\label{p13-10}
G_\Lam(X,Y;T) \equ \int_0^\infty dL \; \e^{-L Y} G_\Lam(X,L;T) \equ \frac{\hW(\bX(T;X))}{\hW(X)}  \; \frac{1}{\bX(T;X) \plu Y}
\eeq
We thus have an ensemble of universes which start start out at $T \equ 0$ with a boundary of length distribution determined 
by the boundary cosmological constant $X$ and which at time $T$ have boundaries with a lengths distribution monitored by 
the boundary cosmological constant $Y$. It is natural to ask about the average length of a spatial universe at time $t$ 
between 0 and $T$. We view $G_\Lam(X,Y;T)$ as the partition function for the ensemble of  universes and then 
the average length at $t$ is defined as 
\beq\label{p13-11}
\la L(t)\ra_{X,Y} = \frac{1}{G_\Lam(X,Y;T)} \int_0^\infty dL  \, G_\Lam(X,L;t) L \, G_\Lam (L,Y;T\mi t).
\eeq
We will  show that 
\beq\label{p13-12}
\la L(t)\ra_{X,Y} = \frac{\hW'(\bX(t;X)) - \hW'(\bX(T;X))}{\hW(\bX(t;X))} +  \frac{\hW(\bX(T;X))}{\hW(\bX(t;X))} \,
 \frac{1}{\bX(T;X) \plu  Y}
 \eeq
 where $\hW'(X)$ denotes the derivative of $\hW(X)$ wrt $X$. 
 
 Before deriving \rf{p13-12}  let us discuss some implication of the formula. First note that we have 
 \beq\label{p13-13}
 \hW(X) = \frac{(X\mi \al)}{\tilde{W}(X)},
 \eeq
 where $\tilde{W}(\al) \neq 0$. This is true both in EDT, CDT and GCDT, just with slightly different\footnote{In all case $\tilde{W}(X)$
 can be viewed as related to the disk amplitude, as discussed in connection with formula \rf{cdt50}. We make here a list:
 \bea
 \tilde{W}(X) &=& \frac{1}{\sqrt{X + \SL}}, \qquad \qquad  \al = \SL/2  \quad   {\rm EDT} \\
 \tilde{W}(X)  &= &\frac{1}{X + \SL}, \qquad  \qquad \quad \al = \SL \qquad  {\rm CDT} \\
  \tilde{W}(X) &= &\frac{1}{\sqrt{(X \plu \al)^2 \mi 2g_s/\al}}, \quad  \al = \al   \qquad {\rm GCDT} 
  \eea} 
  $\al$ and $\tilde{W}(X)$.
 
 The smaller $Y$, the larger we expect length of the 
 exit loop to be, and correspondingly also $\la L(t)\ra_{X,Y}$. In particular a negative $Y$ will try expand the exit loop , to 
 the extent it is possible (such an extension will also typically result in an enlarged area, which is suppressed by the action).
 Let us assume $ Y > -\al$. Now take $T \to \infty$. Recall that $\bX(T;X) \to \al$ for $T\to \infty$ 
 
 \begin{itemize}
 
 \item[(8)] Show that 
 \beq\label{p13-14}
 \la L(t)\ra_{X,Y} =  \frac{\al}{\Sigma^2} + O(\e^{-2\Sigma t}) \quad{\rm for~large~}t.
 \eeq
 In the case of CDT (i.e.\ $g_s \equ 0$) this is just $1/\SL$.

\end{itemize}
We are here considering a situation where $T$ is infinity or very large, and when we are far away from from the entrance 
loop (and by construction very far away from the exit loop) and the average length of the boundary loop is then constant
(and simply $1/\SL$ in the case of CDT). Of course there are fluctuations and the situation is basically the one shown in 
Fig.\ \ref{figcdt11} and captured in eq.\ \rf{cdt35a}. If we choose $Y < - \al$ the system becomes unstable and the exit boundary 
will expand to infinity in a finite time. However, exactly when $Y \equ -\al$ we have a situation where the length of the exit 
boundary expands to infinity when $T \to \infty$. 

\begin{itemize}

\item[(9)] Show that for $Y \equ - \al$ and $T \to \infty$ eq.\ \rf{p13-12} becomes
\beq\label{p13-15}
\la L(t)\ra_{X,Y= -\al} = \frac{1}{\bX(t;X) \mi \al} - \frac{\tilde{W}'(\bX(t;X)}{\tilde{W}(\bX(t;X)},
\eeq
and show that $\la L(t)\ra_{X,Y= -\al}$ grows exponentially like $e^{2\Sigma t}$ for large $t$.
Finally, show that the corrections to \rf{p13-15} if we keep $T$ finite but much larger that $t$ 
is of order $e^{-2\Sigma (T-t)}$.

\end{itemize}

Let us now for simplicity study eq.\ \rf{p13-15} in the CDT case, where $\al \equ \SL$ and $\tilde{W}_\Lam(X) \equ 1/(X+\SL)$.
$\bX(t;X)$ is given by \rf{cdt23}. 

\begin{itemize}

\item[(10)] Show that 
\beq\label{p13-16}
\bX(t;X)) = \SL \, \coth \SL (t+t_0(X)), \qquad  {X} =\SL\, \coth t_0
\eeq
and 
\beq\label{p13-17}
\la L(t)\ra_{X,Y\equ  -\SL} = \frac{1}{\SL}  \sinh 2 \SL (t\plu t_0(X)).
\eeq

\end{itemize}

This shows that if we view $t$ as the geodesic distance from the entrance boundary with boundary cosmological constant  $X$ 
and $\la L(t)\ra_{X,Y\equ  -\SL}$ as the length of the curve a geodesic distance $t$ from the boundary, this average geometry
can be viewed as belonging to the {\it hyperbolic plane}, also called the {\it pseudosphere}. Recall that for a sphere of radius $R$ 
the infinitesimal geodesic distance between points with (spherical) coordinates $(\th,\phi)$ and $(\th \plu d\th,\phi \plu d\phi)$
is given by 
\beq\label{p13-18}
ds^2 = R^2(d\th^2 + \sin^2 \th \, d\phi^2) = dt^2 + R^2 \sin^2 (t/R) \, d\phi^2, \qquad t \equ R \th.
\eeq 
Here $t$ is the geodesic distance on the sphere from the north pole where $\th \equ 0$ to a point with coordinates $(\th,\phi)$.
Also the curve at geodesic distance $t$ from the north pole (curve of fixed latitude $\th$) has the length $\ell(t) =2 \pi R \sin (t/R)$. 
The intrinsic curvature (the Gaussian curvature) is constant  on the sphere, and equal  $R_g = 1/R^2$. 
We obtain the geodesic distance on the pseudo-sphere by 
formally rotating $R \to iR$ in the above line element,
\beq\label{p13-19} 
ds^2 = dt^2 + R^2 \sinh^2 (t/R) \, d\phi^2, \qquad  R_g = - \frac{1}{R^2}, \quad L(t) = 2 \pi R \sinh (t/R).
\eeq
In the $(t,\phi)$ coordinate system, $t$ is also the geodesic distance of point $(t,\phi)$ to the point with coordinate $t \equ 0$
and $L(t)$ the length of the curve of points with geodesic distance $t$ to the point with $t \equ 0$. All points on the pseudo-sphere 
has  intrinsic curvature $R_g \equ  -1/R^2$.  It is now seen that we can view \rf{p13-17} as corresponding to the part 
of the pseudo-sphere where $t \in [t_0(X),\infty[$  if we identify $R \equ 1/2\SL$ and $ \ell(t) \equ \pi \la L(t)\ra_{X,Y\equ  -\SL} $.
It is remarkable that different choices of $X$ lead to the ``same'' pseudo-sphere, with $t\plu  t_0(X)$ being the geodesic distance 
to the ``origin'' (which of course is arbitrary as a point on the pseudo-sphere, like the north-pole being an ``arbitrary'' point
on the sphere ). For $X \to \infty$ the entrance boundary loop will in average contract to a point, and the whole pseudo-sphere is 
covered. The fact that we have to change the length assignment of $L(t)$ relative to $t$ should not be a course of worry. From the 
beginning in the CDT model there was an arbitrariness in the relative length assignment of spatial links and temporal links.

Let us now return to \rf{p13-12} and prove the formula. 

\begin{itemize}

\item[(11)] Use \rf{p13-10} to show that eq.\ \rf{p13-11} can be written as 
\bea\label{p13-21}
\lefteqn{\la L(t)\ra_{X,Y} = \frac{(Y+ \bX(T;X)) \hW(\bX(t;X))}{\hW(\bX(T;X))} \times} \\
 &&\left( -\frac{d}{d \bX(t;X)} \right)
 \left[ \frac{\hW\big(\bX(T\mi t; \bX (t;X))\big)}{\hW(\bX(t;X))} \, \frac{1}{\bX(T \mi t; \bX(t;X)) + Y}\right] \no 
 \eea 
 
 \item[(12)] Show that \rf{p13-2} implies that 
 \beq\label{p13-22}
 \frac{\prt}{ \prt X} \bX(t;X) = \frac{\hW(t;X)}{\hW(X)} 
 \eeq
 
 \item[(13)] Show (if you do not feel it is  trivial) that 
 \beq\label{p13-23}
 \bX(T\mi t; \bX(t;X)) = \bX(T;X)
 \eeq
 
 \item[(14)] Show \rf{p13-12} using \rf{p13-22} in \rf{p13-21}. Note that you are only allowed to use \rf{p13-23} {\it after} 
 having performed the differentiation in \rf{p13-21}.

\end{itemize}

\newpage

\centerline{{\Large \bf Solutions to Problem Sets  1-11}}

\setcounter{equation}{0}
\setcounter{figure}{0}
\renewcommand{\thefigure}{s-1.\arabic{figure}}
 \renewcommand{\theequation}{s-1.\arabic{equation}}

\subsection*{ \hspace{-3mm} Solutions to  Problem Set 1}

\subsubsection*{Gaussian integrals}

\begin{itemize}
    \item[(1)] Switch to polar coordinates. We have
      $$
            \int \e^{-(x^2+y^2)}\, dxdy = \int_0^{2\pi} d\theta \int_0^\infty r dr \, \e^{-r^2} = 2\pi \times \frac{1}{2} \int_0^\infty  dr^2 \, \e^{-r^2} 
            = \pi
      $$
        
    \item[(2)]
        From (1) it is clear that 
        $$
        \int \e^{-x^2} dx = \sqrt{\pi},
        $$
        and thus (by changing integration variable to $y = \sqrt{\frac{a}{2}} \,x$) that 
        $$
        \int \e^{-\frac{a x^2}{2}} dx = \sqrt{\frac{2}{a}} \int \e^{-y^2} dy = \sqrt{\frac{2 \pi}{a}}
        $$
       
    \item[(3)] We write the exponent in vector notation as
        \[ x_i A_{ij} x_j = x^T A x, \quad x \in \mathbb{R}, A \in \mathrm{Sym}_{n \times n} \left(\mathbb{R}\right). \]
        Now any real, symmetric matrix can be decomposed as
        \[ A = Q^T \Lambda Q, \]
        such that $\Lambda$ is diagonal and $Q$ is an orthonormal matrix, meaning that
        \[ Q^{-1} = Q^T, \quad \det{Q} = 1. \]
        Therefore we have
        \[ x^T A x = x^T Q^T \Lambda Q x, \]
        so it is convenient to make the change of variables 
        \[ y = Q x, \quad d^n y  = \det{\left(Q\right)} d^n x = d^n x. \]
        The exponent then simplifies considerably since $\Lambda$ is diagonal. We have
        \[ x^T A x = y^T \Lambda y = \lambda_i y_i^2, \]
        where the $\lambda_i$ are the elements on the diagonal of $\Lambda$ (and therefore, the eigenvalues of $A$).
        Our integral can now be performed for all the $y_i$ separately. We see that
        \begin{align*}
            \int \prod_{i=1}^n dx_i e^{-\frac{1}{2} x_i A_{ij} x_j} = \int \prod_{i=1}^n dy_i e^{-\frac{\lambda_i}{2} y_i^2} = \prod_{i=1}^n \sqrt{\frac{2\pi}{\lambda_i}}
        \end{align*}
        by our result from part (2). Now note that
        \[ \prod_{i=1}^n \lambda_i = \det \Lambda = \det Q^T \det \Lambda \det Q = \det\left(Q^T \Lambda Q\right) = \det A, \]
        which completes the proof.

    \item[(4)]
    We know that  we can decompose a Hermitian matrix $A$ as the product
        \[ A = Q^\dagger \Lambda Q, \]
        again with $\Lambda$ diagonal with {\it real} matrix elements, but now $Q$ is a unitary matrix, meaning that 
        \[Q^{-1} = Q^\dagger, \quad \left|\det Q\right| = 1. \]
       Problem (4) is then reduced to problem (3) by realising that a unitary transformation $w = Q z$ in the
       $n$-dimensional complex vector space becomes an orthogonal transformation in the $2n$-dimensional
       real vector space obtained by writing $z_k = x_k +i y_k,~w_k = u_k + i v_k$ and $w=Q z$ as 
       $$
        \begin{pmatrix} u \\ v \end{pmatrix} = \begin{pmatrix} {\rm Re}\; Q & - {\rm Im} \;Q \\ {\rm Im}\; Q & {\rm Re}\;Q
        \end{pmatrix}
        \begin{pmatrix} x\\ y\end{pmatrix}
        $$
        where 
        $$
        Q^\dagger Q = I_{n\times n}  \implies     
        \begin{pmatrix} {\rm Re}\; Q & - {\rm Im} \;Q \\ {\rm Im}\; Q & {\rm Re}\;Q\end{pmatrix}^T 
        \begin{pmatrix} {\rm Re}\; Q & - {\rm Im} \;Q \\ {\rm Im}\; Q & {\rm Re}\;Q \end{pmatrix} = 
        \begin{pmatrix} I_{n\times n} & 0 \\ 0 & I_{n\times n} \end{pmatrix}
        $$
        By the orthogonal change of variables $(x,y) \to (u,v)$ we obtain
        $$
        z^\dagger A z \to \sum_{k=1}^n \lambda _k (u_k^2 + v_k^2)
        $$
        and as in  (3) we obtain
        $$
            \int \prod_{i=1}^n (dx_i dy_i) e^{-\frac{1}{2} z^*_i A_{ij} z_j} = 
            \int \prod_{i=1}^n (du_i dv_i) e^{-\frac{\lambda_i}{2} (u_i^2+v_i^2)} = 
            \prod_{i=1}^n \Big(\sqrt{\frac{2\pi}{\lambda_i}} \sqrt{\frac{2\pi}{\lambda_i}}\Big) =\frac{(2\pi)^{n}}{\det A}.
            $$

    \item[(5)]
        Use the rules for differentiation:
      \bea\nonumber
            \frac{\partial S(x) }{\partial x_i} &=& 
            \frac{1}{2} \left( \delta_{ij} A_{jk} x_k + x_j A_{p1jk} \delta_{ik} \right)+ b_j \delta_{ij} 
            = A_{ij} x_j + b_i \\
    \frac{\partial^2S(x) }{\partial x_i\partial x_j} &=& A_{ij}.\nonumber
    \eea
       We then find
        $$
         \frac{\partial S(x) }{\partial x_i}\Big|_{x_c} =0 ~~\implies\quad  x_c = -A^{-1} b \quad {\rm and~thus} \quad S(x_c) = - \frac{1}{2} b^T A^{-1} b, \quad 
        $$

Making the expansion $x = x_c \plu \Delta x$, we now compute
  $$
            S(x_c\plu \Delta x)\! = \!S(x_c) + \frac{\partial S(x)}{\partial x_i} \Big|_{x_c}\!\!\! \Delta x_i +
            \frac{1}{2} \frac{\partial^2 S(x)}{\partial x_i\partial x_j} \Big|_{x_c}\!\! \!\Delta x_i \Delta x_j
           \! =\! S(x_c) \plu S(\Delta x, b\equ 0).
  $$       
        Since $x_c$ is a constant vector we have: 
        \[d^n\,\Delta x = d^n\,x.\]
        As a result, we can pull $e^{-S(x_c)}$ out of the integral and obtain
        $$
            \int d^n x\, \e^{-S(x)} =  
          \e^{-S(x_c)} \int d^n \Delta x\; \e^{-S(\Delta x, b=0)} = \e^{-S(x_c)} \int d^n \Delta x \;\e^{-\frac{1}{2} \Delta x^T A \Delta x} 
          $$
          Thus we have obtained the desired result
          $$
             \int d^n x\, \e^{-S(x)}  = \e^{-S(x_c)} \frac{(2\pi)^{\frac{n}{2}}}{\sqrt{\det A}} = 
             \e^{\frac{1}{2} b^TA^{-1}b} \frac{(2\pi)^{\frac{n}{2}}}{\sqrt{\det A}} 
       $$

\item[(6)] We just use the above derived formula with $b =-ik$, then we obtain
$$
{\cal F} ( \e^{- \frac{1}{2} x^T A x}) = \int d^n x \, \e^{- \frac{1}{2} x^T A x} \, \e^{-ik^T x} = \frac{(2\pi)^{\frac{n}{2}}}{\sqrt{\det A}} \;
 \e^{- \frac{1}{2} k^T A^{-1} k}
 $$
 
 \item[(7)]
 
 It is clear from the explicit formula for the Fourier transformed ${\cal F} (f_A)$ that  
 $$
 {\cal F} (f_A) \cdot {\cal F} (f_B) = {\cal F} (f_{A+B})
 $$
Thus we obtain
$$
f_A * f_B = {\cal F}^{-1} \big( {\cal F} (f_A * f_B) \big) =   {\cal F}^{-1} \big( {\cal F} (f_A ) \cdot {\cal F} (f_B) \big) =
{\cal F}^{-1} \big( {\cal F} (f_{A +B}) \big) = f_{A+B}
$$
 
 \end{itemize}
 \vspace{12pt}
 
 \subsubsection*{The free non-relativistic particle}
 
 \begin{itemize}
 
    \item[(8)]
    The integral given depends (by translational invariance)  only on $x'' - x$ 
    since we can shift the integration variable $x' \to x' +x$, and it then becomes a 
    standard convolution with ``external variable'' $x''-x$.  Thus the  formula listed is just the one-dimensional 
    version of our general formula for convolution of Gaussians with $A = ia$ and $B = ib$.
    
    By translational invariance the matrix element will depend only on $x_b- x_a = x_{n+1} -x_0$. We can thus choose 
    $x_0 =0$ and $x_{n+1} = x_b-x_a$. 
    By choosing $a  = \epsilon \hbar/m$ the expression for $\langle x_{n+1} | \hat{{\cal O}}_\epsilon^{n+1} | 0\rangle$ 
    becomes successive convolutions of the gaussian function $f_{ia}$ 
    $$
    \langle x_{n+1} | \hat{{\cal O}}_\epsilon^{n+1} | 0\rangle \equ ( f_{ia} \!*\! f_{ia} \!* \cdots *\! f_{ia}) (x_{n+1}) = f_{i (n+1) a}(x_{n+1}) \equ
    \frac{ \e^{-\frac{x_{n+1}^2}{2 i (n+1) a}}}{\sqrt{2 \pi i (n\plu 1) a} }.
    $$
    Now, using $x_{n+1} \equ x_b\mi x_a $ and  $t = (n+1) \epsilon$ we obtain the desired formula for  
     $\langle x_s | \hat{{\cal O}}_\epsilon^{n+1} | x_0\rangle$.

The result is independent of $n$. The reason is that the Hamiltonian in the case of 
a free particle only depends on the momentum operator. Recall that the $n$-dependence  entered 
because the concept of a path integral was introduced via the Trotter-Kato theorem, where we actually 
changed the operator $e^{-i\epsilon \hat{H} /\hbar}$ into an exponential depending only on the operator $\hat{p}$ and 
another exponential depending only on $\hat{x}$. In this procedure one only got back   $e^{-it \hat{H} /\hbar}$ 
in the limit where $n\to \infty$. However, if $\hat{H}$ only depends on $\hat{p}$ the subdivision in $n$ is 
exact and one is always calculating the matrix element of $\bra{x} e^{-it \hat{H} /\hbar}\ket{y}$ independent of  how many 
subdivisions one makes.

\item[(9)] Again this is a simple application of our Fourier formula for Gaussians $f_{ia} (x)$, with $a = \hbar t/m$.
        
\end{itemize}

\newpage

\setcounter{equation}{0}
\setcounter{figure}{0}
\renewcommand{\thefigure}{s-2.\arabic{figure}}
 \renewcommand{\theequation}{s-2.\arabic{equation}}

\subsection*{Solutions to Problem Set 2}

\begin{itemize}
    \item[1.] The Euler-Lagrange equation  are
   $$
            \frac{\delta L}{\delta x(t)} = \frac{d}{dt} \frac{\delta L}{\delta \dot{x}(t)}, \quad {\rm i.e.} \quad 
            -\omega^2 x(t) = \ddot{x}(t).
    $$
        Clearly, the given solution $x_c(t)$ satisfies this differential equation. To calculate the action one can just insert the 
        solution, or (slightly easier) perform a partial integration
$$
\frac{m}{2}  \int_{t_a}^{t_b} dt  ( \dot{x}^2 \mi \omega^2 x^2) =  
-\frac{m}{2}  \int_{t_a}^{t_b} dt  (\ddot{x} \plu \omega^2 x) \; x \plu  \frac{m}{2}  x  \dot{x} \Big|_{t_a}^{t_b}.
$$
The integrand will be zero if $x(t)$ satisfies the classical equation, and inserting the classical solution in the boundary
term produces the wanted expression for $S[x_{cl}]$.

    \item[2.] We simply insert the decomposition $x_c(t)\plu \Delta x(t)$ in the action:
        \begin{align*}
            S[x(t)] &= S\left[x_c(t)\plu \Delta x(t)\right] \\
            &= \frac{m}{2}\int dt \big( (\dot{x}_c\plu \dot{\Delta x})^2\mi \omega^2 \left(x_c\plu \Delta x\right)^2\big) \\
            &= \frac{m}{2} \int dt \big( \dot{x}_c^2 \plu 2 \dot{x}_c \dot{\Delta x} \plu \dot{\Delta x}^2 
            \mi  \omega^2 x_c\mi 2\omega^2 x_c \Delta x \mi  \omega^2 \Delta x^2 \big)\\
            &= S[x_c] \plu S[\Delta x] \plu  \frac{m}{2} \int dt \big( 2 \dot{x}_c \dot{\Delta x} \mi  2\omega^2 x_c \Delta x \big)\\
            &= S[x_c] \plu  S[\Delta x] \plu \left. m \dot{x}_c \Delta x \right|_{t_a}^{t_b} \plu 
            \frac{m}{2} \int dt \left[ -2 (\ddot{x}_c  \plu \omega^2 x_c )\Delta x \right] \\
            &= S[x_c] \plu S[\Delta x].
        \end{align*}
        Here we performed a partial integration in the fifth line, and used the fact that the variations vanish on the endpoints. Further, the integral also vanishes since $x_c(t)$ satisfies the eom.

\item[3.]
        The ``measure''   $\mathcal{D} x(t)$ is invariant under the decomposition 
        $x(t) \equ x_c(t) \plu\Delta x(t)$. $x_c(t)$ acts like a translation of 
        the ``vector'' $x(t)$, and if we represent $\mathcal{D} x(t)$ as a kind of limit of  $\prod_{i=1}^n dx(t_i)$ we have:
        \[ \mathcal{D}x(t) = \mathcal{D}\Delta x(t), \]
        meaning that we can move the factor $e^{\frac{i}{\hbar} S[x_c(t)]}$ outside the functional integral.

    \item[4.] It is easily checked by direct calculation. Since $y_0 = y_{n+1} = 0$, we can relabel the sum over $y_{i+1}^2$ to $y_i^2$, so that we pick up this term twice. The off-diagonal elements $-1$ provide the cross terms $y_i y_{i+1}$.

    Note that this matrix $A$ is not a unique solution - 
    however, we are looking for a symmetric matrix in order to be able to use our results for Gaussian integrals, and 
    the symmetric matrix is unique.

    \item[5.] We compute the determinant by expanding $D_n$ in the top row:
        \begin{align*}
            D_n &= \left(2-\epsilon^2 \omega^2 \right) D_{n-1} - (-1) 
            \left| 
            \begin{array}{c c c c c} 
                -1 & 2-\omega^2 \epsilon^2 -1 & -1 & & 0 \\
                0 & -1 \ddots & \ddots & \vdots \\
                \vdots & & \ddots & & -1 \\
                0 & 0 & \cdots & -1 & 2-\omega^2 \epsilon^2
            \end{array} 
            \right|\\
            &= \left(2-\epsilon^2 \omega^2 \right) D_{n-1} - D_{n-2}.
        \end{align*}
        The second term in the first line was expanded in the first column. The values of $D_0$ and $D_{-1}$ follow by consistency for the recursion relations for $D_2$ and $D_1$.

\item[6.]        We now find the generating function using the recursion relation:
 \bea
            D(x)-D_0 &=& \sum_{n=1}^\infty D_n x^n =1 + \sum_{n=1}^\infty \big((2-\epsilon^2 \omega^2) D_{n-1} - D_{n-2}\big)x^n
            \nonumber \\
            &=& (2-\epsilon^2 \omega^2) \, x D(x) -(x^2 D(x) -x D_-1) \nonumber
\eea            
       Thus    
  $$
  D(x) \mi 1 =     (2\mi \epsilon^2 \omega^2) \, x D(x) -x^2 D(x)   \quad {\rm or} \quad  D(x) = 
  \frac{1}{1 \mi  (2\mi \epsilon^2 \omega^2) \, x \plu x^2}
  $$

 \item[7.]       In terms of the new variable $\tilde{\om}$, $1 \mi \frac{\omega^2 \epsilon^2}{2} \equ 
 1\mi 2\sin^2 \frac{\tilde{\om} \epsilon}{2} = \cos \tilde{\om} \epsilon$, and therefore
 \bea
            D(x) &=& \frac{1}{1-2x \cos \left(\tilde{\om} \epsilon\right) + x^2} 
            = \frac{1}{\left(\e^{i \tilde{\om} \epsilon}-x\right)\left(e^{-i\tilde{\om}\epsilon}-x\right)} \nonumber \\
    &=&       \frac{1}{\e^{i\tilde{\om} \epsilon} \mi   \e^{-i\tilde{\om} \epsilon}}  \Big(\frac{1}{\left(e^{-i \tilde{\om} \epsilon}-x\right)} -
   \frac{1}{\left(e^{i \tilde{\om} \epsilon}-x\right)}\Big)\nonumber
   \eea    

 \item[8.]    Using $(a \mi x)^{-1} = \sum_{n=0}^\infty x^n/a^{n+1}$ we obtain
 $$
 D_n =  \frac{1}{\e^{i\tilde{\om} \epsilon} \mi   \e^{-i\tilde{\om} \epsilon}} 
 \Big( \e^{i \tilde{\om} \epsilon(n+1)} - \e^{-i \tilde{\om} \epsilon(n+1)}\Big) = 
 \frac{\sin \tilde{\om} \epsilon(n+1)}{\sin \tilde{\om} \epsilon }
 $$
        
\item[9.]        Using that $(n+1) \epsilon = t_b - t_a$, $\tilde{\omega}  = \omega + O( \epsilon)$ and 
        $ \sin \tilde{\omega} \epsilon = \omega \epsilon + O(\epsilon^2)$ we obtain the required formula for 
        the amplitude.  

\item[10.] 
It is readily seen that if we make the replacement $i (t_b-t_a) = \hbar \beta$ and $x_a =x_b =x$ in the formula
for the amplitude, and use that 
$$ 
\frac{ \cosh \omega \beta \hbar -1 }{\sinh \omega \beta \hbar} = 
\tanh (\omega \beta \hbar/2),
$$
we obtain the wanted formula  
$$z(x) = \langle x| e^{-\beta \hat{H}} | x\rangle = \sqrt{ \frac{m \omega}{ 2\pi \hbar \sin \omega \beta \hbar}}
\; e^{- \frac{m\omega}{\hbar} \tanh \frac{\omega \beta \hbar}{2} \; x^2}.
$$
The integral $\int dx \, z(x)$ is a Gaussian integral, which we know how to calculate and  we obtain
$$ 
Z = \int_{-\infty}^{\infty} z(x) =  \sqrt{ \frac{m \omega}{ 2\pi \hbar \sin \omega \beta \hbar} }\;\; 
\sqrt{ \frac{\pi \hbar}{m\omega \tanh  \frac{\omega \beta \hbar}{2} }} = 
\frac{1}{2 \sinh \frac{\omega \beta \hbar}{2} }.
$$

\item[11.]
From the expressions for $z(x)$ and $Z$ we obtain, taking the limit $\beta \to \infty$,
$$
\frac{z(x)}{Z} \to \frac{\sqrt{\frac{m\omega}{2\pi \hbar}} \; e^{-\frac{\hbar \beta \omega}{2}} \; 
e^{- \frac{m \omega}{\hbar} \; x^2}}{e^{-\frac{\hbar \beta \omega}{2}} }= 
\sqrt{\frac{m\omega}{2\pi \hbar}} e^{- \frac{m \omega}{\hbar} \; x^2} = |\psi_0 (x)|^2.
$$        
That the $\beta \to \infty$ limit leads to the square of the ground state wave function should be clear without this detailed 
calculation and is valid for an arbitrary system with a discrete spectrum bounded from below since we have 
$$
Z = \sum_n \langle E_n | e^{ -\beta \hat{H} }| E_n \rangle \to e^{-\beta E_0} +O\big(e^{-\beta E_1}\big)
$$
and similarly
\bea
z(x) &=& \sum_{n,m} \langle x | E_n\rangle \langle E_n |e^{-\beta \hat{H}} | E_m \rangle \langle E_m | x  \rangle =
\sum_{n} \langle x | E_n\rangle e^{-\beta E_n} \langle E_n | x  \rangle \nonumber \\
&\to&  \e^{-\beta E_0 }   |\langle x | E_0\rangle|^2 
+ O\big(\e^{-\beta E_1}\big)\no
\eea

\end{itemize}

\newpage

\setcounter{equation}{0}
\setcounter{figure}{0}
\renewcommand{\thefigure}{s-3.\arabic{figure}}
 \renewcommand{\theequation}{s-3.\arabic{equation}}

\subsection*{Solutions to  Problem Set 3}

\begin{enumerate}

\item 
  \[ \hat{F}(p) = \sum_{x_n} a^D \, \e^{i p \cdot x_n} F(x_n). \]
Prove that
 \[ \hat{F}(p_i) = \hat{F}\left(p_i+\frac{2\pi}{a}e_i\right). \]
Follows trivially from $\frac{2\pi}{a} x_n = 2\pi n$. Thus $\e^{i p_i \cdot x^i_n} = \e^{i (p_i +\frac{2\pi}{a}e_i) x^i_n} $. 

\item
The standard formulas for Fourier series of functions $\hat{F}$ periodic with period $2 \pi$ are   
$$
\hat{F}(q) = \sum_{n} \e^{i q \cdot n} F(n),\quad\quad F(n) = \int_{-\pi}^\pi \frac{d^Dq}{(2\pi)^D} \; 
\e^{-i q \cdot n} \hat{F}(q).
$$
The  formulas in problem 3 are the same, just introducing the dimensionful parameter $a$ (the length of a lattice link).

\item 

One has, from the definition of   $\Delta_L$, 
$$
- a^2 \Delta_L  \e^{-ip x_n }  = \sum_{j=1}^D  \big(2 - \e^{-ip_j a} -\e^{ip_j a}\big) e^{-ip x_n}       
$$
Choosing $x_m=0$ and using that the Fourier transform of $\delta (x_n) $ is 1, we can write
$$
(\Delta_L\plu m^2) \int_{-\frac{\pi}{a}}^{\frac{\pi}{a}} \frac{dp}{(2\pi)^D} e^{-i p x_n} G(p) = 
\int_{-\frac{\pi}{a}}^{\frac{\pi}{a}} \frac{dp}{(2\pi)^D} e^{-i p x_n}  1
$$
or 
$$
\int_{-\frac{\pi}{a}}^{\frac{\pi}{a}} \frac{dp}{(2\pi)^D} e^{-i p x_n} 
\Big( \Big[ \frac{2}{a^2} \sum_{j=1}^D (1\mi \cos(p_ja)) \plu m^2\Big] G(p) \mi1 \Big)=0
$$
valid for all $x_n$, from which we conclude that $\Big( \cdot \Big) =0$.

\item 

$$ 
\Big( \frac{2}{a^2} \sum_{j=1}^D (1-\cos(p_ja)\big) \plu m^2\Big)  \,G(p)  \equ 1~ \Rightarrow~
G(p) \equ  \frac{a^2}{\big(\sum_{j=1}^D 4 \sin^2 \frac{a p_j}{2} \big)\plu m^2 a^2 } 
 $$
 
 \item
 $$
 G(p) = \frac{ 1}{ p^2\plu m^2 \plu O(a^2)} \quad a \to 0.
 $$
 
 \item
 
 It follows from the very definition of the discretized version of $\Delta_L$ that we have the matrix elements:
 $$
 - a^2\left(\Delta_L\right)_{nm} =  2 D \delta_{nm} \mi Q_{nm}
 $$ 
 where $Q_{mn}$ is 1 if $n$ and $m$ label neighbouring sites and zero for all choices of $n$ and $m$.

 \item 
 
 $\Delta_L$ is an operator on $\ell^2(\mathbb{Z})$, the sequences $f(x_n)$ which are  square summable, and
 by Parseval's and Plancherel's theorems Fourier transformation is a unitary map from   $\ell^2(\mathbb{Z})$ to 
 $L^2[-\pi,\pi]$. This map conserves the norm of operators and we can thus analyse the Fourier transformed operator
 which we have already found:
 $$
 -a^2 (\Delta_L + m^2) \to \big(2D + m^2a^2\big)\, I -2\sum_{s3j=1}^D \cos(p_ja) \qquad {\rm by~Fourier~transformation}.
 $$
 Thus it is a simple multiplication operator and we have already found the inverse, namely $G(p)$. We also
 see that the Fourier transform of $Q$ is a multiplication operator 
 $$
 Q \to \hat{Q} =  2   \sum_{s3j=1}^D \cos(p_ja),\quad {\rm i.e} \quad || \hat{Q} \hat{f} || \leq 2D || \hat{f}||.
 $$
 The norm of $\hat{Q}$ is thus less than or equal 2D (in fact it is easily seen to be 2D) 
 and the norm  of $\hat{Q}/(2D + m^2a^2)$ correspondingly less than 1.
 (The norm of a bounded operator $A$ is defined as $\sup_f ||A f||/||f||$). 
 An operator $ I \mi  A$ with $||A|| < 1$ has an inverse operator, which has a convergent expansion in powers
 of $A$ (the Neumann series):
 $$ 
 \frac{1}{I\mi  A} = \sum_{n=0}^\infty A^n
 $$
Writing 
$$
-a^2 (\Delta \plu m^2) = (2D\plu m^2a^2) \Big[ I - \frac{Q}{2D\plu m^2a^2} \Big]
$$ 
leads to the asked for Neumann series:
$$
-  (\Delta \plu m^2)^{-1} = \frac{a^2}{2D \plu  m^2a^2} \; \frac{1}{I \mi  \frac{Q}{2D \plu  m^2a^2}} =  \frac{a^2}{2D \plu m^2a^2} 
\sum_{n=0}^\infty\frac{Q^n}{(2D \plu  m^2a^2)^n}.
$$

\item

\noindent
From the definition of $Q$ it follows that $ (Q^k)_{mn}$ is the number of connected lattice paths of length
$k$ which connect site $m$ to site $n$ and the representation asked for follows.

 \item
 
 Comparing formulas for $G_a(x_n,x_m)$ and  $(- \Delta_L \plu m^2)^{-1}$ as a power series in $Q$ one 
 obtain that
 $$ 
a \, m_0(a) = \log (2D \plu  m^2a^2)
 $$
 
 \item 
 $$
  m_0 (a) = \frac{\log (2D \plu m^2a^2)}{a} = \frac{\log 2D}{a} \plu m^2 a \plu O(a^3).
  $$
 The power dependence on $a$ for the two first terms is universal, but the coefficients are not.
 
 \item
 The number of connected paths on the lattice, starting at a given point $x_n$ and made of $\ell$ links, is 
 $$
 \ell^{2D} = e^{ \ell \log 2D}.
 $$
 
 \end{enumerate}

\noindent 
The formula for $G_a(x_n,x_m)$ can be written as 
$$
G_a(x_n,x_m) = \sum_\ell e^{-m_0 a\ell} {\cal N} (\ell, x_n,x_m),
$$
where ${\cal N} (\ell, x_n,x_m)$ are the number of connected lattice paths from $x_n$ to $x_m$.
It turns out that this number is 
$$
{\cal N} (\ell, x_n,x_m) \propto \frac{1}{\ell^{D/2}} \; \ell^{2D}\quad {\rm for} \quad \ell >> 1.
$$
The contraint that the path should be from $x_n$ to $x_m$ rather the just being an arbitrary path starting at 
$x_n$ only results in a subleading correction to the exponential growth of the number of paths.
It is thus seen that the term in $m_0(a)$ which is divergent for $a \to 0$ is precisely cancelling the exponential growing 
number of paths of length $\ell$, i.e. the {\it entropy} of paths. {\it This will be a universal feature of all the geometric 
systems we will consider: the number of geometric objects of a certain kind will grow exponentially with length, area,
volume or whatever we consider, and in order to have a well defined partition function for these objects, we have 
to adjust the ``bare" coupling constants ({\bf renormalise the coupling constants}) such that this exponential growth 
is cancelled.}

\newpage

\setcounter{equation}{0}
\setcounter{figure}{0}
\renewcommand{\thefigure}{s-4.\arabic{figure}}
 \renewcommand{\theequation}{s-4.\arabic{equation}}

\subsection*{Solutions to Problem Set 4}

\begin{enumerate}
    \item The measure is not affected by this redefinition:
        \[ d S_i = d(\brsh + \dsi) = d \dsi. \]
        We now insert $S_i \equ \brsh \plu \dsi$ into the expression for the partition function:
 \bea
            Z \!\!\!&= \!\!\!&\int \prod_i d S_i  \; 
            \exp \Big[ \mi \sum_i \big(  \knu S_i^2 \plu \lnu S_i^4 \mi \beta h S_i\big) \plu \beta J \sum_{\brak{ij}} S_i S_j \Big] \nonumber \\
            &=&\!\!\! \int  \prod_i d (\dsi) \; \exp \Big[ \mi
            \sum_i \big(\knu \left(\brsh\plu \dsi\right)^2 \plu \lnu (\brsh\plu \dsi)^4\hspace{0.5cm} \nonumber \\
   &&       \hspace{2cm}  \mi  \beta h (\brsh\plu \dsi)\big)  \plu  \beta J \sum_{\brak{ij}} (\brsh\plu \dsi) (\brsh\plu \dsj) \Big] \nonumber
   \eea
        The sum $\brak{ij}$ indicates that we sum  over all the nearest neighbours. 
        Next, we expand  to quadratic order in $\dsi$:
        \begin{align*}
            Z &= \int \prod_i d (\dsi) \;  
            \exp \Big[ \mi \sum_i \Big[  \big\{\knu \brsh^2   \plu \lnu \brsh^4 \mi  \beta h \brsh\big\} +  \\
            & \qquad \quad  \qquad\big\{ 2\knu \brsh  \plu 4 \lnu \brsh^3 \mi \beta h\big\} \dsi 
            \plu \big\{\knu \plu 6 \lnu \brsh^2\big\} \dsi^2\Big]  \\
            &\qquad \quad \qquad \plu \beta J \sum_{\brak{ij}} \left(\brsh^2\plu\brsh (\dsi \plu  \dsj) \plu \dsi \dsj\right)  \Big]\\
            &=\! Z_\mf^V \! \!\int\!\! \prod_i d(\dsi) 
            \exp\! \Big[ \! \mi \sum_i \!\Big[\big\{ 2(\knu \mi DJ\beta) \brsh\plu 4\lnu \brsh^3   \mi \beta h \big\} \dsi
            \\
        &\qquad\qquad \quad   \plu \big\{\knu \mi DJ\beta \plu 6\lnu \brsh^2 \big\}\dsi^2 \Big]   \mi
         \frac{1}{2}\beta J \sum_{\brak{ij}} ( \dsi \mi \dsj)^2\Big]
        \end{align*}
        Here we have defined $Z_\mf$ as the part of $Z$ independent of $\delta S_i$:
        \[ Z_\mf = \exp \left[ - \left(\knu\mi DJ\beta\right) \brsh^2\mi \lnu \brsh^4 \plu \beta h \brsh\right] \]
        and used the fact that $\sum_{\brak{ij}} \brsh = D \sum_i \brsh$ and $\sum_{\brak{ij}}  \dsi = D \sum_i \dsi$
        as well as 
          \[ \sum_{\brak{ij}} \dsi \dsj = - \half \sum_{\brak{ij}}  \left(\dsi \mi  \dsj\right)^2   + D\sum_i\dsi^2  \]
                  If we now let $\brsh$ satisfy
 \beq\label{s4j2}
2 \left( \knu \mi D\beta J\right)\brsh \plu  4\lnu \brsh^3 = \beta h 
 \eeq
        we see that the term linear in $\dsi$ in the exponential drops out. The remaining terms are quadratic in the $\dsi$, which 
        implies that calculating $\brak{\dsi}$ we will indeed obtain $\brak{\dsi} = 0$.

\item

We have 
\beq\label{s4j1}
 \beta f_\mf = (\kappa_0 \mi D J \beta) \left\langle S(h)\right\rangle^2 + \lambda_0 \left\langle S(h)\right\rangle^4 -\beta h \brak{S(h)} 
 \eeq
 and differentiating wrt $h$ we obtain
 \beq\label{s4j3}
 \beta \frac{\prt f_\mf}{\prt h} = \Big[2\left( \knu \mi D\beta J\right)\brsh \plu  4\lnu \brsh^3 \mi  \beta h\Big]
 \frac{\prt \brak{S(h)}}{\prt h} -  \brsh
 \eeq
 and from eq.\ \rf{s4j2} then 
 \beq\label{s4j4}
 \frac{\prt f_\mf}{\prt h} = - \brsh
 \eeq
 
 \item
 Just substitute $h_{\rm eff}$ for $h$ in $f_{\rm free}$.
 
 
 \item  
 For $\kappa_0 > DJ\beta$ the solution is $ \brsh|_{h=0}  =0$, and for $\kappa_0 < DJ\beta$ we have 
 \beq\label{s4j5}
 \brsh|_{h=0} = \sqrt{\frac{2(DJ \beta \mi \kappa_0)}{4\lambda_0}} = c \sqrt{\beta \mi \beta_c}, \qquad c \equ \sqrt{\frac{DJ}{2\lambda_0}},
 ~~~\beta_c \equ \frac{\kappa_0}{DJ}
 \eeq
 
 \begin{figure}[!ht]
\centerline{\includegraphics[width=.5\linewidth, angle = 0]{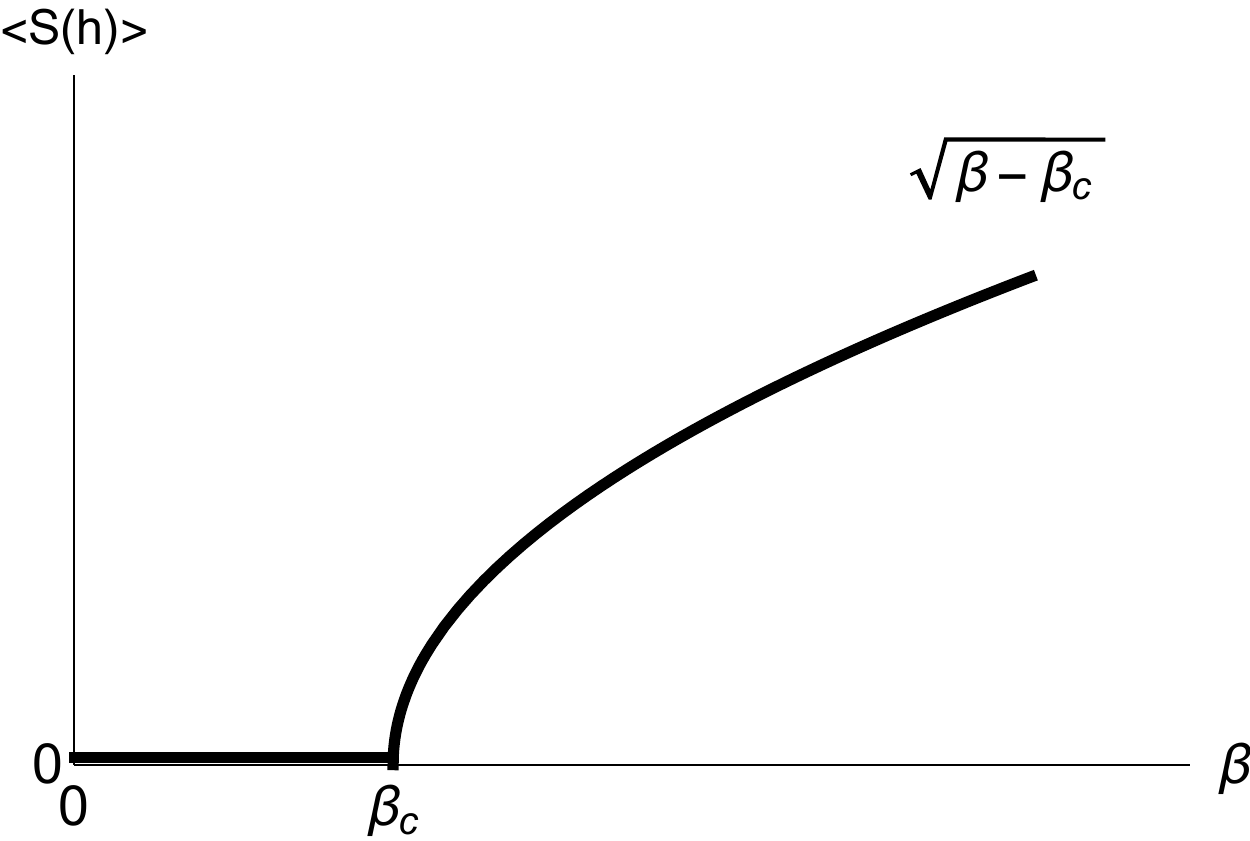}}
\end{figure}

 \item
 Follows from \rf{s4j5}
 
 \item
 
 If we differentiate eq.\ \rf{s4j2} wrt $h$ we obtain
 \beq\label{s4j6}
  2\big[ (\knu-D\beta J) + 6\lnu \brsh^2\big]
 \frac{\prt \brak{S(h)}}{\prt h}  = \beta \nonumber
 \eeq
 i.e.
 \beq
 \frac{\prt \brak{S(h)}}{\prt h}=
 \frac{\beta}{2\big[(\knu\mi D\beta J) \plu 6\lnu \brsh^2\big]} \nonumber
 \eeq
 For $\beta < \beta_c$ we have $\brak{S(h)}|_{h=0} = 0$ and this leads to 
 \beq\label{s4j7}
 \frac{\prt \brak{S(h)}}{\prt h}\Big|_{h=0} = \frac{\beta}{2(\knu \mi D\beta J)}
 \eeq
 while for $\beta > \beta_c$ we have from \rf{s4j5}: $6 \lambda_0 \brsh^2|_{h=0}  = 3(DJ \beta \mi \kappa_0)$ and thus 
\beq\label{s4j8}
 \frac{\prt \brak{S(h)}}{\prt h}\Big|_{h=0} = \frac{\beta}{4(D\beta J \mi \knu)}
 \eeq
 
\item 
This follows since $|\beta\mi \beta_c| = | \beta \mi  \frac{\kappa_0}DJ |$.

\item

Differentiating wrt $J_i$ and $J_j$ in the integral we obtain $\LA x_i x_j\RA$, while differentiating instead 
$e^{s4j_i A_{ij}^{-1}J_j/2}$  (and setting $J=0$) we obtain $A_{ij}^{-1}$. 

\item 

This follows directly from the expression for $H_F(\dsi)$ and the definition of the discrete Laplace operator.

\item

This follows from the expression for $m^2$  given by \rf{p4-mass} and the definition of $T_c$.

\item

Clear again from  from \rf{p4-mass}.

\end{enumerate}

\newpage

\setcounter{equation}{0}
\setcounter{figure}{0}
\renewcommand{\thefigure}{s-5.\arabic{figure}}
 \renewcommand{\theequation}{s-5.\arabic{equation}}

 \subsection*{Solutions to problem set 5}

 \subsubsection*{Rooted planar trees}

 \begin{enumerate}

\item
Assume $h(L) = L^{\gamma-2}$. We have (with notation $ \Delta \mu := \mu\mi  \mu_c$)
$$
z(\mu)= \sum_L e^{-\mu L} \; e^{\mu_c L}L^{\gamma-2} = \Big( \Delta \mu)^{1-\gamma} 
\sum_L \Delta \mu \; e^{ \Delta \mu L } \Big( \Delta \mu L\Big)^{\gamma-2}.
$$ 
Assume now that $\gamma > 1$. When we take the limit $\mu \to \mu_c$ the sum turns into a finite integral
(with value $\Gamma(\gamma-1)$). Thus 
we have for $\gamma >1$:
$$
z(\mu) \to \frac{ \Gamma(\gamma-1)}{ \Delta \mu^{\gamma-1}}, \qquad \Delta \mu \approx \left(1-\frac{g_c}{g}\right), 
\quad {\rm i.e.} \quad z(g) \propto \left(1- \frac{g_c}{g} \right)^{1-\gamma} 
$$
If $0 < \gamma \leq 1$ then differentiate $z(\mu)$ wrt $\mu$. Then the argument
is true for $z'(\mu)$, and integrate to obtain $z(\mu) \equ z(\mu_c) - c \,(\mu\mi \mu_c)^{1-\gamma}$.

Recall that if $z(\mu)$ is the partition function for rooted branched polymers then the susceptibility
is $\chi(\mu) \propto z'(\mu)$ and by definition $\chi(\mu) \sim (\mu - \mu_c)^{-\gamma}$. Thus $\gamma = 1/2$
corresponds to $L^{-3/2}$.

Now let us assume that $z(g) \mi z(g_c) \propto (1\mi g_c/g)^{1 - \gamma}$. We can now Taylor expand and obtain
$$
z(g)  \mi z(g_c)\propto \left( 1- \frac{g_c}{g}\right)^{1 - \gamma} =   \sum_L c_L \left(  \frac{g_c}{g}\right)^L, \quad 
c_L =  \frac{ \Gamma(L\mi (1\mi \gamma))}{\Gamma(L\plu 1) \Gamma(  \gamma-1)}
$$
More precisely we find (writing $s= 1\mi \gamma$ and assuming for simplicity that $s < 0$, but the arguments are also correct for 
$s > 0$)
$$
c_L = \frac{ (-1)^L\;s (s-1) \cdots (s \mi L\plu1)}{L!} = \
\frac{(L\mi s\mi 1) \cdot\cdot (-s)}{L!}\frac{\Gamma(-s)}{\Gamma(-s)} = 
\frac{\Gamma(L\mi s)}{\Gamma(L\plu 1) \Gamma( -s)}.
$$
Here we have used the definition of the Gamma function $s \Gamma(s) = \Gamma(s\plu 1)$, $s! = \Gamma(s\plu 1)$.
Finally the property 
$$
\frac{\Gamma( L\mi s)}{\Gamma(L\plu 1)} \to \frac{1}{L^{s+1}} \quad {\rm for} \quad L \to \infty
$$
leads to the desired power dependence. On can prove the asymptotic behavior using Stirling's formula for the 
asymptotic behavior of the $\Gamma$-function for large argument, as well as $(1-s/L)^L \to e^{-s}$ for $L \to \infty$.

\item
 In order to determine $g_c$, first find $z_c$ by solving $\displaystyle{\frac{dg}{dz}}=0$. We see that
 $$
 g(z) = \frac{1+ \sum_{n=2}^\infty w_n z^{n-1}}{z} = \frac{\sum_{n=0}^\infty z^n}{z}  = \frac{1}{z (1-z)}= \frac{1}{z} + \frac{1}{1-z}.
 $$
 $$
\frac{dg}{dz} = -\frac{1}{z^2}+\frac{1}{(1-z)^2} =0 \quad \Rightarrow \quad z_c = \frac{1}{2} \quad \Rightarrow \quad 
g_c = g(z_c) = 4
$$
Since 
$$
\frac{d^2g}{dz^2}\Big|_{z_c} =  \frac{2}{z^3_c}+\frac{2}{(1-z_c)^3}  > 0,
$$
we know that for $z$ close to $z_c$ we have $g(z) \mi g(z_c) \approx  c (z_c\mi z)^2$, i.e. $z(g) = z_c \mi \tilde{c} \sqrt{g-g_c}$ 
and thus $\gamma \equ 1/2$ and ${\cal N} (L)  \approx 4^L L^{-3/2}$.

\item
$$
g = \frac{1}{z(1\mi z)} \quad\Rightarrow \quad z^2-z+\frac{1}{g} =0\quad  \Rightarrow \quad z = \frac{1- \sqrt{1- \frac{4}{g}}}{2}.
$$

\item Now taking $g(z) = 1/z\plu z$ we could repeat the steps above for $g(z)  \equ 1/(z(1-z))$, but let us instead just solve 
for $z(g)$:
$$
g = 1/z\plu z \Rightarrow z^2-gz +1 =0 \Rightarrow z(g) = \frac{g - \sqrt{g^2 -4}}{2}
$$
It follows that 
$$
g_c = 2,\quad z_c = 1, \quad z(g) -z_c = \sqrt{2} \sqrt{1- \frac{2}{g}} \;\Big( 1+ O( g-g_c) \Big), \quad \gamma = \frac{1}{2}
$$

\item  Now $g(z) = 1/z+z^{n-2}$ and $n>2$ by assumption. 
$$
\frac{dg}{dz}= \frac{-1}{z^2} + (n\mi 2) z^{n-3} ,\qquad \frac{d^2g}{dz^2} = \frac{2}{z^3} + (n\mi 2)(n\mi 3) z^{n-4}.
$$
Thus we find from $g'(z_c) \equ 0$:
$$
z_c = \frac{1}{(n\mi 2)^{\frac{1}{n-1}}}, \qquad g(z_c) = (n\mi 2)^{\frac{1}{n-1}} +  \frac{1}{(n\mi 2)^{\frac{n-2}{n-1}}},  
\qquad g''(z_c) > 0.
$$
This implies as   before the $z(g) \mi z_c \approx c \sqrt{1\mi g_c/g}$ and that $\gamma \equ 1/2$.

We have 
$$
g_c(n) \approx  1+\frac{\ln n}{n}   \quad {\rm for} \quad n\gg 1. 
$$
This choice of weights corresponds to only allowing extremely high branching. This means that for 
$L < n$ the only BP will be the root connected to a vertex of order 1, and for $L < 2n$ there will only be one more BP, namely 
the one where the root is connected to a vertex of order $n$.
 It is thus not surprising that 
exponential growth $g_c^L$ is slow, i.e.\ $g_c(n)$ is close to 1 (although this argument is {\it not} a proof).
\end{enumerate}

Thus we see explicitly in the examples that the critical value of $z_c$ (and therefore $g_c$) will change when choosing different branching weight ratios, but that the power law behavior so far has been universal, corresponding to $\gamma = 1/2$.  We
will study  under what conditions one gets the `standard' value of $\gamma = 1/2$, and what one can do to change it.

\begin{enumerate}\setcounter{enumi}{5}
\item Now $g(z) = 1/z+1$, which we can solve for $z$ as
$$
z(g) = \frac{1}{g-1}  = \frac{1}{g}\; \frac{1}{1 \mi \frac{1}{g} } =  \frac{1}{g}  \sum_{n=0}^\infty \Big(\frac{1}{g}\Big)^{n} = 
\sum_{n=1} (\e^{-\mu})^n, \quad g_c =1, ~~ \gamma = 2.
$$
Recall that we assigned a factor $e^{-\mu}$ for every unit length `link' of the BP, so that this expression for $z$ simply sums over all possible paths without branching (since $w_2 = 1$ and all other weights are 0). Since this BP is not embedded in a target space, the only property of such a BP is its length, and each length is counted exactly once. It is thus not a ``real'' BP and we have a corresponding non-standard $\gamma$ which is obtained for $z_c \equ \infty$.

 \end{enumerate}
    
    \subsubsection*{Criticality of BPs}
    
    \begin{enumerate}\setcounter{enumi}{6}
    
    \item We see that $z$ is an expansion in terms of $e^{-\mu}$, i.e.\ in $1/g$, the first term being $1/g$. 
    Therefore, higher terms in the expansion are higher powers of $1/g$, and in the limit $g \to \infty$ all terms will vanish, giving $z=0$ (assuming we have an absolute convergent power series).
    
    \item We expand $z g(z)$ around the point $z_c$. The first two derivatives are then 
    \[ \frac{d}{dz} (z g(z)) \equ g(z) \plu z g'(z),\quad
    \frac{d^2}{dz^2} (z g(z))  \equ g'(z) \plu g'(z) \plu  z g''(z) \equ 2g'(z) \plu  z g''(z). \]
    Continuing this procedure, we see that 
    \[ \frac{d^n}{dz^n} (z g(z)) = n g^{(n-1)}(z) + z g^{(n)}(z). \]
    The Taylor expansion is then
    $$ 
    \left. (z g(z))\right|_{z_c} = z_c g(z_c) + \sum_{m=1}^\infty \left( m g^{(m-1)}(z_c) + z_c  g^{(m)}(z_c)\right) \frac{(z-z_c)^m}{m!}. 
    $$
    However, using the fact that $z g(z)$ is a polynomial in $z$ of order $n$, we know that all terms with $m > n$ vanish. Furthermore, by the assumption that $g^{(m)}(z_c)=0$ for all $1 \leq m < n$, we see that all terms drop out except the one of order $n$ and the $g^{(0)}$ for $m=1$. The term of order $n$ can be determined from the fact that $g(z)$ is the sum of $1/z$ and a polynomial of order $n-1$. It is clear that $n g^{(n-1)}(z_c) = 0$ (by our assumption), so we only need to compute $z_c g^{(n)}(z_c)$. The $n$th derivative of $g(z)$ can only receive a contribution from the $1/z$ part, since the $n$th derivative of an order $n-1$ polynomial is zero. Therefore
    \bea z g(z)) &=& z_c g(z_c) + g(z_c) (z-z_c) + z_c g^{(n)}(z_c)\frac{(z-z_c)^n}{n!} \nonumber \\
    &= &z g(z_c) +  z_c (-1)^n n! z_c^{-1-n} \frac{(z-z_c)^n}{n!} 
    = z g(z_c) + \left(1-\frac{z}{z_c}\right)^n.\nonumber
    \eea

   $w_2=0 $ and  we have 
   \bea\nonumber
   zg(z) &=& {1 + f(z)} = {1} +w_2z +w_3 z^2 + w_4 z^2 \cdots =  {1} +w_3 z^2+w_4 z^3 \cdots\\
   zg(z) &=& zg(z_c) + \left(1-\frac{z}{z_c}\right)^n  = {1} + \Big( g(z_c)- \frac{n}{z_c}\Big)\,z +  \frac{n(n-1)}{2 z_c^ 2}\,z^2 + \cdots\nonumber
   \eea
   
   where we have just expanded the bracket. Therefore  $z_cg(z_c)\mi  {n}\equ 0$. 

   \item 
  
  We simply find the weights by using  $f(z) = z g(z) \mi 1$ andby  expanding $(1-z/z_c)^n$: the coefficient 
  to $z^{m-1}$ is the weight $w_m$:
  \begin{equation}\label{wm0}
  w_m = \frac{(-1)^{m-1}}{z_c^m}  \; 
  \begin{pmatrix} n \\ m-1 \end{pmatrix}  =  \frac{(-1)^{m-1}}{z_c^{m-1}}  \; 
  \frac{\Gamma(n+1)}{\Gamma(n-m+2)\Gamma(m)},\quad 3 \leq m \leq n+1.
  \end{equation}
  and by definition $w_1 =1$ and $w_2=0$.

  \item We solve $ z_c^n (g \mi g_c) z \equ (z_c \mi z)^n$ iteratively wrt $z$:
  \bea
 z &=& z_c - z_c \,(g \mi g_c)^{1/n} z^{1/n} 
  = z_c - z_c \,(g \mi g_c)^{1/n} \left(z_c \mi z_c (g\mi g_c)^{1/n} z^{1/n}\right)^{1/n} \nonumber \\
  &=& z_c - z_c^{1+1/n} (g\mi g_c)^{1/n} + O((g\mi g_c)^{2/n})\nonumber
  \eea
  We have $z = z_c \mi c\, (g \mi  g_c)^{1-\gamma}$ for the rooted branched polymers. Thus $\gamma = 1\mi 1/n$.
  
  \item 
  If all weights are positive, the second derivative of $g(z)$ is
  $$
  g''(z) = \frac{2}{z^3} + \textrm{positive terms} > 0 \quad {\rm for~all~} z >0.
   $$
  Since $z$ by definition is larger than zero if the weights are positive  and $g(z)$ has a minimum if we assume that
  the weights $w_m \equ 0$ for $m > n \plu 1$, expansion around this minimum at $z_c$ leads to the standard result
 $$
  g(z) = g(z_c) + c_1(z\mi z_c)^2 + \cdots \quad {\rm i.e.} \quad
 z(g) = z_c - c_2  (g\mi g_c)^{1/2} + \cdots
 $$
  Clearly  $\gamma = 1/2$  if and only if  the second derivative of $g(z)$ is  positive at the critical point, and one way to achieve
  this is to have a finite number of branching weights, which in addition are all positive.

  \item Again compute the second derivative of $g(z)$. One can then notice that 
  \[  g''(z) = \frac{f''(z)}{z} -\frac{2}{z} g'(z) \]
 after collecting terms. At the critical point, $g'(z) = 0$, so positivity of $f''(z)$ at the critical point implies positivity of $g''(z)$ there. By our previous discussion, this again leads to $\gamma = 1/2$, provided there {\it is} a critical 
 point. If we add the assumptions that $f(0), f'(0) = 0$ and $f''(x) > 1/x^2$ for large $x$ this is ensured since
 we can write 
 $$
 g'(z) = -\frac{1}{z^2} \;( 1+ f(z) - zf'(z)) =   -\frac{1}{z^2} \Big(1 - \int_0^z dx x f''(x) \Big).
 $$
 The assumptions ensure that $g'(z)$  is negative for small $z$ and positive for large $z$.

  \item The two given examples of functions with the desired property show that not all $w_m$ need to be positive, 
  as long as $f''(z) > 0$ (at $z_c$). Working out the $w_m$ is straightforward from the power series of the functions:
  $$
  w_m = \frac{1}{(m-1)!},~~m>2, \qquad w_{2k-2} = \frac{2^{2k}(2^{2k}\mi 1) B_{2k}}{(2k)!}, ~~w_4 \equ1,~w_6 \equ -\frac{1}{3},\ldots
  $$ 
  The Bernoulli numbers $B_{2k}$ enter in  the weights $w_{2m}$, $m \geq 2$ defined by  the power series of $z^2 \tanh z$,
  and result in  oscillating signs, and a radius of convergence $r = \frac{\pi}{2}$, 
  but the function is perfectly regular along the real axis (there are poles on the imaginary $z$-axis at $\pm i\pi/2$).

  \item The problem of extracting the coeficients $w_m$ from $(1- z/z_c)^s$ is identical to the problem of finding 
  the Taylor coefficient $c_L$ in $(1-g_c/g)^{1-\gamma}$, $s \equ 1 \mi \gamma$, which 
  we addressed in the first question in this problem set. We can then directly use these results with $L \equ m \mi 1$. 
  \bea
   w_m &= & \frac{(-1)^{m-1}}{z_c^{m-1}}\frac{s\cdot(s-1)\cdot(s-2)\cdots (s-(m-1)+1)}{(m-1)!} \nonumber \\
 & =&  \frac{1}{z_c^{m-1}}\frac{(-1)^{m-1}\Gamma(s+1)}{\Gamma(m) \Gamma(s-m+2)} =
   \frac{\Gamma(m-1-s)}{\Gamma(m)\Gamma(-s)}.\label{wm2}
  \eea
 The first expression in the second line is simply the generalization of what we already derived in question 9 for 
 $s$ being an integer $n$, and the second expression is the same as we derived in question 1 with  $L \equ m \mi 1$.    
 The second formula is well suited to find the asymptitic behavior of $w_m$ for large $m$, as we did in question 1, and we find:
 \begin{equation}\label{wm3}
  w_m \to  \frac{(-1)^n}{|\Gamma(-s)|} \; \frac{1}{m^{s+1}}\quad {\rm for}\quad m \to \infty, 
\end{equation}
 The factor $(-1)^n$ comes from $\Gamma(-s)$.
  
  \item
  The oscillating sign of $w_m$ for $m < s \plu 2$ follows from the first expression in the second line of eq.\ \rf{wm2} 
  and the constancy for $m >  s\plu 2$
  follows from the second expression in second line of eq.\ \rf{wm2}.
   
   \item
  We have directly  from (\ref{wm2}) that $w_m$ is positive for $1 < s <2$ and $m > 2$ (by construction $w_1\equ 1$ and 
  $w_2 \equ 0$ since we have fixed $g(z_0) \equ s/z_c$).
  
 \item
 The proof is identical to that in the case $s \equ n$, considered in question 10.

    \end{enumerate}

\newpage

\setcounter{equation}{0}
\setcounter{figure}{0}
\renewcommand{\thefigure}{s-6.\arabic{figure}}
 \renewcommand{\theequation}{s-6.\arabic{equation}}

\subsection*{Solutions to Problem Set 6}

\begin{enumerate}

\item 
The figures are clearly the only possibility if we have the so-called {\it hard dimers}.
That leads to, in terms of  equations 
\beq\label{s6j1}
  z = e^{-\mu} \Big(1+z^2+2z \tz\Big), \qquad \tz = e^{-\mu} \xi \Big(1+z^2\Big)
  \eeq
or the ones given in the problem set. 

\item
Eliminating $\tz$ from \rf{s6j1} leads to 
\beq\label{s6j2}
g = \frac{1\plu z^2}{z} + \frac{2\xi}{g}(1\plu z^2).
\eeq
This is a second order equation in $g$ and we find
\beq\label{s6j3}
g = \half \left[ \frac{1\plu z^2}{z} + \sqrt{\frac{(1\plu z^2)^2}{z^2} + 8 \xi (1\plu z^2)}\right] 
\eeq
where one has to choose the plus sign for square root since we know from \rf{s6j1} that $ g \to 1/z$ for 
$z\to 0$.

\item
Differentiate $g$ given by eq.\ \rf{s6j2} wrt $z$:
\bea
g' (z)& =& -\frac{1}{z^2}+1 + \frac{4\xi z}{g} + F_1(z,g,g'),\\
g''(z) &=& \frac{2}{z^3} + \frac{4 \xi}{g} +  F_2(z,g,g') - \frac{2\xi(1\plu z^2)}{g^2} \; g''(z)\label{s6j4}
\eea
where $F_1$ and $F_2$ are functions which vanish if $g'(z)  =0$. Thus we have for the value of $z$
where $g'(z) = g''(z) = 0$:
\beq\label{s6j5}
0 =  -\frac{1}{z^2}+1 + \frac{4\xi }{g} \,z ,\qquad 0=\frac{1}{z^3} + \frac{2 \xi}{g} ,
\eeq
\item
From this we can find $1/z^2$ and $\xi/g$, and then from \rf{s6j2} the values of $g$ and $\xi$:
\beq\label{s6j6}
\frac{1}{z_c^2} = \frac{1}{3}, \quad \frac{2\xi_c}{g_c} =  \frac{-1}{3\sqrt{3}},\qquad {\rm i.e.} \quad
z_c = \sqrt{3},\quad g_c = \frac{8}{3\sqrt{3}},\quad\xi_c = \frac{-4}{27}.
\eeq
Finally, differentiating $g''(z)$ in \rf{s6j4} one more time and using $g'(z_c) \equ g''(z_c)  \equ 0$ we obtain
\beq\label{s6j7}
g'''(z_c) = - \frac{6}{z_c^4} - \frac{2\xi_c(1+z_c^2)}{g_c^2} \; g'''(z_c)
\eeq
and for the values in \rf{s6j6} we conclude $g'''(z_c) \neq 0$.
\item
We write now $g(z,\xi)$ for the function \rf{s6j3}. As for ordinary BP, for a fixed $\xi > \xi_c$ we have 
now a critical point $z_k(\xi) < z_c$ where $g'_z (z_k,\xi) =0$, and the corresponding $g_k(\xi) = g(z_k,\xi)$.
Thus we have a critical curve $\xi \to (z_k(\xi),g_k(\xi))$ for $\xi > \xi_c$ (see Fig.\ \ref{fig1y}).  Rather than finding the 
parametric form of the curve, we can directly find the form by solving $g'_z(z,\xi) = 0$ from the 
first eq.\ in \rf{s6j5} (finding $\xi/g$) and inserting in \rf{s6j2}:
\beq\label{s6j8}
\frac{2\xi}{g_k} = \frac{1-z_k^2}{2z_k^3},   \quad {\rm i.e.} \quad g_k= (1+z_k^2)\Big(\frac{1}{z_k} + \frac{2\xi}{g_k}\Big)
= \frac{(1+z_k^2)^2}{2z_k^3}.
\eeq
\item
\begin{figure}[!ht]
\centerline{ \includegraphics[height=6cm]{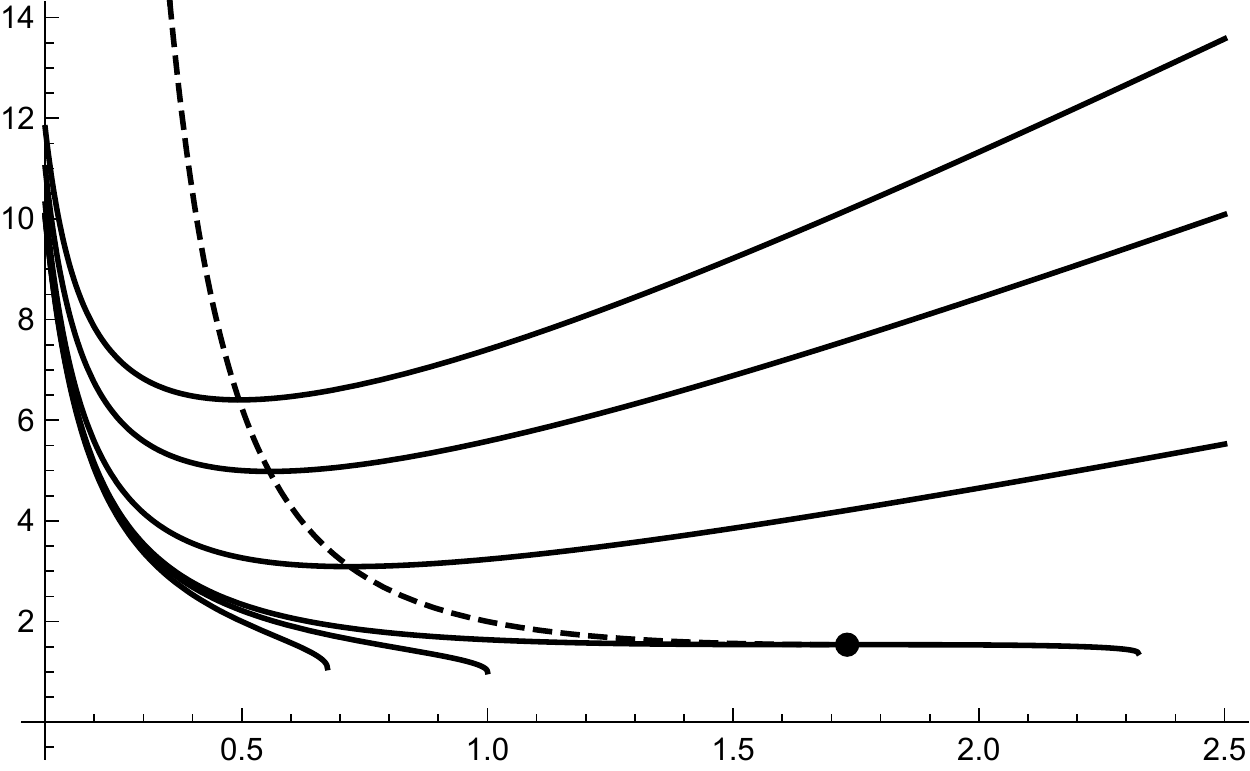}}
\caption{{\small The various curves $z \to g(z,\xi)$ for a number of values of $\xi$ in the range from 10 (top curve)
to -0.4 (bottom curve). The third curve from the bottom corresponds to $\xi=-4/27$ which has the lowest value
of the curve minimum. The dashed curve is the curve of the minima for the curves $g(z,\xi)$ for $\xi > -4/27$, 
i.e.\ the part of the curve $g_k(z_k)$ given by eq.\ \rf{s6j8} until its minimum, i.e.\ for $0<z< \sqrt{3}$. An extended 
range of  $g_k(z_k)$ is shown in Fig.\ \ref{fig2}. The two lowest curves have $\xi < -4/27$ and no local extrema.
For $\xi < -1/8$ the square root in the expression \rf{s6j3} for $g(z,\xi)$ can become negative for sufficient large $z$
and the function is only defined up to this point in the graph.}}
            \label{fig1y}
        \end{figure}

However, the curve $g_k (z_k)$ seemingly continues happily for $z_k > z_c$. This can be understood from 
the explicit solution for $g(z,\xi)$ given in \rf{s6j3}, the behaviour of which are shown in Fig.\ \ref{fig2} for various 
values of $\xi$. We are (from the point of view of physics of BPs )  only interested in the 
first minimum of $g(z,\xi)$ for a given value $\xi$. This first minimum we meet precisely on
the first part of the curve $g_k(z_k)$.
As long as $\xi$ decreases, starting at $\infty$, $z_k(\xi)$ will increase and $g_k(z_k)$ will decrease
{\it until} $dg_k/dz_k=0$. This point is exactly $(z_k,g_k) = (z_c,g_c)$. For $z_k > z_c$ 
 $g_k(z_k)$ will start to increase again. This new point $(z_k,g_k)$ is 
associated with a  second extremum, a local maximum for the function $g(z,\xi)$ for a given value of $\xi$.
From the explicit form of $g(z,\xi)$ one can show that for $\xi \geq -1/8$ there is only one extremum, the 
minimum of   $g(z,\xi)$  at $z_k(\xi)$. For $-4/27 < \xi <-1/8$ there are {\it two} values $z_k$ for which 
$g'_z(z,\xi) =0$.  Fig.\ \ref{fig2} shows how the part of the curve $g_k(z_k)$ for $z_k >z_c$ passes through the maxima 
of the curves $g(z,\xi)$ for $-4/27 < \xi <-1/8$. 
The point $z_k(\xi_c = -4/27) = \sqrt{3}$ is precisely the point where $ dg_k/dz_k=0$. Finally
for $\xi < \xi_c = -4/27$ there is no local minimum, $g'(z,\xi) < 0$ for all $z$ where $g(z,\xi)$ are defined.
\begin{figure}[!ht]

\centerline{ \includegraphics[height=7cm]{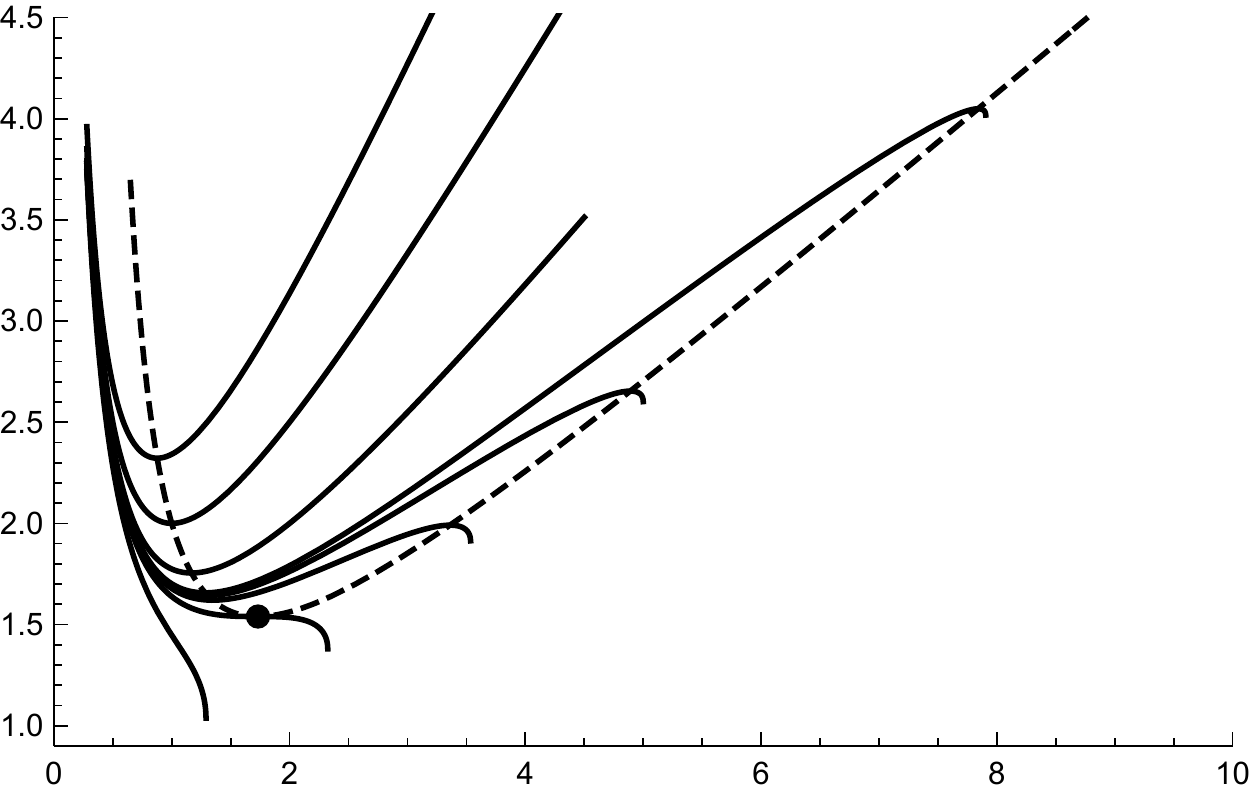}}
\caption{{\small The various curves $z \to g(z,\xi)$ for a number of values of $\xi$. For $\xi \in ]-4/27,-1/8]$
the curves have a local maximum to the right of the local minimum and the curve $g_k(z_k)$ given by eq.\ \rf{s6j8} 
passes through these local maxima for $z_k > \sqrt{3}$, as seen on the figure.}}
            \label{fig2}
        \end{figure}

\item
We can find the function $z_k(\xi)$, or more easily the inverse function $\xi = \xi(z_k)$. From \rf{s6j8}
we obtain 
\beq\label{s6j9}
\xi(z) = \frac{(1-z^4)(1+z^2)}{8 z^6}.
\eeq
\item
We now want to show that 
\beq\label{s6j10}
\frac{d \xi(z)}{dz}\Big|_{z=z_c} =0
\eeq
One can of course directly differentiate \rf{s6j9} and insert $z_c = \sqrt{3}$. Also, taking 
the double derivative one finds that $\xi''(z_c) \neq 0$. 

However it is more in the spirit of critical phenomena to show that the point $z_k^c$ where 
$dg_k (z_k)/dz_k = 0$ is also the point where 
both  $\xi'(z_k) =0$ and $ g'_z(z,\xi)|_{\xi = \xi(z_k)} =  g''_{zz}(z,\xi)|_{\xi = \xi(z_k)} =0$, i.e.\ 
this point, $z_k^c$, is naturally identified with $z_c$ where something special happens, {\it independently
of the specific form of the function $g(z,\xi)$}.
First note that by definition we have
\beq\label{s6j11}
g_k(z) = g(z,\xi(z))\quad {\rm i.e.} \quad \frac{d g_k}{dz} = \frac{\partial g(z,\xi)}{\partial z}\Big|_{\xi(z)} + 
\frac{\partial g(z,\xi)}{\partial \xi}\Big|_{\xi(z)} \frac{d \xi(z)}{dz} 
\eeq
From the very definition of the critical curve $g_k(z_k)$ we have that $ g'_z(z,\xi)|_{\xi = \xi(z)} =0$
everywhere on the curve. Thus we see that there is equivalence between $dg_k (z)/dz = 0$ and $\xi'(z) =0$
unless for some reason  $g'_\xi(z,\xi)|_{\xi = \xi(z)}=0$ (no physics is related to this). To see that 
$dg_k (z)/dz = 0$ implies that $g''_{zz}(z,\xi)|_{\xi = \xi(z)} =0$ expand $g'_z(z,\xi)|_{\xi = \xi(z)} (=0)$
around the point $z_k^c$ where $dg_k (z)/dz = 0$. Let $z = z_k^c + \Delta z$. We then have 
$\xi( z) = \xi(z_k^c) +c (\Delta z)^2$ because $\xi'(z_k^c) =0$ and we obtain
\bea\label{s6j12}
g'_z(z,\xi(z)) &=& g'_z(z_k^c+\Delta z,\xi_k^c + c  (\Delta z)^2) \\
&=& 
g'(z_k^c,\xi_k^c) + g''_{zz}(z_k^c,\xi_k^c) \Delta z +  O((\Delta z)^2). 
\eea
Since, by definition, $g'_z(z,\xi(z))=g'(z_k^c,\xi_k^c)=0$, we conclude that $ g''_{zz}(z_k^c,\xi_k^c) =0$ 
and we have shown that we indeed can identify $z_c$ with $z_k^c$.

\item

The two terms 
\beq\label{s6j13}
\left.\frac{\partial^2 g}{\partial \xi \partial z} \right|_{z_c, \xi_c} \!\!\!\!\!\!\!(z\mi z_c)(\xi\mi \xi_c) +
    \left.\frac{1}{6} \frac{\partial^3 g}{\partial z^3}\right|_{z_c,\xi_c} \!\!\!\!\!\!\!(z\mi z_c)^3 
\eeq
in the expansion (11) of $g(z,\xi)$ around $(z_c,\xi_c)$ both behave like $(\xi\mi \xi_c)^{3/2}$ when one
uses   eq.\ (10) from the problem sheet.

\item
Clear, differentiating twice one obtains the desired behaviour of $g''_k (\xi)$. Recall that 
we have that the free energy $f(\xi) = -\log g_k(\xi)$. Differentiation twice we find the that
singular behaviour of  $f''$ is the same as that of $g_k''$ and we have finally
\beq\label{s6j14}
f''(\xi) \propto \frac{1}{(\xi \mi \xi_c)^{1/2} } \propto (\xi \mi \xi_c)^{\sigma-1} ,\quad {\rm i.e.} \quad \sigma = \half.
\eeq

\end{enumerate}

\newpage

\setcounter{equation}{0}
\setcounter{figure}{0}
\renewcommand{\thefigure}{s-7.\arabic{figure}}
 \renewcommand{\theequation}{s-7.\arabic{equation}}

\subsection*{ \hspace{-4mm} Solutions to Problem Set 7}

\subsubsection*{BPs with infinite Hausdorff dimension}

\begin{enumerate}

\item 
\beq\label{s7jj1}
g(z) = \frac{1+f(z)}{z}
\eeq
$g(z) \to \infty$ for $z\to 0$. The radius of converence for $f(z)$ is $z \equ 1$. If $s \leq 1$  
then the power series of the derivative of $f$ is infinite at $z\equ 1$ ($\sum_n n^{-s}$ is divergent 
for $s \leq 1$). Thus also the derivative of $g(z)$ will go to infinity at $z\equ 1$ for these values 
of $s$. The value of $z$, $z_c$, for which $g(z)$ assumes the minimum is thus $0< z_c < 1$, and 
it is a simple minimum: $g''(z_c) > 0$. This argument does not require $g'(1) \equ \infty$, only 
$g'(1) > 0$, which can be shown to lead to $s < 1.59....$.

\item
We have to cancel an $n^{th}$-order polynomial in $g(z)$ to obtain $g(z) \mi g_c \sim c (1 \mi z)^s$.
This can clearly be done by adding a suitable term $\sum_{k=2}^{n+2} \tilde{w}_k z^{k-1}$ to $f(z)$.

\item
Solve by iteration:
\bea
\Delta z &=& \frac{\Delta g}{c_1} - \frac{c_2}{c_1} (\Delta z)^2 -\frac{c_3}{c_1} (\Delta z)^3 + \cdots \\
(\Delta z)_1 &=& \frac{\Delta g}{c_1}\\
(\Delta z)_{1,2} &=& \frac{\Delta g}{c_1} - \frac{c_2}{c_1} \Big(\frac{\Delta g}{c_1}\Big)^2 \\
(\Delta z)_{1,2,3} &=&\frac{ \Delta g}{c_1} - 
\frac{c_2}{c_1} \left.{\left(\frac{\Delta g}{c_1} - \frac{c_2}{c_1} \Big(\frac{\Delta g}{c_1}\Big)^2\right)^2}\right |_{2,3}
-\frac{c_3}{c_1}  \Big(\frac{\Delta g}{c_1}\Big)^3 
\eea

\item 

We have 
\beq\label{s7jj2a}
G^{ (I)}_\mu (r) = \frac{(1+f(z))^2}{f'(z)}   e^{-m_I(\mu) r},\qquad m_I(\mu) = -\log (e^{-\mu} f'(z)).
\eeq
where $g=e^\mu$. Recall from the notes (differentiating $e^\mu z  =1\plu f(z)$ wrt $\mu$)
\beq\label{s7jj2b}
e^{-\mu}f'(z) = 1 + \frac{z}{z'} = 1 + z \frac{d \mu}{dz}, \quad {\rm i.e.} \quad m_I(\mu) = -\log \Big(1 \plu z \frac{d \mu}{dz}\Big).
\eeq
Thus, if $d\mu/dz \neq 0$ for $z\equ z_c$ we obtain $m_I(\mu_c) \neq 0$.  The reason for this is simply that 
if   $s >2$  then $g(z)$ (or $\mu(z)$) is a decreasing function of $z$ all the way to $z_c \equ 1$ and thus $d\mu/dz < 0$ also
at $z_c$. This is  contrary to the situation for $\gamma \equ 1/2$ 
where $\mu \equ  \mu_c - k (z_c \mi z)^2$  and $d \mu /dz \to 0$ for $z \to z_c$.
\end{enumerate}

\subsubsection*{Ising model coupled to BPs}

\begin{enumerate}

\item Assume the spin of the root is +. 
Depending on whether the first vertex after the root has spin + or spin -, we get factors 
\beq\label{s7jj3}
e^{-\mu} e^{\beta+h} \big(1+ f(Z_+)\big),\qquad {\rm or} \qquad  e^{-\mu} e^{-\beta-h} \big(1+ f(Z_-)\big),
\eeq
and similarly, if the spin of the root is -, we obtain
\beq\label{s7jj4}
e^{-\mu} e^{-\beta+h} \big(1+ f(Z_+)\big),\qquad {\rm or} \qquad  e^{-\mu} e^{+\beta-h} \big(1+ f(Z_-)\big).
\eeq

\item
If $h=0$ then clearly it is consistent to choose $Z_+=Z_-$ and solve the equation for $Z=Z_+=Z_-$. Assuming
there is only one solution to the equations, $Z_-=Z_+ $ is justified. The rest of the questions are easily answered.

\item

\beq\label{s7jj5}
\frac{d}{dZ} \frac{1+f(Z)}{Z}\Big |_{Z_c} =0\quad
 \Longrightarrow \quad \frac{1+f(Z_c)}{Z_c^2} = \frac{f'(Z_c)}{Z_c} .
\eeq

\item

The $\log 2$ comes from the fact that for $\beta =0$ the action $e^{\beta \sum_{<i,j>} \sigma_i \sigma_j}=1$
and in $Z(\beta)$ we thus get $\sum_{\sigma_i} = 2^V$ for the spin contribution. This is precisely the number 
of spin configurations, and the classical entropy is $k_B \log$(number of configurations). 
The entropy density is thus $\log 2$ in our units where $k_B=1$.

\item

The equations are readily obtained by expanding the defining equations to linear order in $h$, writing 
$\mu (\beta,h) = \mu_c(\beta) + \Delta \mu$, where we assume $\Delta \mu = k \cdot h + O(h^2)$, and
then adding and subtracting the resultant equations.

\item
We clearly obtain $\Delta \mu = 0$ from 
$$
 (\Delta Z_+ + \Delta Z_-)\left(e^{\mu_c(\beta)}-2 \cosh \beta f'(Z_c)\right) + 2Z_c e^{\mu_c(\beta)} \Delta \mu = 0
 $$
 since $e^{\mu_c(\beta)}-2 \cosh \beta f'(Z_c) \equ 0$ and this means (since we are expanding only
to linear order in $h$, that $\Delta \mu = O(h^2)$, which implies that $d \Delta \mu/dh \to 0$ for $h\to 0$.
Thus the spontaneous magnetization is zero.

\item 
Using the information that $\Delta Z_+ =0$,  we can write
\bea\label{s7j20}
   &&    \Delta Z_-\left(e^{\mu_c(\beta)}-2 \cosh \beta f'(Z_c)\right) + 2Z_c e^{\mu_c(\beta)} \Delta \mu = 0
            \\
   &&      - \Delta Z_-\left(e^{\mu_c(\beta)} - 2 \sinh \beta f'(Z_c) \right) = 4h(1+f(Z_c)) \sinh \beta. \label{s7j21}
\eea
It follows from the equations that $\Delta Z_- \propto \Delta \mu$
and $\Delta Z_- \propto h$, thus $\Delta \mu  = k \cdot h$, where $k \neq 0$.

\item 
From the definition $ \langle m \rangle= d \Delta \mu/dh\big|_{h=0} = k$ and from \rf{s7j20} and \rf{s7j21} we find explicitly,
using $e^{\mu_c(\beta)} = 2 \cosh \beta \big( 1\plu f(Z_c))/Z_c$ as well as $Z_c =1$:
\beq\label{s7j22} 
 \langle m \rangle = \tanh \beta \; \frac{1 \mi \tanh \beta \frac{f'(1)}{1\plu f(1)}}{1 \mi \frac{f'(1)}{1\plu f(1)}} ,\qquad 
 f(1) \equ {\rm Li}_{s+1}(1),\quad f'(1) \equ {\rm Li}_s (1).
 \eeq
 It is seen that  $\langle m \rangle \to 0$ for $\beta \to 0$ and $\langle m \rangle \to 1$ for $\beta \to \infty$, as one would expect. 

\end{enumerate}

\vspace{6pt}

\subsubsection*{Ising model and dimers}

\begin{enumerate}

\item 
Use the suggested formula for each term in the action and extract $\cosh$-factors.

\item
The lowest order $\tanh \beta$ term is obtained by having a link $\sigma_i\sigma_j \tanh \beta$ anywhere 
on the lattice. However we have to "close" the link with two terms $\sigma_i \tanh h$ and $\sigma_j \tanh h$
in order that the summations over $\sigma_i$ and $\sigma_j$ do not give zero. Similarly when we put down 
two links. If they do not touch (hard dimers) they have to be ``closed'' by four terms of the form
$\sigma_k \tanh h$. A given term $\sigma_{k_1} \cdots \sigma_{k_n} \tanh^n h$ has to meet links
at the  vertices $k_i$ in order that the sum over $\sigma_{k_i}$ does not give zero. Thus $n$ has to 
be even (since links bring an even number of $\sigma$s). Next, the {\it smallest}
number of links one can use is obtained if the links can be put down as hard dimers (and it is $n/2$).
There are many other ways one can dress the  $\tanh^n h$ term with links but they always 
involve more links and thus higher powers of $\tanh \beta$ and thus higher powers of $\beta$.

\item
Follows from the expansion given for $Z(\beta,h)$.

\end{enumerate}

\newpage

\setcounter{equation}{0}
\setcounter{figure}{0}
\renewcommand{\thefigure}{s-8.\arabic{figure}}
 \renewcommand{\theequation}{s-8.\arabic{equation}}

\subsection*{ \hspace{-4mm} Solutions to Problem Set 8}

\subsubsection*{Asymptotic expansions}

\begin{enumerate}

\item  
Write 
\beq
f(x) = x h(x), \qquad B(h)(x) = \sum_{n=0}^\infty (-1)^n x^{n} = \frac{1}{1\plu x}. 
\eeq
\beq
h(x) = \int_0^\infty dt \; \e^{-t} B(h)(xt),\qquad f(x) = x  \int_0^\infty dt \; \e^{-t} \frac{1}{1\plu tx}.
\eeq

\item

Differentiate the formal power series for $f(x)$ to obtain another formal power series
\beq
f'(x) =  \sum_{n=0}^\infty (n\plu 1)!(-1)^n x^{n}
 = \frac{1}{x^2} \Big( -  \sum_{k=0}^\infty k! (-1)^k x^{k+1} +x\Big) 
 =  -\frac{1}{x^2} \; f(x) + \frac{1}{x}\nonumber
\eeq

\item

The general solution to an inhomogeneous linear differential equation:
\beq
f' + a(x) f = b(x), \quad f(x) = \e^{-A(x)} \int^x  \d y\e^{A(y)} \, b(y),\quad A(x) = \int^x \d y \,a(y).\nonumber
\eeq  
Applied to our differential equation we obtain
\beq
f(x) = \e^{1/x} \int^x_0 \d y\, \e^{-1/y} \frac{1}{y} =   
 \int^\infty_0 \d t \, \frac{x\e^{-t}}{1\plu xt}\qquad  \Big[t = \frac{1}{y} - \frac{1}{x}\Big] 
\eeq

\item
Differentiating $e^{1/x} \Ei (-1/x)$ it is easily seen that it satisfies the differential equation. Also, changing
variables as above ($t = \frac{1}{y} - \frac{1}{x}$) it is clear that it is the Borel sum of the original formal 
power series.

\item Successive partial integrations: ($y = 1/x$)
\beq\label{s8j1}
\e^{y} \left(\int_y^\infty \d t  \frac{\e^{-t}}{t} = -\frac{\e^{-t}}{t}\Big|_y^\infty + \frac{\e^{-t}}{t^2}\Big|_y^\infty-
 2\frac{\e^{-t}}{t^3}\Big|_y^\infty+ \cdots + (-1)^{n} n!  \int_y^\infty \d t  \frac{\e^{-t}}{t^{n+1}} \right)
 \eeq

\item
Formally we have from the  power series that $g(x) = -f(-x)$. This clearly leads to both the integral and the 
differential equation. The reason that we have changed the definition of $\Ei(u)$ to $\Ei_c(u)$, is 
that we do not want 0 to be part of the integration interval, since the integral then is ill defined
(one can include it by a so-called principle value prescription, but it is easier to avoid 0)

To find the asymptotic expansion of $e^{-1/x}\Ei_c(1/x)$, we perform the partial integration as in \rf{s8j1} 
\bea\label{s8j2}
\int_{-1/x}^{-c} \d t  \frac{\e^{-t}}{t} &=& -\frac{\e^{-t}}{t}\Big|_{\frac{-1}{x}}^{-c} 
+ \frac{\e^{-t}}{t^2}\Big|_y^\infty-
 2\frac{\e^{-t}}{t^3}\Big|_{\frac{-1}{x}}^{-c}+ \cdots + (-1)^{n} n!  \int_{\frac{-1}{x}}^{-c}  \d t  \frac{\e^{-t}}{t^{n+1}} 
 \nonumber \\
 &=& \e^{1/c} \Big( \frac{1}{c}  +\frac{1}{c^2} + 2! \frac{1}{c^3}  +\cdots \Big)-\e^{1/x}\Big( x +x^2 + 2! x^3 +\cdots \Big)  
 \nonumber 
 \eea
and thus 
$$
-\e^{-1/x} \Ei_c(1/x) = \Big( x +x^2 + 2! x^3 +\cdots \Big)  + F(c) \e^{-1/x}
$$
Note that $F(c)\e^{-1/x}$  is a solution to the homogeneous differential equation. This is why we have a 
solution for any positive $c$ {\it and they all have the same asymptotic expansion since $F(c)\e^{-1/x}$ 
does not contribute to the asymptotic series}.

\end{enumerate}

\newpage

\setcounter{equation}{0}
\setcounter{figure}{0}
\renewcommand{\thefigure}{s-9.\arabic{figure}}
 \renewcommand{\theequation}{s-9.\arabic{equation}}

\subsection*{Solutions to Problem Set 9}

\subsubsection*{Branched polymers with loops}

\begin{enumerate}

\item  
\beq\label{s9j1}
z^2 -gz +1+j =0,\quad {\rm i.e.} \quad z(g,j) = \frac{g}{2} - \sqrt{ \frac{g^2}{4} \mi (j\plu 1)}=  \frac{g}{2} \mi \sqrt{\Del(g,j)}
\eeq
We have to choose the minus sign in front of the square root since $z \to 0$ for $g\to \infty$,
from the very definition of the partition function.

\item

\beq\label{s9j2}
g = \frac{1\plu j \plu  z^2}{z},\qquad  \frac{dg}{dz} = -\frac{1\plu j}{z^2} +1,\qquad \frac{d^2g}{dz^2} = \frac{2(1\plu j)}{z^3}
\eeq
Thus 
\beq\label{s9j3}
 \frac{dg}{dz}=0 ~\Rightarrow ~z_c = \sqrt{1\plu j},\quad g_c = 2 \sqrt{1\plu j}, \quad \Del (g_c,j) = 0,
\eeq
and since $g''(z_c) > 0$ we have $\gm = 1/2$.

\item
From the figure it follows that 
\beq\label{s9j4}
\chi = \frac{1}{g} \plu \frac{ 2z}{g^2 }\plu \frac{ 2^2z^2}{g^3} + \cdots \equ \frac{1}{g \mi 2z} \equ\frac{1}{2\sqrt{\Del}} \equ
\frac{1}{g} \plu 2 \frac{1\plu j}{g^3} \plu  6 \frac{(1\plu j)^2}{g^5}\plu \cdots
\eeq

\item

The expansion in loops of the two first of the hierarchial equations for $Z$ and $\chi^{(k)}$ can be written
\beq\label{xj1}
Z_0 \!+\! \frac{Z_1 }{\Lambda}\! +\!  \frac{ Z_2 }{\Lambda^2}\!+\! \cdots\!\!=\!\!\frac{1}{g} 
\Big(\!1\!+\!g \!+\!\Big[ Z_0 \!+\! \frac{Z_1 }{\Lambda} \!+\!  \frac{ Z_2 }{\Lambda^2}\!+\! \cdots\Big]^2 
\!+\! \frac{1}{\Lambda} \Big[ \chi^{(2)}_0 \!+\! \frac{\chi^{(2)}_1 }{\Lambda}\! +\cdots\Big]\Big)
\eeq
\beq\label{xj2}
 \chi^{(2)}_0\!+\!\frac{ \chi^{(2)}_1}{\Lambda}\!+\!\cdots\!\! =\!\! \frac{1}{g} 
\Big(\!1\!+\! 2 \Big[Z_0 \!+\! \frac{Z_1 }{\Lambda}\! +\!\cdots\Big] \Big[  \chi^{(2)}_0\!+\!\frac{ \chi^{(2)}_1}{\Lambda}\!+\!\cdots\Big]
\!+\! \frac{1}{\Lambda} \Big[ \chi^{(3)}_0\!+\!\cdots\Big] \Big)
\eeq
From this we find
\beq\label{xj3}
Z_1  = \frac{1}{g} \big( 2 Z_0 Z_1 \plu \chi_0^{(2)}\big) \quad \Rightarrow \quad Z_1(g \mi 2Z_0)  = \chi_0^{(2)}\quad
\Rightarrow  Z_1 =  \Big(\chi_0^{(2)}\Big)^2
\eeq
and 
\beq\label{xj4}
\chi_1^{(2)} = \frac{1}{g} \Big( 2Z_0 \chi_1^{(2)}  \plu 2Z_1\chi_0^{(2)}\plu  \chi_0^{(3)}\Big),
\eeq
i.e.
\beq\label{xj5}
\chi_1^{(2)} (g \mi 2Z_0) = 2Z_1\chi_0^{(2)}\plu  \chi_0^{(3)} = 4 \Big( \chi_0^{(2)}\Big)^3\quad  \Rightarrow \quad
\chi_1^{(2)} = 4 \Big( \chi_0^{(2)}\Big)^4.
\eeq
From the $1/\Lambda^2$ term in \rf{xj1} we obtain 
\beq\label{xj6}
Z_2\! =\!  \frac{1}{g} \Big( 2Z_0 Z_2 \!+\! Z_1^2 \! +\! \chi^{(2)}_1\Big) \Rightarrow Z_2 (g \!- \!2Z_0)\! =\!  Z_1^2 \! +\! \chi^{(2)}_1 
\Rightarrow Z_2 \!=\! 5 \Big( \chi_0^{(2)}\Big)^5\!.
\eeq

\item
In any one-loop graph of the kind we discuss, the root is unique, the vertex where the loop starts is 
unique, the shortest path connecting the root to the vertex is unique and the shortest loop-line is unique.
The shortest path between the root and the marked  vertex is dressed with all kind of outgrowths 
and can be used to represent all   BPs where the shortest path between the marked vertices has a fixed length.  
When we then sum over the 
length of these paths  we obtain all BPs connecting the root and the vertex where the 
loop starts, i.e $\chi^{(2)}_0$. Similar arguments apply  to the loop. The vertex where the loop starts, seen from the root,
was unique and can be labelled a new root. We now open the loop by splitting this vertex in two. One 
part is the root, the other vertex will act as the marked vertex in a new  BP, before  forming the loop, 
but now a BP of the kind belonging to $\chi^{(2)}_0$. This makes sense since the vertex we split was of 
order 3 and had thus no $j$ attached. After the split it becomes a root and a vertex of order 1, i.e. precisely 
the two vertices of order 1 which have no $j$ attached in a BP belonging to $\chi^{(2)}_0$. 
The shortest path between these two vertices is exactly the shortest path mentioned before in the loop.
 Summing over such graphs we obtain again all BPs with the marked 
points separated a given distance and summing over the length we obtain again $\chi^{(2)}_0$. In total 
thus $\big(\chi^{(2)}_0\big)^2= 1/(4\Del)$.    
\beq\label{s9j5}
Z_1(g ,j)=\frac{1}{4\Del} = \frac{1}{g^2\mi 4(j\plu 1)}= \frac{1}{g^2} +  4  \frac{s9j\plu 1}{g^2}+  4^2  \frac{(1\plu j)^2}{g^4}\plu  \cdots
\eeq
These are thus 1 one-loop diagram with two lines, 4 one-loop diagrams with 4 lines and 16 one-loop diagrams 
with 6 lines, see figure.
      \begin{figure}[!ht]
\hspace{1cm}\centerline{\includegraphics[width=0.9\linewidth,angle =0]{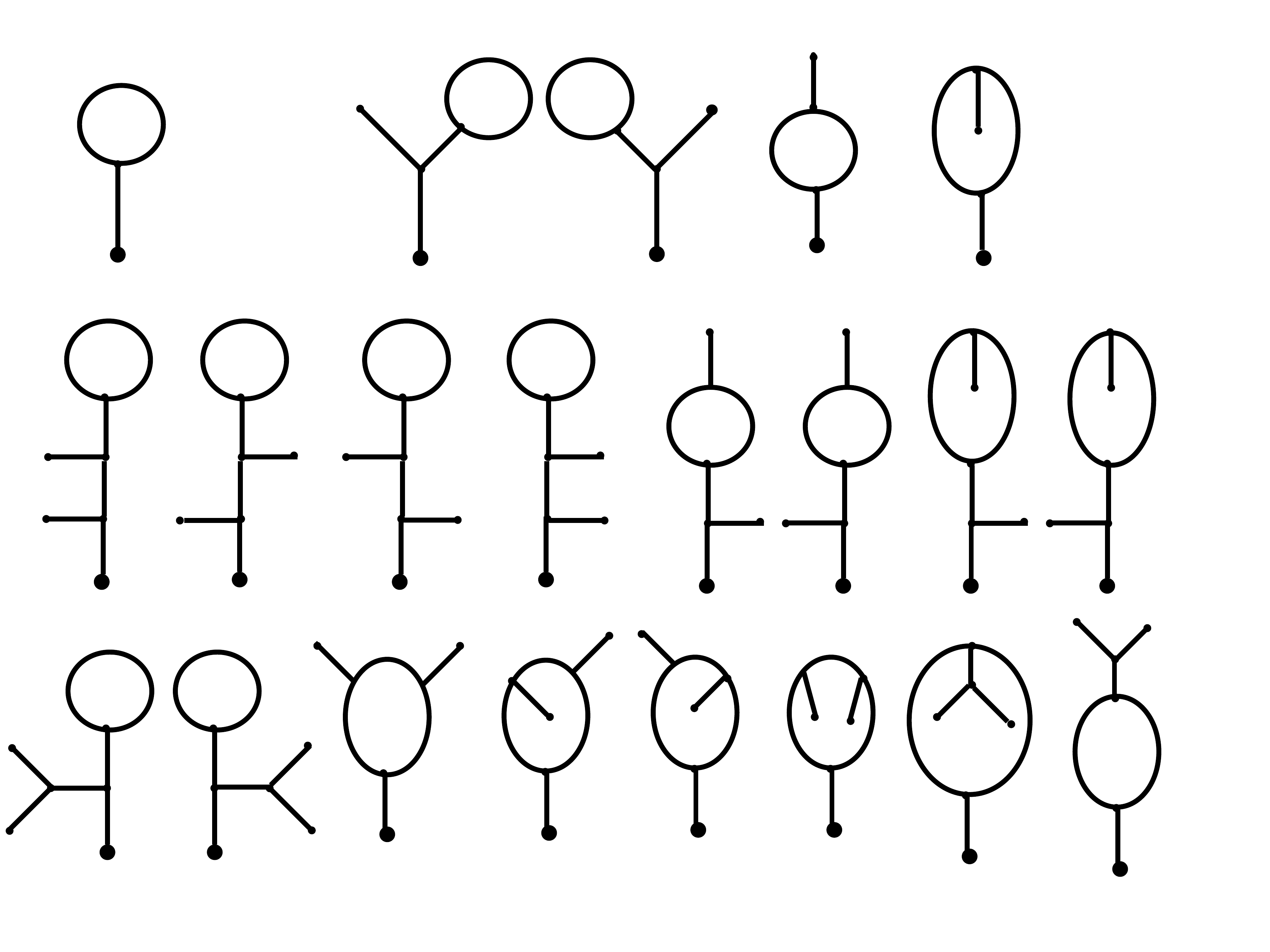}}
\vspace{-1cm}
\caption{{\small The rooted BPs with one loop and two, four and six links,}}
        \end{figure}

\item
First, the 5 ``skeleton'' graphs shown in Fig.\ 5 in the problem sheet are precisely the two-loop graphs 
generated by iterating the graphical Fig.\ 3 in the problem sheet to two loops, assuming that all vertices are of order 
3 except for the root which is of order 1. The graphical iteration is more or less identical to the algebraic iteration we performed
above, which gave us $Z_2$ (eq.\ \rf{xj6}). The middle graph  in Fig.\ 5 corresponds to the term $Z_1^2$, while the four other terms
come from the fact that the one-loop propagator can be decomposed in four components, which according to eq.\ 
\rf{xj4} can be written as $2Z_1\chi_0^{(2)}$ (leading to the two graphs to the left in  in Fig.\ 5) plus $\chi_0^{(3)}$ which
leads to the two graphs to the right in Fig.\ 5 (graphically $\chi_0^{(3)}$ is 2 times a $\phi^3$ vertex connected to three 
external points if we only allow graphs with internal vertices of order 3).

Secondly, whenever one draws a  $\phi^3$ graph where the only 
vertex of order 1 is the root, one can ``extend'' the lines (i.e.\ the links in the graph) to a full BP with two marked 
points, i.e.\ to $\chi^{(2)}_0$ (which we in the following just denote $\chi$).  
Also, given such a $\ell$-loop BP-graph, one can, starting from the root, in a unique 
way identify the vertices which constitute the vertices in a ``skeleton'' $\phi^3$ graph where the only 
vertex of order one is the root. Thus the total contribution is $\chi^L$, where $L$ is the number of 
links in the skeleton $\phi^3$-graph. Let now $G$ be a $\phi^3$-graph with $V_{ex}$ external vertices 
(i.e. vertices of order 1) and $V_I$ internal vertices (i.e.\ vertices of order 3) and $L$ links and $\ell$ loops. 
We then have\footnote{The first equation defines the number of loops in the way we meet them in a Feynman diagram: we 
have to integrate over the momentum for each line (propagators), but for each vertex we have momentum 
conservation, except for allover momentum conservation. The $\ell$ is then the independent momenta we 
have to integrate over, i.e.\ the number of loops associated with the Feynman integral.}
\beq\label{s9j6}
L-(V_I \plu V_{ex}) + 1 =\ell   \quad 3V_I \plu V_{ex} = 2L  \quad {\rm i.e.} \quad 3\ell \mi 1 = L,
\eeq
in the case of tadpoles where $V_{ex} \equ 1$.
Thus the total BP contribution coming from BPs with skeleton graph $G$ with $\ell$ loops is 
$\chi^{3\ell -1}$, and the BP partition function with $\ell$ loops will be the sum over all such skeleton
graphs, i.e. all  ``tadpole'' $\phi^3$-graphs with $\ell$ loop
\beq\label{s9j7}
Z_\ell(g,j) = C_\ell \;\chi^{3\ell -1},\quad C_\ell = \#{\rm ~ tadpole-}\phi^3~{\rm graphs}.
\eeq
It is clear that $Z_\ell = C_\ell \;\chi^{3\ell -1}$ is precisely what we algebraically proved above for $Z_1$ and $Z_2$, 
and it is not too difficult to extend this algebraic proof to all orders in $\ell$.
 
Note that we are not really specifying in a precise way what we mean by the {\it number} of tad-pole $\phi^3$ graphs.
It will not be important for us. The important point is that $Z_\ell \propto \chi^{3\ell -1}$ since this is what determines 
the singular behavior of $Z_\ell(g)$ when $g \to g_c$.

\item
We simply insert 
$$
Z(g,j,\Lambda) = \frac{g}{2} \mi  \sqrt{\Delta}\, F(t),\quad   \Lambda \Delta^{\frac{3}{2}} = \frac{3}{2} t,  \quad
 \frac{d}{dj} = -\frac{d}{d\Delta}= -  \sqrt{\Delta} \,\Lambda \,\frac{d}{dt}
 $$ 
 into the equation 
\beq\label{yj1}
g Z=   (1\plu j)  + Z^2 + \frac{1}{\Lambda}  \frac{d Z}{d j }
\eeq
and obtain
\beq\nonumber
g\Big( \frac{g}{2} - \sqrt{\Delta}\, F\Big) = 1\!+\! j + \Big( \frac{g^2}{4} - g \sqrt{\Delta} \, F + \Delta \, F^2 \Big) + 
\frac{1}{\Lambda}  \Big( \frac{F}{2 \sqrt{\Delta}} + \Lambda \Delta \frac{d F}{dt} \Big)
\eeq
or
\beq\nonumber
\Big(\frac{g^2}{4} - (1\plu j) \Big) = \Delta \, F^2 +\Delta \, \frac{F}{2 \Lambda \Delta^{3/2}} + \Delta\frac{d F}{dt},
\eeq
i.e.
\beq\label{yj3}
1 = F^2 + \frac{F}{3t} +  \frac{d F}{dt}.
\eeq

\item
Inserting the asymptotic expansion in the differential equation leads to a power expansion in $1/t$, where 
the coefficient multiplying  $1/t^{n+1}$ has to be zero. Expressed in terms of the coefficients $c_n$ 
of the asymptotic expansion of $F(t)$ we then obtain the equations:
\bea\label{s9j8}
1& =& c_0^2\\
0 &=&\sum_{k=0}^{n+1} c_k c_{n+1-k} + (\frac{1}{3}-n)c_n = 0, \quad n \geq 0.\label{s9j9}
\eea
The first few equations are (starting with $c_0 =1$)
\bea\nonumber
2 c_1 &=& -\frac{1}{3} \\
2 c_2&=&c_1 -\frac{1}{3} c_1-c_1^2 \nonumber \\
....&...& ......\nonumber \\
2 c_{n+1} &=& nc_n- \Big( \frac{1}{3}c_n + c_1c_n + c_2c_{n-1}\cdots + c_{n-1} c_2 +c_n c_1\Big)   \nonumber
\eea   
The two first equations give $c_1 = -1/6$ and $c_2 = -5/72$. 

\item 
If we can ignore the bracket in the equation for $c_{n+1}$,  a consistent
solution for large $n$ is clearly $c_{n} = - k \Gamma(n) /2^n$, where $k$ is a constant. It is a 
consistent solution for large $n$ up to power corrections $n^{-\alpha}$, since assuming it, one has  
\beq\nonumber
\frac{c_{n+1}}{c_n} = \frac{n}{2} \Big( 1 - \frac{1}{n}\Big[ \frac{1}{3} + \frac{c_1c_n + \cdots + c_nc_1}{c_n}\Big]\Big)
\eeq
and one can check (numerically) that $[\cdot] < 1/3  + 3k$. 

\item

We write the definition of $\gamma_\ell$ as 
$
Z_\ell \sim (g -g_c)^{-\gamma_\ell +1}
$
and since $\Del \sim g-g_c$ for $g \to g_c$ we have (question (5)):
$
Z_{\ell} \sim \Del^{-\frac{3}{2}\ell +\frac{1}{2}},
$
and we conclude $\gamma_\ell = \frac{3}{2}\ell + \frac{1}{2}$.

\item 

We know that the scaling limit of the BPs is universal, independent of the weights $w_3,w_4,\ldots$ as long as 
there are only a finite number of them and they are positive. Thus the susceptibility 
without loops  behaves as $\chi_0(g) \sim 1/\sqrt{g-g_c}$. If we have $w_4,w_5,\ldots$ different from zero 
we can form many more skeleton graphs, involving vertices of order 4,5 etc.. Note also that even if $w_3 =0$
we have not problem constructing  skeleton graphs with $\phi^3$ vertices. Given the number of loops $\ell$,
the question is: which skeleton graph is most singular. Each link in the skeleton graph is 
represented by a BP propagator $\chi_0(g) \sim 1/\sqrt{g-g_c}$, so we simply want the tadpole graphs 
with $\ell$ loops and the maximal number of links $L$. Let $V_n$, $n=3,4,\ldots$ denote the number
of vertices of order $n$. For a  tadpole skeleton graph $G$ with one ``external'' link and one ``external'' vertex, $L$ links
and $\ell$ loops  we have
$$
L = (V_3 +V_4 + \cdots) +\ell,  \qquad   1+3V_3 + 4V_4 + \cdots = 2L.
$$
Thus it is seen that $L$ becomes maximal if all vertices (except the root) are  order 3 vertices.
(if we only have vertices of order $n$:  $ L = (n  \ell +1)/(n-2)$.)

\end{enumerate}

\newpage

\setcounter{equation}{0}
\setcounter{figure}{0}
\renewcommand{\thefigure}{s-10.\arabic{figure}}
 \renewcommand{\theequation}{s-10.\arabic{equation}}

\subsection*{ \hspace{-4mm} Solutions to Problem Set 10}

\subsubsection*{A general even potential V(x)}

\begin{enumerate}

\item  

Contracting the contour as mentioned (using that the contour integral does not change),
and writing $1/ \sqrt{\omega^2\mi a^2} = 1/( \sqrt{\omega\mi a}\sqrt{\omega\plu a})$ we obtain
\bea\label{s10j1}
       \oint_C \frac{d \omega}{2\pi i} \frac{f(\omega)}{\sqrt{\omega^2\mi a^2}} &=&
       \int_{a}^{-a} \frac{dx}{2\pi i} \frac{-i}{\sqrt{a^2\mi x^2}}f(x) \plu  \int_{-a}^{a} \frac{dx}{2\pi i} \frac{i}{\sqrt{a^2\mi x^2}}f(x) 
       \nonumber \\
&=& \int_{-a}^a \frac{dx}{\pi} \frac{f(x)}{\sqrt{a^2\mi x^2}}\; = \int_{-1}^1 \frac{dy}{\pi} \frac{f(ay)}{\sqrt{1\mi y^2}} 
\eea

\item 

The formula for $W(z)$ in the notes can, for an even potential where $c_+ \equ a$ and $c_- \equ -a$,  be written  as 
\beq\label{s10j2}
 \oint_C \frac{d \omega}{2\pi i} \frac{f(\omega)}{\sqrt{\omega^2\mi a^2}} , \quad 
 f(\om) = \frac{V'(\om) \sqrt{z^2\mi a^2} }{2(z \mi \om)} = \frac{(z\plu \om)V'(\om)\sqrt{z^2\mi a^2}}{2(z^2 \mi \om^2)}.
 \eeq
 Thus, for $z$ outside the contour we can directly apply \rf{s10j1}. Since $V'(\om)$ is an odd function, the 
 integral with  $z V'(\om)$ is zero and only the contribution with $\om V'(\om)$ survives, leading to the 
 wanted formula, using  that  $\om V'(\om)$  is an even function.
 
 \item
 Expanding the expression for $W(z)$ in powers of $1/z$ we obtain
 \beq\label{s10j3}
 W(z) = \frac{1}{g}   \int_{0}^a \frac{dx}{\pi} \frac{x \tilde{V}'(x)}{\sqrt{a^2\mi x^2}} \; \frac{1}{z} \plu O\Big(\frac{1}{z^2}\Big)
\eeq
which leads to the determination of $g$ as a function of $a$.

\item

The $t_n$ term will lead to the following integral
\beq\label{s10j4}
  \int_0^a \frac{dx}{\pi} \frac{2n\; x^{2n}}{\sqrt{a^2\mi x^2}} = \frac{2n}{\pi} \int_0^{\frac{\pi}{2}} d\theta \sin^{2n}\theta = 
  \frac{1}{2} \cdot \frac{(2n\mi 1)!!}{(2n\mi 2)!!} 
\eeq
if we set $x =\sin \theta$ and use the hint. We now use 
\bea
&&(2n\mi 1)!! = 2^{n\mi 1} (n\mi \frac{1}{2}) (n\mi \frac{3}{2}) \cdots \frac{1}{2} = 
2^{n\mi 1} \frac{\Gamma(n\plu \frac{1}{2})}{\Gamma(\frac{1}{2})} \nonumber\\
&& (2n\mi 2)!! = 2^{n\mi 2} (n\mi 1) (n\mi 2) \cdots 1 = 2^{n\mi 2} \Gamma(n). \nonumber
\eea
This provides the formula.

\item
We have $B\big(1,\frac{1}{2}\big) \equ 2$ and $B\big(2,\frac{1}{2}\big) \equ 4/3$, and thus
\beq\label{s10j5}
g(a^2) = \frac{1}{4} a^2 \mi \frac{3}{16} a^4 
\eeq
The critical point is where $g'(a^2) \equ0$, i.e.\  $a_c^2 \equ \frac{2}{3}$ and thus $g_c  \equ g(a^2_c) \equ  \frac{1}{12}$.

\item

Assume $t_1 >0$, and the other $t_n \leq 0$ and that $t_n =0$ for $n> N$.  $g'(a^2) =0$ leads to 
\beq\label{s10j6}
 \frac{t_1}{B(1,\frac{1}{2})} + \sum_{n>1} \frac{n t_n}{B(n,\frac{1}{2})} \; (a^2)^{n-1} = 0
 \eeq
 which clearly has only one (positive) $a_c$ solution. Furthermore we have 
 \beq\label{s10j7}
 g''(a^2) =    \sum_{n>1} \frac{n(n\mi 1) t_n}{B(n,\frac{1}{2})} \; (a^2)^{n-2} < 0
 \eeq
 so expanding around $a_c$ we have 
 \beq\label{s10j8}
 g(a^2) = g(a_c^2) + \frac{1}{2} g''(a^2_c) \, (a_c^2\mi a^2)^2 + O((a_c^2\mi a^2)^3),
 \eeq
 
 \item
 
 We know by now that if we have an even polynomial $V(x)$ of order $2m$, then $g(a^2)$ will also 
 be an even polynomial of order $2m$ and we know the relations between $t_n$ and 
 the coefficients $g_n$ in the polynomial 
 \beq\label{s10j9}
 g(a^2) = \sum_{n=1}^m g_n a^{2n}.
 \eeq
 From the assumptions it is clear, by Taylor expanding around $a_c^2$ that $g(a^2)$ can be written 
 as stated. We only need to determine $g(a_c^2)$ and $c$. For this we use the expansion around $a^2 =0$:
 \beq\label{s10j10}
 g(0) \equ 0,\quad g(a^2) \equ g_1 a^2 \plu O(a^4) \equ \frac{t_1}{ B(1,\frac{1}{2})} a^2 \plu  O(a^4) \equ  \frac{t_1}{ 2} a^2 \plu  O(a^4)
 \eeq
 The result now follows ($t_1 =1/2$)
 \beq\label{s10j11}
 g(a_c^2) = c\; a_c^{2m},\qquad   c\; m \, a_c^{2m-2} = \frac{1}{4}.
 \eeq
  
 \item
 
 For $a_c^2 =1$ we have for the coefficient $g_n$ in \rf{s10j9}, expanding $-c (1\mi a^2)^{m}$
 
 \beq\label{s10j12}
 g_n =   \frac{t_n}{ B(n,\frac{1}{2})} = (-1)^{n-1} c \,
 \begin{pmatrix} m \\ n \end{pmatrix}, \quad c = \frac{1}{4m}.
 \eeq
 
 \item 
 
 The formula $\tilde{M}_0(a^2) \equ 2 g(a^2)$ follows directly from the definitions of $\tilde M_0$ and $g$,
 and it is a replacement of $M_0(a^2) \equ 2$ discussed in the notes.
 
 \item 
 
 We know from the definitions that 
 \beq\label{s10j13}
 \tilde M_k (a^2)  \propto \frac{d^k}{d (a^2)^k} \tilde M_0(a^2) \propto \frac{d^k}{d (a^2)^k}  g(a^2)
  \propto \frac{d^k}{d (a^2)^k}  (1\mi a^2)^m 
 \eeq
 which gives the desired result.
 
 \item 
 In problem 5 we saw that the coefficient $c_n$ to $a^{2n}$ in the power expansion of $(1-a^2)^s$ is: 
 $$
 c_n =\frac{\Gamma(n-s)}{\Gamma(-s) \Gamma(n+1)} \propto \frac{1}{n^{s+1}} \quad {\rm for} \quad n \to \infty.
 $$
 $$
 t_n = - \frac{c_n}{4s} \; B(n,1/2)   \propto \frac{1}{n^{s+3/2}} \quad {\rm for} \quad n \to \infty.
 $$

 \item 
 Follows from the definition of $\tilde M_k$ by differentiation wrt $a^2$.
 
 \item
 
 This relation is just as in the notes, the only difference being a factor $1/g$.
 
 \item 
 
 Using the hint, and the relations proven earlier we want to prove that 
 \beq\label{s10j14}
  \left( \tilde M(z) \mi  2(z^2\mi  a^2) \frac{d \tilde M(z)}{da^2} \right) = \tilde M_1.
  \eeq
  We have (using $M_k \equ 0$ for $k >m$)
\bea
 2(z^2-a^2) \frac{d \tilde M(z)}{da^2} & =& -2(z^2\mi a^2) \sum_{k=2}^m (k\mi 1) (z^2\mi a^2)^{k-2}\tilde M_k + \nonumber\\
&& +2(z^2\mi a^2) \sum_{k=1}^{m-1} (k\plu \frac{1}{2} )    (z^2\mi a^2)^{k-1}\tilde M_{k+1}
\nonumber\\
&=& (z^2\mi a^2) \sum_{k=2}^m  (z^2\mi a^2)^{k-2}\tilde M_k = \tilde M (z) \mi \tilde M_1, \nonumber
\eea
i.e.\ the wanted formula.

\end{enumerate}

\newpage

\setcounter{equation}{0}
\setcounter{figure}{0}
\renewcommand{\thefigure}{s-11.\arabic{figure}}
 \renewcommand{\theequation}{s-11.\arabic{equation}}

\subsection*{ \hspace{-4mm} Solutions to Problem Set 11}

\subsubsection*{Multiple Ising spins coupled to 2d quantum gravity}

\begin{enumerate}

\item  
We can sum over the spin configurations in the following way: let $T$ be a triangulation where 
all spins are aligned to the spins on the two boundary triangles who per definition have the same spin (e.g.\ +).
Now take an arbitrary interior link (there are $N_{L(I)}\equ N_L(T)\mi 2$ of these). We either leave the link untouched,
it gives a factor 1, or we can open this link into two, connected to the same vertices and glue 
a new universe with $-$ spin at its boundary to close the surface. In this way we effectively add
a factor  $\e^{-2\bt} G(\mu,\bt)$ to the link. In total we then associate a factor $(1\!+\! \e^{-2\bt} G(\mu,\bt))$ 
with each interior link. In this way we actually perform the sum over allowed spin configurations and 
it leads to the self-consistent equation for $G(\mu,\bt)$. 

Let $T$ be a triangulation with $N_{L(ex)}$ boundary links, $N_{L(I)}$ intrinsic links and $N_T$ triangles.
Then we have 
$$
2    N_{L(I)} \plu  N_{L(ex)} = 3 N_T
$$
which results in the last equation since $ N_{L(ex)} \equ 2$.

\item

The equation should be clear: summing over ${\cal T}^{(2)}(2)$ with exponential weight is by definition $G_0$.

\item 

Just a rearrangement using $G(\mu,\bt)\equ  G_0(\bmu)$.

\item 

Differentiate eq.\ (12) from the problem sheet wrt $\bmu$, using the definition
$$
\chi_0(\bmu) = -\frac{\d G_0(\bmu)}{\d \bmu} .
$$

\item

Differentiate $G(\mu,\bt)\equ  G_0(\bmu)$ using the chain-rule on the rhs and the definition of $\chi$.

\item 

we have 
$$ 
G_0(\bmu) = \sum_T \e^{- \bmu N_T } ,\quad  \chi_0(\bmu)= 
-\frac{\d G_0(\bmu)}{\d \bmu} =\sum_T N_T \;\e^{- \bmu N_T } 
$$
Thus 
$$
\frac{3}{2} \chi_0(\bmu) \mi G_0(\bmu) = \sum_T \Big( \frac{3}{2} N_T\mi 1\Big) \;\e^{- \bmu N_T } 
$$
which is clearly a decreasing function of $\bmu$ (each term is..)

\item

The obvious choice of $\bt_0$ is the value where 
\beq\label{s11-x1}
\e^{2\bt_0} = \frac{3}{2} \chi_0 (\mu_0) \mi  G_0(\mu_0), \quad {\rm i.e.} \quad 
\bt_0 = \frac{1}{2} \log \Big(  \frac{3}{2} \chi_0 (\mu_0) \mi  G_0(\mu_0)\Big).
\eeq
since this is, from above arguments, the largest value the rhs can assume.

Thus we know for sure that if $\bt > \bt_0$ then 
$$
e^{2\bt} \mi \Big( \frac{3}{2} \chi_0 (\bmu) \mi  G_0(\bmu)\Big) > 0 \quad {\rm for}  \quad \bmu \geq \mu_0,
$$
and from $G(\mu,\bt) \equ G_0(\bmu)$ we know that for $\mu > \mu_0(\bt)$ also $\bmu > \mu_0$
(else both sides of the equation could not exist).

\item

The above considerations show that all the way down to $\bmu = \mu_0$ there is a simple 
linear relationship between $\bmu$ and $\mu$ for small changes. The derivatives $\prt \bmu /\prt \mu$ 
and $\prt \mu/ \prt \bmu$ are finite as long as $\bt > \bt_0$ all the way down to and including 
$ \bmu= \mu_0$. Thus the only source of non-analyticity 
in the relation  (16) in the problem sheet can come from $\chi_0(\bmu)$  when $\bmu \to \mu_0$.
Since  $\prt \mu/ \prt \bmu$ is finite at that point and a non-singular function of 
$\chi_0(\bmu)$ and $G_0(\bmu)$,  the non-analyticity  of  $\chi(\mu,\bt)$ must be the same as that of 
$\chi_0(\bmu)$, and thus $\gamma(\bt) \equ \gamma_0$ as long as $\bt > \bt_0$. 

\item
The first of the relations 
$$
\bmu_0(\bt_0 )= \mu_0,\qquad \quad \bmu_0(\bt) > \mu_0\quad {\rm for } \quad \bt < \bt_0.
$$
follows from the definition of $\bt_0$ given above. The other relation is also 
a consequence of that definition: when $\bt < \bt_0$ 
$$
e^{2\bt} =\frac{3}{2} \chi_0 (\bmu) - G_0(\bmu) 
$$
{\it has} a solution $\bmu > \mu_0$ simply because the rhs is an increasing function when $\bmu$ decreases
towards $\mu_0$ and $\bt < \bt_0$

\item 

Eq.\ (20) in the problem sheet clearly implies 
\beq\label{s11-x2}
\mu \mi  \mu_0(\bt) = c\, (\bmu \mi \bmu_0(\bt))^2 + O\big((\bmu \mi \bmu_0(\bt))^3\big).
\eeq
unless, for some reason, the second derivative $\displaystyle{\frac{\prt^2 \mu(\bmu, \bt)}{\prt \bmu^2}=0}$, which we will
assume is not the case. 

\item

Since $\bt < \bt_0$ we have that $\bmu_0 (\bt) > \mu_0$. Thus $\chi_0(\bmu)$ and $G_0(\bmu)$ are
analytic around that point. The source of singularity in $\chi(\mu,\bt)$  in the expression (16) in the 
problem sheet:
$$
\chi(\mu,\bt) = \chi_0 (\bmu(\mu,\bt)) \; \frac{\prt \bmu(\mu.\bt)}{\prt \mu}
$$
can thus not come from $\chi_0(\bmu)$ or $G_0(\bmu)$ and has to  comes from the denominator 
in $\frac{\prt \bmu(\mu.\bt)}{\prt \mu}$, which goes to zero for $\bmu  \to \bmu_0(\bt)$.
However, since $\bmu_0(\bt) > \mu_0$ we can Taylor expand the denominator around   $\bmu_0(\bt)$
and we obtain, using \rf{s11-x2}, the desired result
$$ 
\chi(\mu,\bt) \sim \frac{1}{ \bmu\mi \bmu_0(\bt)} \sim \frac{1}{ \sqrt{\mu\mi \mu_0(\bt)}} \qquad \bt < \bt_0.
$$

\item
we have for $\bmu \to \mu_0$ by definition (see expansion (7) in problem sheet)
\beq\label{s11-x3}
\frac{3}{2} \chi_0(\bmu) \mi  G_0(\bmu) = \frac{3}{2} \chi_0(\mu_0) - G_0(\mu_0) - c\, (\bmu-\mu_0)^{-\gamma_0} +
O(\bmu \mi \mu_0),
 \eeq
 and thus 
 \beq\label{s11-x4}
\e^{2\bt_0} - \Big(\frac{3}{2} \chi_0(\bmu) \mi  G_0(\bmu)\Big)  = c\, (\bmu\mi \mu_0)^{-\gamma_0} + O(\bmu \mi \mu_0).
 \eeq
 This implies 
 \beq\label{s11-21a}
\frac{\prt \mu}{\prt \bmu} \sim (\bmu \mi  \mu_0)^{-\gm_0} + O(\bmu\mi \mu_0), \qquad
\frac{\prt \bmu}{\prt \mu} \sim \frac{1}{(\bmu \mi  \mu_0)^{-\gm_0}}
\eeq

\item 
and by integration of \rf{s11-21a}
\beq\label{s11-22a}
\mu \mi  \mu_0(\bt_0) = c\, (\bmu \mi  \mu_0)^{1-\gm_0}+ O\big((\bmu\mi \mu_0)^2\big)
\eeq

\item 
Finally from 
$$
\chi(\mu,\bt) = \chi_0 (\bmu(\mu,\bt)) \; \frac{\prt \bmu(\mu,\bt)}{\prt \mu}
$$
we obtain, using \rf{s11-21a}
$$
\chi(\mu,\bt_0) \sim \frac{1}{ (\bmu\mi \mu_0)^{-\gm_0}}
$$
and using \rf{s11-22a}
$$
\chi(\mu,\bt_0) \sim \frac{1}{ (\mu\mi \mu_0(\bt_0))^{-\gm_0/(1-\gm_0)}}.
$$
This is the desired result.

\end{enumerate}

\newpage

\setcounter{equation}{0}
\setcounter{figure}{0}
\renewcommand{\thefigure}{s-12.\arabic{figure}}
 \renewcommand{\theequation}{s-12.\arabic{equation}}

\subsection*{ \hspace{-4mm} Solutions to Problem Set 12}

The purpose of this problem set is to derive the the multiloop formulas \rf{5.62}, \rf{5.65} and \rf{5.66} using \rf{5.61}. We will simply 
use the representation \rf{5.52} for the loop insertion operator and act  on the disk function $w(\vg,z)$ written in the form 
\rf{5.49}, using the results \rf{5.51a}-\rf{5.51d}. Let us  for convenience write the two-loop function  \rf{5.62} in the following way
\beq\label{s12-0}
w(\vg,\om,z) = \frac{1}{(z^2\mi \om^2)^2} \left( -2 z\om + 
\frac{2z^2\om^2 \mi c^2(z^2\plu \om^2)}{(z^2\mi c^2)^{1/2} (\om^2\mi c^2)^{1/2}}\right)
\eeq

\begin{itemize}

\item[(1)] {\it Show that }
\beq\label{s12-1}
\frac{2}{\tM_1} \frac{d}{dc^2} \sum_{k=1}^\infty \tM_k (\om^2\mi c^2)^{k-1/2} = - \frac{1}{(\om^2\mi c^2)^{1/2}}.
\eeq
We have
\bea\label{s12-1a}
 \lefteqn{\frac{d}{dc^2} \sum_{k=1}^\infty \tM_k (\om^2\mi c^2)^{k-\oh} \equ 
  \sum_{k=1}^\infty \frac{d \tM_k}{dc^2} (\om^2\mi c^2)^{k-\oh} \plu \tM_k \frac{d  (\om^2\mi c^2)^{k-\oh}}{dc^2}}~~~~~~ \no \\
&=&  \sum_{k=1}^\infty\Big( (k \plu \oh) \tM_{k+1}  (\om^2\mi c^2)^{k-\oh}  \mi  (k\mi \oh) \tM_k  (\om^2\mi c^2)^{k-\frac{3}{2}} \Big)
\no \\
 &=& -\oh \frac{\tM_1}{(\om^2-c^2)^{1/2}}
 \eea
 leading to \rf{s12-1} 

\item[(2)] {\it Show that }
\beq\label{s12-2}
\frac{\prt}{\prt V(z) } \sum_{k=1}^\infty \tM_k (\om^2\mi c^2)^{k-1/2} = 
\frac{d}{dz} \Big[ \Big( \frac{\om^2\mi c^2}{z^2\mi c^2} \Big)^{1/2} \frac{z}{z^2-\om^2}\Big].
\eeq
We have 
\bea\label{s12-2a}
\lefteqn{\sum_{k=1}^\infty \frac{\prt \tM_k}{\prt V(z) }  (\om^2\mi c^2)^{k-\oh } =  \frac{d}{dz} \sum_{k=1}^\infty 
\frac{z (\om^2-c^2)^{k- \oh}}{(z^2-c^2)^{k + \oh}} }\no \\
&=&   \frac{d}{dz} \left(\frac{z (\om^2\mi c^2)^{\oh}}{(z^2\mi c^2)^{\frac{3}{2}}} 
\sum_{l=0}^\infty \Big( \frac{\om^2\mi c^2}{z^2\mi c^2}\Big)^l \right)\no \\
&=&  \frac{d}{dz} \left(\frac{z (\om^2\mi c^2)^{\oh}}{(z^2\mi c^2)^{\frac{3}{2}}} \frac{z^2\mi c^2}{z^2\mi \om^2}\right) =
 \frac{d}{dz}\left(\frac{z (\om^2\mi c^2)^{\oh}}{(z^2\mi c^2)^{\frac{1}{2}}} \frac{1}{z^2\mi \om^2}\right)\no 
 \eea

\item[(3)] {\it Use now \rf{5.51a} to write} 
\beq\label{s12-3}
\frac{d w(\vg,\om) }{d V(z)} = \frac{-2\om z}{(z^2\mi \om^2)^2}  + \frac{c^2}{(z^2\mi c^2)^{3/2} (\om^2\mi c^2)^{1/2}} -   
\frac{d}{dz} \Big[ \Big( \frac{\om^2-c^2}{z^2-c^2} \Big)^{1/2} 
\frac{z}{z^2\mi \om^2}\Big].
\eeq
This is simple consequence of eq.\ \rf{5.51a} and the form of the loop insertion operator given by \rf{5.49}. 

{\it and show that the last two terms, after differentiation, can be reorganized in the following way:}
\beq\label{s12-4}
\frac{1}{(z^2\mi \om^2)^2 }\left(  \frac{(z^2\mi \om^2) \om^2}{(z^2\mi c^2)^{1/2} (\om^2\mi c^2)^{1/2}} + 
\frac{(\om^2\mi c^2) (z^2\plu \om^2)}{(z^2\mi c^2)^{1/2} (\om^2\mi c^2)^{1/2}}   \right) 
\eeq
Here we just have to  perform the differentiation wrt $z$ which leads to the term 
\beq\label{s12-4a}
\frac{(\om^2\mi c^2)^\oh}{(z^2\mi c^2)^{\frac{3}{2}}}  \frac{z^2}{z^2-\om^2}  - \Big(\frac{\om^2\mi c^2}{z^2\mi c^2} \Big)^{\oh}\Big( \frac{1}{z^2-\om^2} - \frac{2z^2}{(z^2-\om^2)^2}\Big) 
  \eeq
  and combining the first term in this expression with the second term in \rf{s12-3} we obtain
  \beq\label{s124b}
  \frac{1}{(z^2\mi c^2)^{\frac{3}{2}} (\om^2\mi c^2)^{\oh}}\left(c^2 \plu  \frac{z^2 (\om^2\mi c^2)}{z^2\mi \om^2}\right) =
  \frac{1}{(z^2\mi c^2)^{\oh} (\om^2\mi c^2)^{\oh}}\; \frac{\om^2}{z^2\mi \om^2}
  \eeq
  This provides us with the first term in \rf{s12-4}. The second term in \rf{s12-4a} is just the second term in \rf{s12-4}.

\item[(4)]
{\it Use the above to prove formula \rf{s12-0}.}

It is just trivial algebra in the numerator of \rf{s12-4} and the use of \rf{5.51a}.

\end{itemize}

We now turn to the proof of the three-loop formula \rf{5.65}. Since the two-loop function only depends on the 
coupling constants $\vg$ via the position of the cut, $c(\vg)$, the loop insertion operator becomes very simple 
in the form \rf{5.49}
when acting on the two-loop function.
\begin{itemize}
\item[(5)] {\it Prove that} 
\beq\label{s12-5}
\frac{d}{dc^2} \left( \frac{2 z^2\om^2 - c^2(z^2\plu \om^2)}{(z^2\mi c^2)^{1/2} (\om^2\mi c^2)^{1/2}}\right)
= \oh \frac{c^2 (z^2\mi \om^2)^2}{(z^2\mi c^2)^{3/2} (\om^2\mi c^2)^{3/2}}
\eeq
Just differentiate and use some simple algebra.

\item[(6)] {\it Use this to prove formula \rf{5.65} for the three-loop function}

We have 
\bea\label{s12-5a}
\lefteqn{w(u,z,\om) = \frac{d}{dV(u)}\, w(z,\om) = \frac{2}{\tM_1 (u^2 \mi c^2)^{\frac{3}{2}}}  \frac{d}{d c^2}  \, w(z,\om)}\\ 
&=& 
\frac{c^2}{\tM_1 (u^2 -c^2)^{\frac{3}{2}}} \frac{1}{(z^2\mi \om^2)^2} 
\frac{d}{dc^2} \left( \frac{2 z^2\om^2 - c^2(z^2\plu \om^2)}{(z^2\mi c^2)^{1/2} (\om^2\mi c^2)^{1/2}}\right)
\no
\eea
and the result now follows from \rf{s12-5}:
\beq\label{s12-5b}
w(u,z,\om) = \frac{1}{2c^2\tM_1} \, \frac{c^2}{(u^2\mi c^2)^{\frac{3}{2}}} \frac{c^2}{(z^2\mi c^2)^{\frac{3}{2}}} 
\frac{c^2}{(\om^2\mi c^2)^{\frac{3}{2}}}
\eeq

\end{itemize}

Let us next prove the 4-loop formula. What we have to show is that
\beq\label{s12-11}
\frac{d}{d V(z)} \frac{f(c)}{\tM_1} = \frac{2}{\tM_1} \frac{d}{dc^2} \frac{ f(c)}{\tM_1 (z^2 \mi c^2)^{3/2}}.
\eeq

\begin{itemize}
\item[(7)] {\it Show that}
\beq\label{s12-11a} 
\frac{\prt \tM_1}{\prt V(z)}  = -\frac{d}{d c^2} \frac{2c^2}{(z^2 \mi c^2)^{3/2}}
\eeq
{\it and use this to show \rf{s12-11}.}

From \rf{5.51b} we have 
\beq\label{s12-11b}
\frac{\prt \tM_1}{\prt V(z)} = \frac{d}{dz} \frac{z}{(z^2\mi c^2)^{3/2}} = -\frac{2z^2\plu c^2}{ (z^2\mi c^2)^{{5}/{2}}} = 
-\frac{d}{d c^2} \frac{2c^2}{(z^2 \mi c^2)^{3/2}} 
\eeq
Next we have 
\bea\label{s12-11d}
\lefteqn{\Big(\frac{\prt}{\prt V(z)} \plu \frac{2 c^2}{\tM_1 (z^2\mi c^2)^{3/2}} \frac{d}{dc^2} \Big) \frac{f(c)}{\tM_1} =} \\
&&- \frac{1}{\tM_1^2}\left( \frac{\prt \tM_1}{\prt V(z)} \plu \Big(  \frac{d}{dc^2} \frac{2c^2}{ (z^2\mi c^2)^{3/2}}\Big)\right) f(c) 
+  \frac{2}{\tM_1} \frac{d}{dc^2} \frac{ f(c)}{\tM_1 (z^2 \mi c^2)^{3/2}} \no 
\eea
and eq.\ \rf{s12-11a} then leads to  formula \rf{s12-11}.

\end{itemize}

Finally, let us turn to the $n$-loop formula, which we have just proven for $n=3,4$. Assume it is correct up to $n\mi 1 \geq 3$.

\begin{itemize}

\item[(7)] {\it prove the following}
\beq\label{s12-6}
\Big[ \frac{d}{d V(z)} , \frac{2}{\tM_1} \frac{d}{dc^2}\Big] =0
\eeq
Using  \rf{5.52} for the loop insertion operator, the commutator can be written
\beq\label{s12-6a}
\Big[ \frac{\prt}{\prt V(z)}\,,\, \frac{2}{\tM_1} \frac{d}{dc^2}\Big] + 
\Big[\frac{c^2}{(z^2\mi c^2)^{\frac{3}{2}}}\frac{2}{\tM_1} \frac{d}{dc^2} \,,\,\frac{2}{\tM_1}  \frac{d}{dc^2}\Big]
\eeq
The first commutator is 
\beq\label{s12-6b}
\left(  \frac{\prt}{\prt V(z)} \frac{2}{\tM_1} \right) \frac{d}{dc^2} = - 
\frac{2}{\tM_1^2} \left(\frac{d}{dz} \frac{z}{(z^2\mi c^2)^{\frac{3}{2}}}\right) \frac{d}{dc^2}
\eeq
The second commutator is (using $[AB,C] \equ A[B,C]\plu [A,C]B$):
\beq\label{s12-6c}
\Big[\frac{c^2}{(z^2\mi c^2)^{\frac{3}{2}}}\,,\, \frac{2}{\tM_1}  \frac{d}{dc^2} \Big]  \frac{2}{\tM_1}  \frac{d}{dc^2} 
=
-\frac{4}{\tM_1^2} \left( \frac{d}{dc^2} \frac{c^2}{(z^2-c^2)^{\frac{3}{2}}} \right)  \frac{d}{dc^2} 
\eeq
Thus the sum of the commutators in \rf{s12-6a} is zero when using  \rf{s12-11a} and we have proven \rf{s12-6}.

\item[(8)] {\it Use this to prove the multiloop formula \rf{5.66}}

Formulas \rf{s12-6} and \rf{s12-11} show that we have
\bea\label{s12-7}
\lefteqn{\frac{d}{d V (z_n)} \Big( \frac{2}{\tM_1} \frac{d}{dc^2}\Big)^{n-4}\; \left(\frac{1}{2c^2 \tM_1} \prod_{k=1}^{n-1} 
\frac{c^2}{(z^2_k\mi c^2)^{3/2}}\right)} \\
&=&  \left( \frac{2}{\tM_1} \frac{d}{dc^2}\right)^{n-4}\!\! \frac{d}{d V (z_n)}\; \left( \frac{1}{2c^2 \tM_1} \prod_{k=1}^{n-1} 
\frac{c^2}{(z^2_k\mi c^2)^{3/2}} \right)\no\\
&=& \left( \frac{2}{\tM_1} \frac{d}{dc^2}\right)^{n-3} \, \left(\frac{1}{2c^2 \tM_1} \prod_{k=1}^{n} 
\frac{c^2}{(z^2_k\mi c^2)^{3/2}} \right) \no 
\eea
i.e.\ the multiloop formula.

\end{itemize}

\newpage

\setcounter{equation}{0}
\setcounter{figure}{0}
\renewcommand{\thefigure}{s-13.\arabic{figure}}
 \renewcommand{\theequation}{s-13.\arabic{equation}}

\subsection*{ \hspace{-4mm} Solutions to Problem Set 13}

\subsubsection*{The characteristic function and the two point function}

\begin{itemize}

\item[(1)] 
\beq\label{s13-1}
\frac{d Y}{dt} = -\hW(Y) \implies {dt} = -\frac{dY}{\hW(Y)} \implies T_2 \mi T_1 = - \int^{Y(T_2)}_{Y(T_1)} \frac{dy}{\hW(y)}.
\eeq 

\item[(2)]
We obtain
\bea\label{s13-4}
\lefteqn{\frac{d\sinh^{-1} \sqrt{F(x)}}{dx} =  \frac{1}{\sqrt{ 1 \plu F(x)}}\; \frac{F'(x)}{2\sqrt{F(x)}}, \qquad 
\Big(F'(x) \equ \frac{-\sqrt{\frac{g_s}{2\al}}\sinh^2\beta }{(x \mi \al)^2}\Big)} \no \\
&=& \frac{-\sqrt{\frac{g_s}{2\al}}\sinh^2\beta /(2(x\mi \al)) }{\sqrt{\cosh^2\frac{\beta}{2} (x\mi \al) + \sinh^2\beta \sqrt{\frac{g_s}{2\al}}}
\ \sqrt{\sinh^2\frac{\beta}{2} (x\mi \al) + \sinh^2\beta \sqrt{\frac{g_s}{2\al}}}} \no \\
&=& 
 \frac{-\sqrt{\frac{g_s}{2\al}}\sinh\beta }{(x\mi \al) \sqrt{(x\mi \al)^2 + 4 \cosh\beta \sqrt{\frac{g_s}{2\al}}(x\mi \al) + 
 4\sinh^2\beta \,{\frac{g_s}{2\al}}}}
 \no\\
 &=& \frac{ -\Sigma}{\sqrt{(x-\al)^2 + 4 \al (x-\al) + 4 \Sigma^2}} =  \frac{ -\Sigma}{\sqrt{(x\plu \al)^2 \mi {\frac{2g_s}{\al}} }} 
\eea
 
\item[(3)] 
 Eq.\ \rf{p13-4} in the Problem Set implies that 
\beq\label{s13-5}
F(\bX(T)) = \sinh^2\big( \Sigma T + \sinh^{-1}\sqrt{ F(X)}\big) 
\eeq
and thus
\bea\label{s13-6}
\bX(T) \mi \al \!\!\!&=&\!\!\! \frac{\sqrt{{g_s}/{2\al}}\;\sinh^2 \beta}{\sinh^2\big( \Sigma T + \sinh^{-1}\sqrt{ F(X)}\big) - \sinh^2 (\beta/2)} 
\hspace{3.5cm}\\
&=&\!\!\! \frac{\Sigma^2}{\sqrt{{g_s}/{2\al}}}\;  \frac{1}{\sinh^2\big( \Sigma T + \sinh^{-1} \sqrt{F(X)}\big) - \sinh^2 (\beta/2)}.\no
\eea

\item[(4)] For $T$ large we have from \rf{s13-6}
\bea\label{s13-7}
\lefteqn{\bX(T) \mi \al \to  \frac{4\Sigma^2}{\sqrt{{g_s}/{2\al}}} \; \e^{-2\Sigma T} \e^{-2 \sinh^{-1} \sqrt{F(X)}} } \non
&=& \frac{4\Sigma^2}{\sqrt{{g_s}/{2\al}}} 
\; \e^{-2\Sigma T} \Big( \cosh  \big( \sinh^{-1} \sqrt{F(X)}\big) \mi \sinh  \big( \sinh^{-1} \sqrt{F(X)}\big) \Big)^2 \non
&=&  \frac{4\Sigma^2}{\sqrt{{g_s}/{2\al}}} \; \e^{-2\Sigma T} \Big(\sqrt{ 1\plu F(X)} \mi \sqrt{F(X)}\Big)^2
\eea

\item[(5)] Let us now take $X \to \infty$. Then 
\beq\label{s13-8a}
\sqrt{F(X)} \to \sinh (\beta/2)\quad {\rm and} \quad  \sinh^{-1} \sqrt{F(X)} \to \beta/2.
\eeq 
\beq\label{s13-8} 
\sinh^2(\Sigma T \plu \frac{\beta}{2}) \mi \sinh^2 \frac{\beta}{2} \equ \frac{\cosh ( 2 \Sigma T \plu \beta) \mi \cosh \beta}{2} \equ
\sinh (\Sigma T \plu \beta) \sinh (\Sigma T)
\eeq 
and we obtain the desired formula by using 
\beq\label{s13-8b}
\sinh(\Sigma T \plu \beta) \equ \sinh \beta \cosh \Sigma T 
\plu \cosh \beta \sinh \Sigma T,
\eeq
 as well as $\sqrt{ \frac{g_s}{2\al}} \sinh \beta = \Sigma$ and $\sqrt{ \frac{g_s}{2\al}} \cosh \beta = \al$.
 
 \item[(6)] For $g_s \to 0$ we have  $\al \equ \Sigma \equ \SL$ and eq. \rf{p13-7} in the Problem Set then reads:
 \beq\label{s13-9}
 \bX(T) = \SL + \frac{2\SL}{ \e^{2\SL T} -1} = \SL \, \coth \SL T,
 \eeq
 which is also \rf{cdt23} for $X \to \infty$.
 
 \item[(7)] The first equation in question (7) in the Problem Set 
 is a trivial consequence of the definitions and that $\hW(X(T)) \equ -\frac{d \bX(T)}{dT}$.
 The second line follows by differentiating $\bX(T)$ (do not do it by hand). Finally $4 (\al^2\mi \Sigma^2) \equ 2g_s/\al$.
 
\end{itemize}

\subsubsection*{The average shape of CDT and GCDT universes}

\begin{itemize}

\item[(8)] By assumption  $Y > -\al$ and for $T\equ \infty$ we have $\bX(T;X) \equ \al$. This implies that
 $\hW(\bX(T)) = \hW(\al) \equ 0$.  Thus eq.\ \rf{p13-12} in the Problem Set becomes
 \beq\label{s13-10}
 \frac{\hW'(\bX(t)) \mi \hW'(\al)}{(\bX(t) \mi \al)/\tilde{W}(\bX(t))} \to -\frac{2 \tilde{W}' (\al)}{\tilde{W} (\al)} \quad {\rm for}  \quad
 t \quad {\rm large}.
 \eeq
 We have here used 
 \beq\label{s13-10a}
 \hW'(x) = \frac{1}{\tilde{W}(x)} - \frac{(x-\al) \tilde{W}'(x)}{\tilde{W}^2(x)}, \quad \hW'(\al) = \frac{1}{\tilde{W}(\al)} -
 \eeq
 \beq\label{s13-10b}
 \hW''(x) = -\frac{2\tilde{W}'(x)}{\tilde{W}^2(x)} - (x \mi \al) \left[ \frac{\tilde{W}''(x) \tilde{W}(x)- 2 (\tilde{W}'(x))^2}{\tilde{W}^3(x)}\right].
 \eeq
 and thus 
 \beq\label{s13-10c}
 \hW'(\bX(t)) \mi \hW'(\al) \equ \hW''(\al) (x \mi \al) + O\big((x\mi \al)^2\big), \quad \hW''(\al) \equ \frac{-2\tilde{W}'(\al)}{\tilde{W}^2(\al)}
 \eeq
 
 For GCDT we have 
 \beq\label{s13-11}
  - \frac{2 \tilde{W}' (x)}{\tilde{W} (x)}\Big|_{x =\al}  = \frac{2 (x \plu \al)}{(x \plu \al)^2 - 2g_s/\al}\Big|_{x =\al} = \frac{\al}{\Sigma^2}.
   \eeq
   og since we have $\bX(t) \equ \al \plu O(\e^{-2\Sigma t})$ this is also the order of the correction to $\la L(t) \ra$. 

\item[(9)] Inserting $Y \equ - \al$ in eq.\ \rf{p13-12} in the Problem Set we obtain
\beq\label{s13-12}
\la L(t)\ra_{X,Y = -\al} = \frac{\hW'(\bX(t)) \mi \hW'(\al)}{{\hW}(\bX(t))} \plu \frac{\tilde{W}(\al)}{\hW(\bX(t))}
= \frac{1}{\bX(t) \mi \al} \mi \frac{\tilde{W}'(\bX(t))}{\tilde{W}(\bX(t))}
\eeq
where we have used \rf{s13-10a}-\rf{s13-10c}.

Now assume we have a finite $T$, but both $t$ and $T$ are large and $T \gg t$. Since in this limit the leading corrections are
$\bX(T) \mi \al \propto \e^{-2\Sigma T}$ and $\bX(t)\mi \al \propto  \e^{-2\Sigma t}$ and since the leading correction to
formula  \rf{p13-16} in The Problem Set when expanding \rf{p13-12} in the Problem Set around $\bX(T\equ \infty) \equ \al$ 
(always assuming $Y \equ -\al$) is of the form 
\beq\label{s13-13a}
\Delta \big(\la L(t)\ra_{X,Y = -\al}\big) = O\left( \frac{\bX(T) \mi \al}{\bX(t)\mi \al}\right)  = O\big( \e^{-2\Sigma (T-t)} \big),
\eeq
we have obtained the desired estimate.

\item[(10)] The CDT solution \rf{cdt23} can be written as
\beq\label{s13-14}
\bX(t;X) = \SL \;\frac{ 1 \plu \e^{-2\SL t}  \frac{X \mi \SL}{X + \SL}}{1 \mi \e^{-2\SL t} \frac{X \mi \SL}{X + \SL}} = 
 \SL \;\frac{ 1 \plu \e^{-2\SL (t+t_0)} }{1 \mi \e^{-2\SL (t+t_0)} } = \SL\, \coth \SL (t\plu t_0)
 \eeq
where $t_0(X)$ is defined by 
\beq\label{s13-15}
\frac{X \mi \SL}{X + \SL} = \e^{-2 \SL t_0}\quad {\rm i.e.} \quad X  = \SL \,\coth \SL\, t_0.
\eeq
Inserting  this $\bX(t;X)$ in \rf{s13-12} and using $\tilde{W}(x) \equ 1/(X\plu \SL)$ we  obtain, using for convenience
the notation $\tilde{t} := t \plu t_0$:
\bea\label{s13-16}
\la L(t)\ra_{X,Y = -\SL}&=& \frac{1}{\SL} \Big(\frac{1}{\coth \SL \tilde{t} \mi 1} + \frac{1}{\coth \SL \tilde{t} \plu 1}\Big)\hspace{2cm}
\non
&= &\!\!\!\frac{1}{\SL} \, \frac{2 \sinh \SL\tilde{t} \cosh \SL \tilde{t}}{\cosh^2\SL\tilde{t} \mi \sinh^2\SL\tilde{t}} 
=\frac{ \sinh 2 \SL \tilde{t}}{\SL}.
\eea

\item[(11)] We have
\bea\label{s13-17}
G(X,Y;T) &=& \frac{\hW(\bX(T;X))}{\hW(X)}\frac{1}{\bX(T;X) \plu Y}, \\
  G(X,L;t) &=&  \frac{\hW(\bX(t;X))}{\hW(X)} \, \e^{-\bX(t;X)L}. \label{s13-17b}
\eea
One can write
\bea\label{s13-17a} 
\lefteqn{\int_0^\infty dL   \,\e^{-\bX(t;X) L}\,L \, G(L,Y;T\mi t)}\non
& =& - \frac{d}{d \bX(t;X)} \int_0^\infty dL  \; \e^{-\bX(t;X) L} G(L,Y;T\mi t)\non
&=&  - \frac{d}{d \bX(t;X)}\,G(\bX(t;X),Y;T\mi t) 
\eea
and we have 
\beq\label{s13-18}
G(\bX(t;X),Y;T\mi t) =\frac{\hW(\bX(T\mi t;\bX(t;X))}{\hW(\bX(t;X))} \, 
\frac{1}{\bX(T\mi t; \bX(t;X ))\plu Y}.
\eeq
Using eqs.\ \rf{s13-17} - \rf{s13-18} in eq.\ \rf{p13-11} one obtains 
\beq\label{s13-19}
\frac{(Y \plu \bX(T;X)) \hW(\bX(t;X))}{\hW(\bX(T;X))}  \frac{d}{d \bX(t;X)} 
 \frac{-\hW(\bX(T\mi t;\bX(t;X))}{\hW(\bX(t;X))(\bX(T\mi t; \bX(t;X ))\plu Y)}.
 \eeq

\item[(12)] We have, differentiating  wrt $X$:
\beq\label{s13-20}
t \equ \int_{\bX(t;X)}^X \frac{dy}{\hW(y)} \implies 0\equ \frac{1}{\hW(X)} - \frac{d \bX(t;X)}{d X} \frac{1}{\hW(\bX(t;X)}.
\eeq 

\item[(13)] When solving the differential equation with the specific boundary condition $\bX(t\equ 0) \equ X$ we can stop 
at any time $t$ where we have reached $\bX(t;X)$, and the continue after the coffee break for the remaining 
$T\mi t$ time, reset to new starting time 0, provided we start out with the value $\bX(t;X)$ we reached at time $t$.
Thus $\bX(T\mi t;\bX(t;X)) \equ \bX(T,X)$, the result we would have obtained in one go, without the coffee break.
It is seen explicitly from our solution
\beq\label{s13-21}
T = \int_{\bX(T;X)}^X \frac{dy}{\hW(y)} = \left[ \int_{\bX(t;X)}^X + \int_{\bX(T\mi t,\bX(t;X))}^{\bX(t;X)}\right] \frac{dy}{\hW(y)} = t + (T-t)
\eeq

\item[(14)] We are now ready to perform the differentiation, obtaining 
\bea
 \lefteqn{\frac{d}{d \bX(t;X)}  \left[\frac{-\hW(\bX(T\mi t;\bX(t;X))}{\hW(\bX(t;X))(Y\plu \bX(T\mi t; \bX(t;X )))}\right] } \non
 &&~~~~~= \frac{ \hW(\bX(T\mi t;X))}{\hW^2(\bX(t;X))} 
 \left[ \frac{\hW'(\bX(t;X)) \mi \hW'(\bX(T\mi t;\bX(t;X)))}{Y \plu \bX(T\mi t;\bX(t;X))} \right.
 \non
 &&\hspace{4.5cm}\left. + \frac{\hW(\bX(T\mi t;X)}{(Y\plu \bX(T\mi t; \bX(t;X )))^2}\right] \label{s13-22}
\eea
In this formula we can now use  $\bX(T\mi t;\bX(t;X)) \equ \bX(T,X)$ and go back and insert \rf{s13-22} in \rf{s13-19}.
This will produce the wanted formula for $\la L(t)\ra_{X,Y}$.

\end{itemize}

  \end{document}